\documentclass[twoside]{report}
\usepackage[T1]{fontenc}
\usepackage{graphics}
\usepackage{floatflt}

\makeatletter

\newcommand{\LyX}{L\kern-.1667em\lower.25em\hbox{Y}\kern-.125emX\@}
\newcommand{\noun}[1]{\textsc{#1}}

\usepackage{a4}
\usepackage{axo}
\usepackage{reporthead}
\usepackage{vmargin}
\input{epsfig.sty}
\def\fnum@table{\tablename~{\bf\thetable}}
\def\fnum@figure{\figurename~{\bf\thefigure}}
\def\tablename{\footnotesize{\bf Table}}
\def\figurename{\footnotesize{\bf Figure}}

\def\be{\begin{equation}}
\def\ee{\end{equation}}

\def\epi{\end{picture}}
\def\emi{\end{minipage}}

\def\QCD{{\rm QCD}}

\makeatother

\begin{document}

\title{\textbf{\Huge Parton-Based }\\
\textbf{\Huge Gribov-Regge Theory}\Huge }
\vspace{1cm}

\author{\textbf{H.J. Drescher\protect\( ^{1}\protect \), M. Hladik\protect\( ^{1,3}\protect \),
S. Ostapchenko\protect\( ^{2,1}\protect \), T. Pierog\protect\( ^{1}\protect \),
}\\
\textbf{and K. Werner}\protect\( ^{1}\protect \)\\
\\
\\
 \textit{\protect\( ^{1}\protect \) SUBATECH, Université de Nantes -- IN2P3/CNRS
-- EMN,  Nantes, France }\\
\textit{\protect\( ^{2}\protect \) Moscow State University, Institute of Nuclear
Physics, Moscow, Russia}\\
\textit{\protect\( ^{3}\protect \) now at SAP AG, Berlin, Germany}}
\vspace{1cm}

\maketitle
\cleardoublepage

\begin{abstract}
We present a new parton model approach for hadron-hadron interactions and, in
particular, for the initial stage of nuclear collisions at very high energies
(RHIC, LHC and beyond). The most important aspect of our approach is a self-consistent
treatment, using the same formalism for calculating cross sections and particle
production, based on an effective, QCD-inspired field theory, where many of
the inconsistencies of presently used models will be avoided.

In addition, we provide a unified treatment of soft and hard scattering, such
that there is no fundamental cutoff parameter any more defining an artificial
border between soft and hard scattering.

Our approach cures some of the main deficiencies of two of the standard procedures
currently used: the Gribov-Regge theory and the eikonalized parton model. There,
cross section calculations and particle production cannot be treated in a consistent
way using a common formalism. In particular, energy conservation is taken care
of in case of particle production, but not concerning cross section calculations.
In addition, hard contributions depend crucially on some cutoff, being divergent
for the cutoff being zero. Finally, in case of several elementary scatterings,
they are not treated on the same level: the first collision is always treated
differently than the subsequent ones. All these problems are solved in our new
approach.

For testing purposes, we make very detailed studies of electron-positron annihilation
and lepton-nucleon scattering before applying our approach to proton-proton
and nucleus-nucleus collisions.

In order to keep a clean picture, we do not consider secondary interactions.
We provide a very transparent extrapolation of the physics of more elementary
interactions towards nucleus-nucleus scattering, without considering any nuclear
effects due to final state interactions. In this sense we consider our model
a realistic and consistent approach to describe the initial stage of nuclear
collisions.
\end{abstract}
\cleardoublepage

\tableofcontents \cleardoublepage

\chapter{Introduction}

The purpose of this paper is to provide the theoretical framework to treat hadron-hadron
scattering and the \textbf{initial stage of nucleus-nucleus collisions} at ultra-relativistic
energies, in particular with view to RHIC and LHC. The knowledge of these initial
collisions is crucial for any theoretical treatment of a possible parton-hadron
phase transition, the detection of which being the ultimate aim of all the efforts
of colliding heavy ions at very high energies.

It is quite clear that coherence is crucial for the very early stage of nuclear
collisions, so a real quantum mechanical treatment is necessary and any attempt
to use a \textbf{transport theoretical parton approach} with incoherent quasi-classical
partons should not be considered at this point. Also \textbf{semi-classical
hadronic cascades} cannot be stretched to account for the very first interactions,
even when this is considered to amount to a string excitation, since it is well
known \cite{wer93} that such longitudinal excitation is simply kinematically
impossible (see appendix \ref{ax-a-2}). There is also the very unpleasant feature
of having to treat formation times of leading and non-leading particles very
different. Otherwise, due to a large gamma factor, it would be impossible for
a leading particle to undergo multiple collisions.

So what are the currently used fully quantum mechanical approaches? There are
presently considerable efforts to describe nuclear collisions via solving \textbf{classical
Yang-Mills equations}, which allows to calculate inclusive parton distributions
\cite{mcl94}. This approach is to some extent orthogonal to ours: here, screening
is due to perturbative processes, whereas we claim to have good reasons to consider
soft processes to be at the origin of screening corrections. 

Provided factorization works for nuclear collisions, on may employ the \textbf{parton
model}, which allows to calculate inclusive cross sections as a convolution
of an elementary cross section with parton distribution functions, with these
distribution function taken from deep inelastic scattering. In order to get
exclusive parton level cross sections, some additional assumptions are needed,
as discussed later.

Another approach is the so-called \textbf{Gribov-Regge theory} \cite{gri68}.
This is an effective field theory, which allows multiple interactions to happen
``in parallel'', with the phenomenological object called ``Pomeron'' representing
an elementary interaction. Using the general rules of field theory, on may express
cross sections in terms of a couple of parameters characterizing the Pomeron.
Interference terms are crucial, they assure the unitarity of the theory. A disadvantage
is the fact that cross sections and particle production are not calculated consistently:
the fact that energy needs to be shared between many Pomerons in case of multiple
scattering is well taken into account when calculating particle production (in
particular in Monte Carlo applications), but energy conservation is not taken
care of for cross section calculations. This is a serious problem and makes
the whole approach inconsistent. Also a problem is the question of how to include
in a consistent way hard processes, which are usually treated in the parton
model approach. Another unpleasant feature is the fact that different elementary
interactions in case of multiple scattering are usually not treated equally,
so the first interaction is usually considered to be quite different compared
to the subsequent ones. 

Here, we present a new approach which we call ``\textbf{Parton-based Gribov-Regge
Theory}'', where we solve some of the above-mentioned problems: we have a consistent
treatment for calculating cross sections and particle production considering
energy conservation in both cases; we introduce hard processes in a natural
way, and, compared to the parton model, we can deal with exclusive cross sections
without arbitrary assumptions. A single set of parameters is sufficient to fit
many basic spectra in proton-proton and lepton-nucleon scattering, as well as
for electron-positron annihilation (with the exception of one parameter which
needs to be changed in order to optimize electron-positron transverse momentum
spectra). 

The basic guideline of our approach is theoretical consistency. We cannot derive
everything from first principles, but we use rigorously the language of field
theory to make sure not to violate basic laws of physics, which is easily done
in more phenomenological treatments (see discussion above).

There are still problems and open questions: there is clearly a problem with
unitarity at very high energies, which should be cured by considering screening
corrections due to so-called triple-Pomeron interactions, which we do not treat
rigorously at present but which is our next project.

\subsubsection*{}

\section{Present Status}

Before presenting new theoretical ideas, we want to discuss a little bit more
in detail the present status and, in particular, the open problems in the parton
model approach and in Gribov-Regge theory.

\subsubsection*{Gribov-Regge Theory}

\begin{figure}[htb]
{\par\centering \resizebox*{!}{0.12\textheight}{\includegraphics{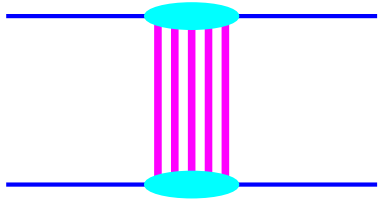}} \par}

\caption{Hadron-hadron scattering in GRT. The thick lines between the hadrons (incoming
lines) represent a Pomeron each. The different Pomeron exchanges occur in parallel.\label{grt}}
\end{figure}

Gribov-Regge theory is by construction a multiple scattering theory. The elementary
interactions are realized by complex objects called ``Pomerons'', who's precise
nature is not known, and which are therefore simply parameterized: the elastic
amplitude \( T \) corresponding to a single Pomeron exchange is given as
\begin{equation}
\label{x}
T(s,t)\sim i\, s^{\alpha _{0}+\alpha 't}
\end{equation}
with a couple of parameters to be determined by experiment. Even in hadron-hadron
scattering, several of these Pomerons are exchanged in parallel, see fig.\ \ref{grt}.
Using general rules of field theory (cutting rules), one obtains an expression
for the inelastic cross section, 
\begin{equation}
\label{x}
\sigma ^{h_{1}h_{2}}_{\mathrm{inel}}=\int d^{2}b\, \left\{ 1-\exp \left( -G(s,b)\right) \right\} ,
\end{equation}
 where the so-called eikonal \( G(s,b) \) (proportional to the Fourier transform
of \( T(s,t) \)) represents one elementary interaction (a thick line in fig.\ \ref{grt}).
One can generalize to nucleus-nucleus collisions, where corresponding formulas
for cross sections may be derived.

In order to calculate exclusive particle production, one needs to know how to
share the energy between the individual elementary interactions in case of multiple
scattering. We do not want to discuss the different recipes used to do the energy
sharing (in particular in Monte Carlo applications). The point is, whatever
procedure is used, this is not taken into account in calculation of cross sections
discussed above. So, actually, one is using two different models for cross section
calculations and for treating particle production. Taking energy conservation
into account in exactly the same way will modify the cross section results considerably,
as we are going to demonstrate later. 

This problem has first been discussed in \cite{bra90}, \cite{abr92}. The authors
claim that following from the non-planar structure  of the corresponding diagrams,
conserving energy and momentum in a consistent way is crucial, and therefore
the incident energy has to be shared between the different elementary interactions,
both real and virtual ones. 

Another very unpleasant and unsatisfactory feature of most ``recipes'' for
particle production is the fact, that the second Pomeron and the subsequent
ones are treated differently than the first one, although in the above-mentioned
formula for the cross section all Pomerons are considered to be identical.

\subsubsection*{The Parton Model}

The standard parton model approach to hadron-hadron or also nucleus-nucleus
scattering amounts to presenting the partons of projectile and target by momentum
distribution functions, \( f_{h_{1}} \) and \( f_{h_{2}} \), and calculating
inclusive cross sections for the production of parton jets with the squared
transverse momentum \( p_{\perp }^{2} \) larger than some cutoff \( Q_{0}^{2} \)
as

\begin{equation}
\label{x}
\sigma ^{h_{1}h_{2}}_{\mathrm{incl}}=\sum _{ij}\int dp_{\perp }^{2}\int dx^{+}\int dx^{-}f^{i}_{h_{1}}(x^{+},p_{\perp }^{2})f_{h_{2}}^{j}(x^{-},p_{\perp }^{2})\frac{d\hat{\sigma }_{ij}}{dp_{\perp }^{2}}(x^{+}x^{-}s)\theta \! \left( p_{\perp }^{2}-Q^{2}_{0}\right) ,
\end{equation}
where \( d\hat{\sigma }_{ij}/dp_{\perp }^{2} \) is the elementary parton-parton
cross section and \( i,j \) represent parton flavors. 

This simple factorization formula is the result of cancelations of complicated
diagrams (AGK cancelations) and hides therefore the complicated multiple scattering
structure of the reaction. The most obvious manifestation of such a structure
is the fact that at high energies (\( \sqrt{s}\gg 10 \) GeV) the inclusive
cross section in proton-(anti)proton scattering exceeds the total one, so the
average number \( \bar{N}^{pp}_{\mathrm{int}} \) of elementary interactions
must be bigger than one:
\begin{equation}
\label{x}
\bar{N}_{\mathrm{int}}^{h_{1}h_{2}}=\sigma ^{h_{1}h_{2}}_{\mathrm{incl}}/\sigma _{\mathrm{tot}}^{h_{1}h_{2}}>1
\end{equation}
 The usual solution is the so-called eikonalization, which amounts to re-introducing
multiple scattering, based on the above formula for the inclusive cross section:
\begin{equation}
\label{x}
\sigma ^{h_{1}h_{2}}_{\mathrm{inel}}(s)=\int d^{2}b\, \left\{ 1-\exp \left( -A(b)\, \sigma ^{h_{1}h_{2}}_{\mathrm{incl}}(s)\right) \right\} =\sum \sigma ^{h_{1}h_{2}}_{m}(s),
\end{equation}
with
\begin{equation}
\label{x}
\sigma ^{h_{1}h_{2}}_{m}(s)=\int d^{2}b\, \frac{\left( A(b)\, \sigma ^{h_{1}h_{2}}_{\mathrm{incl}}(s)\right) ^{m}}{m!}\exp \left( -A(b)\, \sigma ^{h_{1}h_{2}}_{\mathrm{incl}}(s)\right) 
\end{equation}
 representing the cross section for \( m \) scatterings; \( A(b) \) being
the proton-proton overlap function (the convolution of the two proton profiles).
In this way the multiple scattering is ``recovered''. The disadvantage is
that this method does not provide any clue how to proceed for nucleus-nucleus
(\( AB \)) collisions. One usually assumes the proton-proton cross section
for each individual nucleon-nucleon pair of a \( AB \) system. We are going
to demonstrate that this assumption is incorrect.

Another problem, in fact the same one as discussed earlier for the GRT, arises
in case of exclusive calculations (event generation), since the above formulas
do not provide any information on how to share the energy between the many elementary
interactions. The Pythia-method \cite{sjo87} amounts to generating the first
elementary interaction according to the inclusive differential cross section,
then taking the remaining energy for the second one and so on. In this way,
the event generation will reproduce the theoretical inclusive spectrum for hadron-hadron
interaction (by construction), as it should be. The method is, however, very
arbitrary, and is certainly not a convincing procedure for the multiple scattering
aspects of the collisions.

\section{Parton-based Gribov-Regge Theory}

\begin{figure}[htb]
{\par\centering \resizebox*{!}{0.15\textheight}{\includegraphics{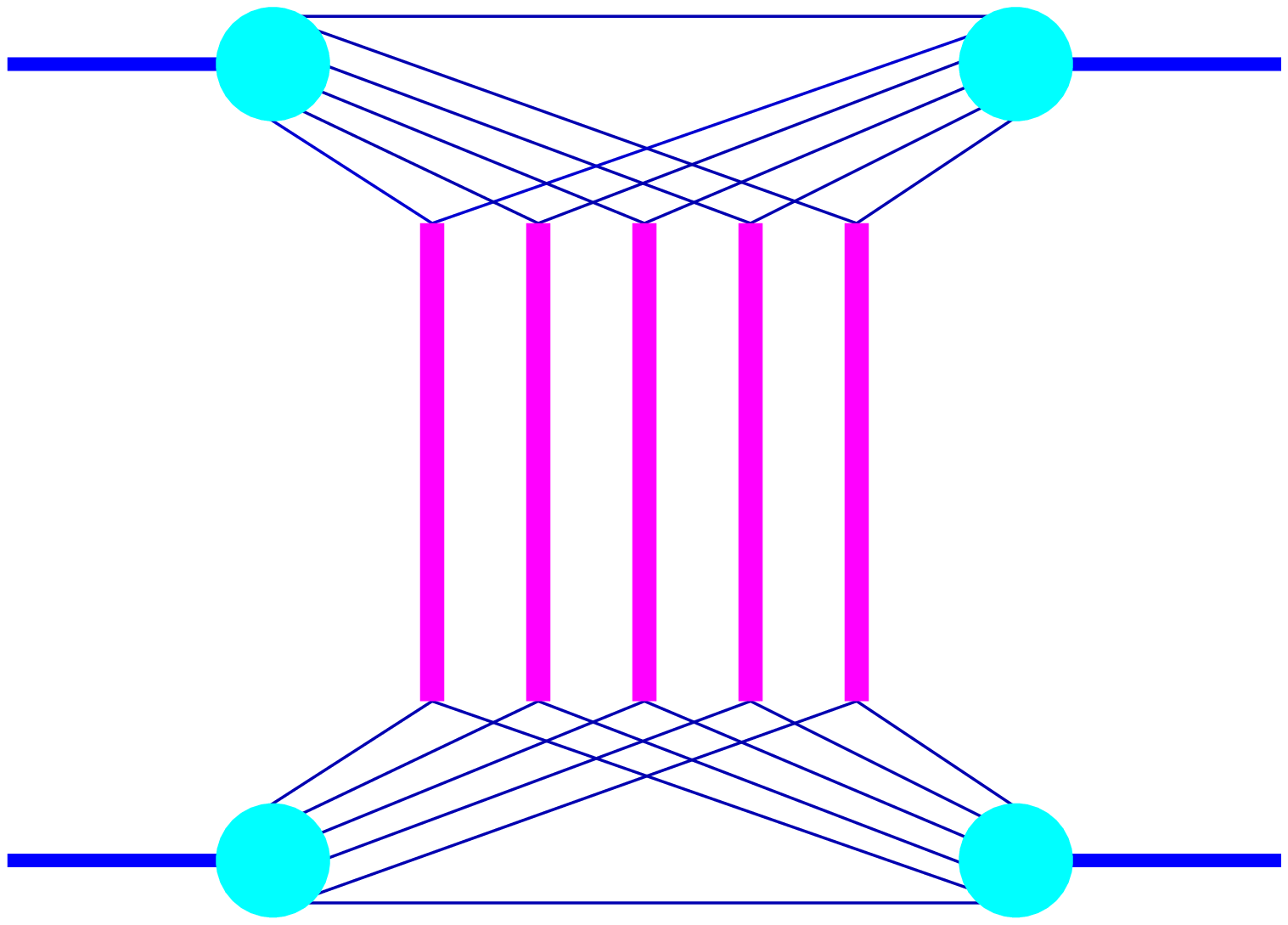}} \par}

\caption{Graphical representation of a contribution to the elastic amplitude of proton-proton
scattering. Here, energy conservation is taken into account: the energy of the
incoming protons is shared among several ``constituents'' (shown by splitting
the nucleon lines into several constituent lines), and so each Pomeron disposed
only a fraction of the total energy, such that the total energy is conserved.\label{grtpp}}
\end{figure}

\begin{figure}[htb]
{\par\centering \resizebox*{!}{0.18\textheight}{\includegraphics{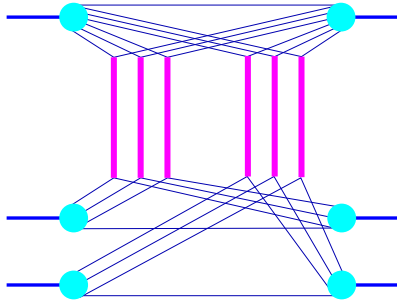}} \par}

\caption{Graphical representation of a contribution to the elastic amplitude of proton-nucleus
scattering, or more precisely a proton interacting with (for simplicity) two
target nucleons, taking into account energy conservation. Here, the energy of
the incoming proton is shared between all the constituents, which now provide
the energy for interacting with two target nucleons.\label{grtppa}}
\end{figure}

In this paper, we present a new approach for hadronic interactions and for the
initial stage of nuclear collisions, which is able to solve several of the above-mentioned
problems. We provide a rigorous treatment of the multiple scattering aspect,
such that questions as energy conservation are clearly determined by the rules
of the field theory, both for cross section and particle production calculations.
In both (!) cases, energy is properly shared between the different interactions
happening in parallel, see fig.\ \ref{grtpp}.
for proton-proton and fig.\ \ref{grtppa} for proton-nucleus collisions (generalization
to nucleus-nucleus is obvious). 
 This is the most important and new aspect of our approach, which we consider
to be a first necessary step to construct a consistent model for high energy
nuclear scattering. 

The elementary interactions, shown as the thick lines in the above figures,
are in fact a sum of a soft, a hard, and a semi-hard contribution, providing
a consistent treatment of soft and hard scattering. To some extend, our approach
provides a link between the Gribov-Regge approach and the parton model, we call
it ``Parton-based Gribov-Regge Theory''. 

There are still many problems to be solved: as we are going to show later, a
rigorous treatment of energy conservation will lead to unitarity problems (increasingly
severe with increasing energy), which is nothing but a manifestation of the
fact that screening corrections will be increasingly important. In this paper
we will employ a ``unitarization procedure'' to solve this problem, but this
is certainly not the final answer. The next step should be a consistent approach,
taking into account both energy conservation and screening corrections due to
multi-Pomeron interactions.

Our approach is realized as the Monte Carlo code \textsc{neXus}, which is nothing
but the direct implementation of the formalism described in this paper. All
our numerical results are calculated with \textsc{neXus} version 2.00. 

\cleardoublepage

\chapter{The Formalism}

We want to calculate cross sections and particle production in a consistent
way, in both cases based on the same formalism, with energy conservation being
ensured. The formalism operates with Feynman diagrams of the QCD-inspired effective
field theory, such that calculations follow the general rules of the field theory.
A graphical representation of a contribution to the elastic amplitude of nucleus-nucleus
scattering (related to particle production via the optical theorem) is shown
in fig.\ \ref{grtppaa}: 
\begin{figure}[htb]
{\par\centering \resizebox*{!}{0.2\textheight}{\includegraphics{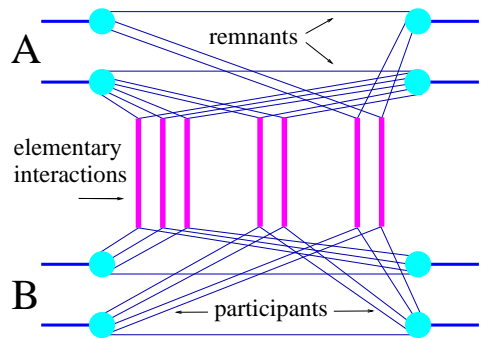}} \par}

\caption{A contribution to the elastic amplitude of a nucleus-nucleus collision, or
more precisely two nucleons from projectile \protect\( A\protect \) interacting
with two nucleons from target \protect\( B\protect \), taking into account
energy conservation. The energy of the incoming nucleons is shared between all
the constituents. \label{grtppaa}}
\end{figure}
here the nucleons are split into several ``constituents'', each one carrying
a fraction of the incident momentum, such that the sum of the momentum fractions
is one (momentum conservation). Per nucleon there are one or several ``participants''
and exactly one ``spectator'' or ``remnant'', where the former ones interact
with constituents from the other side via some ``elementary interaction''
(vertical lines in the figure \ref{grtppaa}). The remnant is just all the rest,
i.e. the nucleon minus the participants. 

After a technical remark concerning profile functions, we are going to discuss
parton-parton scattering, before we develop the multiple scattering theory for
hadron-hadron and nucleus-nucleus scattering.

\section{Profile Functions}

Profile functions play a fundamental role in our formalism, so we briefly sketch
their definition and physical meaning.

\begin{figure}[htb]
{\par\centering \resizebox*{!}{0.15\textheight}{\includegraphics{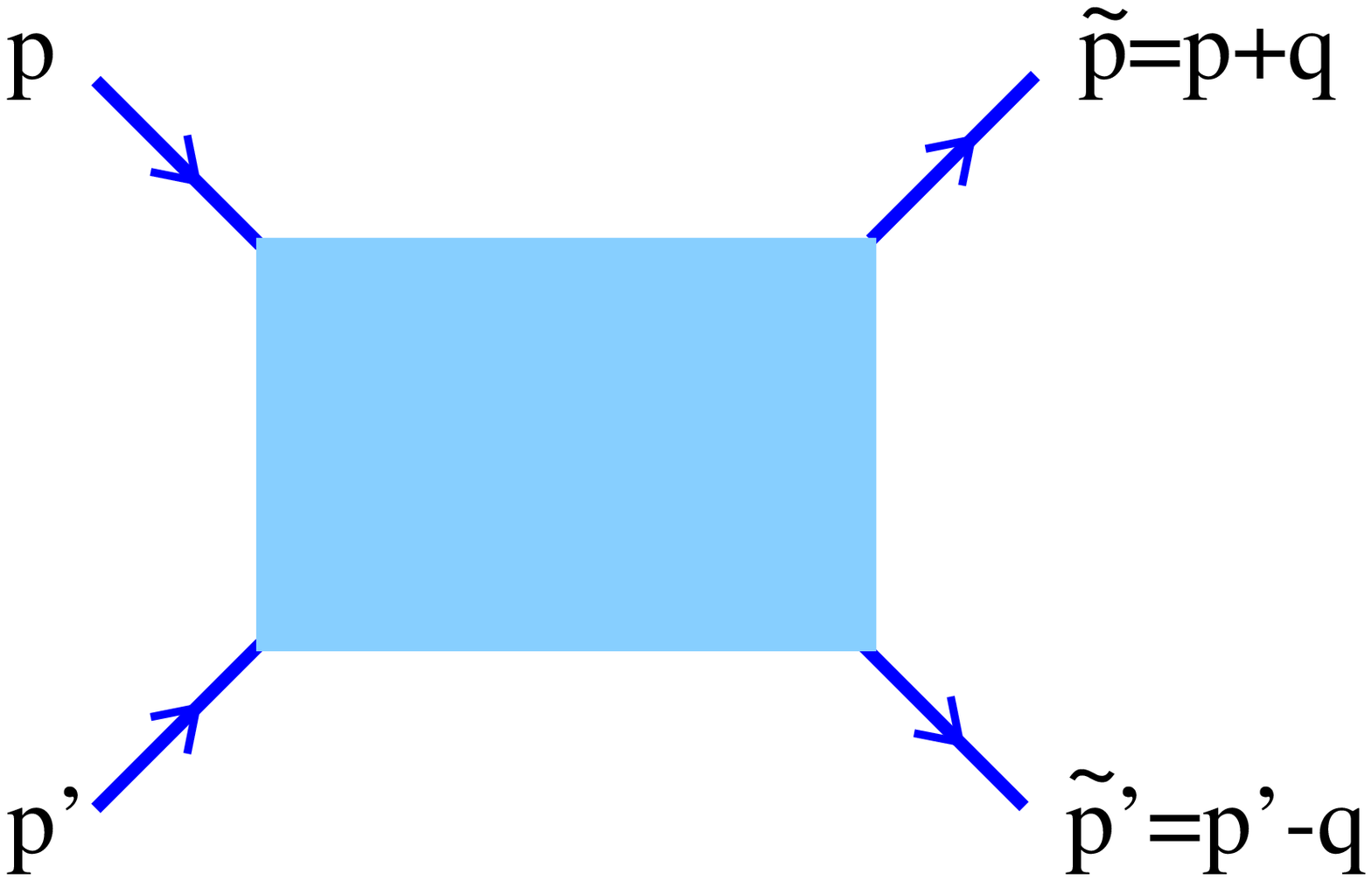}} \par}

\caption{The general elastic scattering amplitude \protect\( T\protect \).\label{fig-elastic}}
\end{figure}

Let \( T \) be the elastic scattering amplitude \( T \) for the two-body scattering
depicted in fig.\ref{fig-elastic}.
The 4-momenta \( p \) and \( p' \) are the ones for the incoming particles
, \( \tilde{p}=p+q \) and \( \tilde{p}'=p'-q \) the ones for the outgoing
particles, and \( q \) the 4-momentum transfer in the process.  We define as
usual the Mandelstam variables \( s \) and \( t \) (see appendix \ref{ax-a-1}).
Using the optical theorem, we may write the total cross section as 
\begin{equation}
\label{eq-2-1-1}
\sigma _{\mathrm{tot}}(s)=\frac{1}{2s}2\mathrm{Im}\, T(s,t=0).
\end{equation}
 We define the Fourier transform \( \tilde{T} \) of \( T \) as 
\begin{equation}
\label{eq-2-1-2}
\tilde{T}(s,b)=\frac{1}{4\pi ^{2}}\int d^{2}q_{\bot }\, e^{-i\vec{q}_{\bot }\vec{b}}\, T(s,t),
\end{equation}
using \( t=-q^{2}_{\bot } \) (see appendix \ref{ax-a-2}), and a so-called
``profile function'' \( G(s,b) \) as
\begin{equation}
\label{eq-2-1-3}
G(s,b)=\frac{1}{2s}2\mathrm{Im}\, \tilde{T}(s,b).
\end{equation}
 One can easily verify that
\begin{equation}
\label{x}
\sigma _{\mathrm{tot}}(s)=\int \! d^{2}b\; G(s,b),
\end{equation}
which allows an interpretation of \( G(s,b) \) to be the probability of an
interaction at impact parameter \( b \). 

In the following, we are working with partonic, hadronic, and even nuclear profile
functions. The central result to be derived in the following sections is the
fact that hadronic and also nuclear profile functions can be expressed in terms
of partonic ones, allowing a clean formulation of a multiple scattering theory.

\section{Parton-Parton Scattering}

We distinguish three types of elementary parton-parton scatterings, referred
to as ``soft'', ``hard'' and ``semi-hard'', which we are going to discuss
briefly in the following. The detailed derivations can be found in appendix
\ref{ax-b-1}.

\subsubsection*{The Soft Contribution }

\begin{figure}[htb]
{\par\centering \resizebox*{!}{0.13\textheight}{\includegraphics{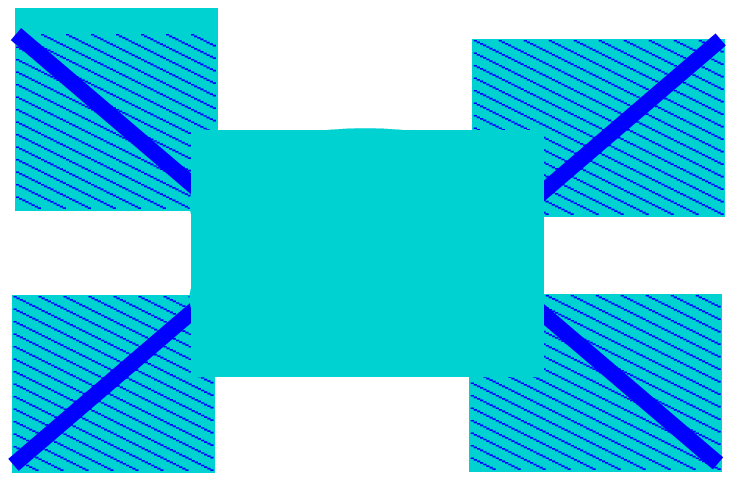}} \par}

\caption{The soft contribution.\label{fig:dsoft}}
\end{figure}
Let us first consider the pure non-perturbative contribution to the process
of the fig.\ \ref{fig-elastic}, where all virtual partons appearing in the
internal structure of the diagram have restricted virtualities \( Q^{2}<Q^{2}_{0} \),
where \( Q_{0}^{2}\sim 1 \) GeV\( ^{2} \) is a reasonable cutoff for perturbative
QCD being applicable. Such soft non-perturbative dynamics is known to dominate
hadron-hadron interactions at not too high energies. Lacking methods to calculate
this contribution from first principles, it is simply parameterized and graphically
represented as a `blob', see fig.\ \ref{fig:dsoft}. It is traditionally assumed
to correspond to multi-peripheral production of partons (and final hadrons)
\cite{afs62} and is described by the phenomenological soft Pomeron exchange
contribution \cite{gri68}:
\begin{equation}
\label{tsoft}
T_{\mathrm{soft}}\! \! \left( \hat{s},t\right) =8\pi s_{0}\eta (t)\, \gamma _{\mathrm{part}}^{2}\, \left( \frac{\hat{s}}{s_{0}}\right) ^{\alpha _{\mathrm{soft}}\! (0)}\exp \! \left( \lambda ^{\! (2)}_{\mathrm{soft}}\! \left( \hat{s}/s_{0}\right) t\right) ,
\end{equation}
with
\begin{equation}
\label{x}
\lambda ^{\! (n)}_{\mathrm{soft}}\! (z)=nR_{\mathrm{part}}^{2}+\alpha '\! _{\mathrm{soft}}\ln \! z,
\end{equation}
where \( \hat{s}=(p+p')^{2} \). The parameters \( \alpha _{\mathrm{soft}}\! (0) \),
\( \alpha '\! _{\mathrm{soft}} \) are the intercept and the slope of the Pomeron
trajectory, \( \gamma _{\mathrm{part}} \) and \( R_{\mathrm{part}}^{2} \)
are the vertex value and the slope for the Pomeron-parton coupling, and \( s_{0}\simeq 1 \)
GeV\( ^{2} \) is the characteristic hadronic mass scale. The so-called signature
factor \( \eta  \) is given as 
\begin{equation}
\label{x}
\eta (t)=i-\cot \frac{\pi \alpha _{_{\rm {P}}}\! (t)}{2}\simeq i.
\end{equation}
 Cutting the diagram of the fig.\ \ref{fig:dsoft} corresponds to the summation
over multi-peripheral intermediate hadronic states, connected via unitarity
to the imaginary part of the amplitude (\ref{tsoft}),
\begin{eqnarray}
\frac{1}{i}\mathrm{disc}_{\hat{s}}\, T_{\mathrm{soft}}\! \! \left( \hat{s},t\right)  & = & \frac{1}{i}\left[ T_{\mathrm{soft}}\! \! \left( \hat{s}+i0,t\right) -T_{\mathrm{soft}}\! \! \left( \hat{s}-i0,t\right) \right] \\
 & = & 2\mathrm{Im}\, T_{\mathrm{soft}}\! \! \left( \hat{s},t\right) \\
 & = & \sum _{n,\mathrm{spins},...}\int d\tau _{n}T_{p,p'\rightarrow X_{n}}T^{*}_{\tilde{p},\tilde{p}'\rightarrow X_{n}},\label{x} 
\end{eqnarray}
where \( T_{p,p'\rightarrow X_{n}} \) is the amplitude for the transition of
the initial partons \( p,p' \) into the \( n \)-particle state \( X_{n} \),
\( d\tau _{n} \) is the invariant phase space volume for the \( n \)-particle
state \( X_{n} \) and the summation is done over the number of particles \( n \)
and over their spins and species, the averaging over initial parton colors and
spins is assumed; \( \mathrm{disc}_{\hat{s}}\, T_{\mathrm{soft}}\! \! \left( \hat{s},t\right)  \)
denotes the discontinuity of the amplitude \( T_{\mathrm{soft}}\! \left( \hat{s},t\right)  \)
on the right-hand cut in the variable \( \hat{s} \).

For \( t=0 \) one obtains via the optical theorem the contribution \( \sigma _{\mathrm{soft}} \)
of the soft Pomeron exchange to the total parton interaction cross section,
\begin{equation}
\label{x}
\sigma _{\mathrm{soft}}\! \left( \hat{s}\right) =\frac{1}{2\hat{s}}2\mathrm{Im}\, T_{\mathrm{soft}}\! \! \left( \hat{s},0\right) =8\pi \gamma _{\mathrm{part}}^{2}\left( \frac{\hat{s}}{s_{0}}\right) ^{\alpha _{\mathrm{soft}}\! (0)-1},
\end{equation}
where \( 2\hat{s} \) defines the initial parton flux.

The corresponding profile function for parton-parton interaction is expressed
via the Fourier transform \( \tilde{T} \) of \( T \) divided by the flux \( 2s \),
\begin{equation}
\label{x}
D_{\mathrm{soft}}\! \left( \hat{s},b\right) =\frac{1}{2\hat{s}}2\mathrm{Im}\tilde{T}_{\mathrm{soft}}(\hat{s},b),
\end{equation}
which gives
\begin{eqnarray}
D_{\mathrm{soft}}(\hat{s},b) & = & \frac{1}{8\pi ^{2}\hat{s}}\int d^{2}q_{\perp }\, \exp \! \left( -i\vec{q}_{\perp }\vec{b}\right) \, 2\mathrm{Im}\, T_{\mathrm{soft}}\! \left( \hat{s},-q^{2}_{\perp }\right) \\
 & = & \frac{2\gamma _{\mathrm{part}}^{2}}{\lambda ^{\! (2)}_{\mathrm{soft}}\! (\hat{s}/s_{0})}\left( \frac{\hat{s}}{s_{0}}\right) ^{\alpha _{\mathrm{soft}}\! (0)-1}\exp \! \left( -\frac{b^{2}}{4\lambda ^{\! (2)}_{\mathrm{soft}}\! (\hat{s}/s_{0})}\right) .\label{soft d} 
\end{eqnarray}
 The external legs of the diagram of fig.\ \ref{fig:dsoft} are ``partonic
constituents'', which are assumed to be quark-antiquark pairs.

\subsubsection*{The Hard Contribution }

\begin{figure}[htb]
{\par\centering \resizebox*{!}{0.18\textheight}{\includegraphics{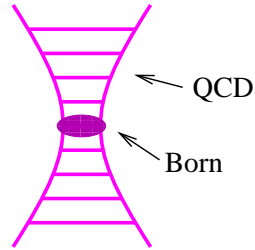}} \par}

\caption{The hard (or val-val) contribution. \label{fig:dval}}
\end{figure}

Let us now consider the other extreme, when all the processes in the `box' of
the fig.\ \ref{fig-elastic} are perturbative, i.e. all internal intermediate
partons are characterized by large virtualities \( Q^{2}>Q^{2}_{0} \). In that
case, the corresponding hard parton-parton scattering amplitude \( T^{jk}_{\mathrm{hard}}\! \! \left( \hat{s},t\right)  \)
(\( j,k \) denote the types (flavors) of the initial partons) can be calculated
using the perturbative QCD techniques \cite{alt82,rey81}, and the intermediate
states contributing to the absorptive part of the amplitude of the fig.\ \ref{fig-elastic}
can be defined in the parton basis. In the leading logarithmic approximation
of QCD, summing up terms where each (small) running QCD coupling constant \( \alpha _{s}(Q^{2}) \)
appears together with a large logarithm \( \ln (Q^{2}/\lambda ^{2}_{\mathrm{QCD}}) \)
(with \( \lambda _{QCD} \) being the infrared QCD scale), and making use of
the factorization hypothesis, one obtains the contribution of the corresponding
cut diagram for \( t=q^{2}=0 \) as the cut parton ladder cross section \( \sigma _{\mathrm{hard}}^{jk}(\hat{s},Q_{0}^{2}) \) \footnote{
Strictly speaking, one obtains the ladder representation for the process only
using axial gauge.
}, which will correspond to the cut diagram of fig.\ \ref{fig:dval}, 
where all horizontal rungs are the final (on-shell) partons and the virtualities
of the virtual \( t \)-channel partons increase from the ends of the ladder
towards the largest momentum transfer parton-parton process (indicated symbolically
by the `blob' in the middle of the ladder):
\begin{eqnarray}
\sigma _{\mathrm{hard}}^{jk}(\hat{s},Q_{0}^{2}) & = & \frac{1}{2\hat{s}}2\mathrm{Im}\, T^{jk}_{\mathrm{hard}}\! \! \left( \hat{s},t=0\right) \nonumber \\
 & = & K\, \sum _{ml}\int dx_{B}^{+}dx_{B}^{-}dp_{\bot }^{2}{d\sigma _{\mathrm{Born}}^{ml}\over dp_{\bot }^{2}}(x_{B}^{+}x_{B}^{-}\hat{s},p_{\bot }^{2})\label{sig-jk-hard} \\
 & \times  & E_{\mathrm{QCD}}^{jm}(x_{B}^{+},Q_{0}^{2},M_{F}^{2})\, E_{\mathrm{QCD}}^{kl}(x_{B}^{-},Q_{0}^{2},M_{F}^{2})\theta \! \left( M_{F}^{2}-Q^{2}_{0}\right) ,\nonumber \label{x} 
\end{eqnarray}
 Here \( d\sigma _{\mathrm{Born}}^{ml}/dp_{\bot }^{2} \) is the differential
\( 2\rightarrow 2 \) parton scattering cross section, \( p_{\bot }^{2} \)
is the parton transverse momentum in the hard process, \( m,l \) and \( x_{B}^{\pm } \)
are correspondingly the types and the shares of the light cone momenta of the
partons participating in the hard process, and \( M_{F}^{2} \) is the factorization
scale for the process (we use \( M_{F}^{2}=p^{2}_{\perp }/4 \)). The `evolution
function' \( E^{jm}_{\mathrm{QCD}}(Q^{2}_{0},M_{F}^{2},z) \) represents the
evolution of a parton cascade from the scale \( Q_{0}^{2} \) to \( M_{F}^{2} \),
i.e.\ it gives the number density of partons of type \( m \) with the momentum
share \( z \) at the virtuality scale \( M_{F}^{2} \), resulted from the evolution
of the initial parton \( j \), taken at the virtuality scale \( Q_{0}^{2} \).
The evolution function satisfies the usual DGLAP equation \cite{alt77} with
the initial condition \( E^{jm}_{\mathrm{QCD}}(Q^{2}_{0},Q_{0}^{2},z)=\delta ^{j}_{m}\; \delta (1-z) \),
as discussed in detail in appendix \ref{ax-time}. The factor \( K\simeq 1.5 \)
takes effectively into account higher order QCD corrections.

In the following we shall need to know the contribution of the uncut parton
ladder \( T_{\mathrm{hard}}^{jk}(\hat{s},t) \) with some momentum transfer
\( q \) along the ladder (with \( t=q^{2} \)). The behavior of the corresponding
amplitudes was studied in \cite{lip86} in the leading logarithmic(\( 1/x \)
) approximation of QCD. The precise form of the corresponding amplitude is not
important for our application; we just use some of the results of \cite{lip86},
namely that one can neglect the real part of this amplitude and that it is nearly
independent on \( t \), i.e. that the slope of the hard interaction \( R_{\mathrm{hard}}^{2} \)
is negligible small, i.e. compared to the soft Pomeron slope one has \( R_{\mathrm{hard}}^{2}\simeq 0 \).
So we parameterize \( T_{\mathrm{hard}}^{jk}(\hat{s},t) \) in the region of
small \( t \) as \cite{rys92}
\begin{equation}
\label{t-ladder}
T_{\mathrm{hard}}^{jk}(\hat{s},t)=i\hat{s}\, \sigma _{\mathrm{hard}}^{jk}(\hat{s},Q_{0}^{2})\: \exp \left( R_{\mathrm{hard}}^{2}\, t\right) 
\end{equation}

The corresponding profile function is obtained by calculating the Fourier transform
\( \tilde{T}_{\mathrm{hard}} \) of \( T_{\mathrm{hard}} \) and dividing by
the initial parton flux \( 2\hat{s} \), 
\begin{equation}
\label{x}
D^{jk}_{\mathrm{hard}}\! \left( \hat{s},b\right) =\frac{1}{2\hat{s}}2\mathrm{Im}\tilde{T}^{jk}_{\mathrm{hard}}(\hat{s},b),
\end{equation}
which gives

\begin{eqnarray}
D^{jk}_{\mathrm{hard}}\left( \hat{s},b\right) =\frac{1}{8\pi ^{2}\hat{s}}\int d^{2}q_{\perp }\, \exp \! \left( -i\vec{q}_{\perp }\vec{b}\right) \, 2\mathrm{Im}\, T_{\mathrm{hard}}^{jk}(\hat{s},-q^{2}_{\perp }) &  & \nonumber \\
=\sigma _{\mathrm{hard}}^{jk}\! \left( \hat{s},Q_{0}^{2}\right) \frac{1}{4\pi R_{\mathrm{hard}}^{2}}\exp \! \left( -\frac{b^{2}}{4R_{\mathrm{hard}}^{2}}\right) , &  & \label{d-val-val} 
\end{eqnarray}

In fact, the above considerations are only correct for valence quarks, as discussed
in detail in the next section. Therefore, we also talk about ``valence-valence''
contribution and we use \( D_{\mathrm{val}-\mathrm{val}} \) instead of \( D_{\mathrm{hard}} \):
\begin{equation}
\label{x}
D^{jk}_{\mathrm{val}-\mathrm{val}}\left( \hat{s},b\right) \equiv D^{jk}_{\mathrm{hard}}\left( \hat{s},b\right) ,
\end{equation}
 so these are two names for one and the same object.

\subsubsection*{The Semi-hard Contribution}

The discussion of the preceding section is not valid in case of sea quarks and
gluons, since here the momentum share \( x_{1} \) of the ``first'' parton
is typically very small, leading to an object with a large mass of the order
\( Q^{2}_{0}/x_{1} \) between the parton and the proton \cite{don94}. Microscopically,
such 'slow' partons with \( x_{1}\ll 1 \) appear as a result of a long non-perturbative
parton cascade, where each individual parton branching is characterized by a
small momentum transfer squared \( Q^{2}<Q^{2}_{0} \) and nearly equal partition
of the parent parton light cone momentum \cite{afs62,bak76}. When calculating
proton structure functions or high-\( p_{t} \) jet production cross sections
that non-perturbative contribution is usually included into parameterized initial
parton momentum distributions at \( Q^{2}=Q^{2}_{0} \). However, the description
of inelastic hadronic interactions requires to treat it explicitly in order
to account for secondary particles produced during such non-perturbative parton
pre-evolution, and to describe correctly energy-momentum sharing between multiple
elementary scatterings. As the underlying dynamics appears to be identical to
the one of soft parton-parton scattering considered above, we treat this soft
pre-evolution as the usual soft Pomeron emission, as discussed in detail in
appendix \ref{ax-b-1}.

\begin{figure}[htb]
{\par\centering \resizebox*{!}{0.18\textheight}{\includegraphics{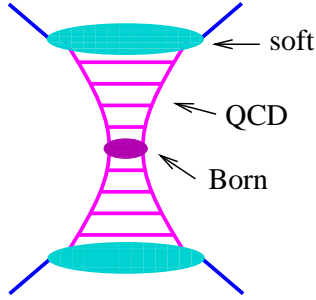}} \par}

\caption{The semi-hard ``sea-sea'' contribution: parton ladder plus ``soft ends''.\label{fig:dsemi}}
\end{figure}

So for sea quarks and gluons, we consider so-called semi-hard interactions between
parton constituents of initial hadrons, represented by a parton ladder with
``soft ends'', see fig.\ \ref{fig:dsemi}. 
As in case of soft scattering, the external legs are quark-antiquark pairs,
connected to soft Pomerons. The outer partons of the ladder are on both sides
sea quarks or gluons (therefore the index ``sea-sea''). The central part is
exactly the hard scattering considered in the preceding section. As discussed
in length in the appendix \ref{ax-b-1}, the mathematical expression for the
corresponding amplitude is given as 
\begin{eqnarray}
iT_{\mathrm{sea}-\mathrm{sea}}(\hat{s},t) & = & \sum _{jk}\int ^{1}_{0}\! \frac{dz^{+}}{z^{+}}\frac{dz^{-}}{z^{-}}\, \mathrm{Im}\, T_{\mathrm{soft}}^{j}\! \! \left( \frac{s_{0}}{z^{+}},t\right) \, \mathrm{Im}\, T_{\mathrm{soft}}^{k}\! \! \left( \frac{s_{0}}{z^{-}},t\right) \, iT_{\mathrm{hard}}^{jk}(z^{+}z^{-}\hat{s},t),\label{x} 
\end{eqnarray}
with \( z^{\pm } \) being the momentum fraction of the external leg-partons
of the parton ladder relative to the momenta of the initial (constituent) partons.
The indices \( j \) and \( k \) refer to the flavor of these external ladder
partons. The amplitudes \( T_{\mathrm{soft}}^{j} \) are the soft Pomeron amplitudes
discussed earlier, but with modified couplings, since the Pomerons are now connected
to the ladder on one side. The arguments \( s_{0}/z^{\pm } \) are the squared
masses of the two soft Pomerons, \( z^{+}z^{-}\hat{s} \) is the squared mass
of the hard piece.

Performing as usual the Fourier transform to the impact parameter representation
and dividing by \( 2\hat{s} \), we obtain the profile function
\begin{equation}
\label{x}
D_{\mathrm{sea}-\mathrm{sea}}\left( \hat{s},b\right) =\frac{1}{2\hat{s}}\, 2\mathrm{Im}\, \tilde{T}_{\mathrm{sea}-\mathrm{sea}}\! \left( \hat{s},b\right) ,
\end{equation}
 which may be written as
\begin{eqnarray}
D_{\mathrm{sea}-\mathrm{sea}}\left( \hat{s},b\right)  & = & \sum _{jk}\int ^{1}_{0}dz^{+}dz^{-}E_{\mathrm{soft}}^{j}\left( z^{+}\right) \, E_{\mathrm{soft}}^{k}\left( z^{-}\right) \, \sigma _{\mathrm{hard}}^{jk}(z^{+}z^{-}\hat{s},Q_{0}^{2})\nonumber \\
 &  & \qquad \times \; \frac{1}{4\pi \, \lambda ^{\! (2)}_{\mathrm{soft}}(1/(z^{+}z^{-}))}\exp \! \left( -\frac{b^{2}}{4\lambda ^{\! (2)}_{\mathrm{soft}}\! \left( 1/(z^{+}z^{-})\right) }\right) \label{d-sea-sea} 
\end{eqnarray}
with the soft Pomeron slope \( \lambda ^{\! (2)}_{\mathrm{soft}} \) and the
cross section \( \sigma _{\mathrm{hard}}^{jk} \) being defined earlier. The
functions \( E_{\mathrm{soft}}^{j}\left( z^{\pm }\right)  \) representing the
``soft ends'' are defined as 
\begin{equation}
\label{x}
\mathrm{E}^{j}_{\mathrm{soft}}(z^{\pm })=\mathrm{Im}\, T_{\mathrm{soft}}^{j}\! \! \left( \frac{s_{0}}{z^{+}},t=0\right) ,
\end{equation}
or explicitly
\begin{eqnarray}
E_{\mathrm{soft}}^{g}\left( z\right)  & = & 8\pi s_{0}\gamma _{\mathrm{part}}\gamma _{g}\, z^{-\alpha _{\mathrm{soft}}\! (0)}\, (1-z)^{\beta _{g}},\label{esoft-g-} \\
E_{\mathrm{soft}}^{q}\left( z\right)  & = & \gamma _{qg}\int ^{1}_{z}\! d\xi \, P_{g}^{q}(\xi )\, E_{\mathrm{soft}}^{g}\left( \frac{z}{\xi }\right) ,\label{esoft-q-} 
\end{eqnarray}
with 
\begin{equation}
\label{x}
\gamma _{qg}\gamma _{g}=w_{\mathrm{split}}\, \tilde{\gamma }_{g},\quad \gamma _{g}=\left( 1-w_{\mathrm{split}}\right) \, \tilde{\gamma }_{g},
\end{equation}
 and 
\begin{equation}
\label{x}
\tilde{\gamma }_{g}=\frac{1}{8\pi s_{0}\gamma _{\mathrm{part}}}\frac{\Gamma \! \left( 3-\alpha _{\mathrm{soft}}\! (0)+\beta _{g}\right) }{\Gamma \! \left( 2-\alpha _{\mathrm{soft}}\! (0)\right) \, \Gamma \! \left( 1+\beta _{g}\right) }
\end{equation}
(see appendix \ref{ax-b-1}). We neglected the small hard scattering slope \( R_{\mathrm{hard}}^{2} \)
compared to the Pomeron slope \( \lambda ^{(2)}_{\mathrm{soft}} \). We call
\( E_{\mathrm{soft}} \) also the `` soft evolution'', to indicate that we
consider this as simply  a continuation of the QCD evolution, however, in a
region where perturbative  techniques do not apply any more. As discussed in
the appendix \ref{ax-b-1}, \( E_{\mathrm{soft}}^{j}\left( z\right)  \) has
the meaning of the momentum distribution of parton \( j \) in the soft Pomeron. 

\begin{figure}[htb]
{\par\centering \resizebox*{!}{0.18\textheight}{\includegraphics{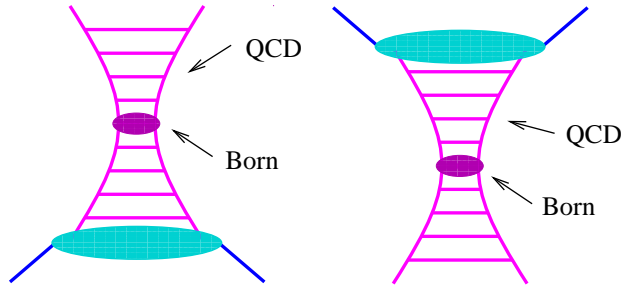}} \par}

\caption{Two ``mixed'' contributions.\label{fig:mixed}}
\end{figure}

Consistency requires to also consider the mixed semi-hard contributions with
a valence quark on one side and a non-valence participant (quark-antiquark pair)
on the other one, see fig.\ \ref{fig:mixed}.

We have
\begin{equation}
\label{x}
iT^{j}_{\mathrm{val}-\mathrm{sea}}(\hat{s},t)=\int ^{1}_{0}\! \frac{dz^{-}}{z^{-}}\sum _{k}\mathrm{Im}\, T_{\mathrm{soft}}^{k}\! \! \left( \frac{s_{0}}{z^{-}},t\right) iT_{\mathrm{hard}}^{jk}\left( z^{-}\hat{s},t\right) \qquad 
\end{equation}
and
\begin{eqnarray}
D^{j}_{\mathrm{val}-\mathrm{sea}}\left( \hat{s},b\right)  & = & \sum _{k}\int ^{1}_{0}\! dz^{-}\, E_{\mathrm{soft}}^{k}\left( z^{-}\right) \, \sigma _{\mathrm{hard}}^{jk}\! \left( z^{-}\hat{s},Q_{0}^{2}\right) \label{d-val-sea} \\
 &  & \qquad \times \; \frac{1}{4\pi \, \lambda ^{\! (1)}_{\mathrm{soft}}(1/z^{-})}\exp \! \left( -\frac{b^{2}}{4\lambda ^{\! (1)}_{\mathrm{soft}}\! \left( 1/z^{-}\right) }\right) \nonumber 
\end{eqnarray}
 where \( j \) is the flavor of the valence quark at the upper end of the ladder
and \( k \) is the type of the parton on the lower ladder end. Again, we neglected
the hard scattering slope \( R^{2}_{\mathrm{hard}} \) compared to the soft
Pomeron slope. A contribution \( D^{j}_{\mathrm{sea}-\mathrm{val}}\left( \hat{s},b\right)  \),
corresponding to a valence quark participant from the target hadron, is given
by the same expression,
\begin{equation}
\label{x}
D^{j}_{\mathrm{sea}-\mathrm{val}}\left( \hat{s},b\right) =D^{j}_{\mathrm{val}-\mathrm{sea}}\left( \hat{s},b\right) ,
\end{equation}
since eq.\ (\ref{d-val-sea}) stays unchanged under replacement \( z^{-}\rightarrow z^{+} \)
and only depends on the total c.m. energy squared \( \hat{s} \) for the parton-parton
system.

\section{Hadron-Hadron Scattering}

\begin{figure}[htb]
{\par\centering \resizebox*{!}{0.15\textheight}{\includegraphics{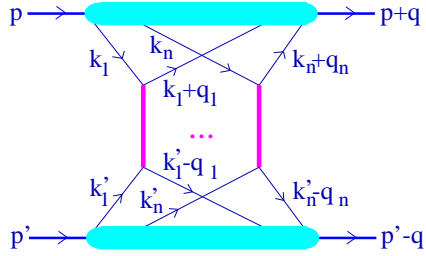}} \par}

\caption{Hadron-hadron interaction amplitude. \label{wave}}
\end{figure}

Let us now consider hadron-hadron interactions (a more detailed treatment can
be found in appendix \ref{ax-b-2}). We ignore first contributions involving
valence quark scatterings. In the general case, the expression for the hadron-hadron
scattering amplitude includes contributions from multiple scattering between
different parton constituents of the initial hadrons, as shown in fig.\ \ref{wave},
 and can be written according to the standard rules \cite{gri68,bak76} as
\begin{eqnarray}
iT_{h_{1}h_{2}}(s,t)=\sum ^{\infty }_{n=1}\frac{1}{n!}\int \! \prod ^{n}_{l=1}\! \left[ \frac{d^{4}k_{l}}{(2\pi )^{4}}\frac{d^{4}k_{l}'}{(2\pi )^{4}}\frac{d^{4}q_{l}}{(2\pi )^{4}}\right] \, N_{h_{1}}^{(n)}\! \left( p,k_{1},\ldots ,k_{n},q_{1},\ldots ,q_{n}\right)  &  & \label{t-h1-h2} \\
\times \; \prod ^{n}_{l=1}\! \left[ iT_{1\mathrm{I}\! \mathrm{P}}\! (\hat{s}_{l},q_{l}^{2})\right] \, N_{h_{2}}^{(n)}\! \left( p',k_{1}',\ldots ,k_{n}',-q_{1},\ldots ,-q_{n}\right) \, (2\pi )^{4}\delta ^{(4)}\! \left( \sum ^{n}_{k=1}\! q_{i}-q\right) , &  & \nonumber 
\end{eqnarray}
with \( t=q^{2} \), \( s=(p+p')^{2}\simeq p^{+}p'^{-} \), with \( p,p' \)
being the 4-momenta of the initial hadrons, and with \( \hat{s}_{l}=(k_{l}+k_{l}')^{2}\simeq k_{l}^{+}k_{l}'^{-} \).
\( T_{1\mathrm{I}\! \mathrm{P}} \) is the sum of partonic one-Pomeron-exchange
scattering amplitudes, discussed in the preceding section, \( T_{1\mathrm{I}\! \mathrm{P}}=T_{\mathrm{soft}}+T_{\mathrm{sea}-\mathrm{sea}} \).
The momenta \( k_{l},k_{l}' \) and \( q_{l} \) denote correspondingly the
4-momenta of the initial partonic constituents (quark-antiquark pairs) for the
\( l \)-th scattering and the 4-momentum transfer in that partial process.
The factor \( 1/n! \) takes into account the identical nature of the \( n \)
re-scattering contributions. \( N_{h}^{(n)}\! \left( p,k_{1},\ldots ,k_{n},q_{1},\ldots ,q_{n}\right)  \)
denotes the contribution of the vertex for \( n \)-parton coupling to the hadron
\( h \). 

As discussed in appendix \ref{ax-b-2}, the hadron-hadron amplitude (\ref{t-h1-h2})
may be written as
\begin{eqnarray}
 &  & iT_{h_{1}h_{2}}(s,t)=8\pi ^{2}s\sum ^{\infty }_{n=1}\frac{1}{n!}\, \int ^{1}_{0}\! \prod ^{n}_{l=1}\! dx_{l}^{+}dx_{l}^{-}\; \prod ^{n}_{l=1}\! \left[ \frac{1}{8\pi ^{2}\hat{s}_{l}}\int \! d^{2}q_{l_{\perp }}\, iT^{h_{1}h_{2}}_{1\mathrm{I}\! \mathrm{P}}\! \left( x_{l}^{+},x_{l}^{-},s,-q_{l_{\perp }}^{2}\right) \right] \nonumber \\
 &  & \times \; F^{h_{1}}_{\mathrm{remn}}\! \! \left( 1-\sum ^{n}_{j=1}\! x_{j}^{+}\right) \, F^{h_{2}}_{\mathrm{remn}}\! \! \left( 1-\sum ^{n}_{j=1}\! x_{j}^{-}\right) \: \delta ^{(2)}\! \left( \sum ^{n}_{k=1}\! \vec{q}_{k_{\perp }}-\vec{q}_{\perp }\right) .\label{t-hadr-hadr} 
\end{eqnarray}
(see eq.\ (\ref{th1h2})), where the partonic amplitudes are defined as \( T^{h_{1}h_{2}}_{1\mathrm{I}\! \mathrm{P}}=T^{h_{1}h_{2}}_{\mathrm{soft}}+T^{h_{1}h_{2}}_{\mathrm{sea}-\mathrm{sea}} \),
with 
\begin{eqnarray}
T^{h_{1}h_{2}}_{\mathrm{soft}/\mathrm{sea}-\mathrm{sea}}\! \left( x^{+},x^{-},s,-q_{\bot }^{2}\right) =T_{\mathrm{soft}/\mathrm{sea}-\mathrm{sea}}\! \left( x^{+}x^{-}s,-q_{\bot }^{2}\right) \, F^{h_{1}}_{\mathrm{part}}(x^{+})\, F^{h_{2}}_{\mathrm{part}}(x^{-}) &  & \nonumber \\
\times \; \exp \! \left( -\left[ R_{h_{1}}^{2}+R_{h_{2}}^{2}\right] q_{\bot }^{2}\right)  &  & \label{thh} 
\end{eqnarray}
representing the contributions of ``elementary interactions plus external legs'';
the functions \( F^{h}_{\mathrm{remn}},F^{h}_{\mathrm{part}} \) are defined
in (\ref{f-part-remn}-\ref{f-remn}) as 
\begin{equation}
\label{f-part-main}
F^{h}_{\mathrm{part}}(x)=\gamma _{h}x^{-\alpha _{\mathrm{part}}},
\end{equation}
\begin{equation}
\label{f-remn-main}
F^{h}_{\mathrm{remn}}\! (x)=x^{\alpha ^{h}_{\mathrm{remn}}}.
\end{equation}
Formula (\ref{t-hadr-hadr}) is also correct if one includes valence quarks
(see appendix \ref{ax-b-2}) , if one defines
\begin{equation}
\label{x}
T_{1\mathrm{I}\! \mathrm{P}}^{h_{1}h_{2}}=T^{h_{1}h_{2}}_{\mathrm{soft}}+T^{h_{1}h_{2}}_{\mathrm{sea}-\mathrm{sea}}+T^{h_{1}h_{2}}_{\mathrm{val}-\mathrm{val}}+T^{h_{1}h_{2}}_{\mathrm{val}-\mathrm{sea}}+T^{h_{1}h_{2}}_{\mathrm{sea}-\mathrm{val}},
\end{equation}
 with the hard contribution
\begin{eqnarray}
T^{h_{1}h_{2}}_{\mathrm{val}-\mathrm{val}}\! \left( x^{+},x^{-},s,-q_{\perp }^{2}\right)  & = & \int ^{x^{+}}_{0}\! dx_{v}^{+}\frac{x^{+}}{x_{v}^{+}}\int ^{x^{-}}_{0}\! dx_{v}^{-}\frac{x^{-}}{x_{v}^{-}}\sum _{j,k}T_{\mathrm{hard}}^{jk}\left( x_{v}^{+}x_{v}^{-}s,-q_{\perp }^{2}\right) \label{x} \\
 &  & \qquad \times \; \bar{F}^{h_{1},j}_{\mathrm{part}}(x_{v}^{+},x^{+}-x^{+}_{v})\, \bar{F}^{h_{2},k}_{\mathrm{part}}(x_{v}^{-},x^{-}-x_{v}^{-})\, \exp \! \left( -\left[ R_{h_{1}}^{2}+R_{h_{2}}^{2}\right] q_{\perp }^{2}\right) ,\nonumber 
\end{eqnarray}
 with the mixed semi-hard ``val-sea'' contribution
\begin{eqnarray}
T^{h_{1}h_{2}}_{\mathrm{val}-\mathrm{sea}}\! \left( x^{+},x^{-},s,-q_{\perp }^{2}\right)  & = & \int ^{x^{+}}_{0}\! dx^{+}_{v}\frac{x^{+}}{x_{v}^{+}}\sum _{j}T_{\mathrm{val}-\mathrm{sea}}^{j}\left( x_{v}^{+}x^{-}s,-q_{\perp }^{2}\right) \label{x} \\
 &  & \qquad \times \; \bar{F}^{h_{1},j}_{\mathrm{part}}(x_{v}^{+},x^{+}-x^{+}_{v})\, F^{h_{2}}_{\mathrm{part}}(x^{-})\, \exp \! \left( -\left[ R_{h_{1}}^{2}+R_{h_{2}}^{2}\right] q_{\perp }^{2}\right) ,\nonumber 
\end{eqnarray}
and with the contribution ``sea-val'' obtained from ``val-sea'' by exchanging
variables,
\[
T^{h_{1}h_{2}}_{\mathrm{sea}-\mathrm{val}}\! \left( x^{+},x^{-},s,-q_{\perp }^{2}\right) =T^{h_{2}h_{1}}_{\mathrm{val}-\mathrm{sea}}\! \left( x^{-},x^{+},s,-q_{\perp }^{2}\right) .\]
Here, we allow formally any number of valence type interactions (based on the
fact that multiple valence type processes give negligible contribution). In
the valence contributions, we have convolutions of hard parton scattering amplitudes
\( T_{\mathrm{hard}}^{jk} \) and valence quark distributions \( \bar{F}^{j}_{\mathrm{part}} \)
over the valence quark momentum share \( x_{v}^{\pm } \) rather than a simple
product, since only the valence quarks are involved in the interactions, with
the anti-quarks staying idle (the external legs carrying momenta \( x^{+} \)
and \( x^{-} \) are always quark-antiquark pairs). The functions \( \bar{F} \)
are given as
\begin{equation}
\label{x}
\bar{F}^{h,i}_{\mathrm{part}}(x_{v},x_{\bar{q}})=N^{-1}\, q^{i}_{\mathrm{val}}(x_{v},Q_{0}^{2})(1-x_{v})^{\alpha _{\mathrm{I}\! \mathrm{R}}-1-\alpha _{\mathrm{remn}}}(x_{\bar{q}})^{-\alpha _{\mathrm{I}\! \mathrm{R}}},
\end{equation}
with the normalization factor

\begin{equation}
\label{x}
N=\frac{\Gamma \! \left( 1+\alpha _{\mathrm{remn}}\right) \, \Gamma \! \left( 1-\alpha _{\mathrm{I}\! \mathrm{R}}\right) }{\Gamma \! \left( 2+\alpha _{\mathrm{remn}}-\alpha _{\mathrm{I}\! \mathrm{R}}\right) },
\end{equation}
where \( q^{i}_{\mathrm{val}} \) is a a usual valence quark distribution function.

\begin{figure}[htb]
{\par\centering \resizebox*{!}{0.15\textheight}{\includegraphics{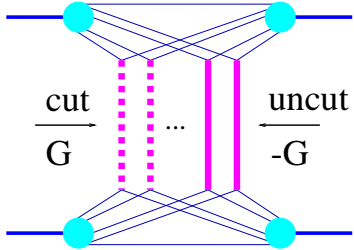}} \par}

\caption{The hadronic profile function \protect\( \gamma \protect \) expressed in terms
of partonic profile functions \protect\( G\equiv G^{h_{1}h_{2}}_{1\mathrm{I}\! \mathrm{P}}\protect \).\label{grtpabppc}}
\end{figure}

The Fourier transform \( \tilde{T}_{h_{1}h_{2}} \) of the amplitude (\ref{t-hadr-hadr})
is given as
\begin{eqnarray}
\frac{i}{2s}\tilde{T}_{h_{1}h_{2}}(s,b) & = & \sum ^{\infty }_{n=1}\frac{1}{n!}\, \int ^{1}_{0}\! \prod ^{n}_{l=1}\! dx_{l}^{+}dx_{l}^{-}\prod ^{n}_{l=1}\frac{i}{2\hat{s}_{l}}\tilde{T}^{h_{1}h_{2}}_{1\mathrm{I}\! \mathrm{P}}(x_{l}^{+},x_{l}^{-},s,b)\nonumber \\
 & \times  & F^{h_{1}}_{\mathrm{remn}}\! \! \left( 1-\sum ^{n}_{j=1}\! x_{j}^{+}\right) \, F^{h_{2}}_{\mathrm{remn}}\! \! \left( 1-\sum ^{n}_{j=1}\! x_{j}^{-}\right) ,\label{thht} 
\end{eqnarray}
with\( \tilde{T}^{h_{1}h_{2}}_{1\mathrm{I}\! \mathrm{P}} \) being the Fourier
transform of \( T^{h_{1}h_{2}}_{1\mathrm{I}\! \mathrm{P}} \). The profile function
\( \gamma  \) is as usual defined as
\begin{equation}
\label{x}
\gamma _{h_{1}h_{2}}(s,b)=\frac{1}{2s}2\mathrm{Im}\tilde{\mathrm{T}}_{h_{1}h_{2}}(s,b),
\end{equation}
which may be evaluated using the AGK cutting rules \cite{abr73},
\begin{eqnarray}
\gamma _{h_{1}h_{2}}(s,b) & = & \sum ^{\infty }_{m=1}\frac{1}{m!}\, \int ^{1}_{0}\! \prod ^{m}_{\mu =1}\! dx_{\mu }^{+}dx_{\mu }^{-}\prod ^{m}_{\mu =1}\frac{1}{2\hat{s}_{\mu }}2\mathrm{Im}\tilde{T}^{h_{1}h_{2}}_{1\mathrm{I}\! \mathrm{P}}(x_{\mu }^{+},x_{\mu }^{-},s,b)\nonumber \\
 & \times  & \sum ^{\infty }_{l=0}\frac{1}{l!}\, \int ^{1}_{0}\! \prod ^{l}_{\lambda =1}\! d\tilde{x}_{\lambda }^{+}d\tilde{x}_{\lambda }^{-}\prod ^{l}_{\lambda =1}\frac{1}{2\hat{s}_{\lambda }}(-2)\mathrm{Im}\tilde{T}^{h_{1}h_{2}}_{1\mathrm{I}\! \mathrm{P}}(\tilde{x}_{\lambda }^{+},\tilde{x}_{\lambda }^{-},s,b)\nonumber \\
 & \times  & F^{h_{1}}_{\mathrm{remn}}\! \! \left( 1-\sum ^{m}_{j=1}\! x_{j}^{+}-\sum ^{l}_{k=1}\! \tilde{x}_{k}^{+}\right) \, F^{h_{2}}_{\mathrm{remn}}\! \! \left( 1-\sum ^{m}_{j=1}\! x_{j}^{-}-\sum ^{l}_{k=1}\! \tilde{x}_{k}^{-}\right) ,\label{gam-agk} 
\end{eqnarray}
where \( 2\mathrm{Im}\tilde{\mathrm{T}}^{h_{1}h_{2}} \) represents a cut elementary
diagram and \( -2\mathrm{Im}\tilde{\mathrm{T}}^{h_{1}h_{2}} \) an uncut one
(taking into account that the uncut contribution may appear on either side from
the cut plane). It is therefore useful to define a partonic profile function
\( G \) via
\begin{eqnarray}
G^{h_{1}h_{2}}_{1\mathrm{I}\! \mathrm{P}}(x_{\lambda }^{+},x_{\lambda }^{-},s,b) & = & \frac{1}{2x_{\lambda }^{+}x_{\lambda }^{-}s}2\mathrm{Im}\, \tilde{T}^{h_{1}h_{2}}_{1\mathrm{I}\! \mathrm{P}}(x_{\lambda }^{+},x_{\lambda }^{-},s,b),\label{gss} 
\end{eqnarray}
which allows to write the integrand of the right-hand-side of eq.\ (\ref{gam-agk})
as a product of \( G \) and \( (-G) \) terms: 
\begin{eqnarray}
\gamma _{h_{1}h_{2}}(s,b) & = & \sum ^{\infty }_{m=1}\frac{1}{m!}\, \int ^{1}_{0}\! \prod ^{m}_{\mu =1}\! dx_{\mu }^{+}dx_{\mu }^{-}\prod ^{m}_{\mu =1}G^{h_{1}h_{2}}_{1\mathrm{I}\! \mathrm{P}}(x_{\mu }^{+},x_{\mu }^{-},s,b)\nonumber \\
 & \times  & \sum ^{\infty }_{l=0}\frac{1}{l!}\, \int ^{1}_{0}\! \prod ^{l}_{\lambda =1}\! d\tilde{x}_{\lambda }^{+}d\tilde{x}_{\lambda }^{-}\prod ^{l}_{\lambda =1}-G^{h_{1}h_{2}}_{1\mathrm{I}\! \mathrm{P}}(\tilde{x}_{\lambda }^{+},\tilde{x}_{\lambda }^{-},s,b)\nonumber \\
 & \times  & F_{\mathrm{remn}}\left( x^{\mathrm{proj}}-\sum _{\lambda }\tilde{x}_{\lambda }^{+}\right) \, F_{\mathrm{remn}}\left( x^{\mathrm{targ}}-\sum _{\lambda }\tilde{x}_{\lambda }^{-}\right) ,\label{gam-agk-g} 
\end{eqnarray}
see fig.\ \ref{grtpabppc}, with
\[
x^{\mathrm{proj}/\mathrm{targ}}=1-\sum x^{\pm }_{\mu }\]
being the momentum fraction of the projectile/target remnant.
This is a very important result, allowing to express the total profile function
\( \gamma _{h_{1}h_{2}} \) via the elementary profile functions \( G^{h_{1}h_{2}}_{1\mathrm{I}\! \mathrm{P}} \). 

Based on the above definitions, we may write the profile function \( G^{h_{1}h_{2}}_{1\mathrm{I}\! \mathrm{P}} \)
as
\begin{equation}
\label{g1p-tot}
G_{1\mathrm{I}\! \mathrm{P}}^{h_{1}h_{2}}=G^{h_{1}h_{2}}_{\mathrm{soft}}+G^{h_{1}h_{2}}_{\mathrm{sea}-\mathrm{sea}}+G^{h_{1}h_{2}}_{\mathrm{val}-\mathrm{val}}+G^{h_{1}h_{2}}_{\mathrm{val}-\mathrm{sea}}+G^{h_{1}h_{2}}_{\mathrm{sea}-\mathrm{val}},
\end{equation}
with 
\begin{equation}
\label{ghh}
G^{h_{1}h_{2}}_{\mathrm{soft}/\mathrm{sea}-\mathrm{sea}}\! \left( x^{+},x^{-},s,b\right) =D^{h_{1}h_{2}}_{\mathrm{soft}/\mathrm{sea}-\mathrm{sea}}\! \left( x^{+}x^{-}s,b\right) \, F^{h_{1}}_{\mathrm{part}}(x^{+})\, F^{h_{2}}_{\mathrm{part}}(x^{-})
\end{equation}
for the soft and semi-hard ``sea-sea'' contribution, with
\begin{eqnarray}
G^{h_{1}h_{2}}_{\mathrm{val}-\mathrm{val}}(x^{+},x^{-},s,b) & = & \int dx_{q_{v}}^{+}dx_{q_{v}}^{-}\sum _{i,j}D^{h_{1}h_{2},ij}_{\mathrm{val}-\mathrm{val}}(x_{q_{v}}^{+}x_{q_{v}}^{-}s,b)\nonumber \\
 &  & \qquad \times \; \bar{F}^{h_{1},i}_{\mathrm{part}}(x_{q_{v}}^{+},x^{+}-x_{q_{v}}^{+})\, \bar{F}^{h_{2},j}_{\mathrm{part}}(x_{q_{v}}^{-},x^{-}-x_{q_{v}}^{-})\label{g-valence} 
\end{eqnarray}
for the hard ``val-val'' contribution, and with
\begin{equation}
\label{g-val-semi}
G^{h_{1}h_{2}}_{\mathrm{val}-\mathrm{sea}}(x^{+},x^{-},s,b)=\int dx_{q_{v}}^{+}\sum _{i}D^{h_{1}h_{2},i}_{\mathrm{val}-\mathrm{sea}}(x_{q_{v}}^{+}x^{-}s,b)\, \bar{F}^{h_{1},i}_{\mathrm{part}}(x_{q_{v}}^{+},x^{+}-x^{+}_{q_{v}})\, F^{h_{2}}_{\mathrm{part}}(x^{-})
\end{equation}
and
\begin{equation}
\label{g-semi-val}
G^{h_{1}h_{2}}_{\mathrm{sea}-\mathrm{val}}(x^{+},x^{-},s,b)=\int dx_{q_{v}}^{-}\sum _{i}D^{h_{2}h_{1},i}_{\mathrm{val}-\mathrm{sea}}(x^{+}x_{q_{v}}^{-}s,b)\, \bar{F}^{h_{2},i}_{\mathrm{part}}(x_{q_{v}}^{-},x^{-}-x^{-}_{q_{v}})\, F^{h_{1}}_{\mathrm{part}}(x^{+}),
\end{equation}
 for the mixed semi-hard ``val-sea'' and ``sea-val'' contributions. For
the soft and ``sea-sea'' contributions, the \( D \)-functions are given as
\begin{eqnarray}
D^{h_{1}h_{2}}_{\mathrm{soft}/\mathrm{sea}-\mathrm{sea}}(\hat{s},b) & = & \int \! d^{2}b'\, D_{\mathrm{soft}/\mathrm{sea}-\mathrm{sea}}(\hat{s},|\vec{b}-\vec{b}'|)\, \frac{1}{4\pi \left( R_{h_{1}}^{2}+R_{h_{2}}^{2}\right) }\, \exp \! \left[ -\frac{b'^{2}}{4\left( R_{h_{1}}^{2}+R_{h_{2}}^{2}\right) }\right] ,\nonumber \label{x} \\
 &  & 
\end{eqnarray}
which means that \( D^{h_{1}h_{2}}_{\mathrm{soft}/\mathrm{sea}-\mathrm{sea}} \)
has the same functional form as \( D_{\mathrm{soft}/\mathrm{sea}-\mathrm{sea}} \),
with \( \lambda ^{(2)}_{\mathrm{soft}}\! (\xi ) \) being replaced by 
\begin{eqnarray}
\lambda ^{h_{1}h_{2}}_{\mathrm{soft}\! }(\xi ) & = & \lambda ^{(2)}_{\mathrm{soft}}(\xi )+R_{h_{1}}^{2}+R_{h_{2}}^{2}\\
 & \simeq  & R_{h_{1}}^{2}+R_{h_{2}}^{2}+\alpha '_{_{\mathrm{soft}}}\ln \! \xi ,\label{x} 
\end{eqnarray}
where we neglected the parton slope \( R_{\mathrm{part}}^{2} \) compared to
the hadron slope \( R_{h}^{2} \).

For the hard contribution, we have correspondingly \( D^{h_{1}h_{2},ij}_{\mathrm{val}-\mathrm{val}}(\hat{s},b) \)
being given in eq.\ (\ref{d-val-val}) with \( R^{2}_{\mathrm{hard}} \) being
replaced by \( R_{h_{1}}^{2}+R_{h_{2}}^{2} \) (neglecting the hard scattering
slope \( R^{2}_{\mathrm{hard}} \) as compared to the hadron Regge slopes \( R_{h_{1}}^{2},R_{h_{2}}^{2} \)).
In case of mixed Pomerons, we have \( D^{h_{1}h_{2},i}_{\mathrm{val}-\mathrm{sea}}(\hat{s},b) \)
given in eq.\ (\ref{d-val-sea}) with \( R_{\mathrm{part}}^{2} \) being replaced
by \( R_{h_{1}}^{2}+R_{h_{2}}^{2} \).

\section{Nucleus-Nucleus Scattering}

We generalize the discussion of the last section in order to treat nucleus-nucleus
scattering. In the Glauber-Gribov approach \cite{gla59,gri69}, the nucleus-nucleus
scattering amplitude is defined by the sum of contributions of diagrams, shown
at fig.\ref{grtppaa}, corresponding to multiple scattering processes between
parton constituents of projectile and target nucleons. Nuclear form factors
are supposed to be defined by the nuclear ground state wave functions. Assuming
the nucleons to be uncorrelated, one can make the Fourier transform to obtain
the amplitude in the impact parameter representation. Then, for given impact
parameter \( \vec{b}_{0} \) between the nuclei, the only formal differences
from the hadron-hadron case will be the possibility for a given nucleon to interact
with a number of nucleons from the partner nucleus as well as the averaging
over nuclear ground states, which amounts to an integration over transverse
nucleon coordinates \( \vec{b}^{A}_{i} \) and \( \vec{b}^{B}_{j} \) in the
projectile and in the target correspondingly. We write this integration symbolically
as 
\begin{equation}
\int dT_{AB}:=\int \prod _{i=1}^{A}d^{2}b^{A}_{i}\, T_{A}(b^{A}_{i})\prod _{j=1}^{B}d^{2}b^{B}_{j}\, T_{B}(b^{B}_{j}),
\end{equation}
 with \( A,B \) being the nuclear mass numbers and with the so-called nuclear
thickness function \( T_{A}(b) \) being defined as the integral over the nuclear
density \( \rho _{A(B)} \), 
\begin{equation}
\label{a.11}
T_{A}(b):=\int dz\, \rho _{A}(\sqrt{b^{2}+z^{2}}).
\end{equation}
For the nuclear densities, we use a parameterization of experimental data of
\cite{mur75},
\begin{equation}
\label{x}
\rho _{A}(r)\propto \left\{ \begin{array}{lcl}
\mathrm{exp}\left\{ -r^{2}/1.72^{2}\right\}  & \mathrm{if} & A=2,\\
\mathrm{exp}\left\{ -r^{2}/(0.9A^{1/3})^{2}\right\}  & \mathrm{if} & 2<A<10,\\
1/\left\{ 1+\mathrm{exp}(\frac{r-0.7A^{0.446}}{0.545})\right\}  & \mathrm{if} & A\geq 10.
\end{array}\right. 
\end{equation}
 It is convenient to use the transverse distance \( b_{k} \) between the two
nucleons from the \( k \)-th nucleon-nucleon pair, i.e.
\begin{equation}
\label{x}
b_{k}=\left| \vec{b}_{0}+\vec{b}^{A}_{\pi (k)}-\vec{b}^{B}_{\tau (k)}\right| ,
\end{equation}
where the functions \( \pi (k) \) and \( \tau (k) \) refer to the projectile
and the target nucleons participating in the \( k^{\mathrm{th}} \) interaction
(pair \( k \)). In order to simplify the notation, we define a vector \( b \)
whose components are the overall impact parameter \( b_{0} \) as well as the
transverse distances \( b_{1},...,b_{AB} \) of the nucleon pairs,
\begin{equation}
\label{x}
b=\{b_{0},b_{1},...,b_{AB}\}.
\end{equation}
Then the nucleus-nucleus interaction cross section can be obtained applying
the cutting procedure to elastic scattering diagrams of fig.\ref{grtppaa} and
written in the form
\begin{equation}
\label{x}
\sigma ^{AB}_{\mathrm{inel}}(s)=\int d^{2}b_{0}\int dT_{AB}\, \gamma _{AB}(s,b),
\end{equation}
where the so-called nuclear profile function \( \gamma _{AB} \) represents
an interaction for given transverse coordinates of the nucleons.

\begin{figure}[htb]
{\par\centering \resizebox*{!}{0.25\textheight}{\includegraphics{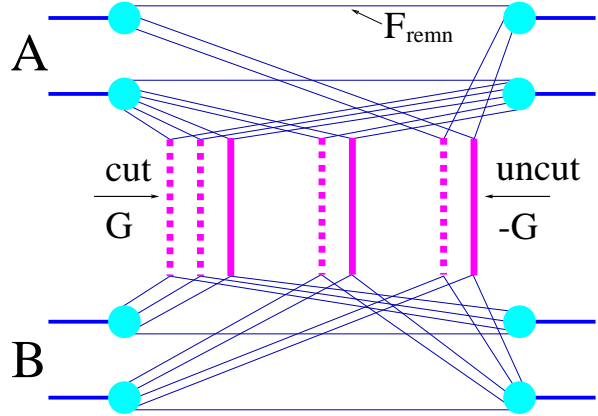}} \par}

\caption{Example for a cut multiple scattering diagram, with cut (dashed lines) and
uncut (full lines) elementary diagrams. This diagram can be translated directly
into a formula for the inelastic cross section (see text).\label{grtppaac}}
\end{figure}

The calculation of the profile function \( \gamma _{AB} \) as the sum over
all cut diagrams of the type shown in fig.\ref{grtppaac} does not differ from
the hadron-hadron case and follows the rules formulated
in the preceding section:

\begin{itemize}
\item For a remnant carrying the light cone momentum fraction \( x \) (\( x^{+} \)
in case of projectile, or \( x^{-} \)in case of target), one has a factor \( F_{\mathrm{remn}}(x) \),
defined in eq.\ (\ref{f-remn-main}). 
\item For each cut elementary diagram (real elementary interaction = dashed vertical
line) attached to two participants with light cone momentum fractions \( x^{+} \)
and \( x^{-} \), one has a factor \( G(x^{+},x^{-},s,b), \) given by eqs.\
(\ref{g1p-tot}-\ref{g-semi-val}). Apart of \( x^{+} \) and \( x^{-} \),
\( G \) is also a function of the total squared energy \( s \) and of the
relative transverse distance \( b \) between the two corresponding nucleons
(we use \( G \) as an abbreviation for \( G_{1\! \mathrm{IP}}^{NN} \) for
nucleon-nucleon scattering).
\item For each uncut elementary diagram (virtual emissions = full vertical line) attached
to two participants with light cone momentum fractions \( x^{+} \) and \( x^{-} \),
one has a factor \( -G(x^{+},x^{-},s,b), \) with the same \( G \) as used
for the cut diagrams.
\item Finally one sums over all possible numbers of cut and uncut Pomerons and integrates
over the light cone momentum fractions.
\end{itemize}
So we find
\begin{eqnarray}
\gamma _{AB}(s,b) & = & \sum _{m_{1}l_{1}}\ldots \sum _{m_{AB}l_{AB}}(1-\delta _{0\Sigma m_{k}})\, \int \, \prod _{k=1}^{AB}\left\{ \prod ^{m_{k}}_{\mu =1}dx_{k,\mu }^{+}dx_{k,\mu }^{-}\, \prod ^{l_{k}}_{\lambda =1}d\tilde{x}_{k,\lambda }^{+}d\tilde{x}_{k,\lambda }^{-}\right\} \nonumber \\
 & \times  & \prod _{k=1}^{AB}\left\{ \frac{1}{m_{k}!}\frac{1}{l_{k}!}\prod _{\mu =1}^{m_{k}}G(x_{k,\mu }^{+},x_{k,\mu }^{-},s,b_{k})\prod _{\lambda =1}^{l_{k}}-G(\tilde{x}_{k,\lambda }^{+},\tilde{x}_{k,\lambda }^{-},s,b_{k})\right\} \nonumber \\
 & \times  & \prod _{i=1}^{A}F_{\mathrm{remn}}\left( x^{+}_{i}-\sum _{\pi (k)=i}\tilde{x}_{k,\lambda }^{+}\right) \prod _{j=1}^{B}F_{\mathrm{remn}}\left( x^{-}_{j}-\sum _{\tau (k)=j}\tilde{x}_{k,\lambda }^{-}\right) \label{gaminel} 
\end{eqnarray}
with 

\begin{eqnarray}
x^{\mathrm{proj}}_{i} & = & 1-\sum _{\pi (k)=i}x_{k,\mu \, ,}^{+}\\
x^{\mathrm{targ}}_{j} & = & 1-\sum _{\tau (k)=j}x_{k,\mu }^{-}\, .\label{x} 
\end{eqnarray}
 The summation indices \( m_{k} \) refer to the number of cut elementary diagrams
and \( l_{k} \) to the number of uncut elementary diagrams, related to nucleon
pair \( k \). For each possible pair \( k \) (we have altogether \( AB \)
pairs), we allow for any number of cut and uncut diagrams. The integration variables
\( x^{\pm }_{k,\mu } \) refer to the \( \mu ^{\mathrm{th}} \) elementary interaction
of the \( k^{\mathrm{th}} \) pair for the cut elementary diagrams, the variables
\( \tilde{x}^{\pm }_{k,\lambda } \) refer to the corresponding uncut elementary
diagrams. The arguments of the remnant functions \( F_{\mathrm{remn}} \) are
the remnant light cone momentum fractions, i.e.\ unity minus the the momentum
fractions of all the corresponding elementary contributions (cut and uncut ones).
We also introduce the variables \( x^{+}_{i} \)and \( x_{j}^{-} \), defined
as unity minus the momentum fractions of all the corresponding cut contributions,
in order to integrate over the uncut ones (see below).

The expression for \( \gamma _{AB}(...) \) sums up all possible numbers of
cut Pomerons \( m_{k} \) with one exception due to the factor \( (1-\delta _{0\Sigma m_{k}}) \):
one does not consider the case of all \( m_{k} \)'s being zero, which corresponds
to ``no interaction'' and therefore does not contribute to the inelastic cross
section. We may therefore define a quantity \( \bar{\gamma }_{AB}(...) \),
representing ``no interaction'', by taking the expression for \( \gamma _{AB}(...) \)
with \( (1-\delta _{0\Sigma m_{k}}) \) replaced by \( (\delta _{0\Sigma m_{k}}) \):
\begin{eqnarray}
\bar{\gamma }_{AB}(s,b) & = & \sum _{l_{1}}\ldots \sum _{l_{AB}}\, \int \, \prod _{k=1}^{AB}\left\{ \prod ^{l_{k}}_{\lambda =1}d\tilde{x}_{k,\lambda }^{+}d\tilde{x}_{k,\lambda }^{-}\right\} \; \prod _{k=1}^{AB}\left\{ \frac{1}{l_{k}!}\, \prod _{\lambda =1}^{l_{k}}-G(\tilde{x}_{k,\lambda }^{+},\tilde{x}_{k,\lambda }^{-},s,b_{k})\right\} \nonumber \label{r} \\
 & \times  & \prod _{i=1}^{A}F^{+}\! \left( 1-\sum _{\pi (k)=i}\tilde{x}_{k,\lambda }^{+}\right) \prod _{j=1}^{B}F^{-}\left( 1-\sum _{\tau (k)=j}\tilde{x}_{k,\lambda }^{-}\right) .\label{rremnant} 
\end{eqnarray}
One now may consider the sum of ``interaction'' and ``no interaction'',
and one obtains easily
\begin{equation}
\label{unitarity}
\gamma _{AB}(s,b)+\bar{\gamma }_{AB}(s,b)=1.
\end{equation}
Based on this important result, we consider \( \gamma _{AB} \) to be the probability
to have an interaction and correspondingly \( \bar{\gamma }_{AB} \) to be the
probability of no interaction, for fixed energy, impact parameter and nuclear
configuration, specified by the transverse distances \( b_{k} \) between nucleons,
and we refer to eq.\ (\ref{unitarity}) as ``unitarity relation''. But we
want to go even further and use an expansion of \( \gamma _{AB} \) in order
to obtain probability distributions for individual processes, which then serves
as a basis for the calculations of exclusive quantities.

The expansion of \( \gamma _{AB} \) in terms of cut and uncut Pomerons as given
above represents a sum of a large number of positive and negative terms, including
all kinds of interferences, which excludes any probabilistic interpretation.
We have therefore to perform summations of interference contributions - sum
over any number of virtual elementary scatterings (uncut Pomerons) - for given
non-interfering classes of diagrams - with given numbers of real scatterings
(cut Pomerons) \cite{abr73}. Let us write the formulas explicitly. We have
 
\begin{eqnarray}
\gamma _{AB}(s,b) & = & \sum _{m_{1}}\ldots \sum _{m_{AB}}(1-\delta _{0\sum m_{k}})\, \int \, \prod _{k=1}^{AB}\left\{ \prod ^{m_{k}}_{\mu =1}dx_{k,\mu }^{+}dx_{k,\mu }^{-}\right\} \nonumber \\
 & \times  & \prod _{k=1}^{AB}\left\{ \frac{1}{m_{k}!}\, \prod _{\mu =1}^{m_{k}}G(x_{k,\mu }^{+},x_{k,\mu }^{-},s,b_{k})\right\} \; \Phi _{AB}\left( x^{\mathrm{proj}},x^{\mathrm{targ}},s,b\right) ,\label{sigmanucl} 
\end{eqnarray}
 where the function \( \Phi  \) representing the sum over virtual emissions
(uncut Pomerons) is given by the following expression
\begin{eqnarray}
\Phi _{AB}\left( x^{\mathrm{proj}},x^{\mathrm{targ}},s,b\right)  & = & \sum _{l_{1}}\ldots \sum _{l_{AB}}\, \int \, \prod _{k=1}^{AB}\left\{ \prod ^{l_{k}}_{\lambda =1}d\tilde{x}_{k,\lambda }^{+}d\tilde{x}_{k,\lambda }^{-}\right\} \; \prod _{k=1}^{AB}\left\{ \frac{1}{l_{k}!}\, \prod _{\lambda =1}^{l_{k}}-G(\tilde{x}_{k,\lambda }^{+},\tilde{x}_{k,\lambda }^{-},s,b_{k})\right\} \nonumber \label{r} \\
 & \times  & \prod _{i=1}^{A}F_{\mathrm{remn}}\left( x_{i}^{\mathrm{proj}}-\sum _{\pi (k)=i}\tilde{x}_{k,\lambda }^{+}\right) \prod _{j=1}^{B}F_{\mathrm{remn}}\left( x^{\mathrm{targ}}_{j}-\sum _{\tau (k)=j}\tilde{x}_{k,\lambda }^{-}\right) .\label{rremnant} 
\end{eqnarray}
 This summation has to be carried out, before we may use the expansion of \( \gamma _{AB} \)
to obtain probability distributions. This is far from trivial, as we are going
to discuss in the next section, but let us assume for the moment that it can
be done. To make the notation more compact, we define matrices \( X^{+} \)
and \( X^{-} \), as well as a vector \( m \), via 
\begin{eqnarray}
X^{+} & = & \left\{ x_{k,\mu }^{+}\right\} ,\\
X^{-} & = & \left\{ x_{k,\mu }^{-}\right\} ,\\
m & = & \{m_{k}\},\label{x} 
\end{eqnarray}
which leads to
\begin{eqnarray}
\gamma _{AB}(s,b) & = & \sum _{m}(1-\delta _{0m})\int dX^{+}dX^{-}\Omega _{AB}^{(s,b)}(m,X^{+},X^{-}),\\
\bar{\gamma }_{AB}(s,b) & = & \Omega _{AB}^{(s,b)}(0,0,0),\label{x} 
\end{eqnarray}
with 
\begin{equation}
\label{omega-ab-bas}
\Omega _{AB}^{(s,b)}(m,X^{+},X^{-})=\prod _{k=1}^{AB}\left\{ \frac{1}{m_{k}!}\, \prod _{\mu =1}^{m_{k}}G(x_{k,\mu }^{+},x_{k,\mu }^{-},s,b_{k})\right\} \; \Phi _{AB}\left( x^{\mathrm{proj}},x^{\mathrm{targ}},s,b\right) .
\end{equation}
This allows to rewrite the unitarity relation eq.\ (\ref{unitarity}) in the
following form,
\begin{equation}
\label{x}
\sum _{m}\int dX^{+}dX^{-}\Omega _{AB}^{(s,b)}(m,X^{+},X^{-})=1.
\end{equation}
This equation is of fundamental importance, because it allows us to interpret
\( \Omega ^{(s,b)}(m,X^{+},X^{-}) \) as probability density of having an interaction
configuration characterized by \( m \), with the light cone momentum fractions
of the Pomerons being given by \( X^{+} \) and \( X^{-} \).

\section{Diffractive Scattering}

We do not have a consistent treatment of diffractive scattering at the moment,
this is left to a future project in connection with a complete treatment of
enhanced diagrams. For the moment, we introduce diffraction ``by hand'': in
case of no interaction in \( pp \) or \( pA \) scattering, we consider the
projectile to be diffractively excited with probability
\begin{equation}
\label{x}
w_{\mathrm{diff}}\frac{\left( 1-\sqrt{\Phi (1,x^{\mathrm{targ}},s,b)}\right) ^{2}}{\Phi (1,x^{\mathrm{targ}},s,b)},
\end{equation}
with a fit parameter \( w_{\mathrm{diff}} \). Nucleus-nucleus scattering is
here (but only here!) considered as composed of \( pA \) or \( Ap \) collisions.

\section{AGK Cancelations in Hadron-Hadron Scattering}

As a first application, we are going to prove that AGK cancelations apply perfectly
in our model\footnote{
We speak here about the contribution of elementary interactions (Pomeron exchanges)
to the secondary particle production; the AGK cancellations do not hold for
the contribution of remnant states (spectator partons) hadronization \cite{bra90}.
}.

As we showed above, the description of high energy hadronic interaction requires
to consider explicitly a great number of contributions, corresponding to multiple
scattering process, with a number of elementary parton-parton interactions happening
in parallel. However, when calculating inclusive spectra of secondary particles,
it is enough to consider the simplest hadron-hadron (nucleus-nucleus) scattering
diagrams containing a single elementary interaction, as the contributions of
multiple scattering diagrams with more than one elementary interaction exactly
cancel each other. This so-called AGK-cancelation is a consequence of the general
Abramovskii-Gribov-Kancheli cutting rules \cite{abr73}.

Let us consider the most fundamental inclusive distribution, where all other
inclusive spectra may be derived from: the distribution \( dn^{h_{1}h_{2}}_{\mathrm{Pom}}/dx^{+}dx^{-} \),
with \( dn^{h_{1}h_{2}}_{\mathrm{Pom}} \) being the number of Pomerons with
light cone momentum fractions between \( x^{+} \) and \( x^{+}+dx^{+} \)and
between \( x^{-} \) and \( x^{-}+dx^{-} \) respectively, at a given value
of \( b \) and \( s \). If AGK cancelations apply, the result for \( dn^{h_{1}h_{2}}_{\mathrm{Pom}}/dx^{+}dx^{-} \)
should coincide with the contribution coming from exactly one elementary interaction
(see eq.\ (\ref{gam-agk-g})):

\begin{equation}
\label{n1pom}
\frac{dn_{\mathrm{Pom}}^{(1)h_{1}h_{2}}}{dx^{+}dx^{-}}(x^{+},x^{-},s,b)=G_{1\mathrm{I}\! \mathrm{P}}^{h_{1}h_{2}}(x^{+},x^{-},s,b)F^{h_{1}}_{\mathrm{remn}}(1-x^{+})F^{h_{2}}_{\mathrm{remn}}(1-x^{-}),
\end{equation}
and the contributions from multiple scattering should exactly cancel. We have
per definition

\begin{eqnarray}
\frac{dn^{h_{1}h_{2}}_{\mathrm{Pom}}}{dx^{+}dx^{-}}(x^{+},x^{-},s,b) & = & \sum ^{\infty }_{m=1}\sum _{l=0}^{\infty }\int \prod ^{m}_{\mu =1}dx^{+}_{\mu }dx^{-}_{\mu }\prod ^{m+l}_{\lambda =m+1}dx^{+}_{\lambda }dx^{-}_{\lambda }\nonumber \\
 & \times  & \frac{1}{m!}\frac{1}{l!}\prod ^{m}_{\mu =1}G_{1\mathrm{I}\! \mathrm{P}}^{h_{1}h_{2}}(x^{+}_{\mu },x^{-}_{\mu },s,b)\prod ^{m+l}_{\lambda =m+1}-G_{1\mathrm{I}\! \mathrm{P}}^{h_{1}h_{2}}(x^{+}_{\lambda },x^{-}_{\lambda },s,b)\nonumber \\
 & \times  & F^{h_{1}}_{\mathrm{remn}}(1-\sum ^{m+l}_{\nu =1}x_{\nu }^{+})F^{h_{2}}_{\mathrm{remn}}(1-\sum ^{m+l}_{\nu =1}x^{-}_{\nu })\label{profile} \\
 & \times  & \sum _{\mu '=1}^{m}\delta (x^{+}-x^{+}_{\mu '})\delta (x^{-}-x_{\mu '}^{-}).\nonumber 
\end{eqnarray}
Due to the symmetry of the integrand in the r.h.s. of eq.\ (\ref{profile})
in the variables \( x^{\pm }_{1},\ldots ,x_{m}^{\pm } \), the sum of delta
functions produces a factor \( mG(x^{+},x^{-},s,b) \), and removes one \( dx^{+}_{\mu }dx_{\mu }^{-} \)
integration. Using
\begin{equation}
\label{x}
m\sum _{m=1}^{\infty }\ldots \frac{1}{m!}\prod _{\mu =1}^{m-1}\ldots =\sum _{m'=0}^{\infty }\ldots \frac{1}{m'!}\prod _{\mu =1}^{m'}\ldots \, ,
\end{equation}
with \( m'=m-1, \) we get
\begin{eqnarray}
\frac{dn^{h_{1}h_{2}}_{\mathrm{Pom}}}{dx^{+}dx^{-}}(x^{+},x^{-},s,b) & = & G_{1\mathrm{I}\! \mathrm{P}}^{h_{1}h_{2}}(x^{+},x^{-},s,b)\frac{1}{n!}\sum _{n=0}^{\infty }\int \prod ^{n}_{\nu =1}dx^{+}_{\nu }dx^{-}_{\nu }\\
 & \times  & \left\{ \sum _{m=0}^{n}\left( \begin{array}{c}
n\\
m
\end{array}\right) \prod ^{m}_{\mu =1}G_{1\mathrm{I}\! \mathrm{P}}^{h_{1}h_{2}}(x^{+}_{\mu },x^{-}_{\mu },s,b)\prod ^{n}_{\lambda =m+1}-G_{1\mathrm{I}\! \mathrm{P}}^{h_{1}h_{2}}(x^{+}_{\lambda },x^{-}_{\lambda },s,b)\right\} \nonumber \\
 & \times  & F^{h_{1}}_{\mathrm{remn}}(1-x^{+}-\sum ^{n}_{\nu =1}x_{\nu }^{+})F^{h_{2}}_{\mathrm{remn}}(1-x^{-}-\sum ^{n}_{\nu =1}x^{-}_{\nu })\, .\nonumber 
\end{eqnarray}
The term in curly brackets \( \left\{ \ldots \right\}  \) is 1 for \( n=0 \)
and zero otherwise, so we get the important final result
\begin{equation}
\label{agk}
\frac{dn^{h_{1}h_{2}}_{\mathrm{Pom}}}{dx^{+}dx^{-}}(x^{+},x^{-},s,b)=\frac{dn_{\mathrm{Pom}}^{(1)h_{1}h_{2}}}{dx^{+}dx^{-}}(x^{+},x^{-},s,b),
\end{equation}
which corresponds to one single elementary interaction; the multiple scattering
aspects completely disappeared, so AGK cancelations indeed apply in our approach.
AGK cancelations are closely related to the factorization formula for jet production
cross section, since as a consequence of eq.\ (\ref{agk}), we may obtain the
inclusive jet cross section in a factorized form as
\begin{equation}
\label{x}
\sigma ^{h_{1}h_{2}}_{\mathrm{jet}}=\sum _{j,k}\int dp_{\perp }^{2}\int dz^{+}\int dz^{-}\, f^{h_{1}}_{j}(z^{+},M_{F}^{2})\, f^{h_{2}}_{k}(z^{-},M_{F}^{2})\, {d\sigma _{\mathrm{Born}}^{jk}\over dp_{\bot }^{2}}\! (z^{+}z^{-}s,p_{\perp }^{2}),
\end{equation}
with \( f^{h_{1}}_{j} \) and \( f^{h_{2}}_{k} \) representing the parton distributions
of the two hadrons.

\section{AGK Cancelations in Nucleus-Nucleus Scattering}

We have shown in the previous section that for hadron-hadron scattering AGK
cancelations apply, which means that inclusive spectra coincide with the contributions
coming from exactly one elementary interaction. For multiple Pomeron exchanges
we have a complete destructive interference, they do not contribute at all.
Here, we are going to show that AGK cancelations also apply for nucleus-nucleus
scattering, which means that the inclusive cross section for \( A+B \) scattering
is \( AB \) times the corresponding inclusive cross section for proton-proton
interaction. 

The inclusive cross section for forming a Pomeron with light cone momentum fractions
\( x^{+} \) and \( x^{-} \) in nucleus-nucleus scattering is given as
\begin{eqnarray}
\frac{d\sigma ^{AB}_{\mathrm{Pom}}}{dx^{+}dx^{-}}(x^{+},x^{-},s) & = & \int d^{2}b_{0}\int dT_{AB}\nonumber \\
 & \times  & \sum _{m_{1}l_{1}}\ldots \sum _{m_{AB}l_{AB}}(1-\delta _{0\sum m_{k}})\int \prod _{k=1}^{AB}\left\{ \prod ^{m_{k}+l_{k}}_{\mu =1}dx_{k,\mu }^{+}dx_{k,\mu }^{-}\right\} \nonumber \\
 & \times  & \prod _{k=1}^{AB}\left\{ \frac{1}{m_{k}!l_{k}!}\prod _{\mu =1}^{m_{k}}G(x_{k,\mu }^{+},x_{k,\mu }^{-},s,b_{k})\prod _{\lambda =m_{k}+1}^{m_{k}+l_{k}}-G(x_{k,\lambda }^{+},x_{k,\lambda }^{-},s,b_{k})\right\} \nonumber \label{sigpom-ab1} \\
 & \times  & \prod _{i=1}^{A}F_{\mathrm{remn}}\left( 1-\sum _{\pi (k)=i}x_{k,\lambda }^{+}\right) \prod _{j=1}^{B}F_{\mathrm{remn}}\left( 1-\sum _{\tau (k)=j}x_{k,\lambda }^{-}\right) \nonumber \\
 & \times  & \sum _{k'=1}^{AB}\sum ^{m_{k}}_{\mu '=1}\delta (x^{+}-x^{+}_{k'\mu '})\delta (x^{-}-x^{-}_{k'\mu '}).\label{sigpom-ab1} 
\end{eqnarray}
 The factor \( (1-\delta _{0\sum m_{k}}) \) makes sure that at least one of
the indices \( m_{k} \) is bigger than zero. Integrating over the variables
appearing in the delta functions, we obtain a factor \( \sum _{k'}G(x^{+},x^{-},s,b_{k})\cdot m_{k'} \)
which may be written in front of \( \sum _{m_{1}l_{1}}... \) . In the following
expression one may rename the integration variables such that the variables
\( x^{+}_{k'm_{k'}} \) and \( x^{-}_{k'm_{k'}} \) disappear. This means for
the arguments of the functions \( F_{\mathrm{remn}} \) that for \( i=\pi (k') \)
and \( j=\tau (k') \) one replaces \( 1 \) by \( 1-x^{+} \) and \( 1-x^{-} \)
respectively. Then one uses the factor \( m_{k'} \) mentioned above to replace
\( m_{k'}! \) by \( (m_{k'}-1)! \). One finally renames \( (m_{k'}-1) \)
by \( m_{k'} \), as a consequence of which one may drop the factor \( (1-\delta _{0\sum m_{k}}) \).
This is crucial, since now we have factors of the form 
\begin{equation}
\label{x}
\sum _{m_{k}=0}^{\infty }\sum _{l_{k}=0}^{\infty }...\frac{1}{m_{k}!l_{k}!}\prod _{\mu =1}^{m_{k}}G(x_{k,\mu }^{+},x_{k,\mu }^{-},s,b)\prod _{\lambda =m_{k}+1}^{m_{k}+l_{k}}-G(x_{k,\lambda }^{+},x_{k,\lambda }^{-},s,b)....
\end{equation}
In this sum only the term for \( m_{k}=0 \) and \( l_{k}=0 \) is different
from zero, namely \( 1 \), and so we get 
\begin{equation}
\label{x}
\frac{d\sigma ^{AB}_{\mathrm{Pom}}}{dx^{+}dx^{-}}(x^{+},x^{-},s,b)=\int d^{2}b_{0}\int dT_{AB}\sum _{k'}G(x^{+},x^{-},s,b_{k'})F_{\mathrm{remn}}(1-x^{+})F_{\mathrm{remn}}(1-x^{-}).
\end{equation}
Using the definition of \( dT_{AB} \), writing \( b_{k'} \) explicitly as
\( |\vec{b}_{0}+\vec{b}^{A}_{\pi (k)}-\vec{b}^{B}_{\tau (k)}| \), we obtain
\begin{eqnarray}
\frac{d\sigma ^{AB}_{\mathrm{Pom}}}{dx^{+}dx^{-}}(x^{+},x^{-},s) & = & AB\int d^{2}b_{0}\int d^{2}b_{+}T(b_{+})\int d^{2}b_{-}T(b_{-})G(x^{+},x^{-},s,|\vec{b}_{0}+\vec{b}_{+}-\vec{b}_{-}|)\nonumber \\
 &  & \qquad \qquad \qquad \qquad \times \; F_{\mathrm{remn}}(1-x^{+})F_{\mathrm{remn}}(1-x^{-}).\label{x} 
\end{eqnarray}
 Changing the order of the integrations, we obtain finally
\begin{equation}
\label{x}
\frac{d\sigma ^{AB}_{\mathrm{Pom}}}{dx^{+}dx^{-}}(x^{+},x^{-},s)=AB\frac{d\sigma ^{pp}_{\mathrm{Pom}}}{dx^{+}dx^{-}}(x^{+},x^{-},s)
\end{equation}
with
\begin{equation}
\label{x}
\frac{d\sigma ^{pp}_{\mathrm{Pom}}}{dx^{+}dx^{-}}(x^{+},x^{-},s)=\int d^{2}b\, G(x^{+},x^{-},s,b)F_{\mathrm{remn}}(1-x^{+})F_{\mathrm{remn}}(1-x^{-}).
\end{equation}
Since any other inclusive cross section \( d\sigma _{\mathrm{incl}}/dq \) may
be obtained from the inclusive Pomeron distribution via convolution, we obtain
the very general result
\begin{equation}
\label{x}
\frac{d\sigma ^{AB}_{\mathrm{incl}}}{dq}(q,s,b)=AB\frac{d\sigma ^{pp}_{\mathrm{incl}}}{dq}(q,s,b),
\end{equation}
 so nucleus-nucleus inclusive cross sections are just \( AB \) times the proton-proton
ones. So, indeed, AGK cancelations apply perfectly in our approach.

\section{Outlook}

What did we achieve so far? We have a well defined model, introduced by using
the language of field theory (Feynman diagrams). We were able to prove some
elementary properties (AGK cancelations in case of proton-proton and nucleus-nucleus
scattering). To proceed further, we have to solve (at least) two fundamental
problems:

\begin{itemize}
\item the sum over virtual emissions has to be performed,
\item tools have to be developed to deal with the multidimensional probability distribution\\
\( \Omega _{AB}^{(s,b)}(m,X^{+},X^{-}) \),
\end{itemize}
both being very difficult tasks. 

Calculating the sum over virtual emissions (\( \Phi _{AB} \)) is not only technically
difficult, there are also conceptual problems. By studying the properties of
\( \Phi _{AB} \), we find that at very high energies the theory is no longer
unitary without taking into account additional screening corrections. In this
sense, we consider our work as a first step to construct a consistent model
for high energy nuclear scattering, but there is still work to be done.

Concerning the multidimensional probability distribution \( \Omega _{AB}^{(s,b)}(m,X^{+},X^{-}) \),
we are going to develop methods well known in statistical physics (Markov chain
techniques), which we also are going to discuss in detail later. So finally,
we are able to calculate the probability distribution \( \Omega _{AB}^{(s,b)}(m,X^{+},X^{-}) \),
and are able to generate (in a Monte Carlo fashion) configurations \( (m,X^{+},X^{-}) \)
according to this probability distribution. 

The two above mentioned problems will be discussed in detail in the following
chapters.

\cleardoublepage

\chapter{Virtual Emissions}

In order to proceed, we need to calculate the sum over virtual emissions, represented
by the function \( \Phi _{AB} \). Understanding the behavior of \( \Phi _{AB} \)
is crucial, since this function is related to \( \bar{\gamma }_{AB} \) and
plays therefore a crucial role in connection with unitarity, the unitarity equation
being given as \( \gamma _{AB}+\bar{\gamma }_{AB}=1 \). By studying the properties
of \( \Phi _{AB} \), we find inconsistencies in the limit of high energies,
in the sense that the individual terms appearing in the unitarity equation are
not necessarily positive. Attempting to understand this unphysical behavior,
we find that any model where AGK cancelations apply (so most of the models used
presently) has to run asymptotically into this problem. So eventually one needs
to construct models, where AGK cancelations are violated, which is going to
be expected when contributions of Pomeron-Pomeron interactions are taken into
consideration. 

As a first phenomenological solution of the unitarity problem, we are going
to ``unitarize'' the ``bare theory'' introduced in the preceding chapter
``by hand'', such that the theory is changed as little as possible, but the
asymptotic problems disappear. The next step should of course be a consistent
treatment including contributions of enhanced Pomeron diagrams.

In the following, we are going to present the calculation of \( \Phi _{AB} \),
we discuss the unitarity problems and the phenomenological solution, as well
as properties of the ``unitarized theory''.

\section{Parameterizing the Elementary Interaction  }

The basis for all the calculations which follow is the function \( G^{NN}_{1\mathrm{I}\! \mathrm{P}} \),
which is the profile function representing a single elementary nucleon-nucleon
(\( NN \)) interaction. For simplicity, we write simply \( G\equiv G^{NN}_{1\mathrm{I}\! \mathrm{P}} \).
This function \( G \) is a sum of several terms, representing soft, semi-hard,
valence, and screening contributions. In case of soft and semi-hard, one has
\( G=(x^{+}x^{-})^{-\alpha _{\mathrm{part}}}D \), where \( D \) represents
the Pomeron exchange and the factor in front of \( D \) the ``external legs'',
the nucleon participants. For the other contributions the functional dependence
on \( x^{+},x^{-} \) is somewhat more complicated, but nevertheless it is convenient
to define a function 
\begin{equation}
\label{x}
D(x^{+},x^{-},s,b)=\frac{G(x^{+},x^{-},s,b)}{(x^{+}x^{-})^{-\alpha _{\mathrm{part}}}}.
\end{equation}
 We obtain \( G \) and therefore \( D \) as the result of a quite involved
numerical calculation, which means that these functions are given in a discretized
fashion. Since this is not very convenient and since the dependence of \( x^{+} \)
, \( x^{-} \), and \( b \) are quite simple, we are going to parameterize
our numerical results and use this analytical expression for further calculations.
This makes the following discussions much easier and more transparent.

\begin{figure}[htb]
{\par\centering \resizebox*{!}{0.4\textheight}{\includegraphics{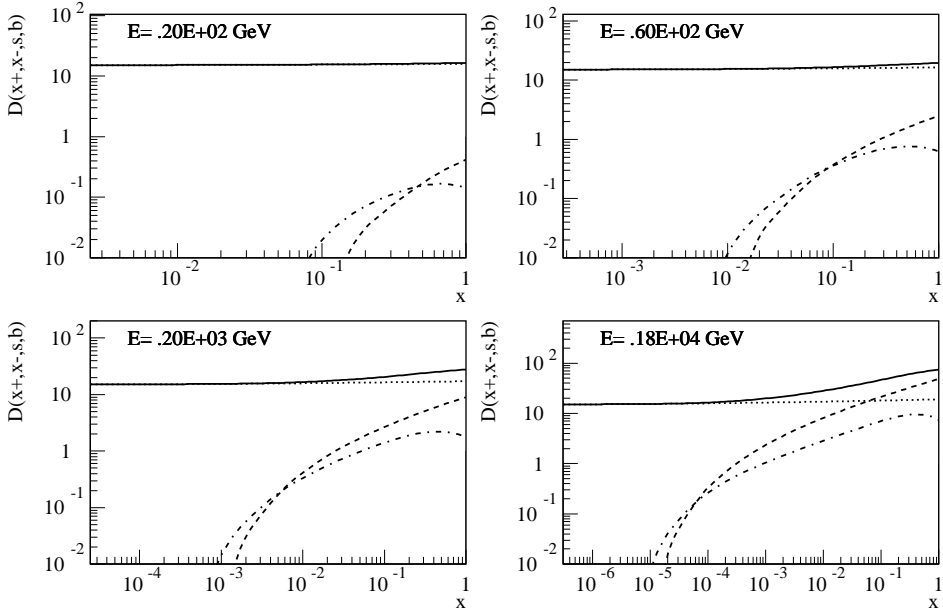}} \par}

\caption{{\small The function \protect\( D\protect \) (solid line) as well as the different
contributions to \protect\( D\protect \) as a function of \protect\( x=x^{+}x^{-}\protect \),
for \protect\( b=0\protect \) , at different energies \protect\( E=\sqrt{s}.\protect \)
We show \protect\( D_{\mathrm{soft}}\protect \) (dotted), \protect\( D_{\mathrm{semi}}\protect \)
(dashed), and \protect\( D_{\mathrm{valence}}\protect \) (dashed-dotted).}
\label{d} }
\end{figure}

\begin{figure}[htb]
{\par\centering \resizebox*{!}{0.4\textheight}{\includegraphics{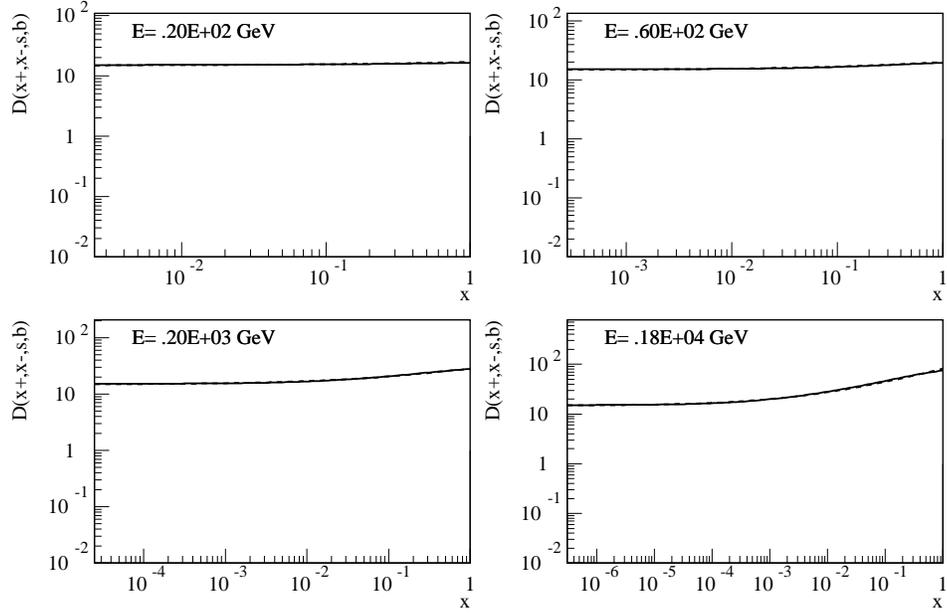}} \par}

\caption{{\small The exact (full) and the parameterized (dashed) function \protect\( D\protect \)
as a function of \protect\( x=x^{+}x^{-}\protect \)}, for \protect\( b=0\protect \),
{\small at different energies \protect\( E=\sqrt{s}.\protect \)}\label{fig:D-x} }
\end{figure}

We first consider the case of zero impact parameter (\( b=0 \)). In fig.\ \ref{d},
we plot the function \( D \) together with the individual contributions as
functions of \( x=x^{+}x^{-} \), for \( b=0 \) and for different values of
\( s \). 
To fit the function \( D \), we make the following ansatz,
\begin{equation}
\label{fit}
D(x^{+},x^{-},s,b=0)=\sum ^{N}_{i=1}\underbrace{\left\{ \alpha _{D_{i}}+\alpha ^{*}_{D_{i}}(x^{+}x^{-})^{\beta _{D_{i}}^{*}}\right\} \left( x^{+}x^{-}s\right) ^{\beta _{D_{i}}}}_{D_{i}}\, ,
\end{equation}
where the parameters may depend on \( s \), and the parameters marked with
a star are non-zero for a given \( i \) only if the corresponding \( \alpha _{D_{i}} \)
is zero. This parameterization works very well, as shown in fig.\ \ref{fig:D-x},
where we compare the original \( D \) function with the fit according to eq.\
(\ref{fit}).

\begin{figure}[htb]
{\par\centering \resizebox*{!}{0.4\textheight}{\includegraphics{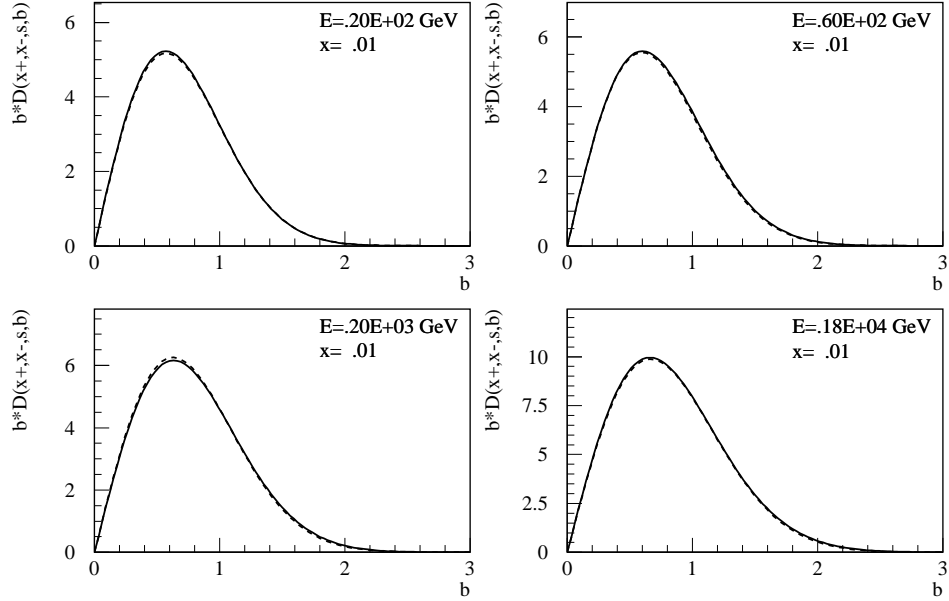}} \par}

\caption{{\small The \protect\( b\protect \)-dependence of \protect\( b\cdot D(x^{+},x^{-},s,b)\protect \)
for a fixed value of \protect\( x=x^{+}x^{-}\protect \), with \protect\( x^{+}=x^{-}\protect \),
at different energies \protect\( E=\sqrt{s}.\protect \) Exact results are represented
as solid lines, the parameterized ones by dashed lines.} \label{fig:D-b}}
\end{figure}

Let us now consider the \( b \)-dependence for fixed \( x^{+} \) and \( x^{-} \).
Since we observe almost a Gaussian shape with a weak (logarithmic) dependence
of the width on \( x=x^{+}x^{-} \), we could make the following ansatz:

\begin{eqnarray}
D(x^{+},x^{-},s,b)=\sum ^{N}_{i=1}\left\{ \alpha _{D_{i}}+\alpha ^{*}_{D_{i}}(x^{+}x^{-})^{\beta _{D_{i}}^{*}}\right\} (x^{+}x^{-}s)^{\beta _{D_{i}}}\exp \left( \frac{-b^{2}}{\delta _{D_{i}}+\epsilon _{D_{i}}\ln (x^{+}x^{-}s)}\right) . &  & 
\end{eqnarray}
However, we can still simplify the parameterization. We have

\begin{eqnarray}
\exp \left( \frac{-b^{2}}{\delta +\epsilon \ln x}\right)  & \approx  & \exp \left( \frac{-b^{2}}{\delta }\left( 1-\frac{\epsilon }{\delta }\ln (x^{+}x^{-}s)\right) \right) \\
 & = & \exp \left( \frac{-b^{2}}{\delta }\right) (x^{+}x^{-}s)^{\gamma b^{2}}
\end{eqnarray}
 So we make finally the ansatz
\begin{equation}
\label{Dfit}
D(x^{+},x^{-},s,b)=\sum ^{N}_{i=1}\left\{ \alpha _{D_{i}}+\alpha ^{*}_{D_{i}}(x^{+}x^{-})^{\beta _{D_{i}}^{*}}\right\} \left( x^{+}x^{-}s\right) ^{\beta _{D_{i}}+\gamma _{D_{i}}b^{2}}e^{-\frac{b^{2}}{\delta _{D_{i}}}},
\end{equation}
which provides a very good analytical representation of the numerically obtained
function \( D \), as shown in fig.\ \ref{fig:D-b}.

\section{Calculating \protect\( \Phi \protect \) for proton-proton collisions}

We first consider proton-proton collisions. To be more precise, we are going
to derive an expression for \( \Phi _{pp} \) which can be evaluated easily
numerically and which will serve as the basis to investigate the properties
of \( \Phi _{pp} \). We have

\begin{eqnarray}
\Phi _{pp}\left( x^{+},x^{-},s,b\right)  & = & \sum _{l=0}^{\infty }\int dx_{1}^{+}dx_{1}^{-}\ldots dx_{l}^{+}dx_{l}^{-}\left\{ \frac{1}{l!}\, \prod _{\lambda =1}^{l}-G(x_{\lambda }^{+},x_{\lambda }^{-},s,b)\right\} \nonumber \label{r} \\
 & \times  & F_{\mathrm{remn}}\left( x^{+}-\sum x_{\lambda }^{+}\right) F_{\mathrm{remn}}\left( x^{-}-\sum x_{\lambda }^{-}\right) .\label{r2i} 
\end{eqnarray}
 with
\begin{eqnarray}
F_{\mathrm{remn}}(x) & = & x^{\alpha _{\mathrm{remn}}}\, \Theta (x)\, \Theta (1-x),\label{Fremn} 
\end{eqnarray}
where \( \Theta (x) \) is the Heavyside function, and
\begin{eqnarray}
G(x_{\lambda }^{+},x_{\lambda }^{-},s,b) & = & (x_{\lambda }^{+}x_{\lambda }^{-})^{-\alpha _{\mathrm{part}}}\, D(x_{\lambda }^{+},x_{\lambda }^{-},s,b).
\end{eqnarray}
Using eq.\ (\ref{Dfit}), we have
\begin{equation}
\label{gi}
G(x_{\lambda }^{+},x_{\lambda }^{-},s,b)=\sum _{i=1}^{N}\underbrace{\alpha _{i}(x_{\lambda }^{+}x_{\lambda }^{-})^{\beta _{i}}}_{G_{i,\lambda }}
\end{equation}
 with 
\begin{eqnarray}
\alpha _{i} & = & \left( \alpha _{D_{i}}+\alpha ^{*}_{D_{i}}\right) \, s^{\left( \beta _{D_{i}}+\gamma _{D_{i}}b^{2}\right) }\, e^{-\frac{b^{2}}{\delta _{D_{i}}}}\, ,\\
\beta _{i} & = & \beta _{D_{i}}+\beta ^{*}_{D_{i}}+\gamma _{D_{i}}b^{2}-\alpha _{\mathrm{part}}\, ,\label{x} 
\end{eqnarray}
with \( \alpha ^{*}_{D_{i}}\neq 0 \) and \( \beta ^{*}_{D_{i}}\neq 0 \) only
if \( \alpha _{D_{i}}=0 \). Using eq.\ (\ref{gi}), we obtain from eq.\ (\ref{r2i})
the following expression
\begin{eqnarray}
\Phi _{pp}(x^{+},x^{-}s,b) & = & \sum ^{\infty }_{r_{1}=0}\ldots \sum ^{\infty }_{r_{N}=0}\frac{1}{r_{1}!}\ldots \frac{1}{r_{N}!}\int \prod ^{r_{1}+...+r_{N}}_{\lambda =1}dx^{+}_{\lambda }dx^{-}_{\lambda }\nonumber \\
 & \times  & \prod _{\rho _{1}=1}^{r_{1}}-G_{1,\rho _{1}}\ldots \prod _{\rho _{N}=r_{1}+...+r_{N-1}+1}^{r_{1}+...+r_{N}}-G_{N,\rho _{N}}\nonumber \\
 & \times  & F_{\mathrm{remn}}(x^{+}-\sum _{\lambda }x^{+}_{\lambda })F_{\mathrm{remn}}(x^{-}-\sum _{\lambda }x^{-}_{\lambda }).
\end{eqnarray}
 Using the fact that the functions \( G_{i,\lambda } \) are separable,
\begin{equation}
\label{x}
G_{i,\lambda }=\alpha _{i}\, (x^{+}_{\lambda })^{\beta _{i}}\, (x^{-}_{\lambda })^{\beta _{i}}\, ,
\end{equation}
one finds finally (see appendix \ref{axd1})

\begin{eqnarray}
\Phi _{pp}(x^{+},x^{-},s,b) & = & x^{\alpha _{\mathrm{remn}}}\sum ^{\infty }_{r_{1}=0}\ldots \sum ^{\infty }_{r_{N}=0}\left\{ \frac{\Gamma (1+\alpha _{\mathrm{remn}})}{\Gamma (1+\alpha _{\mathrm{remn}}+r_{1}\tilde{\beta }_{1}+...+r_{N}\tilde{\beta }_{N})}\right\} ^{2}\nonumber \\
 & \times  & \frac{(-\alpha _{1}x^{\tilde{\beta }_{1}}\Gamma ^{2}(\tilde{\beta }_{1}))^{r}}{r_{1}!}\ldots \frac{(-\alpha _{N}x^{\tilde{\beta }_{N}}\Gamma ^{2}(\tilde{\beta }_{N}))^{t}}{r_{N}!}\label{r2exact} 
\end{eqnarray}
with \( x=x^{+}x^{-} \)and \( \tilde{\beta }_{i}=\beta _{i}+1 \). Since the
sums converge very fast, this expression can be easily evaluated numerically.

\section{Unitarity Problems}

In this section, we are going to present numerical results for \( \Phi _{pp} \),
based on equation (\ref{r2exact}). We will observe an unphysical behavior in
certain regions of phase space, which amounts to a violation of unitarity. Trying
to understand its physical origin, we find that AGK cancelations, which apply
in our model, automatically lead to unitarity violations. This means on the
other hand that a fully consistent approach requires explicit violation of AGK
cancelations, which should occur in case of considering contributions of enhanced
Pomeron diagrams. 

In which way is \( \Phi _{pp} \) related to unitarity? We have shown in the
preceding chapter that the inelastic non-diffractive cross section \( \sigma _{\mathrm{inel}}(s) \)
may be written as
\begin{equation}
\label{x}
\sigma _{\mathrm{inel}}(s)=\int d^{2}b\, \gamma _{pp}(s,b),
\end{equation}
with the profile function \( \gamma _{pp}(s,b) \) representing all diagrams
with at least one cut Pomeron. We defined as well the corresponding quantity
\( \bar{\gamma }_{pp}(s,b) \) representing all diagrams with zero cut Pomerons.
We demonstrated that the sum of these two quantities is one (\ref{unitarity}),
\begin{equation}
\label{prob-cons}
\gamma _{pp}(s,b)+\bar{\gamma }_{pp}(s,b)=1,
\end{equation}
which represents a unitarity relation. The function \( \Phi _{pp} \) enters
finally, since we have the relation
\begin{equation}
\label{x}
\bar{\gamma }_{pp}(s,b)=\Phi _{pp}(1,1,s,b).
\end{equation}
 Based on these formulas, we interpret \( \bar{\gamma }_{pp}(s,b)=\Phi _{pp}(1,1,s,b) \)
as the probability of having no interaction, whereas \( \gamma _{pp}(s,b)=1-\Phi _{pp}(1,1,s,b) \)
represents the probability to have an interaction, at given impact parameter
and energy. Such an interpretation of course only makes sense as long as any
of the \( \gamma  \)'s is positive, otherwise unitarity is said to be violated,
even if the equation (\ref{prob-cons}) still holds.

\begin{figure}[htb]
{\par\centering \resizebox*{!}{0.25\textheight}{\includegraphics{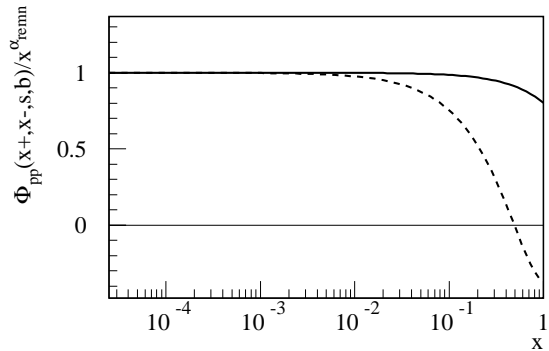}} \par}

\caption{{\small The expression \protect\( \Phi _{pp}(x^{+},x^{-},s,b)\, /\, x^{\alpha _{\mathrm{remn}}}\protect \)
as a function of \protect\( x=x^{+}x^{-}\protect \)for \protect\( b=0\protect \)
(dashed) and for \protect\( b=1.5\protect \) fm (solid curve).\label{r2x}}\small }
\end{figure}

\begin{figure}[htb]
{\par\centering \resizebox*{!}{0.25\textheight}{\includegraphics{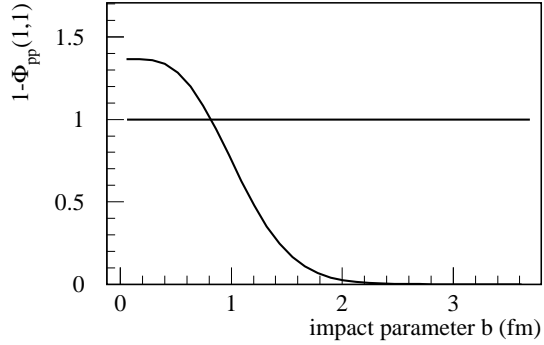}} \par}

\caption{The profile function \protect\( 1-\Phi _{pp}(1,1,s,b)\protect \) as a function
of impact parameter \protect\( b.\protect \) This function should represent
the probability to have an interaction at a given impact parameter. \label{pro1}}
\end{figure}

In fig.\ \ref{r2x}, we plot \( \Phi _{pp} \) as a function of \( x=x^{+}x^{-} \)
for \( \sqrt{s}=200 \) GeV for two different values of \( b \). The curve
for \( b=1.5 \) fm (solid curve) is close to one with a minimum of about 0.8
at \( x=1 \). The \( x \)-dependence for \( b=0 \) fm (dashed curve) is much
more dramatic: the curve deviates from 1 already at relatively small values
of \( x \) and drops finally to negative values at \( x=1. \) 
The values for \( x=1 \) are of particular interest, since \( 1-\Phi _{pp}(1,1,s,b)=\gamma _{pp}(s,b) \)
represents the profile function in the sense that the integration over \( b \)
provides the inelastic non-diffractive cross section. Therefore, in fig.\ \ref{pro1},
we plot the \( b \)-dependence of \( 1-\Phi _{pp}(1,1,s,b) \), which increases
beyond 1 for small values of \( b \), since \( \Phi _{pp} \) is negative in
this region, as discussed above for the case of \( b=0 \) fm. 
 On the other hand, an upper limit of 1 is really necessary in order to assure
unitarity. So the fact that \( \Phi _{pp} \) grows to values bigger than one
is a manifestation of unitarity violation.

In the following, we try to understand the physical reason for this unitarity
problem. We are going to show, that it is intimately related to the fact that
in our approach AGK cancelations are fulfilled, as shown earlier, which means
that any approach where AGK cancelations apply will have exactly the same problem. 

We are going to demonstrate in the following that AGK cancelations imply automatically
unitarity violation. The average light cone momentum taken by a Pomeron may
be calculated from the Pomeron inclusive spectrum \( dn_{\mathrm{Pom}}/dx^{+}dx^{-} \)
as
\begin{equation}
\label{x}
<x^{+}>=\int dx^{+}x^{+}\left\{ \frac{1}{\sigma _{\mathrm{inel}}(s)}\int d^{2}b\int dx^{-}\frac{dn_{\mathrm{Pom}}}{dx^{+}dx^{-}}(x^{+},x^{-},s,b)\right\} .
\end{equation}
If AGK cancelations apply, we have 
\begin{equation}
\label{x}
dn_{\mathrm{Pom}}/dx^{+}dx^{-}=dn^{(1)}_{\mathrm{Pom}}/dx^{+}dx^{-}=G(x^{+},x^{-},s,b)F_{\mathrm{remn}}(x^{+})F_{\mathrm{remn}}(x^{-}),
\end{equation}
 and therefore
\begin{equation}
\label{x}
\frac{dn_{\mathrm{Pom}}}{dx^{+}dx^{-}}(x^{+},x^{-},s,b)=\mu (s)\, (x^{+}x^{-}s)^{\Delta (s)}\, e^{-\frac{b^{2}}{\lambda (s)}}\, f(x^{+})f(x^{-}),
\end{equation}
 where \( \Delta (s) \) is bigger than zero and increases with energy and \( \mu (s) \)
and \( \lambda (s) \) depend weakly (logarithmically) on \( s \), whereas
\( f \) is an energy independent function. We obtain
\begin{equation}
\label{x}
<x^{+}>=\frac{\gamma (s)\, s^{\Delta (s)}}{\sigma _{\mathrm{inel}}(s)},
\end{equation}
where \( \gamma (s) \) depends only logarithmically on \( s \). Since \( <x^{+}> \)
must be smaller or equal to one, we find
\begin{equation}
\label{x}
\sigma _{\mathrm{inel}}(s)\geq \gamma (s)\, s^{\Delta (s)},
\end{equation}
 which violates the Froissard bound and therefore unitarity. This problem is
related to the old problem of unitarity violation in case of single Pomeron
exchange. The solution appeared to be the observation that one needs to consider
multiple scattering such that virtual multiple emissions provide sufficient
screening to avoid the unreasonably fast increase of the cross section. If AGK
cancelations apply (as in our model), the problem comes back by considering
inclusive spectra, since these are determined by single scattering. 

Thus, we have shown that \emph{a consistent application of the eikonal Pomeron
scheme both to interaction cross sections and to particle production calculations
unavoidably leads to the violation of the unitarity}. This problem is not observed
in many models currently used, since there simply no consistent treatment is
provided, and the problem is therefore hidden. The solution of the unitarity
problem requires to employ the full Pomeron scheme, which includes also so-called
enhanced Pomeron diagrams, to be discussed later. The simplest diagram of that
kind - so-called Y-diagram, for example, contributes a negative factor to all
inclusive particle distributions in the particle rapidity region \( y_{0}<y<Y \),
where \( Y \) is the total rapidity range for the interaction and \( y_{0} \)
corresponds to the rapidity position of the Pomeron self-interaction vertex.
Thus, one speaks about breaking of the AGK-cancelations in the sense that one
gets corrections to all inclusive quantities calculated from just one Pomeron
exchange graph \cite{kai86}. In particular, presenting the inclusive Pomeron
distribution \( d\sigma _{\mathrm{Pom}}(x^{+},x^{-})/dx^{+}dx^{-} \) by the
formula (\ref{n1pom}) implies that the functions \( f(x^{\pm }) \) acquire
a dependence on the energy of the interaction \( s \). It is this dependence
which is expected to slow down the energy increase of the Pomeron number and
thus to cure the unitarity problem.

So we think it is mandatory to proceed in the following way: first one needs
to provide a consistent treatment of cross sections and particle production,
which will certainly lead to unitarity problems, and second one has to refine
the theory to solve the unitarity problem in a consistent way, via screening
corrections. The first part of this program is provided in this paper, the second
one will be treated in some approximate fashion later, but a rigorous, self-consistent
treatment of this second has still to be done.

\section{A Phenomenological Solution: Unitarization of \protect\( \Phi \protect \)}

As we have seen in the preceding sections, unitarity violation manifests itself
by the fact that the virtual emission function \( \Phi _{pp} \) appears to
be negative at high energies and small impact parameter for large values of
\( x^{+} \) and \( x^{-} \), particularly for \( x^{+}=x^{-}=1 \). What is
the mathematical origin of these negative values? In eq.\ (\ref{r2exact}),
the sums over \( r_{i} \) contains terms of the form \( (...)^{r_{i}}/r_{i}! \)
and an additional factor of the form
\begin{equation}
\label{x}
\left\{ \frac{\Gamma (1+\alpha _{\mathrm{remn}})}{\Gamma (1+\alpha _{\mathrm{remn}}+r_{1}\tilde{\beta }_{1}+...+r_{N}\tilde{\beta }_{N})}\right\} .
\end{equation}

It is this factor which causes the problem, as it strongly suppresses contributions
of terms with large \( r_{i} \), which are important when the interaction energy
\( s \) increases . Physically it is connected to the reduced phase space in
case of too many virtual Pomerons emitted. By dropping this factor one would
obtain a simple exponential function which is definitely positive.

Our strategy is to modify the scheme such that \( \Phi _{pp} \) stays essentially
unchanged for values of \( s \), \( b \), \( x^{+} \), and \( x^{-} \),
where \( \Phi _{pp} \) is positive and that \( \Phi _{pp} \) is ``corrected''
to assure positive values in regions where it is negative. We call this procedure
``unitarization'', which should not be considered as an approximation, since
one is really changing the physical content of the theory. This is certainly
only a phenomenological solution of the problem, the correct procedure should
amount to taking into account the mentioned screening corrections due to enhanced
Pomeron diagrams, which should provide a ``natural unitarization''. Nevertheless
we consider our approach as a necessary first step towards a consistent formulation
of multiple scattering theory in nuclear (including hadron-hadron) collisions
at very high energies. 

\begin{figure}[htb]
{\par\centering \resizebox*{!}{0.25\textheight}{\includegraphics{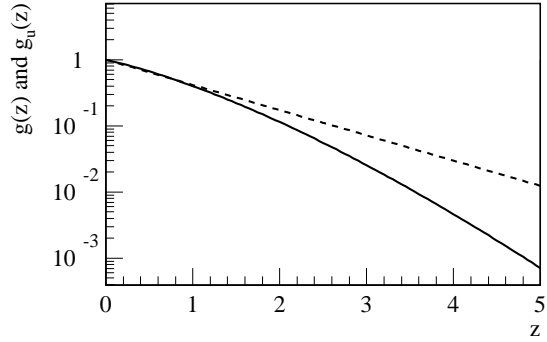}} \par}

\caption{{\small Function \protect\( g(z)\protect \) (full line) and unitarized function
\protect\( g_{\mathrm{u}}(z)\protect \) (dashed).\label{fig:gammag}}\small }
\end{figure}

Let us explain our ``unitarization'' in the following. We define

\begin{equation}
\label{gammaq}
g(z)=\frac{\Gamma (1+\alpha _{\mathrm{remn}})}{\Gamma (1+\alpha _{\mathrm{remn}}+z)}
\end{equation}
such that 
\begin{equation}
\label{x}
g(r_{1}\tilde{\beta }_{1}+...+r_{N}\tilde{\beta }_{N})
\end{equation}
 is the factor causing unitarity problems. This expression should be of the
form 
\[
(...)^{r_{1}}...(...)^{r_{N}},\]
which would make \( \Phi _{AB} \) a well behaved exponential function. In order
to achieve this, the function \( g \) should be an exponential. So we replace
\( g(z) \) by \( g_{\mathrm{e}}(z), \) where the latter function is defined
as

\begin{equation}
\label{gammag}
g_{\mathrm{e}}(z)=e^{-\epsilon _{\mathrm{e}}z},
\end{equation}
where the parameter \( \epsilon _{\mathrm{e}} \) should be chosen such that
\( g(z) \) is well approximated for values of \( z \) between (say) \( 0 \)
and \( 0.5 \) (see fig.\ \ref{fig:gammag}). The index ``e'' refers to ``exponentiation''. 
 So, we replace the factor \( g(r_{1}\tilde{\beta }_{1}+...+r_{N}\tilde{\beta }_{N}) \)
by

\begin{eqnarray}
g_{\mathrm{e}}(r_{1}\tilde{\beta }_{1}+...+r_{N}\tilde{\beta }_{N}) & = & e^{-\epsilon _{\mathrm{e}}(r_{1}\tilde{\beta }_{1}+...+r_{N}\tilde{\beta }_{N})}\\
 & = & \left( e^{-\epsilon _{\mathrm{e}}\tilde{\beta }_{1}}\right) ^{r_{1}}...\left( e^{-\epsilon _{\mathrm{e}}\tilde{\beta }_{N}}\right) ^{r_{N}},\label{gammar} 
\end{eqnarray}
and obtain correspondingly instead of \( \Phi _{pp} \)

\begin{eqnarray}
\Phi _{\mathrm{e}\, _{pp}}(x^{+},x^{-},s,b) & = & x^{\alpha _{\mathrm{remn}}}\sum ^{\infty }_{r_{1}=0}\ldots \sum _{r_{N}=0}^{\infty }\left( e^{-2\epsilon _{\mathrm{e}}\tilde{\beta }_{1}}\right) ^{r_{1}}...\left( e^{-2\epsilon _{\mathrm{e}}\tilde{\beta }_{N}}\right) ^{r_{N}}\nonumber \\
 & \times  & \frac{(-\alpha _{1}x^{\tilde{\beta }_{1}}\Gamma ^{2}(\tilde{\beta }_{1}))^{r_{1}}}{r_{1}!}...\frac{(-\alpha _{N}x^{\tilde{\beta }_{N}}\Gamma ^{2}(\tilde{\beta }_{N}))^{t_{N}}}{r_{N}!}.
\end{eqnarray}
Now the sums can be performed and we get
\begin{eqnarray}
\Phi _{\mathrm{e}\, _{pp}}(x^{+},x^{-},s,b) & = & x^{\alpha _{\mathrm{remn}}}\prod _{i=1}^{N}\exp \left\{ -\alpha _{i}x^{\tilde{\beta }_{i}}\Gamma ^{2}(\tilde{\beta }_{i})e^{-2\epsilon _{\mathrm{e}}\tilde{\beta }_{i}}\right\} 
\end{eqnarray}
which may be written as 
\begin{equation}
\label{x}
\Phi _{\mathrm{e}\, _{pp}}(x^{+},x^{-},s,b)=(x^{+}x^{-})^{\alpha _{\mathrm{remn}}}\exp \left\{ -\tilde{G}(x^{+}x^{-},s,b)\right\} 
\end{equation}
 with

\begin{equation}
\label{G-tilde}
\tilde{G}(x,s,b)=\sum _{i=1}^{N}\tilde{\alpha }_{i}x^{\tilde{\beta }_{i}}
\end{equation}
 with 
\begin{eqnarray}
\tilde{\alpha }_{i} & = & \alpha _{i}\Gamma ^{2}(\tilde{\beta }_{i})e^{-2\epsilon _{\mathrm{e}}\tilde{\beta }_{i}}\\
\tilde{\beta }_{i} & = & \beta _{i}+1,\label{x} 
\end{eqnarray}
where \( \alpha _{i} \) and \( \beta _{i} \) are given as
\begin{eqnarray}
\alpha _{i} & = & \left( \alpha _{D_{i}}+\alpha ^{*}_{D_{i}}\right) \, s^{\left( \beta _{D_{i}}+\gamma _{D_{i}}b^{2}\right) }\, e^{-\frac{b^{2}}{\delta _{D_{i}}}}\, ,\\
\beta _{i} & = & \beta _{D_{i}}+\beta ^{*}_{D_{i}}+\gamma _{D_{i}}b^{2}-\alpha _{\mathrm{part}}\, ,\label{x} 
\end{eqnarray}
with \( \alpha ^{*}_{D_{i}}\neq 0 \) and \( \beta ^{*}_{D_{i}}\neq 0 \) only
if \( \alpha _{D_{i}}\neq 0 \). 

We are not yet done. We modified \( \Phi _{pp} \) such that the new function
\( \Phi _{\mathrm{e}\, _{pp}} \) is surely positive. But what happened to our
unitarity equation? If we replace \( \Phi _{pp} \) by \( \Phi _{\mathrm{e}\, _{pp}} \),
we obtain
\begin{equation}
\label{x}
\bar{\gamma }_{pp}+\gamma _{pp}=\sum _{m=0}^{\infty }\int dX^{+}dX^{-}\Omega _{\mathrm{e}\, _{pp}}^{(s,b)}(m,X^{+},X^{-}),
\end{equation}
with
\begin{equation}
\label{x}
\Omega _{\mathrm{e}\, _{pp}}^{(s,b)}(m,X^{+},X^{-})=\left\{ \frac{1}{m!}\, \prod _{\mu =1}^{m}G(s,x_{\mu }^{+},x_{\mu }^{-},b)\right\} \; \Phi _{\mathrm{e}\, _{pp}}\left( x^{+},x^{-},s,b\right) ,
\end{equation}
where \( x^{+} \) and \( x^{-} \) refer to the remnant light cone momenta.
Since \( \Phi _{\mathrm{e}\, _{pp}} \) is always bigger than \( \Phi _{pp} \)
for small values of \( b \), the sum \( \gamma _{pp}+\bar{\gamma }_{pp} \)
is bigger than one, so the unitarity equation does not hold any more. This is
quite natural, since we modified the virtual emissions without caring about
the real ones. In order to account for this, we define
\begin{equation}
\label{x}
Z(s,b)=\sum _{m=0}^{\infty }\int dX^{+}dX^{-}\left\{ \frac{1}{m!}\, \prod _{\mu =1}^{m}G(s,x_{\mu }^{+},x_{\mu }^{-},b)\right\} \; \Phi _{\mathrm{e}\, _{pp}}\left( x^{+},x^{-},s,b\right) ,
\end{equation}
 which is equal to one in the exact case, but which is different from one if
we use \( \Phi _{\mathrm{e}\, _{pp}} \) instead of \( \Phi _{pp} \). In order
to recover the unitarity equation, we have to ``renormalize'' \( \Phi _{\mathrm{e}\, _{pp}} \),
and we define therefore the ``unitarized'' virtual emission function \( \Phi _{\mathrm{u}\, _{pp}} \)
via
\begin{equation}
\label{x}
\Phi _{\mathrm{u}\, _{pp}}(x^{+},x^{-},s,b)=\frac{\Phi _{\mathrm{e}\, _{pp}}(x^{+},x^{-},s,b)}{Z(s,b)}.
\end{equation}
Now, the unitarity equation holds,
\begin{equation}
\label{x}
\bar{\gamma }_{pp}+\gamma _{pp}=\sum _{m=0}^{\infty }\int dX^{+}dX^{-}\Omega _{\mathrm{u}\, _{pp}}^{(s,b)}(m,X^{+},X^{-})=1\, ,
\end{equation}
with 
\begin{equation}
\label{x}
\Omega _{\mathrm{u}\, _{pp}}^{(s,b)}(m,X^{+},X^{-})=\left\{ \frac{1}{m!}\, \prod _{\mu =1}^{m}G(s,x_{\mu }^{+},x_{\mu }^{-},b)\right\} \; \Phi _{\mathrm{u}\, _{pp}}\left( x^{+},x^{-},s,b\right) 
\end{equation}
being strictly positive, which allows finally the probability interpretation.

\section{Properties of the Unitarized Theory}

\begin{figure}[htb]
{\par\centering \resizebox*{!}{0.25\textheight}{\includegraphics{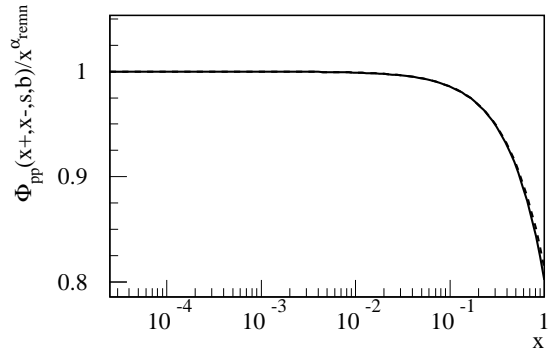}} \par}

\caption{{\small Behavior of \protect\( \Phi _{pp}(x^{+},x^{-},s,b)/(x^{+}x^{-})^{\alpha _{\mathrm{remn}}}\protect \)
as a function of \protect\( x=x^{+}x^{-},\protect \) for the exact function
(solid) and the exponentiated one (dashed curve)} for impact parameter \protect\( b=1.5\protect \)
fm.\label{r2a}}
\end{figure}

\begin{figure}[htb]
{\par\centering \resizebox*{!}{0.25\textheight}{\includegraphics{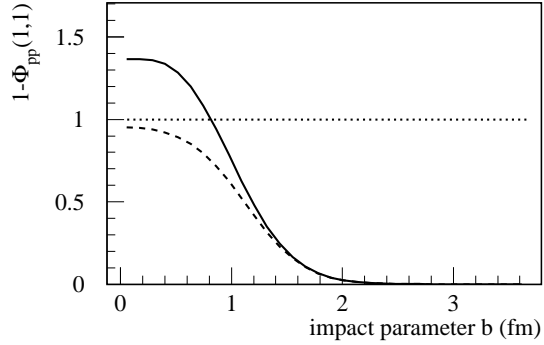}} \par}

\caption{The function \protect\( 1-\Phi _{pp}(1,1,s,b)\protect \) as a function of
impact parameter \protect\( b.\protect \) We show the exact result (solid)
as well as the exponentiated one (dashed curve).\label{pro2}}
\end{figure}

We are now going to investigate the consequences of our unitarization, in other
words, how the results are affected by this modification. In fig.\ \ref{r2a}
we compare the exact and the exponentiated version of the virtual emission function
(\( \Phi _{pp} \) and \( \Phi _{\mathrm{e}\, _{pp}} \)) for a large value
of the impact parameter (\( b=1.5 \)\( \,  \)fm). The exponentiated result
(dashed) is somewhat below the exact one (solid curve), but the difference is
quite small.
The situation is somewhat different in case of zero impact parameter \( b \).
For small values of \( x \) the two curves coincide more or less, however for
\( x=1 \) the exponentiated result (dashed) is well above the exact one (solid
curve). In particular, and this is most important, the dashed curve rests positive
and in this sense corrects for the unphysical behavior (negative values) for
the exact curve.
\begin{figure}[htb]
{\par\centering \resizebox*{!}{0.25\textheight}{\includegraphics{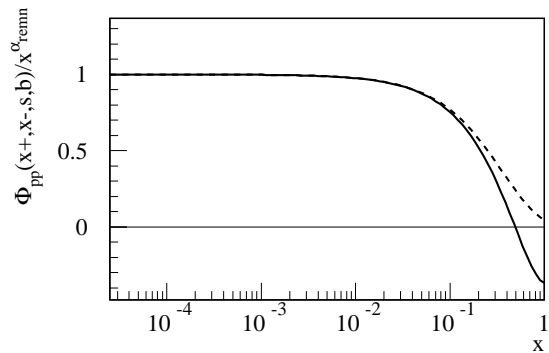}} \par}

\caption{{\small Behavior of \protect\( \Phi _{pp}(x^{+},x^{-},s,b)/(x^{+}x^{-})^{\alpha _{\mathrm{remn}}}\protect \)
as a function of \protect\( x=x^{+}x^{-},\protect \) for the exact function
(solid) and the exponentiated one (dashed curve)} for impact parameter \protect\( b=0\protect \)
fm. \label{r2b}}
\end{figure}
The behavior for \( x=1 \) for different values of \( b \) is summarized in
fig.\( \,  \)\ref{pro2}, where we plot \( 1-\Phi _{pp}(1,1,s,b) \) as a function
of \( b \). We clearly observe that for large \( b \) exact (solid) and exponentiated
(dashed curve) result agree approximately, whereas for small values of \( b \)
they differ substantially, with the exponentiated version always staying below
1, as it should be. So the effect of our exponentiation is essentially to push
the function below 1.

\begin{figure}[htb]
{\par\centering \resizebox*{!}{0.25\textheight}{\includegraphics{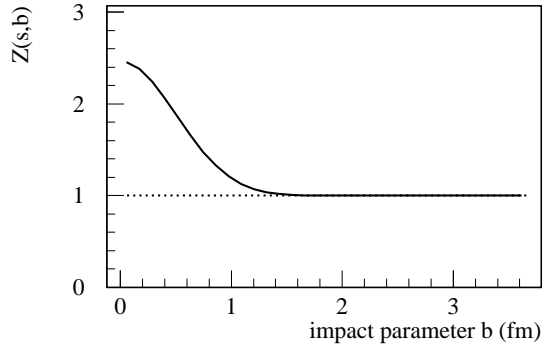}} \par}

\caption{The normalization function \protect\( Z(s,b)\protect \) as a function of the
impact parameter \protect\( b\protect \).\label{norm}}
\end{figure}

Next we calculate explicitly the normalization function \( Z(s,b) \). We have

\begin{eqnarray}
Z(s,b) & = & \sum ^{\infty }_{m=0}\int \prod ^{m}_{\mu =1}dx^{+}_{\mu }dx^{-}_{\mu }\frac{1}{m!}\prod ^{m}_{\mu =1}G(x^{+}_{\mu },x^{-}_{\mu },s,b)\nonumber \\
 &  & \qquad \times \Phi _{\mathrm{e}\, _{pp}}\left( 1-\sum ^{m}_{\nu =1}x_{\nu }^{+},1-\sum ^{m}_{\nu =1}x^{-}_{\nu }\right) ,\label{x} 
\end{eqnarray}
which may be written as 
\begin{equation}
\label{x}
Z(s,b)=\Phi _{\mathrm{e}\, _{pp}}(1,1,s,b)+\int dz^{+}dz^{-}H(z^{+},z^{-})\, \Phi _{\mathrm{e}\, _{pp}}(z^{+},z^{-},s,b),
\end{equation}
with

\begin{eqnarray}
H(z^{+},z^{-}) & = & \sum ^{\infty }_{m=1}\int \prod ^{m}_{\mu =1}dx^{+}_{\mu }dx^{-}_{\mu }\frac{1}{m!}\prod ^{m}_{\mu =1}G(x^{+}_{\mu },x^{-}_{\mu },s,b)\nonumber \\
 & \times  & \delta \left( 1-z^{+}-\sum ^{m}_{\mu =1}x^{+}_{\mu }\right) \, \delta \left( 1-z^{-}-\sum ^{m}_{\mu =1}x^{-}_{\mu }\right) .\label{h1} 
\end{eqnarray}
Using the analytical form of \( G \), we obtain

\begin{eqnarray}
H(z^{+},z^{-}) & =\underbrace{\sum ^{\infty }_{r_{1}=0}...\sum ^{\infty }_{r_{N}=0}}_{r_{1}+...+r_{K}\neq 0} & \frac{\left[ (1-z^{+})(1-z^{-})\right] ^{r_{1}\tilde{\beta }_{1}+...+r_{N}\tilde{\beta }_{N}-1}}{\Gamma (r_{1}\tilde{\beta }_{1}+...+r_{N}\tilde{\beta }_{N})^{2}}\nonumber \\
 & \times  & \frac{(\alpha _{1}\Gamma (\tilde{\beta }_{1})^{2})^{r_{1}}}{r_{1}!}...\frac{(\alpha _{N}\Gamma (\tilde{\beta }_{N})^{2})^{r_{N}}}{r_{N}!}\label{h2} 
\end{eqnarray}
 (see appendix \ref{axd2}). This can be calculated, and after numerically doing
the integration over \( z^{+},z^{-} \), we obtain the normalization function
\( Z(s,b) \), as shown in fig.\ \ref{norm}. 
We observe, as expected, a value close to unity at large values of \( b \),
whereas for small impact parameter \( Z(s,b) \) is bigger than one, since only
at small values of \( b \) the virtual emission function \( \Phi _{pp} \)
has been changed substantially towards bigger values. 

\begin{figure}[htb]
{\par\centering \resizebox*{!}{0.25\textheight}{\includegraphics{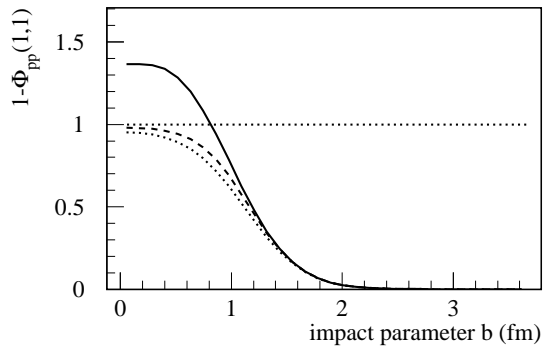}} \par}

\caption{The function \protect\( 1-\Phi _{pp}(1,1,s,b)\protect \) as a function of
the impact parameter \protect\( b.\protect \) We show the exact result \protect\( 1-\Phi _{pp}\protect \)
(solid line) as well as the unitarized one \protect\( 1-\Phi _{\mathrm{u}\, _{pp}}\protect \)
(dashed) and the exponentiated one \protect\( 1-\Phi _{\mathrm{e}\, _{pp}}\protect \)
(dotted).\label{pro3}}
\end{figure}

Knowing \( \Phi _{\mathrm{e}\, _{pp}} \) and \( Z \), we are ready to calculate
the unitarized emission functions \( \Phi _{\mathrm{u}\, _{pp}} \) which finally
replaces \( \Phi _{pp} \) in all formulas for cross section calculations. The
results are shown in fig.\ \ref{pro3}, 
where we plot 1-\( \Phi _{\mathrm{u}\, _{pp}} \) together with \( 1-\Phi _{\mathrm{e}\, _{pp}} \)
and \( 1-\Phi _{pp} \) for both \( x^{+} \) and \( x^{-} \) being one. We
observe that compared to \( 1-\Phi _{\mathrm{e}\, _{pp}} \) the function \( 1-\Phi _{\mathrm{u}\, _{pp}} \)
is somewhat increased at small values of \( b \) due to the fact that here
\( Z(s,b) \) is bigger than one, whereas for large impact parameters there
is no difference.

\begin{figure}[htb]
{\par\centering \resizebox*{!}{0.25\textheight}{\includegraphics{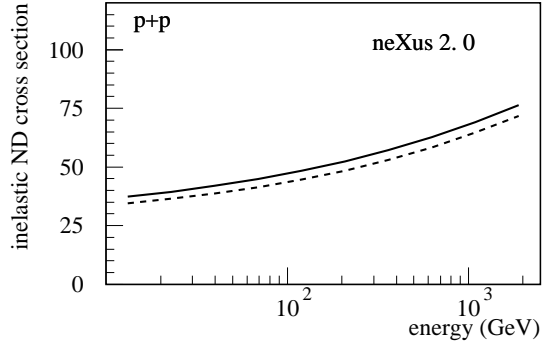}} \par}

\caption{The inelastic non-diffractive cross section as a function of the energy, using
the exact (solid) and the unitarized (dashed) \protect\( \Phi \protect \)-function.\label{sigma}}
\end{figure}

Since the unitarity equation holds, we may integrate \( 1-\Phi _{\mathrm{u}\, _{pp}}(1,1,s,b) \)
over impact parameter, to obtain the inelastic non-diffractive cross section,
\begin{equation}
\label{x}
\sigma _{\mathrm{inel}}(s)=\int d^{2}b\, \left\{ 1-\Phi _{\mathrm{u}\, _{pp}}(1,1,s,b)\right\} ,
\end{equation}
the result being shown in fig.\ \ref{sigma}. Here the exact and the unitarized
result (using \( \Phi _{pp} \) and \( \Phi _{\mathrm{u}\, _{pp}} \) respectively)
are quite close due to the fact that one has a two-dimensional \( b \)-integration,
and therefore the small values of \( b \), where we observe the largest differences,
do not contribute much to the integral. 

\begin{figure}[htb]
{\par\centering \resizebox*{!}{0.25\textheight}{\includegraphics{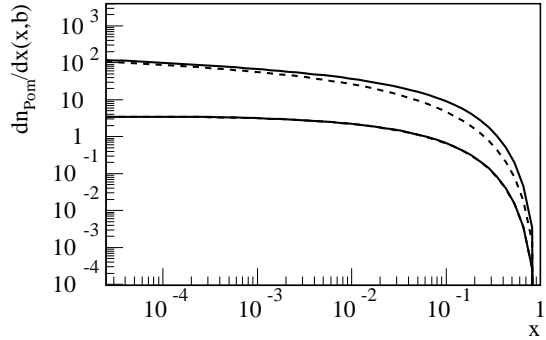}} \par}

\caption{Example of an inclusive spectrum: \protect\( x\protect \)-distribution of
Pomerons. The variable \protect\( x\protect \) is defined as \protect\( x=x^{+}x^{-}\protect \)and
is therefore the squared mass of the Pomeron divided by \protect\( s\protect \).
We show the exact (solid) and the unitarized results (dashed) for \protect\( b=0\protect \)
fm (upper curves) and \protect\( b=1.5\protect \) fm (lower curves). In fact,
for \protect\( b=1.5\protect \) fm the two curves coincide.\label{inc}}
\end{figure}

\begin{figure}[htb]
{\par\centering \resizebox*{!}{0.25\textheight}{\includegraphics{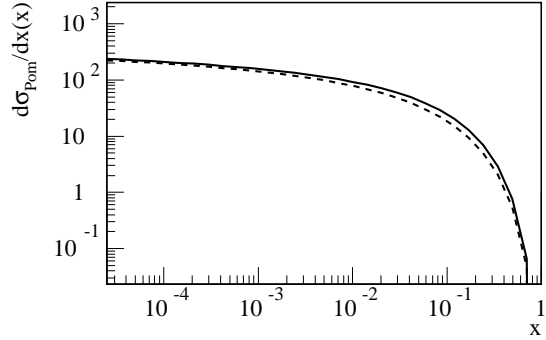}} \par}

\caption{The \protect\( x\protect \)-distribution of Pomerons, averaged over impact
parameter. We show the exact (solid) and the unitarized results (dashed). \label{inci}}
\end{figure}

We now turn to inclusive spectra. We consider the inclusive \( x \)-spectrum
of Pomerons, \( dn_{\mathrm{Pom}}/dx \), where \( x=x^{+}x^{-} \) is the squared
mass of the Pomeron divided by \( s \). In the exact theory, we may take advantage
of the AGK cancelations, and obtain 
\begin{equation}
\label{x}
\frac{dn_{\mathrm{Pom}}}{dx}(x,s,b)=\left. \int _{+\ln \sqrt{x}}^{-\ln \sqrt{x}}\, dy\, \frac{dn_{\mathrm{Pom}}^{(1)}}{dx^{+}dx^{-}}(x^{+},x^{-},s,b)\right| _{x^{+}=\sqrt{x}e^{y},x^{-}=\sqrt{x}e^{-y}},
\end{equation}
where \( dn^{(1)}_{\mathrm{Pom}}/dx^{+}dx^{-} \) is the corresponding inclusive
distribution for one single elementary interaction, which is given in eq.\ (\ref{n1pom}).
The \( y \)-integration can be easily performed numerically, and we obtain
the results shown in fig.\ \ref{inc} as solid curves, the upper one for \( b=0 \)
fm and the lower one for \( b=1.5 \) fm. 
The calculation of the unitarized result is more involved, since now we cannot
use the AGK cancelations any more. We have

\begin{eqnarray}
\frac{dn_{\mathrm{Pom}}}{dx^{+}dx^{-}}(x^{+},x^{-},s,b) & = & \sum ^{\infty }_{m=1}\int \prod ^{m}_{\mu =1}dx^{+}_{\mu }dx^{-}_{\mu }\nonumber \\
 & \times  & \frac{1}{m!}\prod ^{m}_{\mu =1}\left\{ G(x^{+}_{\mu },x^{-}_{\mu },s,b)\right\} \, \Phi _{\mathrm{u}\, _{pp}}(1-\sum _{\mu =1}^{m}x_{\mu }^{+},1-\sum _{\mu =1}^{m}x_{\mu }^{-},s,b)\nonumber \\
 & \times  & \sum _{\mu '=1}^{m}\delta (x^{+}-x^{+}_{\mu '})\delta (x^{-}-x_{\mu '}^{-})\; ,\label{sigma-ine-R} 
\end{eqnarray}
where we used the unitarized version of \( \Phi _{pp} \). We find
\begin{eqnarray}
\frac{dn_{\mathrm{Pom}}}{dx^{+}dx^{-}}(x^{+},x^{-},s,b) & = & G(x^{+},x^{-},s,b)\nonumber \\
 & \times  & \Big [\Phi _{\mathrm{u}\, _{pp}}(1-x^{+},1-x^{-},s,b)\\
 &  & +\int dz^{+}dz^{-}H(z^{+}+x^{+},z^{-}+x^{-})\Phi _{\mathrm{u}\, _{pp}}(z^{+},z^{-},s,b)\Big ],\nonumber \label{x} 
\end{eqnarray}
where \( H \) is defined in eq.\ (\ref{h1}), with the final result given in
eq.\ (\ref{h2}). The integration over \( z^{+},z^{-} \) can now be done numerically.
Expressing \( x^{+} \) and \( x^{-} \) via \( x \) and \( y \) and integrating
over \( y \), we finally obtain \( dn_{\mathrm{Pom}}/dx \), as shown in fig.\ \ref{inc}
(dashed curves). In fig.\ \ref{inci}
we show the b-averaged inclusive spectra, which are given as
\begin{equation}
\label{x}
\frac{1}{\sigma _{\mathrm{inel}}}\int d^{2}b\, \frac{dn_{\mathrm{Pom}}}{dx}(x,s,b)
\end{equation}
 for both, the exact and the unitarized version.

\section{Comparison with the Conventional Approach}

At this point it is noteworthy to compare our approach with the conventional
one \cite{cap94,kai82}. There one neglects the energy conservation effects
in the cross section calculation and sums up virtual Pomeron emissions, each
one taken with the initial energy of the interaction \( s \). We can recover
the conventional approach by simply considering independent (planar) emission
of all the Pomerons, neglecting energy-momentum sharing between them. In the
cross section formulas (\ref{sigmanucl}-\ref{rremnant}) this amounts to perform
formally the convolutions of the Pomeron eikonals \( G(x^{+},x^{-},s,b) \)
with the remnant functions \( F_{\mathrm{remn}}(x^{\pm }) \) for all the Pomerons.
In case of proton-proton scattering, we then get 
\begin{equation}
\label{x}
\Phi _{\mathrm{conv}\, _{pp}}(x^{+},x^{-},s,b)=e^{-\chi (s,b)},
\end{equation}
with
\begin{equation}
\label{eikonal-trad}
\chi (s,b)=\int dx^{+}dx^{-}G(x^{+},x^{-},s,b)\, F_{\mathrm{remn}}(x^{+})\, F_{\mathrm{remn}}(x^{-}),
\end{equation}
 where \( \Phi  \) does not depend on \( x^{+} \) and \( x^{-} \) anymore.
We obtain a unitarity relation of the form
\begin{equation}
\label{x}
\sum _{m=0}^{\infty }\Omega _{m}=1,
\end{equation}
with
\begin{equation}
\label{x}
\Omega _{m}=\frac{\left( \chi (s,b)\right) ^{m}}{m!}e^{-\chi (s,b)}
\end{equation}
representing the probability of having \( m \) cut Pomerons (Pomeron multiplicity
distribution). So in the traditional case, the Pomeron multiplicity distribution
is a Poissonian with the mean value given by \( \chi (s,b) \). As already mentioned
above, that approach is not self-consistent as the AGK rules are assumed to
hold when calculating interaction cross sections but are violated at the particle
production generation. This inconsistency was already mentioned in \cite{bra90},
where the necessity to develop the correct, Feynman diagram-based scheme, was
first argued.

\begin{figure}[htb]
{\par\centering \resizebox*{!}{0.4\textheight}{\includegraphics{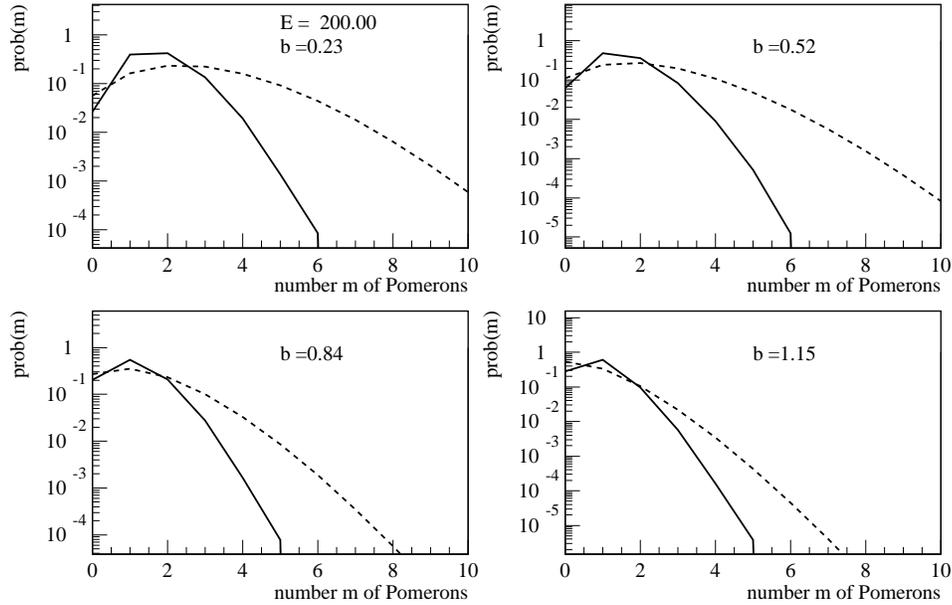}} \par}

\caption{Distribution of the number \protect\( m\protect \) of Pomerons for different
impact parameters. We show the results of a full simulation (solid lines) as
well as the Poissonian distribution obtained by ignoring energy conservation
(dashed line).\label{4-5}}
\end{figure}
The exact procedure is based on the summation over virtual emissions with the
energy-momentum conservation taken into account. This results in the formula
(\ref{r2i}) or, using our parametrization, (\ref{r2exact}) for \( \Phi _{pp}(x^{+},x^{-},s,b) \),
explicitly dependent on the momentum, left after cut Pomerons emission, and
in the formula 
\begin{equation}
\label{sig-inel}
\sigma _{\mathrm{inel}}(s)=\int d^{2}b\left\{ 1-\Phi _{pp}(1,1,s,b)\right\} 
\end{equation}
 for the inelastic cross section; AGK rules are exactly fulfilled both for the
cross sections and for the particle production. But with the interaction energy
increasing the approach starts to violate the unitarity and is no longer self-consistent.

The ''unitarized'' procedure, which amounts to replacing \( \Phi _{pp} \) by
\( \Phi _{\mathrm{u}\, _{pp}} \), allows to avoid the unitarity problems. The
expressions for cross sections and for inclusive spectra are consistent with
each other and with the particle generation procedure. The latter one assures
the AGK cancelations validity in the region, where unitarity problems do not
appear yet (not too high energies or large impact parameters).

In order to see the effect of energy conservation we calculate \( \chi  \)
as given in eq.\ (\ref{eikonal-trad}) with the same parameters as we use in
our approach for different values of \( b \), and we show the corresponding
Pomeron multiplicity distribution in fig.\ \ref{4-5} as dashed lines. We compare
this traditional approach with our full simulation, where energy conservation
is treated properly (solid lines in the figures). One observes a huge difference
between the two approaches. So energy conservation makes the Pomeron multiplicity
distributions much narrower, in other words, the mean number of Pomerons is
substantially reduced. The reason is that due to energy conservation the phase
space of light cone momenta of the Pomeron ends is considerably reduced. 

Of course, in the traditional approach one chooses different parameters in order
to obtain reasonable values for the Pomeron numbers in order to reproduce the
experimental cross sections. But this only ``simulates'' in some sense the
phase space reduction due to energy conservation in an uncontrolled way. 

We conclude that considering energy conservation properly in the cross section
formulas has an enormous effect and cannot be neglected.

\section{Unitarization for Nucleus-Nucleus Scattering}

In this section, we discuss the unitarization scheme for nucleus-nucleus scattering.
The sum over virtual emissions is defined as

\begin{eqnarray}
\Phi _{AB}\left( X^{+},X^{-},s,b\right)  & = & \sum _{l_{1}}\ldots \sum _{l_{AB}}\, \int \, \prod _{k=1}^{AB}\left\{ \prod ^{l_{k}}_{\lambda =1}d\tilde{x}_{k,\lambda }^{+}d\tilde{x}_{k,\lambda }^{-}\right\} \; \prod _{k=1}^{AB}\left\{ \frac{1}{l_{k}!}\, \prod _{\lambda =1}^{l_{k}}-G(\tilde{x}_{k,\lambda }^{+},\tilde{x}_{k,\lambda }^{-},s,b)\right\} \nonumber \label{r} \\
 & \times  & \prod _{i=1}^{A}F_{\mathrm{remn}}\left( x_{i}^{+}-\sum _{\pi (k)=i}\tilde{x}_{k,\lambda }^{+}\right) \prod _{j=1}^{B}F_{\mathrm{remn}}\left( x^{-}_{j}-\sum _{\tau (k)=j}\tilde{x}_{k,\lambda }^{-}\right) .\label{rremnant} 
\end{eqnarray}
 where \( X^{+}=\left\{ x_{1}^{+}\ldots \, x^{+}_{A}\right\}  \), \( X^{-}=\left\{ x_{1}^{-}\ldots \, x^{-}_{B}\right\}  \)
and \( \pi (k) \) and \( \tau (k) \) represent the projectile or target nucleon
linked to pair \( k \). This calculation is very close to the calculation for
proton-proton scattering. Using the expression eq.\ (\ref{gi}) of \( G(\tilde{x}_{k,\lambda }^{+},\tilde{x}_{k,\lambda }^{-},s,b) \),
the definition eq.\ (\ref{Fremn}) of \( F_{\mathrm{remn}}(x) \), one finally
finds

\begin{eqnarray}
 &  & \Phi _{AB}\left( X^{+},X^{-},s,b\right) =\nonumber \\
 &  & \sum _{r_{1,1}\ldots r_{N,1}}\cdots \, \sum _{r_{1,AB}\ldots r_{N,AB}}\, \prod _{k=1}^{AB}\frac{\left( -\alpha _{1}\right) ^{r_{1,k}}}{r_{1,k}!}\ldots \frac{\left( -\alpha _{N}\right) ^{r_{N,k}}}{r_{N,k}!}\label{r2abfinal} \\
 &  & \prod ^{A}_{i=1}\, (x_{i}^{+})^{\alpha _{\mathrm{remn}}}\, \prod _{\pi (k)=i}\left( \Gamma (\tilde{\beta }_{1})(x_{i}^{+})^{\tilde{\beta }_{1}}\right) ^{r_{1,k}}\ldots \left( \Gamma (\tilde{\beta }_{N})(x_{i}^{+})^{\tilde{\beta }_{N}}\right) ^{r_{N,k}}g\left( \sum _{\pi (k)=i}r_{1,k}\tilde{\beta }_{1}+\ldots +r_{N,k}\tilde{\beta }_{N}\right) \nonumber \\
 &  & \prod ^{B}_{j=1}\, (x_{j}^{-})^{\alpha _{\mathrm{remn}}}\, \prod _{\tau (k)=j}\left( \Gamma (\tilde{\beta }_{1})(x_{j}^{-})^{\tilde{\beta }_{1}}\right) ^{r_{1,k}}\ldots \left( \Gamma (\tilde{\beta }_{N})(x_{j}^{-})^{\tilde{\beta }_{N}}\right) ^{r_{N,k}}g\left( \sum _{\tau (k)=j}r_{1,k}\tilde{\beta }_{1}+\ldots +r_{N,k}\tilde{\beta }_{N}\right) \nonumber 
\end{eqnarray}
 (see appendix \ref{axd3}), where the function \( g(z) \) is defined in eq.\
(\ref{gammaq}), and the parameters \( \alpha _{i} \) and \( \tilde{\beta }_{i} \)
are the same ones as for proton-proton scattering.

In case of nucleus-nucleus scattering, we use the same unitarization prescription
as already applied to proton-proton scattering. The first step amounts to replace
the function \( g(z) \), which appears in the final expression of \( \Phi _{AB} \),
by the exponential form \( g_{\mathrm{e}}(z) \). This allows to perform the
sums in eq.\ (\ref{r2abfinal}), and we obtain

\begin{eqnarray}
\Phi _{\mathrm{e}\, _{AB}}\left( X^{+},X^{-},s,b\right)  & = & \prod ^{A}_{i=1}\, (x_{i}^{+})^{\alpha _{\mathrm{remn}}}\, \prod ^{B}_{j=1}\, (x_{j}^{-})^{\alpha _{\mathrm{remn}}}\prod ^{AB}_{k=1}e^{-2\tilde{G}\left( x_{\pi (k)}^{+}.x^{-}_{\tau (k)}\right) }\label{x} 
\end{eqnarray}
 (see appendix \ref{axd3}), where \( \tilde{G}(x) \) is defined in eq.\ (\ref{G-tilde}).
Having modified \( \Phi  \), the unitarity equation
\begin{equation}
\label{x}
\sum _{m}\int dX^{+}dX^{-}\Omega _{\mathrm{e}\, _{AB}}^{(s,b)}(m,X^{+},X^{-})=1
\end{equation}
does not hold any more, since \( \Omega _{\mathrm{e}} \) depends on \( \Phi _{\mathrm{e}} \),
and only the exact \( \Phi  \) assures a correct unitarity relation. So as
in proton-proton scattering, we need a second step, which amounts to renormalizing
\( \Phi _{\mathrm{e}} \). So we introduce a normalization factor
\begin{equation}
\label{x}
Z_{AB}(s,b)=\sum _{m}\int dX^{+}dX^{-}\Omega _{\mathrm{e}\, _{AB}}^{(s,b)}(m,X^{+},X^{-}),
\end{equation}
with \( \Omega _{\mathrm{e}\, _{AB}} \) defined in the same way as \( \Omega  \)
but with \( \Phi _{AB} \) replaced by \( \Phi _{\mathrm{e}\, _{AB}} \), which
allows to define the unitarized \( \Phi _{\mathrm{u}\, _{AB}} \) function as
\begin{equation}
\label{x}
\Phi _{\mathrm{u}\, _{AB}}(x^{+},x^{-},s,b)=\frac{\Phi _{\mathrm{e}\, _{AB}}(x^{+},x^{-},s,b)}{Z_{AB}(s,b)}.
\end{equation}
In this way we recover the unitarity relation,
\begin{equation}
\label{x}
\sum _{m}\int dX^{+}dX^{-}\Omega _{\mathrm{u}\, _{AB}}^{(s,b)}(m,X^{+},X^{-})=1,
\end{equation}
with
\begin{equation}
\label{x}
\Omega _{\mathrm{u}\, _{AB}}^{(s,b)}(m,X^{+},X^{-})=\prod _{k=1}^{AB}\left\{ \frac{1}{m_{k}!}\, \prod _{\mu =1}^{m_{k}}G(s,x_{k,\mu }^{+},x_{k,\mu }^{-},b)\right\} \; \Phi _{\mathrm{u}\, _{AB}}\left( x^{+},x^{-},s,b\right) ,
\end{equation}
and \( \Omega _{\mathrm{u}}^{(s,b)}(m,X^{+},X^{+}) \) may be interpreted as
probability distribution for configurations \( (m,X^{+},X^{+}) \).

\section{Profile Functions in Nucleus-Nucleus Scattering}

In case of nucleus-nucleus scattering, the conventional approach \cite{cap94,kai82}
represents a ``Glauber-type model'', where nucleus-nucleus scattering may
be considered as a sequence of nucleon-nucleon scatterings with constant cross
sections; the nucleons move through the other nucleus along straight line trajectories.
In order to test this picture, we consider all pairs of nucleons, which due
to their distributions inside the nuclei provide a more or less flat \( b \)-distribution. 
\begin{figure}[htb]
{\par\centering \resizebox*{!}{0.25\textheight}{\includegraphics{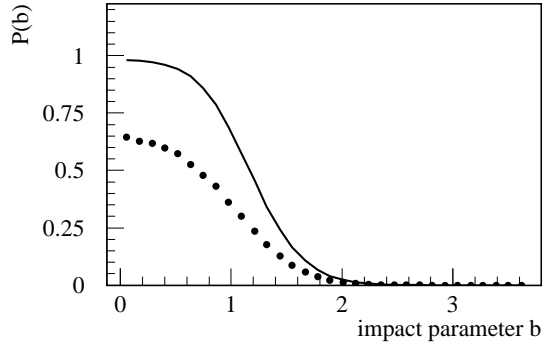}} \par}

\caption{The numerically determined profile function in S+S scattering (points) compared
to the proton-proton profile function (solid curve).\label{4-6}}
\end{figure}
We then simply count, for a given \( b \)-bin, the number of interacting pairs
and then divide by the number of pairs in the corresponding bin. The resulting
distribution, which we call nucleon-nucleon profile function for nucleus-nucleus
scattering, represents the probability density of an interaction of a pair of
nucleons at given impact parameter. This may be compared with the proton-proton
profile function \( 1-\Phi _{\mathrm{u}\, _{pp}}(1,1,s,b) \). In the Glauber
model, these two distributions coincide. As demonstrated in fig.\ref{4-6} for
S+S scattering this is absolutely not the case. The profile function in case
of S+S scattering is considerably reduced as compared to the proton-proton one.
Since integrating the proton-proton profile function represents the inelastic
cross section, one may also define the corresponding integral in nucleus-nucleus
scattering as ``individual nucleon-nucleon cross section''. So we conclude
that this cross section is smaller than the proton-proton cross section. This
is due to the energy conservation, which reduces the number of Pomerons connected
to any nucleon from the projectile and the target and finally affects also the
``individual cross section''.

\section{Inclusive Cross Sections in Nucleus-Nucleus Scattering}

We have shown in the preceding chapter that in the ``bare'' theory AGK cancelations
apply perfectly, which means that nucleus-nucleus inclusive cross sections are
just \( AB \) times the proton-proton ones,
\begin{equation}
\label{x}
\frac{d\sigma ^{AB}_{\mathrm{incl}}}{dq}(q,s,b)=AB\frac{d\sigma ^{pp}_{\mathrm{incl}}}{dq}(q,s,b).
\end{equation}

In the unitarized theory, the results is somewhat different. Unfortunately,
we cannot calculate cross sections analytically any more, so we perform a numerical
calculation using the Markov chain techniques explained later. In order to investigate
the deviation from exact AGK cancelations, we calculate 
\begin{figure}[htb]
{\par\centering \resizebox*{!}{0.25\textheight}{\includegraphics{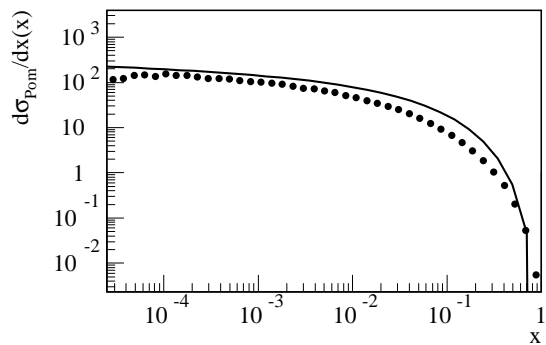}} \par}

\caption{Inclusive cross section of Pomeron production for S+S scattering, divided by
\protect\( AB=32^{2}\protect \) (points), compared to the corresponding proton-proton
cross section (solid line).\label{4-7}}
\end{figure}
\begin{figure}[htb]
{\par\centering \resizebox*{!}{0.25\textheight}{\includegraphics{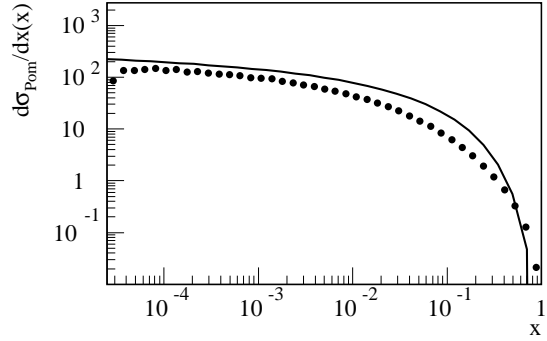}} \par}

\caption{Inclusive cross section of Pomeron production for p+ Au scattering, divided
by \protect\( AB=197\protect \) (points), compared to the corresponding proton-proton
cross section (solid line).\label{4-8}}
\end{figure}
the inclusive nucleus-nucleus cross section for Pomeron production (being the
basic inclusive cross section), divided by \( AB \),
\begin{equation}
\label{x}
\frac{1}{AB}\frac{d\sigma ^{AB}_{\mathrm{Pom}}}{dx}(x,s,b),
\end{equation}
and compare the result with the corresponding proton-proton cross section, see
figs. \ref{4-7} and \ref{4-8}. For large and for small values of \( x \),
we still observe AGK cancelations (the two curves agree), but for intermediate
values of \( x \), the AGK cancelations are violated, the nucleus-nucleus cross
section is smaller than \( AB \) times the nucleon-nucleon one. The effect
is, however, relatively moderate. If one writes the proton-nucleus cross section
as
\begin{equation}
\label{x}
\frac{d\sigma ^{pA}_{\mathrm{Pom}}}{dx}(x,s,b)=A^{\alpha (x)}\frac{d\sigma ^{pp}_{\mathrm{Pom}}}{dx},
\end{equation}
we obtain for \( \alpha (x) \) values between \( 0.85 \) and \( 1 \). So
one may summarize that ``AGK cancelations are violated, but not too strongly''.

\vfill

\cleardoublepage

\chapter{Markov Chain Techniques}

In this chapter we discuss how to deal with the multidimensional probability
distribution \( \Omega _{AB}(K) \) with \( K=\{m,X^{+},X^{-}\} \), where the
vector \( m \) characterizes the type of interaction of each pair of nucleons
(the number of elementary interactions per pair), and the matrices \( X^{+} \),
\( X^{-} \) contain the light cone momenta of all Pomerons (energy sharing
between the Pomerons).

\section{Probability Distributions for Configurations}

In this section we essentially repeat the basic formulas of the preceding chapters
which allowed us to derive probability distributions for interaction configurations
in a consistent way within an effective theory based on Feynman diagrams. 

Our basic formula for the inelastic cross section for a nucleus-nucleus collision
(which includes also as a particular case proton-proton scattering) could be
written in the following form
\begin{equation}
\label{x}
\sigma _{\mathrm{inel}}(s)=\int d^{2}b_{0}\int dT_{AB}\, \gamma _{AB}(s,b_{0},b_{1}\ldots b_{AB}),
\end{equation}
where \( dT_{AB} \) represents the integration over the transverse coordinates
\( b_{i}^{A} \) and \( b_{j}^{B} \) of projectile and target nucleons, \( b_{0} \)
is the impact parameter between the two nuclei, and \( b_{k}=|\vec{b}+\vec{b}_{\pi (k)}^{A}-\vec{b}_{\tau (k)}^{B}| \)
is the transverse distance between the nucleons of \( k^{th} \) pair. Using
the compact notation
\begin{equation}
\label{x}
b=\{b_{k}\},\quad m=\{m_{k}\},\quad X^{+}=\left\{ x_{k,\mu }^{+}\right\} ,\quad X^{-}=\left\{ x_{k,\mu }^{-}\right\} ,
\end{equation}
the function \( \gamma _{AB} \) is given as
\begin{equation}
\label{gam-inel-markov}
\gamma _{AB}(s,b)=\sum _{m}(1-\delta _{0m})\int dX^{+}dX^{-}\Omega _{\mathrm{u}_{AB}}^{(s,b)}(m,X^{+},X^{-}),
\end{equation}
which represents all diagrams with at least one cut Pomeron. One may define
a corresponding quantity \( \bar{\gamma }_{AB} \), which represents the configuration
with exactly zero cut Pomerons. The latter one can be obtained from (\ref{gam-inel-markov})
by exchanging \( 1-\delta _{0m} \) by \( \delta _{0m} \), which leads to
\begin{equation}
\label{x}
\bar{\gamma }_{AB}(s,b)=\Omega ^{(s,b)}_{\mathrm{u}_{AB}}(0,0,0).
\end{equation}
The expression for \( \Omega  \) is given as
\begin{equation}
\label{x}
\Omega _{\mathrm{u}_{AB}}^{(s,b)}(m,X^{+},X^{-})=\prod _{k=1}^{AB}\left\{ \frac{1}{m_{k}!}\, \prod _{\mu =1}^{m_{k}}G(x_{k,\mu }^{+},x_{k,\mu }^{-},s,b_{k})\right\} \; \Phi _{\mathrm{u}_{\mathrm{AB}}}\left( x^{+},x^{-},s,b\right) ,
\end{equation}
with
\begin{eqnarray}
\Phi _{\mathrm{u}_{AB}}(x^{+},x^{-},s,b) & = & \frac{1}{Z_{AB}}\prod _{k=1}^{AB}\exp \left( -\tilde{G}(x^{+}_{\pi (k)}x^{-}_{\tau (k)},s,b_{k})\right) \\
 & \times  & \prod _{i=1}^{A}(x^{+}_{i})^{\alpha _{\mathrm{remn}}}\, \Theta (x_{i}^{+})\, \Theta (1-x_{i}^{+})\, \prod _{j=1}^{B}(x_{j}^{-})^{\alpha _{\mathrm{remn}}}\, \Theta (x_{j}^{-})\, \Theta (1-x_{j}^{-}).\nonumber \label{x} 
\end{eqnarray}
  The arguments of \( \Phi _{\mathrm{u}_{AB}} \) are the momentum fractions
of projectile and target remnants,

\begin{equation}
\label{x}
x_{i}^{+}=1-\sum _{\pi (k)=i}x_{k,\mu \, ,}^{+}\quad x^{-}_{j}=1-\sum _{\tau (k)=j}x_{k,\mu }^{-}\, ,
\end{equation}
where \( \pi (k) \) and \( \tau (k) \) point to the remnants linked to the
\( k^{\mathrm{th}} \) interaction. 

In the following, we perform the analysis for given \( s \) and \( b=(b_{0},b_{1},...,b_{AB}) \),
so we do not write these variables explicitly. In addition, we always refer
to the unitarized functions, so we will also suppress the subscript ``u''.
Furthermore, we suppress the index \( AB \). 

Crucial for our applications is the probability conservation constraint
\begin{equation}
\label{x}
\gamma +\bar{\gamma }=1,
\end{equation}
which may be written more explicitly as 
\begin{equation}
\label{x}
\sum _{m}\int dX^{+}dX^{-}\Omega (m,X^{+},X^{-})=1.
\end{equation}
This allows us to interpret \( \Omega (m,X^{+},X^{-}) \) as the probability
distribution for a configuration \( (m,X^{+},X^{-}) \). For any given configuration
the function \( \Omega  \) can be easily calculated using the techniques developed
in the chapter 2. The difficulty with the Monte Carlo generation of interaction
configurations arises from the fact that the configuration space is huge and
rather nontrivial in the sense that it cannot be written as a product of single
Pomeron contributions. We are going to explain in the next sections, how we
deal with this problem.

\section{The Interaction Matrix}

Since \( \Omega (m,X^{+},X^{-}) \) is a high-dimensional and nontrivial probability
distribution, the only way to proceed amounts to employing dynamical Monte Carlo
methods, well known in statistical and solid state physics. 

\begin{figure}[htb]
{\par\centering \resizebox*{0.3\textwidth}{!}{\includegraphics{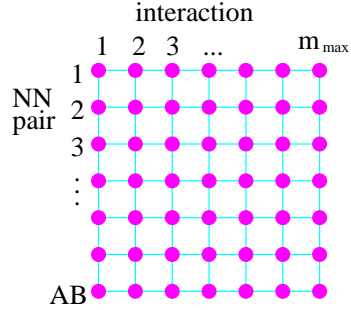}} \par}

\caption{The interaction lattice.\label{lattice}}
\end{figure}

We first need to choose the appropriate framework for our analysis. So we translate
our problem into the language of spin systems \cite{hla98}: we number all nucleon
pairs as 1, 2, ..., \( AB \) and for each nucleon pair \( k \) the possible
elementary interactions as 1,2, ..., \( m_{k\cdot } \) Let \( m_{\mathrm{max}} \)
be the maximum number of elementary interactions per nucleon pair one may imagine.
We now consider a two dimensional lattice with \( AB \) lines and \( m_{\mathrm{max}} \)
columns, see fig.\ \ref{lattice}. Lattice sites are occupied \( \left( =1\right)  \)
or empty \( \left( =0\right)  \), representing an elementary interaction (1)
or the case of no interaction (0), for the \( k^{th} \) pair.
In order to represent \( m_{k} \) elementary interactions for the pair \( k \),
we need \( m_{k} \) occupied cells (1's) in the \( k^{th} \) line. A line
containing only empty cells (0's) represents a pair without interaction. Any
possible interaction may be represented by this ``interaction matrix'' \( M \)
with elements
\begin{equation}
\label{x}
m_{k\mu }\in \left\{ 0,1\right\} .
\end{equation}
 Such an ``interaction configuration'' is exactly equivalent to a spin configuration
of the Ising model. Unfortunately the situation is somewhat more complicated
in case of nuclear collisions: we need to consider the energy available for
each elementary interaction, represented via the momentum fractions \( x^{+}_{k\mu } \)
and \( x^{-}_{k\mu } \). So we have a ``generalized'' matrix \( K \), 
\begin{equation}
\label{x}
K=(M,X^{+},X^{-}),
\end{equation}
representing an interaction configuration, with elements
\begin{equation}
\label{x}
K_{k\mu }=\left( m_{k\mu ,}x^{+}_{k\mu },x_{k\mu }^{-}\right) .
\end{equation}
It is important to note that a number of matrices \( M \) represents one and
the same vector \( m \). In fact, \( m \) is represented by all the matrices
\( M \) with
\begin{equation}
\label{x}
\sum _{\mu =1}^{m_{\mathrm{max}}}m_{k\mu }=m_{k},
\end{equation}
 for each \( k \). Since all the corresponding configurations \( (M,X^{+},X^{-}) \)
should have the same weight, and since there are
\begin{equation}
\label{x}
c=\prod _{k=1}^{AB}\frac{m_{\mathrm{max}}!}{m_{k}!(m_{\mathrm{max}}-m_{k})!}
\end{equation}
 configurations \( (M,X^{+},X^{-}) \) representing the same configuration \( (m,X^{+},X^{-}) \),
the weight for the former is \( c^{-1} \) times the weight for the latter,
so we obtain the following probability distribution for \( K=(M,X^{+},X^{-}) \):
\begin{equation}
\label{x}
\Omega (K)=\prod _{k=1}^{AB}\left\{ \frac{(m_{\mathrm{max}}-m_{k})!}{m_{\mathrm{max}}!}\, \prod _{\mu =1}^{m_{k}}G(s,x_{k,\mu }^{+},x_{k,\mu }^{-},b)\right\} \; \Phi _{\mathrm{u}_{AB}}\left( x^{+},x^{-},s,b\right) ,
\end{equation}
 or, using the expression for \( \Phi _{\mathrm{u}_{AB}} \), 
\begin{eqnarray}
\Omega (K) & = & \frac{1}{Z_{AB}}\prod _{k=1}^{AB}\left\{ \frac{(m_{\mathrm{max}}-m_{k})!}{m_{\mathrm{max}}!}\, \prod _{\mu =1}^{m_{k}}\left\{ G(s,x_{k,\mu }^{+},x_{k,\mu }^{-},b_{k})\right\} \, \exp \left( -\tilde{G}(x^{+}_{\pi (k)}x^{-}_{\tau (k)},s,b_{k})\right) \right\} \nonumber \\
 &  & \times \prod _{i=1}^{A}(x^{+}_{i})^{\alpha _{\mathrm{remn}}}\, \Theta (x_{i}^{+})\, \Theta (1-x_{i}^{+})\, \prod _{j=1}^{B}(x_{j}^{-})^{\alpha _{\mathrm{remn}}}\, \Theta (x_{j}^{-})\, \Theta (1-x_{j}^{-}).\label{x} 
\end{eqnarray}
 The probability conservation now reads
\begin{equation}
\label{x}
\sum _{K}\Omega (K)=\sum _{M}\int dX^{+}dX^{-}\Omega (M,X^{+},X^{-})=1.
\end{equation}
In the following, we shall deal with the ``interaction matrix'' \( K \),
and the probability distribution \( \Omega (K) \).

\section{The Markov Chain Method}

In order to generate \( K \) according to the given distribution \( \Omega \left( K\right)  \),
defined earlier, we construct a Markov chain
\begin{equation}
\label{x}
K^{\left( 0\right) },K^{\left( 1\right) },K^{\left( 2\right) },...K^{\left( t_{\mathrm{max}}\right) }
\end{equation}
 such that the final configurations \( K^{\left( t_{\mathrm{max}}\right) } \)
are distributed according to the probability distribution \( \Omega \left( K\right)  \),
if possible for a \( t_{\mathrm{max}} \) not too large! 

Let us discuss how to obtain a new configuration \( K^{(t+1)}=L \) from a given
configuration \( K^{(t)}=K \). We use Metropolis' Ansatz for the transition
probability
\begin{equation}
\label{x}
p(K,L)=prob\left( K^{\left( t+1\right) }=L\: \right| \left. K^{\left( t\right) }=K\right) 
\end{equation}
 as a product of a proposition matrix \( w(K,L) \) and an acceptance matrix
\( u(K,L) \):
\begin{equation}
\label{x}
p(K,L)=\left\{ \begin{array}{lll}
w(K,L)\, u(K,L) & \mathrm{if} & L\neq K\\
w(K,K)+\sum _{L\neq K}w(K,L)\{1-u(K,L)\} & \mathrm{if} & L=K
\end{array}\right. ,
\end{equation}
where we use 
\begin{equation}
\label{x}
u(K,L)=\min \left( \frac{\Omega (L)}{\Omega (K)}\frac{w(L,K)}{w(K,L)},1\right) ,
\end{equation}
 in order to assure detailed balance. We are free to choose \( w(K,L) \), but
of course, for practical reasons, we want to minimize the autocorrelation time,
which requires a careful definition of \( w \). An efficient procedure requires
\( u(K,L) \) to be not too small (to avoid too many rejections), so an ideal
choice would be \( w\left( K,L\right) =\Omega \left( L\right)  \). This is
of course not possible, but we choose \( w(K,L) \) to be a ``reasonable''
approximation to \( \Omega (L) \) if \( K \) and \( L \) are reasonably close,
otherwise \( w \) should be zero. So we define 
\begin{equation}
\label{x}
w(K,L)=\left\{ \begin{array}{ccc}
\Omega _{0}(L) & \mathrm{if} & d(K,L)\leq 1\\
0 & \mathrm{otherwise} & 
\end{array}\right. ,
\end{equation}
 where \( d(K,L) \) is the number of lattice sites being different in \( L \)
compared to \( K, \) and where \( \Omega ^{(0)} \) is defined by the same
formulas as \( \Omega  \) with one exception : \( \Phi _{\mathrm{u}_{AB}} \)
is replaced by \( 1 \). So we get
\begin{equation}
\label{omega0}
\Omega _{0}(L)\sim \prod _{k=1}^{AB}\left\{ (m_{\mathrm{max}}-m_{k})!\, \prod _{\mu =1}^{m_{k}}G(x_{k,\mu }^{+},x_{k,\mu }^{-},s,b_{k})\right\} .
\end{equation}
The above definition of \( w(K,L) \) may be realized by the following algorithm:

\begin{itemize}
\item choose randomly a lattice site \( (k,\mu ) \),
\item propose a new matrix element \( (m_{k\mu },x^{+}_{k\mu },x^{-}_{k,\mu }) \)
according to the probability distribution \( \rho (m_{k\mu },x^{+}_{k\mu },x^{-}_{k,\mu }) \),
\end{itemize}
where we are going to derive the form of \( \rho  \) in the following. From
eq.\ (\ref{omega0}), we know that \( \rho  \) should be of the form
\begin{equation}
\label{x}
\rho (m,x^{+},x^{-})\sim m_{0}!\, \left\{ \begin{array}{ccc}
G(x^{+},x^{-},s,b) & \mathrm{if} & m=1\\
1 & \mathrm{if} & m=0
\end{array}\right. ,
\end{equation}
where \( m_{0}=m_{\mathrm{max}}-m \) is the number of zeros in the row \( k \).
Let us define \( \bar{m}_{0} \) as the number of zeros (empty cells) in the
row \( k \) not counting the current site \( (k,\mu ) \). Then the factor
\( m_{0}! \) is given as \( \bar{m}_{0}! \) in case of \( m\neq 0 \) and
as \( \bar{m}_{0}!(\bar{m}+1) \) in case of \( m=0 \), and we obtain
\begin{equation}
\label{x}
\rho (m,x^{+},x^{-})\sim (\bar{m}_{0}+1)\delta _{m0}+G(x^{+},x^{-},s,b)\delta _{m1}.
\end{equation}
Properly normalized, we obtain
\begin{equation}
\label{x}
\rho (m,x^{+},x^{-})=p_{0}\, \delta _{m0}+(1-p_{0})\, \frac{G(x^{+},x^{-},s,b)}{\chi }\, \delta _{m1},
\end{equation}
where the probability \( p_{0} \) of proposing no interaction is given as
\begin{equation}
\label{x}
p_{0}=\frac{\bar{m}_{0}+1}{\bar{m}_{0}+1+\chi (s,b)},
\end{equation}
 with \( \chi  \) being obtained by integrating \( G \) over \( x^{+} \)
and \( x^{-} \), 
\begin{equation}
\label{x}
\chi (s,b)=\int ^{1}_{0}dx^{+}dx^{-}G(x^{+},x^{-},s,b).
\end{equation}
Having proposed a new configuration \( L \), which amounts to generating the
values \( m_{k\mu },x_{k\mu }^{+},x_{k\mu }^{-} \) for a randomly chosen lattice
site as described above, we accept this proposal with the probability
\begin{equation}
\label{x}
u(K,L)=\min \left( z_{1}z_{2},1\right) ,
\end{equation}
with 
\begin{equation}
\label{x}
z_{1}=\frac{\Omega (L)}{\Omega (K)},\quad z_{2}=\frac{w(L,K)}{w(K,L)}.
\end{equation}
Since \( K \) and \( L \) differ in at most one lattice site, say \( (k,\mu ) \),
we do not need to evaluate the full formula for the distribution \( \Omega  \)
to calculate \( z_{1} \), we rather calculate 
\begin{equation}
\label{x}
z_{1}=\frac{\Omega ^{k\mu }(L)}{\Omega ^{k\mu }(K)},
\end{equation}
with
\begin{eqnarray}
\Omega ^{k\mu }(K) & = & \rho (m_{k\mu },x_{k\mu }^{+},x_{k\mu }^{-})\exp \left( -\sum _{l\, \mathrm{linked}\, \mathrm{to}\, k}\tilde{G}(x^{+}_{\pi (l)}x^{-}_{\tau (l)},s,b_{l})\right) \nonumber \\
 & \times  & (x^{+}_{\pi (k)})^{\alpha _{\mathrm{remn}}}\Theta (x_{\pi (k)}^{+})\Theta (1-x_{\pi (k)}^{+})\, (x^{-}_{\tau (k)})^{\alpha _{\mathrm{remn}}}\Theta (x_{\tau (k)}^{-})\Theta (1-x_{\tau (k)}^{-}),\label{x} 
\end{eqnarray}
which is technically quite easy. Our final task is the calculation of the asymmetry
\( z_{2} \). In many applications of the Markov chain method one uses symmetric
proposal matrices, in which case this factor is simply one. This is not the
case here. We have
\begin{equation}
\label{x}
z_{2}=\frac{\Omega _{0}(K)}{\Omega _{0}(L)}=\frac{\Omega _{0}^{k\mu }(K)}{\Omega ^{k\mu }_{0}(L)},
\end{equation}
with
\begin{equation}
\label{x}
\Omega _{0}^{k\mu }(K)=\rho (m_{k\mu },x_{k\mu }^{+},x_{k\mu }^{-}),
\end{equation}
which is also easily calculated. So we accept the proposal \( L \) with the
probability \( \min (z_{1}z_{2},1) \), in which case we have \( K^{(t+1)}=L \),
otherwise we keep the old configuration \( K \), which means \( K^{(t+1)}=K \).

\section{Convergence}

\begin{figure}[htb]
{\par\centering \resizebox*{!}{0.6\textheight}{\includegraphics{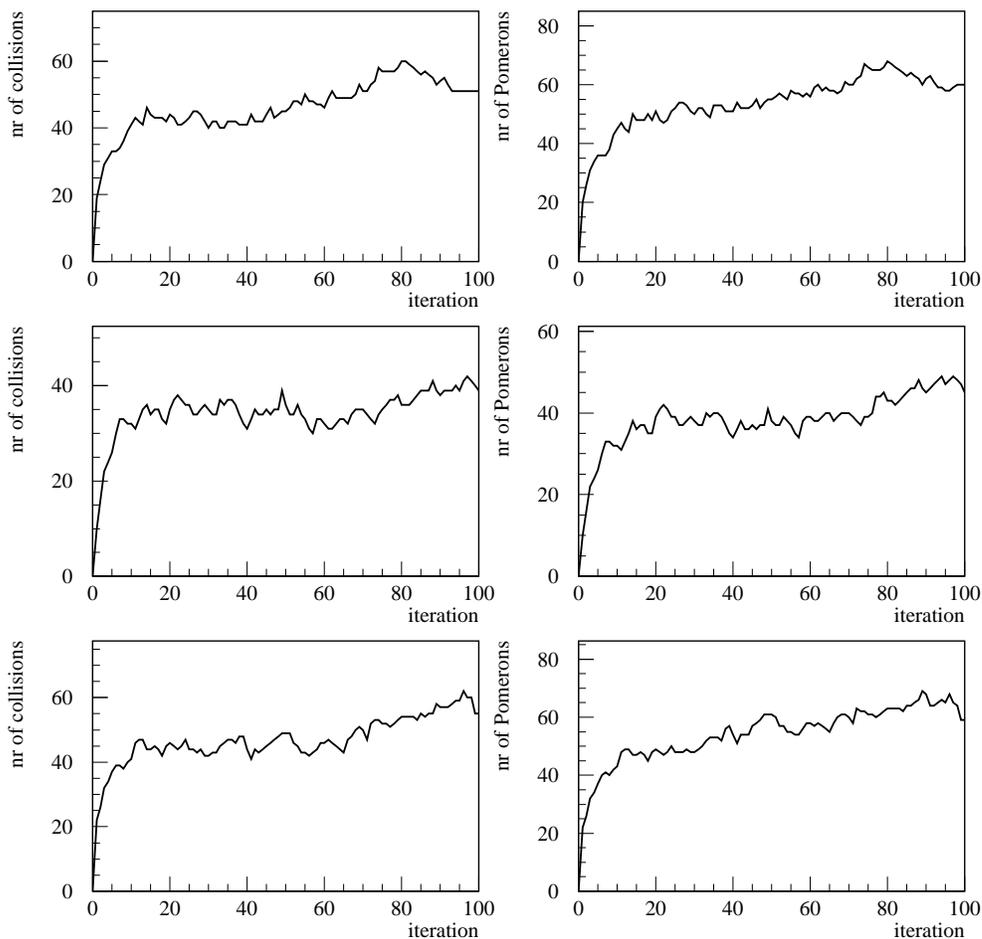}} \par}

\caption{Number of collisions (left) and number of Pomerons (right) as a function of
the iteration step \protect\( t\protect \) (in percent relative to the maximum
number of iterations) for three different S+S events. \label{4-1}}
\end{figure}

A crucial item is the question of how to determine the number of iterations,
which are sufficient to reach the stationary region. In principle one could
calculate the autocorrelation time, or better one could estimate it based on
an actual iteration. One could then multiply it with some ``reasonable number'',
between 10 and 20, in order to obtain the number of iterations. Since this ``reasonable
number'' is not known anyway, we proceed differently. We consider a number
of quantities like the number of binary interactions, the number of Pomerons,
and other observables, and we monitor their values during the iterations. Simply
by inspecting the results for many events, one can quite easily convince oneself
if the numbers of iterations are sufficiently large. As a final check one makes
sure that the distributions of some relevant observables do not change by doubling
the number of iterations. In fig.\ \ref{4-1},
we show the number of collisions (left) and the number of Pomerons (right) as
a function of the iteration step \( t \) for a S+S collision, where the number
of iterations \( t_{\mathrm{max}} \) has been determined according to some
empirical procedure described below. We observe that these two quantities approach
very quickly the stationary region. In order to determine the number \( t_{\mathrm{max}} \)
of iterations for a given reaction \( A+B \), we first calculate the upper
limit for the number of possibly interacting nucleon pairs as the number of
pairs \( k_{\mathrm{max}} \) with a transverse distance smaller than some value
\( b_{\mathrm{max}} \) being defined as
\begin{equation}
\label{x}
1-e^{-\chi (s,b_{\mathrm{max}})}=0.001\: ,
\end{equation}
and we then define 
\begin{equation}
\label{x}
t_{\mathrm{max}}=100\cdot \frac{2}{3}k_{\mathrm{max}\: .}
\end{equation}
Actually, in the real calculations, we never consider sums of nucleon pairs
from 1 to \( AB \), but only from 1 to \( k_{\mathrm{max}} \), because for
the other ones the chance to be involved in an interaction is so small that
one can safely ignore it.

\section{Some Tests for Proton-Proton Scattering}

\begin{figure}[htb]
{\par\centering \resizebox*{!}{0.25\textheight}{\includegraphics{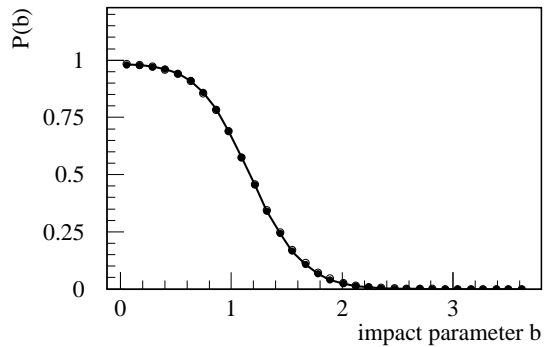}} \par}

\caption{The simulated probability \protect\( P(b)\protect \) (points) of an inelastic
interaction for a proton-proton collision at \protect\( \sqrt{s}=200\protect \)
GeV compared to the profile function \protect\( \gamma _{AB}(s,b)\protect \)
(solid line).\label{4-2}}
\end{figure}

As a first test, we check whether the Monte Carlo procedure reproduces the theoretical
profile function \( \gamma _{\mathrm{inel}} \). So we make a large number of
simulations of proton-proton collisions at a given energy \( \sqrt{s} \), where
the impact parameters are chosen randomly between zero and the earlier defined
maximum impact parameter \( b_{\mathrm{max}} \). We then count simply the number
\( \Delta n_{\mathrm{inel}}(b) \) of inelastic interactions in a given impact
parameter bin \( [b-\Delta b/2,b+\Delta b/2] \), and divide this by the number
\( \Delta n_{\mathrm{tot}}(b) \) of simulations in this impact parameter interval.
Since the total number \( \Delta n_{\mathrm{tot}}(b) \) of simulated configurations
for the given \( b- \)bin splits into the number \( \Delta n_{\mathrm{inel}}(b) \)
of ``interactions'' and the number \( \Delta n_{\mathrm{nonint}}(b) \) of
``non-interactions'', with \( \Delta n_{\mathrm{tot}}(b)= \)\( \Delta n_{\mathrm{inel}}(b)+\Delta n_{\mathrm{nonint}}(b) \),
the result
\begin{equation}
\label{x}
P(b)=\frac{\Delta n_{\mathrm{inel}}(b)}{\Delta n_{\mathrm{tot}}(b)}
\end{equation}
represents the probability to have an interaction at a given impact parameter
\( b \), which should coincide with the profile function 
\begin{equation}
\label{x}
\gamma _{AB}(s,b)=1-\Phi _{\mathrm{u}_{AB}}(1,1,s,b)
\end{equation}
for the corresponding energy. In fig.\ \ref{4-2}, 
 we compare the two quantities for a proton-proton collision at \( \sqrt{s}=200 \)
GeV and we find an excellent agreement, as it should be.

\begin{figure}[htb]
{\par\centering \resizebox*{!}{0.25\textheight}{\includegraphics{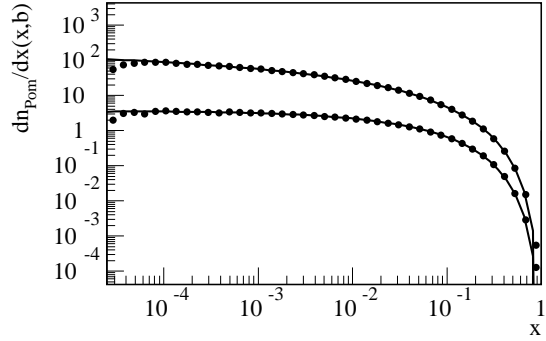}} \par}

\caption{Inclusive \protect\( x\protect \)-distribution of Pomerons. The variable \protect\( x\protect \)
is defined as \protect\( x=x^{+}x^{-}\protect \)and is therefore the squared
mass of the Pomeron divided by \protect\( s\protect \). We show unitarized
analytical results (solid lines) for \protect\( b=0\protect \) fm (upper curves)
and \protect\( b=1.5\protect \) fm (lower curves) and the corresponding simulation
results (points).\label{4-3}}
\end{figure}

\begin{figure}[htb]
{\par\centering \resizebox*{!}{0.25\textheight}{\includegraphics{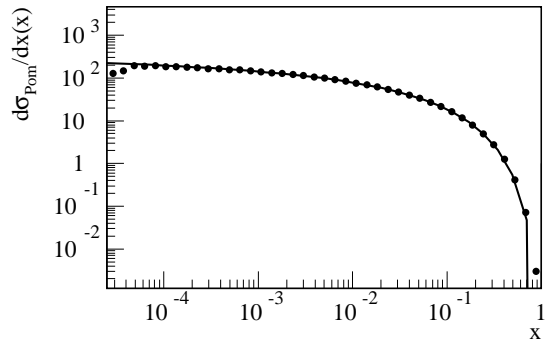}} \par}

\caption{Inclusive cross section of Pomerons versus \protect\( x\protect \). The variable
\protect\( x\protect \) is defined as \protect\( x=x^{+}x^{-}\protect \)and
is therefore the squared mass of the Pomeron divided by \protect\( s\protect \).
We show unitarized analytical results (solid line) and the corresponding simulation
results (points).\label{4-4}}
\end{figure}

Another elementary quantity is the inclusive momentum spectrum of Pomerons.
Pomerons, representing elementary interactions, are characterized by their light
cone momentum fractions \( x^{+} \) and \( x^{-} \), so one might study two
dimensional distributions, or for example a distribution in \( x=x^{+}x^{-} \),
where the second variable \( y=0.5\log (x^{+}/x^{-}) \) is integrated over.
So again we simulate many proton-proton events at a given energy \( \sqrt{s} \)
and we count the number of Pomerons \( \Delta N_{\mathrm{Pom}} \) within a
certain interval \( [x-\Delta x/2,x+\Delta x/2] \) and we calculate
\begin{equation}
\label{x}
\frac{dn^{\mathrm{MC}}_{\mathrm{Pom}}}{dx}=\frac{1}{N_{\mathrm{events}}}\frac{\Delta N_{\mathrm{Pom}}}{\Delta x},
\end{equation}
representing the Monte Carlo Pomeron \( x \)-distribution, which may be compared
with the analytical result calculated earlier, as shown in fig.\ \ref{4-3}. 
The analytical results of course refer to the unitarized theory. Again we find
perfect agreement between Monte Carlo simulations and analytical curves, as
it should be. In fig.\ \ref{4-4},
we compare inclusive Pomeron cross sections (integrated over impact parameter).
Here, the impact parameters are generated randomly between 1 and some \( b_{\mathrm{max}} \),
one counts the number of Pomerons \( \Delta N_{\mathrm{Pom}} \) within a certain
interval of size \( \Delta x \), and one calculates
\begin{equation}
\label{x}
\frac{d\sigma ^{\mathrm{MC}}_{\mathrm{Pom}}}{dx}=\frac{\pi r_{\mathrm{max}}^{2}}{N_{\mathrm{events}}}\frac{\Delta N_{\mathrm{Pom}}}{\Delta x},
\end{equation}
which is compared with the analytical result
\begin{equation}
\label{x}
\frac{d\sigma _{\mathrm{Pom}}}{\mathrm{dx}}(x,s)=\int d^{2}b\, \frac{dn_{\mathrm{Pom}}}{dx}(x,s,b),
\end{equation}
which again show an excellent agreement.

These Pomeron distributions are of particular interest, because they are elementary
distributions based on which other inclusive spectra like a transverse momentum
distribution of pions may be obtained via convolution.

The two examples of this section provide on one hand a check that the numerical
procedures work properly, on the other hand they demonstrate nicely that our
Monte Carlo procedure is a very well defined numerical method to solve a particular
mathematical problem. In simple cases where analytical results exist, they may
be compared with the Monte Carlo results, and they must absolutely agree.

\cleardoublepage

\chapter{Enhanced Pomeron diagrams}

\begin{figure}[htb]
{\par\centering \resizebox*{!}{0.15\textheight}{\includegraphics{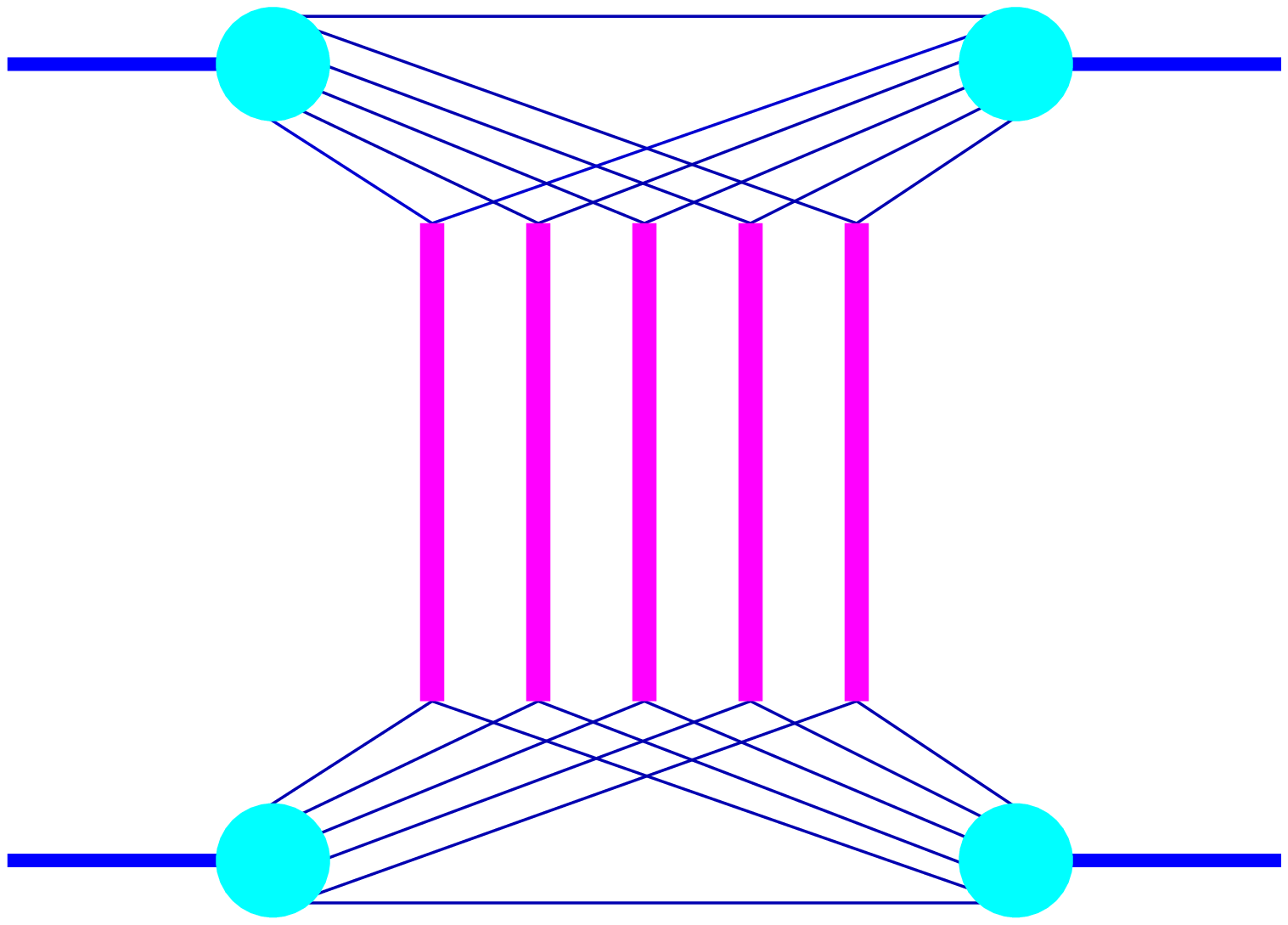}} \par}

\caption{Eikonal type diagrams.\label{eik}}
\end{figure}

\begin{figure}[htb]
{\par\centering \resizebox*{!}{0.17\textheight}{\includegraphics{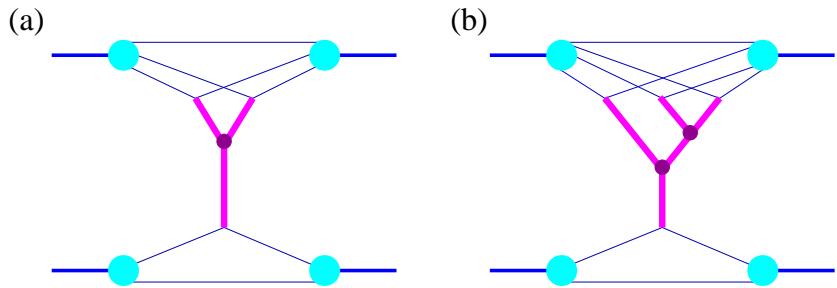}} \par}

\caption{Enhanced diagrams. \label{enh}}
\end{figure}

The eikonal type diagrams shown at fig.\ \ref{eik}, considered in the previous
chapters, correspond to pair-like scatterings between hadron constituents and
form the basis for the description of hadronic interactions at not too high
energies.
However, when the interaction energy increases, the contribution of so-called
enhanced Pomeron diagrams as, for example, the diagrams shown in fig.\ \ref{enh},
become more and more important. 
The latter ones take into account interactions of Pomerons with each other.
The corresponding amplitudes increase asymptotically much faster than the usual
eikonal type contributions considered so far. In this paper, we restrict ourselves
to the lowest order diagrams (\( Y \)-diagrams and inverted \( Y \)-diagrams). 

In the following sections, we discuss the amplitudes corresponding to the lowest
order enhanced diagrams and the modification of the hadronic profile function
in the presence of these diagrams, before we discuss their most important features.

\section{Calculating lowest order enhanced diagrams}

\begin{figure}[htb]
{\par\centering \resizebox*{!}{0.15\textheight}{\includegraphics{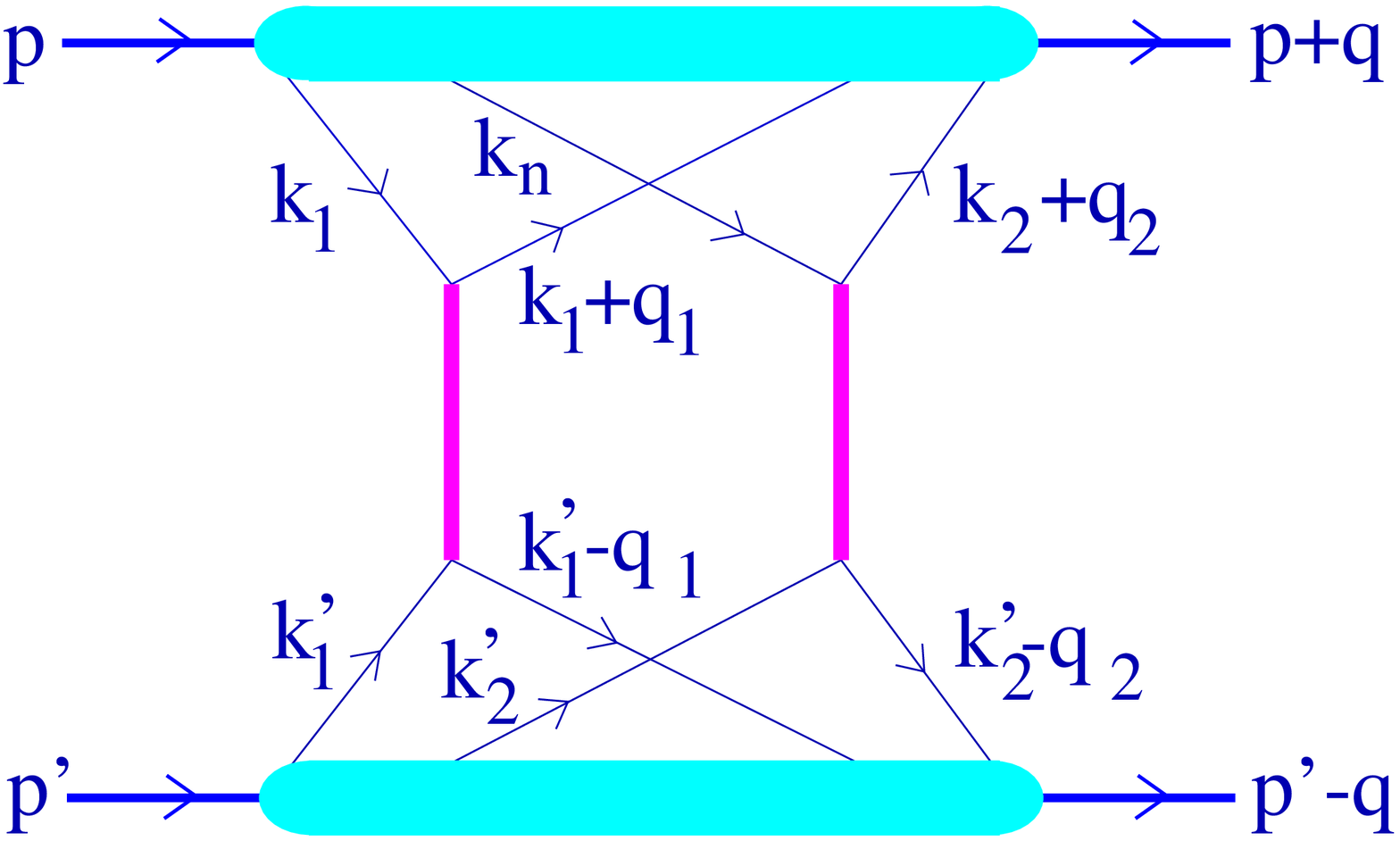}} \par}

\caption{Double Pomeron exchange. \label{2pom}}
\end{figure}

To introduce enhanced type diagrams let us come back to the process of double
soft Pomeron exchange, which is a particular case of the diagram of fig.\ \ref{wave}.
The corresponding contribution to the elastic scattering amplitude is given
in eqs.\ (\ref{t-h1-h2}), (\ref{f-n-x}) with \( n=2 \) and with \( T_{1\mathrm{I}\! \mathrm{P}} \)
being replaced by \( T_{\mathrm{soft}} \): 

\begin{eqnarray}
iT^{(2)}_{h_{1}h_{2}}(s,t) & = & \frac{1}{2}\int \! \frac{d^{4}k_{1}}{(2\pi )^{4}}\frac{d^{4}k_{1}'}{(2\pi )^{4}}\frac{d^{4}k_{2}}{(2\pi )^{4}}\frac{d^{4}k_{2}'}{(2\pi )^{4}}\frac{d^{4}q_{1}}{(2\pi )^{4}}\Theta \! \left( s_{1}^{+}\right) \Theta \! \left( s_{1}^{-}\right) \Theta \! \left( s_{2}^{+}\right) \Theta \! \left( s_{2}^{-}\right) \Theta \! \left( s_{q_{1}}^{+}\right) \Theta \! \left( s_{q_{1}}^{-}\right) \nonumber \label{t-2-elem} \\
 &  & \qquad \times \: \mathrm{disc}_{s_{1}^{+},s_{2}^{+},s_{q_{1}}^{+}}\, N_{h_{1}}^{(2)}\! \left( p,k_{1},k_{2},q_{1},q-q_{1}\right) \nonumber \\
 &  & \qquad \times \; \prod ^{2}_{l=1}\! \left[ iT_{\mathrm{soft}}\! (\hat{s}_{l},q_{l}^{2})\right] \, \mathrm{disc}_{s_{1}^{-},s_{2}^{-},s_{q_{1}}^{-}}\, N_{h_{2}}^{(2)}\! \left( p',k_{1}',k_{2}',-q_{1},-q+q_{1}\right) \label{t-2-soft} 
\end{eqnarray}
 see fig.\ \ref{2pom}.

\begin{figure}[htb]
{\par\centering \resizebox*{!}{0.24\textheight}{\includegraphics{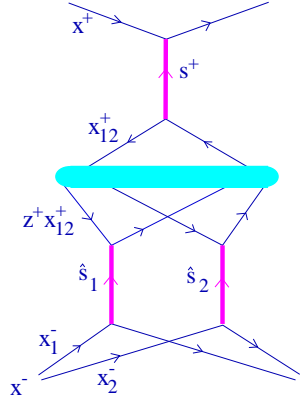}} \par}

\caption{Kinematical variables for the triple Pomeron contribution. \label{3p-pro}}
\end{figure}

We are now interested in the contribution with some of the invariants 
\begin{eqnarray}
s_{1}^{+} & = & (p-k_{1})^{2}\simeq -p^{+}k_{1}^{-},\\
s_{2}^{+} & = & (p-k_{1}-k_{2})^{2}\simeq -p^{+}(k_{1}^{-}+k_{2}^{-}),\\
s_{q_{1}}^{+} & = & (p+q_{1})^{2}\simeq p^{+}q_{1}^{-},\label{x} 
\end{eqnarray}
 being large, implying \( k_{i}^{-},q^{-}_{1} \) to be not too small. As shown
in appendix \ref{ax-c-3}, in that case the above amplitude may be written as
\begin{equation}
\label{t3p-}
iT^{3\mathrm{I}\! \mathrm{P}\! -}_{h_{1}h_{2}}(s,t)=\int ^{1}_{0}\! \! \frac{dx^{+}}{x^{+}}\frac{dx^{-}}{x^{-}}\, F^{h_{1}}_{\mathrm{remn}}\! \left( 1-x^{+}\right) \, F^{h_{2}}_{\mathrm{remn}}\! \! \left( 1-x^{-}\right) \, iT^{h_{1}h_{2}}_{3\mathrm{I}\! \mathrm{P}\! -}(x^{+},x^{-},s,t)
\end{equation}
with
\begin{eqnarray}
 &  & iT_{3\mathrm{I}\! \mathrm{P}\! -}^{h_{1}h_{2}}(x^{+},x^{-},s,t)\: =\: 8\pi ^{2}x^{+}x^{-}s\, \frac{r_{3\mathrm{I}\! \mathrm{P}}}{2}\int ^{x^{+}}_{s_{0}/x^{-}}\! \frac{dx_{12}^{+}}{x_{12}^{+}}\left[ \frac{1}{2s^{+}}\, \mathrm{Im}\, T^{h_{1}}\! \left( x^{+},s^{+},-q_{\perp }^{2}\right) \right] \nonumber \\
 &  & \qquad \qquad \times \; \int dz^{+}\int \! d^{2}q_{1_{\perp }}d^{2}q_{2_{\perp }}\int ^{x^{-}}_{0}\! dx_{1}^{-}dx_{2}^{-}\prod ^{2}_{l=1}\! \left[ \frac{1}{8\pi ^{2}\hat{s}_{l}}\, iT^{h_{2}}\! \left( x_{l}^{-},\hat{s}_{l},-q_{l_{\perp }}^{2}\right) \right] \nonumber \\
 &  & \qquad \qquad \times \; \delta \! (x^{-}-x_{1}^{-}-x_{2}^{-})\, \delta ^{(2)}\! \left( \vec{q}_{\perp }-\vec{q}_{1_{\perp }}-\vec{q}_{2_{\perp }}\right) ,\label{t3p-p} 
\end{eqnarray}
with
\begin{equation}
\label{x}
T^{h}\! \left( x,s,-q_{\perp }^{2}\right) =T_{\mathrm{soft}}^{h}\! \left( x,s,-q_{\perp }^{2}\right) =T_{\mathrm{soft}}\! \left( s,-q_{\perp }^{2}\right) \, F^{h}_{\mathrm{part}}(x)\, \exp \! \left( -R_{h}^{2}\, q_{\perp }^{2}\right) 
\end{equation}
 and
\begin{equation}
\label{x}
\hat{s}_{1}=x_{12}^{+}z^{+}x_{1}^{-}s,\qquad \hat{s}_{2}=x_{12}^{+}(1-z^{+})x_{2}^{-}s,
\end{equation}
 where the following definitions have been used
\begin{eqnarray}
 &  & x^{+}=k^{+}/p^{+},\\
 &  & x_{12}^{+}=k_{12}^{+}/p^{+},\\
 &  & x_{1}^{+}=k_{1}^{+}/p^{+},\\
 &  & x_{12}^{+}-x_{1}^{+}=(k^{+}_{12}-k_{1}^{+})/p^{+}=k_{2}^{+}/p^{+},\\
 &  & z^{+}=k_{1}^{+}/k_{12}^{+}=x_{1}^{+}/x_{12}^{+},\\
 &  & s^{+}=(k-k_{12})^{2}\simeq -k^{+}k_{12}^{-}\simeq s_{0}k^{+}/k^{+}_{12}=s_{0}x^{+}/x^{+}_{12},\label{x} 
\end{eqnarray}
see fig.\ \ref{3p-pro}.
 The sign ``\( - \)'' in ``\( 3\mathrm{I}\! \mathrm{P}\! - \)'' refers
to the Pomeron ``splitting'' towards the target hadron (reversed \( Y \)-diagram);
the lower limit for the integral \( dx_{12}^{+} \) is due to \( x_{12}^{-}\simeq s_{0}/x_{12}^{+}<x^{-} \). 

The triple-Pomeron contribution (\ref{t3p-p}) is by construction expressed
via amplitudes \( T^{h}_{\mathrm{soft}} \) for parton-parton scattering due
to soft Pomeron exchange, each one corresponding to non-perturbative parton
dynamics, characterized by restricted parton virtualities \( Q^{2}<Q_{0}^{2} \).
We can also take into account contributions to the triple-Pomeron diagram from
semi-hard processes, when some part of the parton cascade mediating the scattering
between partons of momenta \( k_{l} \) and \( k_{l}' \) at fig.\ \ref{3pom}
(\( k \) and \( -k_{12} \)) develops in the perturbative region \( Q^{2}>Q_{0}^{2} \).
Then, according to the general discussion of chapter 2, the amplitudes \( T^{h} \)
obtain also contributions from semi-hard sea-type parton-parton scattering \( T^{h}_{\mathrm{sea}-\mathrm{sea}} \)
and from valence quark scattering \( T^{h}_{\mathrm{val}-\mathrm{sea}} \):
\begin{equation}
\label{t-pom-h-sum}
T^{h}=T^{h}_{\mathrm{soft}}+T^{h}_{\mathrm{sea}-\mathrm{sea}}+T^{h}_{\mathrm{val}-\mathrm{sea}},
\end{equation}
with
\begin{equation}
\label{t-pom-h-sea}
T_{\mathrm{soft}/\mathrm{sea}-\mathrm{sea}}^{h}\! \left( x,\hat{s},-q_{\perp }^{2}\right) =T_{\mathrm{soft}/\mathrm{sea}-\mathrm{sea}}\! \left( \hat{s},-q_{\perp }^{2}\right) \, F^{h}_{\mathrm{part}}(x)\, \exp \! \left( -R_{h}^{2}\, q_{\perp }^{2}\right) 
\end{equation}
and 
\begin{eqnarray}
T^{h}_{\mathrm{val}-\mathrm{sea}}\! \left( x,\hat{s},-q_{\perp }^{2}\right) =\int ^{x}_{0}dx_{v}\frac{x}{x_{v}}\sum _{k}T_{\mathrm{val}-\mathrm{sea}}^{k}\left( \frac{x_{v}}{x}\hat{s},-q_{\perp }^{2}\right) \mathrm{exp}\left( -R_{h}^{2}q_{\perp }^{2}\right)  &  & \nonumber \\
\times \; \bar{F}^{h,k}_{\mathrm{part}}(x_{v},x-x_{v}). &  & \label{t-pom-h-valsea} 
\end{eqnarray}

We have to stress again that we do not consider the possibility of Pomeron-Pomeron
coupling in the perturbative region \( Q^{2}>Q_{0}^{2} \). Therefore in our
scheme hard parton processes can only contribute into internal structure of
elementary parton-parton scattering amplitudes but do not influence the triple-Pomeron
coupling. 

A similar contribution \( T_{3\mathrm{I}\! \mathrm{P}\! +}^{h_{1}h_{2}} \)
of the \( Y \)-diagram can be obtained via interchanging \( x^{+}\leftrightarrow x^{-} \)
and \( h_{1}\leftrightarrow h_{2} \): 
\begin{equation}
\label{x}
T_{3\mathrm{I}\! \mathrm{P}\! +}^{h_{1}h_{2}}(x^{+},x^{-},s,t)=T_{3\mathrm{I}\! \mathrm{P}\! -}^{h_{2}h_{1}}(x^{-},x^{+},s,t).
\end{equation}

One can repeat the above derivation for the case of a general soft multiple
scattering process, see eq.\ (\ref{t-h1-h2}), where some of the energy invariants
\( s_{q_{l}}^{+} \), \( s_{q_{l}}^{-} \) are large \cite{bak76}. One then
obtains finally the general multiple scattering expression (\ref{t-hh-part-remn}),
with the contribution of corresponding pairs of Pomerons being replaced by expressions
\begin{equation}
\label{x}
\frac{1}{8\pi ^{2}x^{+}x^{-}s}\int d^{2}q_{\perp }\, iT_{3\mathrm{I}\! \mathrm{P}\pm }^{h_{1}h_{2}}(x^{+},x^{-},s,t),
\end{equation}
 where \( x^{\pm } \) refers to the summary light cone momentum share of the
constituent partons participating in the triple-Pomeron process. So we get
\begin{eqnarray}
 &  & iT_{h_{1}h_{2}}(s,t)=8\pi ^{2}s\sum ^{\infty }_{n=1}\frac{1}{n!}\, \int ^{1}_{0}\! \prod ^{n}_{l=1}\! dx_{l}^{+}dx_{l}^{-}\prod ^{n}_{l=1}\! \left[ \frac{1}{8\pi ^{2}\hat{s}_{l}}\int \! d^{2}q_{l_{\perp }}\, iT_{\mathrm{I}\! \mathrm{P}}^{h_{1}h_{2}}\! \left( \hat{s}_{l},-q_{l_{\perp }}^{2}\right) \right] \nonumber \label{t-h1-h2-3p} \\
 &  & \qquad \qquad \times \; F^{h_{1}}_{\mathrm{remn}}\! \! \left( 1-\sum ^{n}_{j=1}\! x_{j}^{+}\right) \, F^{h_{2}}_{\mathrm{remn}}\! \! \left( 1-\sum ^{n}_{j=1}\! x_{j}^{-}\right) \: \delta ^{(2)}\! \left( \sum ^{n}_{k=1}\! \vec{q}_{k_{\perp }}-\vec{q}_{\perp }\right) .\label{t-h1-h2-3p} 
\end{eqnarray}
with
\begin{equation}
\label{x}
T_{\mathrm{I}\! \mathrm{P}}^{h_{1}h_{2}}=T_{1\mathrm{I}\! \mathrm{P}}^{h_{1}h_{2}}+T^{h_{1}h_{2}}_{3\mathrm{I}\! \mathrm{P}\! -}+T_{3\mathrm{I}\! \mathrm{P}\! +}^{h_{1}h_{2}}
\end{equation}
Here, we allow any number of simple triple-Pomeron diagrams; thus we restrict
ourselves to the contributions of double Pomeron iteration in the \( t \)-channel
rather than to the first order in \( r_{3\mathrm{I}\! \mathrm{P}} \).

The Fourier transform \( \tilde{T} \) of the amplitude (\ref{t-h1-h2-3p})
is given as
\begin{eqnarray}
\frac{i}{2s}\tilde{T}_{h_{1}h_{2}}(s,b) & = & \sum ^{\infty }_{n=1}\frac{1}{n!}\, \int ^{1}_{0}\! \prod ^{n}_{l=1}\! dx_{l}^{+}dx_{l}^{-}\prod ^{n}_{l=1}\frac{i}{2s}\tilde{T}^{h_{1}h_{2}}_{\mathrm{I}\! \mathrm{P}}(x_{l}^{+},x_{l}^{-},s,b)\nonumber \\
 & \times  & F^{h_{1}}_{\mathrm{remn}}\! \! \left( 1-\sum ^{n}_{j=1}\! x_{j}^{+}\right) \, F^{h_{2}}_{\mathrm{remn}}\! \! \left( 1-\sum ^{n}_{j=1}\! x_{j}^{-}\right) ,\label{x} 
\end{eqnarray}
with \( \tilde{T}^{h_{1}h_{2}}_{\mathrm{I}\! \mathrm{P}} \) being the Fourier
transform of \( T^{h_{1}h_{2}}_{\mathrm{I}\! \mathrm{P}} \), 
\begin{equation}
\label{x}
\tilde{T}^{h_{1}h_{2}}_{\mathrm{I}\! \mathrm{P}}=\tilde{T}_{1\mathrm{I}\! \mathrm{P}}^{h_{1}h_{2}}+\tilde{T}^{h_{1}h_{2}}_{3\mathrm{I}\! \mathrm{P}\! -}+\tilde{T}_{3\mathrm{I}\! \mathrm{P}\! +}^{h_{1}h_{2}}\: ,
\end{equation}
 where the Fourier transform \( \tilde{T}^{h_{1}h_{2}}_{3\mathrm{I}\! \mathrm{P}\! -} \)
of the triple Pomeron amplitude \( T^{h_{1}h_{2}}_{3\mathrm{I}\! \mathrm{P}\! -} \)
is given as

\begin{eqnarray}
 &  & \frac{i}{2\hat{s}}\tilde{T}_{3\mathrm{I}\! \mathrm{P}\! -}^{h_{1}h_{2}}(x^{+},x^{-},s,b)\: =\: \frac{r_{3\mathrm{I}\! \mathrm{P}}}{2}\int d^{2}b_{1}\int ^{x^{+}}_{s_{0}/x^{-}}\! \frac{dx_{12}^{+}}{x_{12}^{+}}\left[ \frac{1}{2s^{+}}\, \mathrm{Im}\, \tilde{T}^{h_{1}}\! \left( x^{+},s^{+},|\vec{b}-\vec{b}_{1}|\right) \right] \nonumber \\
 &  & \qquad \qquad \times \; \int dz^{+}\int ^{x^{-}}_{0}\! dx_{1}^{-}dx_{2}^{-}\prod ^{2}_{l=1}\! \left[ \frac{1}{2\hat{s}_{l}}\, i\tilde{T}^{h_{2}}\! \left( x_{l}^{-},\hat{s}_{l},b_{1}\right) \right] \, \delta \! (x^{-}-x_{1}^{-}-x_{2}^{-}),\label{thht3p} 
\end{eqnarray}
with \( \hat{s}=x^{+}x^{-}s \) (and similarly for \( \tilde{T}^{h_{1}h_{2}}_{3\mathrm{I}\! \mathrm{P}\! +} \)).

Here we used
\begin{equation}
\label{x}
\delta ^{(2)}\! \left( \vec{q}_{\perp }-\vec{q}_{1_{\perp }}-\vec{q}_{2_{\perp }}\right) =\frac{1}{4\pi ^{2}}\int \! d^{2}b_{1}\, \exp \! \left( i\left( \vec{q}_{\perp }-\vec{q}_{1_{\perp }}-\vec{q}_{2_{\perp }}\right) \vec{b}_{1}\right) .
\end{equation}

\begin{figure}[htb]
{\par\centering \resizebox*{!}{0.24\textheight}{\includegraphics{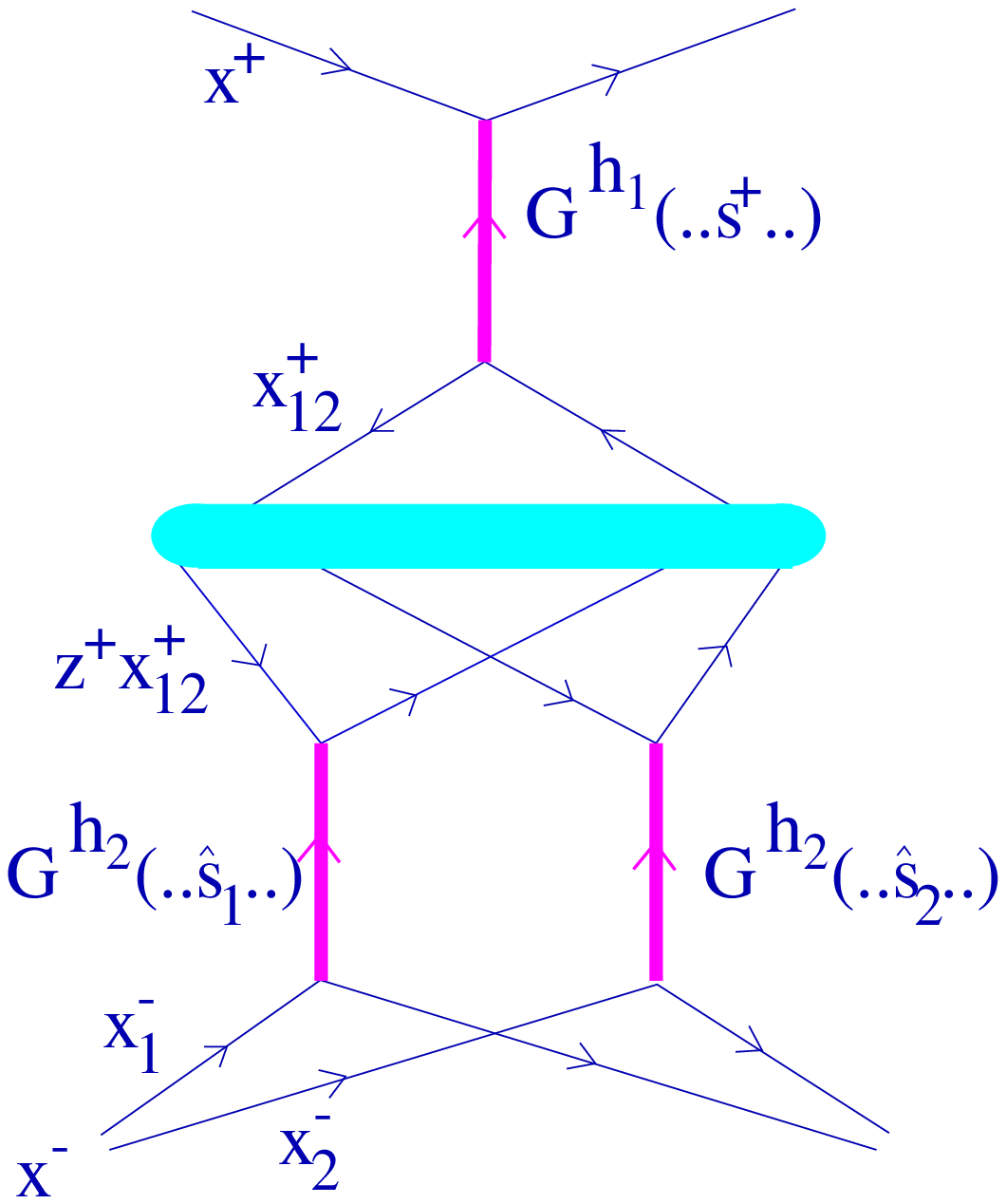}} \par}

\caption{Triple Pomeron contribution. \label{3p-prof}}
\end{figure}

The profile function \( \gamma  \) for hadron-hadron interaction is as usual
defined as
\begin{equation}
\label{x}
\gamma _{h_{1}h_{2}}(s,b)=\frac{1}{2s}2\mathrm{Im}\tilde{\mathrm{T}}_{h_{1}h_{2}}(s,b),
\end{equation}
which may be evaluated using the AGK cutting rules,
\begin{eqnarray}
\gamma _{h_{1}h_{2}}(s,b) & = & \sum ^{\infty }_{m=1}\frac{1}{m!}\, \int \prod ^{m}_{\mu =1}\! dx_{\mu }^{+}dx_{\mu }^{-}\prod ^{m}_{\mu =1}G_{\mathrm{I}\! \mathrm{P}}^{h_{1}h_{2}}(x_{\mu }^{+},x_{\mu }^{-},s,b)\nonumber \\
 & \times  & \sum ^{\infty }_{l=1}\frac{1}{l!}\, \int \prod ^{l}_{\lambda =1}\! d\tilde{x}_{\lambda }^{+}d\tilde{x}_{\lambda }^{-}\prod ^{l}_{\lambda =1}-G_{\mathrm{I}\! \mathrm{P}}^{h_{1}h_{2}}(\tilde{x}_{\lambda }^{+},\tilde{x}_{\lambda }^{-},s,b)\nonumber \\
 & \times  & F_{\mathrm{remn}}\left( x^{\mathrm{proj}}-\sum _{\lambda }\tilde{x}_{\lambda }^{+}\right) \, F_{\mathrm{remn}}\left( x^{\mathrm{targ}}-\sum _{\lambda }\tilde{x}_{\lambda }^{-}\right) ,\label{ghhtot} 
\end{eqnarray}
with \( x^{\mathrm{proj}/\mathrm{targ}}=1-\sum x^{\pm }_{\mu } \) being the
momentum fraction of the projectile/target remnant, and with
\begin{equation}
\label{x}
G_{\mathrm{I}\! \mathrm{P}}^{h_{1}h_{2=}}G_{1\mathrm{I}\! \mathrm{P}}^{h_{1}h_{2}}+G_{3\mathrm{I}\! \mathrm{P}\! -}^{h_{1}h_{2}}+G_{3\mathrm{I}\! \mathrm{P}\! +}^{h_{1}h_{2}},
\end{equation}
 where \( G_{3\mathrm{I}\! \mathrm{P}\! \pm }^{h_{1}h_{2}} \) is twice the
imaginary part of the Fourier transformed triple-Pomeron amplitude \( \tilde{T}_{3\mathrm{I}\! \mathrm{P}\! \pm }^{h_{1}h_{2}} \)
divided by \( 2\hat{s} \),
\begin{equation}
\label{g-3p-h1h2}
G_{3\mathrm{I}\! \mathrm{P}\! \pm }^{h_{1}h_{2}}(x^{+},x^{-},s,b)=\frac{1}{2x^{+}x^{-}s}2\mathrm{Im}\, \tilde{T}_{3\mathrm{I}\! \mathrm{P}\! \pm }^{h_{1}h_{2}}(x^{+},x^{-},s,b),
\end{equation}
 which gives, assuming imaginary amplitudes,
\begin{eqnarray}
G_{3\mathrm{I}\! \mathrm{P}\! -}^{h_{1}h_{2}}(x^{+},x^{-},s,b) & = & -\frac{r_{3\mathrm{I}\! \mathrm{P}}}{8}\, \int d^{2}b_{1}\int ^{x^{+}}_{s_{0}/x^{-}}\! \frac{dx_{12}^{+}}{x_{12}^{+}}\, G^{h_{1}}(x^{+},s^{+},\left| \vec{b}-\vec{b}_{1}\right| )\nonumber \\
 &  & \qquad \times \; \int ^{1}_{0}\! dz^{+}\int ^{x^{-}}_{0}\! dx_{1}^{-}\, G^{h_{2}}(x_{1}^{-},\hat{s}_{1},b_{1})\, G^{h_{2}}(x^{-}-x_{1}^{-},\hat{s}_{2},b_{1}),\label{g-3p} 
\end{eqnarray}
see fig.\ \ref{3p-prof},
 with 
\begin{equation}
\label{x}
s^{+}=s_{0}\frac{x^{+}}{x_{12}^{+}},\quad \hat{s}_{1}=x_{12}^{+}z^{+}x_{1}^{-}s,\quad \hat{s}_{2}=x_{12}^{+}(1-z^{+})(x^{-}-x_{1}^{-})s.
\end{equation}
The functions \( G^{h} \) are defined as
\begin{equation}
\label{gh-th}
G^{h}(x,\hat{s},b)=\frac{1}{2\hat{s}}\, 2\mathrm{Im}\, \tilde{T}^{h}\left( x,\hat{s},b\right) ,
\end{equation}
with \( \tilde{T}^{h} \) being the Fourier transform of \( T^{h} \), which
gives
\begin{equation}
\label{g-h-tot}
G^{h}=G^{h}_{\mathrm{soft}}+G^{h}_{\mathrm{sea}-\mathrm{sea}}+G^{h}_{\mathrm{val}-\mathrm{sea}},
\end{equation}
with
\begin{eqnarray}
 &  & G^{h}_{\mathrm{soft}}\! (x,\hat{s},b)=\frac{2\gamma _{part}}{\lambda ^{h}_{\mathrm{soft}}\! (\frac{\hat{s}}{s_{0}})}\left( \frac{\hat{s}}{s_{0}}\right) ^{\alpha _{_{\mathrm{soft}}}\! (0)-1}\exp \! \left( -\frac{b^{2}}{4\lambda ^{h}_{\mathrm{soft}}\! (\frac{\hat{s}}{s_{0}})}\right) \, F_{\mathrm{part}}(x)\label{g-h-soft-3p} \\
 &  & G^{h}_{\mathrm{sea}-\mathrm{sea}}\! (x,\hat{s},b)=\frac{1}{4\pi }\sum _{jk}\int ^{1}_{0}dz^{+}dz^{-}E_{\mathrm{soft}}^{j}\left( z^{+}\right) \, E_{\mathrm{soft}}^{k}\left( z^{-}\right) \, \sigma _{\mathrm{ladder}}^{jk}(z^{+}z^{-}\hat{s},Q_{0}^{2})\times \\
 &  & \qquad \qquad \qquad \qquad \qquad \times \; \frac{1}{\lambda ^{h}_{\mathrm{soft}}\! \left( 1/z^{+}z^{-}\right) }\exp \! \left( -\frac{b^{2}}{4\lambda ^{h}_{\mathrm{soft}}\! \left( 1/z^{+}z^{-}\right) }\right) \, F_{\mathrm{part}}(x)\nonumber \\
 &  & G^{h}_{\mathrm{val}-\mathrm{sea}}\left( x,\hat{s},b\right) =\frac{1}{4\pi }\sum _{jk}\int ^{x}_{0}dx_{v}\, \int ^{1}_{0}\! dz^{+}\, E_{\mathrm{soft}}^{j}\left( z^{+}\right) \, \sigma _{\mathrm{ladder}}^{jk}\! \left( \frac{x_{v}}{x}z^{+}\hat{s},Q_{0}^{2}\right) \, \label{g-h-val-sea} \\
 &  & \qquad \qquad \qquad \qquad \qquad \times \; \frac{1}{\lambda ^{h}_{\mathrm{soft}}\! \left( 1/z^{+}\right) }\exp \! \left( -\frac{b^{2}}{4\lambda ^{h}_{\mathrm{soft}}\! \left( 1/z^{+}\right) }\right) \bar{F}^{k}_{\mathrm{part}}(x_{v},x-x_{v}),\nonumber \label{x} 
\end{eqnarray}
with 
\begin{equation}
\label{x}
\lambda ^{h}_{\mathrm{soft}}\! \left( z\right) =R_{h}^{2}+\alpha '_{_{\mathrm{soft}}}\ln \! z.
\end{equation}

\section{Cutting Enhanced Diagrams}

\begin{figure}[htb]
{\par\centering \resizebox*{!}{0.15\textheight}{\includegraphics{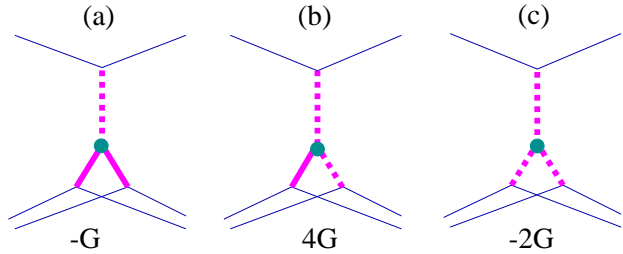}} \par}

\caption{The different cuts of the triple Pomeron inverted \protect\( Y\protect \)-diagram:
none of the lower Pomerons cut (a), one of these Pomerons cut (b), and both
Pomerons cut (c). We also indicate their relations with \protect\( G=G_{3\mathrm{I}\! \mathrm{P}\! -}^{h_{1}h_{2}}\protect \).
\label{3pcuts}}
\end{figure}

To treat particle production, we have to investigate the different cuts of an
enhanced diagram. We consider the inverted \( Y \)-diagram here, the same arguments
apply to the \( Y \)-diagram. We employ the cutting rules to eq.\ (\ref{thht3p}),
\begin{equation}
\label{x}
G_{3\mathrm{I}\! \mathrm{P}\! -}^{h_{1}h_{2}}=\frac{1}{2\hat{s}}2\mathrm{Im}\tilde{T}_{3\mathrm{I}\! \mathrm{P}\! -}^{h_{1}h_{2}}=\left\langle \mathrm{sum}\, \mathrm{over}\, \mathrm{cut}\, \mathrm{diagrams}\right\rangle =\sum _{i}G_{3\mathrm{I}\! \mathrm{P}\! -(i)}^{h_{1}h_{2}},
\end{equation}
where the index \( i \) counts the different cuts. We take into account that
the cutting procedure only influences the two Pomerons exchanged ``in parallel''
in the triple-Pomeron graph (the lower Pomerons) with the third Pomeron already
being cut \cite{bak76}, so that we have three contributions: none of the lower
Pomerons cut (\( i=0 \), diffraction), one of these Pomerons cut (\( i=1 \),
screening), and both Pomerons cut (\( i=2 \), Pomeron-Pomeron fusion), see
fig.\ \ref{3pcuts}. 
 So we have 
\begin{equation}
\label{sumcut}
G_{3\mathrm{I}\! \mathrm{P}\! -}^{h_{1}h_{2}}=G_{3\mathrm{I}\! \mathrm{P}\! -(0)}^{h_{1}h_{2}}+G_{3\mathrm{I}\! \mathrm{P}\! -(1)}^{h_{1}h_{2}}+G_{3\mathrm{I}\! \mathrm{P}\! -(2)}^{h_{1}h_{2}},
\end{equation}
with 
\begin{eqnarray}
G_{3\mathrm{I}\! \mathrm{P}\! -(0)}^{h_{1}h_{2}} & = & \left\{ \frac{i}{2\hat{s}}\tilde{T}_{3\mathrm{I}\! \mathrm{P}\! -}^{h_{1}h_{2}}\right\} \times 2\, ,\nonumber \\
G_{3\mathrm{I}\! \mathrm{P}\! -(1)}^{h_{1}h_{2}} & = & \left\{ \frac{i}{2\hat{s}}\tilde{T}_{3\mathrm{I}\! \mathrm{P}\! -}^{h_{1}h_{2}}\right\} \times (-2)\times 2\times 2\, ,\\
G_{3\mathrm{I}\! \mathrm{P}\! -(2)}^{h_{1}h_{2}} & = & \left\{ \frac{i}{2\hat{s}}\tilde{T}_{3\mathrm{I}\! \mathrm{P}\! -}^{h_{1}h_{2}}\right\} \times 2\times 2\, .\nonumber \label{x} 
\end{eqnarray}
Here, we assumed imaginary amplitudes, and we replace as usual a factor \( i\tilde{T}^{h} \)
in eq.\ (\ref{thht3p}) by \( 2\mathrm{Im}\tilde{T}^{h}=-2i\tilde{T}^{h} \)
for a cut Pomeron, and by \( (i\tilde{T}^{h})^{*}=i\tilde{T}^{h} \) for an
uncut Pomeron being to the left of the cut plane, and by \( i\tilde{T}^{h} \)
for an uncut Pomeron being to the right of the cut plane. Using (see eq.\ (\ref{g-3p-h1h2}))
\begin{equation}
\label{x}
\frac{i}{2\hat{s}}\tilde{T}_{3\mathrm{I}\! \mathrm{P}\! -}^{h_{1}h_{2}}=-\frac{1}{2}G_{3\mathrm{I}\! \mathrm{P}\! -}^{h_{1}h_{2}},
\end{equation}
 we get 
\begin{eqnarray}
G_{3\mathrm{I}\! \mathrm{P}\! -(0)}^{h_{1}h_{2}} & = & -1\times G_{3\mathrm{I}\! \mathrm{P}\! -}^{h_{1}h_{2}}\, ,\nonumber \\
G_{3\mathrm{I}\! \mathrm{P}\! -(1)}^{h_{1}h_{2}} & = & +4\times G_{3\mathrm{I}\! \mathrm{P}\! -}^{h_{1}h_{2}}\, ,\label{factors} \\
G_{3\mathrm{I}\! \mathrm{P}\! -(2)}^{h_{1}h_{2}} & = & -2\times G_{3\mathrm{I}\! \mathrm{P}\! -}^{h_{1}h_{2}}\, ,\nonumber 
\end{eqnarray}
which means that each cut contribution is equal to the profile function, up
to a pre-factor, see fig.\ \ref{3pcuts}. The sum of the three contributions
is \( G_{3\mathrm{I}\! \mathrm{P}\! -}^{h_{1}h_{2}} \), as it should be.

There is a substantial difference between the different cuts of triple-Pomeron
contributions: in the case of both lower Pomerons being cut (\( i=2 \)) all
the momentum of the constituent partons, participating in the process, is transferred
to secondary hadrons produced, whereas for the cut between these Pomerons (\( i=0 \))
only the light cone momentum fractions of the cut Pomeron (\( x^{+} \), \( x_{12}^{-}=s_{0}/x_{12}^{+} \)
in eq.\ (\ref{g-3p})) are available for hadron production, the momentum share
\( x^{-}-x_{12}^{-} \) of the partons, connected to the uncut (virtual) Pomerons
is given back to the remnant state. Correspondingly, the contribution with one
of the two lower Pomerons being cut (\( i=1 \)) defines the screening correction
to the elementary rescattering with the momentum fractions \( x^{+} \), \( x_{1}^{-} \)
(considering the first of the two Pomerons being cut). It is therefore useful
to rewrite the expression for the profile function as
\begin{eqnarray}
\gamma _{h_{1}h_{2}}(s,b) & = & \sum _{m}\frac{1}{m!}\int \prod ^{m}_{\mu =1}dx_{\mu }^{+}dx_{\mu }^{-}d\hat{x}_{\mu }^{+}d\hat{x}_{\mu }^{-}\prod ^{m}_{\mu =1}\hat{G}_{\mathrm{I}\! \mathrm{P}}^{h_{1}h_{2}}(x_{\mu }^{+},x_{\mu }^{-},\hat{x}_{\mu }^{+},\hat{x}_{\mu }^{-},s,b)\nonumber \\
 & \times  & \sum _{l}\frac{1}{l!}\int \prod ^{l}_{\lambda =1}d\tilde{x}_{\lambda }^{+}d\tilde{x}_{\lambda }^{-}\prod ^{l}_{\lambda =1}-G_{\mathrm{I}\! \mathrm{P}}^{h_{1}h_{2}}(\tilde{x}_{\lambda }^{+},\tilde{x}_{\lambda }^{-},s,b)\nonumber \\
 & \times  & F_{\mathrm{remn}}\left( x^{\mathrm{proj}}-\sum _{\mu }\hat{x}_{\mu }^{+}-\sum _{\lambda }\tilde{x}_{\lambda }^{+}\right) \, F_{\mathrm{remn}}\left( x^{\mathrm{targ}}-\sum _{\mu }\hat{x}_{\mu }^{-}-\sum _{\lambda }\tilde{x}_{\lambda }^{-}\right) ,\label{gamenh} 
\end{eqnarray}
 with \( G_{\mathrm{I}\! \mathrm{P}}^{h_{1}h_{2}} \) being defined earlier,
and with
\begin{eqnarray}
\hat{G}_{\mathrm{I}\! \mathrm{P}}^{h_{1}h_{2}}(x_{\mu }^{+},x_{\mu }^{-},\hat{x}_{\mu }^{+},\hat{x}_{\mu }^{-},s,b)=G_{1\mathrm{I}\! \mathrm{P}}\left( x_{\mu }^{+},x_{\mu }^{-},s,b\right) \, \delta \! (\hat{x}_{\mu }^{+})\, \delta \! (\hat{x}_{\mu }^{-})+\sum _{\sigma =\pm }\sum ^{2}_{i=0}\hat{G}_{3\mathrm{I}\! \mathrm{P}\sigma (i)}^{h_{1}h_{2}}(x_{\mu }^{+},x_{\mu }^{-},\hat{x}_{\mu }^{+},\hat{x}_{\mu }^{-},s,b), &  & \nonumber \\
 &  & \label{x} 
\end{eqnarray}
where we used
\begin{equation}
\label{x}
x^{\mathrm{proj}/\mathrm{targ}}=1-\sum x^{\pm }_{\mu }.
\end{equation}
 The variables \( x_{\mu }^{\pm } \) are now the momentum fractions for the
individual cut contributions, which define the energy for the production of
secondary hadrons resulting from a given elementary interaction. The expressions
for the functions \( \hat{G}_{3\mathrm{I}\! \mathrm{P}\sigma (i)}^{h_{1}h_{2}} \)
are obtained from eqs. (\ref{ghhtot}, \ref{g-3p}, \ref{sumcut}, \ref{factors}),
by changing the variables properly. Simplest is the case of all Pomerons being
cut, no changing of variables is necessary, we have simply
\begin{eqnarray}
\hat{G}_{3\mathrm{I}\! \mathrm{P}-(2)}^{h_{1}h_{2}}(x^{+},x^{-},\hat{x}^{+},\hat{x}^{-},s,b) & = & -2G_{3\mathrm{I}\! \mathrm{P}-}^{h_{1}h_{2}}(x^{+},x^{-},s,b)\, \delta \! (\hat{x}^{+})\, \delta \! (\hat{x}^{-})\\
 & = & \frac{r_{3\mathrm{I}\! \mathrm{P}}}{4}\, \int d^{2}b_{1}\int ^{x^{+}}_{s_{0}/x^{-}}\! \frac{dx_{12}^{+}}{x_{12}^{+}}\, G^{h_{1}}(x^{+},s^{+},\left| \vec{b}-\vec{b}_{1}\right| )\nonumber \\
 & \times  & \int ^{1}_{0}\! dz^{+}\int ^{x^{-}}_{0}\! dx_{1}^{-}\, G^{h_{2}}(x_{1}^{-},\hat{s}_{1},b_{1})\, G^{h_{2}}(x^{-}-x_{1}^{-},\hat{s}_{2},b_{1})\nonumber \\
 & \times  & \delta \! (\hat{x}^{+})\, \delta \! (\hat{x}^{-}),\label{g-hat-2} 
\end{eqnarray}
For none of the two lower Pomerons being cut, we rename \( x_{12}^{-} \) into
\( x^{-} \)and \( x^{-}-x_{12}^{-} \) into \( \hat{x}^{-} \), so we get
\begin{eqnarray}
\hat{G}_{3\mathrm{I}\! \mathrm{P}-(0)}^{h_{1}h_{2}}\! (x^{+},x^{-},\hat{x}^{+},\hat{x}^{-},s,b) & = & \frac{r_{3\mathrm{I}\! \mathrm{P}}}{8}\int \! d^{2}b_{1}\, \frac{1}{x^{-}}G^{h_{1}}(x^{+},x^{+}x^{-}s,\left| \vec{b}-\vec{b}_{1}\right| )\nonumber \\
 & \times  & \int ^{1}_{0}\! dz^{+}\int ^{\hat{x}^{-}+x^{-}}_{0}\! dx_{1}^{-}\, G^{h_{2}}(x_{1}^{-},x_{1}^{-}\frac{s_{0}}{x^{-}}z^{+}s,b_{1})\label{g-hat-0} \\
 & \times  & G^{h_{2}}(\hat{x}^{-}+x^{-}-x_{1}^{-},(\hat{x}^{-}+x^{-}-x_{1}^{-})\frac{s_{0}}{x^{-}}(1-z^{+})s,b_{1})\, \delta \! (\hat{x}^{+}).\nonumber \label{x} 
\end{eqnarray}
For one of the two lower Pomerons being cut, we rename \( x_{1}^{-} \) into
\( x^{-} \)and \( x^{-}-x_{1}^{-} \) into \( \hat{x}^{-} \), and we get
\begin{eqnarray}
\hat{G}_{3\mathrm{I}\! \mathrm{P}-(1)}^{h_{1}h_{2}}\! (x^{+},x^{-},\hat{x}^{+},\hat{x}^{-},s,b) & = & -\frac{r_{3\mathrm{I}\! \mathrm{P}}}{2}\int \! d^{2}b_{1}\int ^{x^{+}}_{s_{0}/x^{-}}\! \frac{dx_{12}^{+}}{x_{12}^{+}}\, G^{h_{1}}(x^{+},s^{+},\left| \vec{b}-\vec{b}_{1}\right| )\label{g-hat-1} \\
 & \times  & \int ^{1}_{0}\! dz^{+}\, G^{h_{2}}(x^{-},x_{12}^{+}z^{+}x^{-}s,b_{1})\, G^{h_{2}}(\hat{x}^{-},x_{12}^{+}(1-z^{+})\hat{x}^{-}s,b_{1})\, \delta \! (\hat{x}^{+}).\nonumber \label{x} 
\end{eqnarray}
The contributions \( G_{3\mathrm{I}\! \mathrm{P}\! +(i)}^{h_{1}h_{2}} \) can
be obtained via interchanging \( x^{+}\leftrightarrow x^{-},\hat{x}^{+}\leftrightarrow \hat{x}^{-} \)
and \( h_{1}\leftrightarrow h_{2} \) in the above formulas:
\begin{equation}
\label{x}
G_{3\mathrm{I}\! \mathrm{P}\! +(i)}^{h_{1}h_{2}}(x^{+},x^{-},\hat{x}^{+},\hat{x}^{-},s,b)=G_{3\mathrm{I}\! \mathrm{P}\! -(i)}^{h_{2}h_{1}}(x^{-},x^{+},\hat{x}^{-},\hat{x}^{+},s,b).
\end{equation}
 We define as in the usual eikonal case the virtual emission function 
\begin{eqnarray}
\Phi _{h_{1}h_{2}}\left( x^{\mathrm{proj}},x^{\mathrm{targ}},s,b\right)  & = & \sum _{l}\int \prod ^{l}_{\lambda =1}d\tilde{x}_{\lambda }^{+}d\tilde{x}_{\lambda }^{-}\, \frac{1}{l!}\prod _{\lambda =1}^{l}-G^{h_{1}h_{2}}_{\mathrm{I}\! \mathrm{P}}(\tilde{x}_{\lambda }^{+},\tilde{x}_{\lambda }^{-},s,b)\nonumber \label{r} \\
 & \times  & F_{\mathrm{remn}}\left( x^{\mathrm{proj}}-\sum _{\lambda }\tilde{x}_{\lambda }^{+}\right) \, F_{\mathrm{remn}}\left( x^{\mathrm{targ}}-\sum _{\lambda }\tilde{x}_{\lambda }^{-}\right) ,\label{x} 
\end{eqnarray}
which allows to write the profile functions eq.\ (\ref{gamenh}) as
\begin{eqnarray}
\gamma _{h_{1}h_{2}}(s,b) & = & \sum _{m}\frac{1}{m!}\int \prod ^{m}_{\mu =1}dx_{\mu }^{+}dx_{\mu }^{-}d\hat{x}_{\mu }^{+}d\hat{x}_{\mu }^{-}\prod ^{m}_{\mu =1}\hat{G}_{\mathrm{I}\! \mathrm{P}}^{h_{1}h_{2}}(x_{\mu }^{+},x_{\mu }^{-},\hat{x}_{\mu }^{+},\hat{x}_{\mu }^{-},s,b)\nonumber \\
 & \times  & \Phi _{h_{1}h_{2}}\left( x^{\mathrm{proj}}-\sum _{\mu }\hat{x}_{\mu }^{+},x^{\mathrm{targ}}-\sum _{\mu }\hat{x}_{\mu }^{-},s,b\right) ,\label{x} 
\end{eqnarray}
 which may be approximated as
\begin{eqnarray}
\gamma _{h_{1}h_{2}}(s,b) & = & \sum _{m}\frac{1}{m!}\int \prod ^{m}_{\mu =1}dx_{\mu }^{+}dx_{\mu }^{-}\prod ^{m}_{\mu =1}\widehat{\widehat{G}}_{\mathrm{I}\! \mathrm{P}}^{h_{1}h_{2}}(x_{\mu }^{+},x_{\mu }^{-},x^{\mathrm{proj}},x^{\mathrm{targ}},s,b)\nonumber \\
 &  & \qquad \times \quad \Phi _{h_{1}h_{2}}\left( x^{\mathrm{proj}},x^{\mathrm{targ}},s,b\right) ,\label{gamma-approx} 
\end{eqnarray}
 with
\begin{eqnarray}
\widehat{\widehat{G}}_{\mathrm{I}\! \mathrm{P}}^{h_{1}h_{2}}(x^{+},x^{-},x^{\mathrm{proj}},x^{\mathrm{targ}},s,b)=\frac{1}{F_{\mathrm{remn}}\left( x^{\mathrm{proj}}\right) \, F_{\mathrm{remn}}\left( x^{\mathrm{targ}}\right) }\qquad \qquad \qquad \qquad \qquad  &  & \label{g-hathat} \\
\times \; \int d\hat{x}^{+}d\hat{x}^{-}\hat{G}_{\mathrm{I}\! \mathrm{P}}^{h_{1}h_{2}}(x^{+},x^{-},\hat{x}^{+},\hat{x}^{-},s,b)F_{\mathrm{remn}}\left( x^{\mathrm{proj}}-\hat{x}^{+}\right) \, F_{\mathrm{remn}}\left( x^{\mathrm{targ}}-\hat{x}^{-}\right) . &  & \nonumber \label{x} 
\end{eqnarray}
Based on eq.\ (\ref{gamma-approx}), we proceed as in the eikonal case. We unitarize
the theory by replacing \( \Phi _{h_{1}h_{2}} \) by \( \Phi _{\mathrm{u}\, _{h_{1}h_{2}}} \)
in a complete analogy to the eikonal model. The numerical Markov chain procedures
have to be modified slightly due to the fact that the \( \widehat{\widehat{G}} \)
functions contain \( x^{\mathrm{proj}} \) and \( x^{\mathrm{targ}} \) as arguments.
One can no longer restrict oneself to considering one single site of the interaction
matrix, since changing \( x^{+} \) and \( x^{-} \)modifies as well \( x^{\mathrm{proj}} \)
and \( x^{\mathrm{targ}} \) and affects therefore the other sites as well (in
case of nucleus-nucleus all the sites related to the same projectile and target
nucleon). But this does not pose major problems.

\section{Important Features of Enhanced Diagrams}

\begin{figure}[htb]
{\par\centering \resizebox*{!}{0.15\textheight}{\includegraphics{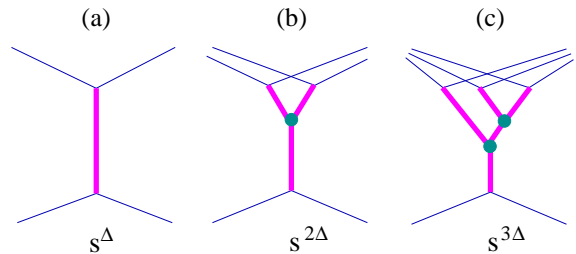}} \par}

\caption{Energy dependence of Pomeron diagrams.\label{enh3}}
\end{figure}

The amplitude corresponding to a single Pomeron exchange as shown at fig.\ \ref{enh3}(a)
behaves as a function of energy approximately as \( s^{\Delta } \), where \( s \)
is the c.m. energy squared for hadron-hadron interaction and \( \Delta  \)
is some effective exponent. At the same moment, the amplitude of the so-called
\( Y \)-diagram shown in fig.\ \ref{enh3}(b) increases asymptotically as \( s^{2\Delta } \)
, as can be seen from eq.\ (\ref{g-3p}). The amplitude corresponding to the
diagram in fig.\ \ref{enh3}(c) behaves as \( s^{3\Delta } \).
This indicates very important property of enhanced diagrams, namely that they
increase with energy much faster than the usual eikonal ones. 

In the following, we discuss exclusively enhanced \( Y \)-diagrams for given
hadron types \( h_{1} \) and \( h_{2} \), and to simplify the notation, we
use simply \( G \) to refer to the corresponding profile function, omitting
all indices referring to the initial hadron types and the Pomeron type.

\begin{figure}[htb]
{\par\centering \resizebox*{!}{0.15\textheight}{\includegraphics{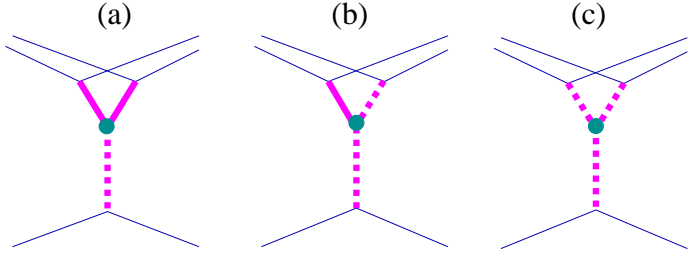}} \par}

\caption{The different cuts of the \protect\( Y\protect \)-diagram. \label{ycut}}
\end{figure}

In order to calculate contributions of enhanced graphs to the total interaction
cross section, one has to consider different cuts of elastic scattering diagrams.
We have
\begin{equation}
\label{sumcut}
G\equiv \frac{1}{2s}2\mathrm{Im}\tilde{T}(s,t=0)=G_{(0)}+G_{(1)}+G_{(2)},
\end{equation}
where \( G_{(i)} \) refers to the different cut diagrams, as shown in fig.\ \ref{ycut}.
We use our convention employed already earlier to plot cut Pomerons as dashed
and uncut ones as full vertical lines. The diagram in fig.\ \ref{ycut}(a) gives
rise to the process of high mass target diffraction (\( G_{(0)} \)), the diagram
in fig.\ \ref{ycut}(b) represents the screening correction to the one cut Pomeron
process (\( G_{(1)} \)), and the diagram in fig \ref{ycut}(c) - the cut Pomeron
fusion process (\( G_{(2)} \)). As discussed in the preceding section, we have
\begin{eqnarray}
G_{(0)} & = & -1G\, ,\nonumber \\
G_{(1)} & = & +4G\, ,\label{factors} \\
G_{(2)} & = & -2G\, ,\nonumber 
\end{eqnarray}
the sum being therefore equal to \( G \), as it should be. Since \( G \) is
negative, the first order contribution to the inelastic cross section is negative,
so another very important property of enhanced diagrams is the suppression of
the increase of the inelastic cross section with energy.

\begin{figure}[htb]
{\par\centering \resizebox*{!}{0.13\textheight}{\includegraphics{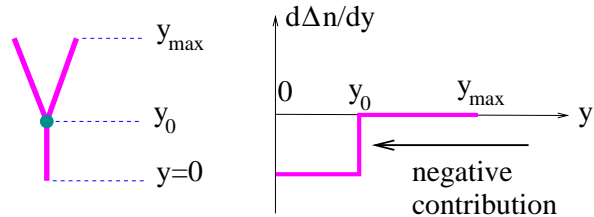}} \par}

\caption{Effect on inclusive spectra.\label{incl}}
\end{figure}

A remarkable feature of enhanced diagrams is connected to their effect on the
inclusive particle spectra. In particular, if one assumes (for a qualitative
discussion) that each cut Pomeron gives rise to a flat rapidity distribution
of secondary hadrons, \( dn/dy=\rho _{0} \), the sum of all three contributions
gives the screening correction to the inclusive particle spectrum as
\begin{equation}
\label{x}
\frac{d\Delta n}{dy}\propto -1G\, \times \, 0\rho _{0}+4G\, \times \, 1\rho _{0}-2G\, \times \, 2\rho _{0}=0,\; \; y>y_{0}
\end{equation}
 and
\begin{equation}
\label{x}
\frac{d\Delta n}{dy}\propto -1G\, \times \, \rho _{0}+4G\, \times \, \rho _{0}-2G\, \times \, \rho _{0}=G\, \rho _{0}<0,\; \; y<y_{0},
\end{equation}
where \( y_{0} \) is the rapidity position of the triple Pomeron vertex, see
fig.\ \ref{incl}. The contribution is negative, because \( G \) is negative.
Thus enhanced diagrams give rise to screening corrections to secondary hadron
spectra only in restricted regions of the kinematical phase space; contributions
of different cuts exactly cancel each other in the region of rapidity space
where two or more Pomerons are exchanged in parallel (in case of the diagram
of fig.\ \ref{incl} for \( y>y_{0} \)) \cite{abr73}. Therefore another important
effect of Pomeron-Pomeron interactions is the modification of secondary hadron
spectra, being mainly suppressed in the fragmentation regions of the interaction:
close to \( y=0 \) from the \( Y \)-diagrams and close to \( y_{\mathrm{max}} \)
for the inverted \( Y \)-diagrams. That explains the great importance of enhanced
diagrams for the solution of the unitarity problems inherent for the pure eikonal
scheme and for the construction of a consistent unitary approach to hadronic
interactions at very high energies (see the discussion in chapter 3).

Another important effect is the considerable increase of fluctuations of hadronic
interactions. Let us for a moment consider the indices indicating Pomeron types:
\( G_{3\mathrm{I}\! \mathrm{P}+} \) for the \( Y \)-diagram as discussed above,
\( G_{3\mathrm{I}\! \mathrm{P}-} \) for the corresponding inverted diagram,
and \( G_{1\mathrm{I}\! \mathrm{P}}(s,x^{+},x^{-}) \) for the normal Pomeron.
The full contribution (so far) is
\begin{equation}
\label{x}
G_{\mathrm{I}\! \mathrm{P}}=G_{1\mathrm{I}\! \mathrm{P}}+G_{3\mathrm{I}\! \mathrm{P}\! -}+G_{3\mathrm{I}\! \mathrm{P}\! +}.
\end{equation}
 Using the fact that \( G_{3\mathrm{I}\! \mathrm{P}\pm } \) can be written
as the sum of the different cut contributions \( G_{3\mathrm{I}\! \mathrm{P}\pm (i)} \),
we get
\begin{eqnarray*}
G_{\mathrm{I}\! \mathrm{P}} & = & G_{1\mathrm{I}\! \mathrm{P}}+\sum _{i}G_{3\mathrm{I}\! \mathrm{P}\! -(i)}+\sum _{i}G_{3\mathrm{I}\! \mathrm{P}\! +(i)},
\end{eqnarray*}
which may be written as
\begin{eqnarray}
G_{\mathrm{I}\! \mathrm{P}} & = & \left\{ G_{1\mathrm{I}\! \mathrm{P}}+G_{3\mathrm{I}\! \mathrm{P}\! -(1)}+G_{3\mathrm{I}\! \mathrm{P}\! +(1)}\right\} +\left\{ G_{3\mathrm{I}\! \mathrm{P}\! -(0)}+G_{3\mathrm{I}\! \mathrm{P}\! +(0)}\right\} +\left\{ G_{3\mathrm{I}\! \mathrm{P}\! -(2)}+G_{3\mathrm{I}\! \mathrm{P}\! +(2)}\right\} .\nonumber \label{x} \\
 &  & 
\end{eqnarray}
This means that we have three contributions: a modified one cut Pomeron exchange,
with probability \( w_{\mathrm{one}} \), the high mass target (see fig.\ \ref{ycut}(a))
and projectile diffraction, with probability \( w_{\mathrm{diff}} \), and the
process of Pomeron fusion of fig.\ \ref{ycut}(c), with probability \( w_{\mathrm{fusion}} \),
where the probabilities are given as
\begin{eqnarray}
\: w_{\mathrm{one}}=\frac{G_{1\mathrm{I}\! \mathrm{P}}+G_{3\mathrm{I}\! \mathrm{P}\! -(1)}+G_{3\mathrm{I}\! \mathrm{P}\! +(1)}}{G_{\mathrm{I}\! \mathrm{P}}},\: w_{\mathrm{diff}}=\frac{G_{3\mathrm{I}\! \mathrm{P}\! -(0)}+G_{3\mathrm{I}\! \mathrm{P}\! +(0)}}{G_{\mathrm{I}\! \mathrm{P}}},\: w_{\mathrm{fusion}}=\frac{G_{3\mathrm{I}\! \mathrm{P}\! -(2)}+G_{3\mathrm{I}\! \mathrm{P}\! +(2)}}{G_{\mathrm{I}\! \mathrm{P}}}. &  & \nonumber \\
 &  & 
\end{eqnarray}
The two latter processes result in correspondingly much smaller and much larger
values of the hadron multiplicity than for the one Pomeron process. 

The problem of consistent treatment of Pomeron-Pomeron interactions was addressed
already in \cite{cap76,kai86}. The number of diagrams which contribute essentially
to the interaction characteristics increases fast with the energy. Therefore,
one has to develop a suitable method to take into account the necessary contributions
to the forward scattering amplitude, the latter one being related via the optical
theorem to the total cross section of the reaction and to the weights for particular
configurations of the interaction (via the ``cutting'' procedure). Such a
scheme is still under development and our current goal was the proper treatment
of some lowest order enhanced diagrams. Thus we proposed a minimal modification
of the standard eikonal scheme, which allowed us to obtain a consistent description
of hadronic interactions in the range of c.m. energies from some ten GeV till
few thousand TeV. Already this minimal scheme allows to achieve partly the goals
mentioned above: the slowing down of the energy increase of the interaction
cross section and the non-AGK-type modification of particle spectra, as well
as the improvement of the description of the multiplicity and inelasticity fluctuations
in hadron-hadron interaction.

An important question exists concerning the nature of the triple-Pomeron coupling.
As discussed above, we used the perturbative treatment for the part of a parton
cascade developing in the region of parton virtualities bigger than some cutoff
\( Q_{0}^{2} \), whereas the region of smaller virtualities is treated phenomenologically,
based on the soft Pomeron. There was an argumentation in \cite{gri83} that
the triple Pomeron coupling is perturbative and therefore can be described on
the basis of the QCD techniques. At the same time, it was shown in \cite{mue86}
that such perturbative coupling, corresponding to non-small parton virtualities,
would result in negligible contribution to the basic interaction characteristics,
in particular, to the proton structure function \( F_{2} \). The latter result
was confirmed experimentally by HERA measurements, where no real shoulder in
the behavior of \( F_{2}(x,Q^{2}) \) (predicted in \cite{gri83}) was found
in the limit \( x\rightarrow 0 \). Another argument in favor of the smallness
of the perturbative Pomeron-Pomeron coupling comes from the HERA diffractive
data, where the proportion of diffractive type events (with a large rapidity
gap in secondary hadron spectra) appeared to be nearly independent on the virtuality
of the virtual photon probe. This implies that that the Pomeron self-interaction
is rather inherent to the non-perturbative initial condition for the QCD evolution
than to the dynamical evolution itself. Therefore we assumed that the Pomerons
interact with each other in the non-perturbative region of parton virtualities
\( Q^{2}<Q_{0}^{2} \) and considered it as the interaction \textbf{between
soft Pomerons}. In our scheme the relatively big value of the soft triple-Pomeron
coupling results in the screening corrections which finally prevent the large
increase of parton densities in the small \( x \) limit and restore the unitarity,
thus leaving a little room for higher twist effects in the perturbative part
of the interaction.

\cleardoublepage

\chapter{Parton Configurations}

In this section, we consider the generation of parton configurations in nucleus-nucleus
(including proton-proton) scattering for a given interaction configuration,
which has already been determined, as discussed above. So, the numbers \( m_{k} \)
of elementary interactions per nucleon-nucleon pair \( k \) are known, as well
as the light cone momentum fractions \( x_{k\mu }^{+} \) and \( x_{k\mu }^{-} \)
of each elementary interaction of the pair \( k \). A parton configuration
is specified by the number of partons, their types and momenta.

\section{General Procedure of Parton Generation}

We showed earlier that the inelastic cross section may be written as

\begin{equation}
\sigma _{\mathrm{inel}}=\int d^{2}b\sum _{m}\int dX^{+}dX^{-}\Omega (m,X^{+},X^{-}),
\end{equation}
 where \( \{m,X^{+},X^{-}\} \) represents an interaction configuration. The
function \( \Omega  \) is known (see eq.\ (\ref{omega-ab-bas})) and is interpreted
as the probability distribution for interaction configurations. For each individual
elementary interaction a term \( G_{1\mathrm{I}\! \mathrm{P}}^{h_{1}h_{2}} \)
appears in the formula for \( \Omega  \), where the function \( G_{1\mathrm{I}\! \mathrm{P}}^{h_{1}h_{2}} \)
itself can be expressed in terms of contributions of different parton configurations.
Namely the \( QCD \) evolution function \( E^{ij}_{\QCD } \), which enters
into the formula for the elementary interaction contribution \( G_{1\mathrm{I}\! \mathrm{P}}^{h_{1}h_{2}} \),
is the solution of a ladder equation, where adding a ladder rung corresponds
to an integration over the momenta of the corresponding resolvable parton emitted.
The complete evolution function is therefore a sum over \( n \)-rung ladder
contributions, where the latter one can be written as an integration over \( n \)
parton momenta. So we have
\begin{equation}
\label{x}
\Omega (m,X^{+},X^{-})=\prod _{k=1}^{AB}\prod _{\mu =1}^{m_{k}}\sum _{\tau =1}^{t}\sum _{\nu =1}^{n_{k\mu \tau }}\int d^{3}p_{k\mu \tau \nu }\: \Psi (\{p_{k\mu \tau \nu }\}),
\end{equation}
where \( t \) is the number of Pomeron types (soft, sea-sea, ...), and \( n_{k\mu \tau } \)
the number of partons for the \( \mu ^{\mathrm{th}} \) interaction of the pair
\( k \) in case of Pomeron type \( \tau  \). We interpret
\begin{equation}
\label{x}
\frac{\Psi (\{p_{k\mu \tau \nu }\})}{\Omega (m,X^{+},X^{-})}
\end{equation}
 as the probability distribution for parton configurations for a given interaction
configuration \( \{m,X^{+},X^{-}\} \). The Monte Carlo method provides a convenient
tool for treating such multidimensional distributions: with \( \Theta  \) being
known (see chapter 2 and the discussion below), one generates parton configurations
according to this distribution. 

We want to stress that the parton generation is also based on the master formula
eq.\ (\ref{omega-ab-bas}), no new elements enter. 

In the following, we want to sketch the generation of parton configurations,
technical details are provided in the next section. Let us consider a particular
elementary interaction with given light cone momentum fractions \( x^{+} \)
and \( x^{-} \) and given impact parameter difference \( b \) between the
corresponding pair of interacting nucleons, for a fixed primary energy squared
\( s \). 

For the sake of simplicity, we discuss here the procedure without the triple-Pomeron
contribution, with the corresponding generalization being done in appendix \ref{ax-c-4}.
We have to start with specifying the type of elementary interaction (soft, semi-hard,
or valence type). The corresponding probabilities are

\begin{eqnarray}
 &  & G^{h_{1}h_{2}}_{\mathrm{soft}}(x^{+},x^{-},s,b)/G_{1\mathrm{I}\! \mathrm{P}}^{h_{1}h_{2}}(x^{+},x^{-},s,b),\nonumber \\
 &  & G^{h_{1}h_{2}}_{\mathrm{sea}-\mathrm{sea}}(x^{+},x^{-},s,b)/G_{1\mathrm{I}\! \mathrm{P}}^{h_{1}h_{2}}(x^{+},x^{-},s,b),\nonumber \\
 &  & G^{h_{1}h_{2}}_{\mathrm{val}-\mathrm{val}}(x^{+},x^{-},s,b)/G_{1\mathrm{I}\! \mathrm{P}}^{h_{1}h_{2}}(x^{+},x^{-},s,b),\label{g-pom-basic} \\
 &  & G^{h_{1}h_{2}}_{\mathrm{val}-\mathrm{sea}}(x^{+},x^{-},s,b)/G_{1\mathrm{I}\! \mathrm{P}}^{h_{1}h_{2}}(x^{+},x^{-},s,b),\nonumber \\
 &  & G^{h_{1}h_{2}}_{\mathrm{sea}-v\mathrm{al}}(x^{+},x^{-},s,b)/G_{1\mathrm{I}\! \mathrm{P}}^{h_{1}h_{2}}(x^{+},x^{-},s,b).\nonumber 
\end{eqnarray}

In the case of a soft elementary interaction, no perturbative parton emission
takes place. Therefore we are left with the trivial parton configuration, consisting
from the initial active partons -- hadron constituents -- to which the Pomeron
is attached.

Let us now consider a semi-hard contribution. We obtain the desired probability
distributions from the explicit expressions for \( G_{\mathrm{sea}-\mathrm{sea}}^{h_{1}h_{2}} \).
For given \( x^{+} \), \( x^{-} \), and \( b \) we have
\begin{eqnarray}
G_{\mathrm{sea}-\mathrm{sea}}^{h_{1}h_{2}}(x^{+},x^{-},s,b) & \propto  & \int _{0}^{1}\! dz^{+}\int _{0}^{1}\! dz^{-}\, \sum _{ij}E_{\mathrm{soft}}^{i}\left( z^{+}\right) \, E_{\mathrm{soft}}^{j}\left( z^{-}\right) \, \sigma _{\mathrm{hard}}^{ij}(z^{+}z^{-}\hat{s},Q_{0}^{2})\label{2g-semi} \\
 & \times  & \; \frac{1}{4\pi \, \lambda ^{h_{1}h_{2}}_{\mathrm{soft}}(1/(z^{+}z^{-}))}\exp \! \left( -\frac{b^{2}}{4\lambda ^{h_{1}h_{2}}_{\mathrm{soft}}\! \left( 1/(z^{+}z^{-})\right) }\right) \, F^{h_{1}}_{\mathrm{part}}(x^{+})\, F^{h_{2}}_{\mathrm{part}}(x^{-})\nonumber 
\end{eqnarray}
 with \( \hat{s}=x^{+}x^{-}s \), \( \lambda ^{h_{1}h_{2}}_{\mathrm{soft}\! }(\xi )=R_{h_{1}}^{2}+R_{h_{2}}^{2}+\alpha '_{_{\mathrm{soft}}}\ln \! \xi , \)
and 
\begin{eqnarray}
\sigma _{\mathrm{hard}}^{ij}(\hat{s},Q_{0}^{2}) & = & K\, \sum _{kl}\int dx_{B}^{+}dx_{B}^{-}dp_{\bot }^{2}{d\sigma _{\mathrm{Born}}^{kl}\over dp_{\bot }^{2}}(x_{B}^{+}x_{B}^{-}\hat{s},p_{\bot }^{2})\nonumber \\
 & \times  & E_{\mathrm{QCD}}^{ik}(x_{B}^{+},Q_{0}^{2},M_{F}^{2})\, E_{\mathrm{QCD}}^{jl}(x_{B}^{-},Q_{0}^{2},M_{F}^{2})\theta \! \left( M_{F}^{2}-Q^{2}_{0}\right) ,\label{sig-ijhard} 
\end{eqnarray}
 representing the perturbative parton-parton cross section, where both initial
partons are taken at the virtuality scale \( Q^{2}_{0} \); we choose the factorization
scale as \( M_{\mathrm{F}}^{2}=p_{\bot }^{2}/4 \). The integrand of eq.\ (\ref{2g-semi})
serves as the probability distribution to generate the momentum fractions \( x_{1}^{\pm }=x^{\pm }z^{\pm } \)
and the flavors \( i \) and \( j \) of the initial partons for the parton
ladder. 

The knowledge of the initial conditions -- the momentum fractions \( x_{1}^{\pm } \)
and the starting virtuality \( Q^{2}_{0} \) for the ``first partons'' as
well as the flavors \( i \) and \( j \) -- allows us to reconstruct the complete
ladder, based on the eq.\ (\ref{sig-ijhard}) and on the evolution equations
(\ref{eqcd-ini}-\ref{ap-res}) for \( E_{\mathrm{QCD}}^{ij} \). To do so,
we generalize the definition of the parton-parton cross section \( \sigma _{\mathrm{hard}}^{ij} \)
to arbitrary virtualities of the initial partons, defining 
\begin{eqnarray}
\sigma _{\mathrm{hard}}^{ij}(\hat{s},Q_{1}^{2},Q_{2}^{2}) & = & K\, \sum _{kl}\int dx_{B}^{+}dx_{B}^{-}dp_{\bot }^{2}{d\sigma _{\mathrm{Born}}^{kl}\over dp_{\bot }^{2}}(x_{B}^{+}x_{B}^{-}\hat{s},p_{\bot }^{2})\label{sigma-ij-12} \\
 & \times  & E_{\mathrm{QCD}}^{ik}(x_{B}^{+},Q_{1}^{2},M_{\mathrm{F}}^{2})\, E_{\mathrm{QCD}}^{jl}(x_{B}^{-},Q_{2}^{2},M_{\mathrm{F}}^{2})\, \Theta \! \left( M_{\mathrm{F}}^{2}-\max \! \left[ Q_{1}^{2},Q_{2}^{2}\right] \right) \nonumber 
\end{eqnarray}
 and 
\begin{eqnarray}
\sigma _{\mathrm{ord}}^{ij}(\hat{s},Q_{1}^{2},Q_{2}^{2}) & = & K\, \sum _{k}\int dx_{B}^{+}dx_{B}^{-}dp_{\bot }^{2}{d\sigma _{\mathrm{Born}}^{kj}\over dp_{\bot }^{2}}(x_{B}^{+}x_{B}^{-}\hat{s},p_{\bot }^{2})\label{sigma-ij-ord} \\
 & \times  & E_{\mathrm{QCD}}^{ik}(x_{B}^{+},Q_{1}^{2},M_{\mathrm{F}}^{2},w^{+})\, \Delta ^{j}(Q_{2}^{2},M_{\mathrm{F}}^{2})\, \Theta \! \left( M_{\mathrm{F}}^{2}-\max \! \left[ Q_{1}^{2},Q_{2}^{2}\right] \right) \nonumber 
\end{eqnarray}
 representing the full ladder contribution (\( \sigma _{\mathrm{hard}} \))
and the contribution, corresponding to the ordering of parton virtualities towards
the end of the ladder, i.e. to the case of parton \( j \), involved into the
highest virtuality Born process (\( \sigma _{\mathrm{ord}} \)).  We calculate
and tabulate \( \sigma _{\mathrm{hard}} \) and \( \sigma _{\mathrm{ord}} \)
initially, so that we can use them via interpolation to generate partons. The
generation of partons is done in an iterative fashion based on  the following
equations:
\begin{eqnarray}
\sigma _{\mathrm{hard}}^{ij}(\hat{s},Q_{1}^{2},Q_{2}^{2}) & = & \sum _{k}\int \frac{dQ^{2}}{Q^{2}}\int d\xi \, \Delta ^{i}(Q_{1}^{2},Q^{2})\, \frac{\alpha _{s}}{2\pi }\, P_{i}^{k}(\xi )\, \sigma _{\mathrm{hard}}^{kj}(\xi \hat{s},Q^{2},Q_{2}^{2})\label{sigma-lad-iter} \\
 & + & \sigma _{\mathrm{ord}}^{ji}(\hat{s},Q_{2}^{2},Q_{1}^{2})\nonumber 
\end{eqnarray}
and

\begin{eqnarray}
\sigma _{\mathrm{ord}}^{ij}(\hat{s},Q_{1}^{2},Q_{2}^{2}) & = & \sum _{k}\int \frac{dQ^{2}}{Q^{2}}\int d\xi \, \Delta ^{i}(Q_{1}^{2},Q^{2})\, \frac{\alpha _{s}}{2\pi }\, P_{i}^{k}(\xi )\, \sigma _{\mathrm{ord}}^{kj}(\xi \hat{s},Q^{2},Q_{2}^{2})\label{sigma-ord-iter} \\
 & + & \sigma _{\mathrm{Born}}^{ij}(\hat{s},Q_{1}^{2},Q_{2}^{2})\nonumber 
\end{eqnarray}
 Here, \( \sigma _{\mathrm{Born}}^{ij} \) gives the contribution of the configuration
without any resolvable emission before the highest virtuality Born process:
\begin{eqnarray}
\sigma _{\mathrm{Born}}^{ij}(\hat{s},Q_{1}^{2},Q_{2}^{2}) & = & K\int dp_{\bot }^{2}\frac{d\sigma _{\mathrm{Born}}^{ij}}{dp_{\bot }^{2}}(\hat{s},p_{\bot }^{2})\label{sigma-ij-born} \\
 & \times  & \Delta ^{i}(Q_{1}^{2},M_{\mathrm{F}}^{2})\, \Delta ^{j}(Q_{2}^{2},M_{\mathrm{F}}^{2})\, \Theta \! \left( M_{\mathrm{F}}^{2}-\max \! \left[ Q_{1}^{2},Q_{2}^{2}\right] \right) \nonumber 
\end{eqnarray}
The procedure is described in detail in the next section.

In the case of elementary interactions involving valence quarks, the method
is almost identical. In that case, the corresponding momentum fractions \( x_{1}^{\pm } \)
are the ones of valence quarks, \( x_{1}^{\pm }=x_{q}^{\pm } \), to be determined
according to the corresponding integrands in the expressions for \( G^{h_{1}h_{2}}_{\mathrm{val}} \),
\( G^{h_{1}h_{2}}_{\mathrm{val}-\mathrm{sea}} \), \( G^{h_{1}h_{2}}_{\mathrm{sea}-\mathrm{val}} \),
see eqs. (\ref{g-valence}-\ref{g-semi-val}). For example, in the case of both
hadron constituents being valence quarks, one generates momentum fractions \( x_{1}^{\pm } \)
and valence quark flavors \( i \), \( j \) with the distribution (up to a
normalization constant)
\begin{equation}
D^{h_{1}h_{2},ij}_{\mathrm{val}-\mathrm{val}}(x_{q_{v}}^{+}x_{q_{v}}^{-}s,b)\, \bar{F}^{h_{1},i}_{\mathrm{part}}(x_{q_{v}}^{+},x^{+}-x_{q_{v}}^{+})\, \bar{F}^{h_{2},j}_{\mathrm{part}}(x_{q_{v}}^{-},x^{-}-x_{q_{v}}^{-}),
\end{equation}
see eqs. (\ref{g-valence}), (\ref{d-val-val}). One then proceeds to generate
parton emissions as discussed above.

\section{Generating the Parton Ladder}

We now discuss in detail the generation of the partons in a ladder, starting
from the initial partons (``leg partons'') with flavors \( i \) and \( j \)
and light cone momentum fractions \( x_{1}^{+} \) and \( x_{1}^{-} \). To
simplify the discussion, we will neglect the effects of finite virtualities
and transverse momenta of initial partons in the kinematical formulas so that
the 4-momenta \( k_{1} \) and \( k_{1}' \) of the two leg partons are purely
longitudinal. In the hadron-hadron (nucleus-nucleus) center of mass frame we
have:
\begin{equation}
\begin{array}{cc}
k_{1}^{+}=x_{1}^{+}\sqrt{s}/2, & \quad k_{1}^{-}=0,\quad k_{1_{\perp }}=0,\\
k_{1}'^{-}=x_{1}^{-}\sqrt{s}/2, & \quad k_{1}'^{+}=0,\quad k_{1_{\perp }}'=0.
\end{array}
\end{equation}
The invariant mass squared of the ladder is \( \hat{s}=(k_{1}+k_{1}')^{2} \). 

One first generates all resolvable partons emitted at one side of the ladder
before the hardest process (for the definiteness we start with the leg parton
\( i \) moving in the forward direction). At each step one decides whether
there is any resolvable emission at the forward end of the ladder before the
hardest process. An emission is done with the probability  
\begin{equation}
\mathrm{prob}(\mathrm{forward}\, \mathrm{emission})=\left( \sigma _{\mathrm{hard}}^{ij}(\hat{s},Q_{1}^{2},Q_{2}^{2})-\sigma _{\mathrm{ord}}^{ji}(\hat{s},Q_{2}^{2},Q_{1}^{2})\right) /\sigma _{\mathrm{hard}}^{ij}(\hat{s},Q_{1}^{2},Q_{2}^{2}).
\end{equation}

In case of an emission, the generation of light cone momentum fraction \( \xi  \)
and momentum transfer squared \( Q^{2} \) for the current parton branching
is done -- up to a normalization constant -- according to the integrand of \( (\sigma _{\mathrm{hard}}^{ij}-\sigma _{\mathrm{ord}}^{ji}) \),
\begin{equation}
\label{f-branch}
\mathrm{prob}(\xi ,Q^{2})\propto \frac{1}{Q^{2}}\, \Delta ^{i}(Q_{1}^{2},Q^{2})\, \frac{\alpha _{s}}{2\pi }\, \sum _{i'}P_{i}^{i'}(\xi )\, \sigma _{\mathrm{hard}}^{i'j}(\xi \hat{s},Q^{2},Q_{2}^{2})
\end{equation}
 see eq.\ (\ref{sigma-lad-iter}). Here the emitted \( s \)-channel parton
gets the share \( 1-\xi  \) of the parent (leg) parton light cone momentum
\( k^{+} \) and the transverse momentum squared \( p^{2}_{\bot }\equiv (1-\xi )Q^{2} \)
. To leading logarithmic accuracy, the initial parton virtuality is neglected
in the branching probability eq.\ (\ref{f-branch}), because of \( Q_{1}^{2}\ll Q^{2} \).
Generating randomly the polar angle \( \varphi  \) for the emission, one reconstructs
the 4-vector \( p \) of the final \( s \)-channel parton as 
\begin{equation}
p^{+}=(1-\xi )k_{1}^{+},\quad p^{-}=p^{2}_{\bot }/((1-\xi )k_{1}^{+}),\quad \vec{p}_{\bot }=\left( \begin{array}{c}
p_{\bot }\cos \varphi \\
p_{\bot }\sin \varphi 
\end{array}\right) .
\end{equation}
 The remaining ladder after the parton emission is now characterized by the
mass squared \( \hat{s}'=(k_{1}+k_{1}'-p)^{2}\simeq \xi \hat{s} \) and the
initial virtualities \( {Q'}^{2}_{1}=Q^{2} \) and \( Q_{2}^{2} \). The flavor
\( i' \) of the new leg parton is generated according to the corresponding
weights in eq.\ (\ref{f-branch}), properly normalized given as 
\begin{equation}
\mathrm{prob}(i')=\frac{P_{i}^{i'}(\xi )\, \sigma _{\mathrm{hard}}^{i'j}(\hat{s}',Q^{2},Q_{2}^{2})}{\sum _{l}P_{i}^{l}(\xi )\, \sigma _{\mathrm{hard}}^{lj}(\hat{s}',Q^{2},Q_{2}^{2})},
\end{equation}
where \( \sigma _{\mathrm{hard}}^{i'j}(\xi \hat{s},Q^{2},Q_{2}^{2}) \) is the
parton cross section (\ref{sigma-ij-12}) for the new ladder. One then renames
\( \hat{s}' \), \( i' \), and \( Q_{1}'^{2} \) into \( \hat{s} \), \( i \),
and \( Q_{1}^{2} \) and repeats the above procedure. 

In case of no forward emission, the generation of all resolvable parton emissions
at the forward side of the ladder has been completed. 

One then proceeds to generate all resolvable parton emissions for the backward
side of the ladder, starting from the original leg parton \( j \) of virtuality
\( Q_{2}^{2}=Q_{0}^{2} \), by using a corresponding recursive algorithm, now
based on eq.\ (\ref{sigma-ord-iter}). On the other end of the ladder, we have
(after renaming) parton \( i \) with the virtuality \( Q_{1}^{2} \). One decides
whether there is any resolvable emission before the hardest process, where the
probability of an emission is given as
\begin{equation}
\mathrm{prob}(\mathrm{backward}\, \mathrm{emission})=\left( \sigma _{\mathrm{ord}}^{ji}(\hat{s},Q_{2}^{2},Q_{1}^{2})-\sigma _{\mathrm{Born}}^{ij}(\hat{s},Q_{1}^{2},Q_{2}^{2})\right) /\sigma _{\mathrm{ord}}^{ji}(\hat{s},Q_{2}^{2},Q_{1}^{2}).
\end{equation}
 In case of an emission, the generation of the fraction \( \xi  \) of the light
cone momentum \( k_{1}'^{-} \), and of the momentum transfer squared \( Q^{2} \)
is done -- up to a normalization constant -- according to the integrand of \( (\sigma _{\mathrm{ord}}^{ji} \)
- \( \sigma _{\mathrm{Born}}^{ij}) \), 
\begin{equation}
\label{f-ord-branch}
\mathrm{prob}(\xi ,Q^{2})\propto \frac{1}{Q^{2}}\Delta ^{j}(Q_{2}^{2},Q^{2})\, \frac{\alpha _{s}}{2\pi }\, \sum _{j'}P_{j}^{j'}(\xi )\, \sigma _{\mathrm{ord}}^{j'i}(\xi \hat{s},Q^{2},Q_{1}^{2}),
\end{equation}
see eq.\ (\ref{sigma-ord-iter}). The flavor \( j' \) of the new leg parton
is defined according to the partial contributions in (\ref{f-ord-branch}).
The generation of resolvable parton emissions is completed when the iterative
procedure stops, with the probability 
\[
1-\mathrm{prob}(\mathrm{backward}\, \mathrm{emission}).\]
 Note, that all parton emissions are simulated in the original Lorentz frame,
where the original leg partons (the initial partons for the perturbative evolution)
are moving along the \( z \)-axis.

The final step is the generation of the hardest \( 2\rightarrow 2 \) parton
scattering process. In the center of mass system of two partons \( i \) and
\( j \) with center-of-mass energy squared \( \hat{s} \), we simulate the
transverse momentum \( p_{\bot }^{2} \) for the scattering within the limits
(given by the condition \( M_{\mathrm{F}}^{2}=p_{\perp }^{2}/4>\max [Q_{1}^{2},Q_{2}^{2}] \))
\begin{equation}
4\max [Q_{1}^{2},Q_{2}^{2}]<p_{\bot }^{2}<\hat{s}/4
\end{equation}
 according to 
\begin{equation}
\mathrm{prob}(p_{\bot }^{2})\propto {d\sigma _{\mathrm{Born}}^{ij}\over dp_{\bot }^{2}}(\hat{s},p_{\bot }^{2})\Delta ^{i}(Q_{1}^{2},p_{\bot }^{2}/4)\, \Delta ^{j}(Q_{2}^{2},p_{\bot }^{2}),
\end{equation}
where the differential parton-parton cross section is
\begin{equation}
\label{sig-born-dif}
{d\sigma _{\mathrm{Born}}^{ij}\over dp_{\bot }^{2}}(\hat{s},p_{\bot }^{2})=\frac{1}{16\pi \hat{s}^{2}\sqrt{1-4p_{\bot }^{2}/\hat{s}}}\sum _{k,l}\left| M^{ij\rightarrow kl}(\hat{s},p_{\bot }^{2})\right| ^{2}
\end{equation}
with \( |M^{ij\rightarrow k,l}(\hat{s},p_{\bot }^{2})|^{2} \) being the squared
matrix elements of the parton subprocesses \cite{owe86}.

Then we choose a particular subprocess \( ij\rightarrow kl \) according to
its contribution to the differential cross section (\ref{sig-born-dif}), and
reconstruct the 4-momenta \( p_{1} \) and \( p_{2} \) of the final partons
in their center of mass system as 
\begin{equation}
\begin{array}{c}
p_{1}^{+}=z\sqrt{\hat{s}},\quad \quad \quad p_{1}^{-}=p_{\bot }^{2}/(z\sqrt{\hat{s}}),\quad \quad \quad \vec{p}_{1\bot }=\left( \begin{array}{c}
p_{\bot }\cos \varphi \\
p_{\bot }\sin \varphi 
\end{array}\right) ,\quad \\
p_{2}^{+}=(1-z)\sqrt{\hat{s}},\quad p_{2}^{-}=p_{\bot }^{2}/((1-z)\sqrt{\hat{s}}),\quad \vec{p}_{2\bot }=\left( \begin{array}{c}
-p_{\bot }\cos \varphi \\
-p_{\bot }\sin \varphi 
\end{array}\right) ,
\end{array}
\end{equation}
 with 
\begin{equation}
z=\frac{1}{2}\left( 1+\sqrt{1-4p_{\bot }^{2}/\hat{s}}\right) 
\end{equation}
 and a random polar angle \( \varphi  \). We finally boost the momenta to the
original Lorentz frame.

\section{The Time-like Parton Cascade}

The above discussion of how to generate parton configurations is not  yet complete:
the emitted partons are in general off--shell and can therefore  radiate further
partons. This so-called time-like radiation is taken into account  using standard
techniques \cite{sjo84}, to be discussed in the following.

The parton emission from an off-shell parton is done using the so-called DGLAP
evolution equations, which describes the process with the leading logarithmic
accuracy. The splitting probability for the initial parton of type \( j \)
is then given as 

\begin{equation}
\label{for:ap}
\frac{d\mathcal{P}}{\mathcal{P}}=-\frac{dQ^{2}}{Q^{2}}\int \! dz\, \sum _{k}\frac{\alpha _{s}(p_{\perp }^{2})}{2\pi }P^{k}_{j}(z),
\end{equation}
 with the usual Altarelli-Parisi splitting functions \( P^{k}_{j}(z) \). Here
\( Q^{2}=Q_{j}^{2} \) is the virtuality of the parent parton \( j \)
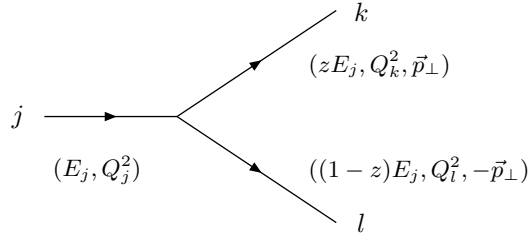
\begin{figure}[htb]
{\par\centering \begin{picture}(200,100) 
\ArrowLine(10,50)(60,50) \Text(0,50)[]{$j$}
\ArrowLine(60,50)(120,10) \Text(130,10)[]{$l$}
\ArrowLine(60,50)(120,90) \Text(130,90)[]{$k$}
\Text(30,30)[]{\small $(E_j,Q_j^2)$} 
\Text(110,70)[l]{\small $(zE_j,Q_k^2,\vec{p}_{\perp})$} 
\Text(110,30)[l]{\small $((1-z)E_j,Q_l^2,-\vec{p}_{\perp})$} 
\end{picture}\par}

\caption{\label{fig:branchement}A branching of a parton \protect\( j\protect \) into
two partons \protect\( k\protect \) and \protect\( l\protect \). The kinematical
variables used to describe the branching are the energies \protect\( E_{i}\protect \),
the virtualities \protect\( Q_{i}^{2}\protect \), and the transverse momenta
\protect\( p_{i_{\perp }}\protect \). }
\end{figure}
and \( z \) is interpreted as the energy fraction carried away by the daughter
parton \( k \). The maximum possible virtuality \( q^{2}_{j\, \mathrm{max}} \)
of the parton \( j \) is given by the virtuality of the parent of \( j \).
One imagines now to decrease the virtuality of \( j \), starting from the maximum
value, such that the DGLAP evolution equations give then the probability \( dP \)
that during a change \( dQ^{2} \) of the virtuality, a parton splits into two
daughter partons \( k \) and \( l \) -- see fig.\ \ref{fig:branchement}.
For the energies of the daughter partons one has 
\begin{equation}
E_{k}=zE_{j},\quad E_{l}=(1-z)E_{j}.
\end{equation}
 Choosing a frame where \( p_{j_{\perp }}=0, \) we have 
\begin{equation}
p_{k_{\perp }}=-p_{l_{\perp }}=p_{\bot },
\end{equation}
 with
\begin{equation}
\label{for:pt2}
p_{\perp }^{2}=\frac{E_{j}^{2}\left( z(1-z)Q_{j}^{2}-zQ_{l}^{2}-(1-z)Q_{k}^{2}\right) -\frac{1}{4}\left( Q_{l}^{2}-Q_{j}^{2}-Q_{k}^{2}\right) ^{2}+Q_{j}^{2}Q_{k}^{2}}{E_{j}^{2}-Q_{j}^{2}}\, .
\end{equation}
 As usual, a cutoff parameter terminates the cascade of parton emissions. We
introduce a parameter \( p^{2}_{\bot \mathrm{fin}} \), which represents the
lower limit for \( p_{\perp }^{2} \) during the evolution, from which we obtain
the lower limit for the virtualities as \( Q_{\min }^{2}=4p_{\bot \mathrm{fin}}^{2} \). 

Let us provide some technical details on the splitting procedure. Using eq.\
(\ref{for:ap}) and applying the rejection method proposed in \cite{sjo84}
we determine the variables \( Q_{j}^{2} \) and \( z \) as well as the flavors
\( k,l \) of the daughter partons -- see appendix \ref{ax-time}. To determine
the 4-momentum \( p_{j} \) of the parent parton we have to distinguish between
three different modes of branching (see figure \ref{fid:troismodes}): 

\begin{enumerate}
\item If \( j \) is the initial parton for the time-like cascade, the energy \( E_{j} \)
is given, and with the obtained value \( Q_{j}^{2} \), we can calculate \( |\vec{p}_{j}|=\sqrt{E^{2}_{j}-Q_{j}^{2}} \).
The direction of \( \vec{p}_{j} \) is obtained from the momentum conservation
constraint for the summary momentum of all partons produced in the current time-like
cascade.
\item If \( j \) is the initial parton for the time-like cascade resulted from the
Born process, together with a partner parton \( j' \), then we know the total
energy \( E_{j}+E_{j'} \) for the two partons. After obtaining the virtualities
\( Q_{j}^{2} \) and \( Q_{j'}^{2} \) from the secondary splittings \( j\rightarrow k,l \)
and \( j'\rightarrow k',l' \) we can use the momentum conservation constraint
in the parton-parton center of mass frame, which gives \( \vec{p}_{j}=-\vec{p}_{j'}=\vec{p} \)
and allows to determine \( |\vec{p}| \). The direction of \( \vec{p} \) can
be chosen randomly.
\item If parton \( j \) is produced in a secondary time-like branching, its energy
\( E_{j}=zE_{\mathrm{parent}} \) as well as the energy of the second ``daughter''
\( E_{j'}=(1-z)E_{\mathrm{parent}} \) are known from the previous branching,
as well as \( Q_{j}^{2} \) and \( Q_{j'}^{2} \) -- from the splittings \( j\rightarrow k,l \)
and \( j'\rightarrow k',l' \), which allows to determine \( \vec{p}_{j_{\perp }}=-\vec{p}_{j'_{\perp }}=\vec{p}_{\perp } \)
using eq.\ (\ref{for:pt2}) in the frame where the parent parton moves along
the \( z \) -axis. In some cases one gets \( p_{_{\perp }}^{2}<0 \); then
the current splitting of the parton \( j \) is rejected and its evolution continues. 
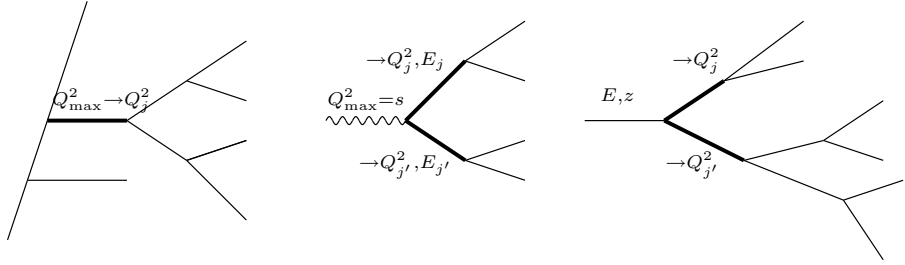
\begin{figure}[htb]
\begin{center}\mbox{\begin{picture}(360,100)(15,45)
\SetScale{1.5}
\SetWidth{0.3}
\Text(50,105)[c]{$\,^{Q^2_{\max}\rightarrow Q^2_j}$}
\Line(10,35)(15,50)\Line(15,50)(40,50) \Line(15,50)(20,65)\Line(20,65)(30,95)

\Line(40,65)(55,75)\Line(55,75)(70,85)
\SetWidth{1} \Line(20,65)(40,65)\SetWidth{0.3} 
\Line(55,75)(70,70) \Line(40,65)(55,55)\Line(55,55)(70,60) \Line(55,55)(70,60)\Line(55,55)(70,40)

\Line(125,80)(140,90) 
\SetWidth{1}\Line(110,65)(125,80) \SetWidth{0.3}
\Line(125,80)(140,75) 
\Photon(90,65)(110,65){1}{6} \Line(125,55)(140,60) 
\Text(150,110)[t]{$\,^{Q^2_{\max}=s}$}
\Text(165,125)[t]{$\,^{\rightarrow Q^2_j,E_j}$}
\Text(165,75)[b]{$ \,^{\rightarrow {Q^2_{j'},E_{j'}}}$}
\SetWidth{1}\Line(110,65)(125,55) \SetWidth{.3}
\Line(125,55)(140,50)

\Text(245,110)[t]{$\,^{E,z}$}
\Text(285,125)[rt]{$\,^{\rightarrow Q^2_j}$}
\Text(285,75)[rb]{$ \,^{\rightarrow {Q^2_{j'}}}$}
\Line(190,75)(210,90) 
\SetWidth{1}\Line(175,65)(190,75) \SetWidth{0.3}
\Line(190,75)(210,80) \Line(155,65)(175,65) \Line(215,60)(230,70) \Line(195,55)(215,60) 
\SetWidth{1}\Line(175,65)(195,55) \SetWidth{0.3}
\Line(215,60)(230,55) \Line(220,45)(235,50) \Line(195,55)(220,45) \Line(220,45)(230,30)
\end{picture}} 
\end{center}

\caption{\label{fid:troismodes}The three branching modes as explained in the text.
The arrows indicate the variables to be determined. }
\end{figure}
 
\end{enumerate}
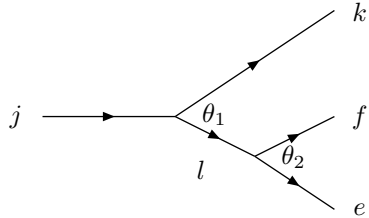
\begin{figure}[htb]
\begin{center}\begin{picture}(200,100) 
\ArrowLine(10,50)(60,50)\Text(0,50)[]{$j$} 
\ArrowLine(60,50)(120,90)\Text(70,30)[]{$l$}
\ArrowLine(60,50)(90,35)\Text(130,90)[]{$k$}
\ArrowLine(90,35)(120,15)\Text(130,15)[]{$e$}
\ArrowLine(90,35)(120,50)\Text(130,50)[]{$f$}
\Text(75,51)[]{$\theta_1$}\Text(105,35)[]{$\theta_2$} 
\end{picture}\end{center}

\caption{\label{fig:classement}Angular ordering in successive branchings.}
\end{figure}

The described leading order algorithm is known to be not accurate enough for
secondary hadron production, in particular it gives too high multiplicities
of secondaries in \( e^{+}e^{-} \)-annihilation. The method can be corrected
if one takes into account the phenomenon of color coherence. The latter one
appears if one considers some higher order corrections to the simplest leading
logarithmic contributions, the latter ones being the basis for the usual Altarelli-Parisi
evolution equations. In the corresponding treatment \cite{dok91} -- so-called
modified leading logarithmic approach -- one essentially recovers the original
scheme for the time-like parton cascading supplemented by the additional condition,
the strict ordering of the emission angles in successive parton branchings.
The appearance of the angular ordering can be explained in a qualitative way:
(see \cite{ell96}): if a transverse wavelength of an emitted gluon (\( f \)
or \( e \) on figure \ref{fig:classement}) is larger than the separation between
the two, this gluon cannot see the color charge of the parent (\( l \)) but
only the total (much smaller) charge of the two partons (\( k+l=j \)) and the
radiation is suppressed. In a Monte-Carlo model this can be easily realized
by imposing the angular ordering condition via rejection \cite{ben87}. Thus,
for each branching we check whether the angular ordering \( \theta _{2}<\theta _{1} \)
is valid, where \( \theta _{1} \) is the angle of the previous branching, and
reject the current splitting otherwise. 

\cleardoublepage

\chapter{Hadronization}

Till now, our discussion concerned exclusively partons, whereas ``in the real
world'' one observes finally hadrons. It is the purpose of this chapter, to
provide the link, i.e. to discuss how to calculate hadron production starting
from partonic configurations, discussed in the previous chapters.

Hadron production is related to the structure of cut Pomerons. A cut Pomeron
is in principle a sum over squared amplitudes of the type \( a+b\: \rightarrow \: \mathrm{hadrons}, \)
integrated over phase space, with \( a \) and \( b \) being the nucleon constituents
involved in the interaction. So far there was no need to talk about the details
of the hadron production, which could be considered to be ``integrated out''.
In case of soft Pomerons, we used a parameterization of the whole object, based
on general asymptotic considerations, which means that all the hadron production
is hidden in the few parameters characterizing the soft Pomeron. In case of
hard Pomerons, we discussed the explicit partonic structure of the corresponding
diagram without talking about hadrons. This is justified based on the assumption
that summing over hadronic final states is identical to summing over partonic
final states, both representing complete sets of states. But although our ignorance
of hadronic states so far was well justified, we finally have to be specific
about the hadronic structure of the cut Pomerons, because these are hadronic
spectra which are measured experimentally, and not parton configurations.

Lacking a rigorous theoretical treatment, we are going to use the same strategy
as we used already for treating the soft Pomeron: we are going to present a
``parameterization'' of the hadronic structure of the cut Pomerons, as simple
as possible with no unnecessary details, in agreement with basic laws of physics
and basic experimental observations. We do not claim at all to understand the
microscopic mechanism, so our parameterization, called ``string model'', should
not be considered as a microscopic hadronization model.

\section{Hadronic Structure of Cut Pomerons}

In order to develop our multiple scattering theory, we used a simple graphical
representation of a cut Pomeron, namely a thick vertical line connecting the
external legs representing nucleon components, 
\begin{figure}[htb]
{\par\centering \resizebox*{!}{0.1\textheight}{\includegraphics{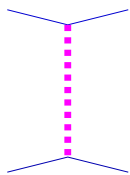}} \par}

\caption{Symbol representing a cut Pomeron.\label{pom}}
\end{figure}
as shown in fig.~\ref{pom}. This simple diagram hides somewhat the fact that
there is a complicated structure hidden in this Pomeron, and the purpose of
this section is to discuss in particular the hadronic content of the Pomeron.

\begin{figure}[htb]
{\par\centering {\huge \resizebox*{!}{0.17\textheight}{\includegraphics{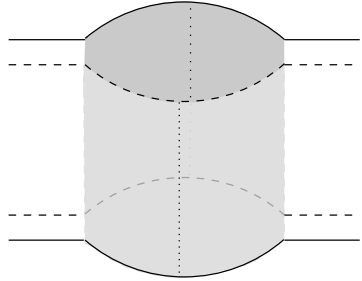}} }\huge \par}

\caption{Cut soft Pomeron represented as a cut cylinder. The grey areas represent unresolved
partons.\label{cyl}}
\end{figure}

Let us start our discussion with the soft Pomeron. Based on Veneziano's topological
expansion one may consider a soft Pomeron as a ``cylinder'' i.e. the sum of
all possible QCD diagrams having a cylindrical topology, see fig.~\ref{cyl}.
As discussed in detail in chapter 2.2, the ``nucleon components'' mentioned
earlier, representing the external legs of the diagram, are always quark-anti-quark
pairs, indicated by a dashed line (anti-quark) and a full line (quark) in fig.
\ref{cyl}. Important for the discussion of particle production are of course
cut diagrams, therefore we show in fig. \ref{cyl} a cut cylinder representing
a cut Pomeron: the cut plane is shown as two vertical dotted lines. Let us consider
the half-cylinder, for example, the one to the right of the cut, representing
an inelastic amplitude. 
\begin{figure}[htb]
{\par\centering \resizebox*{!}{0.2\textheight}{\includegraphics{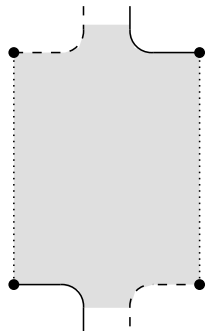}} \par}

\caption{Planar representation of a half-cylinder obtained from cutting a cylinder diagram
(see fig. \ref{cyl}). \label{half-cyl}}
\end{figure}
 
\begin{figure}[htb]
{\par\centering \resizebox*{!}{0.2\textheight}{\includegraphics{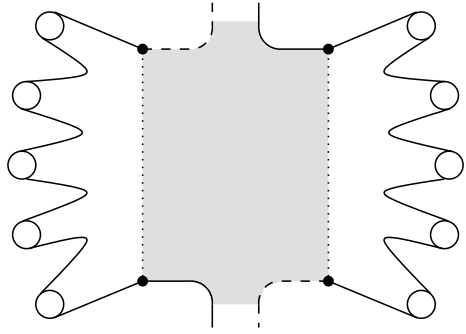}} \par}

\caption{The string model: each cut line (dotted vertical lines) represents a string,
which decays into final state hadrons (circles).\label{string-model}}
\end{figure}
We can unfold this object in order to have a planar representation, as shown
in fig.\ref{half-cyl}. Here, the dotted vertical lines indicate the cuts of
the previous figure, and it is here where the hadronic final state hadrons appear.
Lacking a theoretical understanding of this hadronic structure, we simply apply
a phenomenological procedure, essentially a parameterization. We require the
method to be as simple as possible, with a minimum of necessary parameters.
A solution coming close to these demands is the so-called string model: each
cut line is identified with a classical relativistic string, a Lorentz invariant
string breaking procedure provides the transformation into a hadronic final
state, see fig. \ref{string-model}.

The phenomenological microscopic picture which stays behind this procedure was
discussed in a number of reviews \cite{kai82,cap94, and83}: the string end-point
partons resulted from the interaction appear to be connected by a color field.
With the partons flying apart, this color field is stretched into a tube, which
finally breaks up giving rise to the production of hadrons and to the neutralization
of the color field.

\begin{figure}[htb]
{\par\centering \resizebox*{!}{0.2\textheight}{\includegraphics{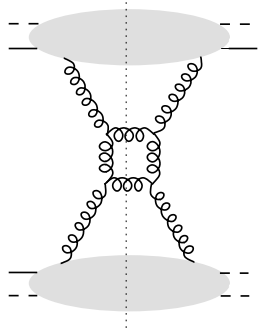}} \par}

\caption{A simple diagram contributing to the semi-hard Pomeron of the ``sea-sea''
type. \label{sea-sea}}
\end{figure}

\begin{figure}[htb]
{\par\centering \resizebox*{!}{0.2\textheight}{\includegraphics{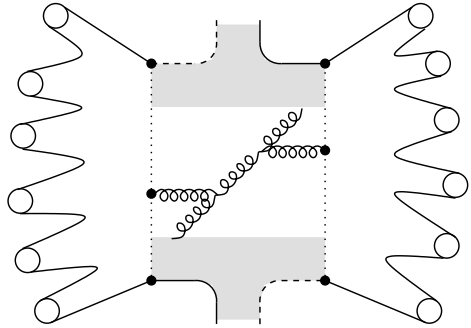}} \par}

\caption{The ``kinky'' string model: the cut line (vertical dotted line) corresponds
to a kinky string, which decays into hadrons (circles). \label{kinky-string-model}}
\end{figure}

We now consider a semi-hard Pomeron of the ``sea-sea'' type, where we have
a hard pQCD process in the middle and a soft evolution at the end, see fig.
\ref{sea-sea}.
 We generalize the picture introduced above for the soft Pomeron. Again, we
assume a cylindrical structure. For the example of fig. \ref{sea-sea}, we have
the picture shown in fig. \ref{sea-sea-cyl}: 
\begin{figure}[htb]
{\par\centering \resizebox*{!}{0.2\textheight}{\includegraphics{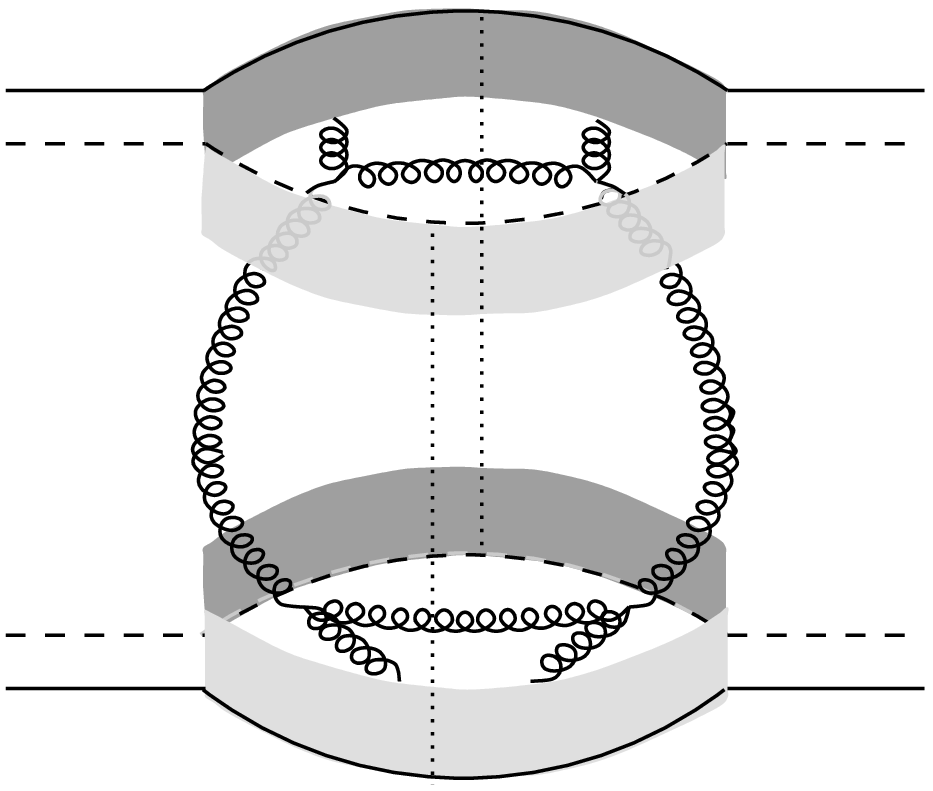}} \( \qquad  \)\( \qquad  \)\resizebox*{!}{0.2\textheight}{\includegraphics{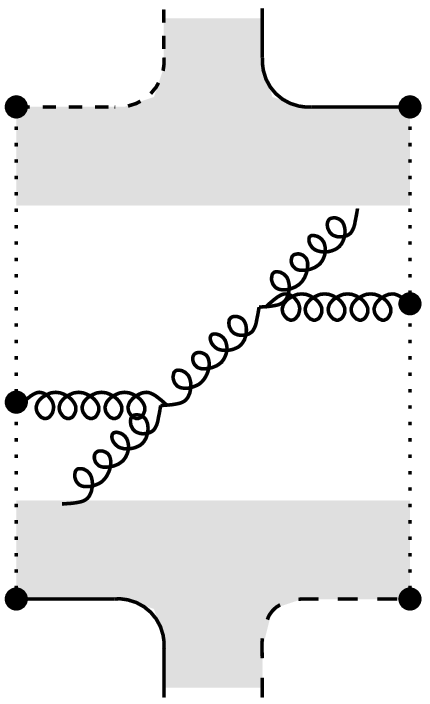}} \par}

\caption{Cylindrical representation of a contribution to the semi-hard Pomeron (left
figure) and planar diagram representing the corresponding half-cylinder (right
figure).\label{sea-sea-cyl}}
\end{figure}
the shaded areas on the cylinder ends represent the soft Pomerons, whereas in
the middle part we draw explicitly the gluon lines on the cylinder surface.
We apply the same procedure as for the soft Pomeron: we cut the diagram and
present a half-cylinder in a planar fashion, see fig. \ref{sea-sea-cyl}. We
observe one difference compared to the soft case: there are three partons (dots)
on each cut line: apart from the quark and the anti-quark at the end, we have
a gluon in the middle. We again apply the string picture, but here we identify
a cut line with a so-called kinky string, where the internal gluons correspond
to internal kinks. The underlying microscopic picture will be presented by three
color-connected partons - the gluon connected by the color field to the quark
and to the anti-quark. The string model provides then a ``parameterization''
of hadron production, see fig. \ref{kinky-string-model}.
The procedure described above can be easily generalized to the case of complicated
parton ladders involving many gluons and quark-anti-quark pairs. One should
note that the treatment of semi-hard Pomerons is just a straightforward generalization
of the string model for soft Pomerons, or one might see it the other way round:
the soft string model is a natural limiting case of the kinky string procedure
for semi-hard Pomerons. 

We now need to discuss Pomerons of valence type. In case of ``valence-valence''
the first partons of the parton ladder are valence quarks, 
\begin{figure}[htb]
{\par\centering \resizebox*{!}{0.2\textheight}{\includegraphics{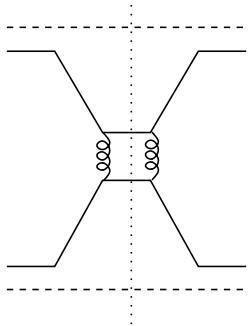}} \par}

\caption{The simplest contribution to the ``valence-valence'' Pomeron. \label{val-val}}
\end{figure}
\begin{figure}[htb]
{\par\centering \resizebox*{!}{0.2\textheight}{\includegraphics{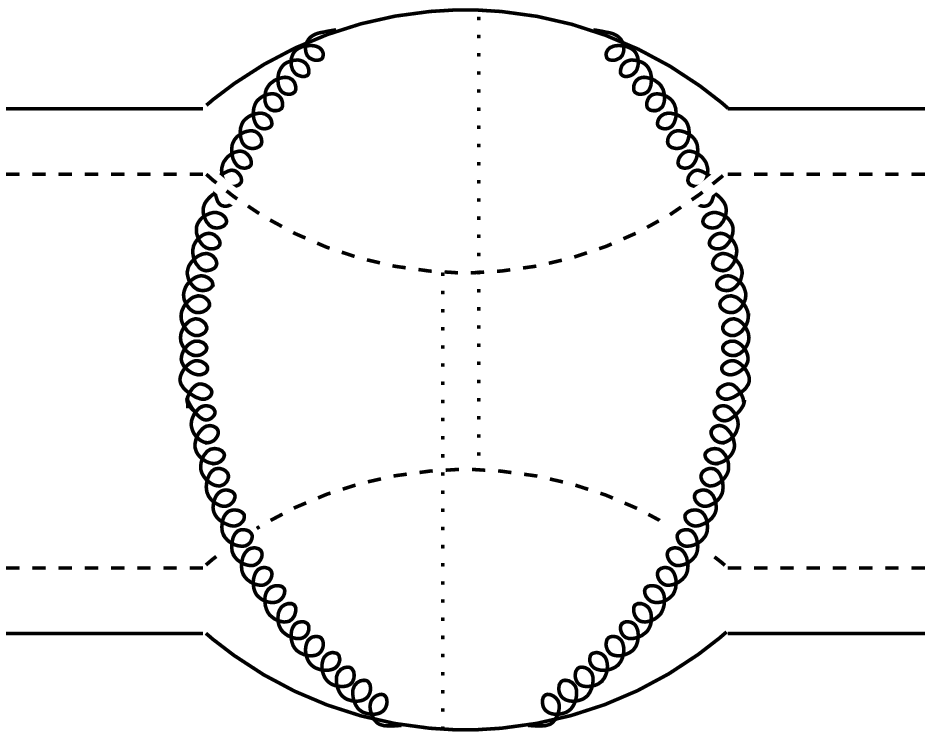}} \( \qquad  \)\resizebox*{!}{0.2\textheight}{\includegraphics{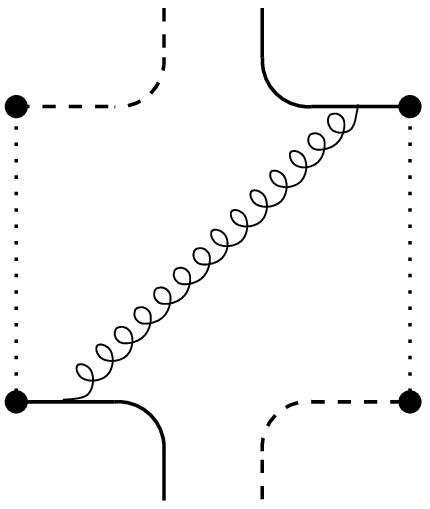}} \par}

\caption{Cylindrical representation of the diagram of fig. \ref{val-val} (left figure)
and planar diagram representing the corresponding half-cylinder (right figure).\label{val-val-cyl}}
\end{figure}
\begin{figure}[htb]
{\par\centering \resizebox*{!}{0.2\textheight}{\includegraphics{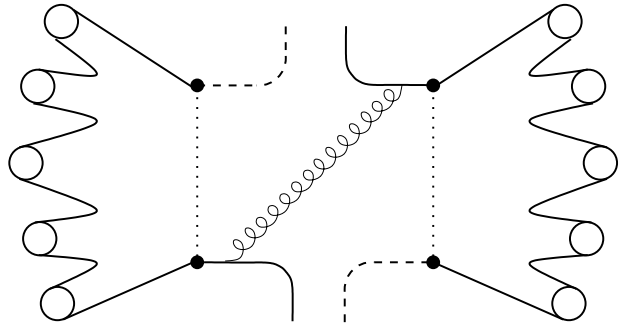}} \par}

\caption{The string model: the cut line (vertical dotted line) corresponds to a string,
which decays into hadrons (circles).\label{string-model-val}}
\end{figure}
there is no soft Pomeron between the parton ladder and the nucleon. The nucleon
components representing the external legs are, as usual, quark-anti-quark pairs,
but the anti-quark plays in fact just the role of a spectator. The simplest
possible interaction is the exchange of two gluons, as shown in fig.\ref{val-val}.
We follow the scheme used for soft Pomerons and ``sea-sea'' type semi-hard
Pomerons: we draw the diagram on a cylinder, see fig. \ref{val-val-cyl}. There
is no soft region, the gluons couple directly to the external partons. We cut
the cylinder, one gluon being to the right and one gluon to the left of the
cut, and then we consider the corresponding half-cylinder presented in a planar
fashion, see fig. \ref{val-val-cyl} (right). Here, we have only internal gluons,
on the cut line we observe just the external partons, the corresponding string
is therefore just an ordinary quark-anti-quark string without internal kinks,
as in the case of the soft Pomerons. We apply the usual string breaking procedure
to obtain hadrons, see fig. \ref{string-model-val}. 

\begin{figure}[htb]
{\par\centering \resizebox*{!}{0.22\textheight}{\includegraphics{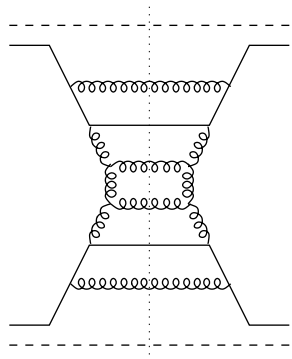}} \par}

\caption{A more complicated contribution to the valence type Pomeron. \label{val-compl}}
\end{figure}

\begin{figure}[htb]
{\par\centering \resizebox*{!}{0.2\textheight}{\includegraphics{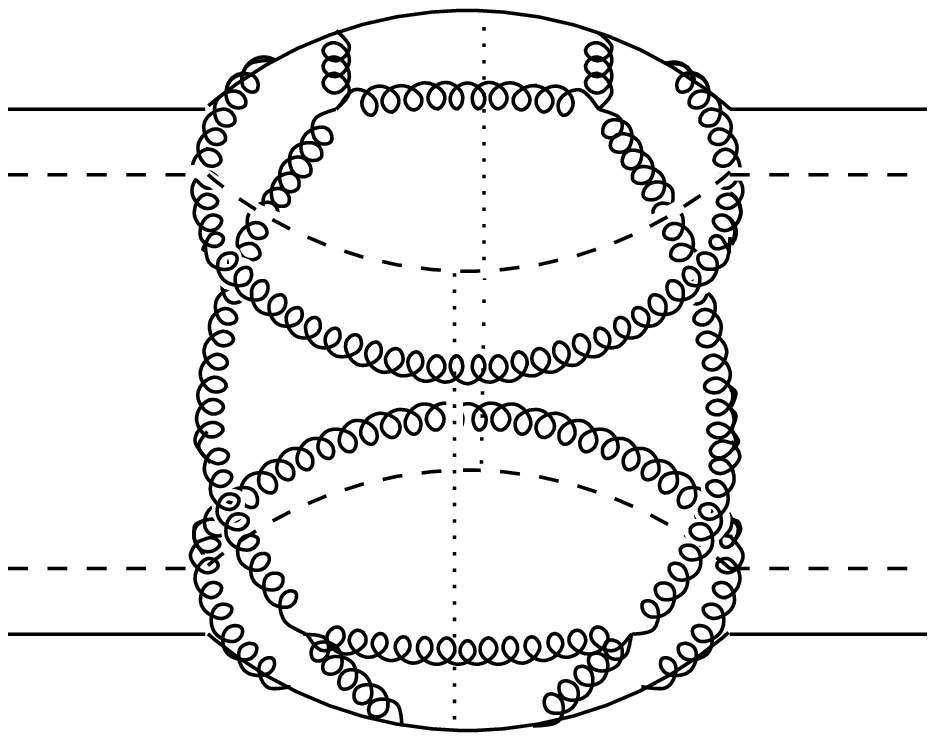}} \( \quad  \)\resizebox*{!}{0.18\textheight}{\includegraphics{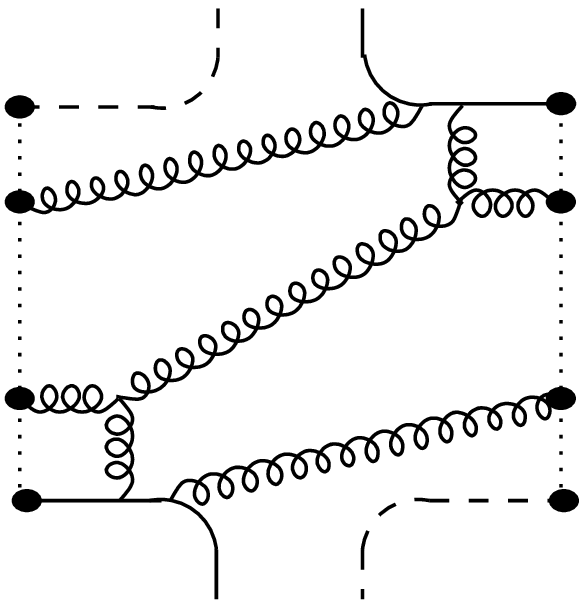}} \par}

\caption{Cylindrical representation of the diagram of fig. \ref{val-compl} (left figure)
and planar diagram representing the corresponding half-cylinder (right figure).
\label{val-compl-cyl}}
\end{figure}

\begin{figure}[htb]
{\par\centering \resizebox*{!}{0.2\textheight}{\includegraphics{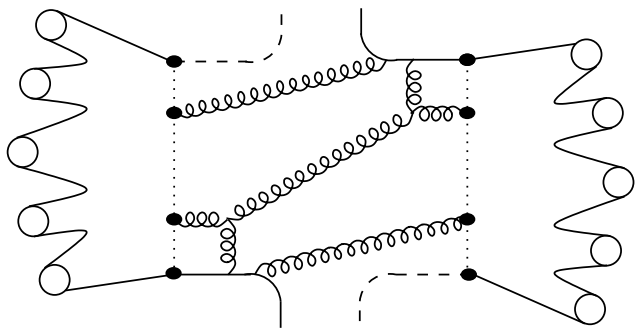}} \par}

\caption{The string model: the cut line (vertical dotted line) corresponds to a string,
which decays into hadrons (circles). \label{string-model-val-compl}}
\end{figure}

Let us consider a more complicated valence-type diagram, as shown in fig. \ref{val-compl}. 
It is again a contribution to the Pomeron of the ``valence-valence'' type:
the external partons of the parton ladders are the valence quarks of the nucleons.
In contrast to the previous example, we have here an emission of s-channel gluons,
traversing the cut. As usual, we present the diagram on the cylinder, as shown
in fig. \ref{val-compl-cyl}, where we also show the corresponding planar half-cylinder.
In addition to internal gluons, we now observe also external ones, presented
as dots on the cut line. As usual, we identify the cut line with a relativistic
kinky string, where each external (s-channel) gluon represents a kink. We then
employ the usual string procedure to produce hadrons, as sketched in fig. \ref{string-model-val-compl}.
\newpage

The general procedure should be clear from the above examples: in any case,
no matter what type of Pomeron, we have the following procedure: 

\begin{enumerate}
\item drawing of a cylinder diagram; 
\item cutting the cylinder;  
\item planar presentation of the half-cylinder; 
\item identification of a cut line with a kinky string;  
\item kinky string hadronization
\end{enumerate}
The last point, the string hadronization procedure, will be discussed in detail
in the following. This work is basically inspired by \cite{mor87,mor90} and
further developed in \cite{dre99t}. The main differences to \cite{mor90}are
that hadrons are directly obtained from strings instead from low mass clusters,
and an intrinsic transverse momentum is added to a string break-up. The main
difference to the Lund-model is to use the area-law instead of a fragmentation
function. 

\clearpage

\section{Lagrange Formalism for Strings}

A string can be considered as a point particle with one additional space-like
dimension. The trajectory in Minkowski space depends on two parameters:

\begin{equation}
x^{\mu }=x^{\mu }(\sigma ,\tau ),\, \, \, \sigma =0...\pi \, ,
\end{equation}
 with \( \sigma  \) being a space-like and \( \tau  \) a time-like parameter.
In order to obtain the equation of motion, we need a Lagrangian. It is obtained
by demanding the invariance of the trajectory with respect to gauge transformations
of the parameters \( \sigma  \) and \( \tau  \). This way we find \cite{wer93}
the Lagrangian of Nambu-Goto:

\begin{equation}
\mathcal{L}=-\kappa \sqrt{(x'\dot{x})^{2}-x'^{2}\dot{x}^{2}}\, ,
\end{equation}
with \( \dot{x}^{\mu }=dx^{\mu }/d\tau  \), \( x^{\mu }\, '=dx^{\mu }/d\sigma  \)
and \( \kappa  \) being the energy density or string tension. With this Lagrangian
we write down the action 
\begin{equation}
S=\int _{0}^{\pi }d\sigma \int _{\tau _{0}}^{\tau _{1}}d\tau \mathcal{L}\, ,
\end{equation}
which leads to the Euler-Lagrange equation:
\begin{equation}
\frac{\partial }{d\tau }\frac{\partial \mathcal{L}}{\partial \dot{x}_{\mu }}+\frac{\partial }{\partial \sigma }\frac{\partial \mathcal{L}}{\partial x_{\mu }'}=0\, ,
\end{equation}
with the initial conditions
\begin{equation}
\frac{\partial \mathcal{L}}{\partial x_{\mu }'}=0,\, \, \, \, \sigma =0,\pi \, ,
\end{equation}
since we have \( \delta x=0 \) for \( \tau =\tau _{0} \) and \( \tau =\tau _{1} \).
This equation can be solved most easily by a partial gauge fixing. We have this
freedom, since the result is independent on the choice of the parameters. This
is done indirectly by imposing the following conditions:
\begin{equation}
\label{for:cordejauge}
\dot{x}^{2}+x'^{2}=0,\, \, \dot{x}x'=0\, .
\end{equation}
The Euler-Lagrange equation gives us a simple solution, the wave equation:
\begin{equation}
\label{for:onde}
\frac{\partial ^{2}x_{\mu }}{\partial \tau ^{2}}-\frac{\partial ^{2}x_{\mu }}{\partial \sigma ^{2}}=0\, ,
\end{equation}
with the following boundary conditions:
\begin{equation}
\label{for:cond_ini1}
\frac{\partial x_{\mu }}{\partial \sigma }=0,\, \, \, \, \sigma =0,\pi .
\end{equation}
 The total momentum of a string is given by \cite{wer93} 
\begin{equation}
\label{for:pstring}
p^{\mu }_{\mathrm{string}}=-\int _{C}\frac{\partial L}{\partial \dot{x}_{\mu }}d\sigma +\frac{\partial L}{\partial x_{\mu }'}d\tau 
\end{equation}
with \( C \) being a curve between the two ends of the string (\( \sigma =0 \)
and \( \sigma =\pi  \)). This gives for (\ref{for:cordejauge}) and for \( d\tau =0 \)
\begin{equation}
\label{for:impcorde}
p_{\mathrm{string}}^{\mu }=\int _{0}^{\pi }\kappa \dot{x}^{\mu }d\sigma .
\end{equation}
 We still have to fix completely the gauge since it has been fixed partially
only. This can be done with the following conditions for the parameter \( \tau  \):
\begin{equation}
n^{\mu }x_{\mu }=\lambda \tau 
\end{equation}

Different choices for \( n \) and \( \lambda  \) are possible, like \( n=(1,-1,0,0) \)
which is called the transverse gauge. We will use \( n=(1,0,0,0) \) which leads
to \( \lambda =E/\pi \kappa  \) and another choice \( \pi =E/\kappa  \) will
identify \( \tau  \) with the time \( x_{0} \), whereas \( E=\int _{0}^{\pi }\kappa \dot{x}_{0}d\sigma  \)
is the total energy of the string. We define ``string units'' via \( \kappa =1 \);
\( \sigma  \) and \( \tau  \) have thereby the dimension of energy and \( \pi =E \).
In ``ordinary'' units, one has \( \kappa =\tilde{\kappa } \) GeV/fm, with
\( \tilde{\kappa } \) being approximately 1, so a length of 1 GeV corresponds
to 1 fm\( /\tilde{\kappa } \) \( \approx  \)1 fm.

The solution of a wave equation is a function which depends on the sum or the
difference of the two parameters \( \sigma  \) and \( \tau  \). As the second
derivative shows up, we have two degrees of freedom to impose the initial conditions
on the space-like extension and the speed of the string at \( \tau =0 \). One
can easily verify that the following Ansatz \cite{mor87,mor90} fulfills the
wave equation (\ref{for:onde}):

\begin{eqnarray}
x^{\mu }(\sigma ,\tau ) & = & \frac{1}{2}\left[ f^{\mu }\left( \sigma +\tau \right) +f^{\mu }\left( \sigma -\tau \right) +\int _{\sigma -\tau }^{\sigma +\tau }g^{\mu }\left( \xi \right) d\xi \right] \label{for:ansatz} \\
f(\sigma ) & = & x^{\mu }(\sigma ,\tau )|_{\tau =0}\\
g(\sigma ) & = & \dot{x}^{\mu }(\sigma ,\tau )|_{\tau =0}
\end{eqnarray}
We identify the function \( f(\sigma ) \) with the initial spatial extension
and \( g(\sigma ) \) with the initial speed of the string at the time \( \tau =0 \).

We will consider here a special class of strings, namely those with \( f=0 \)
(initially point-like) and with a piecewise constant function \( g \), 
\begin{equation}
g(\sigma )=v_{k}\quad \; \mathrm{for}\: E_{k-1}\leq \sigma \leq E_{k},\quad \; 1\leq k\leq n
\end{equation}
 with some integer \( n \). The set \( \{E_{k}\} \) is a partition of the
\( \sigma  \)-range \( [0,E] \),
\begin{equation}
0=E_{0}<E_{1}<...<E_{n-1}<E_{n}=E,
\end{equation}
 and \( \{v_{k}\} \) represents \( n \) constant 4-vectors. Such strings are
called kinky strings, with \( n \) being the number of kinks, and the \( n \)
vectors \( v_{k} \) being called kink velocities. 

In order to use eq.~(\ref{for:ansatz}), we have to extend the function \( g \)
beyond the physical range between \( 0 \) and \( \pi  \). This can be done
by using the boundary conditions, which gives 
\begin{eqnarray}
g(\tau ) & = & g(-\tau ),\\
g(\tau +2\pi ) & = & g(\tau ),
\end{eqnarray}
So \( g \) is a symmetric periodic function, with the period \( 2\pi  \).
This defines \( g \) everywhere, and the eq.~(\ref{for:ansatz}) is the complete
solution of the string equation, expressed in terms of the initial condition
\( g \) (\( f \) is taken to be zero). In case of kinky strings the latter
is expressed in terms of the kink velocities \( \{v_{k}\} \) and the energy
partition \( \{E_{k}\} \).

\section{Identifying Partons and Kinks}

We discussed earlier that a cut Pomeron may be identified with two sequences
of partons of the type
\begin{equation}
q-g-g-...-g-\bar{q},
\end{equation}
representing all the partons on a cut line. We identify such a sequence with
a kinky string, by requiring 
\begin{equation}
\mathrm{parton}=\mathrm{kink},
\end{equation}
which means that we identify the partons of the above sequence with the kinks
of a kinky string, such that the partition of the energy is given by the parton
energies,
\begin{equation}
E_{k}=\mathrm{energy}\, \mathrm{of}\, \mathrm{parton}\, k
\end{equation}
and the kink velocities are just the parton velocities,
\begin{equation}
v_{k}=\frac{\mathrm{momentum}\, \mathrm{of}\, \mathrm{parton}\, k}{E_{k}}.
\end{equation}
We consider massless partons, so that the energy is equal to the absolute value
of the parton momentum.
\begin{figure}[htb]
{\par\centering \resizebox*{!}{10cm}{\includegraphics{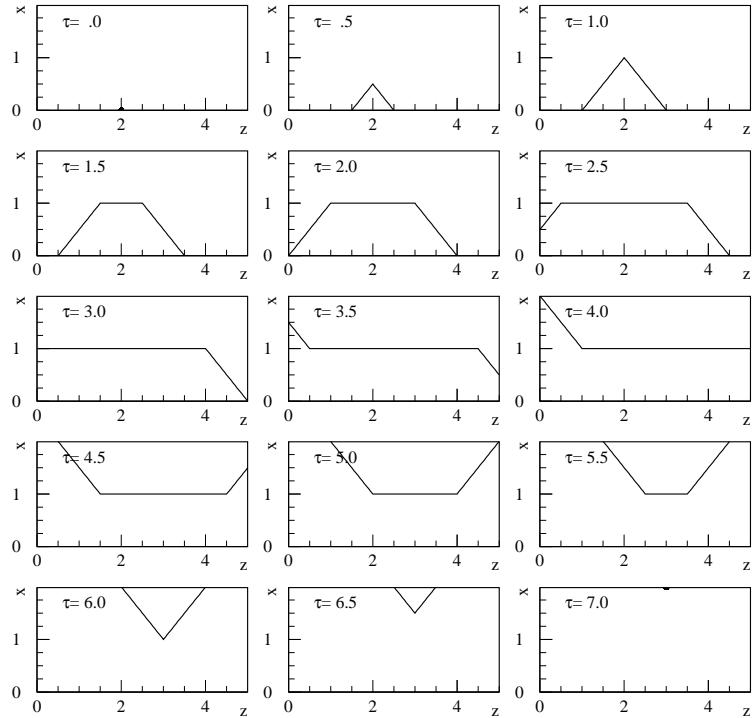}} \par}

\caption{\label{fig:mouvement}Movement of a string with 3 partons on the \protect\( z-x\protect \)
plane: The partons have momenta of \protect\( p_{q}=(p_{x},p_{y},p_{z})=(0,0,-2)\protect \)
GeV/c, \protect\( p_{g}=(2,0,0)\, \textrm{GeV}/\textrm{c}\protect \) and \protect\( p_{\bar{q}}=(0,0,3)\textrm{ GeV}/\textrm{c}\protect \).
The first half-cycle is finished after \protect\( \tau =7\textrm{ GeV}\protect \).}
\end{figure}
Fig.~\ref{fig:mouvement} shows as an example the evolution of a kinky string
representing three partons: a quark, an anti-quark, and a gluon, as a function
of the time \( \tau  \). One sees that the partons start to move along their
original direction with the speed of light. After some time which corresponds
to their energy they take the direction of the gluon. One could say that they
lose energy to the string itself. The gluon loses energy in two directions,
to the quark and to the anti-quark and therefore in half the time. The ends
of the string move continuously with the speed of each of the partons until
the whole string is contracted again in one point. The cycle starts over. 

\begin{figure}[htb]
{\par\centering \resizebox*{!}{10cm}{\includegraphics{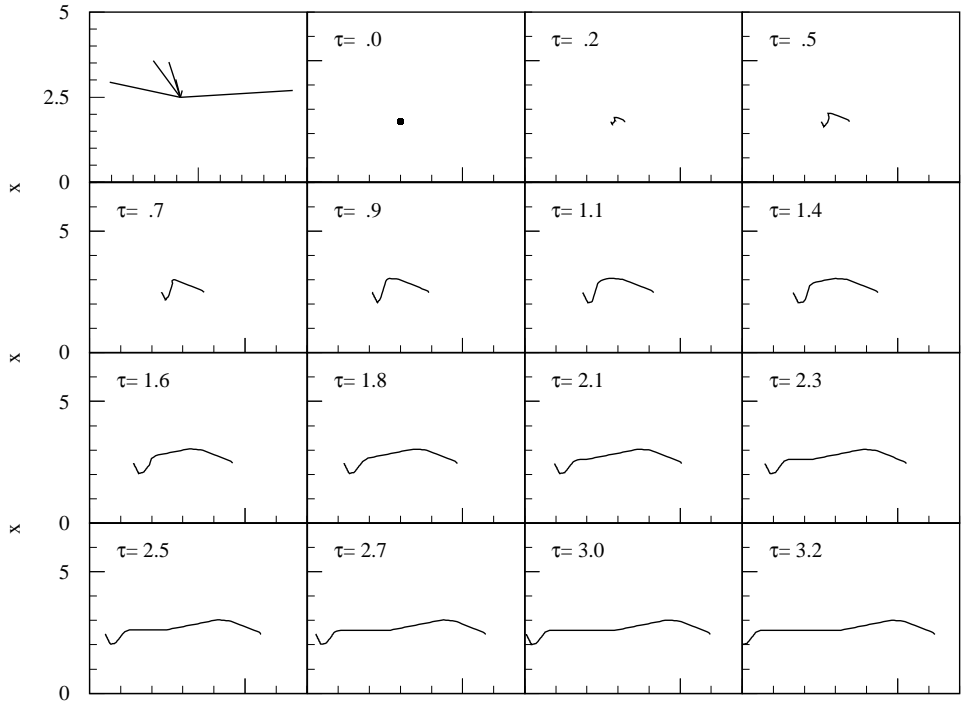}} \par}

\caption{Partons in an \protect\( e^{+}e^{-}\protect \) annihilation event with \protect\( \sqrt{s}=14\textrm{ GeV}\protect \)
in the \protect\( z-x\protect \) plane. The first figure shows the momenta
in the \protect\( p_{z}-p_{x}\protect \) plane\label{fig:kinks_ee}. }
\end{figure}
Another example is shown on fig.~\ref{fig:kinks_ee}, where realistic partons
coming from a simulation of a \( e^{+}e^{-} \) annihilation process at 14 GeV
c.m.s. energy are considered. We will see later how to generate these partons.
We observe 6 partons, 2 quarks and 4 gluons, symbolically displayed in the first
sub-figure. As the total energy is 14 GeV the cycle has a periodicity of 28
\( \textrm{GeV} \). But one sees that the perturbative gluons play an important
role in the beginning of the movement, and later from 2 GeV on, the longitudinal
character dominates. As we will see later, a string breaks typically after \( 1\textrm{ GeV}/\kappa  \)
which gives much importance to the perturbative gluons.

\section{Momentum Bands in Parameter Space }

\begin{figure}[htb]
{\par\centering \resizebox*{!}{0.15\textheight}{\includegraphics{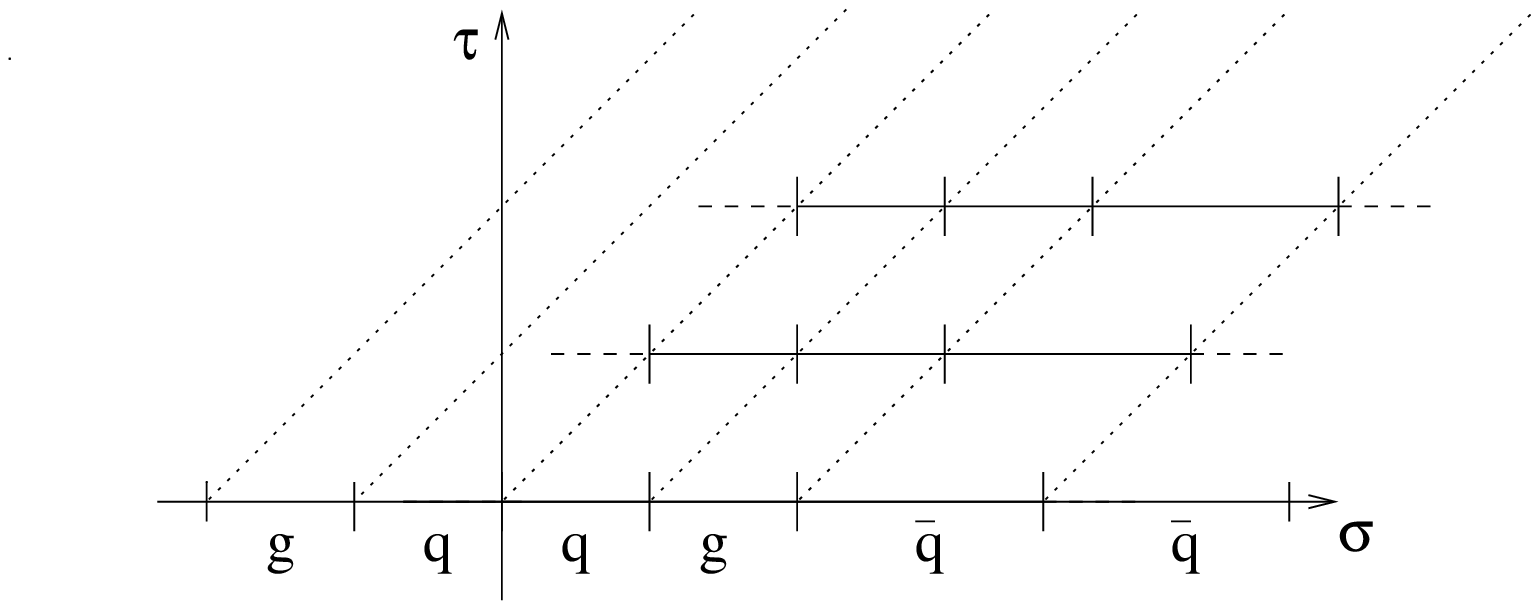}} \par}

\caption{Construction of bands where \protect\( g(\sigma -\tau )\protect \) is constant,
being referred to as R-bands (``right moving bands''). \label{fig:band_plus}}
\end{figure}

\begin{figure}[htb]
{\par\centering \resizebox*{!}{0.15\textheight}{\includegraphics{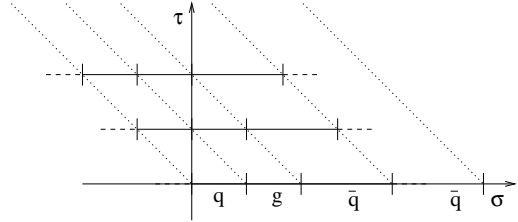}} \par}

\caption{Construction of bands where \protect\( g(\sigma +\tau )\protect \) is constant,
being referred to as L-bands (``left moving bands''). \label{fig:band_minus}}
\end{figure}

As we will see later, it is not necessary for a fragmentation model to know
the spatial extension at each instant. Therefore, we concentrate on a description
in momentum space which simplifies the model even more. By using formula (\ref{for:ansatz})
we can express the derivatives of \( x(\sigma ,\tau ) \) in terms of the initial
conditions \( g(\sigma ) \) as 
\begin{eqnarray}
\dot{x}(\sigma ,\tau ) & = & \frac{1}{2}\left[ g(\sigma +\tau )+g(\sigma -\tau )\right] \label{for:xprime} \\
x'(\sigma ,\tau ) & = & \frac{1}{2}\left[ g(\sigma +\tau )-g(\sigma -\tau )\right] \, \, .\label{for:xdot} 
\end{eqnarray}
 Since the function \( g \) is stepwise constant, we easily identify regions
in the parameter space \( (\sigma ,\tau ) \), where \( g(\sigma +\tau ) \)
is constant or where \( g(\sigma -\tau ) \) is constant, as shown in fig. \ref{fig:band_plus},\ref{fig:band_minus}. 

\begin{figure}[htb]
{\par\centering \resizebox*{!}{0.35\textheight}{\includegraphics{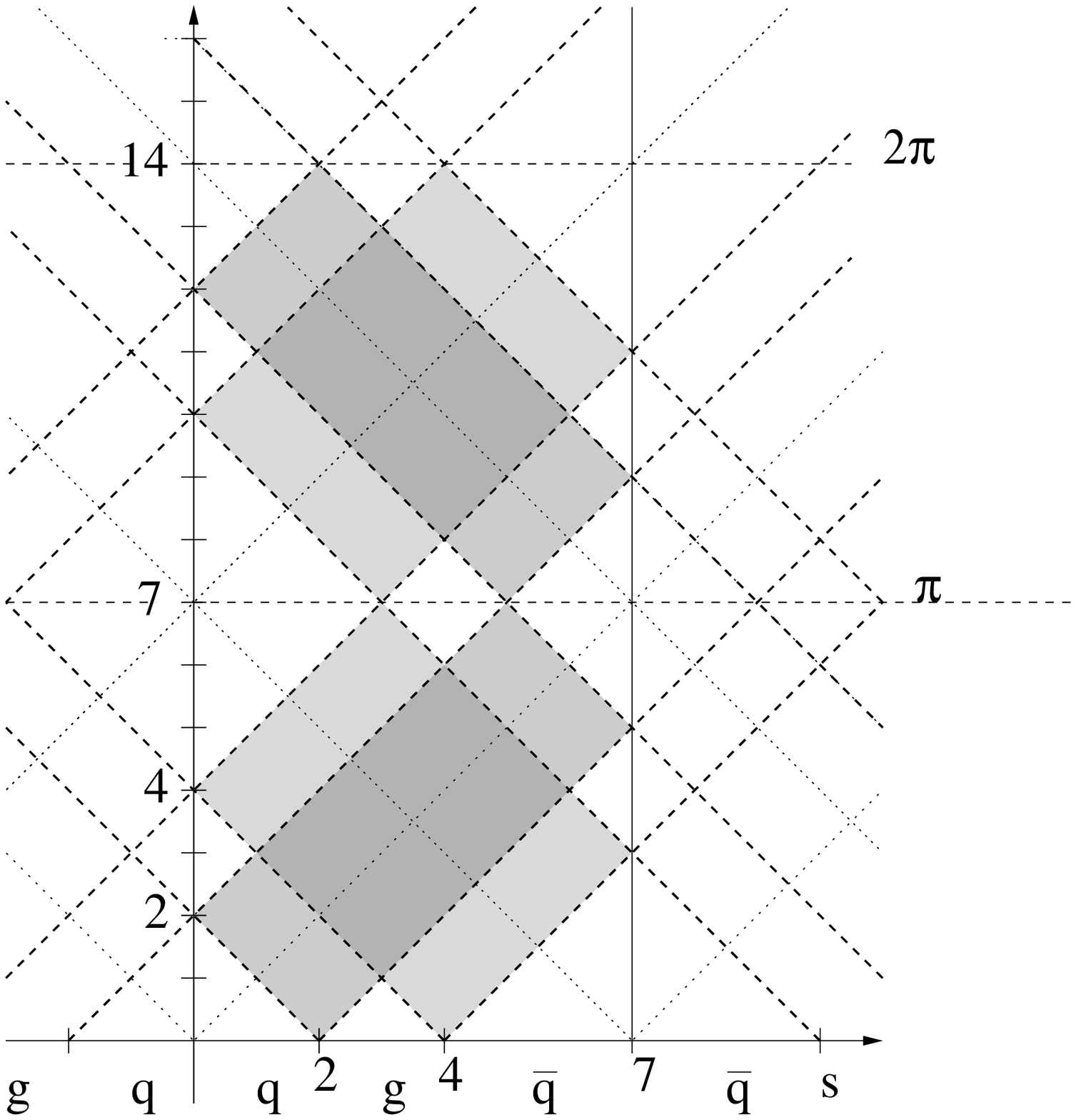}} \par}

\caption{\label{fig:superposition}The superposition of bands where \protect\( g(\sigma +\tau )\protect \)
and \protect\( g(\sigma -\tau )\protect \) are constant (see fig.~\ref{fig:band_plus},\ref{fig:band_minus})
gives regions of constant \protect\( x'\protect \) and \protect\( \dot{x}\protect \). }
\end{figure}

These regions are called momentum bands, more precisely R-bands and L-bands,
being of great importance for the string breaking. If we overlay the two figures
of \ref{fig:band_plus},\ref{fig:band_minus}, we get fig.~\ref{fig:superposition}, 
which allows us to identify regions, where \( g(\sigma +\tau ) \) and \( g(\sigma -\tau ) \)
are constant at the same time, namely the intersections of R-bands and L-bands.
In these areas \( \dot{x} \) and \( x' \) are constant, given as 
\begin{eqnarray}
\dot{x}(\sigma ,\tau ) & = & \frac{1}{2}\left[ v^{-}+v^{+}\right] \label{for:xprime} \\
x'(\sigma ,\tau ) & = & \frac{1}{2}\left[ v^{-}-v^{+}\right] \, \, ,\label{for:xdot} 
\end{eqnarray}
 with \( v^{+} \) and \( v^{-} \) being the velocities of the partons corresponding
to the two intersecting bands.

Rather than considering a \( \sigma  \)-range between \( -\infty  \) and \( +\infty  \),
one may simply consider the physical range from 0 to \( \pi  \), and construct
the bands via reflection. As an example, let us follow the L-band corresponding
to the parton \( i \), starting at \( \tau =0 \). With increasing \( \tau  \)
one reaches at some stage the border \( \sigma =0 \). Here, we have an intersection
with the R-band, corresponding to the same parton \( i \), coming from the
unphysical region \( \sigma <0 \).  We now follow this R-band, which corresponds
to a reflection of the above-mentioned L-band, till we hit the border \( \sigma =\pi  \),
... .

In the regions where \( g(\sigma +\tau ) \) and \( g(\sigma -\tau ) \) have
the same value, corresponding to collinear partons or to an overlap of the momentum
bands of one and the same parton, one finds \( x'=0 \), i.e. there is no spatial
extension in the dependence of the parameter \( \sigma  \). Therefore the coordinates
\( x^{\mu } \) stay unchanged and we recover the speed of the original partons.
In particular, this is the case for the whole string at \( \tau =0 \), due
to \( f=0 \). With the string evolving in time, more and more bands of non
collinear partons overlap, which gives \( x'\neq 0 \); the string is extending
as we have seen in fig.~\ref{fig:mouvement} until \( \tau =7 \).

\section{Area Law}

In order to consider string breaking, we are going to extend the model in a
covariant fashion. We use the method proposed by Artru and Menessier \cite{atr74},
which is based on a simple extension of the decay law of unstable particles,
where the probability \( dP \) to decay within a time interval \( dt \) is
given as
\begin{equation}
dP=\lambda dt\, ,
\end{equation}
with some decay constant \( \lambda  \). For strings, we use the same formula
by replacing the proper time by proper surface in Minkowski space, 
\begin{equation}
dP=\lambda dA\, .
\end{equation}
 By construction, this method is covariant. Since we work in parameter space
it is useful to express this dependence as a function of \( \sigma  \) and
\( \tau  \), 
\begin{equation}
dA=\sqrt{(\dot{x}x')^{2}-x'^{2}\dot{x}^{2}}d\sigma d\tau \, .
\end{equation}
By using the expressions for \( \dot{x} \) and \( x' \) and \( \dot{x}x'=0 \)
and \( g^{2}=0 \) , we find
\begin{eqnarray}
dA & = & \sqrt{-\frac{1}{4}\left( -2g(\sigma +\tau )g(\sigma -\tau )\right) \frac{1}{4}\left( 2g(\sigma +\tau )g(\sigma -\tau )\right) }d\sigma d\tau \\
 & = & \left[ \frac{1}{2}g(\sigma +\tau )g(\sigma -\tau )\right] d\sigma d\tau \\
 & = & \frac{1}{2}\left( 1-\cos \phi \right) d\sigma d\tau \, ,
\end{eqnarray}
with \( \phi  \) being the angle between the partons. Consequently, a string
cannot break at a point where the momentum bands of the same parton overlap,
because in this case the angle \( \phi  \) is zero, which leads to \( dA=0 \).
The maximal contribution is obtained for partons moving in opposite directions. 

\begin{figure}[htb]
{\par\centering \resizebox*{!}{0.27\textheight}{\includegraphics{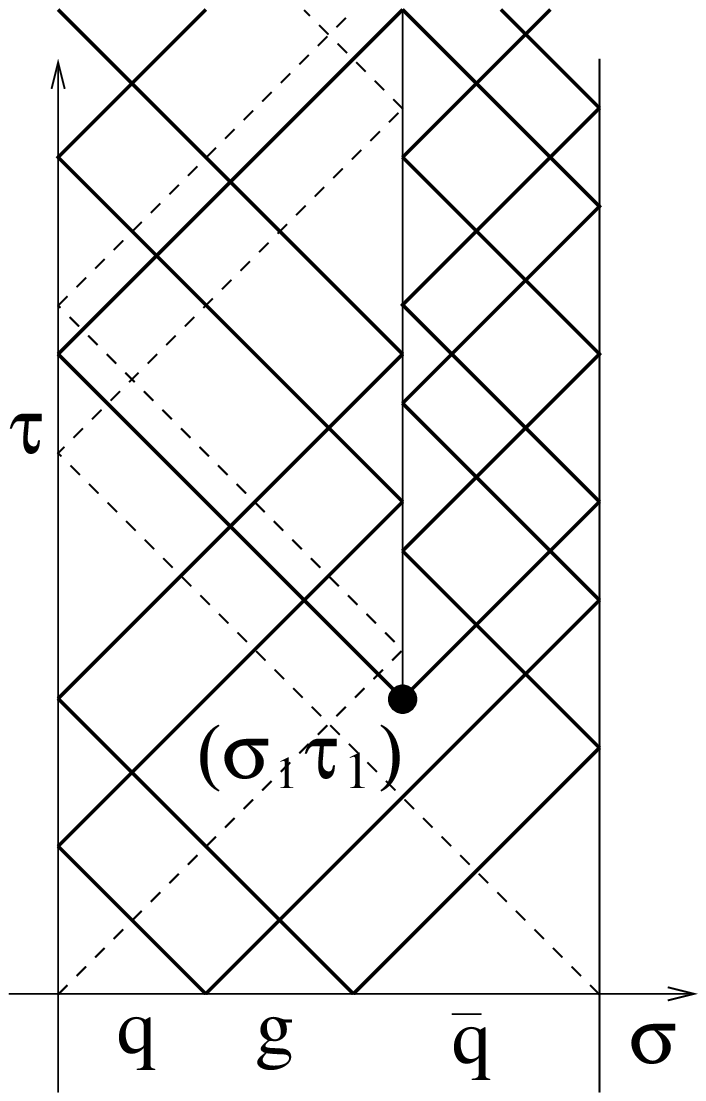}} 
\resizebox*{!}{0.27\textheight}{\includegraphics{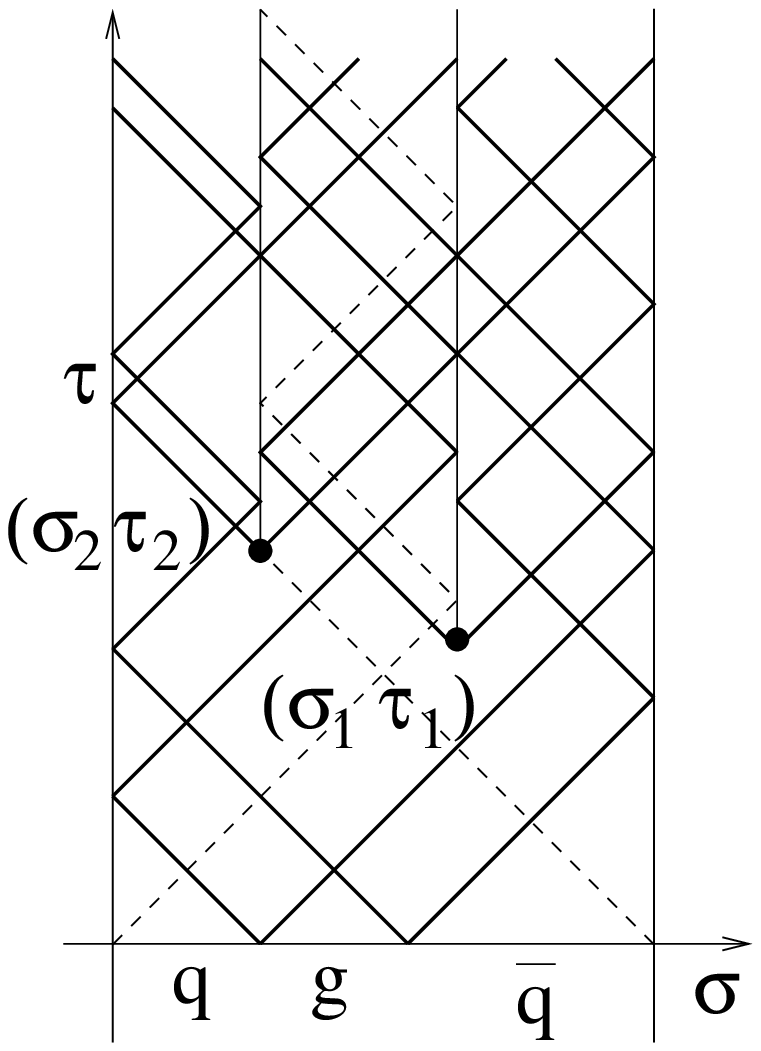}} \par}

\caption{The breaking of a string. The functions \protect\( g\protect \) are found
by imposing the symmetry and periodicity conditions. \label{fig:brisure}}
\end{figure}

We still have to define how a string breaks and how the sub-strings evolve.
At each instant, one knows exactly the momenta of the string by eq.~(\ref{for:xprime})
and (\ref{for:xdot}). The configuration of \( g(\sigma +\tau ) \) and \( g(\sigma -\tau ) \)
at the time \( \tau _{1} \) of the break point is used as initial condition
for the two substrings. The function \( g(\sigma +\tau ) \) is cut into two
pieces between \( 0 \) and \( \sigma _{1} \) and between \( \sigma _{1} \)
and \( \pi  \). The two resulting functions are continued beyond their physical
ranges \( [0,\sigma _{1}] \) and \( [\sigma _{1},\pi ] \) by taking them to
be symmetric and periodic with periods \( 2\sigma _{1} \) and \( 2(\pi -\sigma _{1}) \).
Fig.~\ref{fig:brisure} shows this for a breaking at \( (\sigma _{1},\tau _{1}) \)
and a second break point at \( (\sigma _{2},\tau _{2}) \).  
In principle, the cycle starts over with the two sub-strings breaking each until
the resulting pieces are light enough to form hadrons. However, it is easier
to look for many break-points at once. If they are space-like separated, they
do not interfere with each other. For the coordinates in the parameter space
this translates into the condition 
\begin{equation}
|\sigma _{1}-\sigma _{2}|>|\tau _{1}-\tau _{2}|\, .
\end{equation}

\section{Generating Break Points }

\begin{figure}[htb]
{\par\centering \resizebox*{!}{0.25\textheight}{\includegraphics{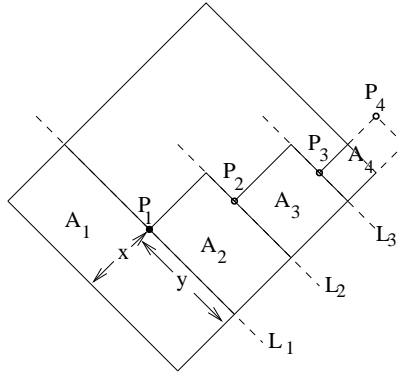}} \par}

\caption{The direct method of searching break points (see text).\label{fig:area1}}
\end{figure}

Having assumed that string breaking occurs according to the area law, 
\begin{equation}
\label{for:surface2}
dP=\lambda dA,
\end{equation}
we now need an algorithm to accomplish this in the framework of the Monte-Carlo
method. The most simple way is to sub-divide a given surface into sufficiently
small pieces and then to decide according to formula (\ref{for:surface2}) if
there is a break point or not. This is what we refer to as the naive method,
which is of course not efficient. We will therefore construct another algorithm
(the direct one) which is based on 
\begin{equation}
P_{0}(A)=e^{-\lambda A}
\end{equation}
being the probability of having no break point within the area \( A \). One
generates surfaces \( A_{1} \), \( A_{2} \), ... according to \( P_{0}(A) \)
as

\begin{equation}
\label{for:A_direct}
A_{i}=-\log (r_{i})/\lambda ,
\end{equation}
with random numbers \( r_{i} \) between \( 0 \) and \( 1 \). The formula
does not say anything about the form of the surfaces \( A_{i} \). Actually,
several choices are possible as long as they do not violate causality. In the
case of a simple string without any gluons, it is easiest to place the surfaces
\( A_{i} \) from left to right such that the break points \( P_{i} \) are
the left upper corners of \( A_{i+1} \), as shown in fig. \ref{fig:area1}:
one first takes \( A_{1} \), which defines the line \( L_{1} \). The first
break point \( P_{1} \) is generated randomly on this line. The next surface
\( A_{2} \) has to be placed in a way that does not violate causality. The
first break point is therefore used as a constraint for the next one, etc. Finally,
if the last surface obtained is too large to be placed on the rest of the string,
the procedure is finished. The advantage of this method is that no break-points
are rejected because the causality principle is obeyed constantly throughout
the whole procedure. 

\begin{figure}[htb]
{\par\centering \resizebox*{!}{6cm}{\includegraphics{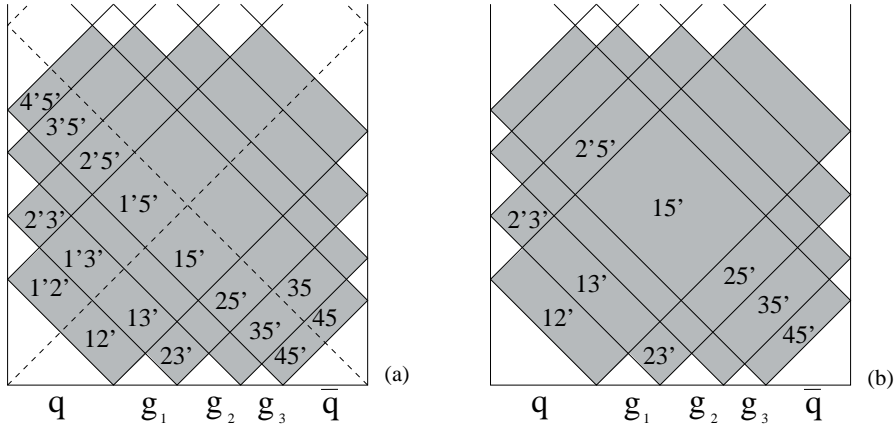}} \par}

\caption{The area \protect\( S_{\mathrm{break}}\protect \) in parameter (\protect\( \sigma ,\tau \protect \))
space, before and after the redefinition of the outer bands.\label{s1}}
\end{figure}

We generalize the method to work for any number of perturbative gluons in the
following way. Since the elementary invariant area \( dA \) is proportional
to the scalar product of the momenta of two partons, we can easily calculate
the area \( A_{ij} \) corresponding to a sub-region \( S_{ij} \) of the \( (\sigma ,\tau ) \)-space,
representing the intersection of the momentum bands of the partons \( i \)
and \( j \). We find
\begin{eqnarray}
A_{ij} & = & \int _{S_{ij}}\left( \frac{1}{2}g(\sigma +\tau )g(\sigma -\tau )\right) d\sigma d\tau =\frac{1}{4}p_{i}\cdot p_{j}\, ,
\end{eqnarray}
with \( p_{i} \) and \( p_{j} \) being the 4-momenta of the two partons. We
now construct the break points in the parameter space rather than in Minkowski
space. One first defines the area in the parameter space of allowed breakpoints
as
\begin{equation}
S_{\mathrm{break}}=\bigcup S_{ij,}
\end{equation}
with the indices running as indicated in fig.~\ref{s1}. To obtain a unique
way of counting the regions, we mark bands which come from a left-moving band
at \( \tau =0 \) with a \( ' \). We further observe that the outer bands \( 1 \)
and \( 1' \) as well as \( 5 \) and \( 5' \), which come from the (anti-)quark-momenta,
are neighboring. It is therefore useful to redefine them as one band \( 1 \)
and \( 5' \) (with double momentum), see fig.~\ref{s1}. 
For each of these sub-areas \( S_{ij} \) the corresponding area in Minkowski
space \( A_{ij} \) is known \( (=p_{i}p_{j}/4) \). One then generates areas
\( A_{1} \), \( A_{2} \), \( A_{3} \), ...(in Minkowski space) according
to eq.~(\ref{for:A_direct}), and places the corresponding areas \( S_{1} \),
\( S_{2} \), \( S_{3} \), ...(in parameter space) into \( S_{\mathrm{break}} \)
from left to right such that the break points \( P_{i} \) are the left upper
corners of \( S_{i+1} \). 

\begin{figure}[htb]
{\par\centering \resizebox*{!}{6cm}{\includegraphics{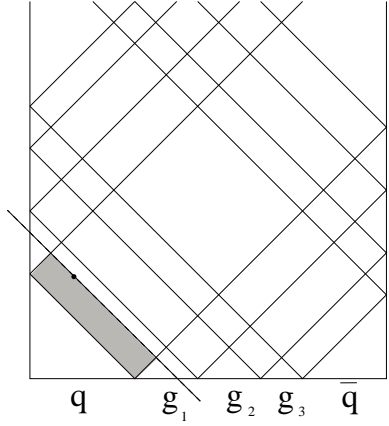}} \par}

\caption{Placing \protect\( S_{1}\protect \) into \protect\( S_{12'}\protect \) in
case of \protect\( A_{1}<A_{12'}\protect \). \label{s2}}
\end{figure}

\begin{figure}[htb]
{\par\centering \resizebox*{!}{6cm}{\includegraphics{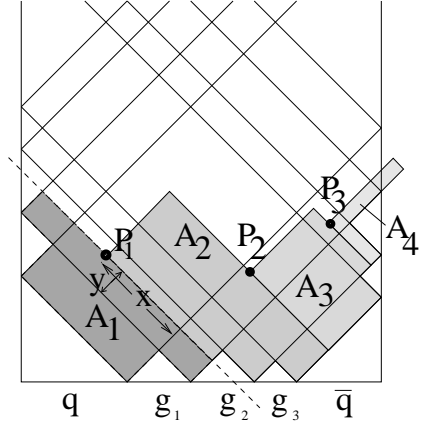}} \par}

\caption{\label{fig:rect_knk}The determination of the break point for the case of a
string with kinks. The regions are weighted according to the scalar product
of their momenta.}
\end{figure}

Let us consider an example of five partons (1,2,3,4,5), see fig. \ref{s2}.
Suppose that we have sampled a surface \( A_{1} \). If it is smaller than the
first region \( A_{12'} \), we determine \( S_{1}=S_{12'}\cdot A_{1}/A_{12'} \)
and we place \( S_{1} \) into the left side of \( S_{12'} \) and generate
the break point \( P_{1} \) randomly on the right upper border of \( S_{12'} \),
see fig. \ref{s2}.
If \( A_{1} \) is greater than \( A_{12'} \), we subtract \( A_{12'} \) from
\( A_{1} \): 
\begin{equation}
A_{1}'=A_{1}-A_{12'}.
\end{equation}
 In the case of the sum of the three areas \( A_{s}=A_{2'3'}+A_{13'}+A_{23'} \)
being greater than \( A_{1}' \), the first coordinate \( x \) of the break
point \( P_{1} \) (see fig. \ref{fig:rect_knk}) 
is determined by 
\begin{equation}
x=\frac{A_{1}'}{A_{s}}.
\end{equation}
 Otherwise we continue the procedure correspondingly. The \( y \) coordinate
is determined as
\begin{equation}
y=\left\{ \begin{array}{cccl}
r\frac{A_{s}}{A_{2'3'}} & \textrm{if} & 0<r<\frac{A_{2'3'}}{A_{s}} & (\textrm{region }S_{2'3'})\\
\left( r-\frac{A_{2'3'}}{A_{s}}\right) \left( \frac{A_{s}}{A_{13'}}\right)  & \textrm{if} & \frac{A_{2'3'}}{A_{s}}<r<\frac{A_{2'3'}+A_{13'}}{A_{s}} & (\textrm{region }S_{13'})\\
\left( r-\frac{A_{2'3'}+A_{13'}}{A_{s}}\right) \left( \frac{A_{s}}{A_{23'}}\right)  & \textrm{if} & \frac{A_{2'3'}+A_{13'}}{A_{s}}<r<1 & (\textrm{region }S_{23'})
\end{array}\right. ,
\end{equation}
 with \( r \) being a random number between \( 0 \) and \( 1 \). This means
that after having determined in which of the regions we find the break point,
it is placed randomly on the world-line which points to the future. After having
obtained the break point \( P_{1} \) we continue the procedure in the same
way by obeying to the principle of causality. The area to sweep over is then
limited by the first break point as shown on fig.~\ref{fig:rect_knk}. 

\begin{figure}[htb]
{\par\centering \resizebox*{!}{0.45\textheight}{\includegraphics{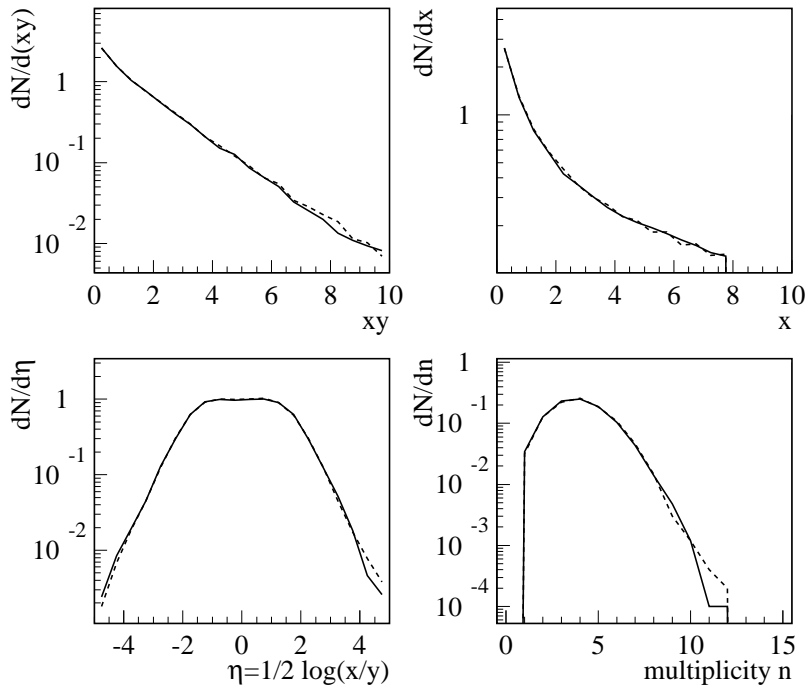}} \par}

\caption{\label{fig:test_knk}Distributions in \protect\( (xy)\protect \), \protect\( x\protect \),
\protect\( \frac{1}{2}\log \left( \frac{x}{y}\right) ,\protect \) and the multiplicity
\protect\( n\protect \) of break points. The full line represents the naive
method, the dashed one the direct one. The coordinates of the break points are
as defined on fig.~\ref{fig:area1}.}
\end{figure}
In fig.~\ref{fig:test_knk}, we apply our hadronization procedure, referred
to as \emph{direct method}, as discussed above, to calculate the distributions
of \( xy \), \( x \), \( \eta =\frac{1}{2}\log \left( \frac{x}{y}\right) , \)
and the multiplicity \( n \) of break points for a quark-anti-quark string
of \( E_{q}=E_{\bar{q}}=8GeV \). We compare our results with the \emph{the
naive method,} where the area of the string is divided into small elements \( \Delta A=8\cdot 8\textrm{ }\mathrm{GeV}^{2}/N^{2} \),
with \( N \) sufficiently large to not change the results any more. In each
of these elements, a break point is found with the probability \( \lambda \Delta A \).
The points which are in the future of another one are rejected. The latter method
is a literal realization of the area-law. As one can easily see in fig.~\ref{fig:test_knk},
the two methods agree within statistical errors.

\section{From String Fragments to Hadrons}

So far, we discussed how to break a string into small pieces, i.e. string fragments
with invariant masses between 0 and about two GeV. In order to identify string
fragments and hadrons, we first have to define the flavors (= quark content)
of the fragments, and then we have to discuss the question of fragment masses.

\subsubsection*{Flavors of String Fragments}

A string as a whole has some flavor, carried by the partons at its two extremities.
Additional flavor is created (by definition) at each break point in the form
of a quark-anti-quark or a diquark-anti-diquark pair of a certain flavor. The
corresponding probabilities are free parameters of the model. In case of quark-anti-quark
formation, we introduce the parameter \( p_{\mathrm{ud}} \), which gives the
probability of flavor \( u \) or \( d \). The probability to get an \( s \)-quark
is therefore \( 1-2p_{\mathrm{ud}} \) which is smaller than \( p_{\mathrm{ud}} \)
because on the larger mass of the \( s \) quark. For diquark-anti-diquark production,
we introduce the corresponding probability \( p_{\mathrm{diquark}} \).

\subsubsection*{Masses of String Fragments}

In the following, we show how to determine the masses of string fragments, characterized
by break points in the parameter space.

\begin{figure}[htb]
{\par\centering \resizebox*{!}{0.27\textheight}{\includegraphics{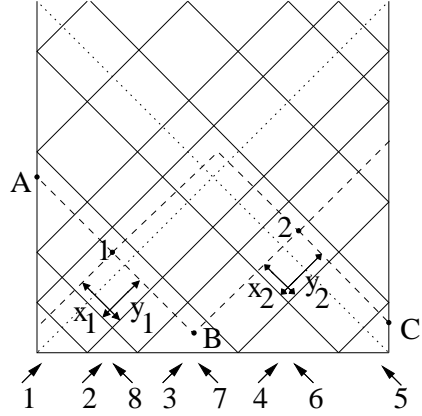}} \par}

\caption{How to calculate the mass of a sub-string. \label{fig:masse1}}
\end{figure}
Fig.~\ref{fig:masse1} shows an example of two break points for a string with
3 inner kinks. The momentum bands and the regions of the their overlaps are
shown: in case of the inner bands, we have three R-bands (2,3,4) and three L-bands
(6,7,8). The bands at the extremities play a special role, since we may have
the corresponding R- and L-band as just one band, due to the fact that one of
the bands is reflected immediately. So, we consider two ``double bands'' (1
and 5). 

The string momentum is given as

\begin{equation}
p_{\mathrm{string}}=\int _{C}\left[ \dot{x}d\sigma +x'd\tau \right] ,
\end{equation}
where \( C \) is an arbitrary curve from one border (\( \sigma =0 \)) to the
other (\( \sigma =\pi  \)) in the parameter space. This leads to
\begin{equation}
p_{\mathrm{string}}=\frac{1}{2}\int _{C}\left( g(\sigma +\tau )+g(\sigma -\tau )\right) d\sigma +\frac{1}{2}\int _{C}\left( g(\sigma +\tau )-g(\sigma -\tau )\right) d\tau 
\end{equation}
The momenta of the bands are by definition
\begin{equation}
p_{(i)}=\left\{ \begin{array}{l}
\frac{1}{2}\int _{\mathrm{band}\, i}g(\sigma -\tau )d\sigma -\frac{1}{2}\int _{\mathrm{band}\, i}g(\sigma -\tau )d\tau \quad \mathrm{if}\, \mathrm{R}\! -\! \mathrm{band}\\
\frac{1}{2}\int _{\mathrm{band}\, i}g(\sigma +\tau )d\sigma +\frac{1}{2}\int _{\mathrm{band}\, i}g(\sigma +\tau )d\tau \quad \mathrm{if}\, \mathrm{L}\! -\! \mathrm{band}
\end{array}\right. ,
\end{equation}
where one integrates along an arbitrary curve from one border of the band to
the other. An important property: an integration path parallel to a band provides
zero contribution. One has to pay attention for the bands at the extremities:
integrating only along \( \tau =0 \) represents only half the band. We have
\begin{equation}
\sum _{i}p_{(i)}=p_{\mathrm{string}.}
\end{equation}
The momenta of the bands are related to the corresponding parton momenta as
\begin{equation}
p_{(i)}=\left\{ \begin{array}{lll}
\frac{1}{2}p_{\mathrm{parton}} & \mathrm{if} & \mathrm{inner}\, \mathrm{band}\\
p_{\mathrm{parton}} & \mathrm{if} & \mathrm{outer}\, \mathrm{band}
\end{array}\right. ,
\end{equation}
which one verifies easily by expressing \( g \) in terms of the parton momenta.
The difference between inner and outer bands is due to the fact that the outer
ones (at the extremities) represent in reality two bands. For the example of
fig. \ref{fig:masse1}, we have 

\begin{eqnarray}
p_{(1)} & = & p_{q}\nonumber \\
p_{(2)}=p_{(8)} & = & \frac{1}{2}p_{g_{1}}\nonumber \\
p_{(3)}=p_{(7)} & = & \frac{1}{2}p_{g_{2}}\\
p_{(4)}=p_{(6)} & = & \frac{1}{2}p_{g_{3}}\nonumber \\
p_{(5)} & = & p_{\bar{q}}\nonumber 
\end{eqnarray}
Summing over the bands, we get
\begin{eqnarray}
\sum _{i=1}^{8}p_{(i)}=p_{q}+p_{g_{1}}+p_{g_{2}}+p_{g_{3}}+p_{\bar{q}}=p_{\mathrm{string}}, & \label{x} 
\end{eqnarray}
 which is the total momentum of the string. 

For a fragment of the string, the momentum is given as 
\begin{equation}
p_{\mathrm{fragm}}=\frac{1}{2}\int _{C'}\left( g(\sigma +\tau )+g(\sigma -\tau )\right) d\sigma +\frac{1}{2}\int _{C'}\left( g(\sigma +\tau )-g(\sigma -\tau )\right) d\tau ,
\end{equation}
where the path of the integration \( C' \) is an arbitrary curve between two
breakpoints, or between one break point and a boundary. One may write
\begin{equation}
p_{\mathrm{fragm}}=p_{\mathrm{R}}+p_{\mathrm{L}},
\end{equation}
 with
\begin{eqnarray}
p_{\mathrm{R}} & = & \frac{1}{2}\int _{C'}g(\sigma -\tau )d\sigma -\frac{1}{2}\int _{C'}g(\sigma -\tau )d\tau ,\\
p_{\mathrm{L}} & = & \frac{1}{2}\int _{C'}g(\sigma +\tau )d\sigma +\frac{1}{2}\int _{C'}g(\sigma +\tau )d\tau ,
\end{eqnarray}
where \( p_{\mathrm{R}} \) and \( p_{\mathrm{L}} \) represent sums of momenta
of R-bands and L-bands. For the example of fig. \ref{fig:masse1}, we can choose
the path \( (1\rightarrow B),\, (B\rightarrow 2) \) for the string fragment
between the break points 1 and 2. Since the first path is parallel to all L-bands
(only R-bands contribute) and the second one is parallel to all R-bands (only
L-bands contribute), we find
\begin{equation}
\label{p-frag}
p_{\mathrm{fragm}\, 1-2}=p_{1\rightarrow B}+p_{B\rightarrow 2}
\end{equation}
 with
\begin{eqnarray}
p_{1\rightarrow B} & = & x_{1}p_{(1)}+p_{(2)}+(1-x_{2})p_{(3)},\\
p_{B\rightarrow 2} & = & (1-y_{1})p_{(7)}^{\mu }+p_{(6)}^{\mu }+y_{2}p_{(5)}^{\mu },
\end{eqnarray}
where the factors \( x_{1} \), \( (1-x_{1}) \), \( y_{1} \), \( (1-y_{1}) \)
represent the fact that the bands at the extremities are only partially integrated
over. The other string fragments are treated correspondingly. For the left string
fragment, we may chose the integration path \( (A\rightarrow 1) \), for the
right one \( (2\rightarrow C) \). So we find for the three string fragments
(referred to as \( a \), \( b \), \( c \)) of fig. \ref{fig:masse1} the
momenta 
\begin{eqnarray}
p_{\mathrm{a}} & = & y_{1}p_{(7)}+p_{(8)}+(1-x_{1})p_{(1)}\label{for:impulsa} \\
p_{\mathrm{b}} & = & x_{1}p_{(1)}+p_{(2)}+(1-x_{2})p_{(3)}+y_{2}p_{(5)}+p_{(6)}+(1-y_{1})p_{(7)}\label{for:impulsb} \\
p_{\mathrm{c}} & = & x_{2}p_{(3)}+p_{(4)}+(1-y_{2})p_{(5)}\, .
\end{eqnarray}
It is easy to verify that the sum of the three sub-strings gives the total momentum
of the string,
\begin{equation}
p_{a}+p_{b}+p_{c}=\sum _{i=1}^{8}p_{(i)}=P_{\mathrm{string}}.
\end{equation}
 The mass squared of the string fragments is finally given as 
\begin{equation}
m^{2}_{\mathrm{fragm}}=p_{\mathrm{fragm}}^{2},
\end{equation}
 for example
\begin{equation}
m_{a}^{2}=2\left( p_{(1)}p_{(8)}-x_{1}p_{(8)}p_{(1)}+y_{1}\left( p_{(7)}p_{(8)}+p_{(7)}p_{(1)}\right) -x_{1}y_{1}p_{(7)}p_{(1)}\right) \, ,
\end{equation}
where we took advantage of the light-cone character of the momenta of the bands
\( (p_{(i)}^{2}=0). \)

\subsubsection*{Determination of Hadrons}

So far, we have determined the flavor \( f \) and the mass \( m \) of each
string fragment. In order to identify string fragments with hadrons, we construct
a mass table, which defines the hadron type as a function of the mass and the
flavor of the fragment. For a given flavor \( f \) of the fragment, we introduce
a sequence \( m^{f}_{1}<m^{f}_{2}<... \) of masses, such that in case of a
fragment mass being within an interval \( [m^{f}_{i-1},m^{f}_{i}] \), one assigns
a certain hadron \( h_{i} \). The masses \( m^{f}_{i} \) are determined by
the masses of the neighboring particles. So we decide for a \( u-\bar{u} \)
pair to be a pion if its mass is between \( 0 \) and \( (140+770)/2=455 \)
MeV. So the particle masses give a natural parameterization. This works, however,
only up to strange flavor. For charm and bottom flavor we choose with a fraction
\( 1:3 \) between pseudo-scalar and vector-mesons.

\subsubsection*{Mass Corrections}

An unrealistic feature of our approach, so far, is the fact that stable particles
are in general off-mass-shell. In order to correct for this, we employ a slight
modification of the break point such that the on-shell mass is imposed. 

\begin{figure}[htb]
{\par\centering \resizebox*{!}{0.27\textheight}{\includegraphics{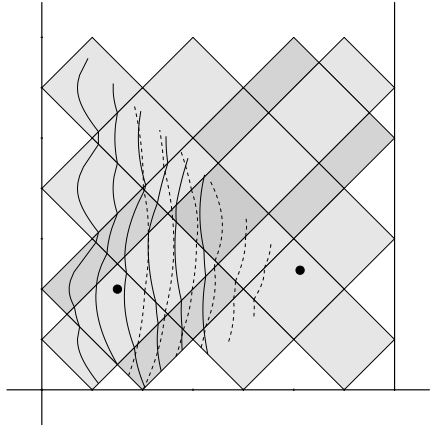}} \par}

\caption{The hyperbolas of constant mass for the two sub-strings if one moves the break-point
on the left side. The solutions of the mass conditions are the crossing of the
hyperbolas.\label{fig:mass2}}
\end{figure}

Let us again consider the example of fig. \ref{fig:masse1}. For a given mass,
the parameters \( x_{1} \) and \( y_{1} \) describe hyperbolas in the regions
of overlapping bands (different ones in different regions). Fig.~\ref{fig:mass2}
shows 
for our example some curves of constant mass for the left sub-string (between
the left side and the first break point). In the same way we find hyperbolas
of constant mass for the right sub-string (between the break points 1 and 2). 

If two neighboring substrings are stable particles, one needs to impose on-shell
masses to both fragments, which amounts to find the intersection of the two
corresponding hyperbolas. 

\begin{figure}[htb]
{\par\centering \resizebox*{!}{0.27\textheight}{\includegraphics{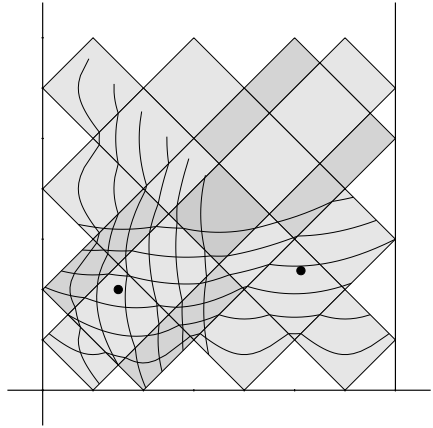}} \par}

\caption{\label{fig:mass3}Lines of constant masses (vertical lines) and constant proper
time (horizontal lines).}
\end{figure}

If one has to modify the break point according to only one mass condition, with
the mass of the second sub-string being still large enough not to represent
a stable hadron, a possible break point must lie on the corresponding hyperbola.
To completely determine the point, we need a second condition. Apart from the
squared mass, another Lorentz invariant variable available is the squared proper
time of the break point, defined as 
\begin{equation}
\Gamma ^{2}=\left( x(\sigma _{\mathrm{break}},\tau _{\mathrm{break}})\right) ^{2}.
\end{equation}
 So the second condition is the requirement that the proper time of the new
break point should coincide with the proper time of the original one. To calculate
the proper time, we use eq.~(\ref{for:ansatz}), to obtain
\begin{equation}
\Gamma ^{2}=\left( \frac{1}{2}\int _{\sigma -\tau }^{\sigma +\tau }g(\xi )\mathrm{d}\xi \right) ^{2}.
\end{equation}
The integration is done in the same way as for the masses, it is a summation
of the momenta of the bands or of fractions of them. In the case of our example,
we find in the region \( (1,7) \)

\begin{eqnarray}
\Gamma ^{2} & = & \frac{1}{4}\left( x_{1}p_{(1)}+p_{(2)}+y_{1}p_{(3)}\right) ^{2}\\
 & = & \frac{1}{2}\left( x_{1}p_{(1)}p_{(2)}+y_{1}p_{(2)}p_{(3)}+x_{1}y_{1}p_{(1)}p_{(3)}\right) \, ,
\end{eqnarray}
which represents again a hyperbola in the parameter space, as shown in fig.~\ref{fig:mass3}. 
So finding a new break point amounts to finding the intersection of the two
curves (hyperbolas) representing constant mass and proper time.

\section{Transverse Momentum}

Inspired by the uncertainty principle, a transverse momentum is generated at
each breaking, which means that 4-vectors \( p_{\bot } \) and \( -p_{\bot } \)
are assigned to the string ends at both sides of the break point. First we choose
the absolute value \( k=|\vec{p}_{\perp }| \) of the transverse momentum according
to the distribution
\begin{equation}
f\left( k\right) \propto e^{-\frac{k}{2p^{\mathrm{break}}_{\perp }}},
\end{equation}
with the parameter \( p_{\perp }^{\mathrm{break}} \) to be fixed. We require
\( p_{\bot } \) to be orthogonal to the momenta \( p_{(i)} \) and \( p_{(j)} \)
of the two intersecting bands where the break point is located. So we have
\begin{eqnarray}
p_{\perp }p_{(i)} & = & 0\\
p_{\perp }p_{(j)} & = & 0\\
p_{\bot }^{2} & = & -k^{2}\, .
\end{eqnarray}
Technically this is most easily done, if we perform a Lorentz-boost into the
center of mass system of the two momenta \( p_{(i)} \) and \( p_{(j)} \) followed
by a rotation such that \( p_{(i)} \) is oriented along the z-axis. One defines
a vector \( p^{_{'}}_{\bot } \), having the components 
\begin{eqnarray}
(p^{_{'}}_{\bot })_{o} & = & 0\\
(p^{_{'}}_{\bot })_{x} & = & k\cos \alpha \\
(p^{_{'}}_{\bot })_{y} & = & k\sin \alpha \\
(p^{_{'}}_{\bot })_{z} & = & 0\, ,
\end{eqnarray}
 \( \alpha  \) being a random angle between \( 0 \) and \( 2\pi  \). The
transformation back to the original system gives the 4-vector \( p_{\bot } \). 

This operation modifies, however, the mass of the string. In order to account
for this, we consider the transverse momentum as an additional band of the string.
It is treated in the same way as the others with the only exception that we
do not look for break points in this region. For our example of fig. \ref{fig:masse1},
we obtain for the left string fragment the momentum
\begin{equation}
p_{\mathrm{a}}=p_{\bot }+y_{1}p_{(7)}+p_{(8)}+(1-x_{1})p_{(1)},
\end{equation}
rather than eq.~(\ref{for:impulsa}). The modification of the coefficients of
the corresponding hyperbola for the mass correction procedure is obvious. In
the case where we have to pass to another region to find a modified break point
for the mass correction, we have to perform a rotation such that the vector
\( p_{\bot } \) is transverse to the two momenta of the new region.

\section{The Fragmentation Algorithm }

In the following, we describe the fragmentation algorithm which is used to obtain
a complete set of particles from one string. 

\begin{enumerate}
\item For a given string, we look for break points. Let \( n \) be the number of
break points. 
\item For each break point, we generate a flavor and a transverse momentum. 
\item We choose one break point by random and calculate the masses of the two neighboring
substrings. 
\item If there is at least one mass in the region of the resonances, we try to modify
the break point as discussed to get exactly this mass. If this is not possible,
we reject (delete) this break point and go to step 3.
\item If the mass of a sub-string is bigger than the upper limit in the mass table,
we fragment this sub-string (go to step 1). 
\end{enumerate}
In this way, we can deal in an elegant manner with the kinematical constraints.
Often, break points are rejected when a sampled transverse momentum is too high,
which results in a negative mass squared for a final particle. In this case
we look for another break point with another transverse momentum until a valid
configuration is found.

\cleardoublepage

\chapter{Parameters\label{parameters}}

We discuss in this chapter the parameters of the model, how they are determined,
and also their values. Parameter fixing is done in a systematic way, starting
with the hadronization parameters and the ones determining the time-like cascade,
before considering parton-parton-scattering and hadron-hadron scattering.

\section{Hadronization  }

The breaking probability \( p_{\mathrm{break}} \) is the essential parameter
in the hadronization model to determine the multiplicity and the form of the
rapidity distribution. For \( \gamma p,\, pp,\, pA,\, AA \) , we use a a fixed
value, fitted to reproduce the pion multiplicity in \( \gamma p \) scattering.
For \( e^{+}e^{-} \)annihilation, this parameter is considered to be \( Q \)
dependent as 
\begin{equation}
\label{pbreak}
p_{\mathrm{break}}(Q)=0.14+\frac{7.16\mathrm{GeV}}{Q}-\frac{111.1\mathrm{GeV}^{2}}{Q^{2}}+\frac{855\mathrm{GeV}^{3}}{Q^{3}}
\end{equation}
 in the region \( 14\leq Q\leq 91.2 \) GeV, and to be constant outside this
interval.
\begin{figure}[htb]
{\par\centering \resizebox*{0.6\columnwidth}{!}{\includegraphics{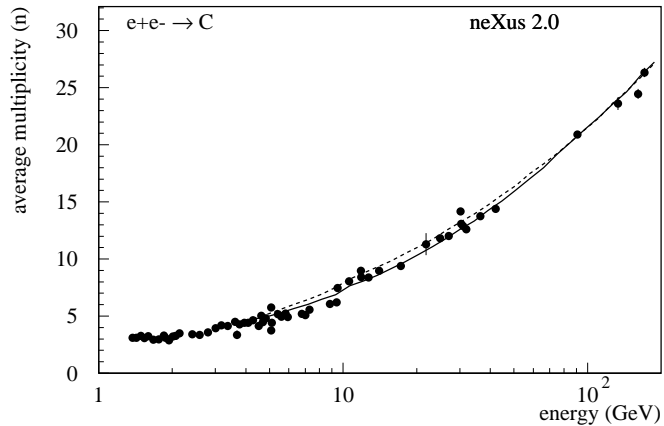}} \par}

\caption{\label{fig:muleng}Charged particles multiplicity as a function of energy.
The full line is for \protect\( p_{\mathrm{break}}\protect \) parameterized,
the dashed one for \protect\( p_{\mathrm{break}}=0.21\protect \).}
\end{figure}
\begin{figure}[htb]
{\par\centering \resizebox*{0.5\columnwidth}{!}{\includegraphics{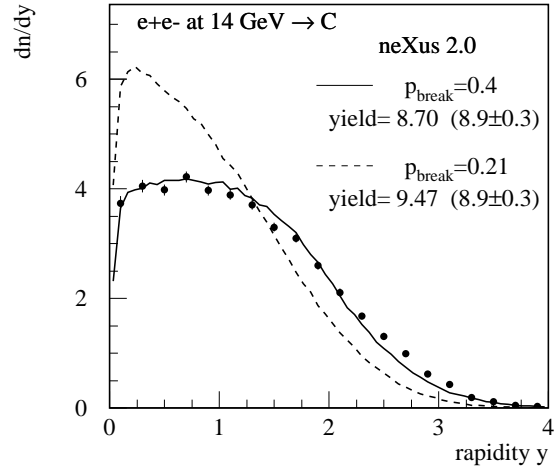}} \par}

\caption{\label{fig:ee14pbreak}Rapidity distributions for the parameters \protect\( p_{\mathrm{break}}=\protect \)
0.4 (full line) and \protect\( p_{\mathrm{break}}=0.21\protect \) (dashed).}
\end{figure}
\begin{figure}[htb]
{\par\centering \resizebox*{0.9\columnwidth}{!}{\includegraphics{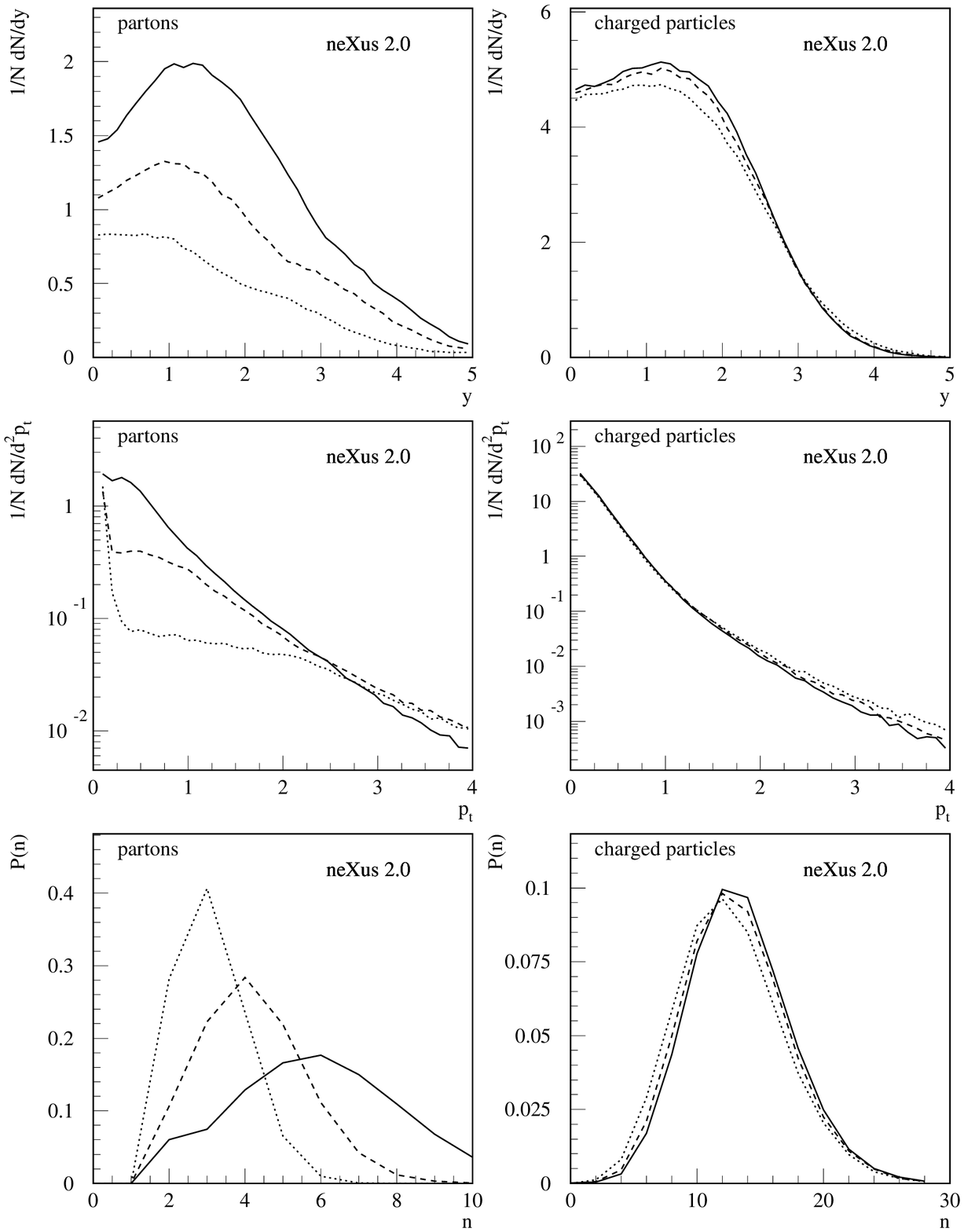}} \par}

\caption{\label{fig:partonsandcharged}Rapidity, transverse momentum and multiplicity
distributions for partons (left column) and for charged particles (right column).
The curves correspond to the parameters \protect\( q^{2}_{\mathrm{fin}}=0.25,\, 1.0,\, 2.0\protect \)
(full, dashed, dotted). Only parton distributions exhibit strong sensitivity
to the value of \protect\( q^{2}_{\mathrm{fin}}\protect \).}
\end{figure}
Figure \ref{fig:muleng} shows the total multiplicity of charged particles in
\( e^{+}e^{-} \) annihilation as a function of energy. The solid line corresponds
to the \( Q^{2} \) dependent \( p_{\mathrm{break}} \), the dashed line is
for \( p_{\mathrm{break}}=0.21 \) (the value for \( Q=91 \)GeV. In fig.~\ref{fig:ee14pbreak},
we show the corresponding rapidity distributions. The effect of making \( p_{\mathrm{break}} \)
Q-dependent shows up more in the shape of the rapidity distribution, not in
the total multiplicity. The parameter \( p^{\mathrm{break}}_{\bot } \), which
determines the transverse momenta of the partons at a string break, is determined
by investigating transverse momentum spectra of charged particles. A value of
0.50 GeV provides the best fit to data concerning \( \gamma p,\, pp,\, pA,\, AA \),
whereas for \( e^{+}e^{-} \) a value of 0.35 GeV is more favorable. The parameter
\( p_{\mathrm{ud}} \) affects strongly kaon production, we use \( p_{\mathrm{ud}}=0.44 \)
adjusted to the multiplicity of kaons. Baryon production is determined by \( p_{\mathrm{diquark}} \),
we use 0.08 adjusted to proton production in \( e^{+}e^{-} \)-annihilation.
In table \ref{hadr-par}, we give the complete list of hadronization parameters,
with their default values.
\begin{table}[htb]
{\centering \begin{tabular}{|l|l|l|}
\hline 
name&
value&
meaning\\
\hline 
\( p_{\mathrm{break}} \)&
eq. (\ref{pbreak}) for \( e^{+}e^{-} \)&
break probability\\
&
0.40 for \( \gamma p,\, pp,\, pA,\, AA \)&
\\
\hline 
\( p^{\mathrm{break}}_{\perp } \)&
0.35 GeV for \( e^{+}e^{-} \) &
mean transverse momentum\\
&
0.50 GeV for \( \gamma p,\, pp,\, pA,\, AA \)&
\( \quad  \)at break\\
\hline 
\( p_{\mathrm{ud}} \)&
0.44&
probability of \( u \)-\( \bar{u} \) or \( d \)-\( \bar{d} \) \\
\hline 
\( p_{\mathrm{diquark}} \)&
0.08&
probability of \( qq \)-\( \bar{q}q \)\\
\hline 
\end{tabular}\par}

\caption{Hadronization parameters.\label{hadr-par}}
\end{table}
So we use absolutely the same parameters for all the reactions \( \gamma p,\, pp,\, pA,\, AA \).
A perfect fit for \( e^{+}e^{-} \)requires a modification of two parameters,
\( p_{\mathrm{break}} \) and \( p^{\mathrm{break}}_{\perp } \).

\section{Time-Like Cascade}

Let us discuss the parameters which determine the time-like cascade, all fixed
via studying \( e^{+}e^{-} \) annihilation. For the pQCD parameter, we use
the usual leading order value \( \Lambda =0.2\textrm{ GeV} \). We have also
a technical parameter \( q_{\mathrm{fin}}^{2} \), which determines the lower
mass limit of partons in the time-like cascade. In figure \ref{fig:partonsandcharged},
we analyze how certain spectra depend on this parameter. We show rapidity, transverse
momentum and multiplicity distributions for partons and for charged particles
for an \( e^{+}e^{-} \) annihilation at 34 GeV. We show results for different
values of \( q^{2}_{\mathrm{fin}} \), namely \( 0.25 \) GeV\( ^{2} \), \( 1.0 \)
GeV\( ^{2} \)and \( 4.0 \) GeV\( ^{2} \), which means a lower mass limit
of \( 1 \) GeV, \( 2 \) GeV or \( 4 \) GeV, respectively. One sees that only
parton distributions are sensitive to the choice of this parameter, whereas
the corresponding charged particle spectra exhibit rather week dependence on
it. This can be explained from the fact that decreasing \( q^{2}_{\mathrm{fin}} \)
mainly results in production of additional partons with transverse momenta \( p^{2}_{\perp }\sim q^{2}_{\mathrm{fin}} \)
and such soft collinear partons hardly affect the fragmentation procedure, which
is in this sense ``infrared stable''. The number \( N_{f} \) of active flavors
is taken to be 5 for \( e^{+}e^{-} \) annihilation, whereas for \( \gamma p,\, pp,\, pA,\, AA \)
we use for the moment \( N_{f}=3 \). In table \ref{casc-par}
\begin{table}[htb]
{\centering \begin{tabular}{|l|l|l|}
\hline 
name&
value&
meaning\\
\hline 
\hline 
\( \Lambda  \)&
0.2 GeV&
pQCD parameter\\
\hline 
\( q^{2}_{\mathrm{fin}} \)&
0.25 GeV\( ^{2} \)&
transverse momentum cutoff\\
\hline 
\( N_{f} \) &
5 for \( e^{+}e^{-} \)&
active flavors\\
&
3 for \( \gamma p,\, pp,\, pA,\, AA \)&
\\
\hline 
\end{tabular}\par}

\caption{Time-like cascade parameters.\label{casc-par}}
\end{table}
 we show the cascade parameters and their default values.

\section{Parton-Parton Scattering}

There is first of all the parameter \( Q_{0} \) which defines the borderline
between soft and hard processes, where one has to choose a reasonable value
(say between 1 and 2 GeV\( ^{2} \)). 

Then we have a couple of parameters characterizing the soft Pomeron: the intercept
\( \alpha _{\mathrm{soft}}\! (0) \) and the slope \( \alpha '\! _{\mathrm{soft}} \)
of the Pomeron trajectory, the vertex value \( \gamma _{\mathrm{part}} \) and
the slope \( R_{\mathrm{part}} \) for the Pomeron-parton coupling, and the
characteristic hadronic mass scale \( s_{0} \). We have two parameters, \( \beta _{g} \)
and \( w_{\mathrm{split}} \), characterizing the coupling between the soft
Pomeron and the parton ladder. Whereas for \( s_{0} \) one just chooses some
``reasonable'' value and \( R_{\mathrm{part}} \) is taken to be zero, one
fixes the other parameters by trying to get a good fit for the total cross section
and the slope parameter for proton-proton scattering as a function of the energy
as well as the structure function \( F_{2} \) of deep inelastic lepton-proton
scattering. 

Concerning the hard scattering part, the resolutions scale \( p^{2}_{\bot \mathrm{res}} \)
and the K-factor \( K \) are fixed such that the standard parton evolution
is reproduced.

Finally, we have the triple Pomeron coupling weight \( r_{3\mathrm{I}\! \mathrm{P}} \),
which is fixed by as well checking the energy dependence of the proton-proton
total cross section.

\begin{table}[htb]
{\centering \begin{tabular}{|l|l|l|}
\hline 
name&
value&
meaning\\
\hline 
\hline 
\( Q^{2}_{0} \)&
1.5 GeV\( ^{2} \)&
soft virtuality cutoff\\
\hline 
\hline 
\( s_{0} \)&
1 GeV\( ^{2} \)&
soft mass scale\\
\hline 
\( \alpha _{\mathrm{soft}}\! (0) \)&
1.054&
Pomeron intercept\\
\hline 
\( \alpha '\! _{\mathrm{soft}} \)&
0.21 GeV\( ^{-2} \)&
Pomeron slope\\
\hline 
\( \gamma _{\mathrm{part}} \)&
1.11GeV\( ^{-1} \)&
Pomeron-parton coupling vertex\\
\hline 
\( R_{\mathrm{part}} \)&
0&
Pomeron-parton coupling slope\\
\hline 
\( \beta _{g} \)&
0.5&
Pomeron-ladder coupling parameter\\
\hline 
\( w_{\mathrm{split}} \)&
0.15&
Pomeron-ladder coupling parameter\\
\hline 
\( p^{2}_{\bot \mathrm{res}} \)&
0.25 GeV\( ^{2} \)&
resolutions scale\\
\hline 
\( K \)&
1.5&
K-factor\\
\hline 
\hline 
\( r_{3\mathrm{I}\! \mathrm{P}} \)&
0.0096 GeV\( ^{-1} \)&
triple Pomeron coupling constant\\
\hline 
\end{tabular}\par}

\caption{Parton-parton scattering parameters.\label{par-par}}
\end{table}
The values of these parameters are shown in table \ref{par-par}.

\section{Hadron-Hadron Scattering}

Let us first discuss the parameters related to the partonic wave function of
the hadron \( h \) (for numerical applications we only consider nucleons: \( h=N \)).
The transverse momentum distribution is characterized by the hadronic Regge
radius squared \( R^{2}_{N} \), the longitudinal momentum distributions are
given in terms of two exponents, \( \alpha _{\mathrm{remn}}^{N} \) and \( \alpha _{\mathrm{part}} \).
The latter one is taken to be independent of the hadron type as \( 2\alpha _{\mathrm{I}\! \mathrm{R}}(0)-1 \),
with the usual Reggeon intercept \( \alpha _{\mathrm{I}\! \mathrm{R}}(0)=1/2. \)
The parameter \( R_{N} \) also affects the proton-proton total cross sections,
whereas \( \alpha _{\mathrm{remn}}^{N} \) can be determined by investigating
baryon spectra (but it also influences the total cross section).

There are several more parameters, which have not been mentioned so far: the
remnant excitation probability \( p_{\mathrm{remn}.\mathrm{ex}} \) and the
exponent \( \alpha _{\mathrm{remn}.\mathrm{ex}} \), which gives a remnant mass
distribution as 
\begin{equation}
\left( M^{2}\right) ^{-\alpha _{\mathrm{remn}.\mathrm{ex}}}.
\end{equation}
The minimum string mass \( m_{\mathrm{string}\, \mathrm{min}} \) assures that
Pomerons with string masses less than this minimal mass are ignored. The partons
defining the string ends are assumed to have transverse momenta according a
Gaussian distribution with a mean value \( p_{\bot \mathrm{SE}} \). Diffractive
scattering is assumed to transfer transverse momentum according to a Gaussian
distribution, with a mean value \( p_{\bot \mathrm{diff}} \). The parameter
\( m_{\mathrm{string}\, \mathrm{min}} \) is taken to be just slightly bigger
than two times the pion mass to allow at least string fragmentation into two
pions. The other parameters can be fixed by comparing with experimental inclusive
spectra. 

\begin{table}[htb]
{\centering \begin{tabular}{|l|l|l|}
\hline 
name&
value&
meaning\\
\hline 
\hline 
\( R^{2}_{N} \)&
2 GeV\( ^{-2} \)&
parton-hadron coupling slope\\
\hline 
\( \alpha _{\mathrm{part}} \)&
0&
participant exponent\\
\hline 
\( \alpha _{\mathrm{remn}}^{N} \)&
1.5&
remnant exponent\\
\hline 
\hline 
\( p_{\mathrm{remn}.\mathrm{ex}} \)&
0.45&
remnant excitation probability \\
\hline 
\( \alpha _{\mathrm{remn}.\mathrm{ex}} \)&
1.4&
remnant excitation exponent\\
\hline 
\( m_{\mathrm{string}\, \mathrm{min}} \)&
0.29 GeV&
minimal string mass\\
\hline 
\( p_{\bot \mathrm{SE}} \)&
0.21 GeV&
mean \( p_{\bot } \) for string ends\\
\hline 
\( p_{\bot \mathrm{diff}} \)&
0.35 GeV&
mean \( p_{\bot } \) for diffractive scattering\\
\hline 
\end{tabular}\par}

\caption{hadron-hadron scattering parameters (for the case of nucleons).\label{hadr-par}}
\end{table}
In table \ref{hadr-par}, we show the numerical values of the parameters.

\cleardoublepage

\chapter{Testing Time-like Cascade and Hadronization: Electron-Positron Annihilation}

Electron-positron annihilation is the simplest possible system to test the time-like
cascade as well as the model of fragmentation, since the decay of a virtual
photon in electron-positron annihilation gives a quark and an anti-quark, both
emitting a cascade of time-like partons, which finally hadronize.

Electron-positron annihilation is therefore used to test both the time-like
cascade and the hadronization model, and in particular to fix parameters. 

The simulation of an electron-positron annihilation event can be divided into
three different stages: 

\begin{enumerate}
\item The annihilation into a virtual photon or a \( Z \) boson and its subsequent
decay into a quark-anti-quark pair (the basic diagram). 
\item The evolution of the quark and the anti-quark into on-shell partons by radiation
of perturbative partons (time-like cascade).
\item The transition of the partonic system into hadrons via a fragmentation model
(hadronization).
\end{enumerate}
These stages are discussed in the following sections. 

After having described the three stages of electron-positron annihilation, we
will be able to test the model against numerous data available. We will show
comparisons with experimental results at low energies at PETRA (DESY), by the
TASSO collaboration \cite{alt84}. The center-of-mass energies are 14, 22 and
34 GeV. Higher energies are reached at LEP, where we compare especially with
results for \( 91.2 \) GeV, the \( Z^{0} \) mass, where a big number of events
has been measured. 

By comparing with data, we will be able to fix the essential parameters of the
hadronization model, namely \( p_{\mathrm{break}} \), \( p_{\mathrm{ud}} \),
\( p_{\mathrm{diquark}} \) and \( p_{\bot }^{\mathrm{break}} \). The free
parameters in the parton cascade are the pQCD scaling parameter \( \Lambda  \)
and \( q_{\mathrm{fin}}^{2} \), representing the minimum transverse momentum
for a branching in the cascade. For the pQCD parameter, we use the usual leading
order value \( \Lambda =0.2\textrm{ GeV} \). The influence of the technical
parameter \( q^{2}_{\mathrm{fin}} \) has been investigated in detail. 
\newpage

\section{The Basic Diagram}

\begin{figure}[htb]
\begin{center}\begin{picture}(190,100) 
\ArrowLine(60,50)(10,10)\Text(10,30)[l]{$e^+$}
\ArrowLine(10,90)(60,50)\Text(10,70)[l]{$e^-$}
\Photon(60,50)(120,50){2}{6}\Text(90,60)[]{$\gamma^*,Z^{0*}$}
\ArrowLine(120,50)(180,90)\Text(170,70)[l]{$q$}
\ArrowLine(180,10)(120,50)\Text(170,30)[l]{$\bar{q}$}
\end{picture}\end{center}

\caption{\label{fig:branchement}Electron-positron annihilation.}
\end{figure}
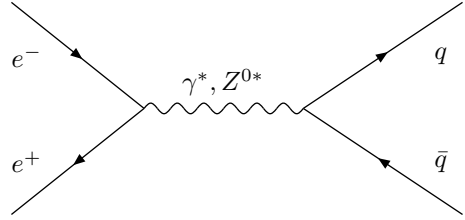
The first order differential cross section for the process 
\begin{equation}
\, e^{+}\, e^{-}\, \rightarrow \, \gamma ^{*}\, \mathrm{or}\, Z\, \rightarrow \, q\, \bar{q}\, 
\end{equation}
 is given as \cite{ell96}

\begin{eqnarray}
\frac{d\sigma }{d\cos \theta } & = & \frac{\pi \alpha ^{2}}{2s}\left[ \begin{array}{c}
\\
\\

\end{array}\left( 1+\cos ^{2}\theta \right) \left\{ q_{f}^{2}-2q_{f}V_{e}V_{f}\chi _{1}(s)+\left( A_{e}^{2}+V_{e}^{2}\right) \left( A_{f}^{2}+V_{f}^{2}\right) \chi _{2}(s)\right\} \right. \nonumber \\
 &  & +\left. \cos \theta \left\{ -4q_{f}A_{e}A_{f}\chi _{1}(s)+8A_{e}V_{e}A_{f}V_{f}\chi _{2}(s)\right\} \begin{array}{c}
\\
\\

\end{array}\right] \label{for:eecross} 
\end{eqnarray}
 with
\begin{eqnarray}
\chi _{1}(s) & = & \kappa \frac{s(s-M_{Z}^{2})}{(s-M_{Z}^{2})^{2}+\Gamma _{Z}^{2}M_{Z}^{2}},\\
\chi _{2}(s) & = & \kappa ^{2}\frac{s^{2}}{(s-M_{Z}^{2})^{2}+\Gamma _{Z}^{2}M_{Z}^{2}},
\end{eqnarray}
where \( \kappa  \) is given as 
\[
\kappa =\frac{\sqrt{2}G_{F}M_{Z}^{2}}{16\pi \alpha }.\]
Here, \( \alpha  \) is the fine structure constant, \( q_{f} \) the quark
flavor, \( M_{z} \) the mass of the \( Z \) boson, \( \Gamma _{Z} \) its
decay width. The vector and axial coupling factors are 
\begin{equation}
V_{f}=T_{f}^{3}-2q_{f}\sin ^{2}\theta _{W},\quad \, \, A_{f}=T_{f}^{3}
\end{equation}
 and 
\begin{equation}
T_{f}^{3}=\left\{ \begin{array}{lll}
\frac{1}{2} & \mathrm{if} & f=\nu ,u,c\\
-\frac{1}{2} & \mathrm{if} & f=e,d,s,b
\end{array}\right. .
\end{equation}
 \( \theta _{W} \) is the Weinberg mixing angle. The Fermi constant 
\begin{equation}
G_{F}=\frac{\sqrt{2}g^{2}_{W}}{8M_{W}^{2}}
\end{equation}
 is expressed via the weak coupling \( g^{2}_{W}=4\pi \alpha /\sin ^{2}\theta _{W} \)
and the W-boson mass \( M_{w} \). At low energies, \( s\ll M_{Z}^{2} \), we
recover the well known formula 
\begin{equation}
\sigma =\frac{4\pi \alpha ^{2}}{3s}q_{f}^{2}\, .
\end{equation}
The factors \( \chi _{1}(s) \) and \( \chi _{2}(s) \) correspond to the intermediate
\( Z \)-boson state and to the photon-Z-boson interference, respectively. Formula
(\ref{for:eecross}) can now be used to generate an initial quark-anti-quark
pair.

\section{The Time-like Parton Cascade and String Formation}

\begin{figure}[htb]
{\par\centering \resizebox*{!}{0.2\textheight}{\includegraphics{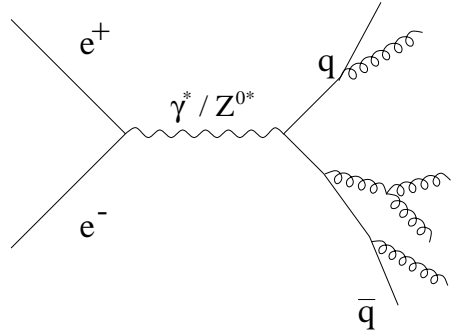}} \par}

\caption{Time-like parton cascade.\label{cascade} }
\end{figure}

If one considers a quark and an anti-quark coming from the decay of a virtual
photon or Z-boson to be on-shell, this amounts to a lowest order treatment.
At high energies, a perturbative correction has to be done. Since it is difficult
to calculate the higher order Feynman diagrams exactly, one uses the so-called
DGLAP evolution equations, which describes the evolution of a parton system
with leading logarithmic accuracy. This amounts to successively emitting partons
(time-like parton cascade, see fig.~\ref{cascade}).
 This has been discussed in detail in connection with the parton production
in proton-proton or nucleus-nucleus collisions. We use exactly the same method
here. Even for determining three momenta of the primary quark and anti-quark,
we do not need any new input, since this corresponds exactly to the case of
the time-like cascade of the two partons involved in the Born scattering in
hadronic collisions.

The fact that we use one and the same procedure for the time-like parton cascade
in all the different reactions, allows us to test elements of hadronic interactions
in a much simpler context of elementary electron-positron interactions.

\begin{figure}[htb]
{\par\centering (a){\Large \( \qquad \qquad \qquad \qquad  \)}(b)\( \qquad \qquad \qquad \qquad \qquad  \)\\
\( \,  \)\hfill{}\resizebox*{!}{0.2\textheight}{\includegraphics{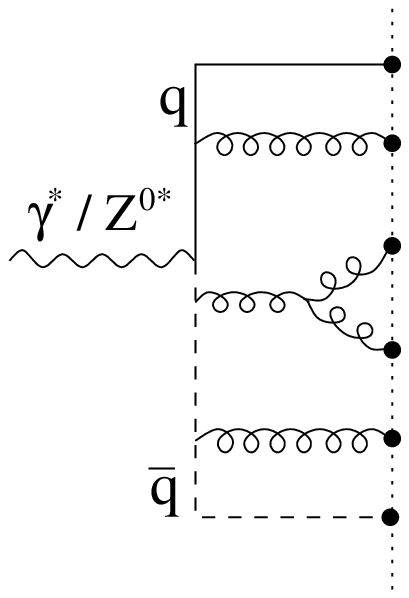}} \hfill{}
\resizebox*{!}{0.2\textheight}{\includegraphics{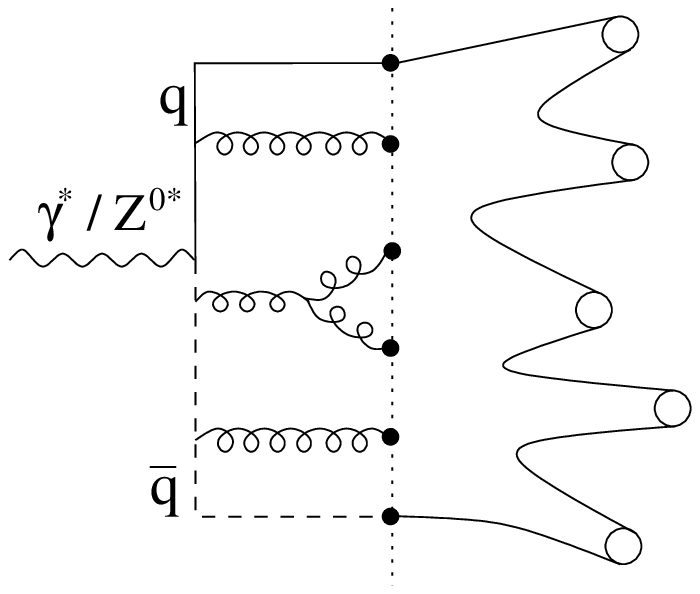}} \hfill{}\par}

\caption{\label{fig:fluxdecouleur}Forming strings for an \protect\( e^{+}e^{-}\protect \)
event.}
\end{figure}
The final step is the hadronization of the above-mentioned parton configuration.
Here we use the string model, as in case of \( pp \) or \( \gamma p \) scattering.
The string formation in \( e^{+}e^{-} \) is much simpler than in proton-proton
scattering, where the cut Pomeron is represented as a cylinder. The structure
of an \( e^{+}e^{-} \) event is planar in the sense that the whole event can
be represented on a plane. So we simply plot the diagram on a plane with only
one cut line. In fig.~\ref{fig:fluxdecouleur}(a), we present a half-plane (on
one side of the cut) for the amplitude shown in fig.~\ref{cascade}. The dotted
line represents the cut. There are a couple of partons crossing the cut, indicated
by dots. As in the case of proton-proton or photon-proton scattering, we identify
the cut line as a kinky relativistic string, with the partons representing the
kinks. So in our example, we have a kinky string with six kinks, two external
ones and four internal ones. We then apply the usual hadronization procedure,
discussed earlier in detail, in order to calculate hadron production from a
fragmenting string, see fig.~\ref{fig:fluxdecouleur}(b).

\section{Event Shape Variables }

\begin{figure}[htb]
{\par\centering \resizebox*{0.9\columnwidth}{!}{\includegraphics{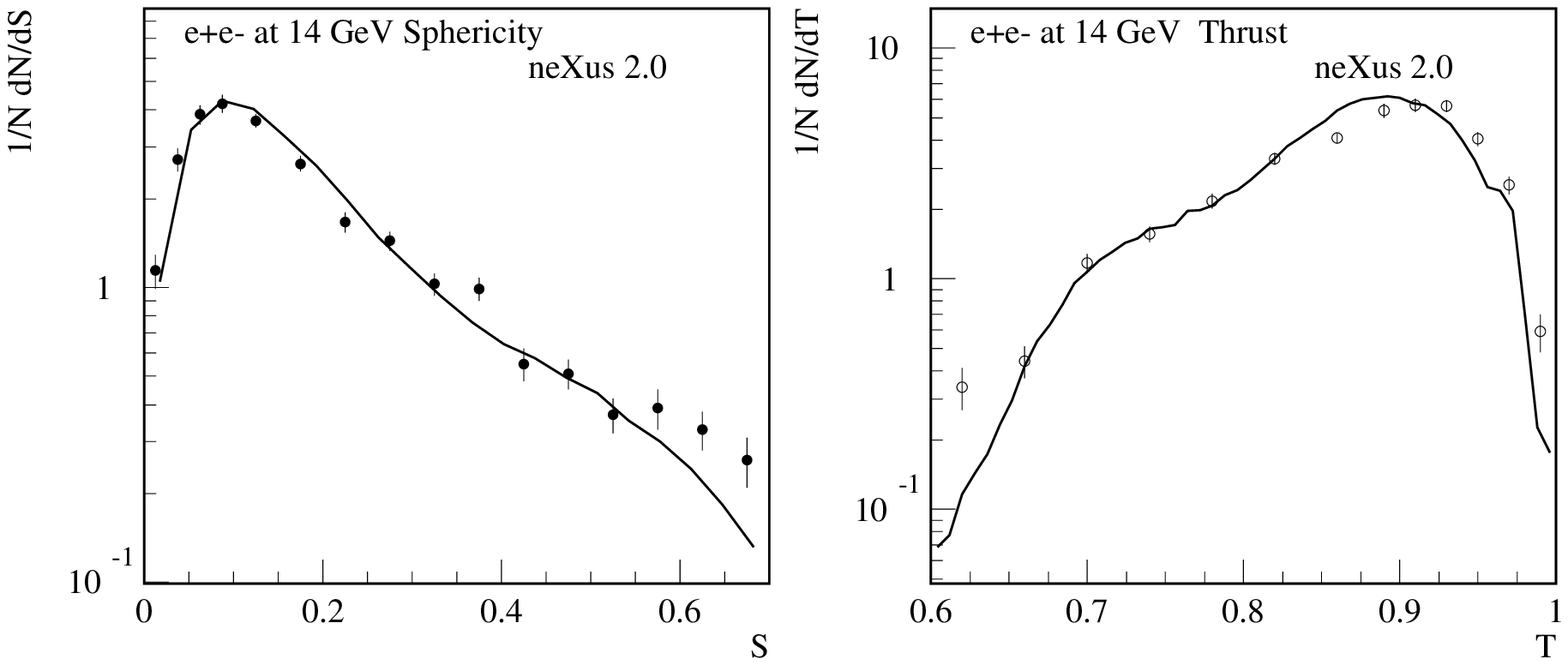}} \par}

{\par\centering \resizebox*{0.9\columnwidth}{!}{\includegraphics{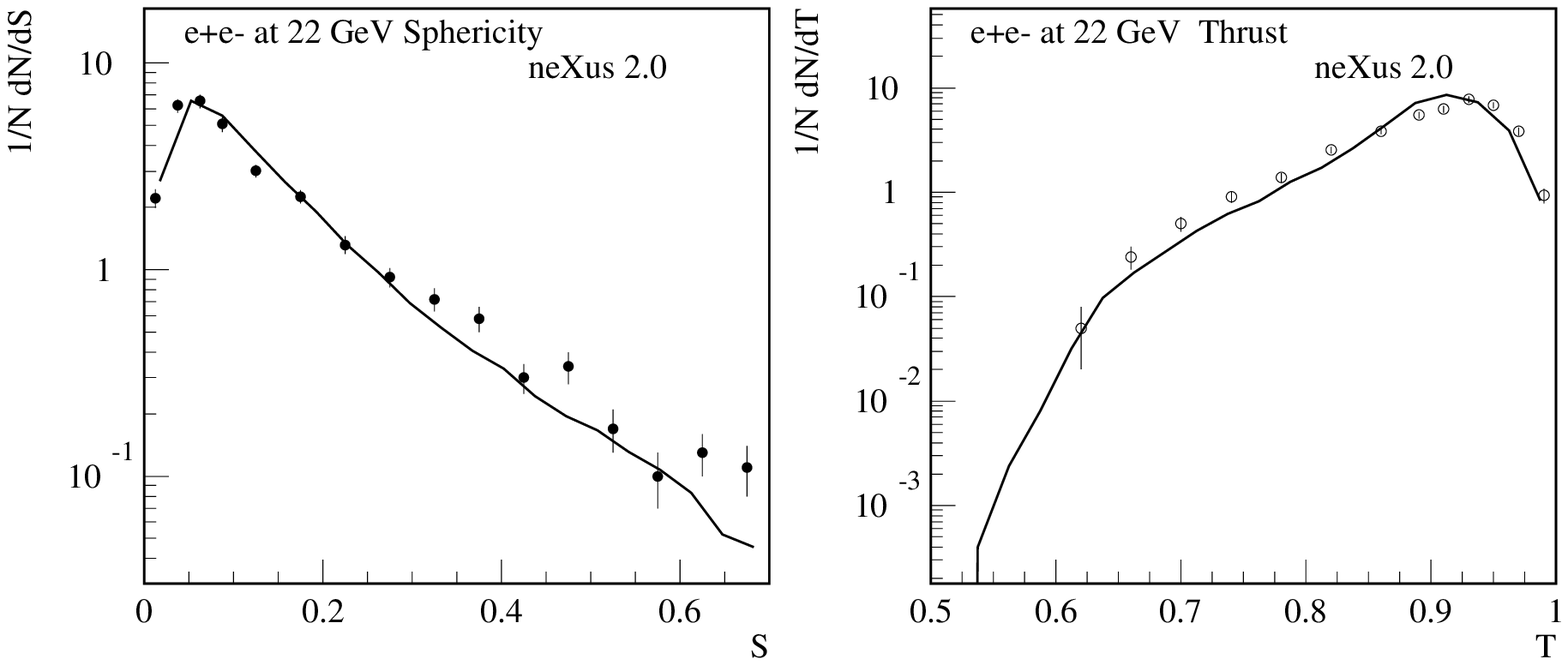}} \par}

{\par\centering \resizebox*{0.9\columnwidth}{!}{\includegraphics{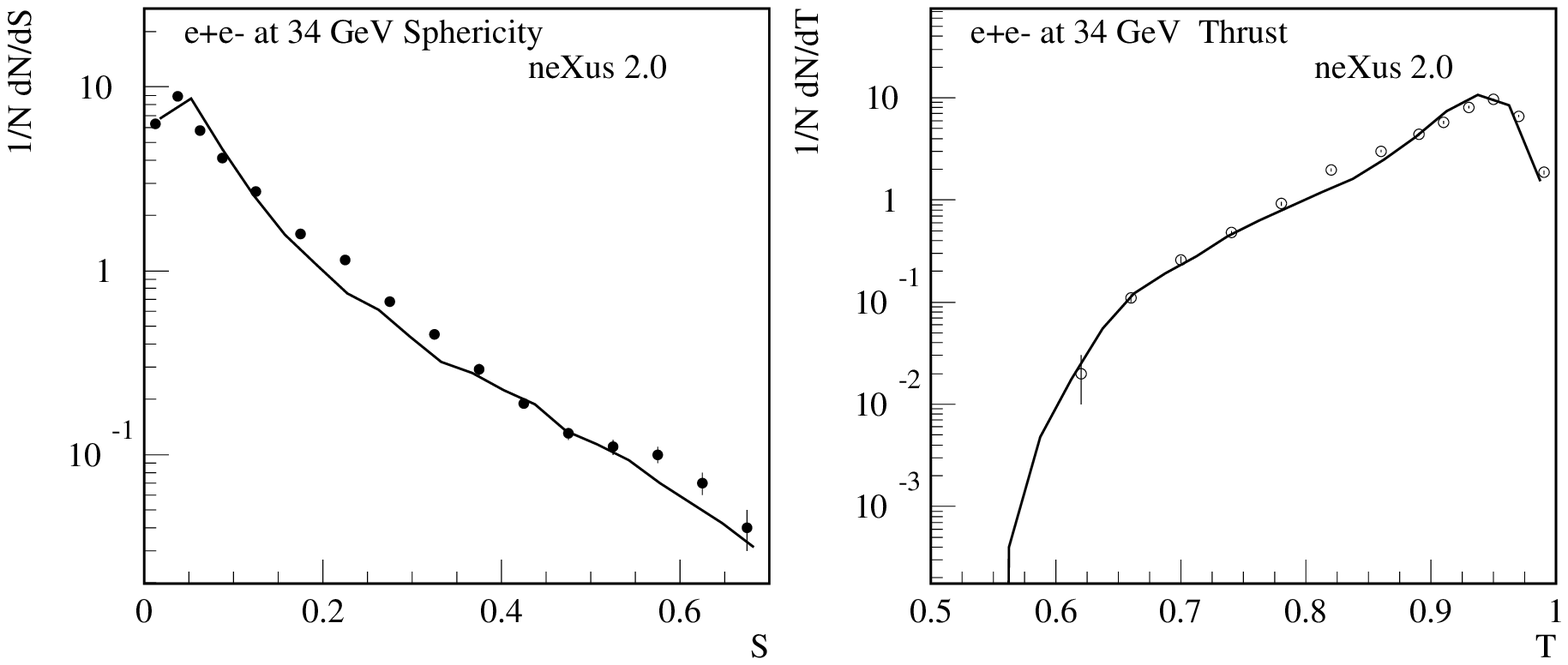}} \par}

\caption{\label{fig:ST14-34}The sphericity and thrust for the energies 14 GeV, 22 GeV
and 34 GeV.}
\end{figure}
 
\begin{figure}[htb]
{\par\centering \resizebox*{1\columnwidth}{!}{\includegraphics{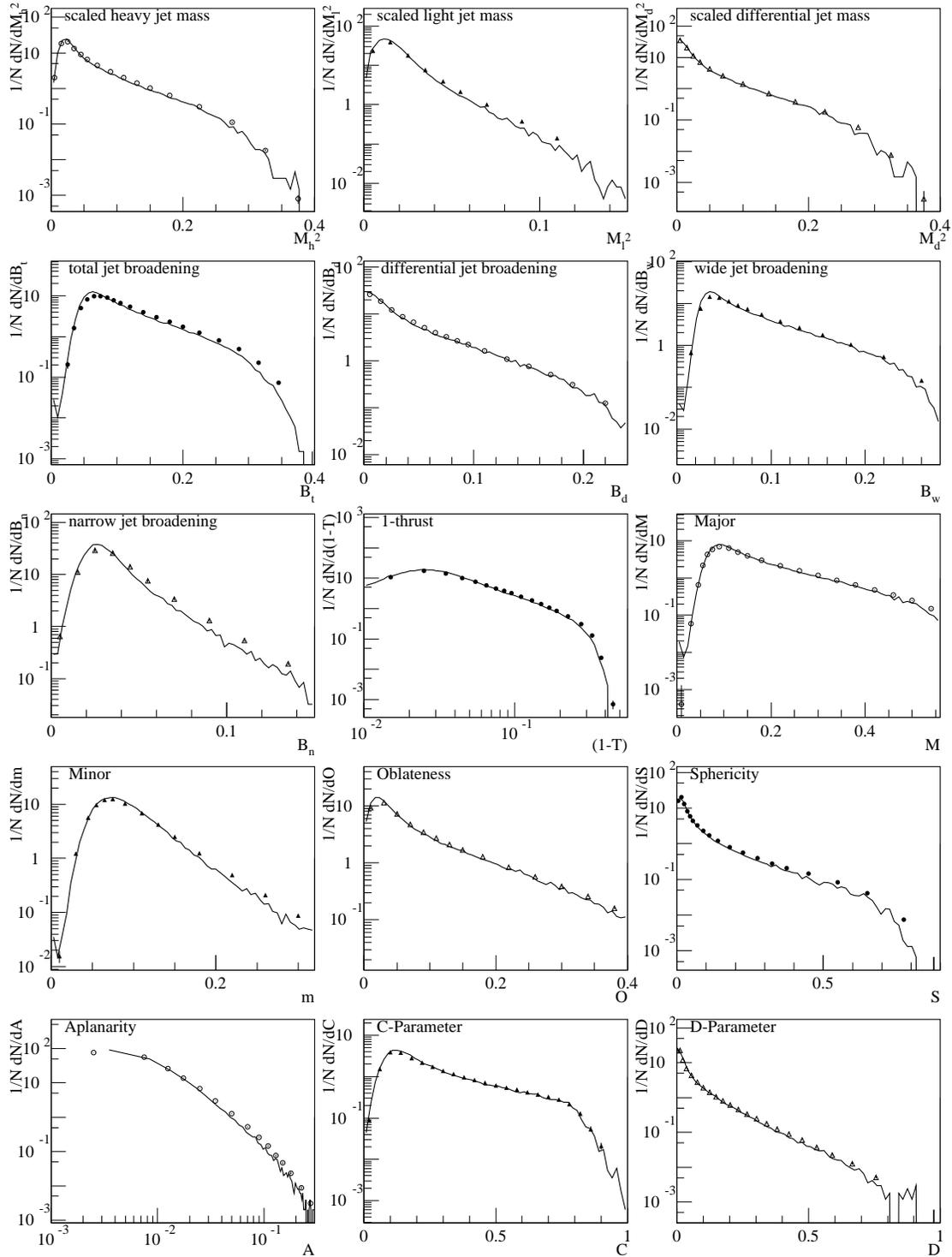}} \par}

\caption{Event shape variables at 91.2 GeV. The data (dots) are from the DELPHI collaboration
\cite{abr96}. \label{91}}
\end{figure}
\begin{figure}[htb]
{\par\centering \resizebox*{1\columnwidth}{!}{\includegraphics{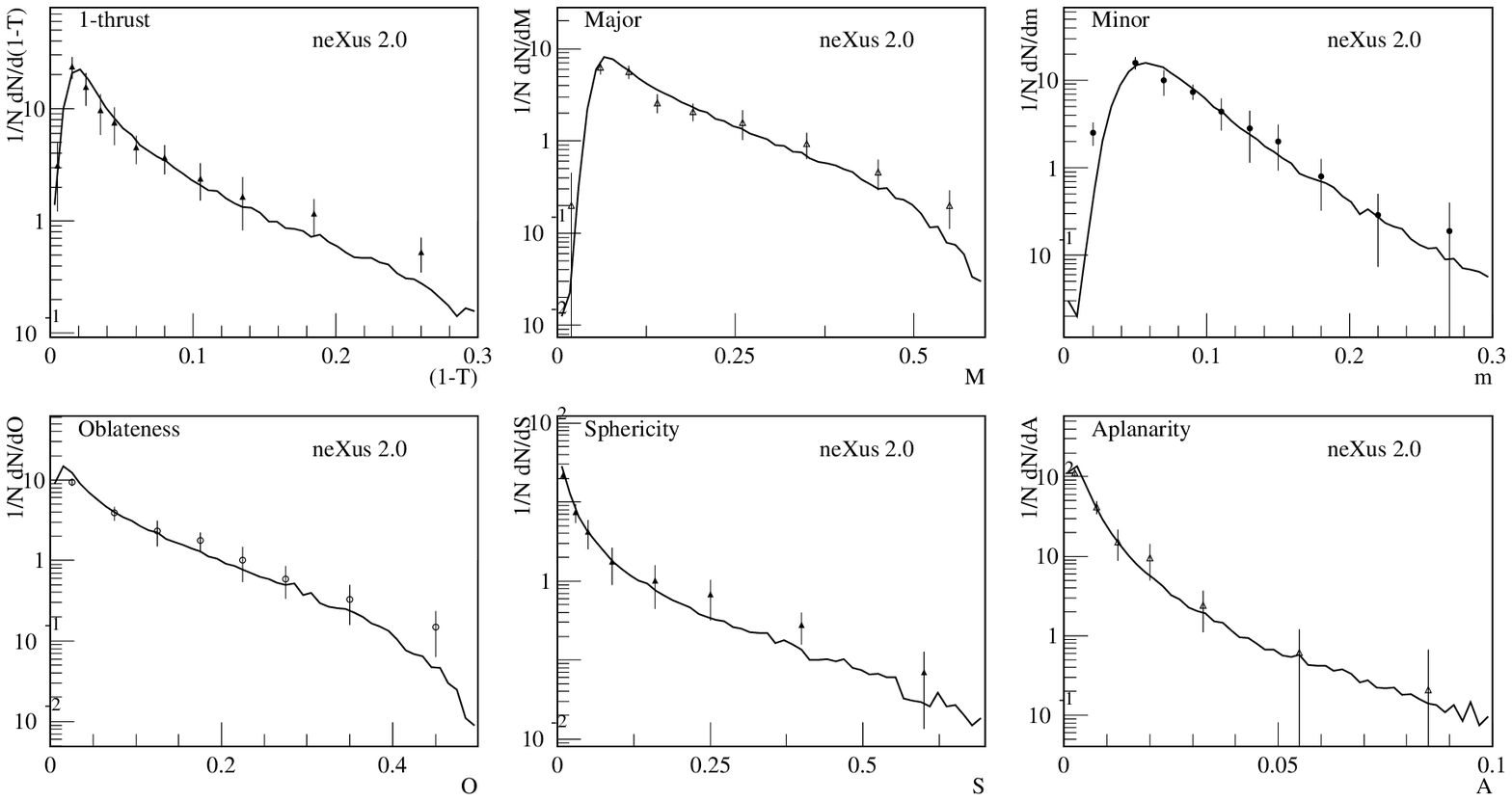}} \par}

\caption{Event shape variables for 133 GeV. The data (dots) are from OPAL collaboration
\cite{ale96}.\label{133}}
\end{figure}
\begin{figure}[htb]
{\par\centering \resizebox*{1\columnwidth}{!}{\includegraphics{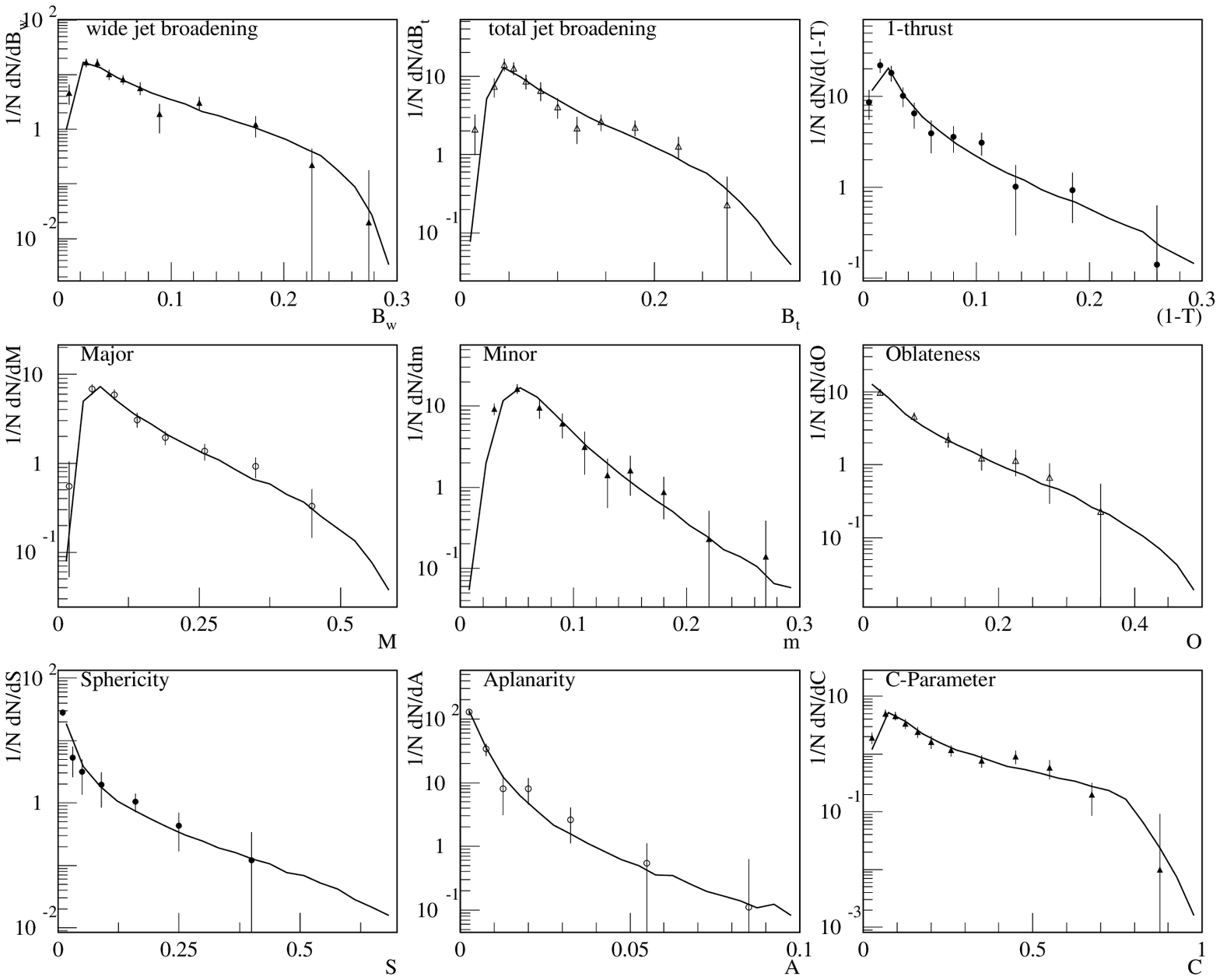}} \par}

\caption{Event shape variables for 161 GeV. The data (dots) are from OPAL collaboration
\cite{ale96}.\label{161}}
\end{figure}
\begin{figure}[htb]
{\par\centering \resizebox*{1\columnwidth}{!}{\includegraphics{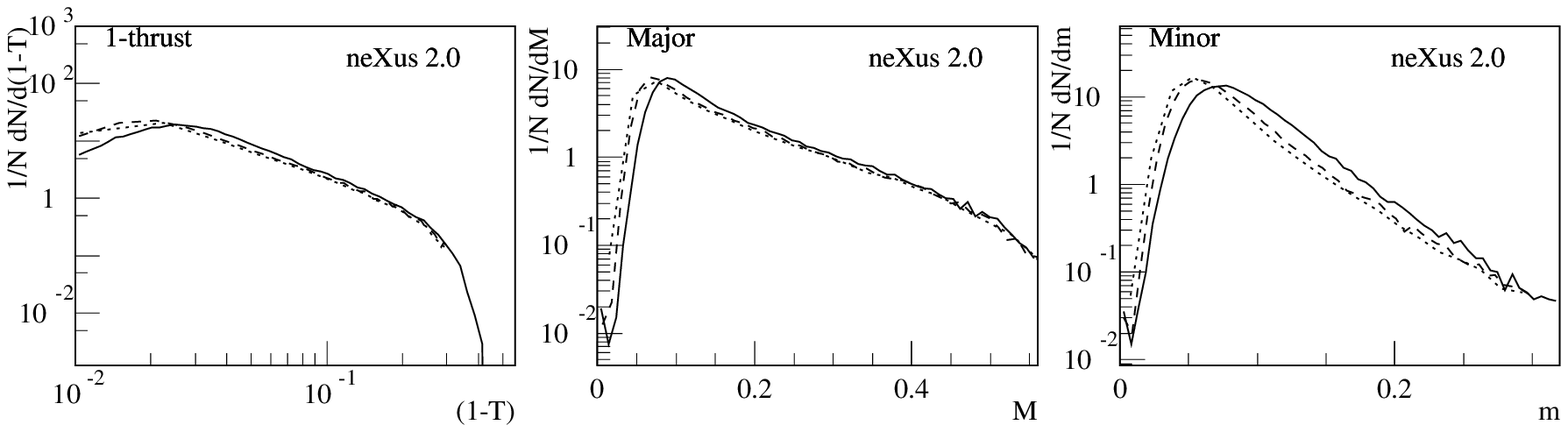}} \par}

\caption{\label{fig:tmm}Thrust, minor and major for 91 GeV (full line), 133 GeV (dashed)
and 161 GeV (dotted). }
\end{figure}
We start our presentation of results by considering the so-called event-shape
variables, which describe the form of an \( e^{+}e^{-} \) event in general.
For example, one is interested in knowing whether the particle momenta are essentially
aligned along a certain axis, distributed isotropically over the phase space,
or lying more or less in a plane. In the figs.~\ref{fig:ST14-34} to \ref{fig:tmm},
we are going to compare our calculated distributions of several event-variables
with data. Let us first discuss the different event-shape variables, one after
the other.

\subsubsection{Sphericity}

The sphericity is defined by the eigenvalues of the sphericity tensor, 
\begin{equation}
S^{\alpha \beta }=\sum _{i}(p_{i})^{\alpha }(p_{i})^{\beta }\, \, ,
\end{equation}
where \( i \) sums all particles and \( p^{\alpha } \) is the particle four-momentum.
One finds three eigenvalues \( \lambda _{i} \) with \( \lambda _{1}<\lambda _{2}<\lambda _{3} \)
and \( \lambda _{1}+\lambda _{2}+\lambda _{3}=1 \). The sphericity is then
defined as 
\begin{equation}
S=\frac{3}{2}(\lambda _{1}+\lambda _{2})\, .
\end{equation}
For a perfectly isotropic event, one finds \( \lambda _{1}=\lambda _{2}=\lambda _{3}=1/3 \)
and therefore \( S=1 \). An event oriented along one axis gives \( S=3/2(0+0)=0 \)
. 

To test whether an event has planar geometry, one defines the aplanarity 
\begin{equation}
A=\frac{3}{2}\lambda _{1\, .}
\end{equation}
For events in a plane we will find \( \lambda _{1}=A=0 \). The maximum of this
value is \( A=3/2\cdot 1/3=1/2 \) for an isotropic event, since the eigenvalues
are ordered. 

The three eigenvectors \( \overrightarrow{v}_{1,2,3} \) of the matrix \( S^{\alpha \beta } \)
can be used to define a coordinate system.

\subsubsection{C and D Parameters}

The C-parameter is defined by 
\begin{equation}
C=3\left( \lambda _{1}\lambda _{2}+\lambda _{1}\lambda _{3}+\lambda _{2}\lambda _{3}\right) ,
\end{equation}
with \( \lambda _{1,2,3} \) being the eigenvalues of the tensor 
\begin{equation}
M^{\alpha \beta }=\frac{\sum _{i}\frac{p_{i}^{\alpha }p^{\beta }_{i}}{|p_{i}|}}{\sum _{i}|p_{i}|}\, .
\end{equation}
The \( D \) -parameter is 
\begin{equation}
D=27\, \lambda _{1}\lambda _{2}\lambda _{3}\, .
\end{equation}

These values measure the multiple jet-structure of events. For small values
of C two of the eigenvalues are close to zero, we have a two-jet event. If one
of the three eigenvalues is close to zero, the D-parameter is approaching zero
as well, we have at least a planar event.

\subsubsection{Thrust}

The thrust of an event is defined as
\begin{equation}
T=\max _{\vec{n}}\frac{\sum _{j}\left| \vec{n}\cdot \vec{p}_{j}\right| }{\sum _{j}\left| \vec{p}_{j}\right| }\, .
\end{equation}
The vector \( \vec{n}_{\mathrm{thrust}} \), which maximizes this expression,
defines the thrust axis. A two-jet event will give a thrust value of 1 and a
thrust axis along the two jets. An isotropic event gives \( T\sim 1/2 \).

One can repeat the same algorithm with the imposed condition \( \vec{n}\perp \vec{n}_{\mathrm{thrust}} \);
this gives an expression for the major \( M \) with the axis \( \vec{n}_{\mathrm{Major}} \). 

A third variable, the minor \( m \), is obtained by evaluating the above expression
with \( \vec{n}\perp \vec{n}_{\mathrm{thrust}} \) and \( \vec{n}\perp \vec{n}_{\mathrm{Major}} \),
the axis being already given. Each of these values describes the extension of
the event perpendicular to the thrust axis. Similar values for \( M \) and
\( m \) describe therefore a cylindrical event. For this, the oblateness is
defined as \( O=M-m \) which, as the aplanarity, describes a cylindrical event
for \( O=0 \) an a planar event for higher values.

\subsubsection{Jet Broadening}

In each hemisphere, the sum of the transverse momenta of the particles relative
to the thrust axis is divided by the sum of the absolute values of the momenta.
\begin{equation}
B_{\pm }=\frac{\sum _{\pm \vec{p}.\vec{n}_{\mathrm{thrust}}>0}|\vec{p}_{i}\times \vec{n}_{\mathrm{thrust}}|}{2\sum _{i}|\vec{p}_{i}|}
\end{equation}
The greater \( B_{+} \) is, the greater is the mean transverse momentum. One
defines in addition the following variables,
\begin{equation}
B_{\mathrm{wide}}=\max (B_{+},B_{-}),\, \, \, B_{\mathrm{narrow}}=\min (B_{+},B_{-}),
\end{equation}
\begin{equation}
B_{\mathrm{total}}=B_{+}+B_{-},\, \, \, \, B_{\mathrm{diff}}=|B_{+}-B_{-}|\, ,
\end{equation}
to compare jet broadening in both hemispheres. For small \( B_{\mathrm{wide}} \),
one finds a longitudinal event, \( B_{\mathrm{diff}} \) measures the asymmetry
between the two hemispheres.

\subsubsection{Heavy Jet or Hemisphere Mass}

The variable \( M^{2}_{\mathrm{h}} \) is defined as
\begin{equation}
M_{\mathrm{h}}^{2}=\max \left( \left( \sum _{\vec{p}_{i}\dot{\cdot }\vec{n}_{\mathrm{thrust}}>0}p_{i}\right) ^{2},\left( \sum _{\vec{p}_{i}\dot{\cdot }\vec{n}_{\mathrm{thrust}}<0}p_{i}\right) ^{2}\right) .
\end{equation}
 and corresponds to the maximal invariant mass squared of the hemispheres. The
corresponding formula with ``\( \min  \)'' instead of ``\( \max  \)''
defines the variable \( M_{\mathrm{l}} \). One defines as well \( M_{\mathrm{diff}}=M_{\mathrm{h}}-M_{\mathrm{l}} \).
Usually one analyzes distributions of \( \frac{M_{\mathrm{h}}^{2}}{E_{\mathrm{vis}}^{2}} \),
\( \frac{M_{\mathrm{l}}^{2}}{E_{\mathrm{vis}}^{2}} \) or \( \frac{M_{\mathrm{diff}}^{2}}{E_{\mathrm{vis}}^{2}} \)
, which describe the squared masses normalized to the visible energy.

\subsubsection*{Some Comments}

Fig.~\ref{fig:ST14-34} shows the distributions of sphericity and thrust for
the lower energies 14, 22 et 34 GeV. Even though one expects an increasing contribution
of perturbative gluons, which is confirmed by the inclusive hadron distributions
(see next section), the events are more longitudinal at higher energies, corresponding
to the values of thrust close to 1 and to the sphericity \( S\simeq 0 \) .
This can be explained by the fact that the leading quarks dominate the event
shape. The results for higher energies (figs.~\ref{91}, \ref{133}, \ref{161})
confirm the above statements. 

In figs.~\ref{91}, \ref{133}, \ref{161}, we show also the distributions of
other event variables, like heavy jet mass, Major, etc. In general, our model
describes quite well all these event shape variables.

\section{The Charged Particle Distributions }

We will now consider the distributions of charged particles, which by definition
contain all the particles with a decay time smaller than \( 10^{-9} \)s, i.e.
the spectra contain, for example, products of decay of \( K^{0}_{\mathrm{short}} \),
while \( K^{\pm } \) are considered stable. The decay products of strange baryons
are also included in the distributions. 
\begin{figure}[htb]
{\par\centering \resizebox*{0.9\columnwidth}{!}{\includegraphics{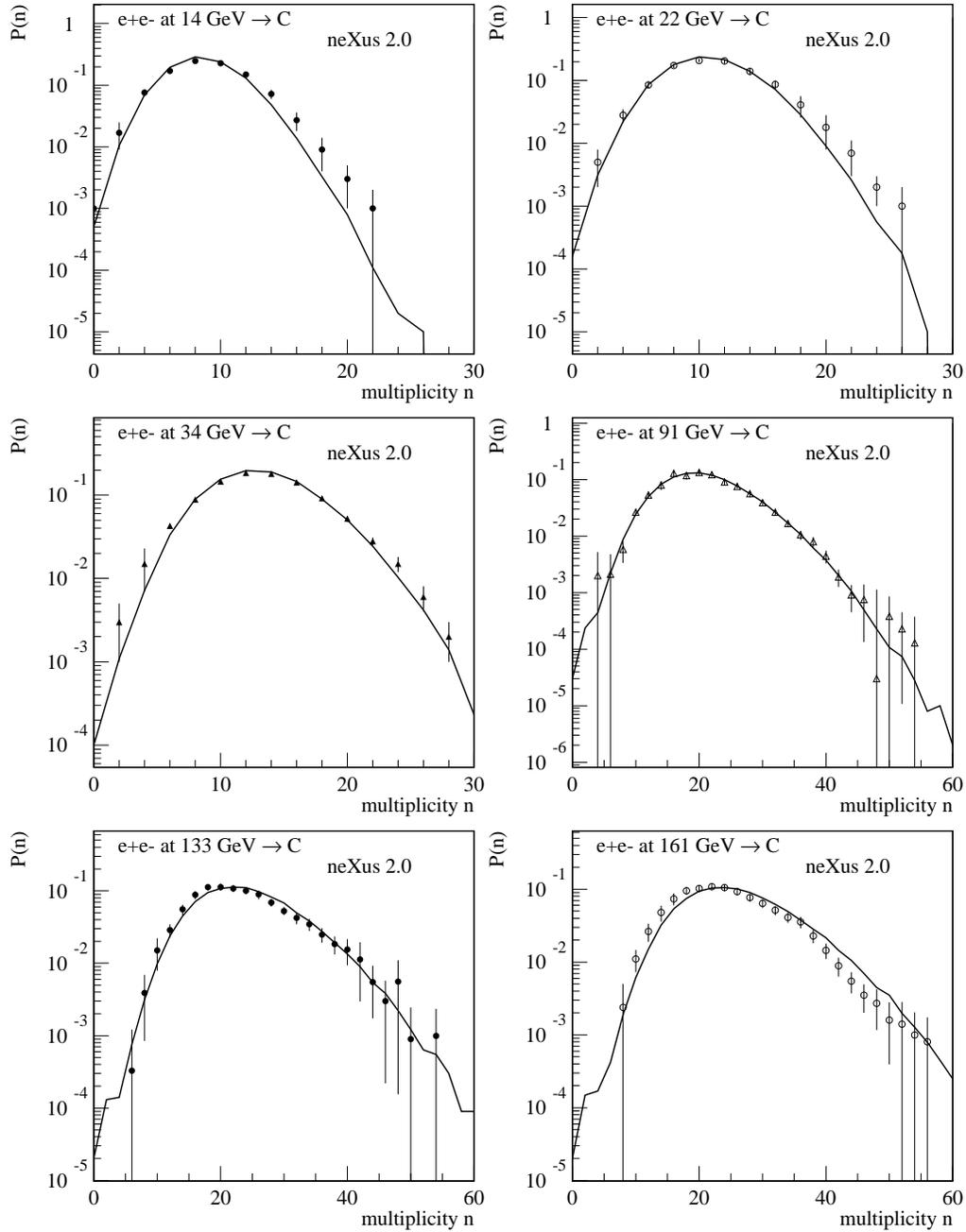}} \par}

\caption{\label{fig:mul_C}Multiplicity distributions for charged particles.}
\end{figure}
\begin{figure}[htb]
{\par\centering \resizebox*{0.9\columnwidth}{!}{\includegraphics{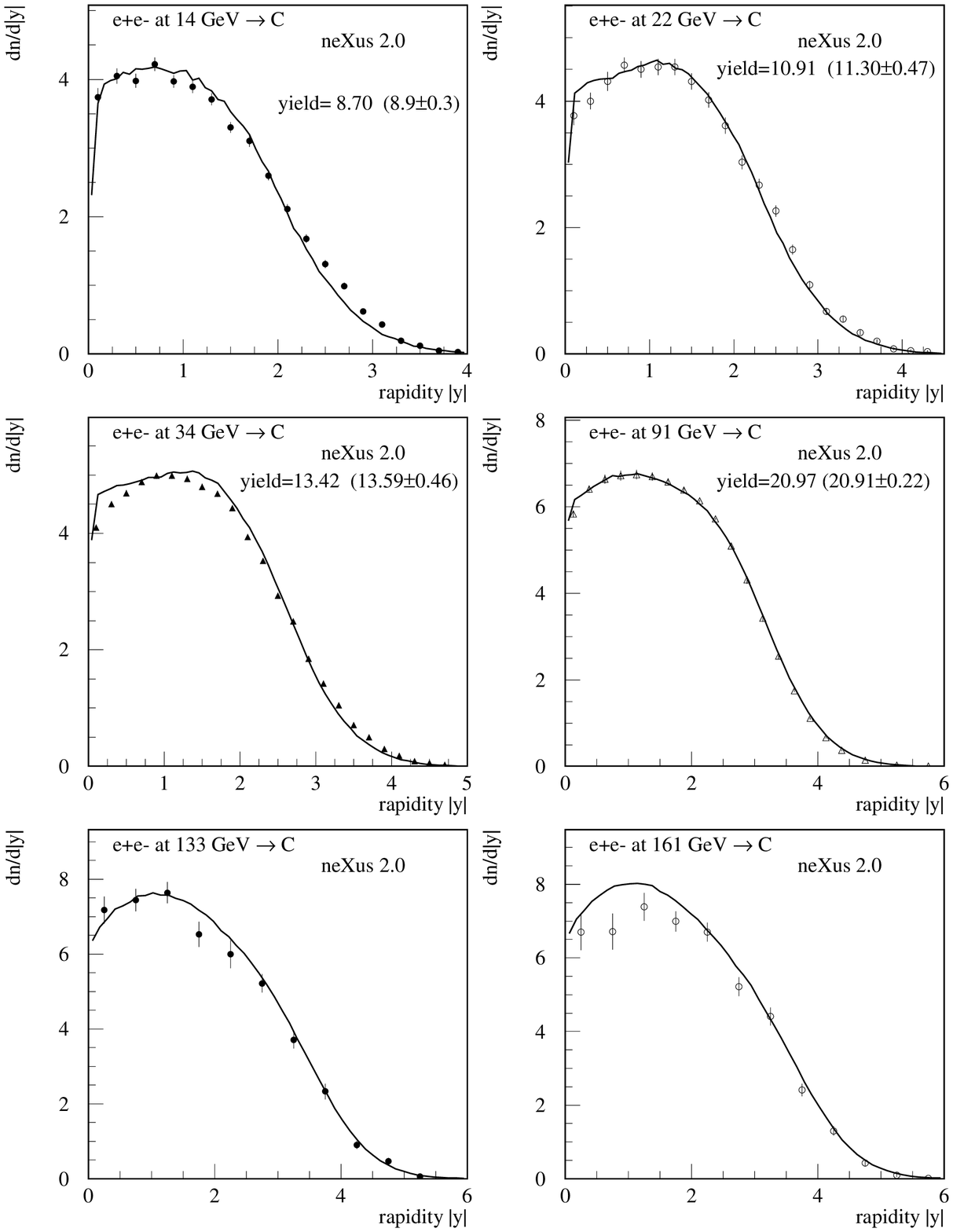}} \par}

\caption{\label{fig:rap_C}Rapidity distributions for charged particles.}
\end{figure}

In fig.~\ref{fig:mul_C}, we plot multiplicity distributions of charged particles
for three different energies, where one observes an obvious increase of the
multiplicity with energy. In fig.~\ref{fig:rap_C}, we show the distributions
of the absolute value of the rapidity for the energies 14 GeV, 22 GeV, 34 GeV,
91.2 GeV, 133 GeV and 161 GeV. The rapidity is defined as \( y=0.5\ln ((E+p_{z})/(E-p_{z})) \),
where the variable \( p_{z} \) may be defined along the thrust axis or along
the sphericity axis. For both, multiplicity and rapidity distributions, the
theoretical curves agree well with the data. 

The multiplicity increases faster than \( <n_{\mathrm{ch}}>=a+b\ln s \) as
a function of \( s \) \cite{wu84}, which is due to the fact that the maximal
height of the rapidity distribution increases with energy, as seen in fig.~\ref{fig:rap_C}.
This comes from radiated gluons, leading to kinky strings, since a flat string
without gluons shows an increasing width but a constant height, as one can see
in 
\begin{figure}[htb]
{\par\centering \resizebox*{0.9\columnwidth}{!}{\includegraphics{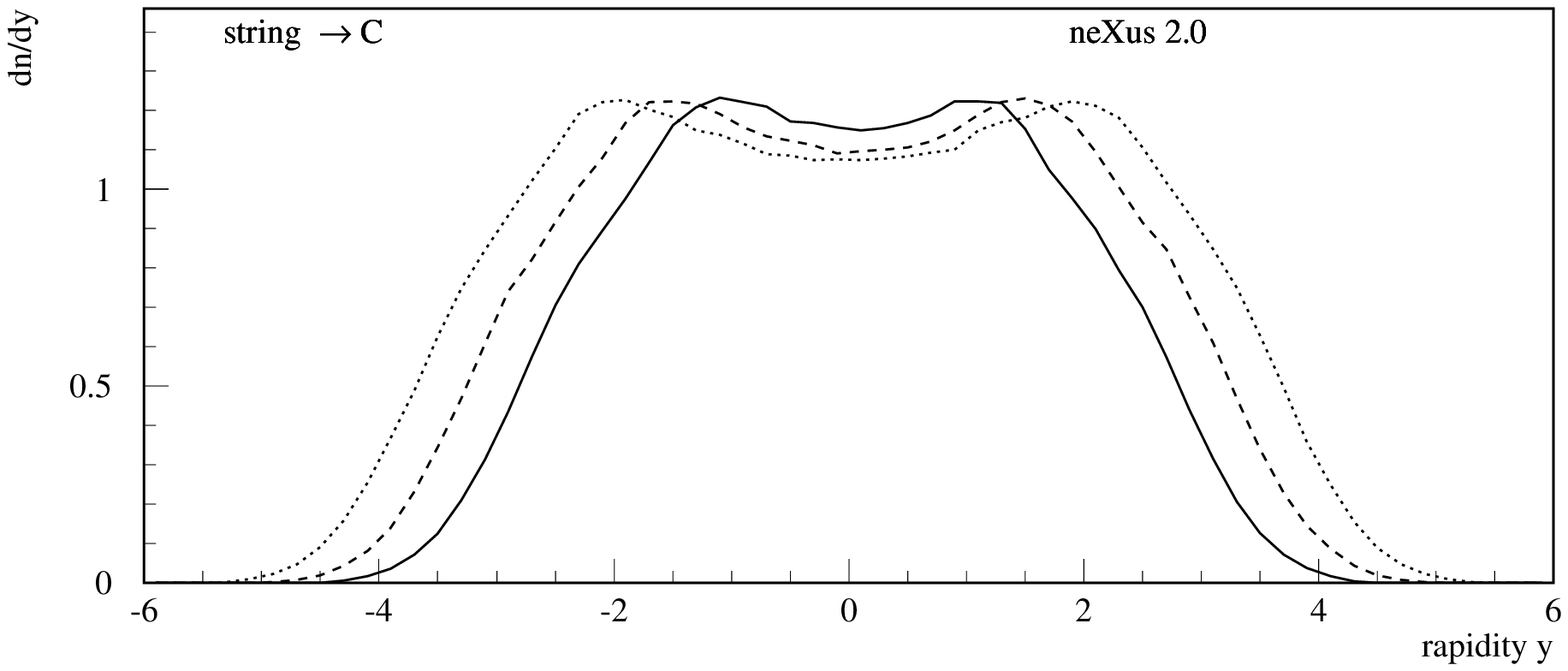}} \par}

\caption{\label{fig:corderap}Rapidity distribution of charged particles for flat strings
at 14 GeV (full), 22 GeV (dashed ) and 34 GeV (dotted). The height of the distribution
does not change, but its width does.}
\end{figure}
fig.~\ref{fig:corderap}. Here, the rapidity distributions of charged particles
are plotted for the fragmentation of a flat \( d-\bar{d} \) string for the
energies 14, 22 and 34 GeV. One observes that the width of the distributions
increases whereas its height does not change, giving rise to a proportionality
to \( \ln s \). Additional hard gluons with non-collinear momenta increase
the multiplicity in the mid-rapidity region.

\begin{figure}[htb]
{\par\centering \resizebox*{0.9\columnwidth}{!}{\includegraphics{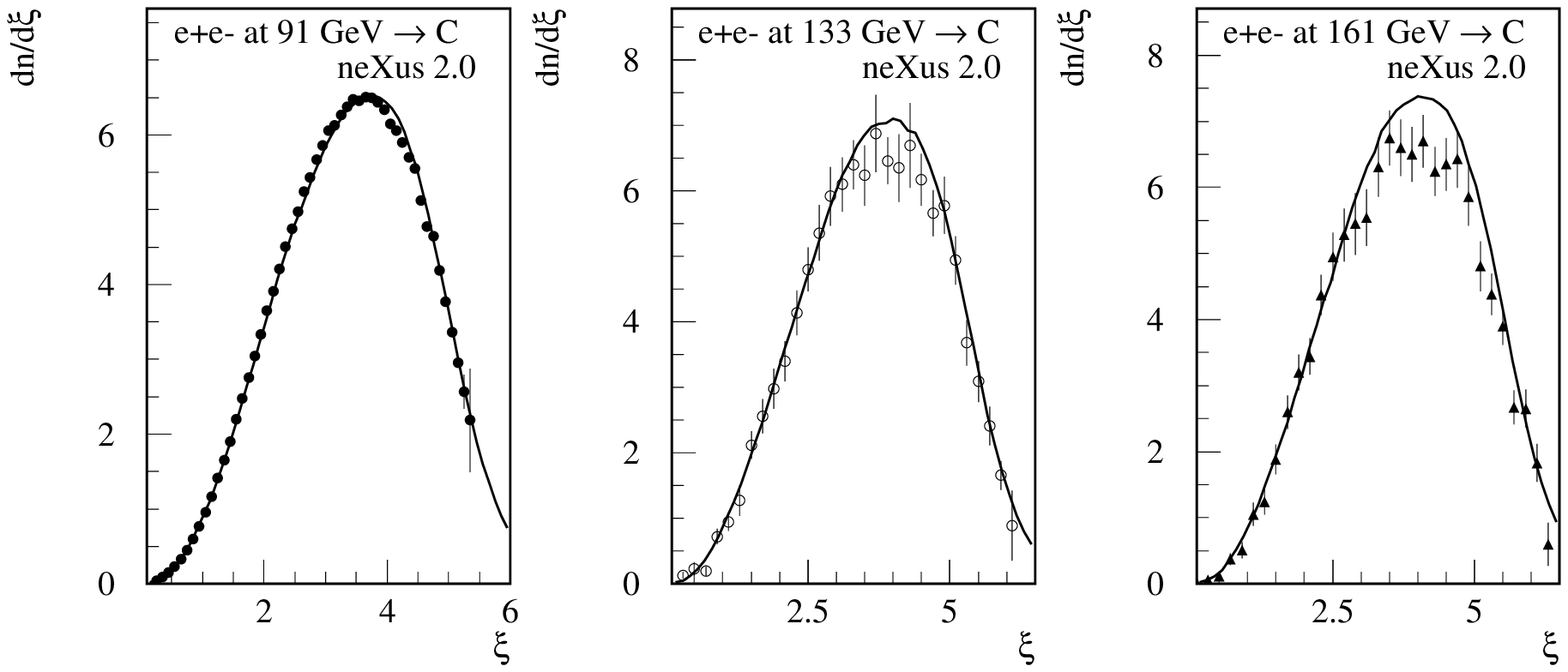}} \par}

\caption{\label{fig:xi_C}\protect\( \xi =-\ln x_{p}\protect \) distributions of charged
particles at the energies 91 GeV, 133 GeV and 161 GeV. }
\end{figure}

Rather than the rapidity, one may consider the scaled momentum \( x_{\mathrm{p}}=2|\vec{p}_{i}|/E \)
or the scaled energy \( x_{\mathrm{E}}=2E_{i}/E \), as well as the ``rapidity-like''
variable \( \xi =-\ln x_{\mathrm{p}} \). Concerning the \( \xi  \) distributions
shown in fig.~\ref{fig:xi_C}, 
\begin{figure}[htb]
{\par\centering \resizebox*{0.9\columnwidth}{!}{\includegraphics{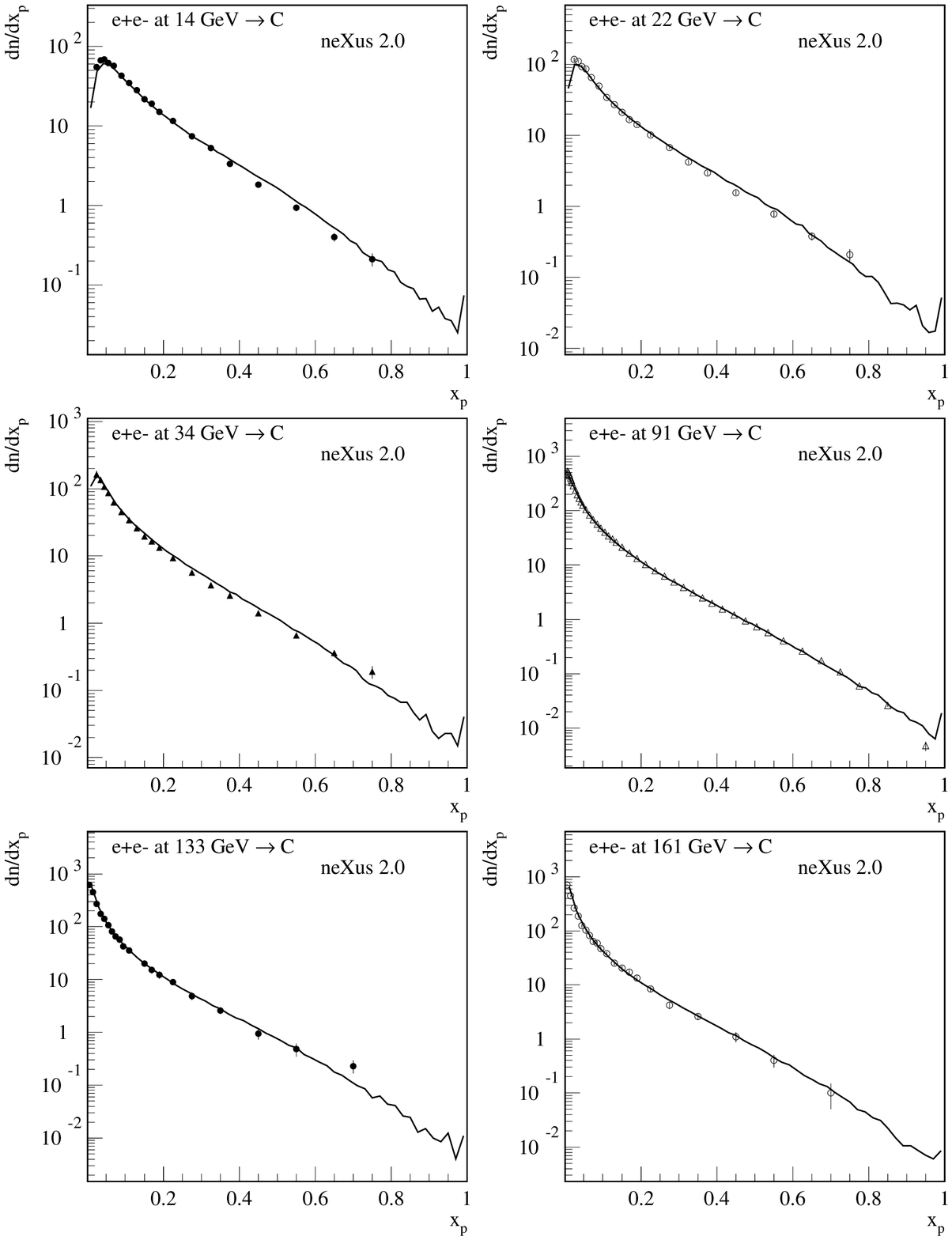}} \par}

\caption{\label{fig:xp_C}\protect\( x_{p}\protect \) distributions of charged particles.}
\end{figure}
 one sees that the value \( \xi _{\mathrm{max}} \) corresponding to the maximum
of the curves increases with energy. The \( x_{\mathrm{p}} \) distributions
(see fig.~\ref{fig:xp_C}) show the development of a more and more pronounced
peak at \( x_{\mathrm{p}} \) close to zero, with increasing energy. 

Having discussed in detail the variables describing the longitudinal phase space,
we now turn to transverse momentum,
\begin{figure}[htb]
{\par\centering \resizebox*{0.8\columnwidth}{!}{\includegraphics{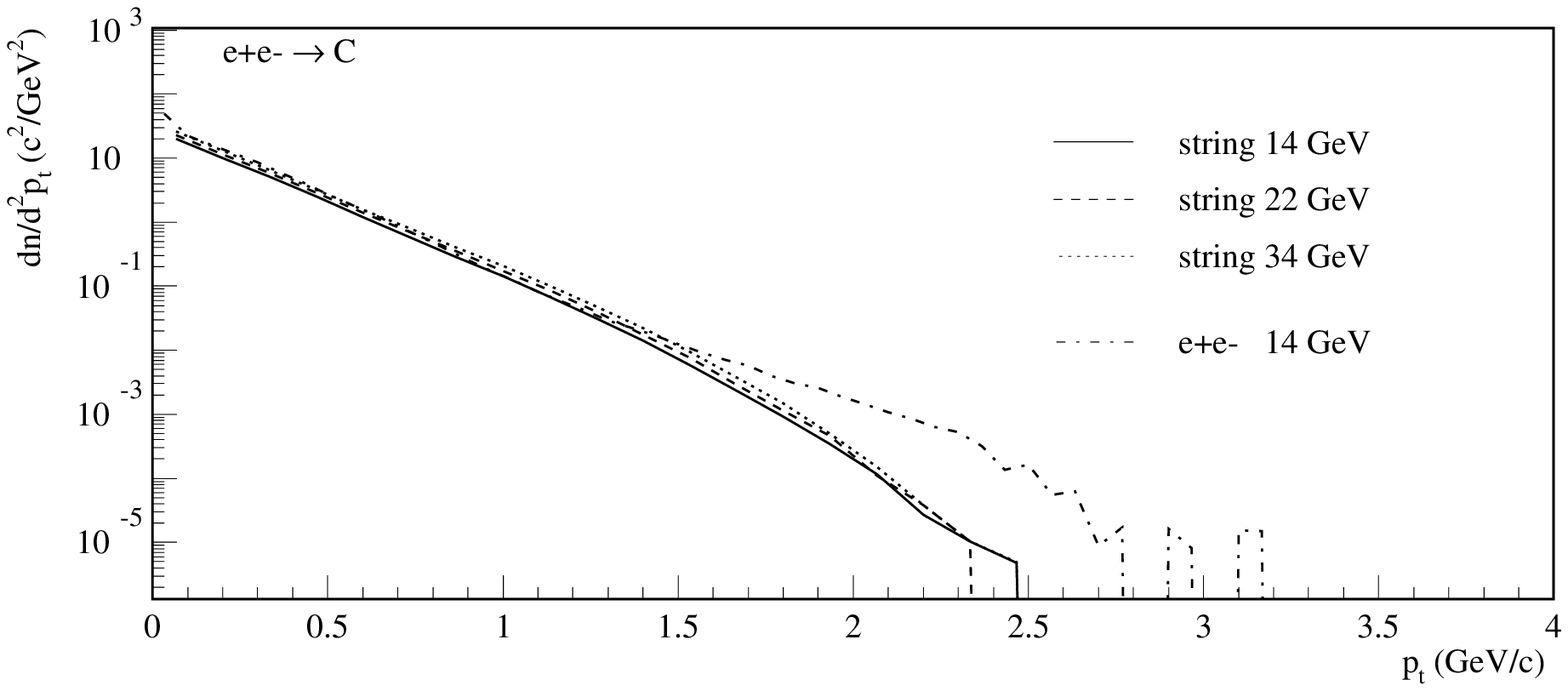}} \par}

\caption{\label{fig:ptcorde}Transverse momentum spectra of charged particles for a
flat string at energies 14, 22 and 34 GeV together with a simulation for \protect\( e^{+}e^{-}\protect \)
at 14 GeV.}
\end{figure}
\begin{figure}[htb]
{\par\centering \resizebox*{0.62\columnwidth}{!}{\includegraphics{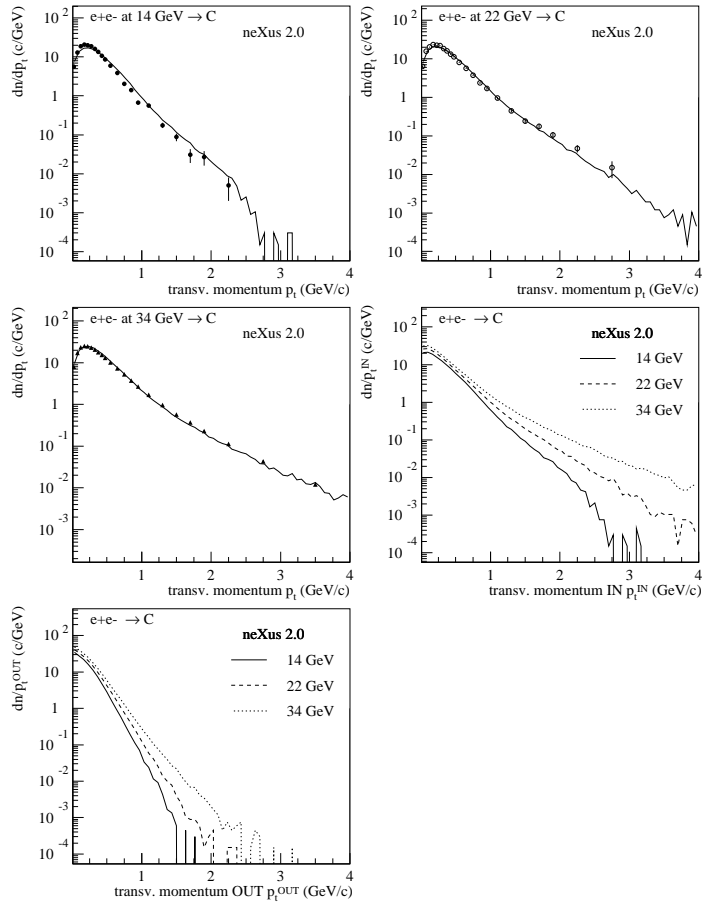}} \par}

\caption{\label{fig:pt14-34}Transverse momentum spectra of charged particles at 14,
22 and 34 GeV. The spectra of \protect\( p_{\perp }^{\mathrm{out}}\protect \)
are steeper than the ones of \protect\( p_{\perp }^{\mathrm{in}}\protect \). }
\end{figure}
\begin{figure}[htb]
{\par\centering \resizebox*{0.9\columnwidth}{!}{\includegraphics{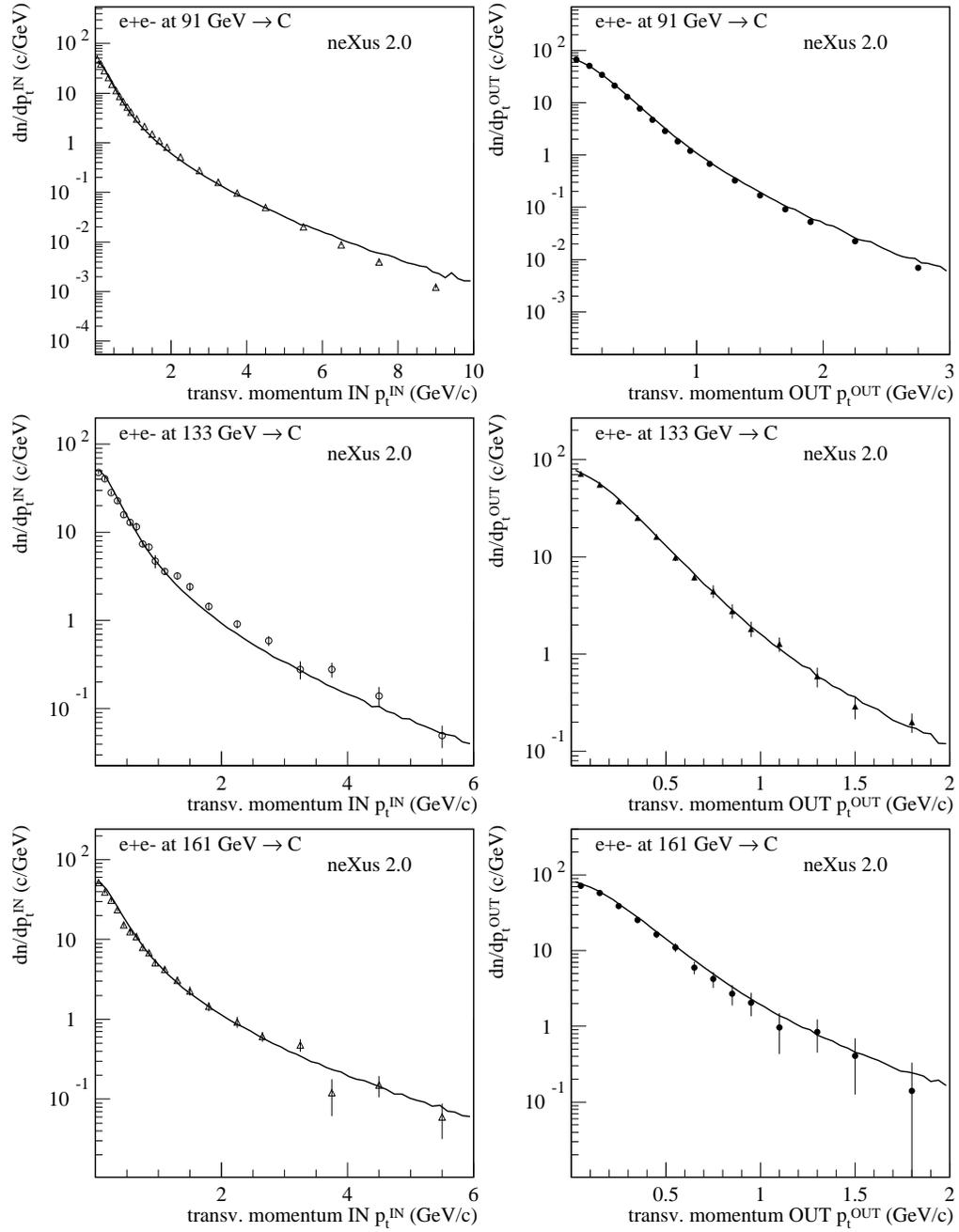}} \par}

\caption{\label{fig:pt91-161}Transverse momentum spectra of charged particles at 91,
133 and 161 GeV.}
\end{figure}
which can be defined according the the sphericity axis or to the thrust axis.
One writes 
\begin{equation}
p^{\mathrm{in}}_{\perp }=\left\{ \begin{array}{ccc}
|\vec{v}_{2}\cdot \vec{p}| & \textrm{for sphericity} & \\
|\vec{n}_{\mathrm{major}}\cdot \vec{p}| & \textrm{for thrust} & 
\end{array}\right. ,
\end{equation}
and
\begin{equation}
p^{\mathrm{out}}_{\perp }=\left\{ \begin{array}{ccc}
|\vec{v}_{3}\cdot \vec{p}| & \textrm{for sphericity} & \\
|\vec{n}_{\mathrm{minor}}\cdot \vec{p}| & \textrm{for thrust} & 
\end{array}\right. ,
\end{equation}
 There are mainly two ``sources'' of transverse momentum. The first one is
the transverse momentum created at each string break. The second one is the
transverse momentum from hard gluon radiation, which can be much larger than
the first one. So we find large values of \( p_{\perp } \) in the event plane,
and smaller ones out of the event plane. Here the event plane is essentially
defined by the direction of the hardest gluon emitted.

Let us have a look at transverse momenta of charged particles coming from a
string decay for different energies (see fig.~\ref{fig:ptcorde}). As expected,
the curves show the same behavior. In the same figure, we show the results for
an \( e^{+}e^{-} \) annihilation at 14 GeV. Already at 14 GeV the influence
of parton radiation is important. Fig.~\ref{fig:pt14-34} shows the results
for 14-34 GeV . Our results agree well with data from the TASSO collaboration
\cite{alt84}. One can see how transverse momenta increase with the energy as
an indication of more hard gluon radiation: the \( p_{\perp }^{\mathrm{out}} \)
distributions change little, whereas the \( p_{\perp }^{\mathrm{in}} \) distributions
get much harder at high energies as compared to lower ones.

\clearpage

\section{Identified Particles}

\begin{figure*}[htb]
{\par\centering \resizebox*{0.9\columnwidth}{!}{\includegraphics{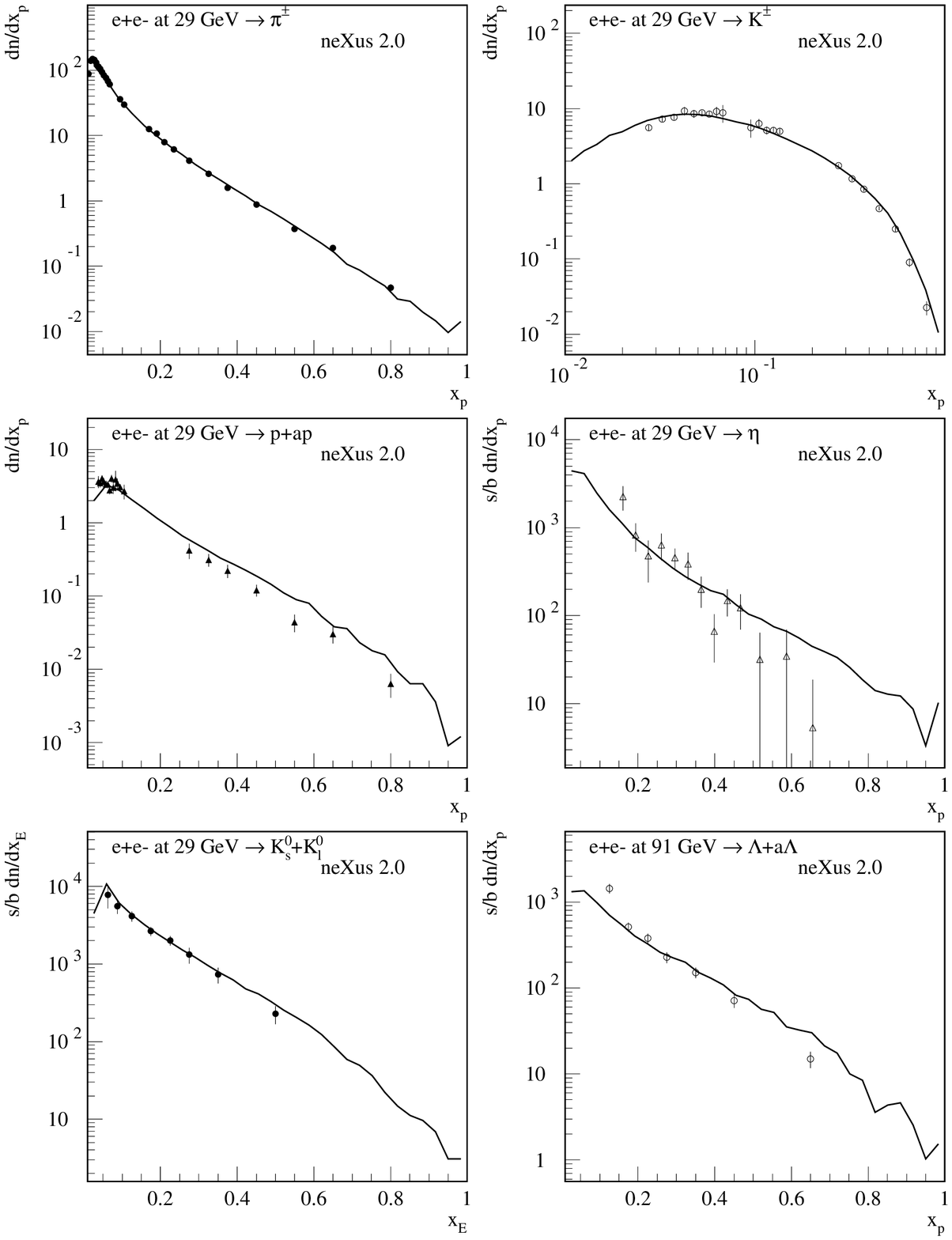}} \par}

\caption{\label{fig:ee29a}Longitudinal momentum fraction distributions for different
identified hadrons at \protect\( \sqrt{s}=29\protect \) GeV.}
\end{figure*}
\begin{figure*}[htb]
{\par\centering \resizebox*{0.9\columnwidth}{!}{\includegraphics{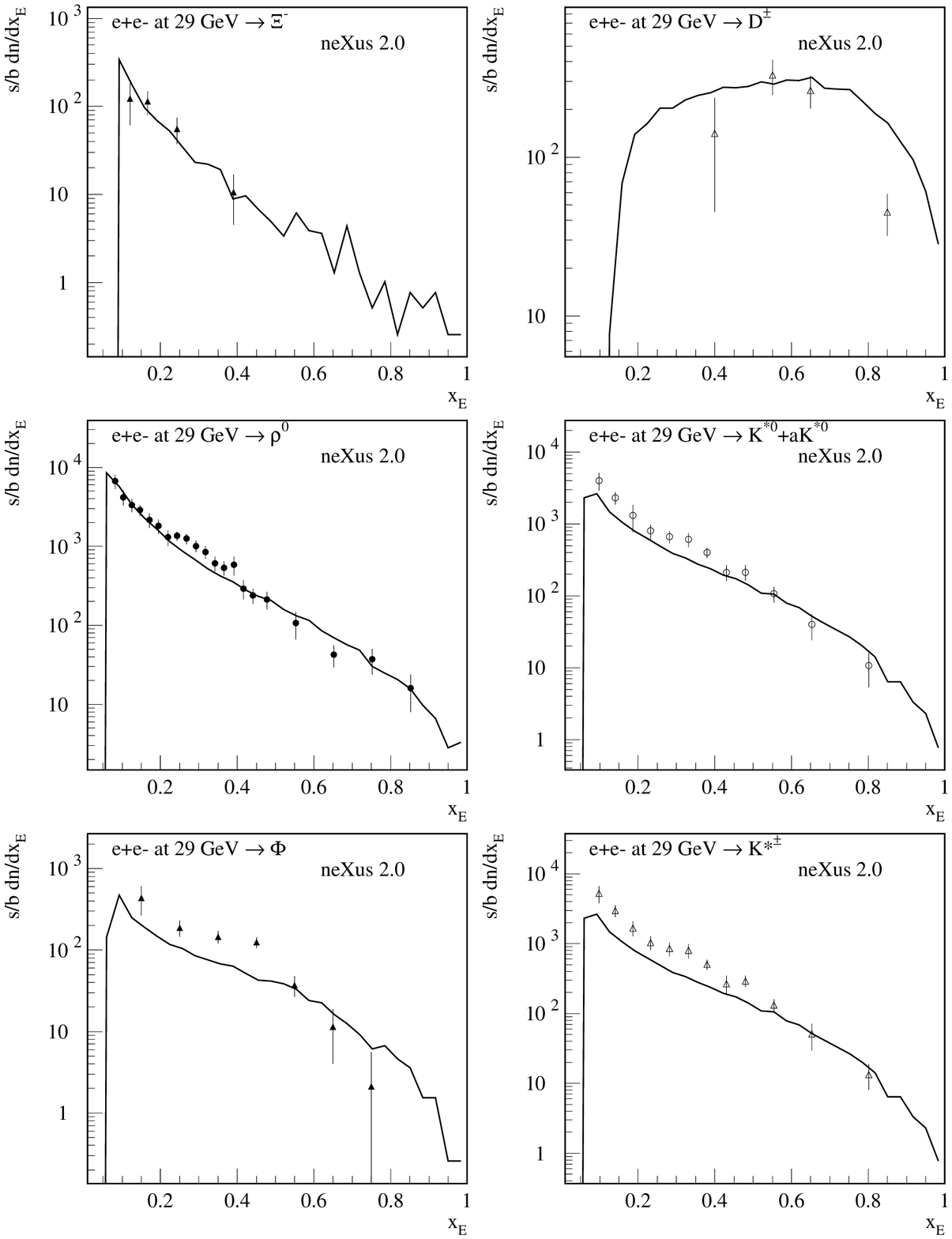}} \par}

\caption{\label{fig:ee29b}Longitudinal momentum fraction distributions for different
identified hadrons at \protect\( \sqrt{s}=29\protect \) GeV.}
\end{figure*}
\begin{figure}[htb]
{\par\centering \resizebox*{0.9\columnwidth}{!}{\includegraphics{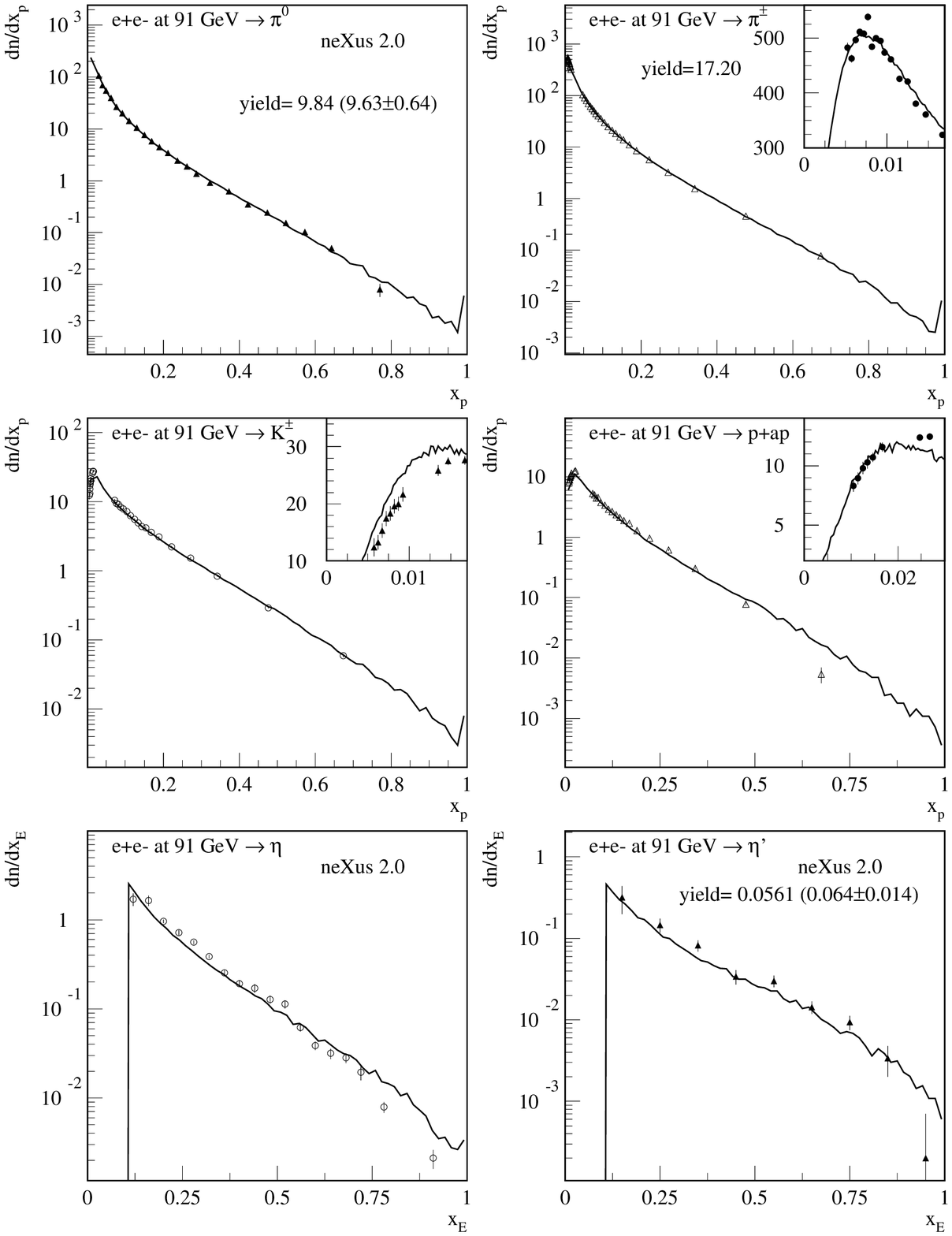}} \par}

\caption{\label{91xa}Longitudinal momentum fraction distributions for different identified
hadrons at \protect\( \sqrt{s}=91\protect \) GeV.}
\end{figure}
\begin{figure}[htb]
{\par\centering \resizebox*{0.9\columnwidth}{!}{\includegraphics{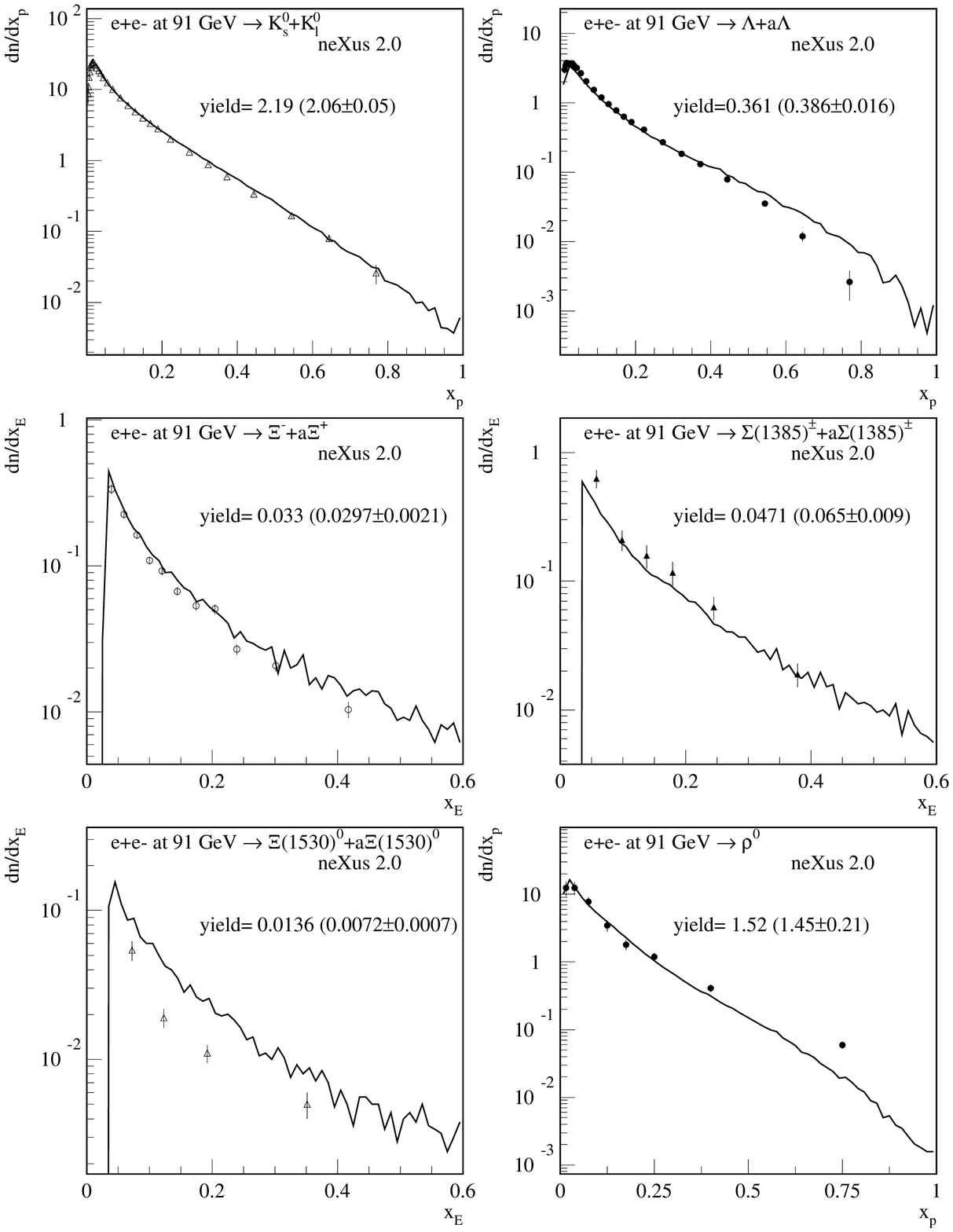}} \par}

\caption{\label{91xb}Longitudinal momentum fraction distributions for different identified
hadrons at \protect\( \sqrt{s}=91\protect \) GeV.}
\end{figure}
\begin{figure}[htb]
{\par\centering \resizebox*{0.9\columnwidth}{!}{\includegraphics{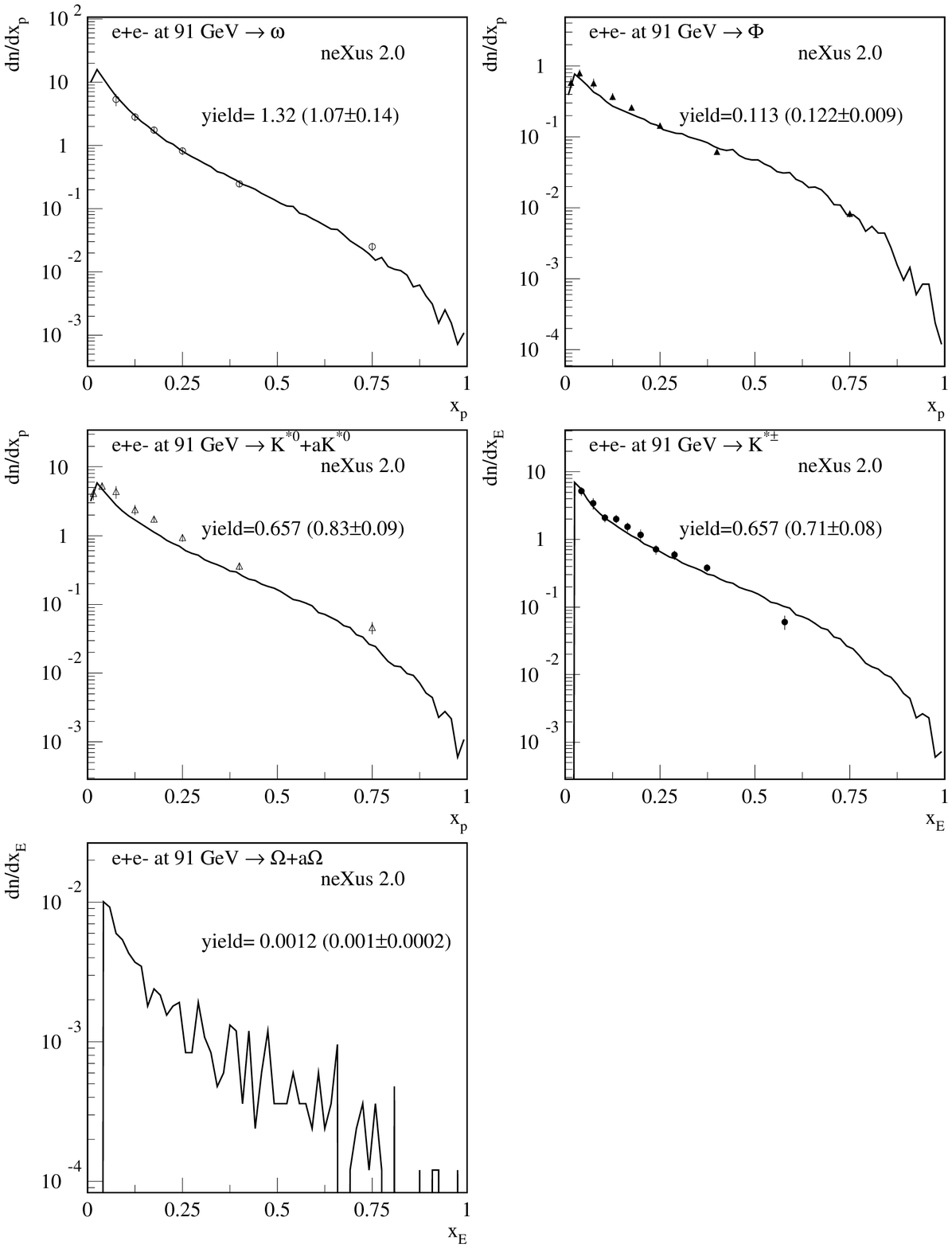}} \par}

\caption{\label{91xc}Longitudinal momentum fraction distributions for different identified
hadrons at \protect\( \sqrt{s}=91\protect \) GeV.}
\end{figure}
In this section we consider inclusive spectra of identified hadrons. This provides
a crucial test of the fragmentation model and allows to fix the two hadronization
parameters \( p_{\mathrm{ud}} \) et \( p_{\mathrm{diquark}} \). The first
one gives the probability to find a pair \( u-\bar{u} \) or \( d-\bar{d} \),
fixing so the strangeness probability to be \( (1-2p_{\mathrm{ud}}) \). The
parameter \( p_{\mathrm{diquark}} \) determines the multiplicity of baryons. 

Let us look at spectra at 29 GeV obtained at SLAC (\cite{aih84}, \cite{kle87})
(figs.~\ref{fig:ee29a}, \ref{fig:ee29b}) and at 91 GeV at LEP (figs.~\ref{91xa},
\ref{91xb}, \ref{91xc}). Since the total multiplicity is dominated by small
\( x \)-values, we show this regions separately for some figures. The results
are in general quite good, however, \( K^{*} \)'s are underestimated. 

For charmed particles, there is no production from the string decay due to the
large mass of the \( c-\bar{c} \) pair. The corresponding probability \( p_{\mathrm{charm}} \)
is taken to be zero. Charmed quarks come therefore directly from the decay of
the virtual photon as well as from perturbative parton cascade. This explains
as well the drop of \( D \) spectrum at small \( x_{p} \). 

\clearpage

\section{Jet Rates}

Jet multiplicities play an important role in \( e^{+}e^{-} \) physics since
their measurements proved the validity of perturbative QCD. 
\begin{figure}[htb]
{\par\centering \resizebox*{9cm}{!}{\rotatebox{-0.5}{\includegraphics{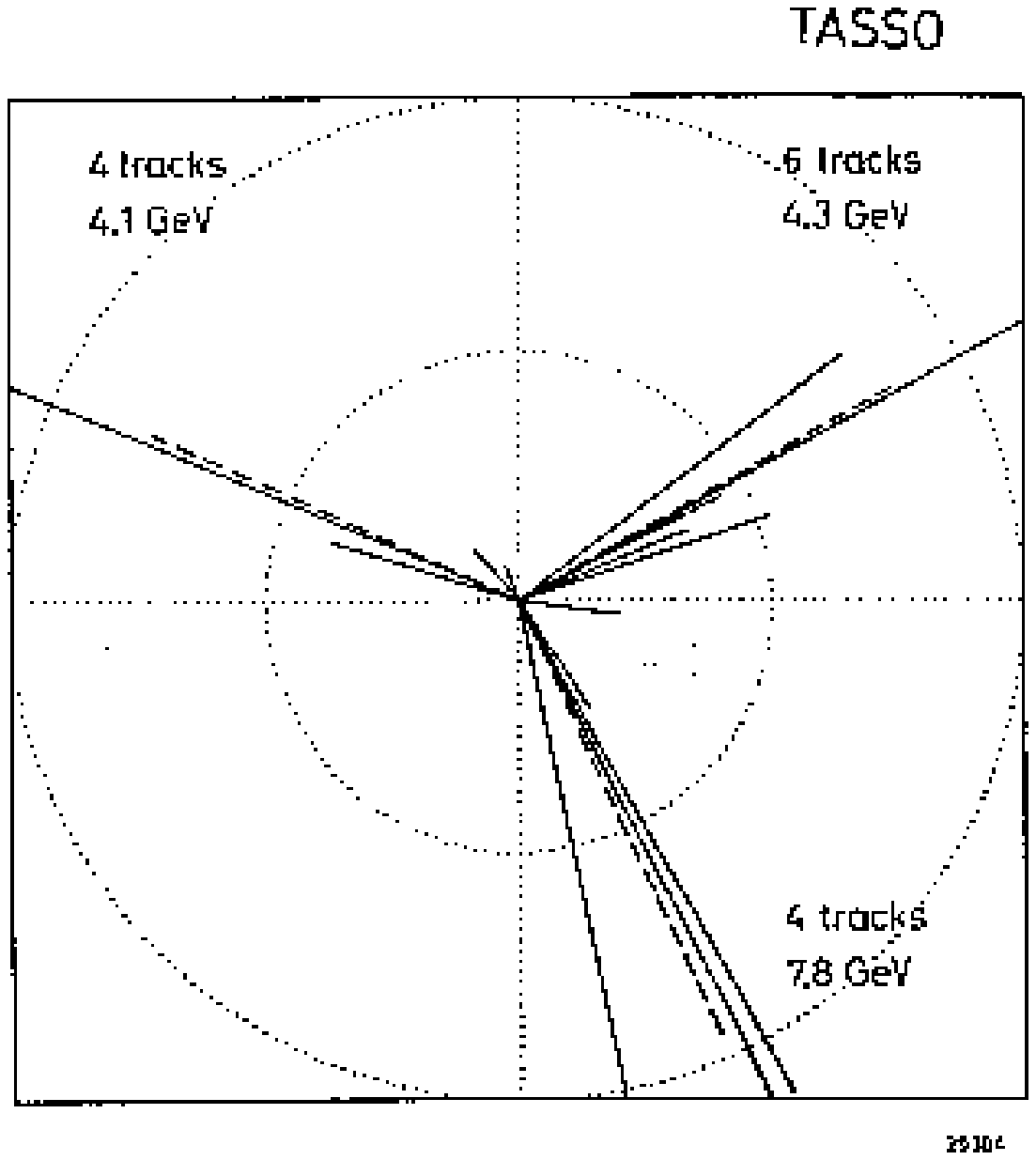}}} \par}

\caption{\label{fig:3jets}The first 3-jet event \cite{wu84}.}
\end{figure}
\begin{figure}[htb]
{\par\centering \resizebox*{0.75\columnwidth}{!}{\includegraphics{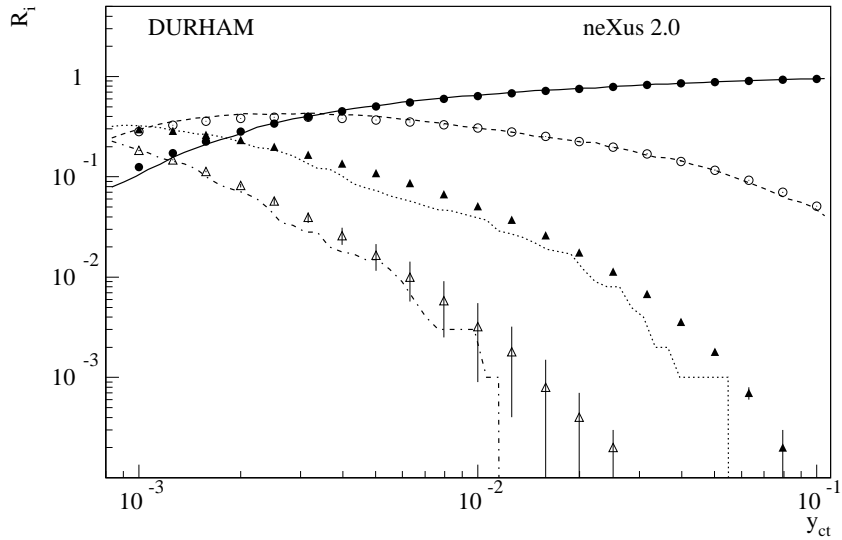}} \par}

\caption{\label{fig:jet91}Jet rates at 91.2 GeV calculated with the DURHAM algorithm
as a function of \protect\( y_{\mathrm{cut}}\protect \). The rates for 2, 3,
4 and 5 jets are shown.}
\end{figure}
\begin{figure}[htb]
{\par\centering \resizebox*{0.9\columnwidth}{!}{\includegraphics{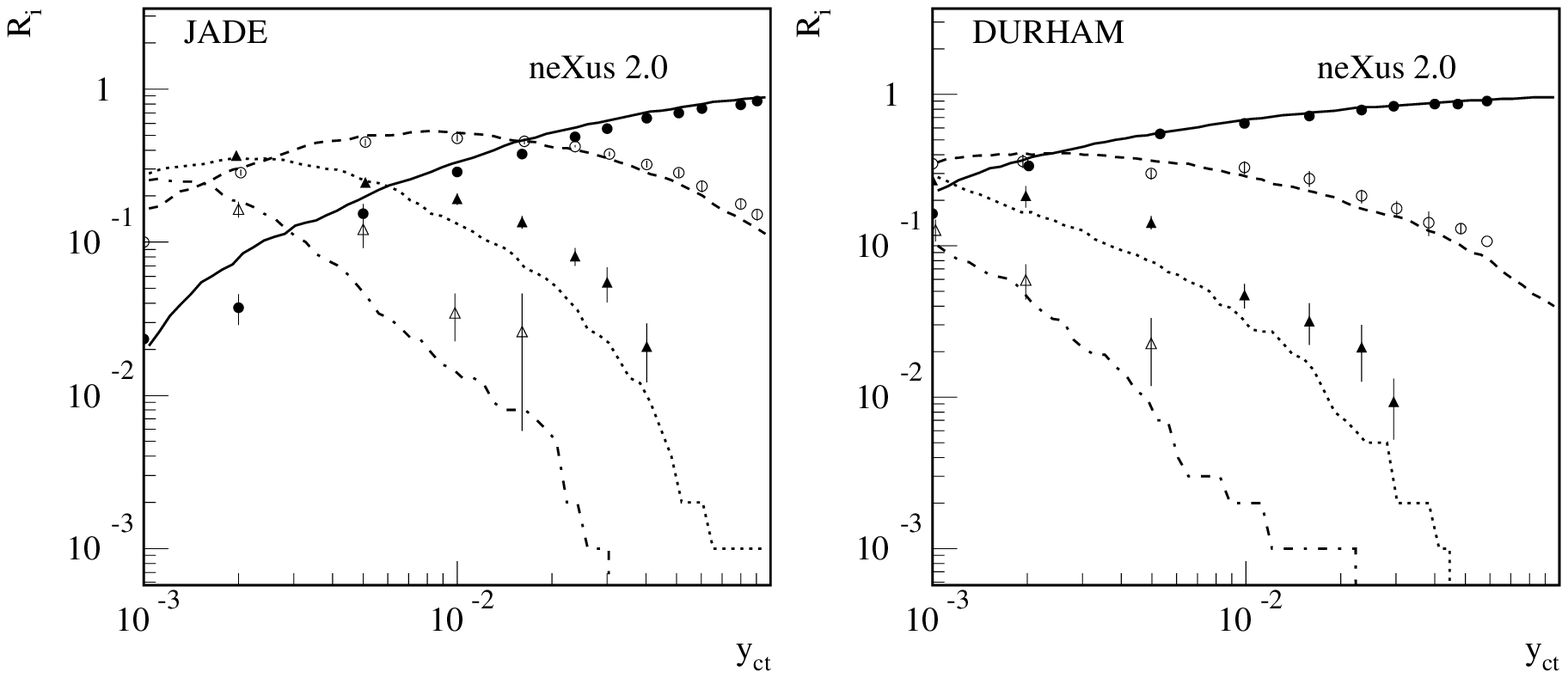}} \par}

{\par\centering \resizebox*{0.9\columnwidth}{!}{\includegraphics{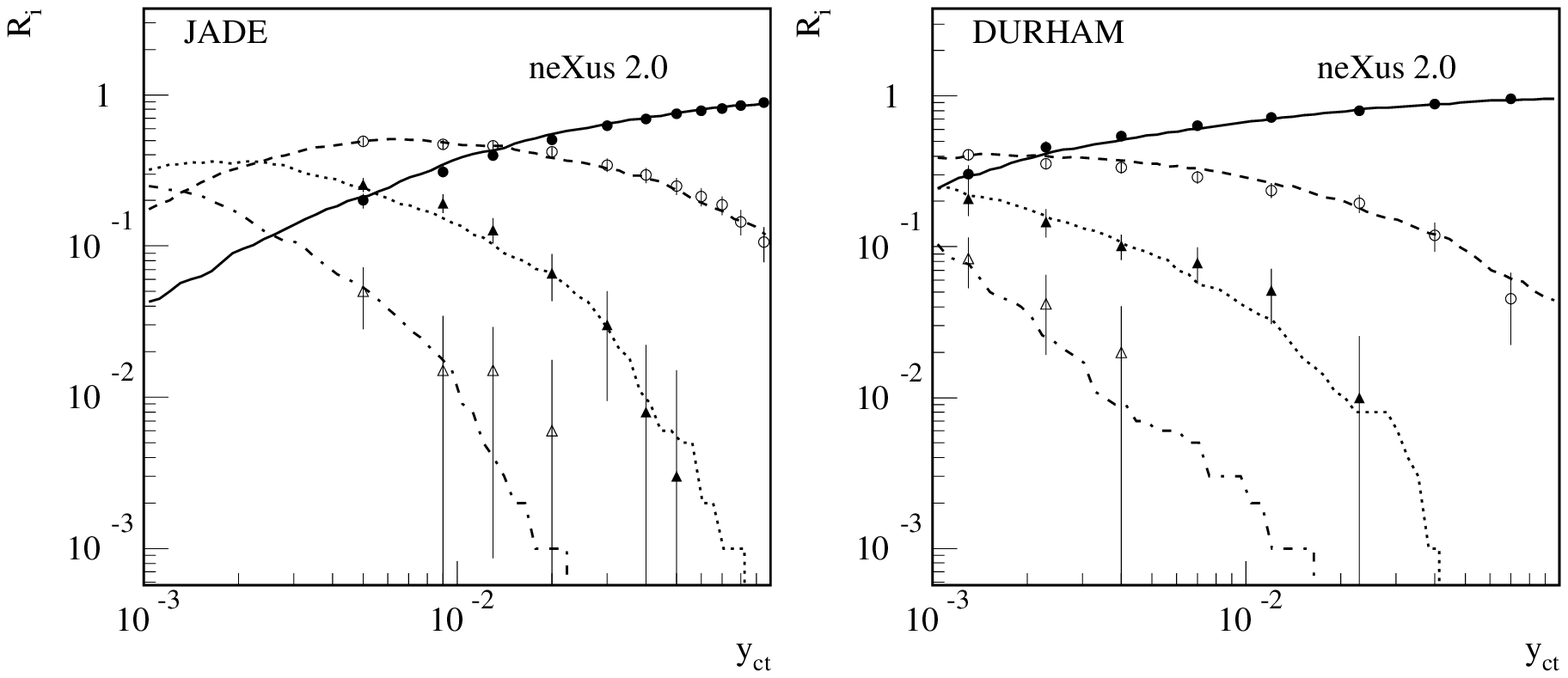}} \par}

\caption{\label{fig:jet133-161}Jet rates at 133 GeV GeV (top) et 161 GeV (bottom) for
the two algorithms DURHAM and JADE. }
\end{figure}
The first 3-jet event was found in 1979. Fig.~\ref{fig:3jets} shows this historical
event. 

There are several methods to determine the number of jets in an event, but they
are all based on some distance \( y_{ij} \) between two particles \( i \)
and \( j \) in momentum space, something like an invariant mass. For the JADE
algorithm one defines \cite{bar86}
\begin{equation}
y_{ij}=\frac{2E_{i}E_{j}(1-\cos \theta _{ij})}{E^{2}_{vis}}\, ,
\end{equation}
 \( \theta _{ij} \) being the angle between the two particles, and for the
algorithm DURHAM one has \cite{cat91}
\begin{equation}
y_{ij}=\frac{2\min (E^{2}_{i},E^{2}_{j})(1-\cos \theta _{ij})}{E^{2}_{vis}}\, \, .
\end{equation}
 \( E_{vis} \) is the total visible energy of all the particles which contribute
to the jet finding. The algorithm works as follows: one determines the pair
with the lowest distance \( y \) and replaces it with one pseudo-particle having
the sum of the momenta of the two particles \( i \) and \( j \) : \( p^{\mu }=p_{i}^{\mu }+p_{j}^{\mu } \).
One repeats this until all pairs of pseudo-particles have a distance greater
than \( y_{\mathrm{cut}} \). The number of jets is then the total number of
pseudo-particles. Of course, this depends on the choice of \( y_{\mathrm{cut}} \).
Therefore one displays often the jet multiplicity distribution as a function
of \( y_{\mathrm{cut}} \) . 

Let us compare the jet rates for different energies. Fig.~\ref{fig:jet91} shows
the jet rates for 91.2 GeV, fig.~\ref{fig:jet133-161} for 133 and 161 GeV.
The greater is \( y_{\mathrm{cut}} \) the smaller is the number of jets. 

\cleardoublepage

\chapter{Testing the Semi-hard Pomeron: Photon-Proton Scattering}

It is well known that both photon-proton (\( \gamma ^{*}p \)) scattering and
hard processes in proton-proton (\( pp \)) collisions can be treated on the
basis of perturbative QCD, using the same evolution equations. In both cases,
the perturbative partons are finally coupled softly to the proton(s). This provides
a very useful consistency check: any model for (semi)hard proton-proton collisions
should be applied to photon-proton scattering, where a wealth of data exists,
mainly from deep inelastic electron-proton scattering (DIS). In particular the
soft coupling to the protons is not calculable from first principles, so photon-proton
scattering provides a nice opportunity to test the scheme.

Let us discuss the relation between \( \gamma ^{*}p \) and \( pp \) scattering. 
\begin{figure}[htb]
{\par\centering \resizebox*{!}{0.14\textheight}{\includegraphics{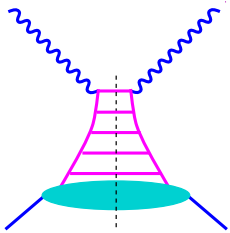}} \par}

\caption{The  cut diagram representing photon-proton (\protect\( \gamma ^{*}p\protect \))
scattering. \label{fig:ddh}}
\end{figure}
In figure \ref{fig:ddh}, we show the cut diagram (integrated squared amplitude),
representing a contribution to photon-proton  scattering: a photon couples to
a quark of the  proton, where this quark represents the last one in a ``cascade''
of partons  emitted from the nucleon. In the leading logarithmic approximation
(LLA) the virtualities of the partons are  ordered such that the largest one
is close to the photon \cite{rey81,alt82}.
\begin{figure}[htb]
{\par\centering \resizebox*{!}{0.18\textheight}{\includegraphics{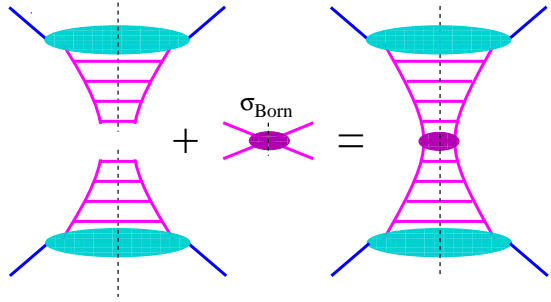}} \par}

\caption{The universality hypothesis implies that the upper and the lower part of the
Pomeron diagram are identical to the photon-proton diagram. \label{cpomdis}}
\end{figure}
Comparing the cut diagram for \( \gamma ^{*}p \) (figure \ref{fig:ddh}) with
the cut diagram representing a semi-hard elementary proton-proton scattering
(figure \ref{cpomdis}), we see immediately that the latter one is essentially
made of two \( \gamma ^{*}p \) diagrams, glued together by a Born process.
So, understanding \( \gamma ^{*}p \) implies  understanding an elementary nucleon-nucleon
interaction as well. Actually, probably everybody agrees with this statement,
 which is nothing but the factorization hypothesis, proved in QCD \cite{mue81},
and the standard  procedure to calculate  inclusive cross sections in proton-proton
scattering  amounts to using input from the DIS structure functions. But one
can profit  much more from studying \( \gamma ^{*}p \), for example, concerning
the production of hadrons. Not being calculable from first principles, the hadronization
of parton configurations is a delicate issue in any model for proton-proton
(or  nucleus-nucleus) scattering. So studying \( \gamma ^{*}p \) provides an
excellent possibility to ``gauge'' the hadronization procedure, such that
there is no freedom left on the level  of nucleon-nucleon (or nucleus-nucleus)
scattering. 

The simple picture, depicted at the fig.\ \ref{fig:ddh}, is correct for large
virtualities, but it fails when the photon virtuality becomes small. In that
case a virtual photon behaves to a large extent as a hadron and is characterized
by some parton content instead of interacting with a proton just as a point-like
object. 
\begin{figure}[htb]
{\par\centering \resizebox*{!}{0.18\textheight}{\includegraphics{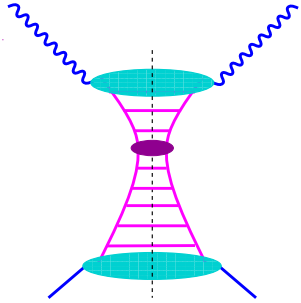}} \par}

\caption{The  cut diagram representing resolved photon-proton (\protect\( \gamma ^{*}p\protect \))
scattering. \label{fig:ddr}}
\end{figure}
Then the contribution of so-called resolved photon interactions - see fig.\
\ref{fig:ddr} - is important and has to be taken into account properly for
the description of hadron production in DIS. Only then one may deduce the parton
momentum distributions of the proton from the measured virtual photon-proton
cross section \( \sigma ^{\gamma ^{*}p} \).

\section{Kinematics }

\begin{figure}[htb]
{\par\centering \resizebox*{!}{0.14\textheight}{\includegraphics{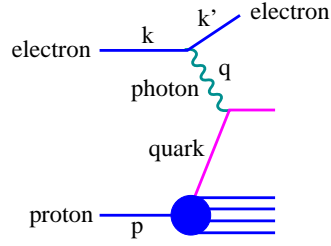}} \par}

\caption{Kinematics of electron-proton scattering.\label{kine}}
\end{figure}

In the following, we consider photon-proton collisions in the context of electron-proton
scattering. We first recall the basic kinematic variables, see fig.\ \ref{kine}.
We use standard conventions: \( k \), \( k' \), and \( p \) are the four-momenta
of  incoming and outgoing lepton and the target nucleon, \( q=k-k' \) is the
four-momentum of the photon. 
Then the photon virtuality is \( Q^{2}=-q^{2} \), and one defines the Bjorken
\( x \)-variable as 
\begin{equation}
x_{\mathrm{B}}=\frac{Q^{2}}{2(pq)}.
\end{equation}
 The \( y \)-variable is defined as 
\begin{equation}
y=\frac{(pq)}{(pk)}=\frac{2(pq)}{s}=\frac{Q^{2}}{x_{\mathrm{B}}s},
\end{equation}
which gives the energy fraction of the photon relative to the incident electron
in the proton rest frame. In the above formula we used 
\begin{equation}
s=(k+p)^{2},
\end{equation}
being the total center-of-mass squared energy and neglected the proton and electron
masses, \( p^{2}\simeq 0 \), \( k^{2}\simeq 0 \). For the center-of-mass squared
energy of photon-proton interaction, we use 
\begin{equation}
\hat{s}=(p+q)^{2},
\end{equation}
 and we finally define the variable 
\begin{equation}
\tilde{s}=\hat{s}+Q^{2}=2\, (pq)=ys,
\end{equation}
which allows to write
\begin{equation}
x_{\mathrm{B}}=\frac{Q^{2}}{\tilde{s}}.
\end{equation}
 It is often convenient to take \( \tilde{s} \) and \( Q^{2} \) as the basic
kinematical variables instead of \( x_{\mathrm{B}} \) and \( Q^{2} \).

\section{Cross Sections}

The differential cross section for deep inelastic electron-proton scattering
in the one-photon approximation can be written as \cite{ell96}

\begin{equation}
\label{sigmadis-dif1}
\frac{d\sigma ^{ep}\! \left( s,x_{B},Q^{2}\right) }{dQ^{2}dx_{\mathrm{B}}}=\frac{\alpha }{\pi Q^{2}x_{\mathrm{B}}}\left[ L_{T}^{\gamma }(y)\, \sigma ^{\gamma ^{*}p}_{T}\! (\tilde{s},Q^{2})+L_{L}^{\gamma }(y)\, \sigma ^{\gamma ^{*}p}_{L}\! (\tilde{s},Q^{2})\right] ,
\end{equation}
where \( \sigma ^{\gamma ^{*}p}_{T(L)}(\tilde{s},Q^{2}) \) are the cross sections
for interactions of transversely (longitudinally) polarized photons of virtuality
\( Q^{2} \) with a proton, \( \alpha  \) is the fine structure constant and
factors \( L_{T(L)}^{\gamma ^{*}}(y) \) define the flux of transversely (longitudinally)
polarized photons,

\begin{equation}
L_{T}^{\gamma }(y)=\frac{1+(1-y)^{2}}{2}\; \; \: \: \: L_{L}^{\gamma }(y)=(1-y).
\end{equation}
 The cross sections \( \sigma ^{\gamma ^{*}p}_{T(L)}(\tilde{s},Q^{2}) \) are
related to the proton structure functions \( F_{2} \), \( F_{L} \), describing
the proton structure as seen by a virtual photon probe, as
\begin{eqnarray}
F_{2}(x_{\mathrm{B}},Q^{2}) & = & \frac{Q^{2}}{4\pi ^{2}\alpha }\left[ \sigma ^{\gamma ^{*}p}_{T}\! \left( \tilde{s},Q^{2}\right) +\sigma ^{\gamma ^{*}p}_{L}\! \left( \tilde{s},Q^{2}\right) \right] \label{f2-sigma} \\
F_{L}(x_{\mathrm{B}},Q^{2}) & = & \frac{Q^{2}}{4\pi ^{2}\alpha }\sigma ^{\gamma ^{*}p}_{L}\! \left( \tilde{s},Q^{2}\right) \label{fl-sigma} 
\end{eqnarray}

To the leading logarithmic accuracy one has to take into account a number of
processes, contributing to \( \sigma ^{\gamma ^{*}p}_{T(L)}(\tilde{s},Q^{2}) \).
In the case of transverse polarization of the photon, we have three contributions:
the direct coupling of the virtual photon \( \gamma ^{*} \) to a light quark
from the proton (``light''), the direct coupling to a charm quark (``charm''),
and finally we have a ``resolved'' contribution, i.e.
\begin{equation}
\label{sigma-t}
\sigma ^{\gamma ^{*}p}_{T}(\tilde{s},Q^{2})=\sigma ^{\gamma ^{*}p}_{T(\mathrm{light})}(\tilde{s},Q^{2})+\sigma ^{\gamma ^{*}p}_{T(\mathrm{charm})}(\tilde{s},Q^{2})+\sigma ^{\gamma ^{*}p}_{T(\mathrm{resolved})}(\tilde{s},Q^{2}).
\end{equation}
The latter one is becoming essential at small \( Q^{2} \) and large \( \tilde{s} \).
For our study, contributions of beauty and top quarks can be neglected. The
longitudinal photon cross section receives leading order contributions only
from the direct \( \gamma ^{*} \)-coupling to a charm quark. So we have
\begin{equation}
\label{sigma-l}
\sigma ^{\gamma ^{*}p}_{L}(\tilde{s},Q^{2})=\sigma ^{\gamma ^{*}p}_{L(\mathrm{charm})}(\tilde{s},Q^{2}).
\end{equation}
 Again, contributions of beauty and top quarks can be neglected.

Let us list the different contributions. The leading order light quark-\( \gamma ^{*} \)
coupling contribution (``light'') can be expressed via the quark momentum
distributions in the proton \( f_{q/p}\! \left( x_{B},Q^{2}\right)  \) as
\begin{equation}
\label{sigma-light}
\sigma ^{\gamma ^{*}p}_{T(\mathrm{light})}(\tilde{s},Q^{2})=\frac{4\pi ^{2}\alpha }{Q^{2}}\sum _{i\in \{u,d,s,\bar{u},\bar{d},\bar{s}\}}e_{i}^{2}\, x_{B}\, f_{i/p}\! \left( x_{B},Q^{2}\right) ,
\end{equation}
where \( e_{q}^{2} \) is the quark \( q \) electric charge squared.

The contributions of heavy quarks can be taken into account via photon-gluon
fusion (PGF) process \cite{glu94},

\begin{equation}
\label{sigma-cc}
\sigma ^{\gamma ^{*}p}_{T/L(\mathrm{charm})}(\tilde{s},Q^{2})=e_{c}^{2}\int \! dx^{-}\, dp_{\bot }^{2}\frac{d\sigma ^{\gamma ^{*}g\rightarrow c\bar{c}}_{T/L}\! \left( x^{-}\tilde{s},Q^{2},p_{\bot }^{2}\right) }{dp_{\bot }^{2}}f_{g/p}\! \left( x^{-},M^{2}_{F}\right) ,
\end{equation}
where \( f_{g/p}\! \left( x^{-},M^{2}_{F}\right)  \) is the gluon momentum
distribution in the proton at the factorization scale \( M^{2}_{F} \), and
where the photon-gluon cross section in lowest order is given as

\begin{equation}
\label{sig-born-cc}
\frac{d\sigma ^{\gamma ^{*}g\rightarrow c\bar{c}}_{T/L}\! \left( \tilde{s},Q^{2},p_{\bot }^{2}\right) }{dp_{\bot }^{2}}=\frac{1}{16\pi \tilde{s}\hat{s}\sqrt{1-4(p_{\bot }^{2}+m_{c}^{2})/\hat{s}}}\left| M_{T/L}^{\gamma ^{*}g\rightarrow c\bar{c}}(\tilde{s},Q^{2},p_{\bot }^{2})\right| ^{2}
\end{equation}
with \( \hat{s}=\tilde{s}-Q^{2} \), and with the corresponding matrix elements
squared given as
\begin{eqnarray}
\left| M_{T}^{\gamma ^{*}g\rightarrow c\bar{c}}(\tilde{s},Q^{2},p_{\bot }^{2})\right| ^{2} & = & \pi \alpha \alpha _{s}\left[ \left( \frac{t'}{u'}+\frac{u'}{t'}\right) \frac{Q^{4}+\hat{s}^{2}}{\tilde{s}^{2}}\right. \label{m-cc-t} \\
 & + & \left. \frac{2m^{2}_{c}\tilde{s}^{2}}{t'^{2}u'^{2}}\left( Q^{2}-2m^{2}_{c}\right) +\frac{4m^{2}_{c}}{t'u'}\left( \tilde{s}-2Q^{2}\right) \right] \nonumber \\
\left| M_{L}^{\gamma ^{*}g\rightarrow c\bar{c}}(\tilde{s},Q^{2},p_{\bot }^{2})\right| ^{2} & = & 8\pi \alpha \alpha _{s}\left( \frac{\hat{s}Q^{2}}{\tilde{s}^{2}}-\frac{m^{2}_{c}Q^{2}}{t'u'}\right) \label{m-cc-l} 
\end{eqnarray}
 Here, the variables \( t',u' \) are expressed via standard Mandelstam variables
for parton-parton scattering as \( t'=t-m_{c}^{2} \), \( u'=u-m_{c}^{2} \).
According to \cite{glu94}, the factorization scale \( M^{2}_{F} \) has to
be chosen irrespectively to the photon virtuality \( Q^{2} \) to assure the
perturbative stability of the result; we use \( M^{2}_{F}=p_{\bot }^{2}+m^{2}_{c} \),
which coincides at large \( \hat{s} \) with the virtuality (off-shellness)
of the intermediate \( t \)-channel \( c \)-quark \( |t'| \), with \( m_{c}=1.6 \)
GeV being the \( c \)-quark mass.

In addition, at small \( Q^{2} \) and large \( \tilde{s} \), the contribution
of resolved photon processes becomes important for the production of parton
jets of transverse momenta \( p^{2}_{\bot }>Q^{2} \). Here,
\begin{eqnarray}
\sigma ^{\gamma ^{*}p}_{T(\mathrm{resolved})}(\tilde{s},Q^{2}) & = & \int dx^{+}\, dx^{-}\, dp_{\bot }^{2}\: \sum _{i,j}\frac{d\sigma _{\mathrm{Born}}^{ij}\! \left( x^{+}x^{-}\hat{s},p_{\bot }^{2}\right) }{dp_{\bot }^{2}}\label{sigma-res} \\
 & \times  & f_{i/\gamma ^{*}}\! \left( x^{+},M^{2}_{\gamma },Q^{2}\right) \; f_{j/p}\! \left( x^{-},M^{2}_{p}\right) \; \theta \! \left( p_{\bot }^{2}-Q^{2}\right) \nonumber 
\end{eqnarray}
where \( \hat{s}=\tilde{s}-Q^{2} \) is the c.m. energy squared for \( \gamma ^{*} \)-proton
interaction, \( d\sigma _{\mathrm{Born}}^{ij}/dp_{\bot }^{2} \) is the differential
partonic cross section, \( f_{i/\gamma ^{*}} \) is the parton momentum distribution
in the photon, \( M_{p}^{2} \), \( M_{\gamma }^{2} \) are the factorization
scales for the proton and photon correspondingly. As in hadron-hadron scattering,
we use \( M_{p}^{2}=p^{2}_{\bot }/4 \), whereas for the photon we take \( M_{\gamma }^{2}=4p^{2}_{\bot } \).
This requires some more explanation. To the leading order accuracy, the factorization
scales are rather undefined as the difference between the results, obtained
for different scale choices, is due to higher order corrections. The scheme
would be scale independent only after summing up all order contributions both
in the structure functions and in the partonic cross section. High \( p_{\bot } \)
jet production in \( \gamma ^{*} \)-proton interaction is known to obtain essential
contributions from next to leading order (NLO) direct processes \cite{pot97}.
As it was shown in \cite{pot97}, the sum of the leading order resolved \( \gamma ^{*} \)-proton
cross section and the NLO direct one exhibits a remarkable independence on the
scale \( M_{\gamma }^{2} \) for the production of parton jets with \( p_{\bot }>Q \),
where \( Q \) is the photon virtuality. So our strategy is to choose \( M_{\gamma }^{2} \)
such that it allows to represent effectively the full contribution by the leading
order resolved cross section.

\section{Parton Momentum Distributions }

The cross sections mentioned in the preceding section are all expressed via
the so-called parton distribution functions \( f_{i/a} \), representing the
momentum fraction distribution of parton \( i \) inside particle \( a \) (proton
or photon). In this section, we are going to discuss these distribution functions.

We are first discussing parton distribution functions of the proton. They are
represented by the hadronic part of the photon-proton diagram, i.e.\ the diagram
without the external photon line. As mentioned before, this diagram is also
a building block of one of the the elementary diagrams of \( pp \) scattering,
and one can therefore repeat literally the argumentation of the chapter 2. In
\( pp \) scattering, we have (apart of the soft one) four contributions, since
on each side the parton ladder couples to the nucleon either via a soft Pomeron
or it connects directly to a valence quark. In addition, there is a triple Pomeron
diagram. Corresponding we have here three contributions, referred to as ``sea'',
``triple Pomeron'', and ``valence'' .

\subsubsection*{The Sea Contribution}

\begin{figure}[htb]
{\par\centering \resizebox*{!}{0.1\textheight}{\includegraphics{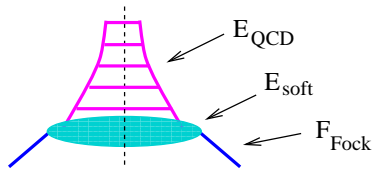}} \par}

\caption{The diagram corresponding to the parton distribution functions. \label{fpom}}
\end{figure}

The sea contribution contains the perturbative parton cascade, described as
a parton ladder with strictly ordered virtualities, and the non-perturbative
soft block, dominated by the soft Pomeron asymptotics, see fig. \ \ref{fpom}. 
We obtain the momentum distribution \( f^{1\mathrm{I}\! \mathrm{P}}_{i/p(\mathrm{sea})}\left( x,M^{2}_{F}\right)  \)
of the parton \( i \) at the virtuality scale \( M_{F}^{2} \) in the proton
as the convolution of three distributions (see fig.\ \ref{fpom}): the inclusive
parton Fock state distribution in the proton \( \tilde{F}_{p}^{(1)}(x_{0}) \)
(see eq.\ (\ref{f-part-remn})),
\begin{equation}
\label{fp-(1)}
\tilde{F}_{p}^{(1)}(x_{0})=F^{p}_{\mathrm{remn}}(1-x_{0})\, F^{p}_{\mathrm{part}}(x_{0}),
\end{equation}
 the distribution for the parton momentum share in the soft Pomeron \( E_{\mathrm{soft}}^{j}(z) \)
(see eqs.\ (\ref{esoft-g}-\ref{esoft-q})), and the QCD evolution function
\( E^{ji}_{\mathrm{QCD}}\left( z,Q^{2}_{0},M_{F}^{2}\right)  \): 

\begin{eqnarray}
f^{1\mathrm{I}\! \mathrm{P}}_{i/p(\mathrm{sea})}\! \left( x,M^{2}_{F}\right)  & = & \sum _{j}\int ^{1}_{x}\! \frac{dx_{0}}{x_{0}}\int ^{x_{0}}_{x}\! \frac{dx_{1}}{x_{1}}\nonumber \\
 &  & \qquad \times \; \tilde{F}_{p}^{(1)}(x_{0})\; E_{\mathrm{soft}}^{j}\! \left( \frac{x_{1}}{x_{0}}\right) \; E^{ji}_{\mathrm{QCD}}\! \left( \frac{x}{x_{1}},Q^{2}_{0},M_{F}^{2}\right) ,\label{f1p-sea} 
\end{eqnarray}
 see fig.\ \ref{fpom}. This equation may be written as
\begin{equation}
f^{1\mathrm{I}\! \mathrm{P}}_{i/p(\mathrm{sea})}\! \left( x,M^{2}_{F}\right) =\sum _{j}\int ^{1}_{x}\! \frac{dx_{1}}{x_{1}}\, \varphi ^{1\mathrm{I}\! \mathrm{P}}_{j/p(\mathrm{sea})}\! \left( x_{1}\right) \, E^{ji}_{\mathrm{QCD}}\! \left( \frac{x}{x_{1}},Q^{2}_{0},M_{F}^{2}\right) ,
\end{equation}
where \( \varphi ^{1\mathrm{I}\! \mathrm{P}}_{j/p(\mathrm{sea})}\! \left( x_{1}\right)  \)
by construction corresponds to the distribution at the initial scale \( Q_{0}^{2} \),
\begin{equation}
\label{sf-q0-pom}
\varphi ^{1\mathrm{I}\! \mathrm{P}}_{j/p(\mathrm{sea})}\! \left( x_{1}\right) =\int ^{1}_{x_{1}}\! \frac{dx_{0}}{x_{0}}\, \tilde{F}_{p}^{(1)}\! (x_{0})\, E_{\mathrm{soft}}^{j}\! \left( \frac{x_{1}}{x_{0}}\right) 
\end{equation}
Here we use the same expressions for \( F^{p}_{\mathrm{remn}}(x) \), \( F^{p}_{\mathrm{part}}(x) \),
and \( E_{\mathrm{soft}}^{j}(z) \) as in the case of proton-proton or nucleus-nucleus
scattering, see eqs.\ (\ref{f-part}-\ref{f-remn}), (\ref{esoft-g}-\ref{esoft-q}).

\subsubsection*{The Triple Pomeron Contribution}

\begin{figure}[htb]
{\par\centering \resizebox*{!}{0.2\textheight}{\includegraphics{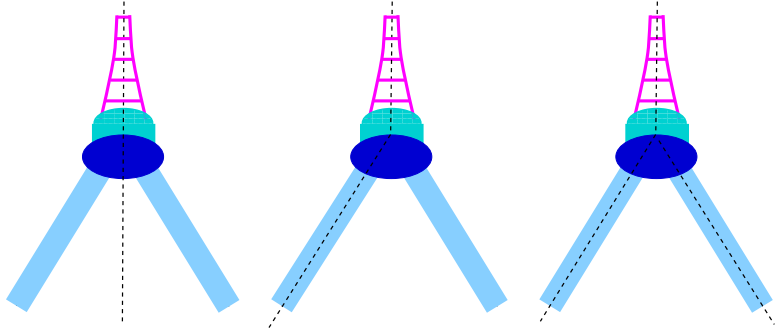}} \par}

\caption{The triple Pomeron contribution.\label{fpomtri}}
\end{figure}

We have to take also into account triple-Pomeron contributions \( f^{3\mathrm{I}\! \mathrm{P}}_{i/p(\mathrm{sea})}\left( x,M^{2}_{F}\right)  \)
to gluon and sea quark momentum distributions. The latter ones are defined by
the diagrams of fig.\ \ref{3pcuts} with the upper cut Pomeron being replaced
by the ``half'' of the sea-sea type semihard Pomeron which consists from a
soft Pomeron coupled to the triple-Pomeron vertex and the parton ladder describing
the perturbative parton evolution from the initial scale \( Q_{0}^{2} \) to
the final scale \( M_{F}^{2} \) , see fig. \ref{fpomtri}.
The lower legs are (un)cut Pomerons. 

Let us consider the part \( \Delta \sigma _{pp(\mathrm{sea}-\mathrm{sea})}^{3\mathrm{I}\! \mathrm{P}-}(s) \)
of the contribution of cut triple-Pomeron diagram 
\[
\Delta \sigma _{pp}^{3\mathrm{I}\! \mathrm{P}-}(s)=\frac{1}{2s}2\mathrm{Im}T^{3\mathrm{I}\! \mathrm{P}\! -}_{pp}(s,t=0),\]
 corresponding to the semihard sea-sea type parton-parton scattering in the
upper Pomeron, which is according to eqs.\ (\ref{t3p-}-\ref{t3p-p}) given
as
\begin{eqnarray}
\Delta \sigma _{pp(\mathrm{sea}-\mathrm{sea})}^{3\mathrm{I}\! \mathrm{P}-}(s) & = & -\frac{r_{3\mathrm{I}\! \mathrm{P}}}{2}\mathrm{Im}\left[ \int ^{1}_{0}\! \! dx^{+}\int ^{1}_{0}\! \! dx_{1}^{-}\int ^{1-x_{1}^{-}}_{0}\! \! dx_{2}^{-}\, F^{p}_{\mathrm{remn}}\! \left( 1-x^{+}\right) \, F^{p}_{\mathrm{remn}}\! \! \left( 1-x_{1}^{-}-x_{2}^{-}\right) \right. \nonumber \\
 &  & \times 8\pi ^{2}\int \! \frac{dx_{12}^{-}}{x_{12}^{-}}\left[ \frac{1}{2s^{+}}\, \mathrm{Im}\, T^{p}_{\mathrm{sea}-\mathrm{sea}}\! \left( x^{+},s^{+},0\right) \right] \nonumber \\
 &  & \times \left. \int dz^{+}\int \! d^{2}q_{1_{\perp }}\prod _{l=1}^{2}\! \left[ \frac{1}{8\pi ^{2}\hat{s}_{l}}\, iT^{p}\! \left( x_{l}^{-},\hat{s}_{l},-q_{1_{\perp }}^{2}\right) \right] \right] ,\label{sig-3p-semi} 
\end{eqnarray}
where we used \( x_{12}^{-}=s_{0}/(x_{12}^{+}s) \), \( s^{+}=x^{+}x^{-}_{12}s \),
\( \hat{s}_{1}=z^{+}s_{0}x_{1}^{-}/x_{12}^{-} \), \( \hat{s}_{2}=(1-z^{+})s_{0}x_{2}^{-}/x_{12}^{-} \),
and \( T^{p} \), \( T^{p}_{\mathrm{sea}-\mathrm{sea}} \) are defined in (\ref{t-pom-h-sum}-\ref{t-pom-h-valsea}).
Applying the AGK cutting rules, the contribution (\ref{sig-3p-semi}) can be
written as a sum of three terms,
\begin{equation}
\Delta \sigma _{pp(\mathrm{sea}-\mathrm{sea})}^{3\mathrm{I}\! \mathrm{P}-}(s)=\Delta \sigma _{pp(\mathrm{sea}-\mathrm{sea})}^{3\mathrm{I}\! \mathrm{P}-(0)}(s)+\Delta \sigma _{pp(\mathrm{sea}-\mathrm{sea})}^{3\mathrm{I}\! \mathrm{P}-(1)}(s)+\Delta \sigma _{pp(\mathrm{sea}-\mathrm{sea})}^{3\mathrm{I}\! \mathrm{P}-(2)}(s),
\end{equation}
 corresponding to the diffractive type semihard interaction, screening correction
to the usual semihard interaction, and double cut Pomeron contribution, with
the weights
\begin{eqnarray}
\Delta \sigma _{pp(\mathrm{sea}-\mathrm{sea})}^{3\mathrm{I}\! \mathrm{P}-(0)}(s) & =-1 & \, \times \; \Delta \sigma _{pp(\mathrm{sea}-\mathrm{sea})}^{3\mathrm{I}\! \mathrm{P}-}(s)\\
\Delta \sigma _{pp(\mathrm{sea}-\mathrm{sea})}^{3\mathrm{I}\! \mathrm{P}-(1)}(s) & =+4 & \, \times \; \Delta \sigma _{pp(\mathrm{sea}-\mathrm{sea})}^{3\mathrm{I}\! \mathrm{P}-}(s)\\
\Delta \sigma _{pp(\mathrm{sea}-\mathrm{sea})}^{3\mathrm{I}\! \mathrm{P}-(2)}(s) & =-2 & \, \times \; \Delta \sigma _{pp(\mathrm{sea}-\mathrm{sea})}^{3\mathrm{I}\! \mathrm{P}-}(s)
\end{eqnarray}
 Making use of (\ref{t-pom-h-sea}), (\ref{sig-sea-sea}), (\ref{sig-jk-hard}),
we have
\begin{eqnarray}
\frac{1}{2s^{+}}\mathrm{Im}\, T^{p}_{\mathrm{sea}-\mathrm{sea}}(x^{+},s^{+},0) & = & \frac{1}{2}F^{p}_{\mathrm{part}}(x^{+})\, \sum _{jk}\int ^{1}_{0}\! \frac{dz_{1}^{+}}{z_{1}^{+}}\frac{dz_{1}^{-}}{z_{1}^{-}}\, E_{\mathrm{soft}}^{j}\left( z_{1}^{+}\right) \, E_{\mathrm{soft}}^{k}\left( z_{1}^{-}\right) \nonumber \\
 & \times  & \sum _{ml}\int ^{1}_{z_{1}^{+}}\! dz_{B}^{+}\int ^{1}_{z_{1}^{-}}\! dz_{B}^{-}\int \! dp_{\bot }^{2}\, E_{\mathrm{QCD}}^{jm}(z_{B}^{+}/z_{1}^{+},Q_{0}^{2},M_{F}^{2})\, E_{\mathrm{QCD}}^{kl}(z_{B}^{-}/z_{1}^{-},Q_{0}^{2},M_{F}^{2})\nonumber \\
 & \times  & K\, {d\sigma _{\mathrm{Born}}^{ml}\over dp_{\bot }^{2}}(z_{B}^{+}z_{B}^{-}s^{+},p_{\bot }^{2})\, \theta \! \left( M_{F}^{2}-Q^{2}_{0}\right) 
\end{eqnarray}
 Now, using (\ref{fp-(1)}-\ref{f1p-sea}), we can rewrite (\ref{sig-3p-semi})
as
\begin{eqnarray}
\Delta \sigma _{pp(\mathrm{sea}-\mathrm{sea})}^{3\mathrm{I}\! \mathrm{P}-}(s) & = & \sum _{ml}\int \! dx^{+}_{B}dx^{-}_{B}dp_{\bot }^{2}\; f^{1\mathrm{I}\! \mathrm{P}}_{m/p(\mathrm{sea})}\! \left( x^{+}_{B},M^{2}_{F}\right) \, f^{3\mathrm{I}\! \mathrm{P}}_{l/p(\mathrm{sea})}\! \left( x^{-}_{B},M^{2}_{F}\right) \nonumber \\
 & \times  & K\, {d\sigma _{\mathrm{Born}}^{ml}\over dp_{\bot }^{2}}(x_{B}^{+}x_{B}^{-}s,p_{\bot }^{2})\, \theta \! \left( M_{F}^{2}-Q^{2}_{0}\right) ,\label{sig-3p-semi-fact} 
\end{eqnarray}
 where we denoted 
\[
x^{+}_{B}=x^{+}z^{+}_{B},\quad x^{-}_{B}=z^{-}_{B}\frac{s_{0}}{x_{12}^{+}s},\]
and defined 
\begin{equation}
f^{3\mathrm{I}\! \mathrm{P}}_{i/p(\mathrm{sea})}\! \left( x,M^{2}_{F}\right) =\sum _{j}\int ^{1}_{x}\! \frac{dx_{1}}{x_{1}}\, \varphi ^{3\mathrm{I}\! \mathrm{P}}_{j/p(\mathrm{sea})}\! \left( x_{1}\right) \, E^{ji}_{\mathrm{QCD}}\! \left( \frac{x}{x_{1}},Q^{2}_{0},M_{F}^{2}\right) ,
\end{equation}
with
\begin{eqnarray}
\varphi ^{3\mathrm{I}\! \mathrm{P}}_{j/p(\mathrm{sea})}\! \left( x\right)  & = & -\frac{r_{3\mathrm{I}\! \mathrm{P}}}{2}\, \int ^{1}_{x}\! \frac{dx_{12}}{x_{12}}\, E_{\mathrm{soft}}^{j}\! \left( \frac{x}{x_{12}}\right) \, \int \! \! dx_{1}dx_{2}\, F^{p}_{\mathrm{remn}}\! \! \left( 1-x_{1}-x_{2}\right) \nonumber \\
 & \quad \times  & 4\pi ^{2}\int \! dz\int \! d^{2}q_{\perp }\, \mathrm{Im}\left[ \prod ^{2}_{l=1}\! \left[ \frac{1}{8\pi ^{2}\hat{s}_{l}}\, iT^{p}\! \left( x_{l},\hat{s}_{l},-q_{\perp }^{2}\right) \right] \right] \nonumber \\
 & = & -\frac{r_{3\mathrm{I}\! \mathrm{P}}}{8}\, \int ^{1}_{x}\! \frac{dx_{12}}{x_{12}}\, E_{\mathrm{soft}}^{j}\! \left( \frac{x}{x_{12}}\right) \, \int \! \! dx_{1}dx_{2}\, F^{p}_{\mathrm{remn}}\! \! \left( 1-x_{1}^{-}-x_{2}^{-}\right) \nonumber \\
 & \quad \times  & \int \! d^{2}b\int \! dz\, G^{p}(x_{1},\hat{s}_{1},b)\, G^{p}(x_{2},\hat{s}_{2},b),\label{phi-3p-sea} 
\end{eqnarray}
with \( \hat{s}_{1}=z\, s_{0}x_{1}/x_{12} \), \( \hat{s}_{2}=(1-z)s_{0}x_{2}/x_{12} \),
where \( G^{p}(x,\hat{s},b) \) is given in (\ref{g-h-tot}-\ref{g-h-val-sea}
).

Now, replacing in (\ref{sig-3p-semi-fact}) the interaction with the projectile
parton \( m \) by the interaction with a virtual photon probe of virtuality
\( M_{F}^{2} \) or by the interaction with a hypotetical probe which couples
directly to a gluon, we immediately see that \( f^{3\mathrm{I}\! \mathrm{P}}_{i/p(\mathrm{sea})}\! \left( x,M^{2}_{F}\right)  \)
defines the (negative) contribution of the triple-Pomeron diagram to parton
structure functions.

\subsubsection*{The Valence Contribution}

\begin{figure}[htb]
{\par\centering \resizebox*{!}{0.1\textheight}{\includegraphics{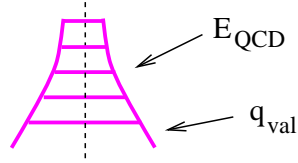}} \par}

\caption{The diagram corresponding to the ``valence'' contribution to the parton distribution
functions.\label{fig:ddhhv}}
\end{figure}

The third contribution, referred to as ``valence'', amounts to the case where
a valence quark is the first parton of the ladder, see fig. \ref{fig:ddhhv}. 
Here, the soft pre-evolution, governed by the secondary Reggeon, is typically
short and therefore not considered explicitly. One simply uses the parameterized
input for valence quark momentum distributions at the initial scale \( Q_{0}^{2} \),
\begin{equation}
\label{sf-dir}
\varphi _{i/p(\mathrm{val})}(x_{1})=q^{i}_{\mathrm{val}}(x_{1},Q_{0}^{2})
\end{equation}
 with the Gluck-Reya-Vogt parameterization for \( q^{i}_{\mathrm{val}}(x,Q_{0}^{2}) \)
\cite{glu95}.

\subsubsection{The Complete Proton Distribution Function}

The total parton distribution in the proton at the initial scale \( Q_{0}^{2} \)
is then defined as 
\begin{equation}
\label{sf-tot}
\varphi _{i/p}(x_{1})=\varphi _{i/p(\mathrm{sea})}(x_{1})+\varphi _{i/p(\mathrm{val})}(x_{1}),
\end{equation}
 with
\begin{equation}
\label{sf-sea-tot}
\varphi _{i/p(\mathrm{sea})}(x_{1})=\varphi ^{1\mathrm{I}\! \mathrm{P}}_{i/p(\mathrm{sea})}(x_{1})+\varphi ^{3\mathrm{I}\! \mathrm{P}}_{i/p(\mathrm{sea})}(x_{1})
\end{equation}
which results for an arbitrary scale \( M^{2}_{F} \) in

\begin{equation}
\label{sf-mf}
f_{i/p}\left( x,M^{2}_{F}\right) =\sum _{j}\int \! \frac{dx_{1}}{x_{1}}\varphi _{i/p}\left( x_{1}\right) E^{ji}_{\mathrm{QCD}}\left( \frac{x}{x_{1}},Q^{2}_{0},M_{F}^{2}\right) .
\end{equation}
For \( M^{2}_{F}=Q_{0}^{2} \), the semi-hard contribution is a function which
peaks at very small values  of \( x \) and then decreases monotonically towards
zero for \( x=1 \). The valence contribution, on the  other hand, has a maximum
at large values of \( x \) and goes towards zero for small  values of \( x \).
For moderate values of \( M^{2}_{F} \), the precise form of \( f \) depends
crucially on the  exponent for the Pomeron-nucleon coupling \( \alpha _{\mathrm{part}} \),
whereas for large \( M^{2}_{F} \) it is mainly defined by the QCD evolution
and depends weekly on the initial conditions at the scale \( Q_{0}^{2} \).

\subsubsection*{The Photon Distribution Functions}

To calculate the resolved photon cross section (\ref{sigma-res}), one needs
also to know parton momentum distributions of a virtual photon \( f_{i/\gamma ^{*}}\left( x,M^{2}_{\gamma },Q^{2}\right)  \).
According to \cite{glu99}, \( f_{i/\gamma ^{*}} \) gets contributions both
from vector meson states of the photon and from perturbative point-like photon
splitting into a quark-anti-quark pair,
\begin{equation}
f_{i/\gamma }\! \left( x,M^{2}_{\gamma },Q^{2}\right) =f^{\mathrm{VDM}}_{i/\gamma }\! \left( x,M^{2}_{\gamma },Q^{2}\right) +f^{\mathrm{point}}_{i/\gamma }\! \left( x,M^{2}_{\gamma },Q^{2}\right) .
\end{equation}

For the former one, one has \cite{glu99}
\begin{equation}
\label{f-vdm}
f^{\mathrm{VDM}}_{i/\gamma }\! \left( x,M^{2}_{\gamma },Q^{2}\right) =\eta (Q^{2})\, \alpha \left[ G^{2}_{i}\, f_{i/\pi ^{0}}\! \left( x,M^{2}_{\gamma }\right) +\frac{1}{2}\delta _{i}\, \left( G^{2}_{u}-G^{2}_{d}\right) \, f_{s/\pi ^{0}}\! \left( x,M^{2}_{\gamma }\right) \right] ,
\end{equation}
where the function \( \eta  \) is given as
\begin{equation}
\eta (Q^{2})=\left( 1+Q^{2}/m^{2}_{\rho }\right) ^{-2},
\end{equation}
 with \( m^{2}_{\rho }=0.59 \) GeV\( ^{2} \) and with
\begin{equation}
\begin{array}{llll}
\delta _{u}=-1, & \delta _{d}=1, & \delta _{s}=0, & \delta _{g}=0,\\
G^{2}_{u}=0.836,\;  & G^{2}_{d}=0.250,\;  & G^{2}_{s}=0.543, & \; G^{2}_{g}=0.543.
\end{array}
\end{equation}
 Pion structure functions \( f_{i/\pi ^{0}} \) are defined in the same way
as proton ones, namely

\begin{equation}
\label{sf-mf-pi}
f_{i/\pi ^{0}}\! \left( x,M^{2}_{F}\right) =\sum _{j}\int \! \frac{dx_{1}}{x_{1}}\, \varphi _{i/\pi ^{0\! }}\left( x_{1}\right) \, E^{ji}_{\mathrm{QCD}}\! \left( \frac{x}{x_{1}},Q^{2}_{0},M_{F}^{2}\right) ,
\end{equation}
 with
\begin{equation}
\label{sf-tot-pi}
\varphi _{i/\pi ^{0}}(x_{1})=\varphi _{i/\pi (\mathrm{sea})}(x_{1})+\varphi _{i/\pi ^{0}(\mathrm{val})}(x_{1}),
\end{equation}
 Here \( \varphi _{i/\pi ^{0}(\mathrm{val})} \) is a parameterized initial
distribution for the valence component from \cite{glu97} 
\begin{equation}
\varphi _{i/\pi ^{0}(\mathrm{val})}(x_{1})=q^{i}_{\mathrm{val}/\pi ^{0}}(x_{1},Q_{0}^{2}),
\end{equation}
 and the sea component \( \varphi _{j/\pi (\mathrm{sea})} \) is given by the
formulas (\ref{sf-tot}-\ref{sf-sea-tot}), (\ref{sf-q0-pom}), (\ref{phi-3p-sea})
, (\ref{sf-dir}) with the subscript \( p \) being replaced by \( \pi  \)
and using the appropriate parameters \( \alpha ^{\pi }_{\mathrm{remn}} \),
\( \gamma _{\pi } \) in \( F^{\pi }_{\mathrm{part}},\, F^{\pi }_{\mathrm{remn}} \),
but keeping all other parameters, characterizing the Pomeron trajectory, unchanged
compared to proton case. 

The point-like contribution \( f^{\mathrm{point}}_{i/\gamma } \) is given as
a convolution of the photon splitting into a quark-anti-quark pair (with the
Altarelli-Parisi splitting function \( P^{\gamma \rightarrow q\bar{q}}(z)=N_{c}/2\, (z^{2}+(1-z)^{2}) \),
\( N_{c}=3 \) being the number of colors), followed by the QCD evolution of
a (anti-)quark from the initial virtuality \( q^{2} \) of the splitting till
the scale \( M_{\gamma }^{2} \)

\begin{eqnarray}
f^{\mathrm{point}}_{i/\gamma }\left( x,M^{2}_{\gamma },Q^{2}\right)  & = & \langle e_{q}^{2}\rangle \int \! \frac{dq^{2}}{q^{2}}\int _{x}^{1}\! \frac{dx_{\gamma }}{x_{\gamma }}\, \frac{\alpha }{2\pi }\, P^{\gamma \rightarrow q\bar{q}}(x_{\gamma })\nonumber \\
 & \times  & \sum _{j\in \{u,d,s,\bar{u},\bar{d},\bar{s}\}}\, E^{ji}_{\mathrm{QCD}}\left( \frac{x}{x_{\gamma }},q^{2},M_{\gamma }^{2}\right) \, \Theta \! \left( q^{2}-\max \! \left[ Q_{0}^{2},x_{\gamma }Q^{2}\right] \right) 
\end{eqnarray}
with \( x_{\gamma } \) being the share of the virtual photon light cone momentum
taken by the (anti-)quark and with
\begin{equation}
\langle e_{q}^{2}\rangle =\frac{1}{3}\left( e_{u}^{2}+e_{d}^{2}+e_{s}^{2}\right) 
\end{equation}
 being the average light quark charge squared. Here we have chosen the initial
scale for the QCD evolution of a \( t \)-channel (anti-)quark to be equal to
its initial virtuality \( q^{2}=x_{\gamma }Q^{2}+p^{2}_{\perp }/(1-x_{\gamma }) \),
with \( p_{\perp }^{2} \) being the transverse momentum squared for the splitting
in the \( \gamma ^{*}p \) center of mass system \cite{ros98}.

\section{The Structure Function \protect\( F_{2}\protect \)}

\begin{figure}[htb]
{\par\centering \resizebox*{!}{0.9\textheight}{\includegraphics{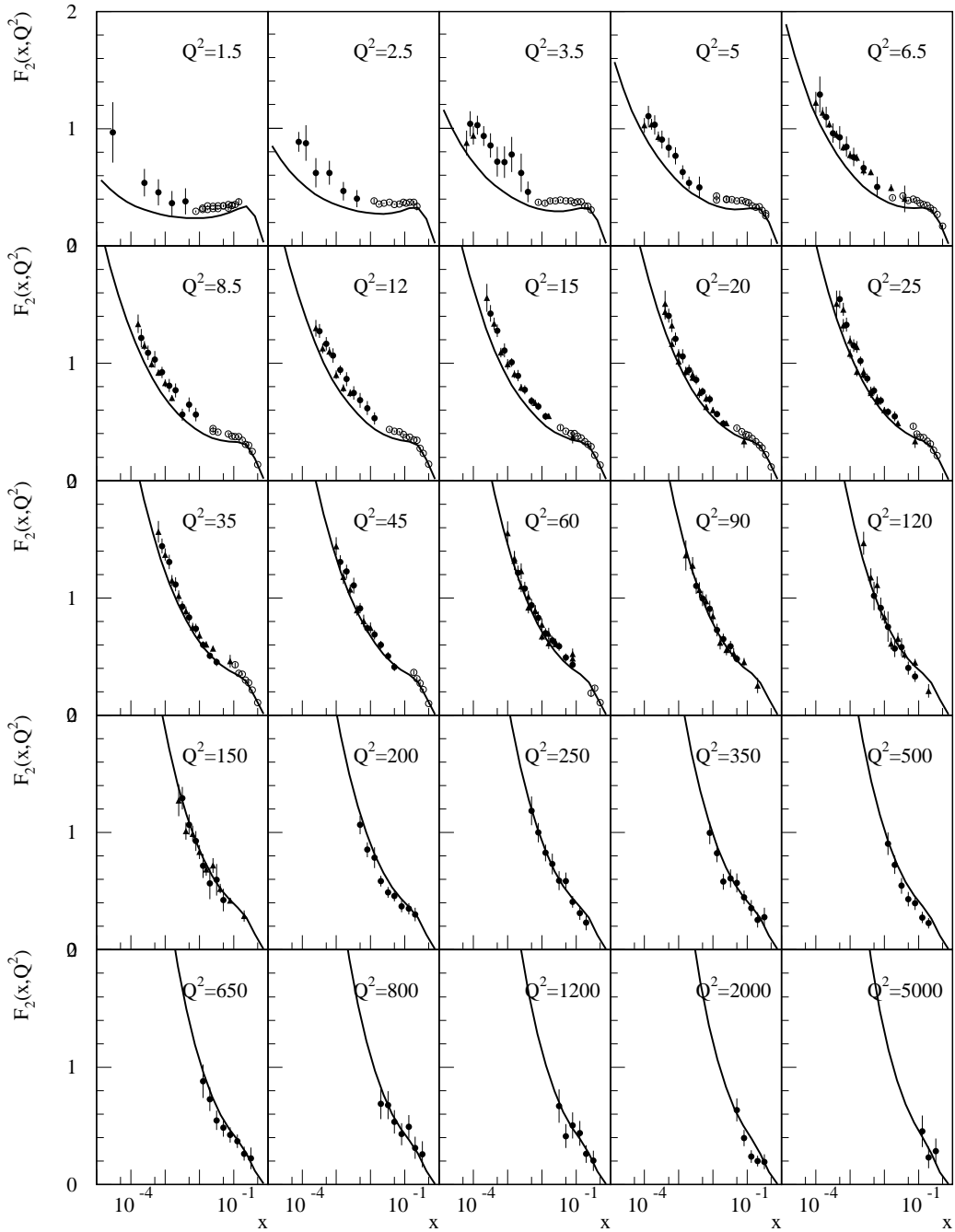}} \par}

\caption{The structure function \protect\( F_{2}\protect \) for different values of
\protect\( Q^{2}\protect \) together with experimental data from H1 \cite{h1-96a},
ZEUS \cite{zeus96}and NMC \cite{nmc95}.\label{ep1}}
\end{figure}
\begin{figure}[htb]
{\par\centering \resizebox*{!}{0.9\textheight}{\includegraphics{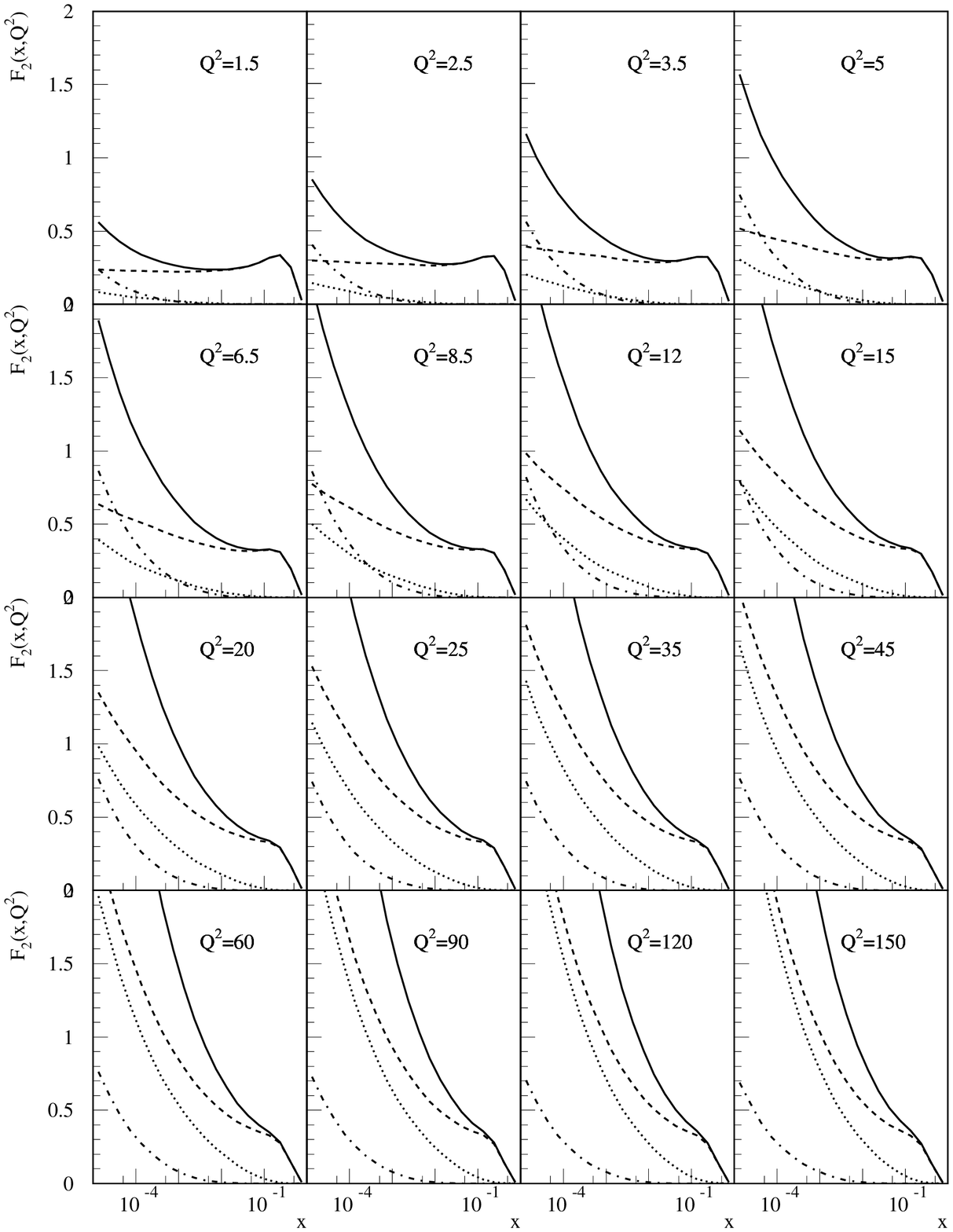}} \par}

\caption{Same as fig. \ref{ep1}, but here we show separately the ``direct-light''
contribution (dashed), the ``direct-charm'' contribution (dotted), and the
``resolved'' contribution (dashed-dotted).\label{ep1b}}
\end{figure}
We have now all elements to calculate the structure function \( F_{2} \), based
on the formula (\ref{f2-sigma}), with all leading order contributions to \( \sigma ^{\gamma ^{*}p}_{T(L)} \)
as given in eqs. (\ref{sigma-t}-\ref{sigma-res}) and with the parton momentum
distributions in the proton and photon as discussed in the preceding section.
The results for  \( F_{2}(x,Q^{2}) \) are shown in fig.\ \ref{ep1} together
with experimental data from H1 \cite{h1-96a}, ZEUS \cite{zeus96} and NMC \cite{nmc95}.
The parameters affecting the results for \( F_{2} \) are actually the same
ones which affect parton-parton and hadron-hadron scattering. So we fix them
in order to have an overall good fit for \( F_{2} \) and at the same time the
energy dependence of the total cross section and of the slope parameter. It
is possible to get a reasonable agreement, which is of course not perfect due
to the fact that enhanced diagrams are only treated to lowest order. In fig.\ \ref{ep1b},
we show separately the direct light quarks contribution (\ref{sigma-light}),
as well as the ones of charm quarks (\ref{sigma-cc}) and of resolved photons
(\ref{sigma-res}), for \( Q^{2}= \)1.5 GeV\( ^{2} \), 5 GeV\( ^{2} \), and
25 GeV\( ^{2} \). It is easy to see that the resolved photon cross section
contributes significantly to \( F_{2}(x,Q^{2}) \) for small photon virtualities
\( Q^{2} \).

\section{Parton Configurations: Basic Formulas}

In order to have a coherent approach, we base our treatment of particle production
on exactly the same formulas as derived earlier for the cross sections. To be
more precise, we take the formulas for \( d\sigma /dx_{B}dQ^{2} \) as a basis
for treating particle production, which means first of all parton production. 

The differential cross section for lepton-nucleon scattering is given in eq.\
(\ref{sigmadis-dif1}). Using (\ref{eqcd-ini}), (\ref{eqcd-n-rungs}), the
\( QCD \) evolution function \( E^{ij}_{\QCD } \), which enters into the formulas
for the cross sections \( \sigma ^{\gamma ^{*}p}_{T(L)}(\tilde{s},Q^{2}) \),
can be expanded as a sum over \( n \)-rung ladder contributions, where the
latter ones can be written as an integration over the momenta \( p_{1},p_{2},\ldots ,p_{n} \)
of \( n \) resolvable final partons. Introducing a multidimensional variable
\begin{equation}
\label{p2}
P=\{p_{1},p_{2},\ldots ,p_{n}\},
\end{equation}
 and considering the symbol \( \sum _{P} \) representing \( \sum _{n}\int dP_{n} \),
with \( dP_{n} \) being the invariant phase space volume for \( n \)-parton
state, we may write
\begin{equation}
\label{p3}
{d\sigma _{lp}\over dx_{B}\, dQ^{2}}=\sum _{P}\sigma (x_{B},Q^{2},P).
\end{equation}
After normalization, \( \sigma (x_{B},Q^{2},P) \) may be interpreted as the
probability distribution for a parton configuration \( P \) for given values
of \( x_{B} \) and \( Q^{2} \). The Monte Carlo method provides a convenient
tool for treating such multidimensional distributions: with \( \sigma (x_{B},Q^{2},P) \)
being known (see preceding sections), one generates parton configurations \( P \)
according to this distribution. In addition to \( x_{B} \), \( Q^{2} \), and
\( P \), additional variables occur, specifying a particular contribution to
the DIS cross section. One essentially follows the structure of the formula
for the cross section. Let us discuss the procedure to generate parton configurations
in detail. 

We start with some useful definitions. Using the relation (\ref{eqcd-ini})
for the evolution function \( E^{ij}_{\QCD } \), any parton momentum distribution
at the scale \( Q^{2} \) can be decomposed into two contributions, corresponding
to the case of no resolvable emission in the range of virtualities between \( Q_{0}^{2} \)
and \( Q^{2} \) and to at least one resolvable emission:
\begin{equation}
f_{j}(x,Q^{2})=f_{j}(x,Q_{0}^{2})\, \Delta ^{j}(Q_{0}^{2},Q^{2})+\sum _{i}\int _{x}^{1-\epsilon }\frac{dz}{z}\bar{E}^{ij}_{\QCD }\! \left( z,Q^{2}_{0},Q^{2}\right) \: f_{i}\! \left( \frac{x}{z},Q^{2}_{0}\right) 
\end{equation}

\begin{figure}[htb]
{\par\centering \resizebox*{!}{0.4\textheight}{\includegraphics{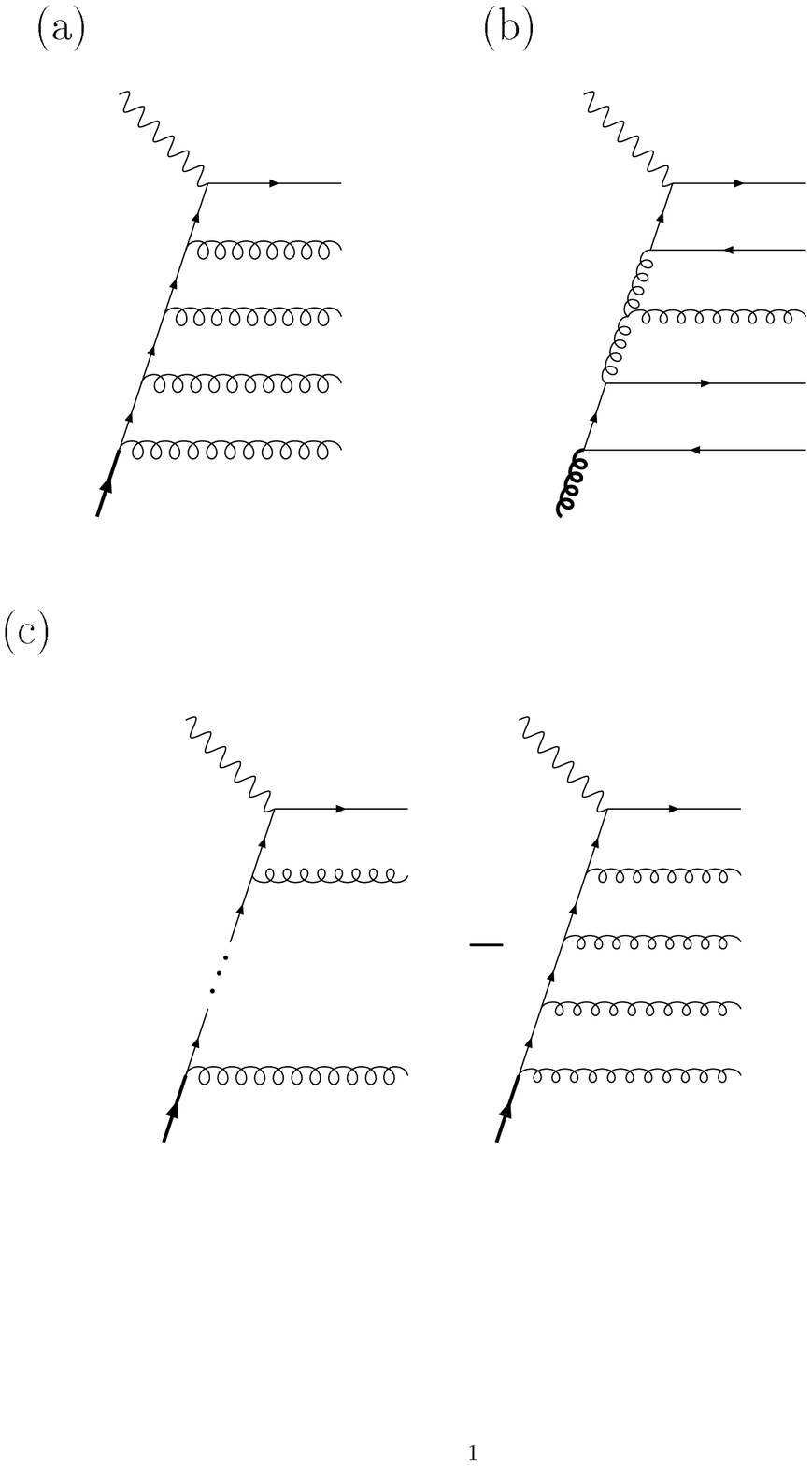}} \par}

\caption{\label{fig:sigdissns}The non-singlet contribution (a), the gluon contribution
(b), and the singlet contribution (c).}
\end{figure}

In case of ``\( i \)'' and ``\( j \)'' being quarks, we split the sum
\( \sum _{j}\bar{E}^{ij}_{\QCD } \) into two components,
\begin{equation}
\sum _{j}\bar{E}^{ij}_{\QCD }=\bar{E}_{\mathrm{NS}}+\bar{E}_{\mathrm{S}}\qquad (i,j=\mathrm{quark}s),
\end{equation}
 with the so-called non-singlet evolution \( \bar{E}_{\mathrm{NS}} \), where
only gluons are emitted as final \( s \)-channel partons, and the singlet evolution
\( \bar{E}_{\mathrm{S}} \) representing all the other contributions, see fig.\ \ref{fig:sigdissns}a,c. 
The non-singlet evolution (compare with eq. (\ref{ap-res})) satisfies the evolution
equation
\begin{eqnarray}
 &  & \bar{E}_{\mathrm{NS}}\left( x,Q^{2}_{0},Q^{2}\right) =\label{eqcd-ns} \\
 &  & \quad \int ^{Q^{2}}_{Q_{0}^{2}}\! \frac{dQ_{1}^{2}}{Q_{1}^{2}}\left[ \int _{x}^{1-\epsilon }\! \frac{dz}{z}\, \frac{\alpha _{s}}{2\pi }\, P_{q}^{q}\! (z)\: \bar{E}_{\mathrm{NS}}\! \left( \frac{x}{z},Q^{2}_{0},Q_{1}^{2}\right) \, \Delta ^{q}(Q_{1}^{2},Q^{2})+\frac{\alpha _{s}}{2\pi }\, P_{q}^{q}\! (x)\, \Delta ^{q}(Q_{0}^{2},Q^{2})\right] ,\nonumber \label{eqcd-ns} 
\end{eqnarray}
and the singlet one
\begin{eqnarray}
\bar{E}_{\mathrm{S}}\left( x,Q^{2}_{0},Q^{2}\right) =\int ^{Q^{2}}_{Q_{0}^{2}}\! \frac{dQ_{1}^{2}}{Q_{1}^{2}}\int ^{1-\epsilon }_{x}\! \frac{dz}{z}\, \frac{\alpha _{s}}{2\pi }\left[ P^{q}_{q}\! (z)\, \bar{E}_{\mathrm{S}}\! \left( \frac{x}{z},Q^{2}_{0},Q_{1}^{2}\right) \right. \qquad \qquad \qquad  &  & \nonumber \\
\left. +\; 2n_{f}\, P^{q}_{g}\! (z)\, \bar{E}^{qg}_{\QCD }\! \left( \frac{x}{z},Q^{2}_{0},Q_{1}^{2}\right) \right] \, \Delta ^{q}(Q_{1}^{2},Q^{2}), &  & \label{eqcd-s} 
\end{eqnarray}
with \( n_{f}=3 \) being the number of active quark flavors.

Now it is convenient to define parton level cross sections, corresponding to
different contributions to the deep inelastic scattering process and to different
partons, entering the perturbative evolution at the initial scale \( Q_{0}^{2} \),
see fig.\ \ref{fig:sigdissns}. Essentially, we include into the cross sections
the perturbative part of parton evolution, whereas the initial conditions, given
by parton momentum densities at the initial scale \( Q_{0}^{2} \), are factorized
out. The non-singlet and singlet contributions to the direct (light) photon-quark
interaction are defined as
\begin{equation}
\label{sigma-qgamns}
\sigma ^{\gamma ^{*}q}_{T(\mathrm{NS})}(\tilde{s},Q^{2},Q_{0}^{2})=\frac{4\pi ^{2}\alpha }{\tilde{s}}\bar{E}_{\mathrm{NS}}\left( \frac{Q^{2}}{\tilde{s}},Q^{2}_{0},Q^{2}\right) 
\end{equation}
\begin{equation}
\label{sigma-qgams}
\sigma ^{\gamma ^{*}q}_{T(\mathrm{S})}(\tilde{s},Q^{2},Q_{0}^{2})=\frac{4\pi ^{2}\alpha }{\tilde{s}}\bar{E}_{\mathrm{S}}\left( \frac{Q^{2}}{\tilde{s}},Q^{2}_{0},Q^{2}\right) 
\end{equation}
 For the direct (light) photon-gluon interaction, we define 
\begin{equation}
\label{sigma-ggam}
\sigma _{T(\mathrm{light})}^{\gamma ^{*}g}(\tilde{s},Q^{2},Q_{0}^{2})=\frac{4\pi ^{2}\alpha }{\tilde{s}}\bar{E}^{gq}_{\QCD }\left( \frac{Q^{2}}{\tilde{s}},Q^{2}_{0},Q^{2}\right) 
\end{equation}
 The photon-parton charm production cross section is defined as 
\begin{equation}
\label{sigma-ccgam}
\sigma ^{\gamma ^{*}i}_{T/L(\mathrm{charm})}(\tilde{s},Q^{2},Q_{0}^{2})=\int \! dx^{-}\int \! dp_{\bot }^{2}\, E^{ig}_{\QCD }\! \left( x^{-},Q^{2}_{0},M_{F}^{2}\right) \, \frac{d\sigma ^{\gamma ^{*}g\rightarrow c\bar{c}}_{T/L}\! \left( x^{-}\tilde{s},Q^{2},p_{\bot }^{2}\right) }{dp_{\bot }^{2}},
\end{equation}
where the parton (\( i \)) may be a quark or a gluon. Finally, we define the
parton-parton cross section for resolved processes similarly to (\ref{sigma-ij-12})
as
\begin{eqnarray}
\sigma _{T(\mathrm{resolved})}^{ij}(\hat{s},Q^{2},q^{2},Q_{0}^{2}) & = & \int \! dx^{+}dx^{-}\int \! dp_{\bot }^{2}\, \sum _{k,l}\frac{d\sigma _{\mathrm{Born}}^{kl}\left( x^{+}x^{-}\hat{s},p_{\bot }^{2}\right) }{dp_{\bot }^{2}}\label{sigma-ij} \\
 & \times  & E^{ik}_{\QCD }\! \left( x^{+},q^{2},M_{\gamma }^{2}\right) \, E^{jl}_{\QCD }\! \left( x^{_{-}},Q^{2}_{0},M_{p}^{2}\right) \, \theta \! \left( p_{\bot }^{2}-Q^{2}\right) .\nonumber 
\end{eqnarray}
Based on the above partial photon-parton cross sections, we define now the total
photon-parton cross sections, summed over the quark flavors of the quark coupling
to the photon with the appropriate squared charge (\( e^{2} \)) factor. We
obtain for the photon-gluon cross section

\begin{eqnarray}
\sigma ^{\gamma ^{*}g}_{T}(\tilde{s},Q^{2},Q_{0}^{2}) & = & \langle e_{q}^{2}\rangle \, \sigma _{T(\mathrm{light})}^{\gamma ^{*}g}(\tilde{s},Q^{2},Q_{0}^{2})+e_{c}^{2}\, \sigma ^{\gamma ^{*}g}_{T(\mathrm{charm})}(\tilde{s},Q^{2},Q_{0}^{2})\nonumber \\
 & + & \sum _{j}\int \! dx_{\gamma }\, f^{\mathrm{VDM}}_{j/\gamma }\! \left( x_{\gamma },Q_{0}^{2},Q^{2}\right) \, \sigma _{T(\mathrm{resolved})}^{jg}(x_{\gamma }(\tilde{s}-Q^{2}),Q^{2},Q_{0}^{2},Q_{0}^{2})\label{sig-gamma-g} \\
 & + & \langle e_{q}^{2}\rangle \, \int \! \frac{dq^{2}}{q^{2}}\int \! dx_{\gamma }\, \frac{\alpha }{2\pi }\, P^{\gamma \rightarrow q\bar{q}}(x_{\gamma })\nonumber \\
 & \times  & \sum _{j\in \{u,d,s,\bar{u},\bar{d},\bar{s}\}}\sigma _{T(\mathrm{resolved})}^{jg}(x_{\gamma }\tilde{s}-q^{2},Q^{2},q^{2},Q_{0}^{2})\, \Theta \! \left( q^{2}-\max \! \left[ Q_{0}^{2},x_{\gamma }Q^{2}\right] \right) .\nonumber 
\end{eqnarray}
The photon-quark cross section for a quark with flavor ``\( i \)'' is given
as

\begin{eqnarray}
\sigma ^{\gamma ^{*}i}_{T}(\tilde{s},Q^{2},Q_{0}^{2}) & = & e_{i}^{2}\, \sigma _{T(\mathrm{NS})}^{\gamma ^{*}q}(\tilde{s},Q^{2},Q_{0}^{2})+\langle e_{q}^{2}\rangle \, \sigma _{T(\mathrm{S})}^{\gamma ^{*}q}(\tilde{s},Q^{2},Q_{0}^{2})\nonumber \\
 & + & e_{c}^{2}\, \sigma ^{\gamma ^{*}g}_{T(\mathrm{charm})}\! \left( \tilde{s},Q^{2},Q_{0}^{2}\right) \label{sig-gamma-q} \\
 & + & \sum _{j}\int \! dx_{\gamma }\, f^{\mathrm{VDM}}_{j/\gamma }\! \left( x_{\gamma },Q_{0}^{2},Q^{2}\right) \, \sigma _{T(\mathrm{resolved})}^{ji}(x_{\gamma }(\tilde{s}-Q^{2}),Q^{2},Q_{0}^{2},Q_{0}^{2})\nonumber \\
 & + & \langle e_{q}^{2}\rangle \, \int \! \frac{dq^{2}}{q^{2}}\int \! dx_{\gamma }\, \frac{\alpha }{2\pi }\, P^{\gamma \rightarrow q\bar{q}}(x_{\gamma })\nonumber \\
 & \times  & \sum _{j\in \{u,d,s,\bar{u},\bar{d},\bar{s}\}}\sigma _{T(\mathrm{resolved})}^{ji}(x_{\gamma }\tilde{s}-q^{2},Q^{2},q^{2},Q_{0}^{2})\, \Theta \! \left( q^{2}-\max \! \left[ Q_{0}^{2},x_{\gamma }Q^{2}\right] \right) \nonumber 
\end{eqnarray}
The longitudinal photon-parton cross section for a parton of flavor \( i \)
(quark or gluon) is finally given as
\begin{equation}
\label{sig-gamma-i-l}
\sigma ^{\gamma ^{*}i}_{L}(\tilde{s},Q^{2},Q_{0}^{2})=e_{c}^{2}\, \sigma ^{\gamma ^{*}i}_{L(\mathrm{charm})}(\tilde{s},Q^{2},Q_{0}^{2}).
\end{equation}
 Finally, we may express the photon-proton cross sections in terms of the above
photon-parton cross sections and the parton distribution functions \( \varphi (x)=f(x,Q^{2}_{0}) \)
at the scale \( Q^{2}_{0} \). The transverse cross section is given as
\begin{eqnarray}
\sigma _{T}^{\gamma ^{*}p}(\tilde{s},Q^{2}) & = & \int \! dx_{1}\, \varphi _{g/p}\left( x_{1}\right) \, \sigma _{T}^{\gamma ^{*}g}(x_{1}\tilde{s},Q^{2},Q_{0}^{2})\label{sig-tra} \\
 & + & \sum _{i\in \{u,d,s,\bar{u},\bar{d},\bar{s}\}}\left( \frac{4\pi ^{2}\alpha }{Q^{2}}\, e_{i}^{2}\, x_{B\, }\varphi _{i/p}(x_{B})\, \Delta ^{q}(Q_{0}^{2},Q^{2})+\int \! dx_{1}\, \varphi _{i/p}(x_{1})\, \sigma ^{\gamma ^{*}i}_{T}(x_{1}\tilde{s},Q^{2},Q_{0}^{2})\right) ,\nonumber \label{x} 
\end{eqnarray}
the longitudinal cross section is given as
\begin{eqnarray}
\sigma _{L}^{\gamma ^{*}p}(\tilde{s},Q^{2}) & = & \int \! dx_{1}\, \varphi _{g/p}\! \left( x_{1}\right) \, \sigma _{L}^{\gamma ^{*}g}(x_{1}\tilde{s},Q^{2},Q_{0}^{2})\label{sig-long} \\
 & + & \sum _{i\in \{u,d,s,\bar{u},\bar{d},\bar{s}\}}\int \! dx_{1}\, \varphi _{i/p}(x_{1})\, \sigma ^{\gamma ^{*}i}_{L}(x_{1}\tilde{s},Q^{2},Q_{0}^{2}),\nonumber 
\end{eqnarray}
with the quark momentum distributions being a sum of two terms,
\begin{equation}
\varphi _{i/p}(x_{1})=\varphi _{i/p(\mathrm{sea})}(x_{1})+\varphi _{i/p(\mathrm{val})}(x_{1}),
\end{equation}
see eqs. (\ref{sf-q0-pom}-\ref{sf-tot}). 

The above formulas together with eq.\ (\ref{sigmadis-dif1}) serve as the basis
to generate all main variables for the description of deep inelastic scattering.
After modeling \( Q^{2} \) and \( x_{B} \) we simulate types (a valence quark,
a sea quark, or a gluon) and kinematical characteristics for first partons,
entering the perturbative evolution (at the initial scale \( Q_{0}^{2}) \),
and then, for given initial conditions, generate corresponding parton configurations,
based on the particular structure of perturbative cross sections (\ref{sigma-qgamns}-\ref{sigma-ij}).
The detailed description of this procedure is given in the next sections.

The triple Pomeron contributions are included here in the definition of the
parton distribution \( \varphi _{i/p(\mathrm{sea})} \). At HERA energies, the
triple Pomeron contribution is dominated by the process where the two Pomerons
exchanged in parallel are soft ones, and therefore no additional parton production
needs to be considered in that case.

\section{Generating Initial Conditions for the Perturbative Evolution}

We start with the generation of the kinematical variables \( x_{B} \) and \( Q^{2} \)
according to the differential cross section eq.\ (\ref{sigmadis-dif1}) together
with the explicit form for the photon-proton cross sections eqs.\ (\ref{sig-tra}-\ref{sig-long}).
Then we choose an interaction with the transverse or with the longitudinal polarization
component of the photon, with the corresponding weights 
\begin{equation}
L_{T/L}^{\gamma ^{*}}(y)\, \sigma ^{\gamma ^{*}p}_{T/L}(\tilde{s},Q^{2})/\left( L_{T}^{\gamma ^{*}}(y)\, \sigma ^{\gamma ^{*}p}_{T}(\tilde{s},Q^{2})+L_{L}^{\gamma ^{*}}(y)\, \sigma ^{\gamma ^{*}p}_{L}(\tilde{s},Q^{2})\right) .
\end{equation}
After that, we consider virtual photon-proton interaction for given photon virtuality
\( Q^{2} \) and polarization (\( T \), \( L \)), and for given c.m.\ energy
squared \( \hat{s}=\tilde{s}-Q^{2} \) for the interaction; we use the photon-proton
center of mass system.

Let us first consider the case of transverse photon polarization. We have to
choose between ``sea'' and ``valence '' contribution, where the latter one
is chosen with the probability 
\begin{eqnarray}
\mathrm{prob}(\mathrm{val})=\frac{1}{\sigma ^{\gamma ^{*}p}_{T}(\tilde{s},Q^{2})}\left\{ \frac{4\pi ^{2}\alpha }{Q^{2}}\sum _{i\in \{u,d\}}e_{i}^{2}\, x_{B}\, \varphi _{i/p(\mathrm{val})}(x_{B})\, \Delta ^{q}(Q_{0}^{2},Q^{2})\right.  & + & \nonumber \label{1-wdir} \\
\left. +\; \sum _{i\in \{u,d\}}\int \! dx_{1}\, \varphi _{i/p(\mathrm{val})}(x_{1})\, \sigma ^{\gamma ^{*}i}_{T}(x_{1}\tilde{s},Q^{2},Q_{0}^{2})\right\}  &  & \label{w-dir} 
\end{eqnarray}
and the former one with the probability \( 1-\mathrm{prob}(\mathrm{val}) \),
which is the sum of the contributions, corresponding to a gluon or a sea quark
from the proton being the first ladder parton,
\begin{eqnarray}
\mathrm{prob}(\mathrm{sea}) & = & \frac{1}{\sigma ^{\gamma ^{*}p}_{T}(\tilde{s},Q^{2})}\left\{ \frac{4\pi ^{2}\alpha }{Q^{2}}\sum _{i\in \{u,d,s,\bar{u},\bar{d},\bar{s}\}}e_{i}^{2}\, x_{B}\, \varphi _{i/p(\mathrm{sea})}(x_{B})\, \Delta ^{q}(Q_{0}^{2},Q^{2})\right. +\nonumber \label{1-wdir} \\
 &  & \left. +\; \sum _{i\in \{g,u,d,s,\bar{u},\bar{d},\bar{s}\}}\int \! dx_{1}\, \varphi _{i/p(\mathrm{sea})}(x_{1})\, \sigma ^{\gamma ^{*}i}_{T}(x_{1}\tilde{s},Q^{2},Q_{0}^{2})\right\} .\label{1-wdir} 
\end{eqnarray}

The next step consists of defining the type (flavor) of the first ladder parton
and its momentum share \( x_{1} \) in the proton. Here one has to distinguish
two possible parton configurations for the interaction: parton cascading without
any resolvable parton emission in the ladder (corresponding to a valence or
a sea quark of the proton scattered back in the Breit frame), with the relative
weights given by the first term in the curly brackets in eqs. (\ref{w-dir},
\ref{1-wdir}), i.e. 
\begin{equation}
\mathrm{prob}(\mathrm{no}\, \mathrm{emission})=\frac{\frac{4\pi ^{2}\alpha }{Q^{2}}\sum _{i}e_{i}^{2}\, x_{B}\, \varphi _{\mathrm{i}/p(\mathrm{val}/\mathrm{sea})}(x_{B})\, \Delta ^{q}(Q_{0}^{2},Q^{2})}{\mathrm{prob}(\mathrm{val}/\mathrm{sea})\, \sigma ^{\gamma ^{*}p}_{T}(\tilde{s},Q^{2})},
\end{equation}
 and the configurations with at least one resolvable parton emission, with the
weight \( 1-\mathrm{prob}(\mathrm{no}\, \mathrm{emission}) \). In the case
of no resolvable emission we have \( x_{1}=x_{B}=Q^{2}/\tilde{s} \) and the
flavor of the quark is generated according to the weights 
\begin{equation}
\left\{ \begin{array}{lll}
e_{i}^{2}\, \varphi _{\mathrm{i}/p(\mathrm{val})}(x_{B}) & \mathrm{if} & \mathrm{val}\\
e_{i}^{2}\, \varphi _{i/p(\mathrm{sea})}(x_{B}) & \mathrm{if} & \mathrm{sea}
\end{array}\right. .
\end{equation}
 Then we are left with a trivial parton configuration. For a valence quark contribution
it consists of an anti-quark, moving along the original proton direction, and
the quark, scattered back. 

In the case of at least one resolvable emission, one generates the type (flavor)
\( i \) and the light cone momentum fraction \( x_{1} \) of  the first parton
of the QCD cascade according to the distributions, given by the expressions
in the curly brackets in (\ref{w-dir}), (\ref{1-wdir}),
\begin{equation}
\mathrm{prob}(i,x_{1})=\left\{ \begin{array}{lll}
\delta ^{i}_{q}\, \varphi _{\mathrm{i}/p(\mathrm{val})}(x_{1})\, \sigma ^{\gamma ^{*}i}_{T}(x_{1}\tilde{s},Q^{2},Q_{0}^{2}) & \mathrm{if} & \mathrm{val}\\
\varphi _{i/p(\mathrm{sea})}(x_{1})\, \sigma ^{\gamma ^{*}i}_{T}(x_{1}\tilde{s},Q^{2},Q_{0}^{2}) & \mathrm{if} & \mathrm{sea}
\end{array}\right. ,
\end{equation}
where \( \delta ^{i}_{q} \) is zero if \( i=g \) and otherwise one.

Then one chooses between different types of interactions, contributing to the
photon-parton cross section, according to their partial weights in eqs. (\ref{sig-gamma-g},
\ref{sig-gamma-q}). The direct photon-parton contribution is chosen with the
weight 

\begin{equation}
\mathrm{prob}(\mathrm{direct})=\left\{ \begin{array}{lll}
\left[ e_{i}^{2}\, \sigma _{T(\mathrm{NS})}^{\gamma ^{*}q}+\langle e_{q}^{2}\rangle \, \sigma _{T(\mathrm{S})}^{\gamma ^{*}q}+e_{c}^{2}\, \sigma ^{\gamma ^{*}q}_{T(\mathrm{charm})}\right] /\sigma ^{\gamma ^{*}i}_{T} & \mathrm{if} & i=\mathrm{quark}\\
\left[ \langle e_{q}^{2}\rangle \, \sigma _{T(\mathrm{light})}^{\gamma ^{*}g}+e_{c}^{2}\, \sigma ^{\gamma ^{*}g}_{T(\mathrm{charm})}\right] /\sigma ^{\gamma ^{*}g}_{T} & \mathrm{if} & i=\mathrm{gluon}
\end{array}\right. .
\end{equation}
 For the resolved photon contributions, the weight is given by the last two
terms in eqs. (\ref{sig-gamma-g}, \ref{sig-gamma-q}). The probability for
the VDM part is 
\begin{equation}
\mathrm{prob}(\mathrm{VDM})=\sum _{j}\int \! dx_{\gamma }\, f^{\mathrm{VDM}}_{j/\gamma }\! \left( x_{\gamma },Q_{0}^{2},Q^{2}\right) \, \sigma _{T(\mathrm{resolved})}^{ji}(x_{\gamma }(x_{1}\tilde{s}-Q^{2}),Q^{2},Q_{0}^{2},Q_{0}^{2})/\sigma ^{\gamma ^{*}i}_{T},
\end{equation}
where \( x_{1} \) is the light cone momentum share and \( i \) is the type
(flavor) of the first ladder parton on the proton side (already determined),
whereas the probability for the point-like resolved contribution is 
\begin{eqnarray}
\mathrm{prob}(\mathrm{point}) & = & \langle e_{q}^{2}\rangle \, \int \! \frac{dq^{2}}{q^{2}}\int \! dx_{\gamma }\, \frac{\alpha }{2\pi }\, P^{\gamma \rightarrow q\bar{q}}(x_{\gamma })\\
 & \times  & \sum _{j\in \{u,d,s,\bar{u},\bar{d},\bar{s}\}}\sigma _{T(\mathrm{resolved})}^{ji}(x_{\gamma }x_{1}\tilde{s}-q^{2},Q^{2},q^{2},Q_{0}^{2})\, \Theta \! \left( q^{2}-\max \! \left[ Q_{0}^{2},x_{\gamma }Q^{2}\right] \right) /\sigma ^{\gamma ^{*}i}_{T}.\nonumber 
\end{eqnarray}

In case of a direct light contribution, one has to generate the configuration
for a parton ladder, strictly ordered in parton virtualities towards the virtual
photon. The method is quite analogous to the one of section 2 in chapter 5 and
is described in the next section. 

In case of a resolved contribution, we need to define the initial conditions
for the other end of the parton ladder, on the photon side, as well as the parton
type (a quark of some flavor or a gluon) and the share of the light cone momentum
fraction \( x_{\gamma } \), taken by the parton from the photon. For the direct
resolved contribution, corresponding to the point-like photon splitting into
a quark-anti-quark pair, the flavor \( j \), the share \( x_{\gamma } \),
and the virtuality \( q^{2} \) of the (anti-)quark, being the first ladder
parton, are generated according to
\begin{equation}
\mathrm{prob}(j,x_{\gamma },q^{2})\sim \frac{1}{q^{2}}\, \frac{\alpha }{2\pi }\, P^{\gamma \rightarrow q\bar{q}}(x_{\gamma })\, \sigma _{T(\mathrm{resolved})}^{ji}(x_{\gamma }x_{1}\tilde{s}-q^{2},Q^{2},q^{2},Q_{0}^{2})\, \Theta \! \left( q^{2}-\max \! \left[ Q_{0}^{2},x_{\gamma }Q^{2}\right] \right) \, \delta ^{j}_{q}
\end{equation}
 For the VDM contribution, the first ladder parton is taken at the initial virtuality
\( Q_{0}^{2} \) and its type \( j \) and momentum share \( x_{\gamma } \)
are chosen according to the distribution

\begin{equation}
\mathrm{prob}(j,x_{\gamma })\sim f^{\mathrm{VDM}}_{j/\gamma }\! \left( x_{\gamma },Q_{0}^{2},Q^{2}\right) \, \sigma _{T(\mathrm{resolved})}^{ji}(x_{\gamma }(x_{1}\tilde{s}-Q^{2}),Q^{2},Q_{0}^{2},Q_{0}^{2}),
\end{equation}
with the VDM parton momentum distributions in the photon defined in (\ref{f-vdm}).
In both cases for resolved photon interactions, the simulation of parton configurations,
corresponding to the ladder of given mass squared \( \hat{s}' \) (\( \hat{s}'=x_{\gamma }x_{1}\tilde{s}-q^{2} \)
for the direct resolved contribution and \( \hat{s}'=x_{\gamma }(x_{1}\tilde{s}-Q^{2}) \)
for the VDM one), and of given types and virtualities of the leg partons, is
done exactly in the same way as for proton-proton (nucleus-nucleus) interactions,
as described in the chapter 5. The only difference comes from the presence of
two different scales \( M_{p}^{2} \), \( M_{\gamma }^{2} \) and the cutoff
\( p_{\perp }^{2}>Q^{2} \) in the parton-parton cross section \( \sigma _{T(\mathrm{resolved})}^{ij} \)
for resolved DIS processes as given in eq. (\ref{sigma-ij}), when compared
to the cross section in eq. (\ref{sigma-ij-12}).

In the case of interaction with the longitudinal photon component, the procedure
simplifies considerably, as one only has to consider direct photon-parton interactions
via the parton-gluon fusion process. One starts by choosing between the coupling
of the parton ladder to a valence quark (``val'') or to a soft Pomeron (``sea''),
the weights are

\begin{eqnarray}
\mathrm{prob}(\mathrm{val}) & = & \frac{\int dx_{1}\, \sum _{i}\varphi _{i/p(\mathrm{val})}(x_{1})\, \sigma ^{\gamma ^{*}i}_{L}(x_{1}\tilde{s},Q^{2},Q_{0}^{2})}{\sigma ^{\gamma ^{*}p}_{L}(\tilde{s},Q^{2})};\label{w-dir-l} \\
\mathrm{prob}(\mathrm{sea}) & = & \frac{\int dx_{1}\, \sum _{i}\varphi _{i/p(\mathrm{sea})}(x_{1})\, \sigma ^{\gamma ^{*}i}_{L}(x_{1}\tilde{s},Q^{2},Q_{0}^{2})}{\sigma ^{\gamma ^{*}p}_{L}(\tilde{s},Q^{2})}.\label{1-wdir-l} 
\end{eqnarray}
The weight for the parton type (flavor) \( i \) and the distribution for the
light cone momentum fraction \( x_{1} \) of  the first parton of the QCD cascade
is given by the integrands of (\ref{w-dir-l}-\ref{1-wdir-l}). The final step
amounts to generating the configuration for the parton ladder, strictly ordered
in parton virtualities towards the virtual photon, with the largest momentum
transfer parton process of parton-gluon fusion type, as discussed in the next
section.

\section{Generating the Ladder Partons}

In this section we describe the procedure to generate parton configurations,
corresponding to direct photon-parton interaction with at least one resolvable
emission in the parton cascade. In that case, a parton ladder is strictly ordered
in parton virtualities towards the virtual photon and the ladder cross section
is given as a sum of the contributions eqs.\ (\ref{sigma-qgamns}-\ref{sigma-ccgam})
in the case of transverse photon polarization, 
\begin{equation}
\label{sig-q-dir}
\sigma _{T(\mathrm{direct})}^{\gamma ^{*}i}(\tilde{s},Q^{2},Q_{0}^{2})=\left\{ \begin{array}{lll}
e_{i}^{2}\, \sigma _{T(\mathrm{NS})}^{\gamma ^{*}q}(\tilde{s},Q^{2},Q_{0}^{2})+\langle e_{q}^{2}\rangle \, \sigma _{T(\mathrm{S})}^{\gamma ^{*}q}(\tilde{s},Q^{2},Q_{0}^{2}) &  & \\
+e_{c}^{2}\, \sigma ^{\gamma ^{*}q}_{T(\mathrm{charm})}(\tilde{s},Q^{2},Q_{0}^{2}) & i=q, & \\
\langle e_{q}^{2}\rangle \, \sigma ^{\gamma ^{*}g}(\tilde{s},Q^{2},Q_{0}^{2})+e_{c}^{2}\, \sigma ^{\gamma ^{*}g(c\bar{c})}_{T}(\tilde{s},Q^{2},Q_{0}^{2}) & i=g, & 
\end{array}\right. 
\end{equation}
or by the cross section eq.\ (\ref{sigma-ccgam}) for the longitudinal one,
\begin{equation}
\label{sig-i-l-dir}
\sigma _{L(\mathrm{direct})}^{\gamma ^{*}i}(\tilde{s},Q^{2},Q_{0}^{2})=e_{c}^{2}\, \sigma ^{\gamma ^{*}i}_{T(\mathrm{charm})}(\tilde{s},Q^{2},Q_{0}^{2}).
\end{equation}
 All the photon-parton cross sections are expressed in terms of the QCD evolution
functions \( \bar{E}_{\mathrm{QCD}} \). Using the explicit representation eqs.\
(\ref{eqcd-n-rungs}-\ref{ap-1-res}) for \( \bar{E}_{\QCD } \), one can rewrite
the recursive relations eqs.\ (\ref{ap-res}), (\ref{eqcd-ns}-\ref{eqcd-s})
in a form, such that the first (lowest virtuality) emission in the ladder is
treated explicitly, multiplied by a weight factor, given by the contribution
of the rest of the ladder (the sum of any number of additional resolvable emissions),

\begin{eqnarray}
\bar{E}^{ij}_{\mathrm{QCD}}\left( x,Q^{2}_{0},Q^{2}\right)  & = & \int ^{Q^{2}}_{Q_{0}^{2}}\! \frac{dQ_{1}^{2}}{Q_{1}^{2}}\sum _{k}\int _{x}^{1-\epsilon }\! \frac{dz}{z}\, \frac{\alpha _{s}}{2\pi }\, P_{i}^{k}\! (z)\, \Delta ^{i}(Q_{0}^{2},Q_{1}^{2})\, \bar{E}^{kj}_{\mathrm{QCD}}\! \left( \frac{x}{z},Q^{2}_{1},Q^{2}\right) \nonumber \label{xxx} \\
 & + & \int ^{Q^{2}}_{Q_{0}^{2}}\! \frac{dQ_{1}^{2}}{Q_{1}^{2}}\, \Delta ^{i}(Q_{0}^{2},Q_{1}^{2})\, \Delta ^{j}(Q_{1}^{2},Q^{2})\, \frac{\alpha _{s}}{2\pi }\, P_{i}^{j}\! (x)\label{eqcd-inv} \\
\bar{E}_{NS}\left( x,Q^{2}_{0},Q^{2}\right)  & = & \int ^{Q^{2}}_{Q_{0}^{2}}\! \frac{dQ_{1}^{2}}{Q_{1}^{2}}\int _{x}^{1-\epsilon }\! \frac{dz}{z}\, \frac{\alpha _{s}}{2\pi }\, P_{q}^{q}\! (z)\, \Delta ^{q}(Q_{0}^{2},Q_{1}^{2})\, \bar{E}_{NS}\! \left( \frac{x}{z},Q^{2}_{1},Q^{2}\right) \nonumber \\
 & + & \Delta ^{q}(Q_{0}^{2},Q^{2})\, \int ^{Q^{2}}_{Q_{0}^{2}}\! \frac{dQ_{1}^{2}}{Q_{1}^{2}}\, \frac{\alpha _{s}}{2\pi }\, P_{q}^{q}\! (x)\label{eqcd-ns-inv} \\
\bar{E}_{S}\left( x,Q^{2}_{0},Q^{2}\right)  & = & \int ^{Q^{2}}_{Q_{0}^{2}}\! \frac{dQ_{1}^{2}}{Q_{1}^{2}}\sum _{k}\int _{x}^{1-\epsilon }\! \frac{dz}{z}\, \Delta ^{q}(Q_{0}^{2},Q_{1}^{2})\, \frac{\alpha _{s}}{2\pi }\, \left[ P^{q}_{q}\! (z)\, \bar{E}_{\mathrm{S}}\! \left( \frac{x}{z},Q^{2}_{1},Q^{2}\right) \right. \nonumber \\
 & + & \left. P^{g}_{q}\! (z)\, \bar{E}^{gq}_{\QCD }\! \left( \frac{x}{z},Q^{2}_{1},Q^{2}\right) \right] \label{eqcd-s-inv} 
\end{eqnarray}
 With the help of eqs.\ (\ref{eqcd-inv}-\ref{eqcd-s-inv}), (\ref{sigma-qgamns}-\ref{sigma-ccgam}),
one can obtain the recursive relations for the cross sections eqs.\ (\ref{sig-q-dir},
\ref{sig-i-l-dir}) for an arbitrary virtuality \( Q^{2}_{1} \) of the initial
parton \( i \),
\begin{eqnarray}
\sigma _{T/L(\mathrm{direct})}^{\gamma ^{*}i}(\tilde{s},Q^{2},Q_{1}^{2}) & = & \sum _{j}\int ^{Q^{2}}_{Q_{1}^{2}}\! \frac{d\tilde{Q}^{2}}{\tilde{Q}^{2}}\, \Delta ^{i}(Q_{1}^{2},\tilde{Q}^{2})\int \! dz\, \frac{\alpha _{s}}{2\pi }\, P_{i}^{j}(z)\, \sigma _{T/L(\mathrm{direct})}^{\gamma ^{*}j}(z\tilde{s},Q^{2},\tilde{Q}^{2})+\nonumber \\
 & + & \sigma _{T/L(\mathrm{direct}\, 2\rightarrow 2)}^{\gamma ^{*}i}(\tilde{s},Q^{2},Q_{1}^{2}),\label{sig-tl-dir} 
\end{eqnarray}
where \( \sigma _{T/L(\mathrm{direct}\, 2\rightarrow 2)}^{\gamma ^{*}i} \)
represents the contribution of parton configurations with only one resolvable
parton emission in the ladder, or with the photon-gluon fusion process without
any additional resolvable parton emissions,
\begin{eqnarray}
\sigma _{T(\mathrm{direct}\, 2\rightarrow 2)}^{\gamma ^{*}i}(\tilde{s},Q^{2},Q_{1}^{2}) & = & \frac{4\pi ^{2}\, \alpha \, e_{i}^{2}}{\tilde{s}}\, E^{(1)qq}_{\QCD }\! \left( \frac{Q^{2}}{\tilde{s}},Q^{2}_{1},Q^{2}\right) \, \Theta \! (Q^{2}-Q_{1}^{2})\quad (i=\mathrm{quark})\\
\sigma _{T(\mathrm{direct}\, 2\rightarrow 2)}^{\gamma ^{*}g}(\tilde{s},Q^{2},Q_{1}^{2}) & = & \frac{4\pi ^{2}\, \alpha \, \langle e_{q}^{2}\rangle }{\tilde{s}}\, E^{(1)gq}_{\QCD }\! \left( \frac{Q^{2}}{\tilde{s}},Q^{2}_{1},Q^{2}\right) \, \Theta \! (Q^{2}-Q_{1}^{2})\nonumber \\
 & + & e_{c}^{2}\int \! dp_{\bot }^{2}\, \frac{d\sigma ^{\gamma ^{*}g\rightarrow c\bar{c}}_{T}\left( \tilde{s},Q^{2},p_{\bot }^{2}\right) }{dp_{\bot }^{2}}\, \Delta ^{g}(Q_{1}^{2},M_{F}^{2})\\
\sigma _{L(\mathrm{direct}\, 2\rightarrow 2)}^{\gamma ^{*}i}(\tilde{s},Q^{2},Q_{1}^{2}) & = & \delta ^{g}_{i}\, e_{c}^{2}\int \! dp_{\bot }^{2}\, \frac{d\sigma ^{\gamma ^{*}g\rightarrow c\bar{c}}_{L}\left( \tilde{s},Q^{2},p_{\bot }^{2}\right) }{dp_{\bot }^{2}}\, \Delta ^{g}(Q_{1}^{2},M_{F}^{2})\label{sig-2-2-l} 
\end{eqnarray}
 It is noteworthy that to the leading logarithmic accuracy one should not use
in \( \sigma _{T(\mathrm{direct}\, 2\rightarrow 2)}^{\gamma ^{*}i} \) the Born
process matrix elements \( d\sigma ^{\gamma ^{*}g\rightarrow q\bar{q}}_{T}/dp_{\bot }^{2} \)
and \( d\sigma ^{\gamma ^{*}q\rightarrow gq}_{T}/dp_{\bot }^{2} \); the contribution
of just one resolvable parton emission is proportional to \( E^{(1)iq}_{\QCD } \),
eq.\ (\ref{ap-1-res}), defined by the corresponding Altarelli-Parisi kernel
\( P_{i}^{q}(z) \).

The formulas (\ref{sig-tl-dir}-\ref{sig-2-2-l}) allow us to generate the cascade
of partons, corresponding to the direct \( \gamma ^{*} \)-quark (gluon) interaction
of energy squared \( \hat{s}=\tilde{s}-Q^{2} \) and photon virtuality \( Q^{2} \),
starting  from an initial parton with a flavor \( i \), taken at a scale \( Q^{2}_{1}=Q_{0}^{2} \).
We use an iterative procedure, similar to the one of chapter 5. At each step
one checks whether there is any additional resolvable parton emission before
the last one , with the probability 
\begin{equation}
\mathrm{prob}(\mathrm{forward}\, \mathrm{emission})=\frac{\sigma _{T/L(\mathrm{direct})}^{\gamma ^{*}i}(\tilde{s},Q^{2},Q_{1}^{2})-\sigma _{T/L(\mathrm{direct}\, 2\rightarrow 2)}^{\gamma ^{*}i(2\rightarrow 2)}(\tilde{s},Q^{2},Q_{1}^{2})}{\sigma _{T/L(\mathrm{direct})}^{\gamma ^{*}i}(\tilde{s},Q^{2},Q_{1}^{2})}
\end{equation}

In case of an emission, the flavor \( j \) of the new ladder leg parton, the
light cone momentum fraction \( z \), taken from the parent parton, and the
virtuality \( \tilde{Q}^{2} \) are generated according to  the integrand of
\( \sigma _{T/L(\mathrm{direct})}^{\gamma ^{*}i}-\sigma _{T/L(\mathrm{direct}\, 2\rightarrow 2)}^{\gamma ^{*}i} \)
in (\ref{sig-tl-dir}), 
\begin{equation}
\label{f-dis-branch}
\mathrm{prob}(j,z,\tilde{Q}^{2})\sim \frac{1}{\tilde{Q}^{2}}\, \Delta ^{i}(Q_{1}^{2},\tilde{Q}^{2})\, \frac{\alpha _{s}}{2\pi }\, P_{i}^{j}(z)\, \sigma _{T/L(\mathrm{direct})}^{\gamma ^{*}j}(z\tilde{s},Q^{2},\tilde{Q}^{2}).
\end{equation}
The process is repeated for the new ladder of energy squared \( \hat{s}'=z\tilde{s}-Q^{2} \),
with the initial parton \( i'=j \), and with virtuality \( Q_{1}'^{2}=\tilde{Q}^{2} \)
and so on. At each step one decides about emission or not, which finally terminates
the iteration.

Having done the iterative parton emission, we finally generate the last (highest
virtuality) resolvable parton emission or the photon-gluon fusion process (if
\( i'=g \)) in the photon-parton center-of-mass system. The photon-gluon fusion
(PGF) process is chosen for \( i'=g \) with the probability 
\begin{equation}
\mathrm{prob}(\mathrm{PGF})=\frac{1}{\sigma _{T(\mathrm{direct}\, 2\rightarrow 2)}^{\gamma ^{*}g}(\tilde{s}',Q^{2},\tilde{Q}^{2})}e_{c}^{2}\int \! dp_{\bot }^{2}\, \frac{d\sigma ^{\gamma ^{*}g\rightarrow c\bar{c}}_{T}\left( \tilde{s}',Q^{2},p_{\bot }^{2}\right) }{dp_{\bot }^{2}}\, \Delta ^{g}(\tilde{Q}^{2},M_{F}^{2})
\end{equation}
for the transverse photon polarization and always for the longitudinal photon
polarization. For \( i'=q \), we have \( \mathrm{prob}(\mathrm{PGF})=0 \).
In case of PGF, with our choice \( M_{F}^{2}=m_{c}^{2}+p_{\bot }^{2} \), we
generate the transverse momentum squared of final charm quarks in the region
\( \tilde{Q}^{2}-m^{2}_{c}<p_{\bot }^{2}<\frac{1}{4}\hat{s}-m_{c}^{2} \) according
to 
\begin{equation}
\mathrm{prob}(p_{\bot }^{2})\sim \frac{d\sigma ^{\gamma ^{*}g\rightarrow c\bar{c}}_{T}\left( \tilde{s}',Q^{2},p_{\bot }^{2}\right) }{dp_{\bot }^{2}}\Delta ^{g}(\tilde{Q}^{2},M_{F}^{2})
\end{equation}
 In case of no PGF, we generate the momentum transfer squared for the last resolvable
parton emission in the range \( \tilde{Q}^{2}<Q'^{2}<Q^{2} \) according to
\begin{equation}
\mathrm{prob}(Q'^{2})\sim \frac{1}{Q'^{2}}\Delta ^{i}(\tilde{Q}^{2},Q'^{2})\; \Delta ^{q}(Q'^{2},Q^{2})\; \frac{\alpha _{s}}{2\pi }P_{i}^{q}\! \left( z'\right) ,
\end{equation}
with 
\begin{equation}
z'=1-\left( 1-\frac{Q'^{2}}{\tilde{s}}\right) \left( 1-\frac{Q^{2}}{\tilde{s}}\right) ,
\end{equation}
 and find the parton transverse momentum squared as \( p^{2}_{\bot }=Q'^{2}(1-z') \).

We then reconstruct final parton 4-momenta in their center of mass system with
a random polar angle for the transverse momentum vector \( \vec{p}_{\bot } \)
and boost them to the original Lorentz frame. This completed the description
of the algorithm to generate parton  configurations, based on exactly the same
formulas as for calculatins of \( F_{2} \) before.

The above discussion of how to generate parton configurations is not  yet complete:
the emitted partons are in general off--shell and can therefore  radiate further
partons. This so called time-like radiation is taken into account  using standard
techniques \cite{sjo84}, as discussed already in chapter 5.

\section{Hadron Production}

\begin{figure}[htb]
{\par\centering \resizebox*{!}{0.2\textheight}{\includegraphics{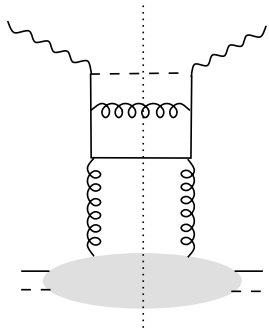}} \par}

\caption{A diagram contributing to the photon-proton cross section.\label{light-sea}}
\end{figure}

\begin{figure}[htb]
{\par\centering \resizebox*{!}{0.2\textheight}{\includegraphics{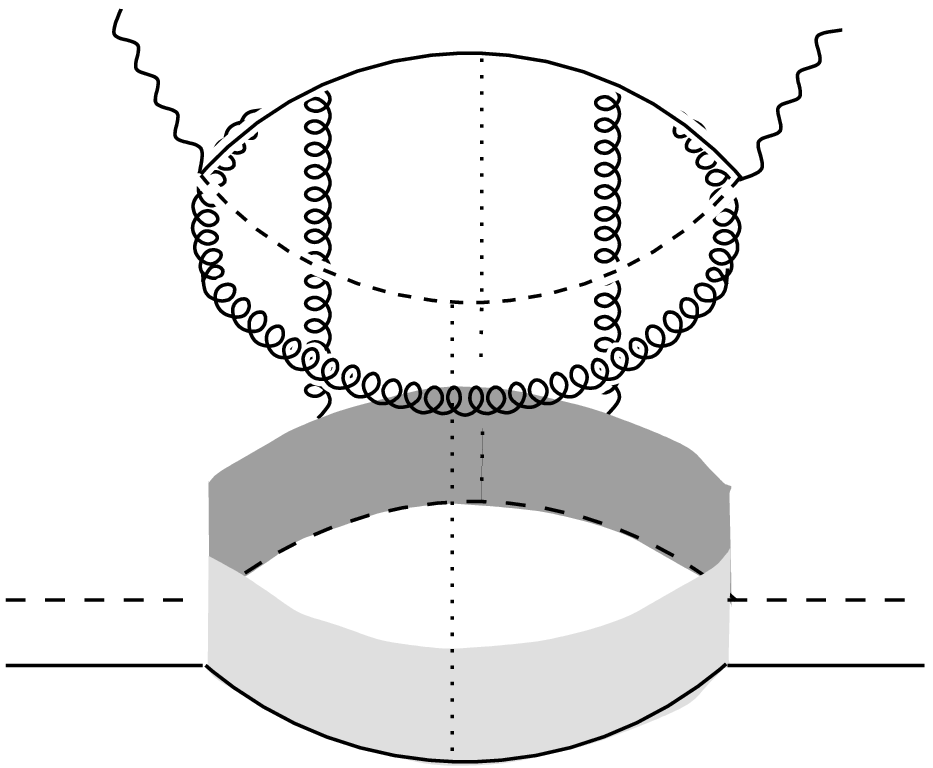}} \( \quad  \)\resizebox*{!}{0.2\textheight}{\includegraphics{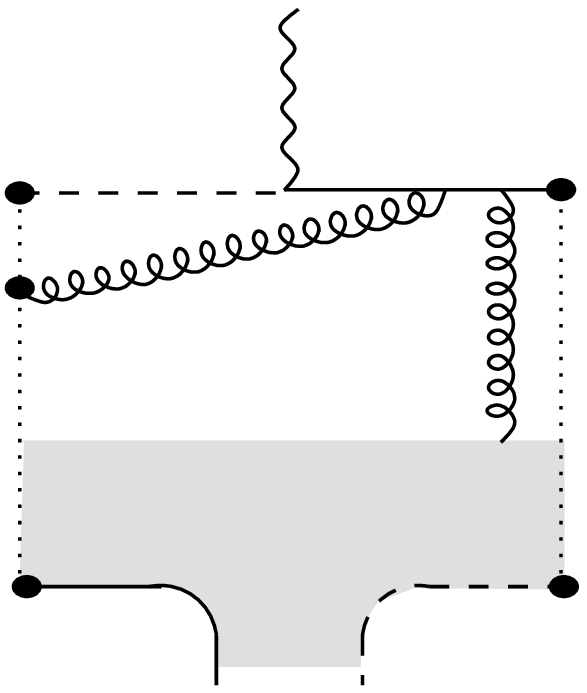}} \par}

\caption{Cylindrical representation of the diagram of fig. \ref{light-sea} (left figure)
and planar diagram representing the corresponding half-cylinder (right figure).
\label{light-sea-cyl}}
\end{figure}

\begin{figure}[htb]
{\par\centering \resizebox*{!}{0.2\textheight}{\includegraphics{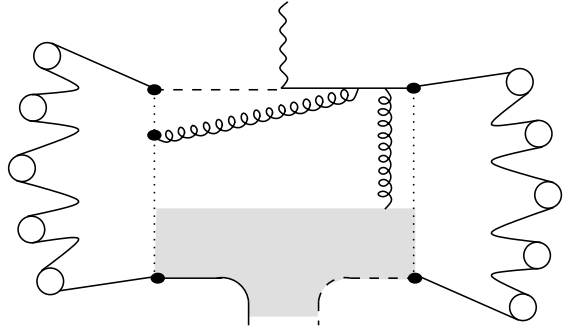}} \par}

\caption{The string model: the cut line (vertical dotted line) corresponds to a string,
which decays into hadrons (circles). \label{string-model-light-sea}}
\end{figure}

For the hadronization, we use exactly the same philosophy and even the same
procedure as in case of proton-proton (\( pp \)) scattering. Hadronization
is not considered as a dynamical procedure, rather we consider the hadronic
states as being integrated out in the considerations of cross section calculations
of the preceding sections. Hadronization means simply a phenomenological procedure
to explicitly reintroduce these hadronic states. The procedure employed for
\( pp \) scattering and to be used here as well is as follows:

\begin{enumerate}
\item drawing a cylinder diagram; 
\item cutting the cylinder;  
\item planar presentation of half-cylinder; 
\item identification of cut line with kinky string;  
\item kinky string hadronization (as explained in chapter 6).
\end{enumerate}
We are going to explain the steps (1-4) for a concrete example of a diagram
contributing to photon-proton scattering, where the photon interacts directly
with a light quark (contribution ``light''), and where the first parton of
the ladder on the proton side couples to the proton via a soft Pomeron (contribution
``sea''), as shown in fig.\ \ref{light-sea}.
The external legs on the lower (proton) side are a quark (full line) and an
anti-quark (dashed), representing together the proton constituent participating
in the interaction. In fig.\ \ref{light-sea-cyl}(left), we show the result
of plotting the diagram on a cylinder. The shaded area on the lower part of
the cylinder indicates the soft Pomeron, a complicated non-resolved structure,
where we do not specify the microscopic content. The two space-like gluons emerge
out of this soft structure. The cut is represented by the two vertical dotted
lines on the cylinder. We now consider one of the two half-cylinders, say the
left one, and we plot it in a planar fashion, as shown in fig.\ \ref{light-sea-cyl}(right).
We observe one internal gluon, and one external one, appearing on the cut line.
We now identify the two cut lines with kinky strings such that a parton on the
cut line corresponds to a kink: we have one kinky string with one internal kink
(gluon) in addition to the two end kinks, and we have one flat string with just
two end kinks, but no internal one. The strings are then hadronized according
to the methods explained in chapter 6, see fig.\ \ref{string-model-light-sea}. 

In principle, we have also triple Pomerons contributing to the event topology.
However, due to AGK cancellations, such contributions to the inclusive spectra
cancel each other in the kinematical region where the two Pomerons are in parallel.
Therefore, the average charteristics are correctly described by considering
the simple cylinder-type topology corresponding to one Pomeron exchange.

\section{Results}

We are now capable to simulate events from deep inelastic scattering. When fixing
the parameters, we found that all the ones found in \( e^{+}e^{-} \) can be
kept with the exception of the mean transverse momentum of string breaking \( p_{\perp }^{f} \)and
the break probability \( p_{\mathrm{break}} \), see the discussion in chapter
\ref{parameters}. 

\begin{figure}[htb]
{\par\centering \resizebox*{0.7\columnwidth}{!}{\includegraphics{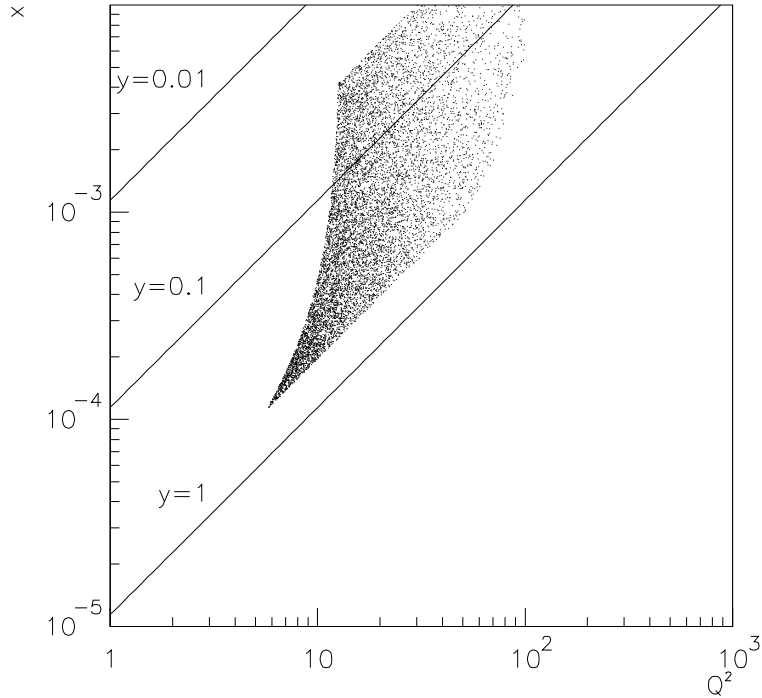}} \par}

\caption{\label{fig:disq2x}Distribution of the simulated events if the \protect\( Q^{2}-x\protect \)
plane.}
\end{figure}

We show results from \( ep \) scattering and compare to the data of the experiments
accomplished at HERA. Electrons of an energy 26.7 GeV collide with protons of
820 GeV, which gives a center of mass energy 296 GeV. We made the analysis for
the kinematical region \( 10^{-4}<x<10^{-2} \) and \( 10<Q^{2}<100\textrm{ GeV}^{2} \).
The distribution of the events calculated using our model is shown on figure
\ref{fig:disq2x}.

\begin{figure}[htb]
{\par\centering \resizebox*{0.9\columnwidth}{!}{\includegraphics{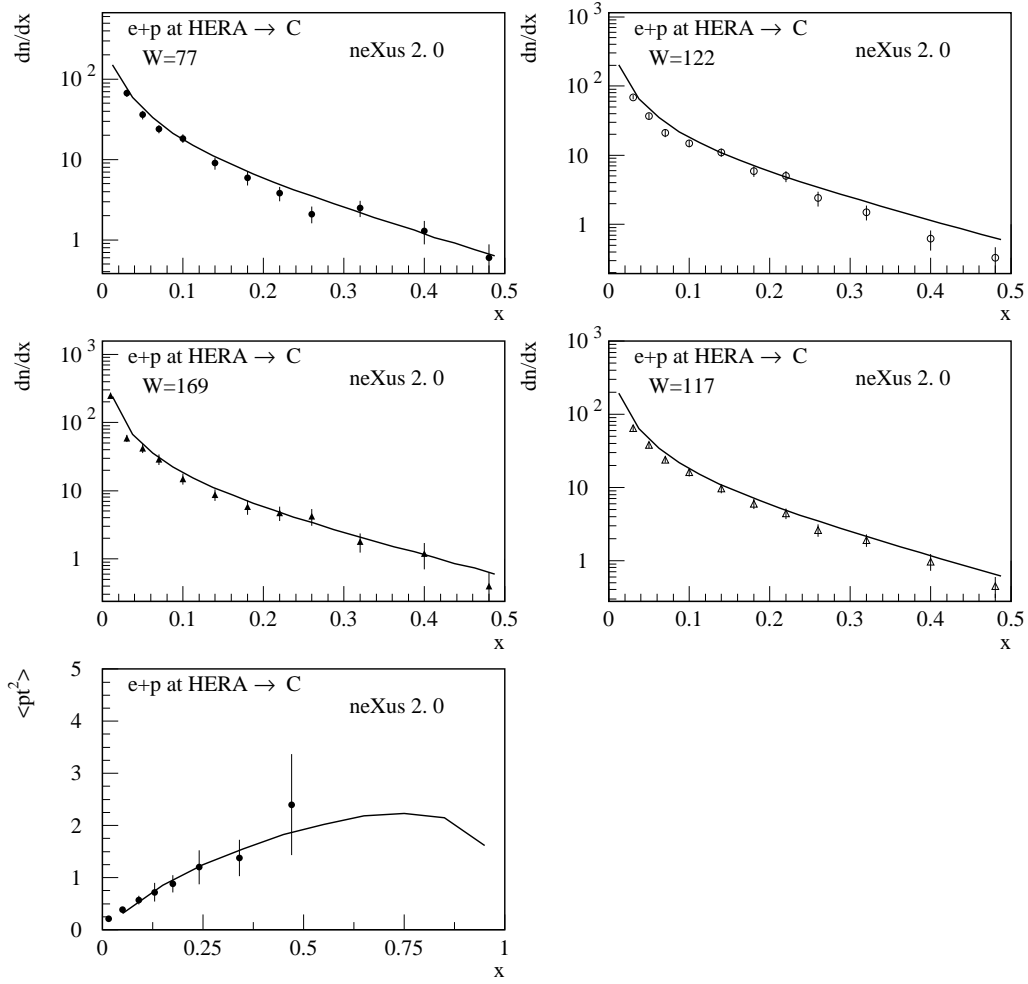}} \par}

\caption{\label{fig:hera-1} \protect\( x_{F}\protect \) distribution of charged particles
for different values of \protect\( W\protect \): 50-100 GeV, 100-150 GeV, 150-200
GeV and for the total \protect\( W\protect \) region of 50-200 GeV. The last
diagram shows mean \protect\( p_{\perp }^{2}\protect \) as a function of \protect\( x_{F}\protect \).
All the variables are defined in the hadronic center of mass system. The data
are from H1 collaboration \cite{h1-94}.}
\end{figure}
\begin{table}[htb]
\vspace{0.3cm}
{\centering \begin{tabular}{|c|c|c|c|c|c|}
\hline 
bin &
\( x/10^{-3} \)&
\( Q^{2}/\textrm{GeV}^{2} \)&
\( <x>/10^{-3} \)&
\( <Q^{2}>/\textrm{GeV}^{2} \)&
\( <W^{2}>/\textrm{GeV}^{2} \)\\
\hline 
\hline 
0&
0.1-10&
5-50&
1.14&
18.3&
24975\\
\hline 
\hline 
1&
0.1-0.2&
5-10&
0.16&
7.7&
45296\\
\hline 
2&
0.2-0.5&
6-10&
0.29&
8.8&
31686\\
\hline 
\hline 
3&
0.2-0.5&
10-20&
0.37&
13.1&
36893\\
\hline 
4&
0.5-0.8&
10-20&
0.64&
14.0&
22401\\
\hline 
5&
0.8-1.5&
10-20&
1.1~~&
14.3&
13498\\
\hline 
6&
1.5-4.0&
10-20&
2.1~~&
15.3&
7543\\
\hline 
\hline 
7&
0.5-1.4&
20-50&
0.93&
28.6&
32390\\
\hline 
8&
1.4-3.0&
20-50&
2.1&
31.6&
16025\\
\hline 
9&
3.0-10&
20-50&
4.4&
34.7&
8225\\
\hline 
\end{tabular}\par}\vspace{0.3cm}

\caption{\label{tab:herabins}Bins in \protect\( x\protect \) and \protect\( Q^{2}\protect \)
for the figures \ref{fig:hera-1}-\ref{fig:hera-6}.}
\end{table}
 
\begin{figure}[htb]
{\par\centering \resizebox*{0.9\columnwidth}{!}{\includegraphics{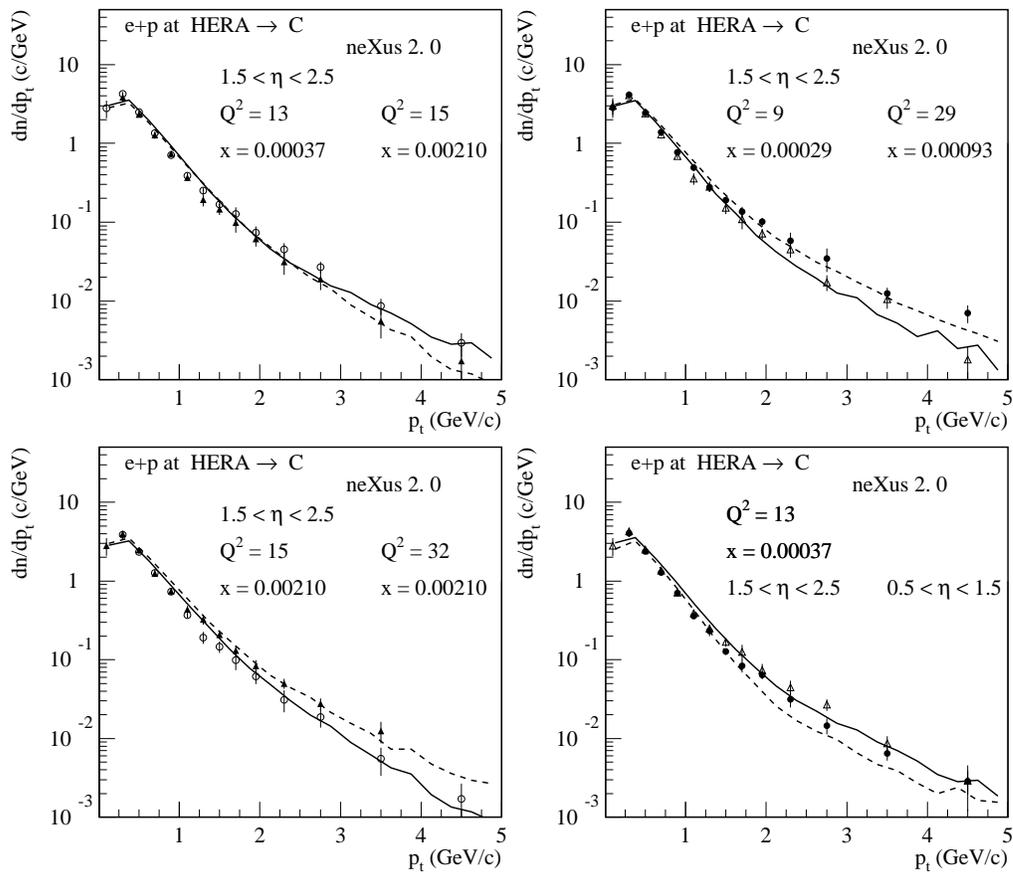}} \par}

\caption{\label{fig:hera-2} \protect\( p_{\perp }\protect \) distribution of charged
particles for different kinematic regions. The values of the cuts are indicated
on the figures (column 1 - full line, column 2 - dashed line ).}
\end{figure}
\begin{figure}[htb]
{\par\centering \resizebox*{!}{0.85\textheight}{\includegraphics{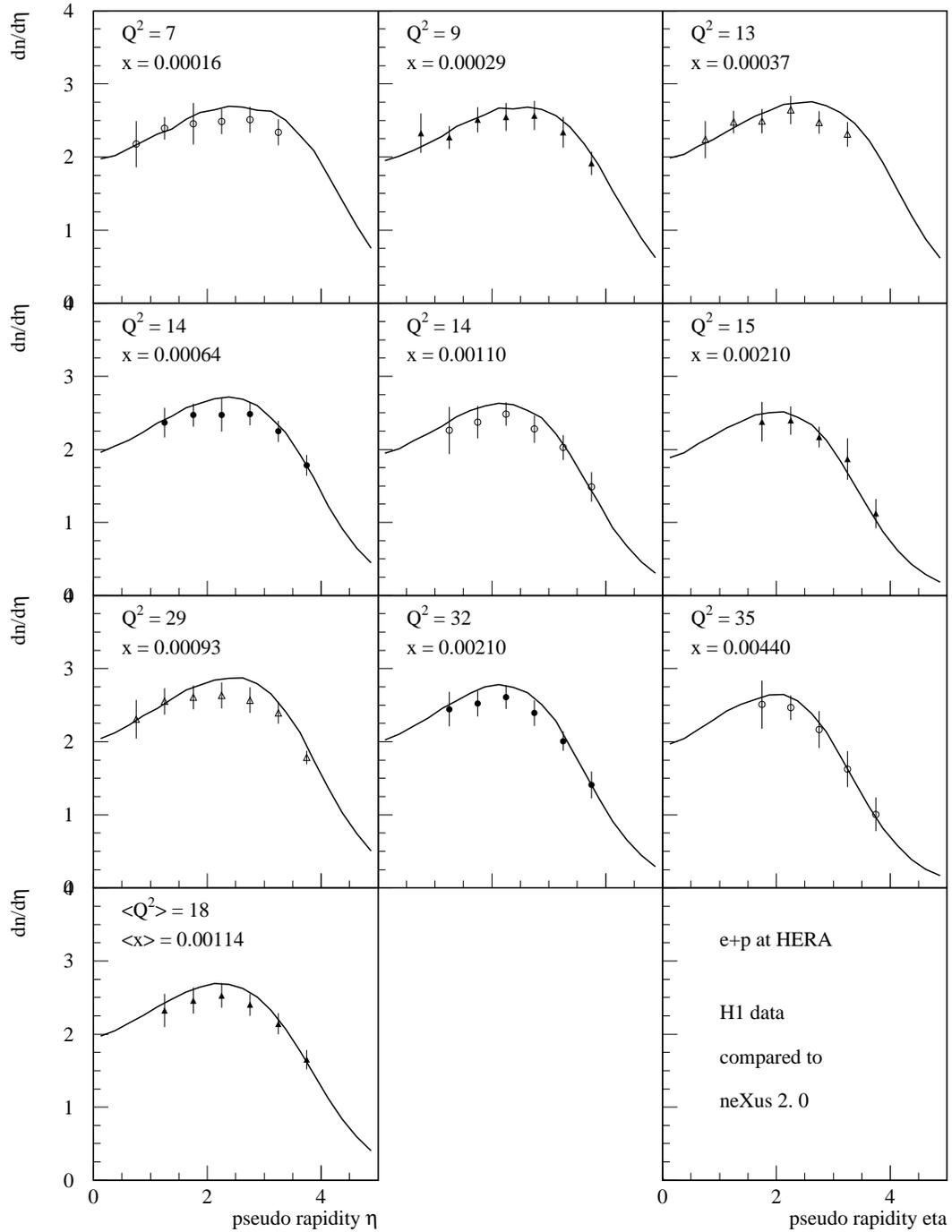}} \par}

\caption{\label{fig:hera-3}Pseudo rapidity distribution of charged particles for the
bins indicated in table \ref{tab:herabins}. The lower left diagram represents
no cut results. }
\end{figure}
\begin{figure}[htb]
{\par\centering \resizebox*{!}{0.85\textheight}{\includegraphics{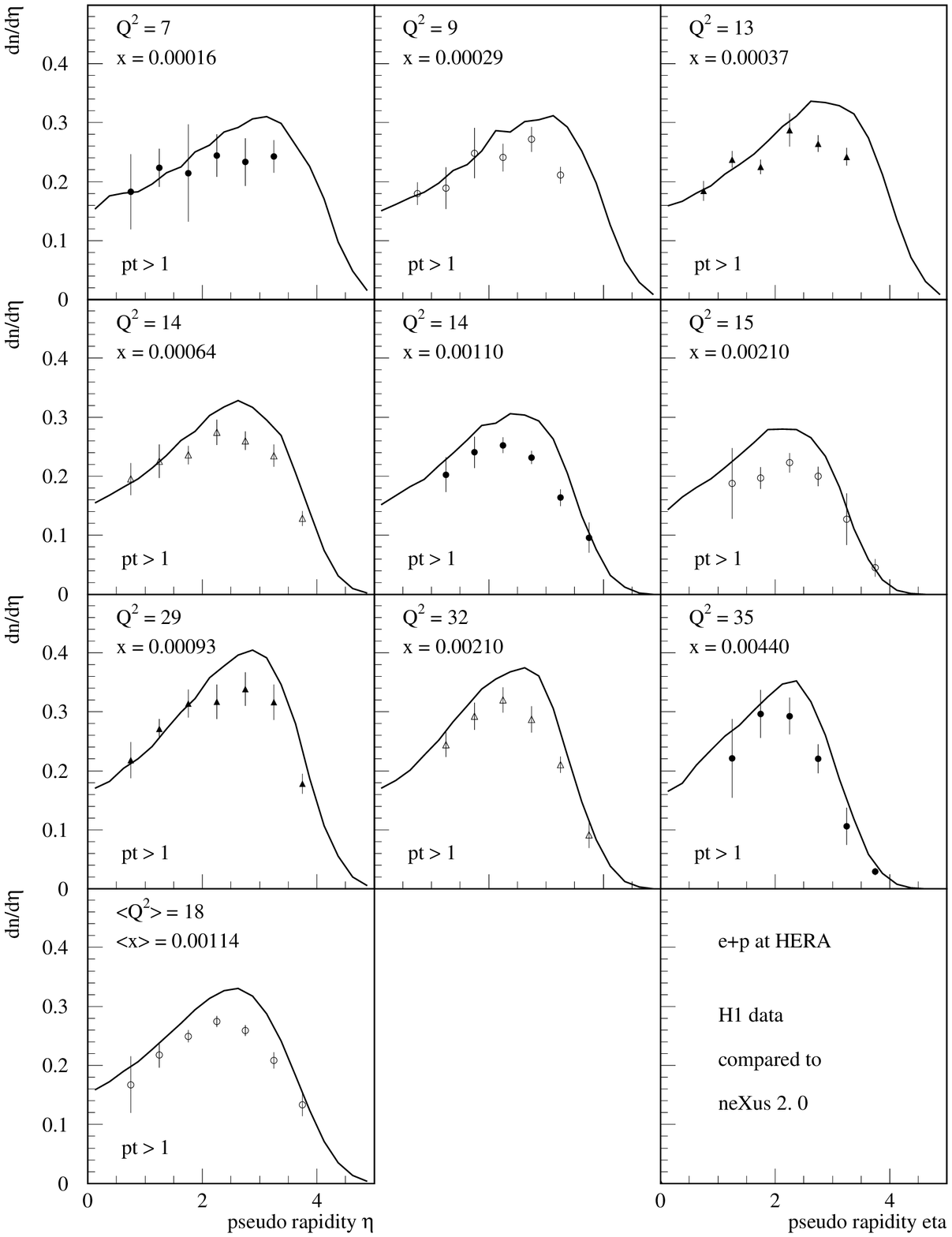}} \par}

\caption{\label{fig:hera-4}The same as on figure \ref{fig:hera-3}, but with an additional
cut \protect\( p_{\perp }>1\textrm{ GeV}\protect \).}
\end{figure}
\begin{figure}[htb]
{\par\centering \resizebox*{!}{0.85\textheight}{\includegraphics{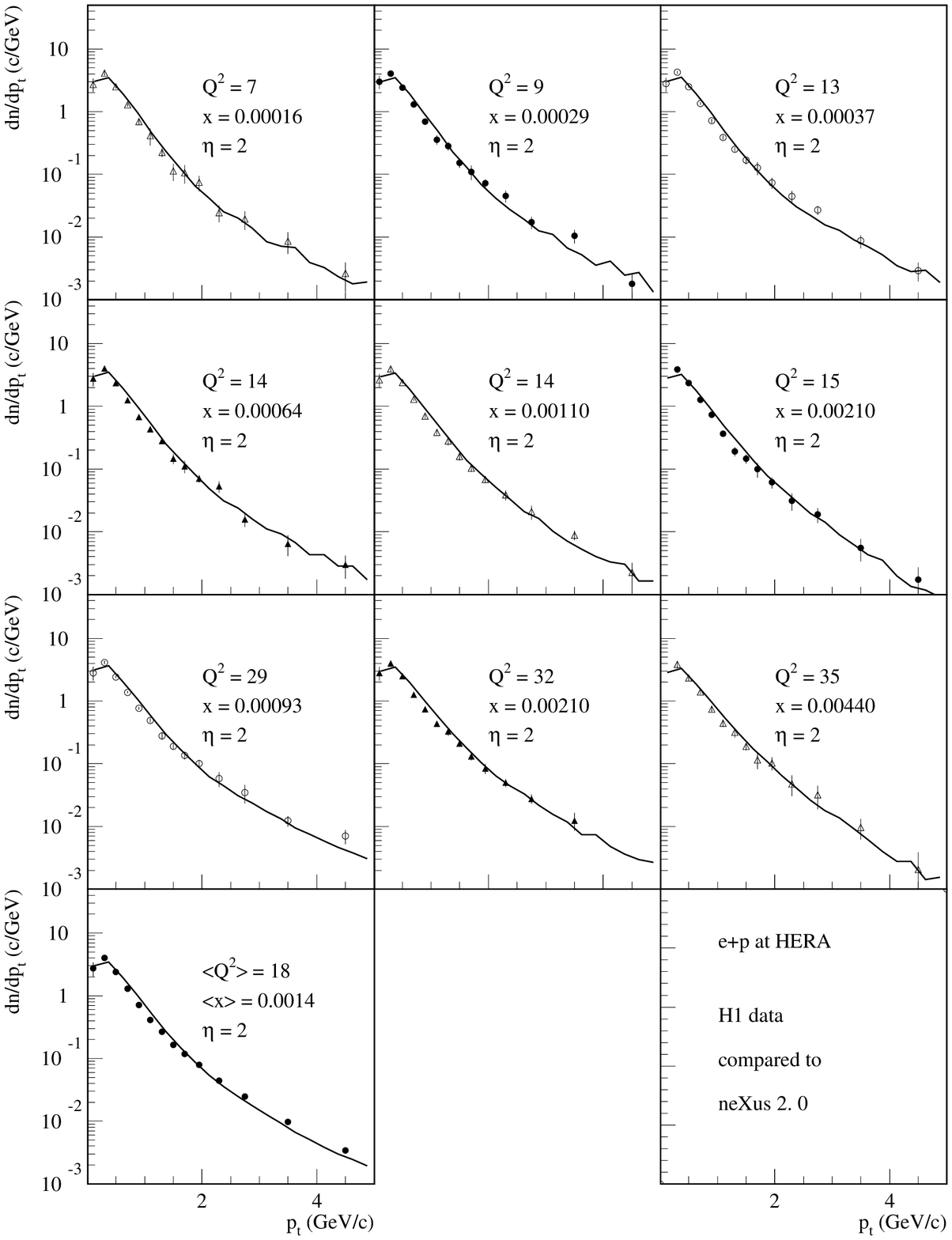}} \par}

\caption{\label{fig:hera-5}Transverse momentum of charged particles for \protect\( 1.5<\eta <2.5\protect \).}
\end{figure}
\begin{figure}[htb]
{\par\centering \resizebox*{!}{0.85\textheight}{\includegraphics{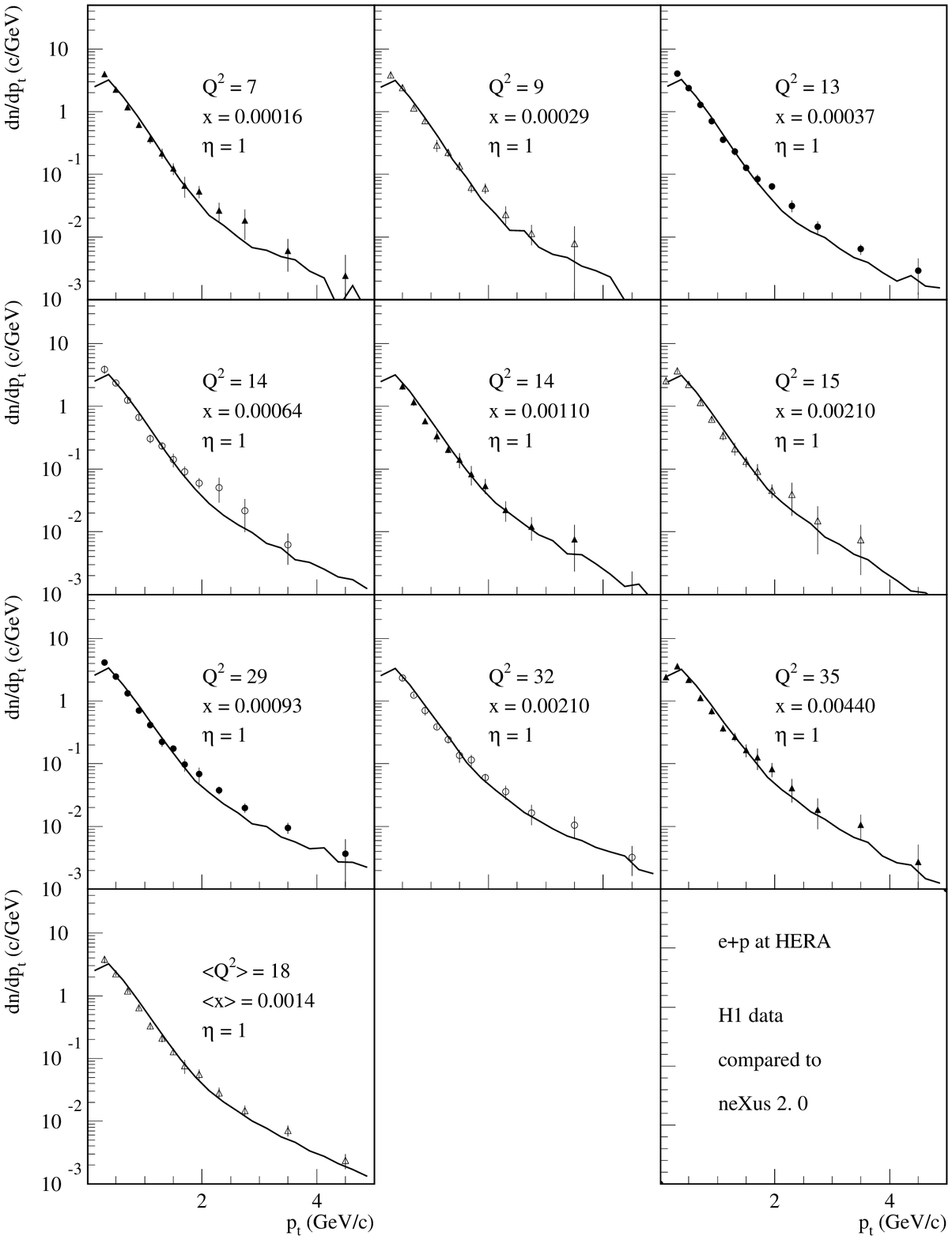}} \par}

\caption{\label{fig:hera-6}Transverse momentum of charged particles for \protect\( 0.5<\eta <1.5\protect \).}
\end{figure}
We recall the principal variables 
\begin{equation}
x\equiv x_{\mathrm{B}}=\frac{Q^{2}}{2(pq)}=\frac{Q^{2}}{ys};\, y=\frac{Q^{2}}{xs},
\end{equation}
which gives straight lines for \( y=\textrm{const}. \)\ in fig.\ \ref{fig:disq2x}.
The sharp borders are due to imposing the experimental cuts \( 0.03<y \), \( E'>12\textrm{ GeV} \)
and \( \theta _{e}>7.5 \). On figure \ref{fig:hera-1}, we plot charged particle
distributions for different values of \( W=\hat{s}=2(pq)-Q^{2} \). The particle
spectra look very similar for different \( W \) values; the distributions decrease
rapidly with \( x_{F} \). The dependence of the average \( p_{\perp }^{2} \)on
\( x_{F} \) shows an overall good agreement with the data from H1 collaboration
\cite{h1-94}.

In table \ref{tab:herabins} the bins in \( x \) and \( Q^{2} \) are given
for the experimental data points on figures \ref{fig:hera-1}-\ref{fig:hera-6}
(see \cite{h1-96b}). The bin 0 is the sum of all the others. First, we compare
the \( p_{\perp } \) distribution in the photon-proton center of mass system
for the different bins - fig.\ \ref{fig:hera-2}. Fig.\ \ref{fig:hera-2}(a)
shows the comparison of the results for low and high \( x \) values for the
values of \( Q^{2}\simeq 10-20\textrm{ GeV}^{2} \)(bins 6 and 3). We find harder
spectra for smaller \( x \), which is the consequence of the larger kinematical
space (in \( x \)) for the initial state radiation. Next, on figures \ref{fig:hera-2}(b,c),
the spectra are compared for two different values of \( Q^{2} \) and either
the energy \( W\sim Q^{2}/x \) being fixed (bins 2 and 7, fig.\ \ref{fig:hera-2}(b)),
or for a given value of \( x \) (bins 6 and 8, figure \ref{fig:hera-2}(c)).
The spectra in \( p_{\perp } \) are always harder for larger \( Q^{2} \),
which is now the consequence of the larger kinematical space in \( p^{2}_{\perp } \)
for the initial state radiation. Two cuts in pseudo-rapidity \( \eta  \) for
bin 3 are considered on figure \ref{fig:hera-2}(d). We see a harder distribution
for \( 1.5<\eta <2.5 \). Around mid-rapidity, where \( \eta  \) is maximal,
one finds higher transverse momenta as this region is dominated by the contribution
of the largest virtuality photon process. 

Let us now consider pseudo-rapidity distributions of charged particles. Figures
\ref{fig:hera-3}, \ref{fig:hera-4} show the \( \eta  \)-distributions for
the 9 bins of table \ref{tab:herabins} for different values of \( Q^{2} \)
and \( x \). On figure \ref{fig:hera-4}, a cut for \( p_{\perp }>1\textrm{ GeV} \)
has been made to extract the contribution of hard processes. The latter one
results in approximately 10\% of the total hadron multiplicity. At fig.\ \ref{fig:hera-4}
we find fewer particles at small \( \eta  \) which is the consequence of smaller
kinematical space (in \( x \)) for the initial state radiation and the reduced
influence of the largest virtuality photon process. 

The transverse momenta for all particles are generally well described by the
model - figs.\ \ref{fig:hera-5}, \ref{fig:hera-6}. The fact that we find harder
spectra for higher \( Q^{2} \) and lower \( x \) is best seen for smaller
values of \( \eta  \) - see fig.\ \ref{fig:hera-6}.\\

\cleardoublepage

\chapter{Results for Proton-Proton Scattering}

In this section we are going to discuss our results for proton-proton interactions
in the energy range between roughly 10 and 2000 GeV, which represents the range
of validity of our approach. The lower limit is a fundamental limitation due
to the fact that our approach requires hadron production to start after the
primary interactions are finished, which is no longer fulfilled at low energies.
The upper limit is due to the fact that above 2000 GeV higher order screening
corrections need to be taken into account.

\section{Energy dependence}

\begin{figure}[htb]
{\par\centering \resizebox*{!}{0.25\textheight}{\includegraphics{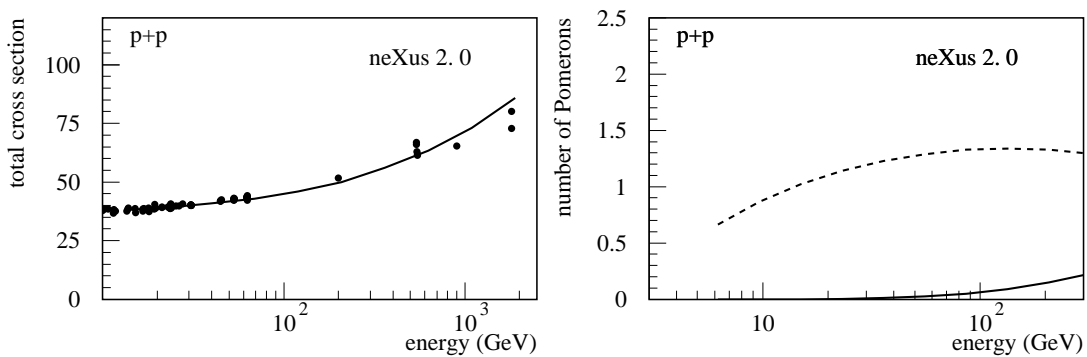}} \par}

\caption{The total cross section as a function of the energy \protect\( \sqrt{s}\protect \)
(left figure): the full line is the simulation, the points represent data. Pomeron
numbers as a function of the energy \protect\( \sqrt{s}\protect \) (right):
soft (dashed) and semi-hard (solid line) Pomerons. \label{pp-1}}
\end{figure}

We first consider the energy dependence of some elementary  quantities in \( pp \)
scattering in the above mentioned energy range. In fig.\ \ref{pp-1}, the results
for the total cross section are shown.
The cross section is essentially used to fit the soft Pomeron parameters. Also
shown in the figure is the energy dependence of the number of soft and semi-hard
Pomerons. Over the whole energy range shown in the figure, soft physics is dominating.
So for example at RHIC, soft physics dominates by far. 

\begin{figure}[htb]
{\par\centering \resizebox*{!}{0.45\textheight}{\includegraphics{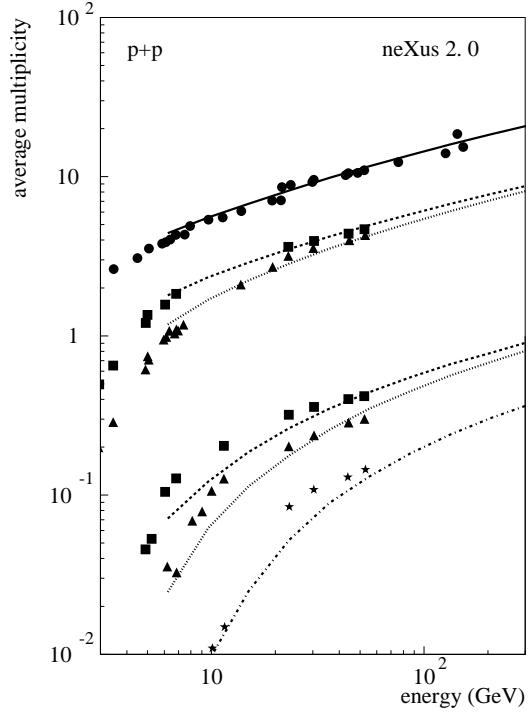}} \par}

\caption{The average multiplicities of different hadron species, as a function of the
energy \protect\( \sqrt{s}\protect \). From top to bottom: all charged particles,
\protect\( \pi ^{+}\protect \), \protect\( \pi ^{-}\protect \), K\protect\( ^{+},\protect \)
K\protect\( ^{-},\protect \) \protect\( \bar{p}\protect \). The full lines
are simulations, the points represent data (from \cite{gia79}).\label{pp-4b}}
\end{figure}

\begin{figure}[htb]
{\par\centering \resizebox*{!}{0.25\textheight}{\includegraphics{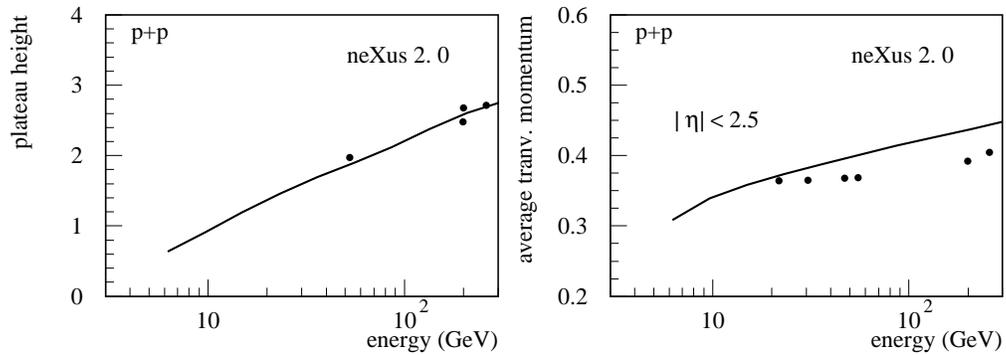}} \par}

\caption{Pseudo-rapidity plateau \protect\( dn/d\eta (0)\protect \) and mean squared
transverse momentum \protect\( <p_{t}^{2}>\protect \) as a function of the
energy \protect\( \sqrt{s}\protect \). The full lines are simulations, the
points represent data.\label{pp-3}}
\end{figure}

Our results for hadron production are based on the Pomeron parameters, defined
from the cross sections fitting, and on the fragmentation procedure, adjusted
on the basis of \( e^{+}e^{-} \)data. In fig.\ \ref{pp-4b}, average multiplicities
of different hadron species are given as a function of the energy. In fig.\
 \ref{pp-3}, we show the energy dependence of the pseudo-rapidity plateau \( dn_{C}/d\eta (0) \)
and of the mean squared transverse momentum \( <p_{t}^{2}> \).

\section{Charged Particle and Pion Spectra}

\begin{figure}[htb]
{\par\centering \resizebox*{!}{0.25\textheight}{\includegraphics{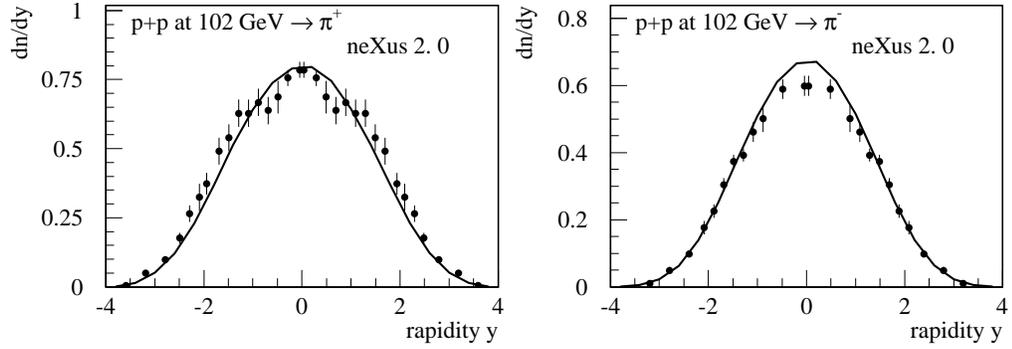}} \par}

\caption{Rapidity distributions of pions at 100 GeV. The full lines are simulations,
the points represent data. \label{pp-4}}
\end{figure}

\begin{figure}[htb]
{\par\centering \resizebox*{!}{0.45\textheight}{\includegraphics{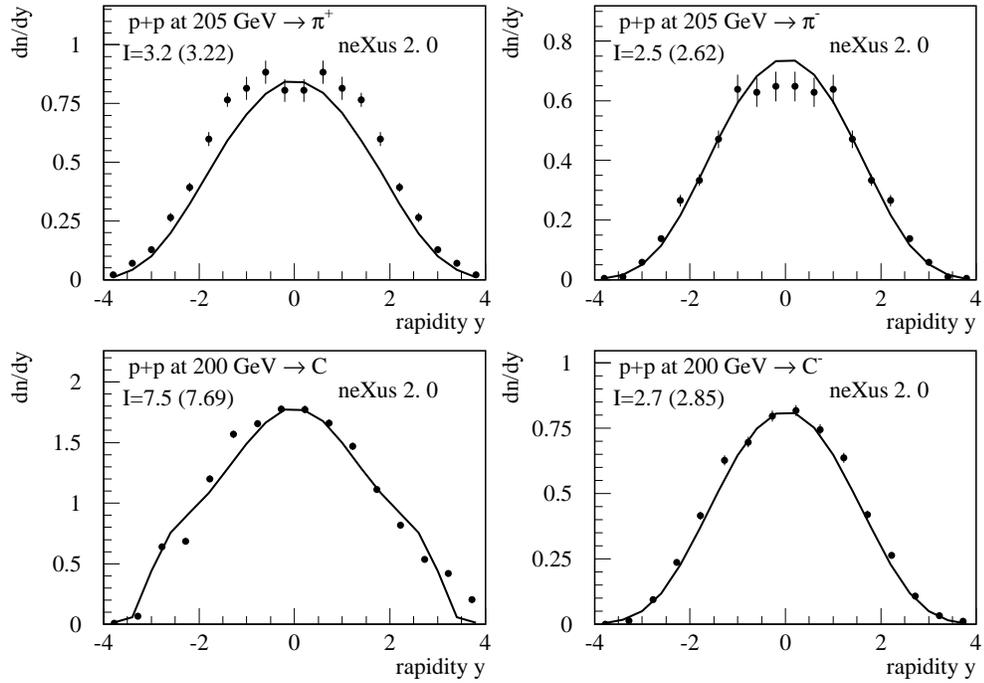}} \par}

\caption{Pseudo-rapidity distributions of pions (\protect\( \pi ^{+},\pi ^{-}\protect \))
and charged particles (all charged and negatively charged) at 200 GeV. The full
lines are simulations, the points represent data.\label{pp-5}}
\end{figure}

\begin{figure}[htb]
{\par\centering \resizebox*{!}{0.45\textheight}{\includegraphics{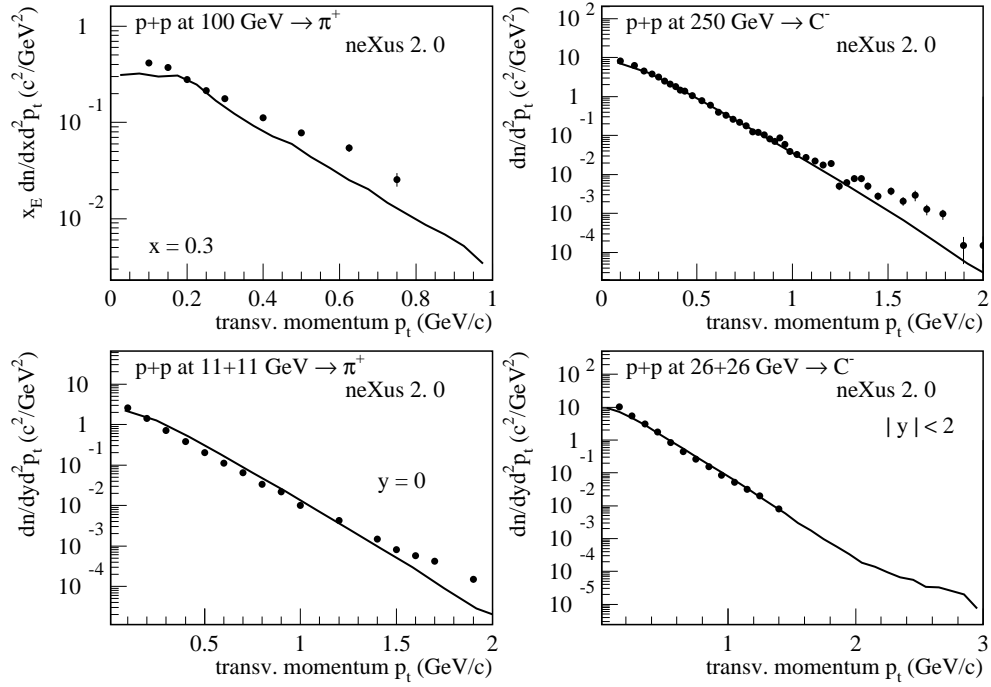}} \par}

\caption{Transverse momentum distributions of pions or negatively charged particles
at different energies. The full lines are simulations, the points represent
data.\label{pp-8}}
\end{figure}

\begin{figure}[htb]
{\par\centering \resizebox*{!}{0.3\textheight}{\includegraphics{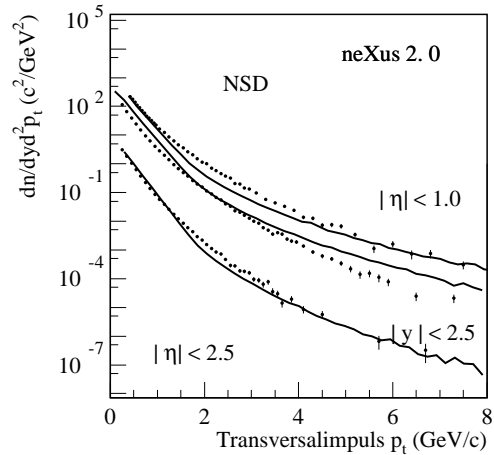}} \par}

\caption{Transverse momentum distributions of charged particles at (from bottom to top)
200, 900, and 1800 GeV (cms). The full lines are simulations, the points represent
data.\label{pp-9}}
\end{figure}

In fig.\ \ref{pp-4}, we present rapidity distributions of pions at 100 GeV,
in fig.\ \ref{pp-5} rapidity distributions of pions and charged particles at
200 GeV. The values following the Symbol ``I='' represent the integrals, i.e.
the average multiplicity. The first number is the simulation, the second number
(in brackets) represents data.
\newpage
\begin{figure}[htb]
{\par\centering \resizebox*{!}{0.25\textheight}{\includegraphics{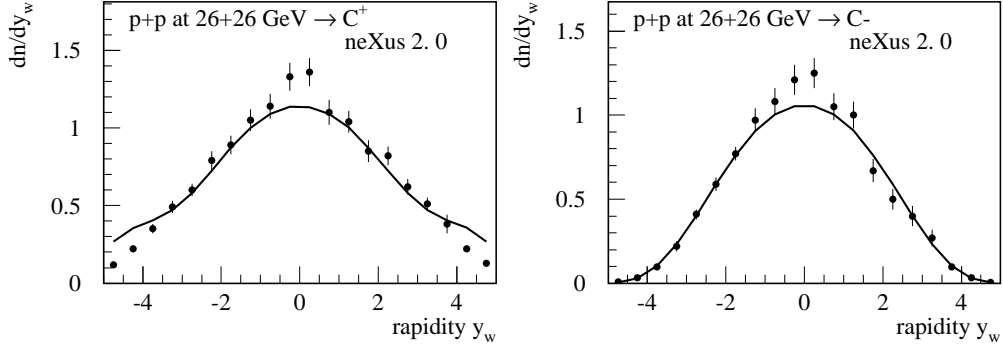}} \par}

\caption{Pseudo-rapidity distributions of positively and negatively charged particles
at 53 GeV (cms). The full lines are simulations, the points represent data.\label{pp-6}}
\end{figure}
\begin{figure}[htb]
{\par\centering \resizebox*{!}{0.25\textheight}{\includegraphics{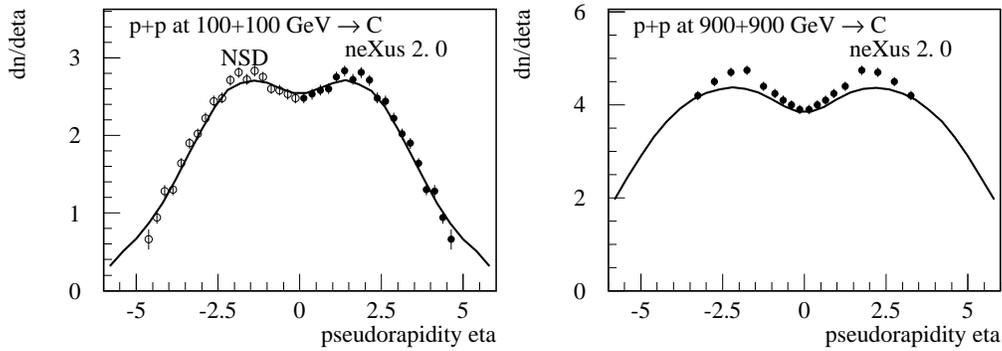}} \par}

\caption{Pseudo-rapidity distributions of charged particles at 200 and 1800 GeV (cms).
The full lines are simulations, the points represent data.\label{pp-7}}
\end{figure}
In fig.\ \ref{pp-6}, we show rapidity distributions for positively and negatively
charged particles at 53 GeV (cms), where we adopted also for the simulations
the experimental definition of the rapidity by always taking the pion mass.
In fig.\ \ref{pp-7}, we show pseudo-rapidity distributions of charged particles
at 200 and 1800 GeV (cms). In figs. \ref{pp-8} and \ref{pp-9}, we finally
show transverse momentum spectra at different energies between 100 GeV (lab)
and 1800 GeV (cms).

\clearpage

\section{Proton spectra}

\begin{figure}[htb]
{\par\centering \resizebox*{!}{0.45\textheight}{\includegraphics{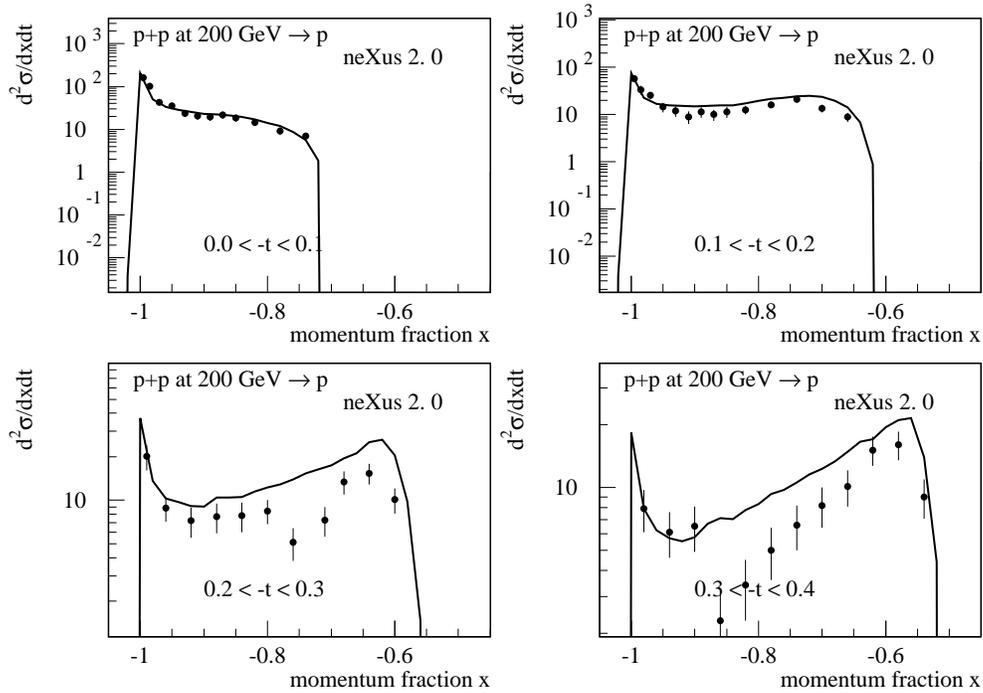}} \par}

\caption{Longitudinal momentum fraction distributions of protons for different values
of \protect\( t\protect \) at 200 GeV. The full lines are simulations, the
points represent data.\label{pp-10}}
\end{figure}

\begin{figure}[htb]
{\par\centering \resizebox*{!}{0.25\textheight}{\includegraphics{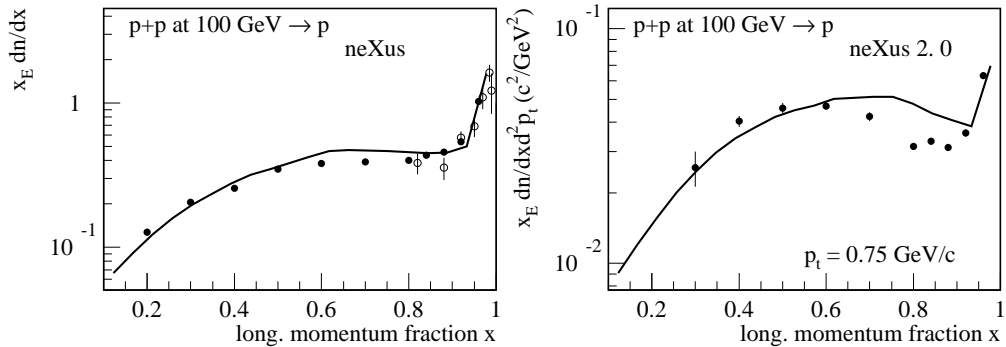}} \par}

\caption{Longitudinal momentum fraction distribution of protons at 100 GeV, integrated
over \protect\( p_{t}\protect \) (left) and for \protect\( p_{t}=0.75\protect \)
GeV/c (right). The full lines are simulations, the points represent data.\label{pp-11}}
\end{figure}

\begin{figure}[htb]
{\par\centering \resizebox*{!}{0.24\textheight}{\includegraphics{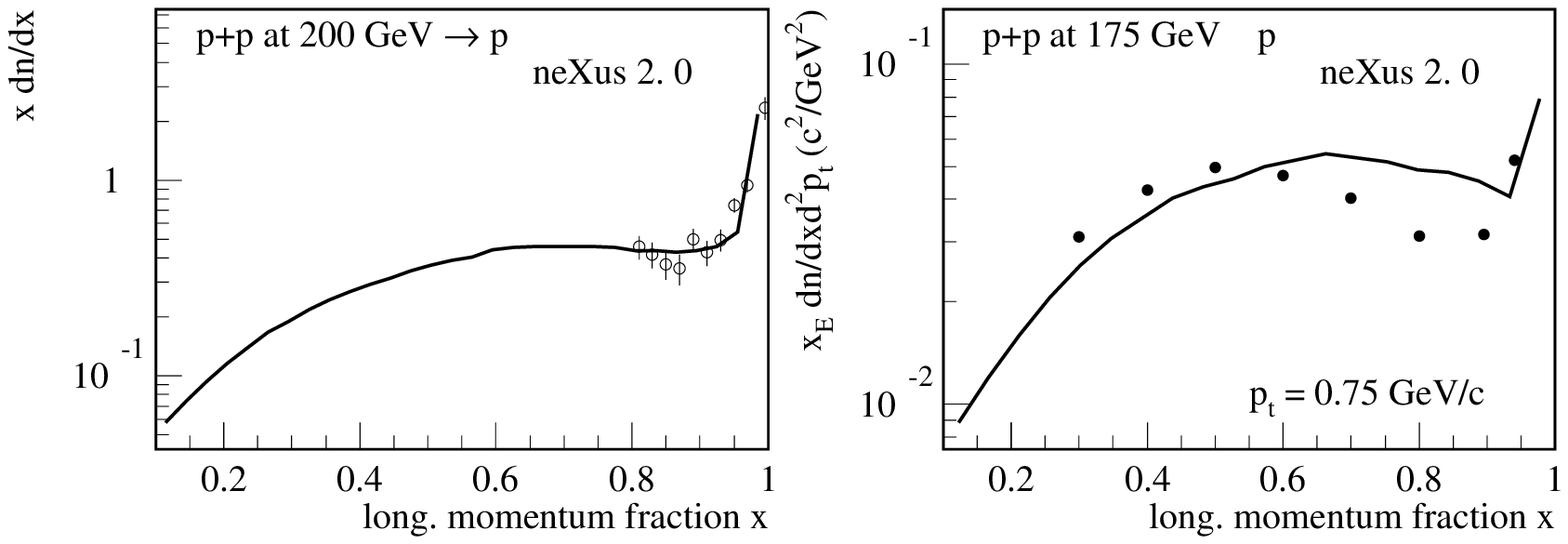}} \par}

\caption{Longitudinal momentum fraction distributions of protons at 200 GeV, integrated
over \protect\( p_{t}\protect \) (left) and at 175 GeV, for \protect\( p_{t}=0.75\protect \)
GeV/c (right). The full lines are simulations, the points represent data.\label{pp-12}}
\end{figure}

\begin{figure}[htb]
{\par\centering \resizebox*{!}{0.24\textheight}{\includegraphics{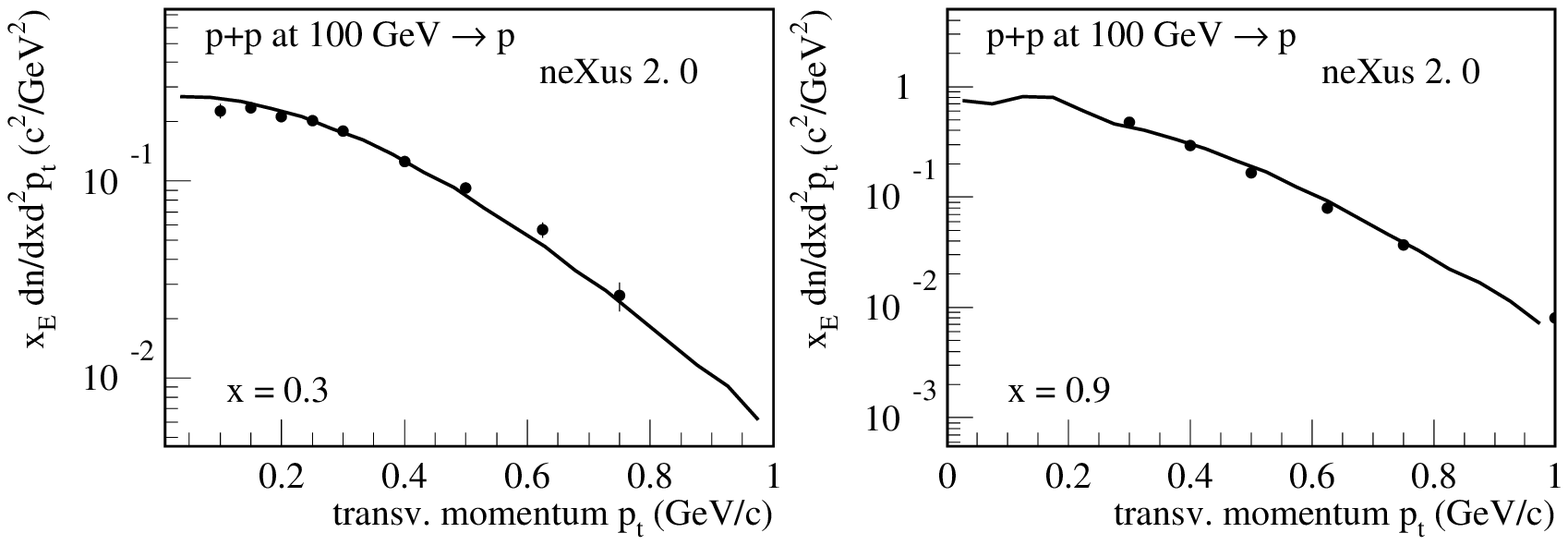}} \par}

\par\vspace{-0.8cm}\par

{\par\centering \resizebox*{!}{0.43\textheight}{\includegraphics{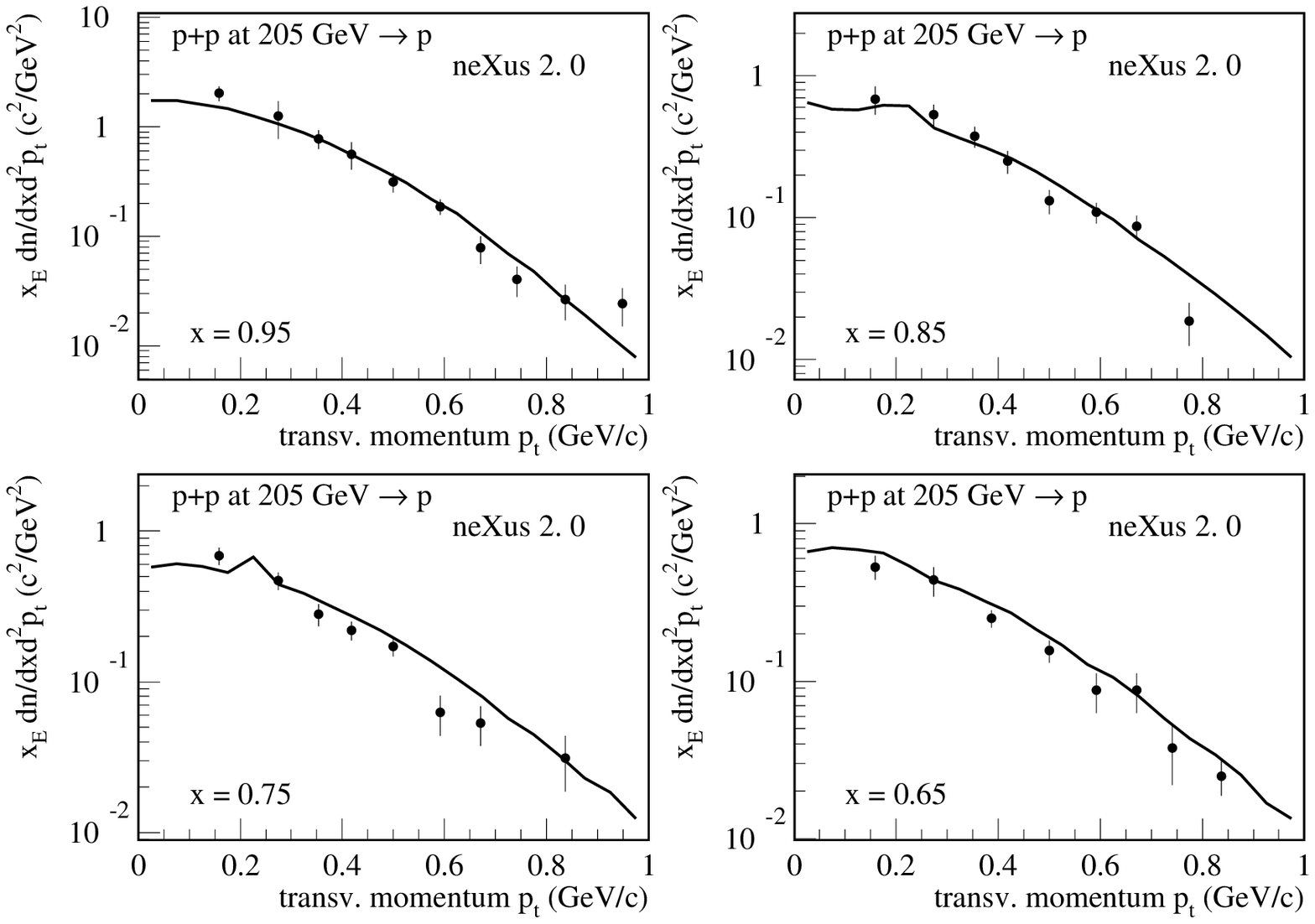}} \par}

\caption{Transverse momentum spectra of protons for different values of the longitudinal
momentum fraction \protect\( x\protect \) at 100 and 205 GeV. The full lines
are simulations, the points represent data.\label{pp-13}}
\end{figure}

In fig.\ \ref{pp-10}, we plot longitudinal momentum fraction distributions
of protons for different values of \( t \) at 200 GeV, in figs. \ref{pp-11}
and \ref{pp-12} as well longitudinal momentum fraction distributions at 100-200
GeV, for given values of \( p_{t} \) or integrated over \( p_{t} \). In fig.\ \ref{pp-13},
we show transverse momentum spectra of protons for different values of the longitudinal
momentum fraction \( x \) at 100 and 205 GeV.

\clearpage

\section{Strange particle spectra}

\begin{figure}[htb]
{\par\centering \resizebox*{!}{0.45\textheight}{\includegraphics{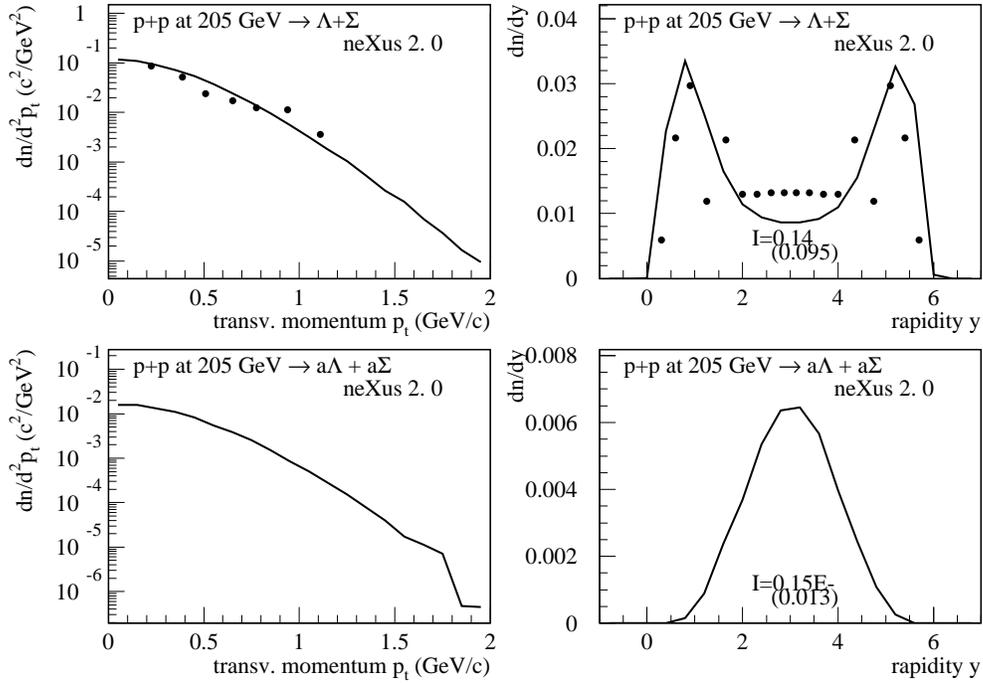}} \par}

\caption{Distributions of transverse momentum (left) and rapidity (right) of lambdas
plus neutral sigmas (upper) and of anti-lambdas plus neutral anti-sigmas (lower)
at 205 GeV. The full lines are simulations, the points represent data.\label{pp-14}}
\end{figure}

\begin{figure}[htb]
{\par\centering \resizebox*{!}{0.25\textheight}{\includegraphics{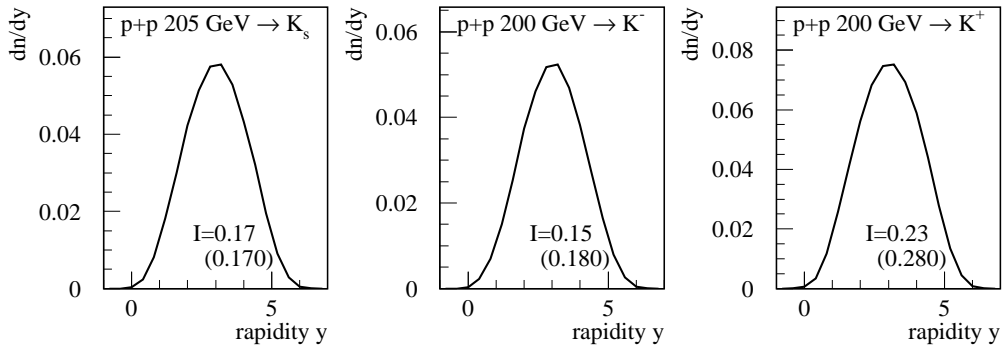}} \par}

\caption{Rapidity distributions of kaons (\protect\( K_{s}\protect \), \protect\( K^{-}\protect \),
\protect\( K^{+}\protect \)) at 205 GeV. The numbers represent \protect\( y\protect \)-integrated
results: the first number is the simulated one, the number in brackets the experimental
one.\label{pp-15}}
\end{figure}

In figs. \ref{pp-14} and \ref{pp-15}, we show transverse momentum and rapidity
spectra of lambdas (including \( \Sigma _{0} \)), anti-lambdas (including \( \bar{\Sigma }_{0} \)),
and kaons. The numbers represent the integrals, i.e. the average multiplicity.
The first number is the simulation, the second number (in brackets) represents
data.

\cleardoublepage

\chapter{Results for Collisions Involving Nuclei}

It is well known that secondary interactions play an important role in collisions
involving nuclei. Nevertheless, in this report, we do no want to consider any
rescattering procedure, we just present bare \noun{neXus} simulations. This
seems to us the most honest way to present results.

\section{Proton-Nucleus Scattering}

\begin{figure}[htb]
{\par\centering \resizebox*{!}{0.45\textheight}{\includegraphics{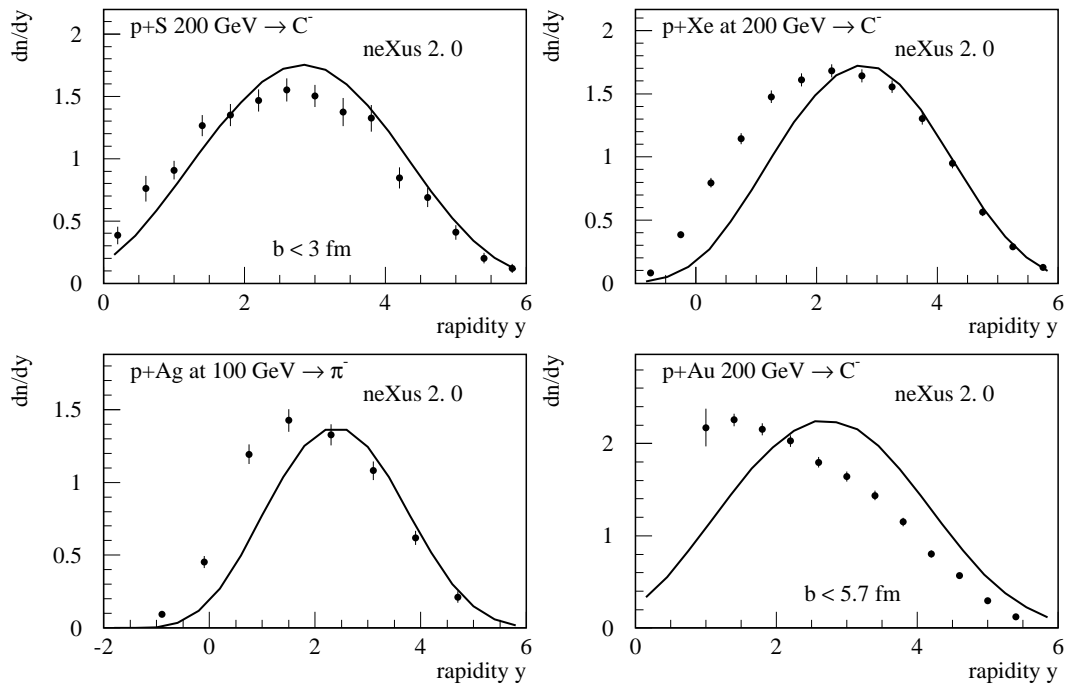}} \par}

\caption{Rapidity distributions of charged particles or negative pions for different
\protect\( p+A\protect \) collisions at 200 GeV (lab).\label{1}}
\end{figure}

\begin{figure}[htb]
{\par\centering \resizebox*{!}{0.45\textheight}{\includegraphics{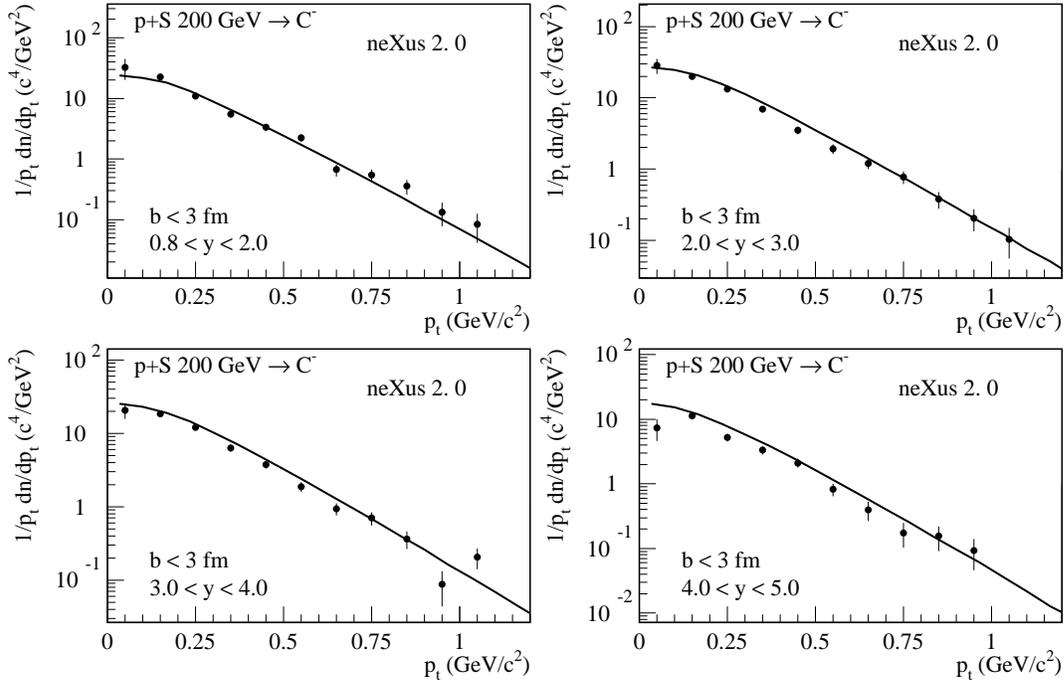}} \par}

\caption{Transverse momentum distributions of negatively charged particles for different
\protect\( p+S\protect \) collisions at 200 GeV (lab).\label{2}}
\end{figure}

\begin{figure}[htb]
{\par\centering \resizebox*{!}{0.42\textheight}{\includegraphics{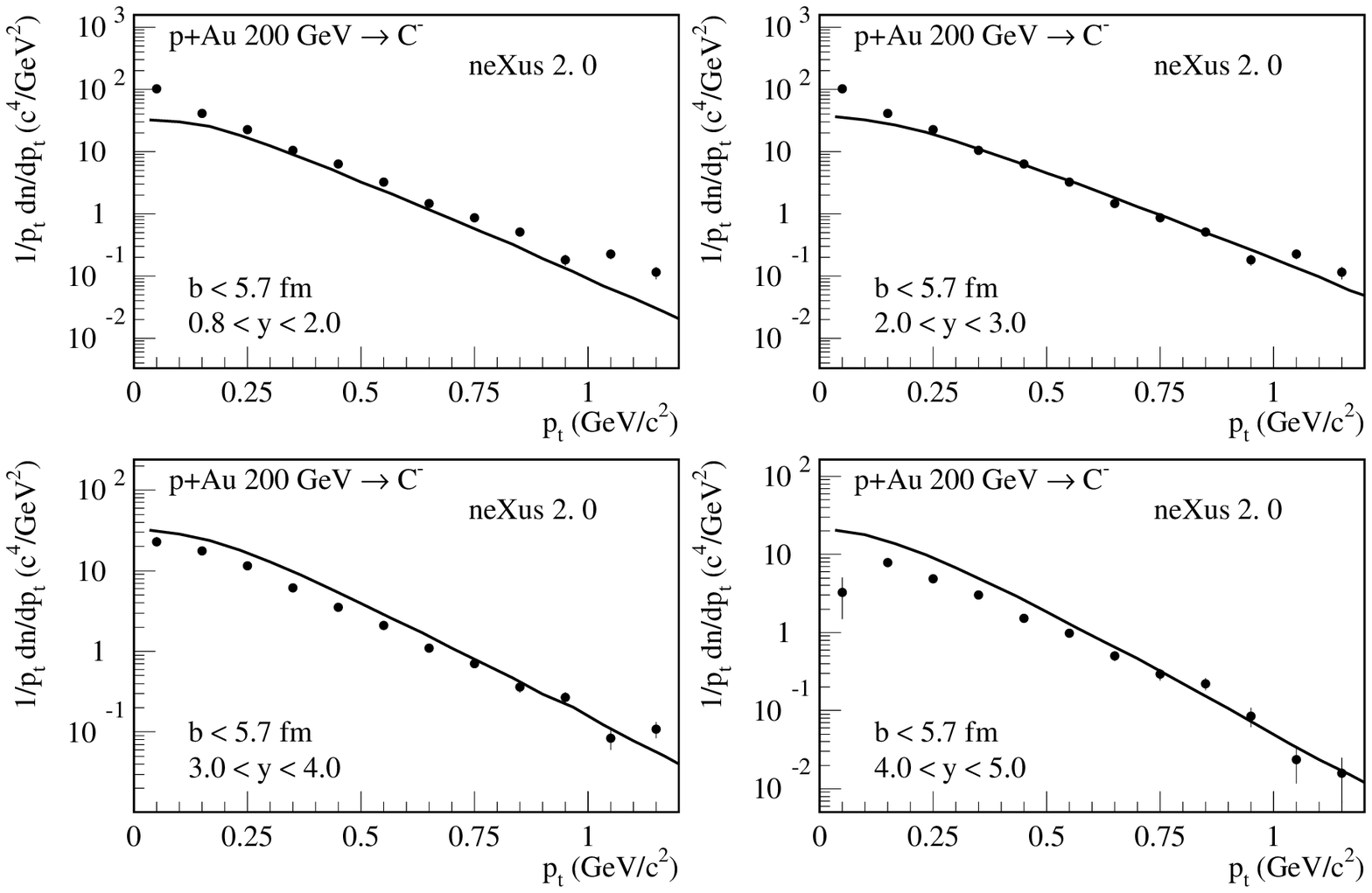}} \par}

\caption{Transverse momentum distributions of negatively charged particles for different
\protect\( p+Au\protect \) collisions at 200 GeV (lab).\label{3}}
\end{figure}

\noindent In fig.\ \ref{1}, we show rapidity spectra of negatively charged
particles for different target nuclei. Missing particles in the backward region
are certainly due to rescattering. In the forward region, the model works well,
except for \( p+ \)Au, which represents the heaviest target, but in addition
one has here a centrality trigger, in contrast to the other reactions. Here,
we expect some reduction due to nuclear screening effects. The transverse momentum
spectra are well reproduced in case of \( p+S \), as shown in fig.\ \ref{2},
whereas for \( p+Au \) one sees some deviations for small values of \( p_{t} \),
see fig.\ \ref{3}.

\begin{figure}[htb]
{\par\centering \resizebox*{!}{0.42\textheight}{\includegraphics{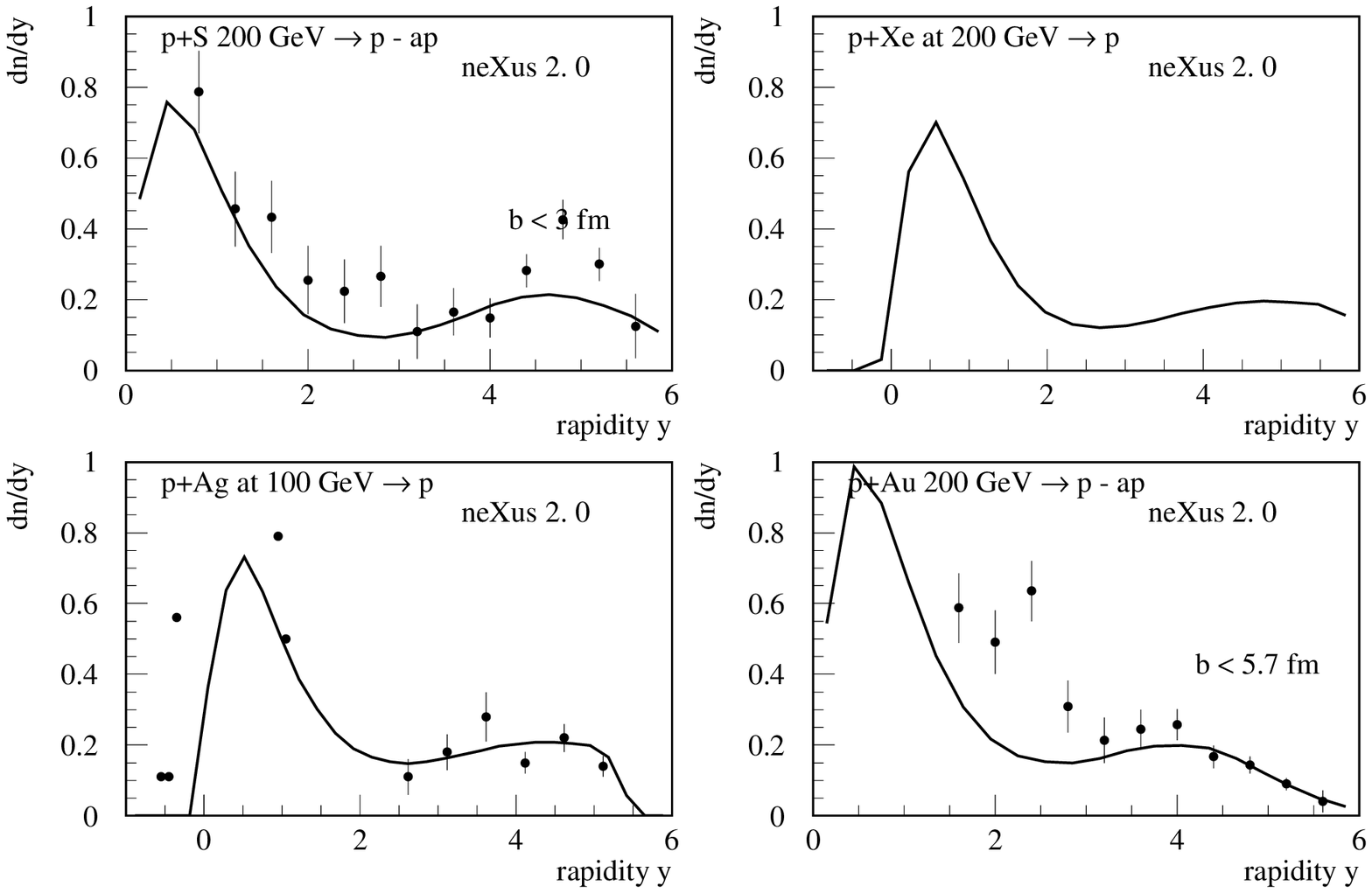}} \par}

\caption{Rapidity distributions of net protons for different \protect\( p+A\protect \)
collisions at 200 GeV (lab).\label{4}}
\end{figure}

\begin{figure}[htb]
{\par\centering \resizebox*{!}{0.21\textheight}{\includegraphics{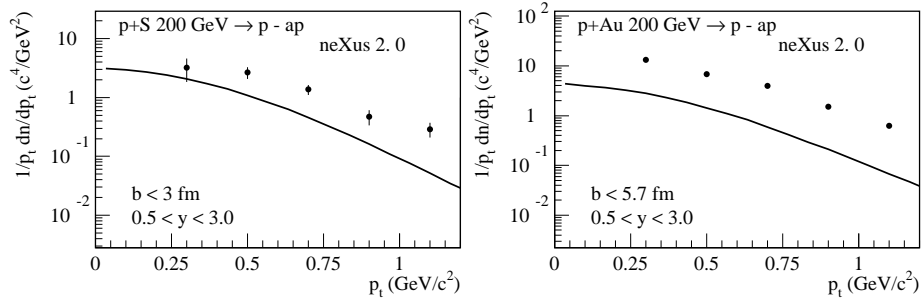}} \par}

\caption{Transverse momentum spectra of net protons for different \protect\( p+A\protect \)
collisions at 200 GeV (lab).\label{5}}
\end{figure}

Let us turn to proton spectra. In fig.\ \ref{4}, we show rapidity spectra of
net protons (protons minus anti-protons) for different target nuclei. Since
secondary interactions are not considered, we are missing the pronounced peak
around rapidity zero (not visible in the figure, since we have chosen the range
for the y-axis to be {[}0,1{]}). Apart of the target fragmentation region, the
model works well. The transverse momentum spectra shown in fig.\ \ref{5} refer
to the target fragmentation region, and therefore the absolute number is too
small, whereas the shape of the spectra is quite good.

\begin{figure}[htb]
{\par\centering \resizebox*{!}{0.21\textheight}{\includegraphics{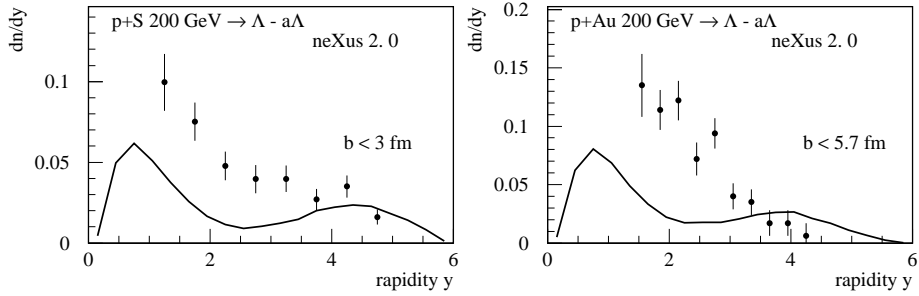}} \par}

\caption{Rapidity distributions of net lambdas for different \protect\( p+A\protect \)
collisions at 200 GeV (lab).\label{6}}
\end{figure}

\begin{figure}[htb]
{\par\centering \resizebox*{!}{0.4\textheight}{\includegraphics{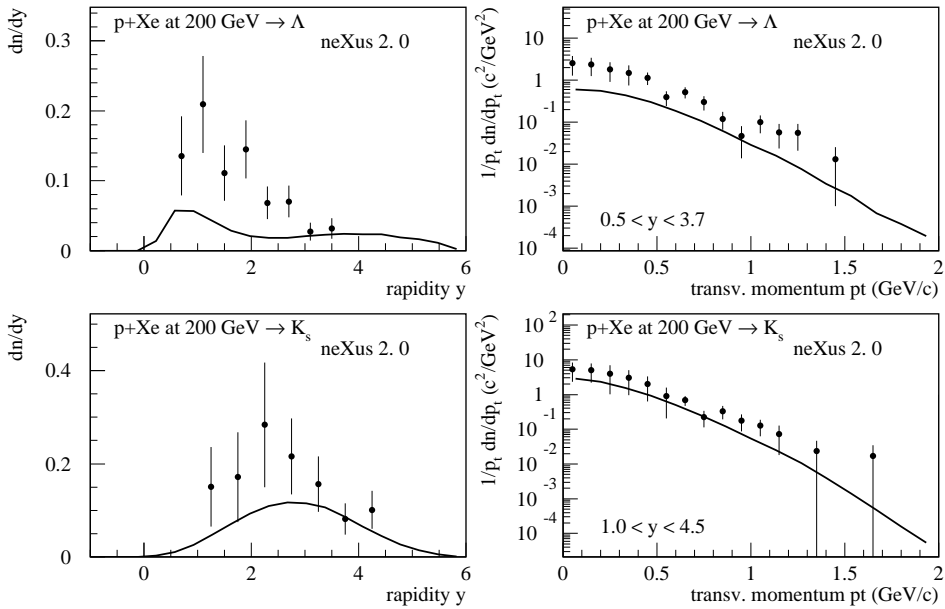}} \par}

\caption{Rapidity and \protect\( p_{t}\protect \) spectra of net lambdas and K\protect\( _{s}\protect \)
for \protect\( p+Xe\protect \) collisions at 200 GeV (lab).\label{7}}
\end{figure}

Strange particle spectra are shown in figs. \ref{6} and \ref{7}. Whereas the
simulations agree with the data for the kaons, the spectra are largely underestimated
in case of lambdas, in particular in the backward region, where rescattering
plays a dominant role.
\cleardoublepage

\section{Nucleus-Nucleus Scattering}

\begin{figure}[htb]
{\par\centering \resizebox*{!}{0.45\textheight}{\includegraphics{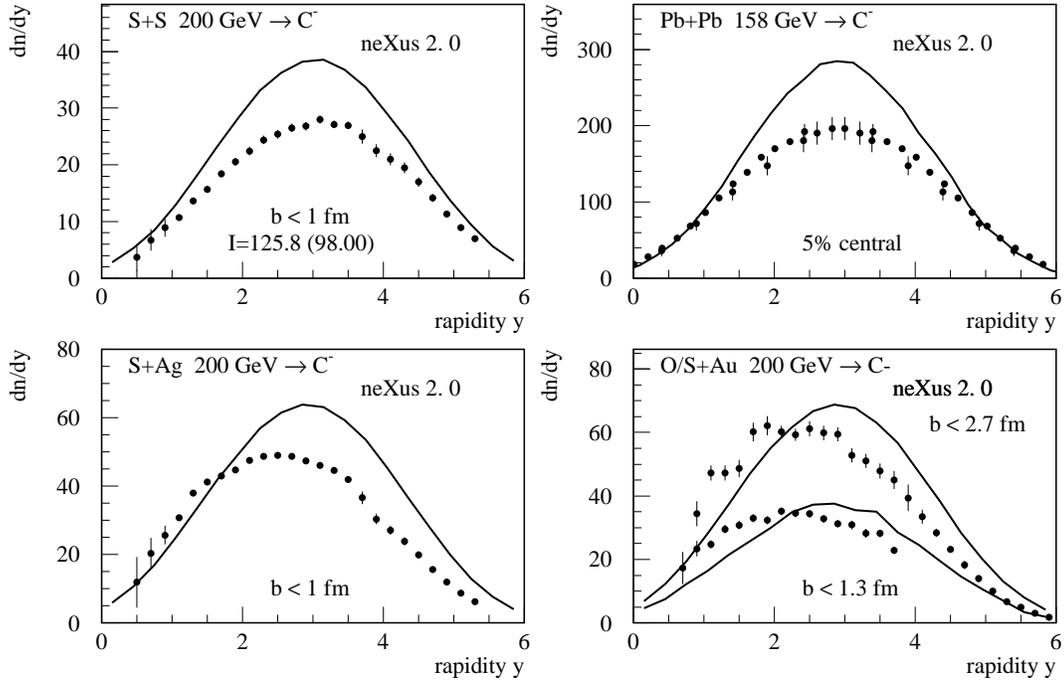}} \par}

\caption{Rapidity distributions of negatively charged particles for different \protect\( A+B\protect \)
collisions at 200 GeV (lab).\label{8}}
\end{figure}

\begin{figure}[htb]
{\par\centering \resizebox*{!}{0.45\textheight}{\includegraphics{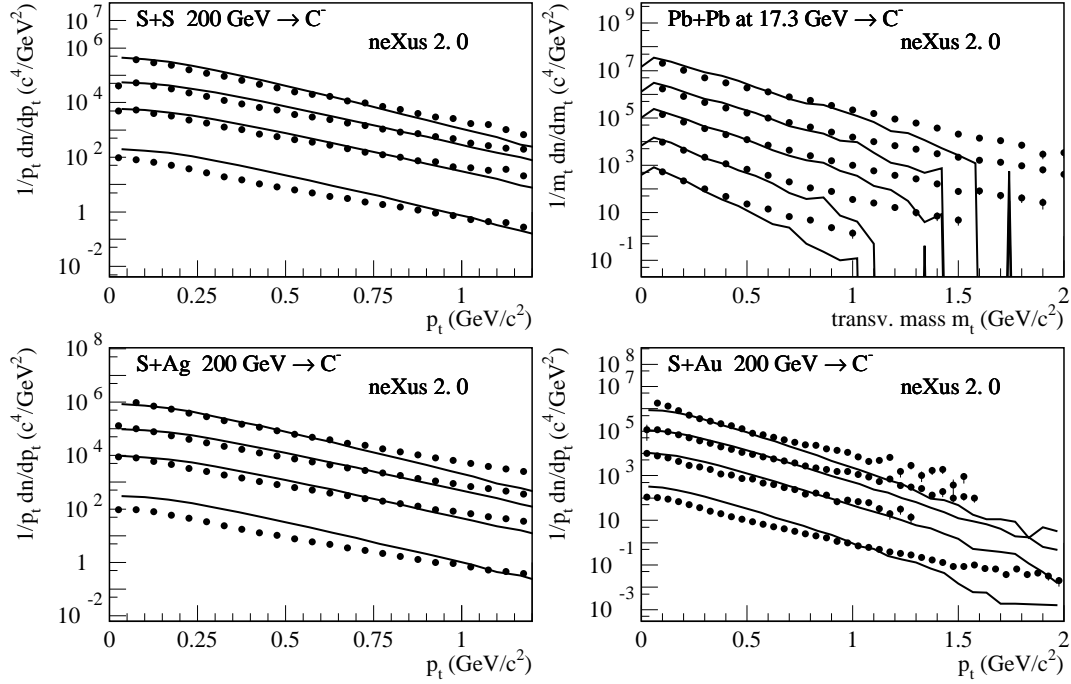}} \par}

\caption{Transverse momentum or transverse mass spectra of negatively charged particles
in several rapidity windows (see text) for different \protect\( A+B\protect \)
collisions at 200 GeV (lab). \label{9}}
\end{figure}

Again, we show results of the bare \textsc{neXus} model, without any secondary
interactions. Considering rapidity distributions of negatively charged particles,
as shown in fig.\ \ref{8}, we observe a strong excess at central rapidities
compared to the data. Rescattering will partly cure this, but not completely.
For asymmetric systems like for example S+Ag, we observe in addition a missing
asymmetry in the shape of the rapidity spectrum, in other words, the simulated
spectrum is too symmetric. This is not surprising, since in our approach AGK
cancelations apply, which make \( A+B \) spectra identical to the \( p+p \)
ones, up to a factor. Rescattering will not cure this, since it acts essentially
at central rapidities. But we expect another important effect effect due to
additional screening effects coming from contributions of ehnanced Pomeron diagrams.
In fig.\ \ref{9}, we show transverse momentum or transverse mass spectra of
negatively charged particles for different \( A+B \) collisions at 200 GeV
(lab). Several rapidity windows are shown; from top to bottom: 3.15-3.65, 3.65-4.15,
4.15-4.65, 4.65-5.15, 5.15-5.65 in case of Pb+Pb and 0.8-2, 2-3, 3-4, 4-4.4
for the other reactions. The lowest curves are properly normalized, the next
ones are multiplied by ten, etc. We again observe an excess at certain rapidities,
as already discussed above. In addition, in particular for heavy systems, the
slopes are too steep, there is clearly some need of secondary interactions to
``heat up'' the system. This is consistent with the fact that the multiplicity
is too high: collective motion should reduce the multiplicity but instead increase
the transverse energy per particle.
\newpage

\begin{figure}[htb]
{\par\centering \resizebox*{!}{0.45\textheight}{\includegraphics{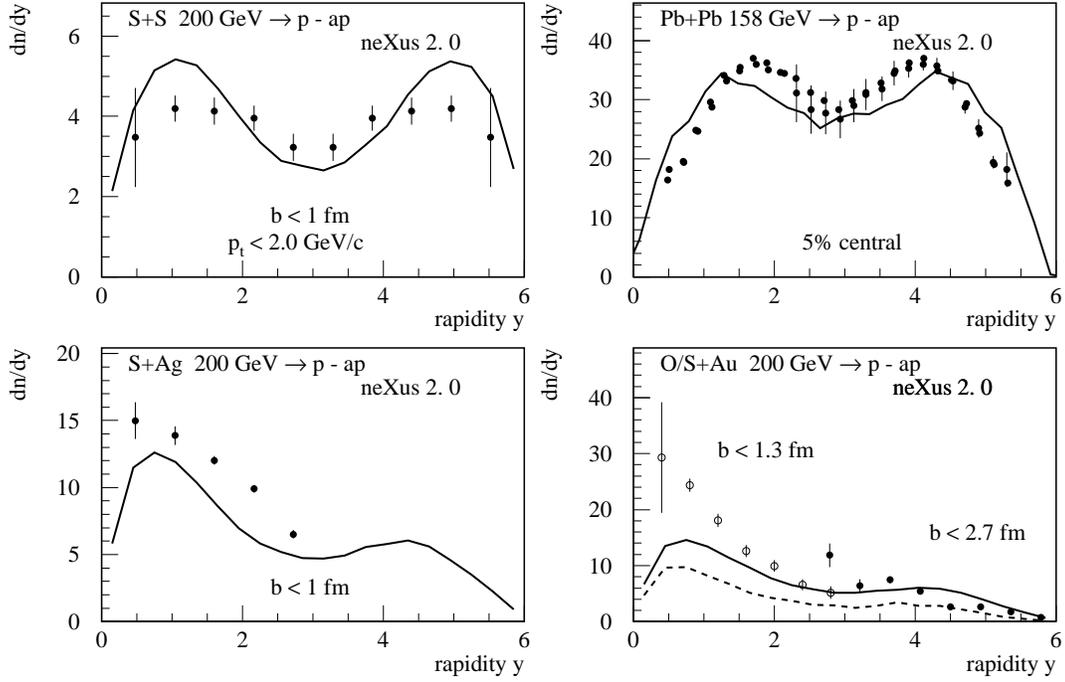}} \par}

\caption{Rapidity distributions of net protons for different \protect\( A+B\protect \)
collisions at 200 GeV (lab).\label{10}}
\end{figure}

\begin{figure}[htb]
{\par\centering \resizebox*{!}{0.45\textheight}{\includegraphics{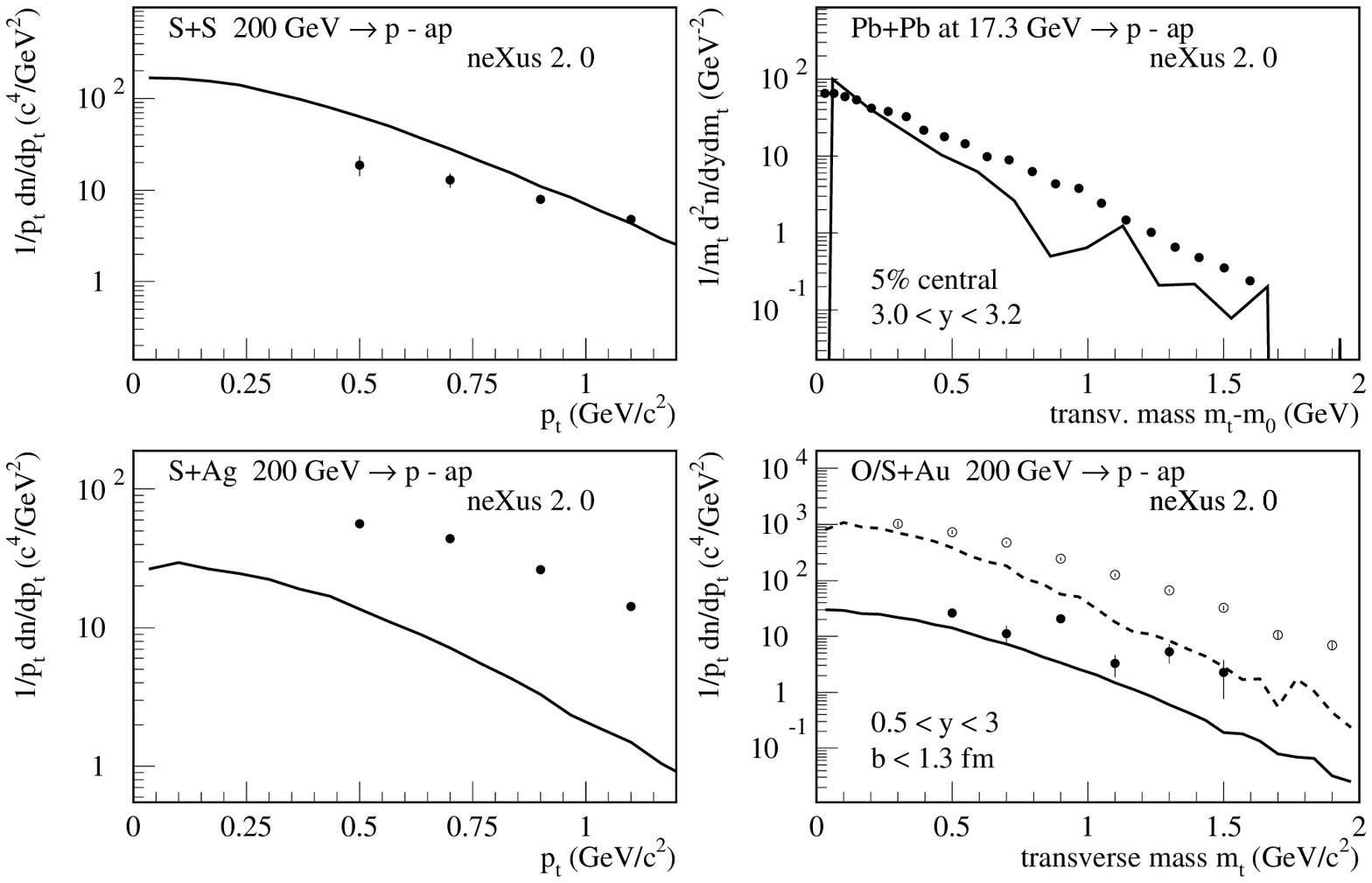}} \par}

\caption{\protect\( p_{t}\protect \) distributions of net protons for different \protect\( A+B\protect \)
collisions at 200 GeV (lab).\label{11}}
\end{figure}

In fig.\ \ref{10}, we show rapidity distributions of net protons (protons minus
anti-protons) for different \( A+B \) collisions at 200 GeV. For the asymmetric
systems one observes clearly the effect of missing target nucleons, which should
be cured by rescattering. The simulated results for symmetric systems are close
to the data, rescattering does not contribute much here. But as for pion production,
we expect also some changes due to screening effects. Transverse momentum spectra,
as shown in fig.\ \ref{11}, show a similar behavior as for pions but even more
pronounced: the theoretical spectra are much too steep, in particular for heavy
systems.

\begin{figure}[htb]
{\par\centering \resizebox*{!}{0.7\textheight}{\includegraphics{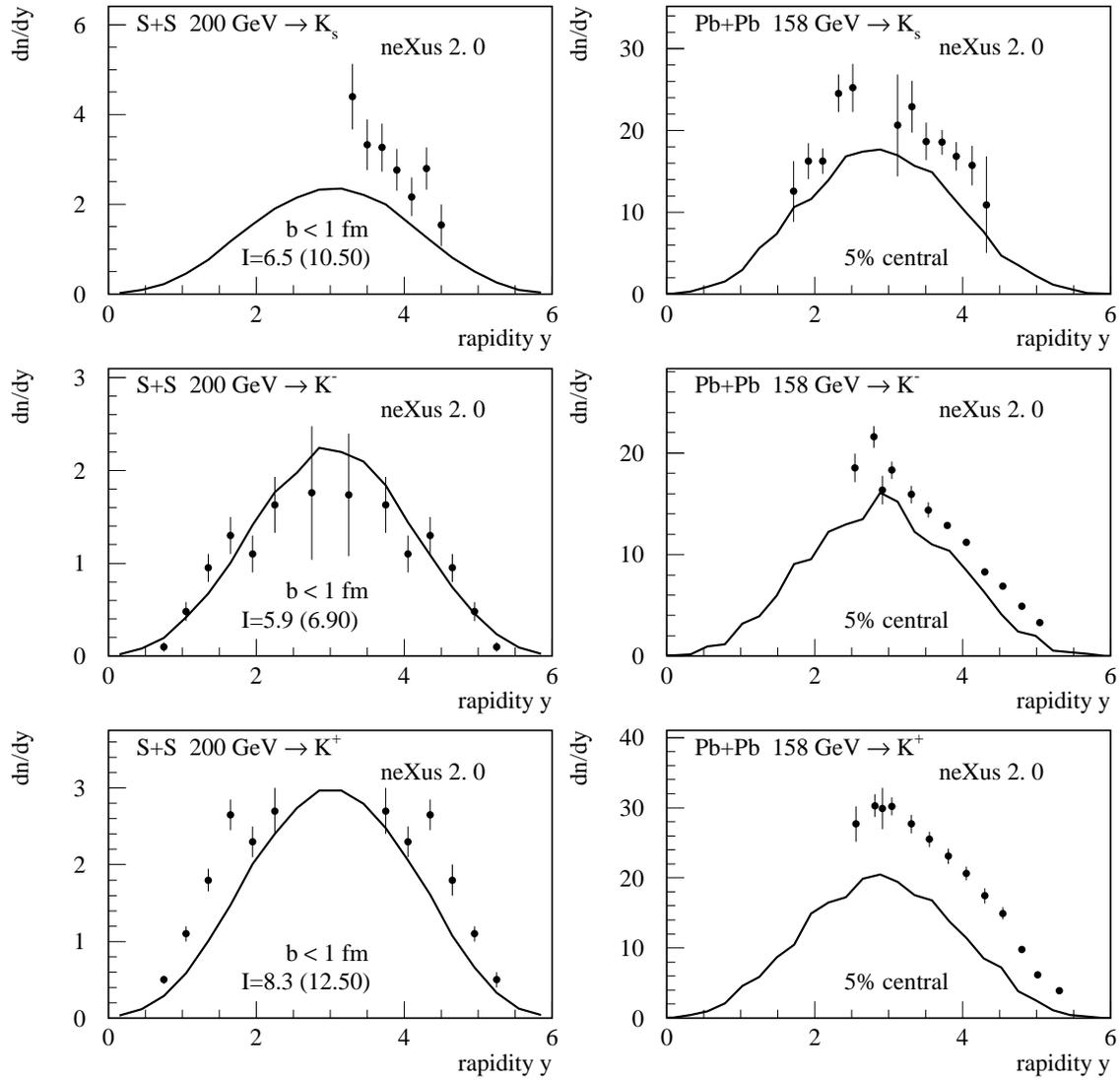}} \par}

\caption{Rapidity distributions of kaons for S+S and Pb+Pb.\label{12}}
\end{figure}

\begin{figure}[htb]
{\par\centering \resizebox*{!}{0.7\textheight}{\includegraphics{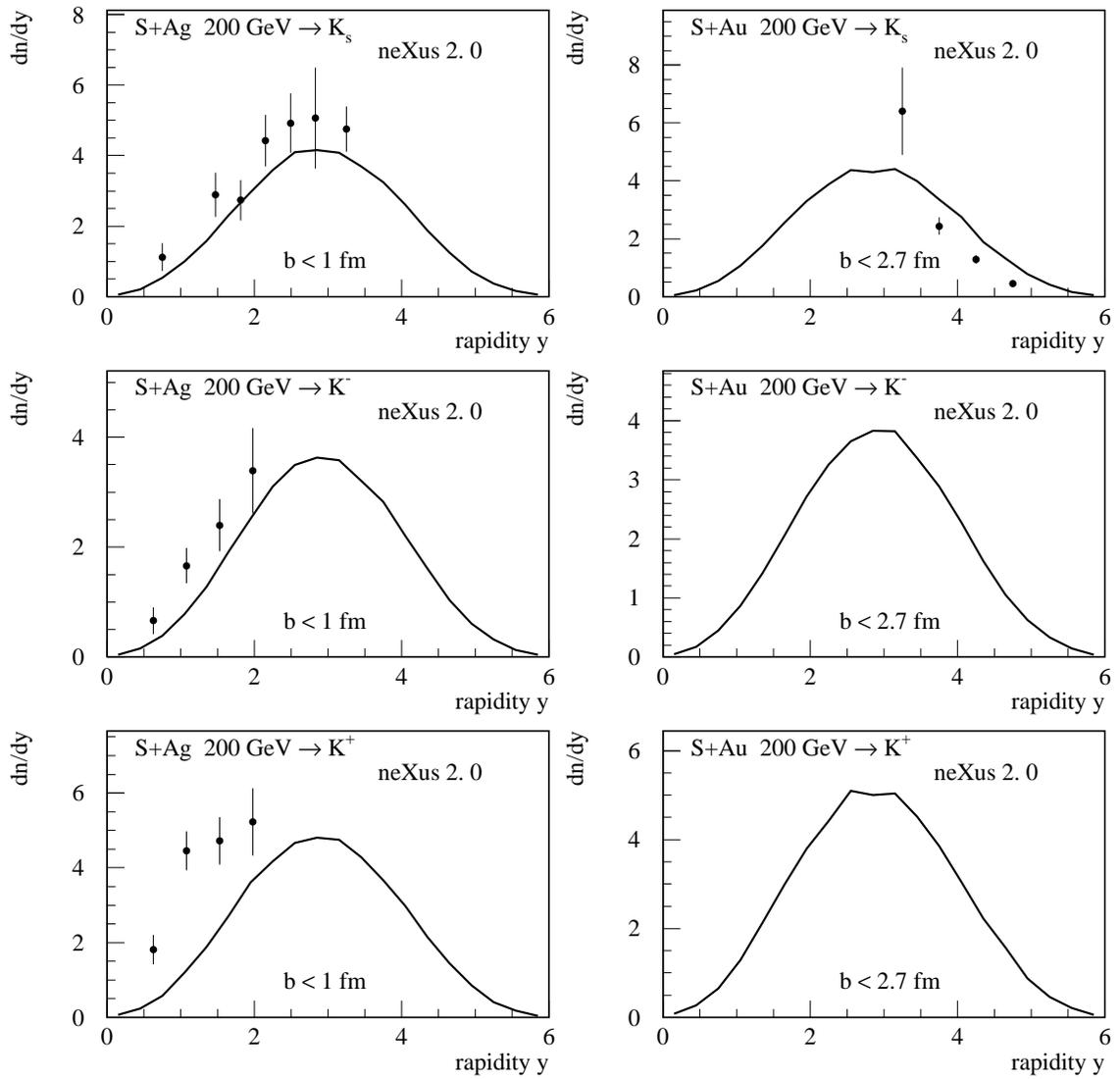}} \par}

\caption{Rapidity distributions of kaons for S+Ag and S+Au.\label{13}}
\end{figure}

\begin{figure}[htb]
{\par\centering \resizebox*{!}{0.7\textheight}{\includegraphics{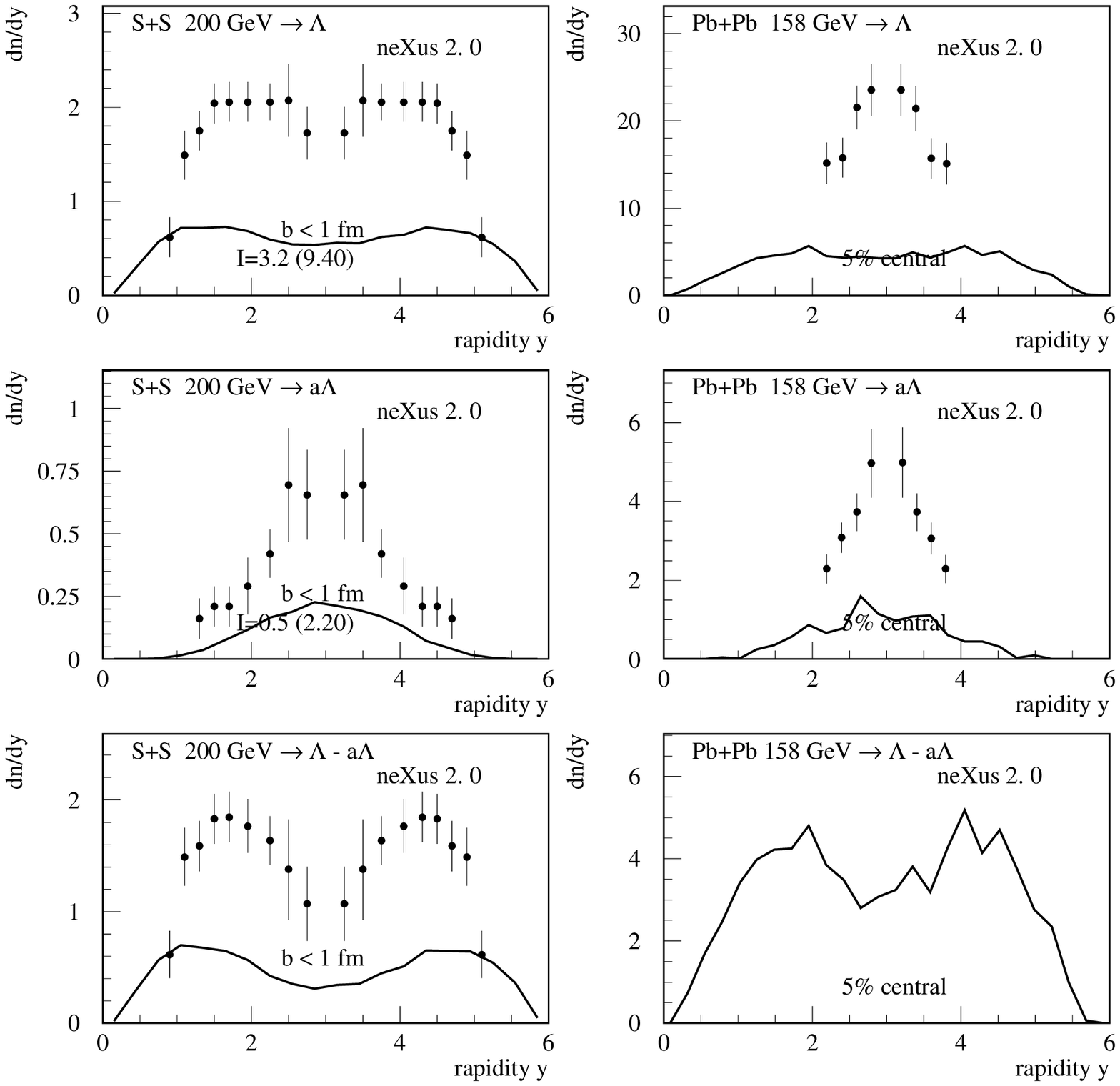}} \par}

\caption{Rapidity distributions of (from top to bottom) lambdas, anti-lambdas, and net
lambdas for S+S and Pb+Pb.\label{14}}
\end{figure}

\begin{figure}[htb]
{\par\centering \resizebox*{!}{0.7\textheight}{\includegraphics{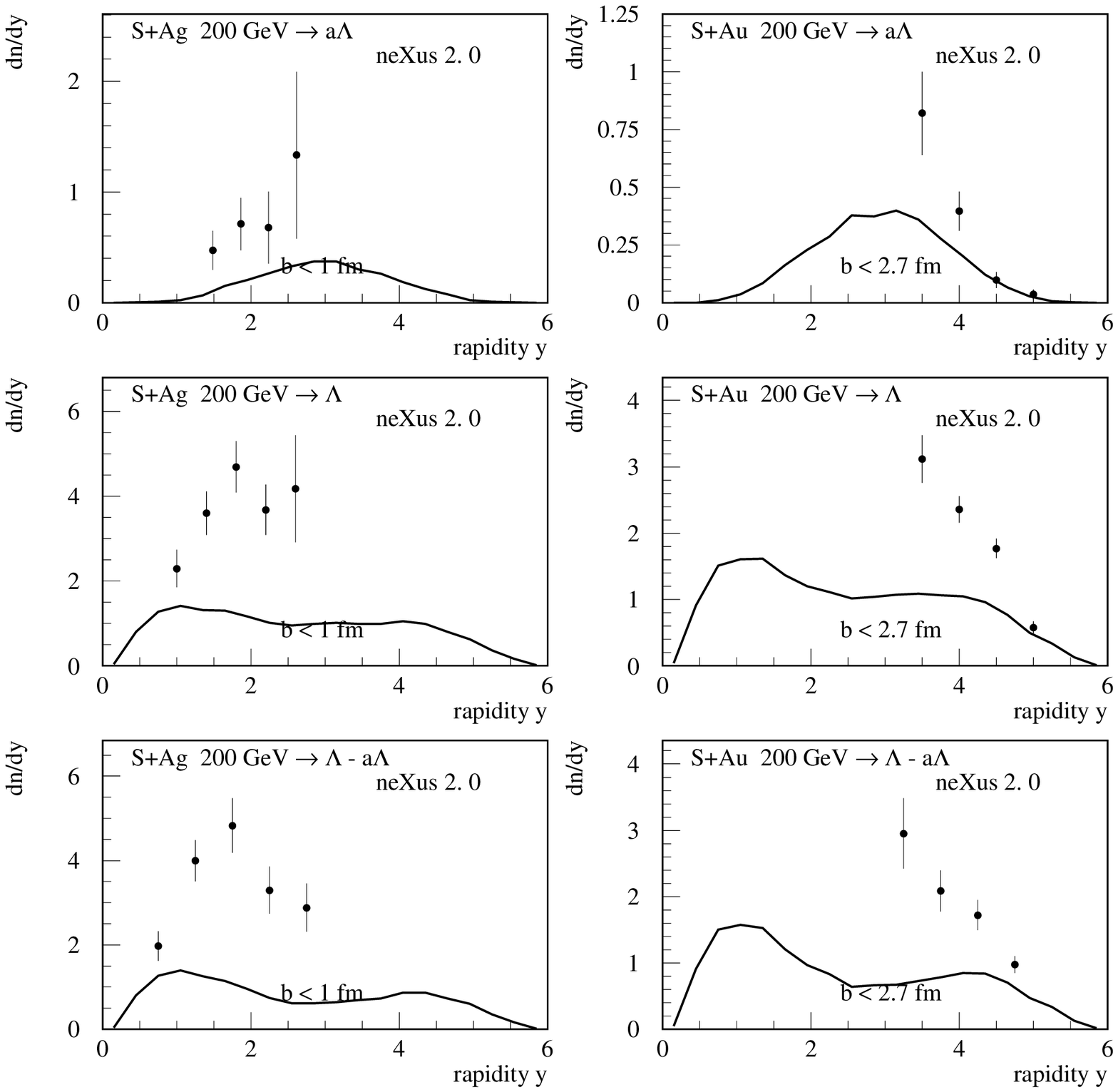}} \par}

\caption{Rapidity distributions of (from top to bottom) lambdas, anti-lambdas, and net
lambdas for S+Ag and S+Au.\label{15}}
\end{figure}

In fig.\ \ref{12}-\ref{15}, we show rapidity spectra of strange particles.
\( \mathrm{K}^{-} \) seem to be correct, whereas \( \mathrm{K}^{+} \) are
in general somewhat too low compared to the data. Lambdas and anti-lambdas are
way too low. For these particles, rescattering has to provide most of the particles
which are finally observed.

\cleardoublepage

\chapter{Summary}

We presented a new approach for hadronic interactions and for the initial stage
of nuclear collisions, which solves several conceptual problems of certain classes
of models, which are presently widely used in order to understand experimental
data. 

The main problem of these models is the fact that energy is not conserved in
a consistent fashion: the fact that energy needs to be shared between many elementary
interactions in case of multiple scattering is well taken into account when
calculating particle production, but energy conservation is not taken care of
in cross section calculations. Related to this problem is the fact that different
elementary interactions in case of multiple scattering are usually not treated
equally, so the first interaction is usually considered to be quite different
compared to the subsequent ones. 

We provided a rigorous treatment of the multiple scattering aspect, such that
questions as energy conservation are clearly determined by the rules of field
theory, both for cross section and particle production calculations. In both
cases, energy is properly shared between the different interactions happening
in parallel. This is the most important and new aspect of our approach, which
we consider to be a first necessary step to construct a consistent model for
high energy nuclear scattering. 

Another important aspect of our approach is the hypothesis that particle production
is a universal process for all the elementary interactions, from \( e^{+}e^{-} \)
annihilation to nucleus-nucleus scattering. That is why we also carefully study
\( e^{+}e^{-} \) annihilation and deep inelastic scattering. This allows to
control reasonably well for example the hadronization procedure, which is not
treatable theoretically from first principles.

This work has been funded in part by the IN2P3/CNRS (PICS 580) and the Russian
Foundation of Basic Researches (RFBR-98-02-22024). 

\cleardoublepage

\begin{appendix}

\chapter{Kinematics of Two Body Collisions}

\section{Conventions\label{ax-a-1}}

\begin{figure}[htb]
{\par\centering \resizebox*{!}{0.15\textheight}{\includegraphics{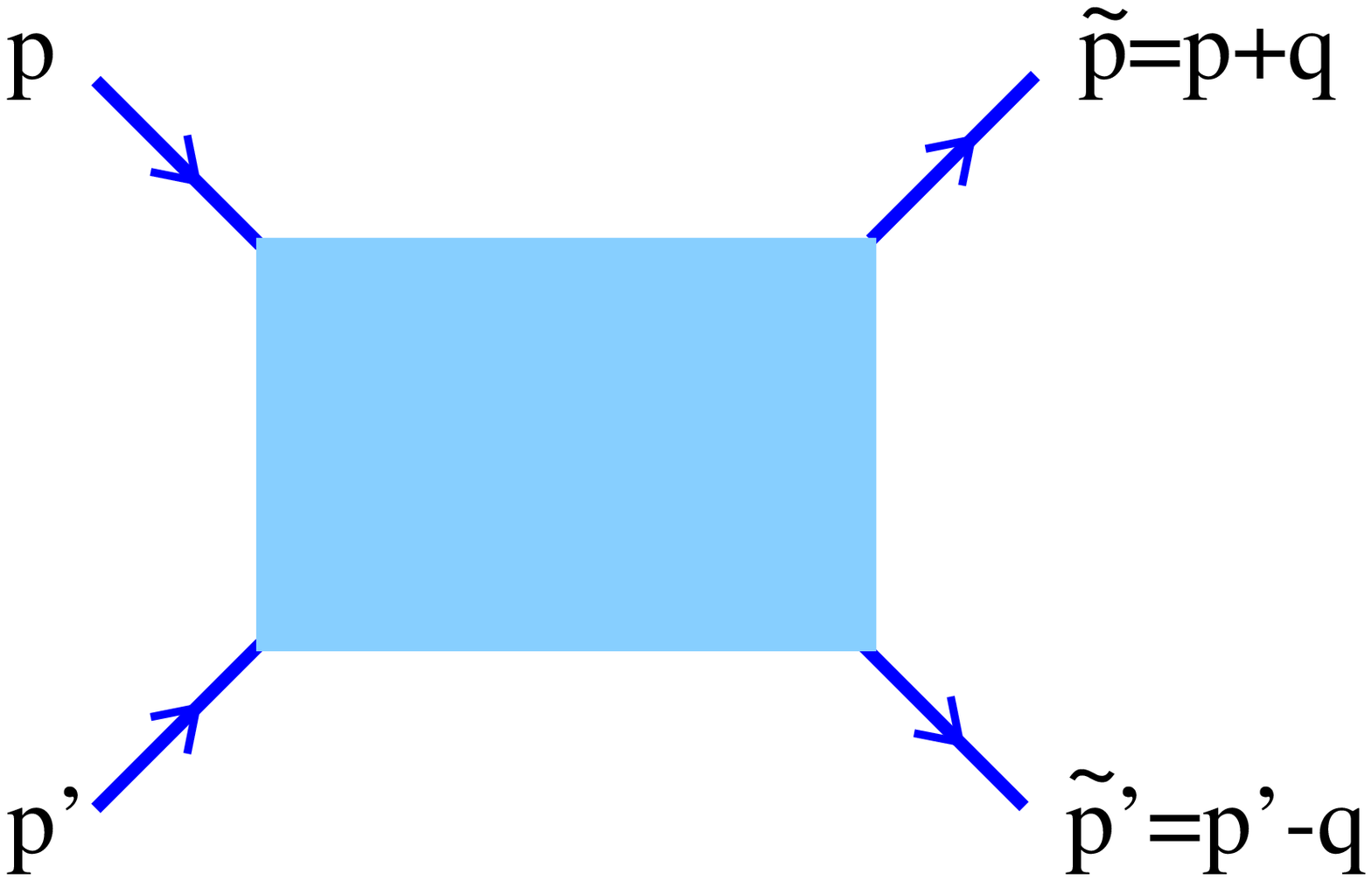}} \par}

\caption{Two body kinematics.\label{xa1}}
\end{figure}

We consider a scattering of a projectile \( P \) on a target \( T \) (hadron-hadron
or parton-parton). We define the incident 4-momenta to be \( p \) and \( p' \)
and the transferred momentum \( q \), so that the outgoing momenta are \( \tilde{p}=p+q \)
and \( \tilde{p}'=p'-q \), see fig.\ \ref{xa1}.
We define as usual the Mandelstam variables 
\begin{equation}
s=(p+p')^{2},\qquad t=(\tilde{p}-p)^{2}.
\end{equation}
Usually, we employ light cone momentum variables, connected to the energy and
\( z \)-component of the particle momentum, as 
\begin{equation}
p^{\pm }=p_{0}\pm p_{z}\, ,
\end{equation}
 and we denote the particle 4-vector as 
\begin{equation}
p=(p^{+},p^{-},\vec{p}_{\perp }).
\end{equation}

\section{Proof of the Impossibility of Longitudinal Excitations\label{ax-a-2} }

Here, we present a mathematical proof of the well known fact that longitudinal
excitations are impossible at high energies. Consider a collision between two
hadrons \( h \) and \( h' \) which leads to two hadrons \( \tilde{h} \) and
\( \tilde{h}' \), 
\begin{equation}
h(p)+h'(p')\rightarrow \tilde{h}(\tilde{p})+\tilde{h}'(\tilde{p}')
\end{equation}
 with four-momenta \( p \), \( p' \), \( \tilde{p} \), \( \tilde{p}' \).
As usual, we define \( s=(p+p')^{2} \) and \( t=(\tilde{p}-p)^{2} \). For
the following, we consider always the limit \( s\rightarrow \infty  \) and
ignore terms of the order \( p^{2}/s \). We expand \( q=\tilde{p}-p=p'-\tilde{p}' \)
as 
\begin{equation}
q=\alpha p+\beta p'+q_{\bot }\, ,
\end{equation}
and obtain 
\begin{equation}
\alpha =\frac{2qp'}{s},\; \beta =\frac{2qp}{s},
\end{equation}
 where we used 
\begin{equation}
p^{2}=0,\, \, p'^{2}=0,\, \, pq_{\bot }=p'q_{\bot }=0,\, \, s=2pp'.
\end{equation}
 We get 
\begin{equation}
q=\frac{q^{2}}{s}(p-p')+q_{\bot },
\end{equation}
 having used 
\begin{equation}
q=\tilde{p}-p=p'-\tilde{p}',\, \, p^{2}=p'^{2}=0,\, \, p'\tilde{p}'=p\tilde{p}=-q^{2}/2.
\end{equation}
 This proves 
\begin{equation}
q=q_{\bot }
\end{equation}
 for \( s\rightarrow \infty  \) and limited \( q^{2} \). In other words, momentum
transfer is purely transversal at high energies.

\cleardoublepage

\chapter{Partonic Interaction Amplitudes}

\section{Semi-hard Parton-Parton Scattering\label{ax-b-1}}

\begin{figure}[htb]
{\par\centering \resizebox*{!}{0.18\textheight}{\includegraphics{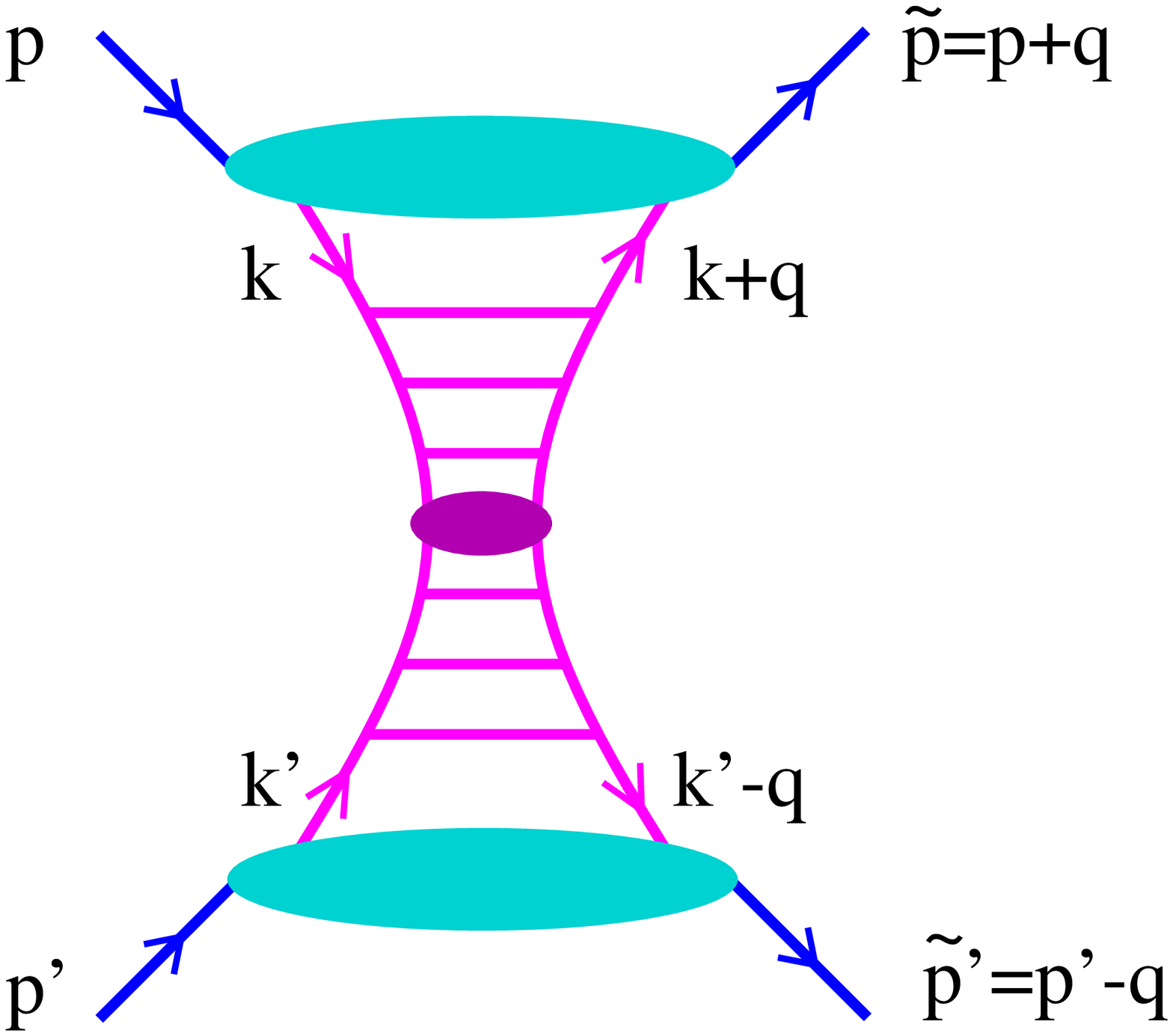}} \par}

\caption{Semi-hard contribution \protect\( T_{\mathrm{sea}-\mathrm{sea}}\protect \).
\label{xb4}}
\end{figure}

Let us derive the mathematical expression corresponding to the contribution
of so-called semi-hard parton-parton scattering, see fig.\ \ref{xb4}.
Here \( p,p' \) are the 4-momenta of the constituent partons participating
in the process. We denote by \( k,k' \) the 4-momenta of the first partons
entering the perturbative evolution, i.e. the initial partons for the perturbative
parton cascade (characterized by parton virtualities \( Q^{2}>Q^{2}_{0} \)).
Further, we define light cone momentum fractions

\begin{equation}
x^{+}=\frac{p^{+}}{p_{0}^{+}},\quad x^{-}=\frac{p'^{-}}{p_{0}^{-}},\quad x_{1}^{+}=\frac{k^{+}}{p_{0}^{+}},\quad x_{1}^{-}=\frac{k'^{-}}{p_{0}^{-}},
\end{equation}
with \( p_{0}^{\pm } \) being the total light-cone momenta for the interaction.

At high energies, the dominant contribution to the process comes from the kinematical
region where these partons are slow, i.e. \( x_{1}^{\pm }\ll x^{\pm } \), so
that a relatively small contribution of the perturbative parton cascade (of
the ladder part of the diagram of the fig.\ \ref{xb4}) is compensated by the
large density of such partons, resulted from the soft pre-evolution \cite{gri83,wer97}.
Since the initial partons \( k,k' \) are gluons or sea quarks (contrary to
valence quarks) we talk about ``sea-sea'' contribution.

Let us first consider the case where the intermediate partons \( k,k' \) are
gluons. Then the amplitude for the diagram of fig.\ \ref{xb4} can be written
as
\begin{eqnarray}
iT^{gg}_{\mathrm{sea}-\mathrm{sea}}(\hat{s},t) & = & \int \frac{d^{4}k}{(2\pi )^{4}}\frac{d^{4}k'}{(2\pi )^{4}}\sum _{\lambda \lambda '\gamma \gamma '\delta \delta '\tau \tau '}\: iM^{g}_{\mathrm{soft}}\! \! \left( p,-k,p+q,-k-q\right) _{\lambda \gamma }\label{sig-elem} \\
 & \times  & D_{\lambda \delta }^{g}\! \! \left( k^{2}\right) \: D_{\gamma \tau }^{g}\! \! \left( (k+q)^{2}\right) iM^{gg}_{\mathrm{hard}}\! \! \left( k,k',k+q,k'-q,Q_{0}^{2}\right) _{\delta \tau \delta '\tau '}\nonumber \\
 & \times  & D_{\lambda '\delta '}^{g}\! \! \left( k'^{2}\right) \: D_{\gamma '\tau '}^{g}\! \! \left( (k'-q)^{2}\right) \: iM^{g}_{\mathrm{soft}}\! \! \left( p',-k',p'-q,-k'+q\right) _{\lambda '\gamma '}\, ,\nonumber 
\end{eqnarray}
 where the amplitude \( M^{gg}_{\mathrm{hard}}\! \! \left( k,k',k+q,k'-q,Q_{0}^{2}\right) _{\delta \tau \delta '\tau '} \)
represents the perturbative contribution of the parton ladder with the initial
partons of momenta \( k,k' \) and with the momentum transfer along the ladder
\( q \) (hard parton-parton scattering), the amplitude \( M^{g}_{\mathrm{soft}}\! \left( p,k,p+q,k-q\right) _{\lambda \gamma } \)
corresponds to the non-perturbative soft interaction between the constituent
parton with the 4-momentum \( p \) and the gluon with the 4-momentum \( k, \)
and \( D_{\lambda \delta }\! \! ^{g}(k) \) is the non-perturbative gluon propagator,
\( D_{\lambda \delta }\! \! ^{g}(k)=i\tilde{D}^{g}\! \! \left( k^{2}\right) \, \varepsilon _{\lambda \delta }(k) \)
with \( \varepsilon _{\lambda \delta }(k) \) being the usual gluon polarization
tensor in the axial gauge; \( \lambda ,\gamma ,\delta ,\ldots  \) denotes symbolically
the combination of color and Lorentz indexes for the intermediate gluons.

As by construction partons of large virtualities \( Q^{2}>Q^{2}_{0} \) can
only appear in the parton ladder part of the diagram of the fig.\ \ref{xb4},
we assume that the integral \( \int d^{4}k \) converges in the region of restricted
virtualities \( k^{2}\sim -s_{0} \) with \( s_{0}\simeq 1 \) GeV\( ^{2} \)
being the typical hadronic mass scale, i.e. the combination 
\begin{equation}
\label{m}
M^{g}_{\mathrm{soft}}\! \! \left( p,-k,p+q,-k-q\right) _{\lambda \gamma }D_{\lambda \delta }^{g}\! \! \left( k^{2}\right) \: D_{\gamma \tau }^{g}\! \! \left( (k+q)^{2}\right) 
\end{equation}
 drops down fast with increasing \( \left| k^{2}\right|  \); this implies that
the transverse momentum \( k_{\perp } \) is also restricted to the region \( k_{\perp }\leq s_{0} \).
Similar arguments apply for \( k' \). Further we make the usual assumption
that longitudinal polarizations in the gluon propagators \( D_{\lambda \delta }^{g}\! \! \left( k^{2}\right)  \)
are canceled in the convolution with the soft contribution \( M^{g}_{\mathrm{soft}} \)
even for finite gluon virtualities \( k^{2} \) \cite{don94}. Finally we assume
that in the considered limit \( x_{1}^{\pm }\ll x^{\pm } \) the amplitude \( M^{g}_{\mathrm{soft}}\! \left( p,k,p+q,k-q\right) _{\lambda \gamma } \)
is governed by the non-perturbative soft Pomeron exchange between the constituent
parton \( p \) and the gluon \( k \), which implies in particular that it
has the singlet structure in the color and Lorentz indexes:
\begin{equation}
\label{M-delta}
M^{g}_{\mathrm{soft}}\! \left( p,k,p+q,k-q\right) _{\lambda \gamma }\sim \delta _{\gamma }^{\lambda }.
\end{equation}
 Then, for small momentum transfer \( q \) in the process of fig.\ \ref{xb4}
the intermediate gluons of momenta \( k,k',k+q,k'-q \) can be considered as
real (on-shell) ones with respect to the perturbative parton evolution in the
ladder, characterized by large momentum transfers \( Q^{2}>Q_{0}^{2} \). Then
we obtain
\begin{eqnarray}
\frac{1}{K^{2}_{g}}\sum _{\lambda \lambda '\delta \delta '\tau \tau '}M^{gg}_{\mathrm{hard}}\! \! \left( k,k',k+q,k'-q,Q_{0}^{2}\right) _{\delta \tau \delta '\tau '} &  & \label{mhard-tlad} \\
\times \; \varepsilon _{\lambda \delta }(k)\varepsilon _{\delta \tau }(k+q)\varepsilon _{\lambda '\delta '}(k')\varepsilon _{\delta '\tau '}(k'-q) & \simeq  & T_{\mathrm{hard}}^{gg}(\hat{s}_{\mathrm{hard}},q^{2},Q_{0}^{2}),\nonumber 
\end{eqnarray}
where the averaging over the spins and the colors of the initial gluons \( k,k' \)
is incorporated in the factor \( K_{g}^{2} \), \( T_{\mathrm{hard}}^{gg}(\hat{s}_{\mathrm{hard}},t,Q_{0}^{2}) \)
is defined in (\ref{t-ladder}), and \( \hat{s}_{\mathrm{hard}}=(k+k')^{2}\simeq k^{+}k'^{-}=x_{1}^{+}x_{1}^{-}s \).

Now, using (\ref{M-delta}-\ref{mhard-tlad}) we can rewrite (\ref{sig-elem})
as
\begin{eqnarray}
iT^{gg}_{\mathrm{sea}-\mathrm{sea}}(\hat{s},t) & = & \int \frac{dk^{+}dk^{-}d^{2}k_{\perp }}{2(2\pi )^{4}}\frac{dk'^{+}dk'^{-}d^{2}k_{\perp }'}{2(2\pi )^{4}}iT_{\mathrm{hard}}^{gg}(\hat{s}_{\mathrm{hard}},q^{2},Q_{0}^{2})\nonumber \\
 & \times  & \left[ -i\sum _{\lambda }M^{g}_{\mathrm{soft}}\! \! \left( p,-k,p+q,-k-q\right) _{\lambda \lambda }\, \tilde{D}^{g}\! \! \left( k^{2}\right) \, \tilde{D}^{g}\! \! \left( (k+q)^{2}\right) \right] \label{tsemi-tlad} \\
 & \times  & \left[ -i\sum _{\lambda '}M^{g}_{\mathrm{soft}}\! \! \left( p',-k',p'-q,-k'+q\right) _{\lambda '\lambda '}\, \tilde{D}^{g}\! \! \left( k'^{2}\right) \, \tilde{D}^{g}\! \! \left( (k'-q)^{2}\right) \right] .\nonumber 
\end{eqnarray}
 It is convenient to perform separately the integrations over \( k^{-},k'^{+},k_{\perp },k_{\perp }' \)
keeping in mind that the only dependence on those variables in the eq.\ (\ref{tsemi-tlad})
appears in the non-perturbative contributions in the square brackets. Let us
consider the first of those contributions, corresponding to the upper soft blob
at fig.\ \ref{xb4} together with the intermediate gluon propagators (for the
lower blob the derivation is identical). Being described by the soft Pomeron
exchange, the amplitude \( M^{g}_{\mathrm{soft}}\! \! \left( p,-k,p+q,-k-q\right) _{\lambda \lambda } \)
is an analytical function of the energy invariants \( \hat{s}_{\mathrm{soft}}=(p-k)^{2} \)
\( \simeq -p^{+}k^{-} \) and \( \hat{u}_{\mathrm{soft}}=(p+k+q)^{2} \) \( \simeq p^{+}k^{-} \)with
the singularities in the complex \( k^{-} \)-plane, corresponding to the values
\( \hat{s}_{\mathrm{soft}}=s'-i0 \) for all real \( s' \) which are greater
than some threshold value \( s_{\mathrm{thr}} \) for the Pomeron asymptotics
to be applied, as well as for \( \hat{u}_{\mathrm{soft}}=u'-i0 \), where \( u'>u_{\mathrm{thr}} \)
with some threshold value \( u_{\mathrm{thr}} \) \cite{bak76}. Thus one has
the singularities in the variable \( k^{-} \)in the upper half of the complex
plane at \( k^{-}\simeq (-s'-i0)/p^{+} \) and in the lower half of the complex
plane at \( k^{-}\simeq (u'-i0)/p^{+} \). Then one can use the standard trick
to rotate the integration contour \( C \) in the variable \( k^{-} \)such
that the new contour \( C' \) encloses the left-hand singularities in \( k^{-} \),
corresponding to the right-hand singularities in the variable \( \hat{s}_{\mathrm{soft}} \)
\cite{bak76}. Then one ends up with the integral over the discontinuity of
the amplitude \( M^{g}_{\mathrm{soft}} \) on the left-hand cut in the variable
\( k^{-} \), which is up to a minus sign equal to the discontinuity on the
right-hand cut in \( \hat{s}_{\mathrm{soft}} \): 
\begin{eqnarray}
 &  & \int ^{+\infty }_{-\infty }\! dk^{-}\left[ \sum _{\lambda }M^{g}_{\mathrm{soft}}\! \! \left( p,-k,p+q,-k-q\right) _{\lambda \lambda }\, \tilde{D}^{g}\! \! \left( k^{2}\right) \, \tilde{D}^{g}\! \! \left( (k+q)^{2}\right) \right] \nonumber \\
 &  & =\int ^{-s_{\mathrm{thr}}/k^{+}}_{-\infty }\! dk^{-}\, \mathrm{disc}_{\hat{s}_{\mathrm{soft}}}\left[ \sum _{\lambda }M^{g}_{\mathrm{soft}}\! \! \left( p,-k,p+q,-k-q\right) _{\lambda \lambda }\, \tilde{D}^{g}\! \! \left( k^{2}\right) \, \tilde{D}^{g}\! \! \left( (k+q)^{2}\right) \right] \\
 &  & =\int ^{-s_{\mathrm{thr}}/k^{+}}_{-\infty }\! dk^{-}\, 2i\, \mathrm{Im}\left[ \sum _{\lambda }M^{g}_{\mathrm{soft}}\! \! \left( p,-k,p+q,-k-q\right) _{\lambda \lambda }\, \tilde{D}^{g}\! \! \left( k^{2}\right) \, \tilde{D}^{g}\! \! \left( (k+q)^{2}\right) \right] .
\end{eqnarray}
 Now, using \( \int dk^{-}=\int dk^{2}/k^{+} \) and recalling our assumption
that the integral over \( k^{2} \) gets dominant contribution from the region
\( k^{2}\sim -s_{0} \), we may write
\begin{eqnarray}
 & \int \frac{dk^{-}d^{2}k_{\perp }}{(2\pi )^{4}}\mathrm{Im}\left[ \sum _{\lambda }M^{g}_{\mathrm{soft}}\! \! \left( p,-k,p+q,-k-q\right) _{\lambda \lambda }\, \tilde{D}^{g}\! \! \left( k^{2}\right) \, \tilde{D}^{g}\! \! \left( (k+q)^{2}\right) \right]  & \nonumber \\
 & =\frac{1}{k^{+}}\mathrm{Im}\, T_{\mathrm{soft}}^{g}\! \! \left( \hat{s}_{\mathrm{soft}},q^{2}\right)  & 
\end{eqnarray}
with
\begin{equation}
\hat{s}_{\mathrm{soft}}=s_{0}\frac{p^{+}}{k^{+}}=s_{0}\frac{x^{+}}{x_{1}^{+}}.
\end{equation}
The integrations over \( k^{2} \) in the vicinity of \( k^{2}=-s_{0} \) and
over \( k_{\perp }\leq s_{0} \) are supposed to just contribute to the redetermination
of the Pomeron-gluon coupling of the usual soft Pomeron amplitude, and therefore
we parameterize the amplitude \( T_{\mathrm{soft}}^{g} \) as (compare with
eq.\ (\ref{tsoft}))
\begin{equation}
\label{t-parton-gluon}
T_{\mathrm{soft}}^{g}\! \! \left( \hat{s},t\right) =8\pi s_{0}\eta (t)\, \gamma _{\mathrm{part}}\gamma _{g}\, \left( \frac{\hat{s}}{s_{0}}\right) ^{\alpha _{_{\rm {P}}}\! (0)}\exp \! \left( \lambda ^{(1)}\! _{\mathrm{soft}}(\hat{s}/s_{0})\, t\right) \, \left( 1-\frac{s_{0}}{\hat{s}}\right) ^{\beta _{g}},
\end{equation}
with
\begin{equation}
\lambda ^{(1)}\! _{\mathrm{soft}}\! (z)=R_{\mathrm{part}}^{2}+\alpha '\! _{\mathrm{soft}}\ln \! z,
\end{equation}
where we used \( \gamma _{g} \) for the Pomeron-gluon coupling and we neglected
the radius of the Pomeron-gluon vertex assuming that the coupling is local in
the soft Pomeron; the factor \( \left( 1-s_{0}/\hat{s}\right) ^{\beta _{g}} \)
is included to ensure that the Pomeron has sufficiently large mass, which is
the necessary condition for applying Regge description for the soft evolution.
As we shall see below the parameter \( \beta _{g} \) determines the gluon momentum
distribution in the Pomeron at \( x_{1}^{\pm }\rightarrow x^{\pm } \).

Finally, using the above results we obtain
\begin{eqnarray}
iT^{gg}_{\mathrm{sea}-\mathrm{sea}}(\hat{s},t) & = & \int \frac{dk^{+}}{k^{+}}\frac{dk'^{-}}{k'^{-}}\mathrm{Im}\, T_{\mathrm{soft}}^{g}\! \! \left( \hat{s}_{\mathrm{soft}},t\right) \, \mathrm{Im}\, T_{\mathrm{soft}}^{g}\! \! \left( \hat{s}'\! _{\mathrm{soft}},t\right) \, iT_{\mathrm{hard}}^{gg}(\hat{s}_{\mathrm{hard}},t,Q_{0}^{2})\\
 & = & \int ^{1}_{0}\frac{dz^{+}}{z^{+}}\frac{dz^{-}}{z^{-}}\, \mathrm{Im}\, T_{\mathrm{soft}}^{g}\! \! \left( \frac{s_{0}}{z^{+}},t\right) \, \mathrm{Im}\, T_{\mathrm{soft}}^{g}\! \! \left( \frac{s_{0}}{z^{-}},t\right) \, iT_{\mathrm{hard}}^{gg}(z^{+}z^{-}\hat{s},t,Q_{0}^{2}),
\end{eqnarray}
where the following definitions have been used:
\begin{equation}
z^{\pm }=\frac{x_{1}^{\pm }}{x^{\pm }},\quad \hat{s}_{\mathrm{soft}}=s_{0}\frac{x^{+}}{x_{1}^{+}}=\frac{s_{0}}{z^{+}},\quad \hat{s}'\! _{\mathrm{soft}}=s_{0}\frac{x^{-}}{x_{1}^{-}}=\frac{s_{0}}{z^{-}},\quad \hat{s}_{\mathrm{hard}}=x_{1}^{+}x_{1}^{-}s=z^{+}z^{-}\hat{s}.
\end{equation}

In the case of the intermediate parton \( k \) being a (anti-)quark, we assume
that it originates from local gluon splitting in the soft Pomeron. Thus we neglect
the slope of the (non-perturbative) gluon-quark vertex. Using the usual Altarelli-Parisi
kernel \( P_{g}^{q}(z)=\frac{1}{2}(z^{2}+(1-z)^{2}) \) for the gluon light
cone momentum partition between the quark and the anti-quark, we get the corresponding
amplitude \( T^{qg}_{\mathrm{sea}-\mathrm{sea}}(\hat{s},t) \) as
\begin{equation}
\label{t-semi-qg}
iT^{qg}_{\mathrm{sea}-\mathrm{sea}}(\hat{s},t)=\int ^{1}_{0}\! \frac{dz^{+}}{z^{+}}\frac{dz^{-}}{z^{-}}\, \mathrm{Im}\, T_{\mathrm{soft}}^{q}\! \! \left( \frac{s_{0}}{z^{+}},t\right) \, \mathrm{Im}\, T_{\mathrm{soft}}^{g}\! \! \left( \frac{s_{0}}{z^{-}},t\right) \, iT_{\mathrm{hard}}^{qg}(z^{+}z^{-}\hat{s},t,Q_{0}^{2}),
\end{equation}
where the imaginary part of the amplitude \( \mathrm{Im}\, T_{\mathrm{soft}}^{q} \)
for the soft Pomeron exchange between the constituent parton and the quark \( q \)
\( \in  \) \( \{u,d,s,\bar{u},\bar{d},\bar{s}\} \) is expressed via \( \, \mathrm{Im}\, T_{\mathrm{soft}}^{g} \)
as
\begin{equation}
\label{t-pom-q}
\mathrm{Im}\, T_{\mathrm{soft}}^{q}\! (\hat{s}_{\mathrm{soft}},t)=\gamma _{qg}\int \! d\xi \, P_{g}^{q}(\xi )\, \mathrm{Im}\, T_{\mathrm{soft}}^{g}\! (\xi \, \hat{s}_{\mathrm{soft}},t),
\end{equation}
with \( \gamma _{qg} \) representing the quark-gluon vertex value and \( \xi  \)
being the ratio of the quark and the parent gluon light cone momentum, \( \xi =k^{+}/k_{g}^{+} \);
the mass squared of the Pomeron between the initial constituent parton and the
gluon is then 
\begin{equation}
(p-k_{g})^{2}\simeq s_{0}\frac{p^{+}}{k_{g}^{+}}=\xi \, \hat{s}_{\mathrm{soft}}.
\end{equation}

The full amplitude for the semi-hard scattering is the sum of the different
terms discussed above, i.e.
\begin{eqnarray}
iT_{\mathrm{sea}-\mathrm{sea}}(\hat{s},t) & = & \sum _{jk}iT^{jk}_{\mathrm{sea}-\mathrm{sea}}(\hat{s},t)\nonumber \\
 & = & \sum _{jk}\int ^{1}_{0}\! \frac{dz^{+}}{z^{+}}\frac{dz^{-}}{z^{-}}\, \mathrm{Im}\, T_{\mathrm{soft}}^{j}\! \! \left( \frac{s_{0}}{z^{+}},t\right) \, \mathrm{Im}\, T_{\mathrm{soft}}^{k}\! \! \left( \frac{s_{0}}{z^{-}},t\right) \, iT_{\mathrm{hard}}^{jk}(z^{+}z^{-}\hat{s},t,Q_{0}^{2}),\label{t-semi-full} 
\end{eqnarray}
where \( j,k \) denote the types (flavors) of the initial partons for the perturbative
evolution (quarks or gluons).

The discontinuity of the amplitude \( T_{\mathrm{sea}-\mathrm{sea}}(\hat{s},t) \)
on the right-hand cut in the variable \( \hat{s} \) defines the contribution
of the cut diagram of the fig.\ \ref{xb4}. Cutting procedure amounts here to
replace the hard parton-parton scattering amplitude \( iT_{\mathrm{hard}}^{jk}(\hat{s}_{\mathrm{hard}},t,Q_{0}^{2}) \)
in (\ref{t-semi-full}) by \( 2\mathrm{Im}\, T_{\mathrm{hard}}^{jk}(\hat{s}_{\mathrm{hard}},t,Q_{0}^{2}) \),
whereas the contributions of the soft parton cascades \( \mathrm{Im}\, T_{\mathrm{soft}}^{j} \)
stay unchanged as they are already defined by the discontinuities in the corresponding
variables \( \hat{s}_{\mathrm{soft}} \) and \( \hat{s}'\! _{\mathrm{soft}} \).
So the cut diagram contribution is just \( 2\mathrm{Im}\, T_{\mathrm{sea}-\mathrm{sea}}(\hat{s},t) \).
At \( t=0 \) it defines the cross section for the semi-hard parton-parton scattering:
\begin{equation}
\label{sig-sea-sea}
\sigma _{\mathrm{sea}-\mathrm{sea}}\! \left( \hat{s}\right) =\frac{1}{2\hat{s}}2\mathrm{Im}\, T_{\mathrm{sea}-\mathrm{sea}}(\hat{s},0)=\sum _{jk}\int ^{1}_{0}dz^{+}dz^{-}E_{\mathrm{soft}}^{j}\left( z^{+}\right) \, E_{\mathrm{soft}}^{k}\left( z^{-}\right) \, \sigma _{\mathrm{hard}}^{jk}(z^{+}z^{-}\hat{s},Q_{0}^{2}),
\end{equation}
where we used the explicit expressions (\ref{t-ladder}), (\ref{t-parton-gluon}),
(\ref{t-pom-q}) for \( T_{\mathrm{hard}}^{jk} \), \( T_{\mathrm{soft}}^{j} \),
and the functions \( E_{\mathrm{soft}}^{j} \) are defined as
\begin{eqnarray}
E_{\mathrm{soft}}^{g}\left( z\right)  & = & 8\pi s_{0}\gamma _{\mathrm{part}}\gamma _{g}\, z^{-\alpha _{\mathrm{soft}}\! (0)}\, (1-z)^{\beta _{g}}\label{esoft-g} \\
E_{\mathrm{soft}}^{q}\left( z\right)  & = & \gamma _{qg}\int ^{1}_{z}\! d\xi \, P_{g}^{q}(\xi )\, E_{\mathrm{soft}}^{g}\left( \frac{z}{\xi }\right) .\label{esoft-q} 
\end{eqnarray}

It is easy to see that \( E_{\mathrm{soft}}^{j}\left( z\right)  \) has the
meaning of the momentum distribution of parton \( j \) at the scale \( Q_{0}^{2} \)
for an elementary interaction, i.e. parton distribution \textbf{in the soft
Pomeron}. Introducing the gluon splitting probability \( w_{\mathrm{split}} \)
and the coupling \( \tilde{\gamma }_{g} \) via 
\begin{equation}
\gamma _{qg}\gamma _{g}=w_{\mathrm{split}}\, \tilde{\gamma }_{g},\quad \gamma _{g}=\left( 1-w_{\mathrm{split}}\right) \, \tilde{\gamma }_{g},
\end{equation}
 the light cone momentum conservation reads 
\begin{equation}
1=\int _{0}^{1}\! dz\, \sum _{j}z\, E_{\mathrm{soft}}^{j}\left( z\right) =8\pi s_{0}\gamma _{\mathrm{part}}\tilde{\gamma }_{g}\int _{0}^{1}\! dz\, z^{1-\alpha _{\mathrm{soft}}\! (0)}\, (1-z)^{\beta _{g}}
\end{equation}
and therefore
\begin{equation}
\tilde{\gamma }_{g}=\frac{1}{8\pi s_{0}\gamma _{\mathrm{part}}}\frac{\Gamma \! \left( 3-\alpha _{\mathrm{soft}}\! (0)+\beta _{g}\right) }{\Gamma \! \left( 2-\alpha _{\mathrm{soft}}\! (0)\right) \, \Gamma \! \left( 1+\beta _{g}\right) }
\end{equation}

\section{Parton Evolution\label{ax-evol}}

In this appendix, we discuss the properties of the evolution function \( E_{\mathrm{QCD}} \),
describing the perturbative evolution of partons.

The evolution function \( E^{jm}_{\mathrm{QCD}}\left( z,Q^{2}_{0},Q^{2}\right)  \)
satisfies the usual DGLAP equation
\begin{equation}
\frac{dE^{jm}_{\mathrm{QCD}}\left( Q^{2}_{0},Q^{2},x\right) }{d\ln Q^{2}}=\sum _{k}\int _{x}^{1}\frac{dz}{z}\frac{\alpha _{s}}{2\pi }\tilde{P}_{k}^{m}\! (z)\: E^{jk}_{\mathrm{QCD}}\left( \frac{x}{z},Q^{2}_{0},Q^{2}\right) 
\end{equation}
with the initial condition \( E^{jm}_{\mathrm{QCD}}\left( Q^{2}_{0},Q_{0}^{2},z\right) =\delta ^{m}_{j}\; \delta (1-z). \)
Here \( \tilde{P}_{k}^{m}\! (z) \) are the usual Altarelli-Parisi splitting
functions, regularized at \( z\rightarrow 1 \) by the contribution of virtual
emissions.
\begin{figure}[htb]
{\par\centering \resizebox*{0.4\textwidth}{!}{\includegraphics{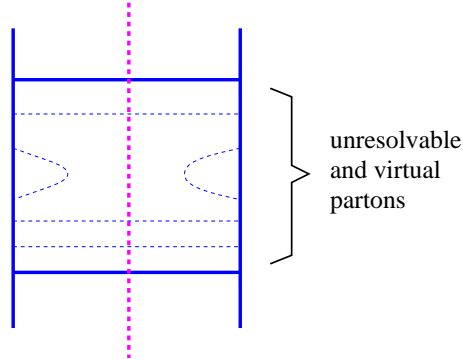}} \par}

\caption{Unresolvable emissions are summed over, providing the so-called Sudakov form
factor.\label{suda}}
\end{figure}

One can introduce the concept of ``resolvable'' parton emission, i.e. an emission
of a final (\( s \)-channel) parton with a finite share of the parent parton
light cone momentum \( (1-z)>\epsilon =p_{\bot \textrm{res }}^{2\textrm{ }}/Q^{2} \)
(with finite relative transverse momentum \( p_{\bot }^{2}= \) \( Q^{2}(1-z) \)
\( >p_{\bot \mathrm{res}\textrm{ }}^{2\textrm{ }} \)) \cite{web84} and use
the so-called Sudakov form factor, corresponding to the contribution of any
number of virtual and unresolvable emissions (i.e. emissions with \( (1-z)<\epsilon  \))
- see fig.\ \ref{suda}. 
\begin{equation}
\Delta ^{k}(Q_{0}^{2},Q^{2})=\exp \left\{ \int _{Q_{0}^{2}}^{Q^{2}}{dq^{2}\over q^{2}}\int _{1-\epsilon }^{1}dz\, {\alpha _{s}\over 2\pi }\, \tilde{P}_{k}^{k}(z)\right\} 
\end{equation}
 This can also be interpreted as the probability of no  resolvable  emission
between \( Q_{0}^{2} \) and \( Q^{2} \).

Then \( E^{jm}_{\mathrm{QCD}} \) can be expressed via \( \bar{E}^{jm}_{\mathrm{QCD}} \),
corresponding to the sum of any number (but at least one) resolvable emissions,
allowed by the kinematics:
\begin{equation}
\label{eqcd-ini}
E^{jm}_{\mathrm{QCD}}\left( z,Q^{2}_{0},Q^{2}\right) =\delta ^{m}_{j}\; \delta (1-z)\, \Delta ^{j}(Q_{0}^{2},Q^{2})+\bar{E}^{jm}_{\mathrm{QCD}}\left( z,Q^{2}_{0},Q^{2}\right) ,
\end{equation}
where \( \bar{E}^{jm}_{\mathrm{QCD}}\left( z,Q^{2}_{0},Q^{2}\right)  \) satisfies
the integral equation
\begin{eqnarray}
\bar{E}^{jm}_{\mathrm{QCD}}\left( x,Q^{2}_{0},Q^{2}\right)  & = & \int ^{Q^{2}}_{Q_{0}^{2}}\frac{dQ_{1}^{2}}{Q_{1}^{2}}\left[ \sum _{k}\int _{x}^{1-\epsilon }\frac{dz}{z}\frac{\alpha _{s}}{2\pi }P_{k}^{m}\! (z)\: \bar{E}^{jk}_{\mathrm{QCD}}\left( \frac{x}{z},Q^{2}_{0},Q_{1}^{2}\right) +\right. \nonumber \\
 & + & \left. \Delta ^{j}(Q_{0}^{2},Q_{1}^{2})\frac{\alpha _{s}}{2\pi }P_{j}^{m}\! (x)\right] \Delta ^{m}(Q_{1}^{2},Q^{2})\label{ap-res} 
\end{eqnarray}
 Here \( P_{j}^{k}\! (z) \) are the Altarelli-Parisi splitting functions for
real emissions, i.e. without \( \delta  \)-function and regularization terms
at \( z\rightarrow 1 \). 
\begin{figure}[htb]
{\par\centering \resizebox*{!}{0.18\textheight}{\includegraphics{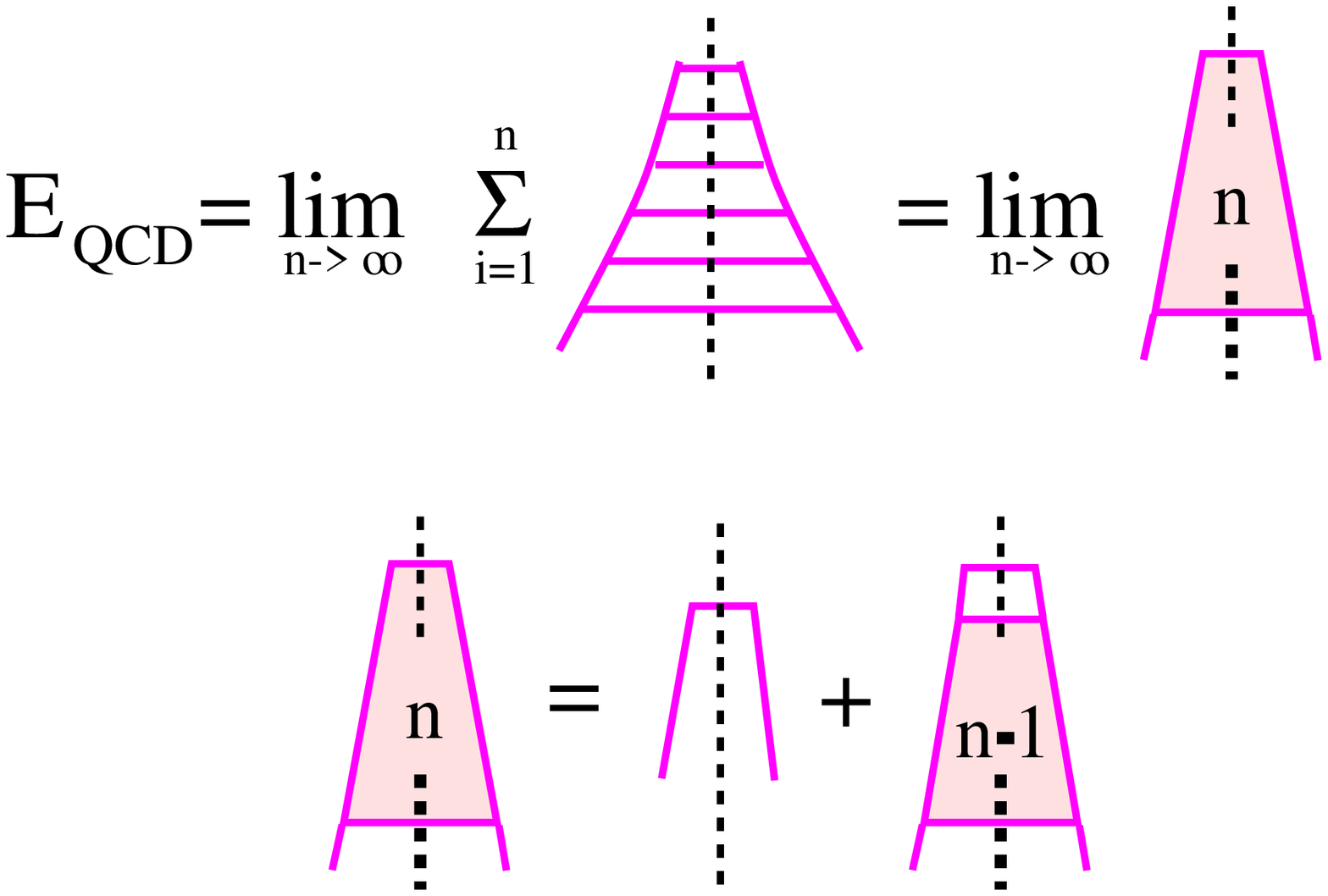}} \par}

\caption{The calculation of \protect\( \bar{E}_{\mathrm{QCD}}\protect \).\label{evol}}
\end{figure}

Eq.\ (\ref{ap-res}) can be solved iteratively, expressing \( \bar{E}^{jm}_{\mathrm{QCD}} \)
as the contribution of at most \( n \) (\( n\rightarrow \infty  \)) resolvable
emissions (of an ordered ladder with at most \( n \) ladder rungs) - see fig.\ \ref{evol}:
\begin{equation}
\label{eqcd-n-rungs}
\bar{E}^{jm}_{\QCD }(Q_{0}^{2},Q^{2},x)=\lim _{n\to \infty }E^{(n)jm}_{\QCD }(Q_{0}^{2},Q^{2},x),
\end{equation}
 with
\begin{eqnarray}
E_{\QCD }^{(n)jm}\left( x,Q^{2}_{0},Q^{2}\right)  & = & \int ^{Q^{2}}_{Q_{0}^{2}}\frac{dQ_{1}^{2}}{Q_{1}^{2}}\left[ \sum _{k}\int _{x}^{1-\epsilon }\frac{dz}{z}\frac{\alpha _{s}}{2\pi }P_{k}^{m}\! (z)\: E_{\QCD }^{(n-1)jk}\left( \frac{x}{z},Q^{2}_{0},Q_{1}^{2}\right) \right] \nonumber \label{ap-n-res} \\
 & \times  & \Delta ^{m}(Q_{1}^{2},Q^{2})+E_{\QCD }^{(1)jm}\left( x,Q^{2}_{0},Q^{2}\right) \label{ap-n-res} \\
E_{\QCD }^{(1)jm}\left( x,Q^{2}_{0},Q^{2}\right)  & = & \int ^{Q^{2}}_{Q_{0}^{2}}\frac{dQ_{1}^{2}}{Q_{1}^{2}}\; \Delta ^{j}(Q_{0}^{2},Q_{1}^{2})\; \Delta ^{m}(Q_{1}^{2},Q^{2})\; \frac{\alpha _{s}}{2\pi }P_{j}^{m}\! (z)\label{ap-1-res} 
\end{eqnarray}
So the procedure amounts to only  considering resolvable emissions, but to multiply
each propagator with \( \Delta ^{j} \), which is the reason for the appearance
of \( \Delta ^{j} \) in eqs.\ (\ref{ap-res}), (\ref{ap-n-res}-\ref{ap-1-res}).

\section{Time-Like Parton Splitting\label{ax-time}}

We discuss here the algorithm for Monte Carlo generation of time-like parton
emission on the basis of the eq.\ (\ref{for:ap}). 

The standard procedure is to apply the Monte Carlo rejection method \cite{sjo84}.
We consider the splitting of a parton \( j \) with a maximal virtuality \( Q^{2}_{j\, \mathrm{max}} \)
given by the preceding process. For the proposal function, we define the limits
in \( z \) for given \( Q_{j}^{2} \) using an approximate formula
\begin{equation}
p_{\perp }^{2}\simeq z(1-z)Q_{j}^{2}-zQ_{l}^{2}-(1-z)Q_{k}^{2}
\end{equation}
instead of the exact one, eq.\ (\ref{for:pt2}), with the lowest possible virtualities
for daughter partons \( Q_{k}^{2}=Q_{l}^{2}=p_{\bot \mathrm{fin}}^{2} \), which
gives 
\begin{equation}
z_{\min /\max }(Q_{j}^{2})=\frac{1}{2}\pm \frac{1}{2}\sqrt{1-\frac{4p^{2}_{\bot \mathrm{fin}}}{Q_{j}^{2}}}\, .
\end{equation}
 Further, we define upper limits,
\begin{eqnarray}
\bar{P}^{g}_{g}(z) & = & 3\left\{ \frac{1}{z}+\frac{1}{1-z}\right\} ,\nonumber \\
\bar{P}^{q}_{g}(z) & = & \frac{N_{f}}{2},\\
\bar{P}^{g}_{q}(z) & = & \frac{4}{3}\frac{2}{1-z},\nonumber 
\end{eqnarray}
 for the splitting functions, 
\begin{eqnarray}
 &  & \frac{1}{2}P^{g}_{g}(z)=3\frac{(1-z(1-z))^{2}}{z(1-z)},\nonumber \\
 &  & \frac{1}{2}\sum _{i\in \{u,d,s,\bar{u},\bar{d},\bar{s}\}}P^{i}_{g}(z)=\frac{N_{f}}{2}(z^{2}+(1-z)^{2}),\\
 &  & P^{g}_{q}(z)=\frac{4}{3}\frac{1+z^{2}}{1-z},\nonumber 
\end{eqnarray}
with \( N_{f} \) being the number of active quark flavors. Integrating these
three functions \( \bar{P}^{k}_{j}(z) \) over \( z \) from \( z_{\mathrm{min}}=z_{\mathrm{min}}(Q_{j}^{2}) \)
to \( z_{\mathrm{max}}=z_{\mathrm{max}}(Q^{2}_{j}) \) as
\begin{equation}
\label{ijk-pjk}
I^{k}_{j}(Q_{j}^{2})=\int ^{z_{\max }}_{z_{\min }}\! dz\, \bar{P}^{k}_{j}\! (z),
\end{equation}
 one obtains
\begin{eqnarray}
I^{g}_{g}(Q_{j}^{2}) & = & 3\left( \ln \left( \frac{z_{\max }}{z_{\min }}\right) +\ln \left( \frac{1-z_{\min }}{1-z_{\max }}\right) \right) \label{for:propalt1} \\
I^{q}_{g}(Q_{j}^{2}) & = & \frac{n_{f}}{2}\left( z_{\max }-z_{\min }\right) \label{for:propalt2} \\
I^{q}_{q}(Q_{j}^{2}) & = & \frac{8}{3}\ln \left( \frac{1-z_{\min }}{1-z_{\max }}\right) \, .\label{for:propalt3} 
\end{eqnarray}
Defining \( I_{j}(Q_{j}^{2}) \) as
\begin{equation}
\label{i-j-ii}
I_{j}(Q_{j}^{2})=\left\{ I^{g}_{g}(Q_{j}^{2})+I^{q}_{g}(Q_{j}^{2})\right\} \, \delta ^{g}_{j}+I^{q}_{q}(Q_{j}^{2})\, \delta ^{q}_{j}\, ,
\end{equation}
 we propose the value \( Q_{j}^{2} \) based upon the probability distribution
\begin{equation}
f_{j}(Q_{j}^{2})=-\frac{\alpha _{\max }}{2\pi }\, I_{j}(Q_{j}^{2})\, \frac{1}{Q_{j}^{2}},
\end{equation}
 with \( \alpha _{\mathrm{max}}=\alpha _{s}(p_{\bot \mathrm{min}}^{2})=\alpha _{s}(p_{\bot \mathrm{fin}}^{2}) \).
The flavor \( k \) of the daughter parton is then chosen according to partial
contributions \( I^{k}_{j}(Q_{j}^{2}) \) in (\ref{i-j-ii}), and the value
of \( z \) according to the functions \( \bar{P}^{k}_{j}(z) \). 

The proposed values for \( Q_{j}^{2}=Q^{2} \), \( k \), and \( z \) are accepted
according to the probability 
\begin{equation}
\frac{\alpha _{s}(p_{\bot }^{2})}{\alpha _{\max }}\, w_{j}^{k}\, ,
\end{equation}
 with \( p^{2}_{\perp }=z(1-z)Q_{j}^{2} \) and
\begin{eqnarray}
w_{g}^{g}= & (1-z(1-z))^{2}, & \\
w_{g}^{q}= & z^{2}+(1-z)^{2}, & \\
w_{q}^{g}= & (1+z^{2})/2\, .
\end{eqnarray}
Otherwise, the proposal is rejected and one looks for another splitting with
\( Q^{2}_{j\, \mathrm{max}}=Q_{j}^{2} \).

\cleardoublepage

\chapter{Hadron-Hadron Amplitudes\label{ax-b-2}}

In this appendix, we discuss the hadron-hadron scattering amplitude \( T_{h_{1}h_{2}} \),
where \( h_{1} \) and \( h_{2} \) represent any pair of hadrons.

\section{Neglecting Valence Quark Scatterings}

We start with the general expression for hadron-hadron scattering amplitude,
eq.\ (\ref{t-h1-h2}),
\begin{eqnarray}
iT_{h_{1}h_{2}}(s,t)=\sum ^{\infty }_{n=1}\frac{1}{n!}\int \! \prod ^{n}_{l=1}\! \left[ \frac{d^{4}k_{l}}{(2\pi )^{4}}\frac{d^{4}k_{l}'}{(2\pi )^{4}}\frac{d^{4}q_{l}}{(2\pi )^{4}}\right] \, N_{h_{1}}^{(n)}\! \left( p,k_{1},\ldots ,k_{n},q_{1},\ldots ,q_{n}\right)  &  & \nonumber \\
\times \; \prod ^{n}_{l=1}\! \left[ iT_{1\mathrm{I}\! \mathrm{P}}\! (\hat{s}_{l},q_{l}^{2})\right] \, N_{h_{2}}^{(n)}\! \left( p',k_{1}',\ldots ,k_{n}',-q_{1},\ldots ,-q_{n}\right) \, (2\pi )^{4}\delta ^{(4)}\! \left( \sum ^{n}_{k=1}\! q_{i}-q\right) , &  & 
\end{eqnarray}
 with \( t=q^{2} \), \( s=(p+p')^{2}\simeq p^{+}p'^{-} \), and with \( p,p' \)
being the 4-momenta of the initial hadrons. We consider for simplicity identical
parton constituents (neglecting valence quark scatterings) and take \( T_{1\mathrm{I}\! \mathrm{P}} \)
to be the sum of the soft Pomeron exchange amplitude (see eq.\ (\ref{tsoft}))
and the semi-hard sea-sea scattering amplitude (see eq.\ (\ref{t-sea-sea})):
\begin{equation}
T_{1\mathrm{I}\! \mathrm{P}}\! (\hat{s}_{l},q^{2}_{l})=T_{\mathrm{soft}}(\hat{s}_{l},q^{2}_{l})+T_{\mathrm{sea}-\mathrm{sea}}(\hat{s}_{l},q^{2}_{l}),
\end{equation}
 with \( \hat{s}_{l}=(k_{l}+k_{l}')^{2}\simeq k_{l}^{+}k_{l}'^{-} \). The momenta
\( k_{l},k_{l}' \) and \( q_{l} \) denote correspondingly the 4-momenta of
the initial partons for the \( l \)-th scattering and the 4-momentum transfer
in that partial process. The factor \( 1/n! \) takes into account the identical
nature of the \( n \) scattering contributions. \( N_{h}^{(n)}\! \left( p,k_{1},\ldots ,k_{n},q_{1},\ldots ,q_{n}\right)  \)
denotes the contribution of the vertex for \( n \)-parton coupling to the hadron
\( h \). 

We assume that the initial partons \( k_{l},k_{l}' \) are characterized by
small virtualities \( k^{2}_{l}\sim -s_{0} \), \( k_{l}'^{2}\sim -s_{0} \),
and therefore by small transverse momenta \( k^{2}_{l_{\perp }}<s_{0} \), \( k_{l_{\perp }}'^{2}<s_{0} \),
so that the general results of the analysis made in \cite{gri68,bak76} are
applicable for the hadron-parton vertices \( N_{h}^{(n)} \). Using
\begin{equation}
d^{4}k_{l}=\frac{1}{2}dk_{l}^{+}dk_{l}^{-}d^{2}k_{l_{\perp }},\qquad d^{4}q_{l}=\frac{1}{2}dq_{l}^{+}dq_{l}^{-}d^{2}q_{l_{\perp }},
\end{equation}
 we can perform the integrations over \( k^{-}_{l},k_{l_{\perp }},q_{l}^{-} \)
and \( k_{l}'^{+},k_{l_{\perp }}',q_{l}^{+} \) separately for the upper and
the lower vertexes by making use of 
\begin{equation}
q^{2}_{l}\simeq -q_{l_{\perp }}^{2},\qquad k_{l}^{-},q_{l}^{-}\ll k_{l}'^{-},\qquad k_{l}'^{+},q_{l}^{+}\ll k_{l}^{+},
\end{equation}
 as well as the fact that the integrals \( dk_{l}^{+},dk_{l}'^{-} \) are restricted
by the physical region 
\begin{equation}
0<k_{l}^{+}\leq p^{+},\qquad \sum _{l}k_{l}^{+}\leq p^{+},
\end{equation}
 (similar for \( k_{l}'^{-} \)) \cite{bak76}. We shall consider explicitly
the upper vertex as for the lower one the derivation is identical. The integrals
over \( k^{-}_{l},q_{l}^{-} \) are defined by the discontinuities of the analytic
amplitude \( N_{h}^{(n)} \) with respect to the singularities in the corresponding
energy invariants 
\begin{eqnarray}
 & s_{1}^{+}=(p-k_{1})^{2}\simeq -p^{+}k_{1}^{-}, & \nonumber \\
 & ... & \\
 & s_{n}^{+}=(p-k_{1}-\cdots -k_{n})^{2}, & \nonumber 
\end{eqnarray}
 and 
\begin{eqnarray}
 & s_{q_{1}}^{+}=(p+q_{1})^{2}\simeq p^{+}q_{1}^{-}, & \nonumber \\
 & ... & \\
 & s_{q_{n-1}}^{+}=(p+q_{1}+\cdots +q_{n-1})^{2}. & \nonumber 
\end{eqnarray}
 As the processes corresponding to large values of \( s_{l}^{+},s_{q_{l}}^{+} \)
need an explicit treatment (the so-called enhanced diagrams, see chapter 5),
we only get contributions from the region of large \( k_{l}^{+}\sim p^{+} \),
so that 
\begin{equation}
s_{l}^{+}\simeq -p^{+}(k_{1}^{-}+\cdots +k_{l}^{-})\sim s_{0}\left( \frac{p^{+}}{k_{1}^{+}}+\cdots +\frac{p^{+}}{k_{l}^{+}}\right) <M_{0}^{2},
\end{equation}
 where \( M_{0}^{2} \) is some minimal mass for the Pomeron asymptotics to
be applied. The similar argument holds for the momenta \( q_{l}^{-} \), such
that 
\begin{equation}
s_{q_{l}}^{+}\simeq p^{+}(q_{1}^{-}+\cdots +q_{l}^{-})<M_{0}^{2}.
\end{equation}
 Using 
\begin{equation}
dq_{l}^{-}=\frac{ds_{q_{l}}^{+}}{p^{+}},\qquad dk_{l}^{-}=\frac{dk_{l}^{2}}{k_{l}^{+}},
\end{equation}
 we get
\begin{eqnarray}
 &  & \int \! \prod ^{n}_{l=1}\! \left[ \frac{d^{4}k_{l}}{(2\pi )^{4}}\right] \, \prod ^{n-1}_{l=1}\! \left[ \frac{dq_{l}^{-}}{2\pi }\right] \, N_{h_{1}}^{(n)}\! \left( p,k_{1},\ldots ,k_{n},q_{1},\ldots ,q_{n}\right) \nonumber \\
 &  & \qquad =\int \! \prod ^{n}_{l=1}\! \left[ \frac{dk^{2}_{l}dk_{l}^{+}d^{2}k_{l_{\perp }}}{2(2\pi )^{4}k_{l}^{+}}\Theta \! \left( s_{l}^{+}\right) \right] \prod ^{n-1}_{l=1}\! \left[ \frac{ds_{q_{l}}^{+}}{2\pi p^{+}}\Theta \! \left( s_{q_{l}}^{+}\right) \right] \, \nonumber \\
 &  & \qquad \qquad \qquad \times \; disc_{s_{1}^{+},\ldots ,s_{n}^{+},s_{q_{1}}^{+},\ldots ,s_{q_{n-1}}^{+}}\, N_{h_{1}}^{(n)}\! \left( p,k_{1},\ldots ,k_{n},q_{1},\ldots ,q_{n}\right) \Theta \! \left( 1-\sum ^{n}_{j=1}\! x_{j}^{+}\right) \nonumber \\
 &  & \qquad \equiv \frac{1}{\left( p^{+}\right) ^{n-1}}\int ^{1}_{0}\! \prod ^{n}_{l=1}\! \frac{dx_{l}^{+}}{x_{l}^{+}}\, F_{h_{1}}^{(n)}\! \left( x_{1}^{+},\ldots x_{n}^{+},q_{1_{\perp }}^{2},\ldots ,q_{n_{\perp }}^{2}\right) \, \Theta \! \left( 1-\sum ^{n}_{j=1}\! x_{j}^{+}\right) .\label{f-n-x} 
\end{eqnarray}
 The only difference of the formula eq.\ (\ref{f-n-x}) from the traditional
expression for \( F_{h}^{(n)} \) in \cite{gri68,bak76} is the fact that we
keep explicitly the integrations over the light cone momentum shares of the
constituent partons \( x_{l}^{+}=k_{l}^{+}/p^{+} \). Further we assume that
the dependences on the light cone momentum fractions \( x_{l}^{+} \) and on
the momentum transfers along the Pomerons \( q_{l}^{2}\simeq -q_{l_{\perp }}^{2} \)
factorize in \( F_{h}^{(n)} \), and we use the Gaussian parameterization for
the latter one,
\begin{equation}
\label{factoriz}
F_{h}^{(n)}\! \left( x_{1}^{+},\ldots x_{n}^{+},q_{1_{\perp }}^{2},\ldots ,q_{n_{\perp }}^{2}\right) =\tilde{F}_{h}^{(n)}\! \left( x_{1}^{+},\ldots x_{n}^{+}\right) \, \exp \! \left( -R_{h}^{2}\sum ^{n}_{j=1}\! q_{j_{\perp }}^{2}\right) ,
\end{equation}
where the parameter \( R_{h}^{2} \) is known as the hadron Regge slope \cite{bak76}.
Based on the above discussion and a corresponding treatment of the lower part
of the diagram, eq.\ (\ref{sig-elem}) can be rewritten as
\begin{eqnarray}
iT_{h_{1}h_{2}}(s,t)=\sum ^{\infty }_{n=1}\frac{1}{n!}\int \! \prod ^{n-1}_{l=1}\! \left[ \frac{d^{2}q_{l_{\perp }}}{8\pi ^{2}s}\right] \, \int ^{1}_{0}\! \prod ^{n}_{l=1}\! \frac{dx_{l}^{+}}{x_{l}^{+}}\frac{dx_{l}^{-}}{x_{l}^{-}}\; \tilde{F}_{h_{1}}^{(n)}\! \left( x_{1}^{+},\ldots x_{n}^{+}\right) \, \tilde{F}_{h_{2}}^{(n)}\! \left( x_{1}^{-},\ldots x_{n}^{-}\right)  &  & \nonumber \label{t-hh-fact} \\
\times \; \prod ^{n}_{l=1}\! \left[ iT_{1\mathrm{I}\! \mathrm{P}}\! \left( \hat{s}_{l},-q_{l_{\perp }}^{2}\right) \, \exp \! \left( -\left[ R_{h_{1}}^{2}+R_{h_{2}}^{2}\right] q_{l_{\perp }}^{2}\right) \right] \, \Theta \! \left( 1-\sum ^{n}_{j=1}\! x_{j}^{+}\right) \, \Theta \! \left( 1-\sum ^{n}_{j=1}\! x_{j}^{-}\right)  &  & \label{t-hh-fact} 
\end{eqnarray}

\begin{figure}[htb]
{\par\centering \resizebox*{!}{0.1\textheight}{\includegraphics{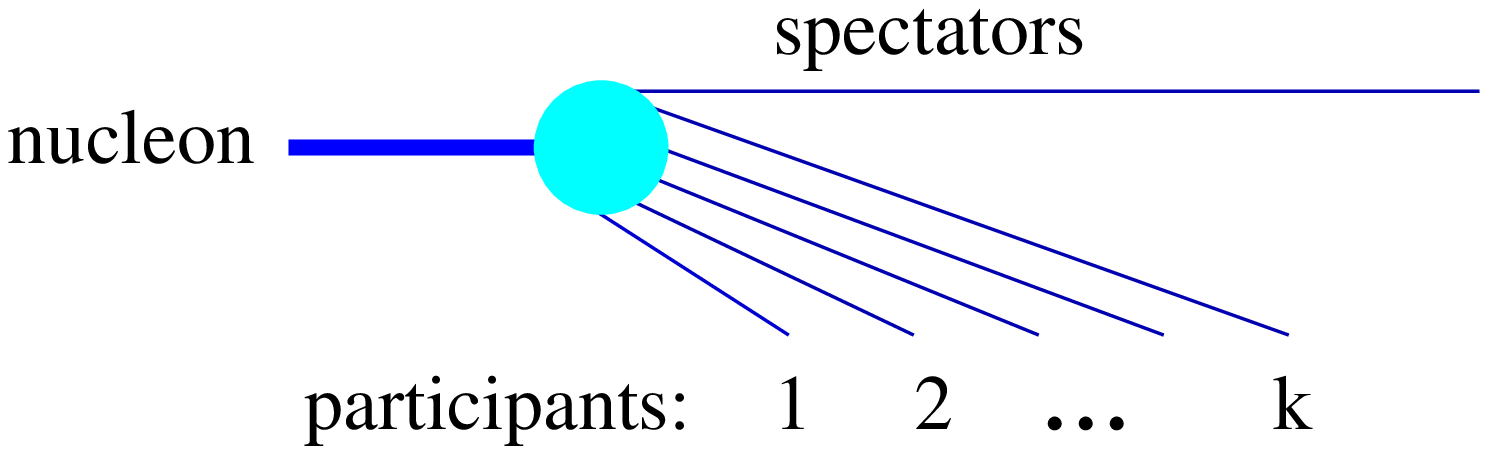}} \par}

\caption{Nucleon Fock state.\label{fig:fock}}
\end{figure}

The formula (\ref{t-hh-fact}) can be also obtained using the parton momentum
Fock state expansion of the hadron eigenstate \cite{abr92} 
\begin{equation}
|h\rangle =\sum ^{\infty }_{k=1}\frac{1}{k!}\int ^{1}_{0}\! \prod ^{k}_{l=1}\! dx_{l}\, f_{k}\! \left( x_{1},\ldots x_{k}\right) \, \delta \! \left( 1-\sum ^{k}_{j=1}\! x_{j}\right) \, a^{+}\! (x_{1})\cdots a^{+}\! (x_{k})\left| 0\right\rangle ,
\end{equation}
where \( f_{k}\! \left( x_{1},\ldots x_{k}\right)  \) is the probability amplitude
for the hadron \( h \) to consist of \( k \) constituent partons with the
light cone momentum fractions \( x_{1},\ldots ,x_{k} \) and \( a^{+}\! \left( x\right)  \)
is the creation operator for a parton with the fraction \( x \). \( f_{k}\! \left( x_{1},\ldots x_{k}\right)  \)
satisfies the normalization condition
\begin{equation}
\label{normalization}
\sum ^{\infty }_{k=1}\frac{1}{k!}\int ^{1}_{0}\! \prod ^{k}_{l=1}\! dx_{l}\, \left| f_{k}\! \left( x_{1},\ldots x_{k}\right) \right| ^{2}\, \delta \! \left( 1-\sum ^{k}_{j=1}\! x_{j}\right) =1
\end{equation}
 Then, for the contribution of \( n \) pair-like scatterings between the parton
constituents of the projectile and target hadrons one obtains eq.\ (\ref{t-hh-fact}),
as shown in \cite{abr92}, with 
\begin{equation}
\frac{1}{n!}\tilde{F}_{h}^{(n)}\! \left( x_{1},\ldots x_{n}\right) =\sum ^{\infty }_{k=n}\frac{1}{k!}\frac{k!}{n!\, (k-n)!}\int ^{1}_{0}\! \prod ^{k}_{l=n+1}\! \! dx_{l}\; \left| f_{k}\! \left( x_{1},\ldots x_{k}\right) \right| ^{2}\, \delta \! \left( 1-\sum ^{k}_{j=1}\! x_{j}\right) 
\end{equation}
 representing the ``inclusive'' momentum distributions of \( n \) ``participating''
parton constituents, involved in the scattering process. From the normalization
condition (\ref{normalization}) follows the momentum conservation constraint
\begin{equation}
\label{mom-cons}
\int ^{1}_{0}\! dx\, x\tilde{F}_{h}^{(1)}\! \left( x\right) =1
\end{equation}

To get further simplifications, we assume that \( \tilde{F}_{h_{1}(h_{2})}^{(n)}\! \left( x_{1},\ldots x_{n}\right)  \)
can be represented in a factorized form as a product of the contributions \( F^{h}_{\mathrm{part}}(x_{l}) \),
depending on the momentum shares \( x_{l} \) of the ``participating'' or
``active'' parton constituents, and on the function \( F^{h}_{\mathrm{remn}}\! \left( 1-\sum ^{n}_{j=1}x_{j}\right)  \),
representing the contribution of all ``spectator'' partons, sharing the remaining
share \( 1-\sum _{j}x_{j} \) of the initial light cone momentum (see fig.\ \ref{fig:fock}).
So we have 
\begin{equation}
\label{f-part-remn}
\tilde{F}_{h}^{(n)}\! \left( x_{1},\ldots x_{n}\right) =\prod ^{n}_{l=1}\! F^{h}_{\mathrm{part}}(x_{l})\, \; F^{h}_{\mathrm{remn}}\! \left( 1-\sum ^{n}_{j=1}x_{j}\right) 
\end{equation}
 The participating parton constituents are assumed to be quark-anti-quark pairs
(not necessarily of identical flavors), such that the baryon numbers of the
projectile and of the target are conserved. So we have \( x=x_{q}+x_{\bar{q}} \)
with \( x_{q} \) and \( x_{\bar{q}} \) being the light-cone momentum fractions
of the quark and the anti-quark. The function \( F^{h}_{\mathrm{part}} \) may
thus be written as
\begin{equation}
F^{h}_{\mathrm{part}}(x)=\int \! dx_{q}dx_{\bar{q}}\; \bar{F}^{h}_{\mathrm{part}}(x_{q},x_{\bar{q}})\; \delta (x-x_{q}-x_{\bar{q}}).
\end{equation}
 In case of soft or semi-hard Pomerons, \( \bar{F}^{h}_{\mathrm{part}} \) is
taken as a product of two asymptotics \( x_{i}^{-\alpha _{q}},\; i=q,\bar{q} \),
so we have
\begin{equation}
\label{f-part}
F^{h}_{\mathrm{part}}(x)=\gamma _{h}x^{-\alpha _{\mathrm{part}}},
\end{equation}
with \( \alpha _{\mathrm{part}}=2\alpha _{q}-1 \). The parameter \( \alpha _{q} \)
defines the probability to slow down the constituent (``dressed'') (anti-)quark;
it is related to the Regge intercept of the \( q\bar{q} \)-trajectory \cite{kai84}:
\( \alpha _{q}=\alpha _{\mathrm{I}\! \mathrm{R}}\! (0)\simeq 0.5 \). The remnant
function \( F^{h}_{\mathrm{remn}} \) defines the probability to slow down the
initial hadron quark configuration; it is assumed to be of the form
\begin{equation}
\label{f-remn}
F^{h}_{\mathrm{remn}}\! (x)=x^{\alpha ^{h}_{\mathrm{remn}}},
\end{equation}
with an adjustable parameter \( \alpha ^{h}_{\mathrm{remn}} \). Using (\ref{f-part-remn}-\ref{f-remn}),
the eq.\ (\ref{t-hh-fact}) can be rewritten as
\begin{eqnarray}
 &  & iT_{h_{1}h_{2}}(s,t)=8\pi ^{2}s\sum ^{\infty }_{n=1}\frac{1}{n!}\, \int ^{1}_{0}\! \prod ^{n}_{l=1}\! dx_{l}^{+}dx_{l}^{-}\; \prod ^{n}_{l=1}\! \left[ \frac{1}{8\pi ^{2}\hat{s}_{l}}\int \! d^{2}q_{l_{\perp }}\, iT_{1\mathrm{I}\! \mathrm{P}}^{h_{1}h_{2}}\! \left( x_{l}^{+},x_{l}^{-},s,-q_{l_{\perp }}^{2}\right) \right] \nonumber \label{t-hh-part-remn} \\
 &  & \qquad \times \; F^{h_{1}}_{\mathrm{remn}}\! \! \left( 1-\sum ^{n}_{j=1}\! x_{j}^{+}\right) \, F^{h_{2}}_{\mathrm{remn}}\! \! \left( 1-\sum ^{n}_{j=1}\! x_{j}^{-}\right) \: \delta ^{(2)}\! \left( \sum ^{n}_{k=1}\! \vec{q}_{k_{\perp }}-\vec{q}_{\perp }\right) .\label{th1h2} 
\end{eqnarray}
with
\begin{equation}
T_{1\mathrm{I}\! \mathrm{P}}^{h_{1}h_{2}}\! \left( x_{l}^{+},x_{l}^{-},s,-q_{l_{\perp }}^{2}\right) =T_{1\mathrm{I}\! \mathrm{P}}\! \left( \hat{s}_{l},-q_{l_{\perp }}^{2}\right) \, F^{h_{1}}_{\mathrm{part}}(x^{+}_{l})\, F^{h_{2}}_{\mathrm{part}}(x^{-}_{l})\, \exp \! \left( -\left[ R_{h_{1}}^{2}+R_{h_{2}}^{2}\right] q_{l_{\perp }}^{2}\right) 
\end{equation}
representing the contributions of ``elementary interactions plus external legs''.

\section{Including Valence Quark Hard Scatterings }

To include valence quark hard scatterings one has to replace the inclusive parton
momentum distributions 
\begin{equation}
\frac{1}{n!}\tilde{F}_{h}^{(n)}\! \left( x_{1},\ldots x_{n}\right) 
\end{equation}
 in (\ref{t-hh-fact}) by the momentum distributions 
\begin{equation}
\frac{1}{n_{v}!\, (n-n_{v})!}\tilde{F}_{h}^{(n,n_{v})i_{1},\ldots i_{n_{v}}}\! \left( x_{v_{1}},\ldots ,x_{v_{n_{v}}},x_{n_{v}+1},\ldots ,x_{n}\right) ,
\end{equation}
corresponding to the case of \( n_{v} \) partons being valence quarks with
flavors \( i_{1},\ldots ,i_{n_{v}} \) (taken at the virtuality scale \( Q_{0}^{2} \))
and other \( n-n_{v} \) partons being usual non-valence participants (quark-anti-quark
pairs). One has as well to take into account different contributions for scatterings
between a pair of valence quarks or between a valence quark and a non-valence
participant. In particular, for a single hard scattering on a valence quark
we have to use
\begin{equation}
\tilde{F}_{h}^{(1,1)}\! \left( x_{v}\right) =q^{i}_{\mathrm{val}}(x_{\nu }),
\end{equation}
 where \( q^{i}_{\mathrm{val}} \) is the momentum distribution of a valence
quark of flavor \( i \) at the scale \( Q^{2}_{0} \). In order to conserve
the initial hadron baryon content and to keep the simple factorized structure
(\ref{f-part-remn}), we associate a ``quasi-spectator'' anti-quark to each
valence quark interaction, defining the joint contribution \( \bar{F}_{\mathrm{part}}^{i}\left( x_{v},x_{\bar{q}}\right)  \)
of the valence quark with the flavor \( i \) and the anti-quark. Thus we have
\begin{eqnarray}
 &  & \tilde{F}_{h}^{(n,n_{v})i_{1},\ldots ,i_{n_{v}}}\! \left( x_{v_{1}},\ldots ,x_{v_{n_{v}}},x_{n_{v+1}},\ldots ,x_{n}\right) \label{f-fact-val} \\
 &  & \qquad =\prod ^{n_{v}}_{l=1}\! \left[ \int ^{1}_{x_{v_{l}}}\! dx_{l}\, \bar{F}_{\mathrm{part}}^{h,i_{l}}\left( x_{v_{l}},x_{l}-x_{v_{l}}\right) \right] \; \prod ^{n}_{m=n_{v}+1}\! F^{h}_{\mathrm{part}}(x_{m})\; F^{h}_{\mathrm{remn}}\left( 1-\sum ^{n}_{k=1}x_{k}\right) ,\nonumber \label{f-fact-val} 
\end{eqnarray}
where \( x_{l} \) is the sum of the momentum fractions of \( l^{\mathrm{th}} \)
valence quark and the corresponding anti-quark; we allow here formally any number
of valence quark participants (based on the fact that multiple valence type
processes give negligible contribution to the scattering amplitude). By construction
the integral over \( x_{\bar{q}} \) of the function \( \tilde{F}_{h}^{_{(1,1)i}}\! \left( x_{v},x_{\bar{q}}\right)  \)
gives the inclusive momentum distribution of the valence quark \( i \). Thus
the function \( \bar{F}^{i}_{\mathrm{part}} \) has to meet the condition 
\begin{equation}
\int ^{1-x_{v}}_{0}dx_{\bar{q}}\, \bar{F}_{\mathrm{part}}^{h,i}(x_{v},x_{\bar{q}})\; F^{h}_{\mathrm{remn}}\! \left( 1-x_{v}-x_{\bar{q}}\right) =q^{i}_{\mathrm{val}}(x_{v},Q_{0}^{2})
\end{equation}
 Assuming that the anti-quark momentum distribution behaves as \( (x_{\bar{q}})^{-\alpha _{R}} \),
and using the above-mentioned parameterization for \( F^{h}_{\mathrm{remn}} \),
we get 
\begin{equation}
\bar{F}^{h,i}_{\mathrm{part}}(x_{v},x_{\bar{q}})=N^{-1}\, q^{i}_{\mathrm{val}}(x_{v},Q_{0}^{2})(1-x_{v})^{\alpha _{\mathrm{I}\! \mathrm{R}}-1-\alpha _{\mathrm{remn}}}(x_{\bar{q}})^{-\alpha _{\mathrm{I}\! \mathrm{R}}},
\end{equation}
with the normalization factor

\begin{equation}
N=\frac{\Gamma \! \left( 1+\alpha _{\mathrm{remn}}\right) \, \Gamma \! \left( 1-\alpha _{\mathrm{I}\! \mathrm{R}}\right) }{\Gamma \! \left( 2+\alpha _{\mathrm{remn}}-\alpha _{\mathrm{I}\! \mathrm{R}}\right) }.
\end{equation}
 Now we can write the normalization condition (\ref{mom-cons}) for ``active''
(participating in an interaction) partons as
\begin{equation}
\int ^{1}_{0}\! dx\, xF^{h}_{\mathrm{part}}\! \left( x\right) \, F^{h}_{\mathrm{remn}}\! \left( 1-x\right) +\sum _{i}\int ^{1}_{0}\! x_{v}dx\int ^{x}_{0}\! dx_{v}\, \bar{F}_{\mathrm{part}}^{h,i}(x_{v},x-x_{v})\; F^{h}_{\mathrm{remn}}\! \left( 1-x\right) =1,
\end{equation}
which gives
\begin{equation}
\gamma _{h}=(1-\langle x_{v}\rangle )\frac{\Gamma \! \left( 3-\alpha _{\mathrm{part}}+\alpha ^{h}_{\mathrm{remn}}\right) }{\Gamma \! \left( 2-\alpha _{\mathrm{part}}\right) \, \Gamma \! \left( 1+\alpha ^{h}_{\mathrm{remn}}\right) },
\end{equation}
with 
\begin{equation}
\langle x_{v}\rangle =\sum _{i}\int ^{1}_{0}\! x_{v}dx_{v}\, q^{i}_{\mathrm{val}}(x_{v},Q_{0}^{2})
\end{equation}
 being the average light cone momentum fraction carried by valence quarks.

\section{Enhanced Diagrams\label{ax-c-3}}

In this appendix, we demonstrate how triple Pomeron contributions appear naturally
in the Gribov-Regge formalism under certain kinematical conditions, and we derive
a formula for the hadron-hadron scattering amplitude in this case.

\begin{figure}[htb]
{\par\centering \resizebox*{!}{0.15\textheight}{\includegraphics{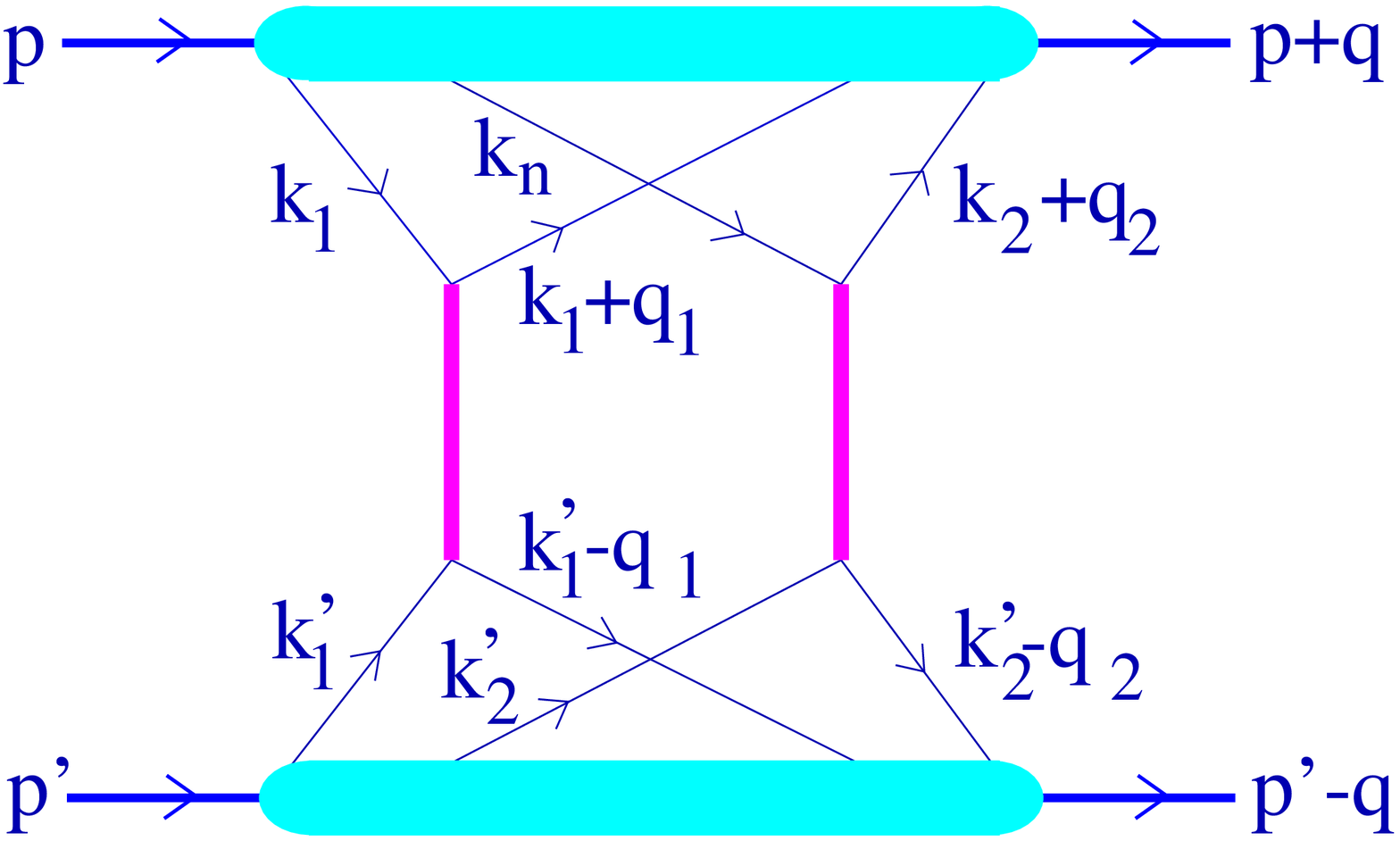}} \par}

\caption{Double Pomeron exchange. \label{2pomx}}
\end{figure}

To introduce enhanced type diagrams let us come back to the process of double
soft Pomeron exchange, which is a particular case of the diagram of fig \ref{wave}.
The corresponding contribution to the elastic scattering amplitude is given
in eqs.\ (\ref{t-h1-h2}), (\ref{f-n-x}) with \( n=2 \) and with \( T_{1\mathrm{I}\! \mathrm{P}} \)
being replaced by \( T_{\mathrm{soft}} \):

\begin{eqnarray}
iT^{(2)}_{h_{1}h_{2}}(s,t) & = & \frac{1}{2}\int \! \frac{d^{4}k_{1}}{(2\pi )^{4}}\frac{d^{4}k_{1}'}{(2\pi )^{4}}\frac{d^{4}k_{2}}{(2\pi )^{4}}\frac{d^{4}k_{2}'}{(2\pi )^{4}}\frac{d^{4}q_{1}}{(2\pi )^{4}}\Theta \! \left( s_{1}^{+}\right) \Theta \! \left( s_{1}^{-}\right) \Theta \! \left( s_{2}^{+}\right) \Theta \! \left( s_{2}^{-}\right) \Theta \! \left( s_{q_{1}}^{+}\right) \Theta \! \left( s_{q_{1}}^{-}\right) \nonumber \label{t-2-elem} \\
 &  & \qquad \times \: \mathrm{disc}_{s_{1}^{+},s_{2}^{+},s_{q_{1}}^{+}}\, N_{h_{1}}^{(2)}\! \left( p,k_{1},k_{2},q_{1},q-q_{1}\right) \nonumber \\
 &  & \qquad \times \; \prod ^{2}_{l=1}\! \left[ iT_{\mathrm{soft}}\! (\hat{s}_{l},q_{l}^{2})\right] \, \mathrm{disc}_{s_{1}^{-},s_{2}^{-},s_{q_{1}}^{-}}\, N_{h_{2}}^{(2)}\! \left( p',k_{1}',k_{2}',-q_{1},-q+q_{1}\right) \label{t-2-elem} 
\end{eqnarray}
 see fig.\ \ref{2pomx}.

\begin{figure}[htb]
{\par\centering \resizebox*{!}{0.2\textheight}{\includegraphics{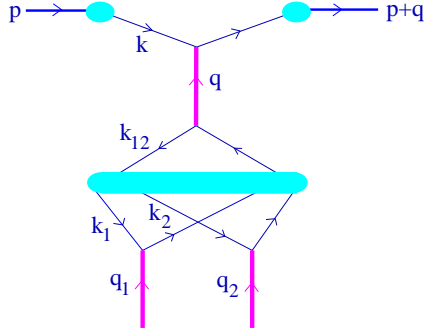}} \par}

\caption{Three Pomeron coupling. \label{3pom}}
\end{figure}

We are now interested in the contribution with some of the invariants 
\begin{eqnarray}
s_{1}^{+} & = & (p-k_{1})^{2}\simeq -p^{+}k_{1}^{-},\nonumber \\
s_{2}^{+} & = & (p-k_{1}-k_{2})^{2}\simeq -p^{+}(k_{1}^{-}+k_{2}^{-}),\\
s_{q_{1}}^{+} & = & (p+q_{1})^{2}\simeq p^{+}q_{1}^{-},\nonumber 
\end{eqnarray}
 being large, implying \( k_{i}^{-},q^{-}_{1} \) to be not too small. As was
shown in \cite{bak76}, one can restrict the integration region to \( s_{i}^{+}\leq s_{q_{1}}^{+} \),
because \( s_{i}^{+}\gg s_{q_{1}}^{+} \), (\( k_{i}^{-}\gg q_{1}^{-} \) )
either correspond to processes with exchange of more than two Pomerons or to
the Pomeron self-coupling, the latter one just renormalizing the Pomeron amplitude.
Then from 
\begin{equation}
k_{i}^{2},\: k_{i}'^{2},\: (k_{1}'-q_{1})^{2},\: (k_{1}+q_{1})^{2},\: (k_{2}-q_{1}-q)^{2}\sim -s_{0}
\end{equation}
 it follows that 
\begin{equation}
k_{i}^{-}\sim -\frac{s_{0}}{k_{i}^{+}},\quad k_{i}'^{+}\sim -\frac{s_{0}}{k_{i}'^{-}},\quad q_{1}^{+}\sim -\frac{s_{0}}{k_{1}'^{-}},\quad q_{1}^{-}\sim \frac{s_{0}}{k_{1}^{+}},\frac{s_{0}}{k_{2}^{+}}
\end{equation}
 and correspondingly \( k_{1}^{+}\sim k_{2}^{+} \) and the invariants \( s_{1}^{+},s_{2}^{+},s_{q_{1}}^{+} \)
are of the same order. The vertex \( N_{h}^{(2)} \) for large \( s_{q_{1}}^{+} \)
can be described by the soft Pomeron asymptotics and we may write \( disc_{s_{1}^{+},s_{2}^{+},s_{q_{1}}^{+}}\, N_{h_{1}}^{(2)}\! \left( p,k_{1},k_{2},q_{1},q-q_{1}\right)  \)
as a product of the Pomeron-hadron coupling \( N_{h}^{(1)}\! \left( p,k,q\right)  \),
twice imaginary part of the soft Pomeron exchange amplitude \( 2\mathrm{Im}\, T_{\mathrm{soft}}\! \left( (k-k_{12})^{2},q^{2}\right)  \)
with \( k_{12}=k_{1}+k_{2} \), and a term \( V^{\rm {3P}}\! \left( k_{12},k_{1},k_{2},q,q_{1},q-q_{1}\right)  \)
describing the coupling of the three Pomerons\footnote{
The triple Pomeron vertex is assumed to have nonplanar structure, corresponding
to having the two lower Pomerons ``in parallel''; the planar triple-Pomeron
vertex would correspond to subsequent emission of these Pomerons and gives no
contribution in the high energy limit \cite{bor91}.
}. So we get
\begin{eqnarray}
 &  & \int \! \frac{d^{4}k_{1}}{(2\pi )^{4}}\frac{d^{4}k_{2}}{(2\pi )^{4}}\frac{dq^{-}_{1}}{2\pi }\Theta \! \left( s_{1}^{+}\right) \Theta \! \left( s_{2}^{+}\right) \Theta \! \left( s_{q_{1}}^{+}\right) \, disc_{s_{1}^{+},s_{2}^{+},s_{q_{1}}^{+}}\, N_{h_{1}}^{(2)}\! \left( p,k_{1},k_{2},q_{1},q-q_{1}\right) \nonumber \\
 &  & \qquad =\int \! \frac{d^{4}k_{1}}{(2\pi )^{4}}\frac{d^{4}k_{2}}{(2\pi )^{4}}\frac{dq^{-}_{1}}{2\pi }\Theta \! \left( s_{1}^{+}\right) \Theta \! \left( s_{2}^{+}\right) \Theta \! \left( s_{q_{1}}^{+}\right) \frac{d^{4}k}{(2\pi )^{4}}\frac{d^{4}k_{12}}{(2\pi )^{4}}(2\pi )^{4}\delta \! (k_{12}-k_{1}-k_{2})\nonumber \\
 &  & \qquad \qquad \qquad \, \times \; N_{h_{1}}^{(1)}\! \left( p,k,q\right) \, 2\mathrm{Im}\, T_{\mathrm{soft}}\! \left( (k-k_{12})^{2},q^{2}\right) \, V^{3\mathrm{I}\! \mathrm{P}}\! \left( k_{12},k_{1},k_{2},q,q_{1},q-q_{1}\right) \nonumber \\
 &  & \qquad =\int \! \frac{dk^{+}dk^{2}d^{2}k_{\perp }}{2k^{+}(2\pi )^{4}}\Theta \! \left( s_{0}^{+}\right) \, disc_{s_{0}^{+}}\, N_{h_{1}}^{(1)}\! \left( p,k,q\right) \, \int \! \frac{dk_{12}^{+}dk_{12}^{2}d^{2}k_{12_{\perp }}}{2k_{12}^{+}(2\pi )^{4}}\, 2\mathrm{Im}\, T_{\mathrm{soft}}\! \left( (k-k_{12})^{2},q^{2}\right) \nonumber \\
 &  & \qquad \qquad \qquad \times \; \int \! \frac{dk_{1}^{+}dk_{1}^{2}d^{2}k_{1_{\perp }}}{2k_{1}^{+}(2\pi )^{4}}\frac{d(k_{12}-k_{1}-q_{1}-q)^{2}}{2\pi (k_{12}^{+}-k_{1}^{+})}\Theta \! \left( s_{1}^{+}\right) \Theta \! \left( s_{2}^{+}\right) \Theta \! \left( s_{q_{1}}^{+}\right) \nonumber \label{n3p-v3p} \\
 &  & \qquad \qquad \qquad \times \; V^{3\mathrm{I}\! \mathrm{P}}\! \left( k_{12},k_{1},k_{12}-k_{1},q,q_{1},q-q_{1}\right) \label{n3p-v3p} 
\end{eqnarray}
 see fig.\ \ref{3pom},
 with \( s_{0}^{+}=(p-k)^{2}\simeq -p^{+}k^{-} \) and \( dq_{1}^{-}=\frac{1}{k_{2}^{+}}d(k_{2}-q_{1}-q)^{2} \).
Now we can perform the integrations over \( k^{2},k_{12}^{2},k_{1}^{2},(k_{2}-q_{1}-q)^{2} \)
assuming their convergence in the region \( k_{i}^{2}\sim -s_{0} \), as well
as over \( k_{\perp },k_{12_{\perp }},k_{1_{\perp }}\leq s_{0} \) to transform
eq.\ (\ref{n3p-v3p}) to the form
\begin{eqnarray}
\int \! dx^{+}\, F^{h_{1}}_{\mathrm{part}}(x^{+})\, F^{h_{1}}_{\mathrm{remn}}\! \left( 1-x^{+}\right) \, \exp \! \left( -R_{h_{1}}^{2}q_{\perp }^{2}\right) \, \int \! \frac{dx_{12}^{+}}{x_{12}^{+}}\, \frac{1}{2s^{+}}\, \mathrm{Im}\, T_{\mathrm{soft}}\! \left( s^{+},-q_{\perp }^{2}\right)  & \times  & \nonumber \\
\times \; \int \! \frac{dz^{+}}{x_{1}^{+}\, (x_{12}^{+}-x_{1}^{+})\, p^{+}}\, \tilde{V}^{3\mathrm{I}\! \mathrm{P}}\! \left( -q^{2}_{\perp },-q^{2}_{1_{\perp }},-\left( \vec{q}_{\perp }-\vec{q}_{1_{\perp }}\right) ^{2}\right)  &  & \label{3p-vertex-fin} 
\end{eqnarray}
with 
\begin{eqnarray}
 &  & x^{+}=k^{+}/p^{+},\\
 &  & x_{12}^{+}=k_{12}^{+}/p^{+},\\
 &  & x_{1}^{+}=k_{1}^{+}/p^{+},\\
 &  & x_{12}^{+}-x_{1}^{+}=(k^{+}_{12}-k_{1}^{+})/p^{+}=k_{2}^{+}/p^{+},\\
 &  & z^{+}=k_{1}^{+}/k_{12}^{+}=x_{1}^{+}/x_{12}^{+},\\
 &  & s^{+}=(k-k_{12})^{2}\simeq -k^{+}k_{12}^{-}\simeq s_{0}k^{+}/k^{+}_{12}=s_{0}x^{+}/x^{+}_{12}.
\end{eqnarray}
We used (see eq.\ (\ref{f-n-x}-\ref{f-part-remn}))
\begin{equation}
\int \! \frac{dk^{2}d^{2}k_{\perp }}{2(2\pi )^{4}}\Theta \! \left( s_{0}^{+}\right) \, disc_{s_{0}^{+}}\, N_{h}^{(1)}\! \left( p,k,q\right) =F^{h}_{\mathrm{part}}(x^{+})\, F^{h}_{\mathrm{remn}}\! \left( 1-x^{+}\right) \, \exp \! \left( -R_{h}^{2}q_{\perp }^{2}\right) 
\end{equation}
and the definition
\begin{eqnarray}
\tilde{V}^{3\mathrm{I}\! \mathrm{P}}\! \left( -q^{2}_{\perp },-q^{2}_{1_{\perp }},-(\vec{q}_{\perp }-\vec{q}_{1_{\perp }})^{2}\right)  & = & s_{0}\int \! \frac{dk_{12}^{2}d^{2}k_{12_{\perp }}}{(2\pi )^{4}}\frac{dk_{1}^{2}d^{2}k_{1_{\perp }}}{(2\pi )^{4}}\frac{d\left[ (k_{12}-k_{1}-q_{1}-q)^{2}\right] }{2\pi }\\
 &  & \times \; \Theta \! \left( s_{1}^{+}\right) \Theta \! \left( s_{2}^{+}\right) \Theta \! \left( s_{q_{1}}^{+}\right) \, V^{3\mathrm{I}\! \mathrm{P}}\! \left( k_{12},k_{1},k_{12}-k_{1},q,q_{1},q-q_{1}\right) \: .\nonumber 
\end{eqnarray}
We have also taken into account the fact that the triple Pomeron vertex \( V^{3\mathrm{I}\! \mathrm{P}} \)
has a scalar structure, and we therefore suppose that it can only depend on
the invariants 
\begin{equation}
k_{i}^{2},\; (k_{1}+q_{1})^{2},\; (k_{2}-q_{1}-q)^{2}\sim -s_{0},\; q^{2}\simeq -q_{\perp }^{2},\; q_{1}^{2}\simeq -q_{1_{\perp }}^{2},\; q_{2}^{2}\simeq -(q_{\perp }-q_{1_{\perp }})^{2}
\end{equation}
 and on the partition of the light cone momentum \( k_{12}^{+} \) between the
two lower Pomerons, 
\[
z^{+}=k_{1}^{+}/k_{12}^{+}=(p'+k_{1})^{2}/(p'+k_{12})^{2}.\]
Furthermore, we did assume the flat distribution in \( z^{+} \) in order to
obtain \( k_{1}^{+}\sim k_{2}^{+}\sim k_{12}^{+}/2 \) (see the discussion above).
We use the Gaussian parameterization for the \( q_{i}^{2} \)-dependence of
\( \tilde{V}^{3\mathrm{I}\! \mathrm{P}} \),
\begin{equation}
\label{3p-param}
\tilde{V}^{3\mathrm{I}\! \mathrm{P}}\! \left( -q^{2}_{\perp },-q^{2}_{1_{\perp }},-q_{2_{\perp }}^{2}\right) \equiv r_{3\mathrm{I}\! \mathrm{P}}\, \exp \! \left( -R_{3\mathrm{I}\! \mathrm{P}}^{2}\left[ q^{2}_{\perp }+q_{1_{\perp }}^{2}+q_{2_{\perp }}^{2}\right] \right) ,
\end{equation}
where \( r_{3\mathrm{I}\! \mathrm{P}} \) is the triple-Pomeron coupling and
\( R_{3\mathrm{I}\! \mathrm{P}}^{2} \) is the slope of the triple-Pomeron vertex.
The slope \( R_{3\mathrm{I}\! \mathrm{P}}^{2} \) is known to be small and will
be neglected in the following.

Now, using (\ref{t-2-elem}-\ref{3p-param}) for the upper vertex and doing
the usual treatment (\ref{f-n-x}-\ref{f-part-remn}) of the lower one, we get
for the triple Pomeron amplitude
\begin{equation}
\label{t3p-x}
iT^{3\mathrm{I}\! \mathrm{P}\! -}_{h_{1}h_{2}}(s,t)=\int ^{1}_{0}\! \! \frac{dx^{+}}{x^{+}}\frac{dx^{-}}{x^{-}}\, F^{h_{1}}_{\mathrm{remn}}\! \left( 1-x^{+}\right) \, F^{h_{2}}_{\mathrm{remn}}\! \! \left( 1-x^{-}\right) \, iT^{h_{1}h_{2}}_{3\mathrm{I}\! \mathrm{P}\! -}(x^{+},x^{-},s,t)
\end{equation}
with
\begin{eqnarray}
 &  & iT_{3\mathrm{I}\! \mathrm{P}\! +}^{h_{1}h_{2}}(x^{+},x^{-},s,t)\: =\: 8\pi ^{2}x^{+}x^{-}s\, \frac{r_{3\mathrm{I}\! \mathrm{P}}}{2}\int ^{x^{+}}_{s_{0}/x^{-}}\! \frac{dx_{12}^{+}}{x_{12}^{+}}\left[ \frac{1}{2s^{+}}\, \mathrm{Im}\, T^{h_{1}}\! \left( x^{+},s^{+},-q_{\perp }^{2}\right) \right] \nonumber \\
 &  & \qquad \qquad \times \; \int dz^{+}\int \! d^{2}q_{1_{\perp }}d^{2}q_{2_{\perp }}\int ^{x^{-}}_{0}\! dx_{1}^{-}dx_{2}^{-}\prod ^{2}_{l=1}\! \left[ \frac{1}{8\pi ^{2}\hat{s}_{l}}\, iT^{h_{2}}\! \left( x_{l}^{-},\hat{s}_{l},-q_{l_{\perp }}^{2}\right) \right] \nonumber \\
 &  & \qquad \qquad \times \; \delta \! (x^{-}-x_{1}^{-}-x_{2}^{-})\, \delta ^{(2)}\! \left( \vec{q}_{\perp }-\vec{q}_{1_{\perp }}-\vec{q}_{2_{\perp }}\right) ,\label{t3p-px} 
\end{eqnarray}
with
\begin{equation}
T^{h}\! \left( x,\hat{s},-q_{\perp }^{2}\right) =T_{\mathrm{soft}}^{h}\! \left( x,\hat{s},-q_{\perp }^{2}\right) =T_{\mathrm{soft}}\! \left( \hat{s},-q_{\perp }^{2}\right) \, F^{h}_{\mathrm{part}}(x)\, \exp \! \left( -R_{h}^{2}\, q_{\perp }^{2}\right) 
\end{equation}
 and
\begin{equation}
\hat{s}_{1}=x_{12}^{+}z^{+}x_{1}^{-}s,\qquad \hat{s}_{2}=x_{12}^{+}(1-z^{+})x_{2}^{-}s.
\end{equation}
 The sign ``\( - \)'' in ``\( 3\mathrm{I}\! \mathrm{P}\! - \)'' refers
to the Pomeron ``splitting'' towards the target hadron (reversed \( Y \)-diagram);
the lower limit for the integral \( dx_{12}^{+} \) is due to \( x_{12}^{-}\simeq s_{0}/x_{12}^{+}<x^{-} \).

\section{Parton Generation for Triple-Pomeron Diagrams\label{ax-c-4}}

The inclusion of the triple-Pomeron contributions results only in slight modification
of the standard procedure. Now the full contribution of an elementary interaction
is
\begin{equation}
G_{1\mathrm{I}\! \mathrm{P}}^{h_{1}h_{2}}(x^{+},x^{-},s,b)+\sum _{\sigma }\sum ^{2}_{i=0}\widehat{\widehat{G}}_{3\mathrm{I}\! \mathrm{P}\! \sigma (i)}^{h_{1}h_{2}}(x^{+},x^{-},x^{\mathrm{proj}},x^{\mathrm{targ}},s,b),
\end{equation}
where \( x^{\mathrm{proj}},x^{\mathrm{targ}} \) are the corresponding remnant
light cone momentum fractions, and with the contributions of different cuts
of triple-Pomeron diagrams being obtained from eq. (\ref{g-hathat}), together
with eqs. (\ref{g-hathat}), (\ref{g-hat-2}-\ref{g-hat-1}) as 
\begin{eqnarray}
\widehat{\widehat{G}}_{3\mathrm{I}\! \mathrm{P}-(0)}^{h_{1}h_{2}}\! (x^{+},x^{-},x^{\mathrm{proj}},x^{\mathrm{targ}},s,b) & = & \frac{r_{3\mathrm{I}\! \mathrm{P}}}{8}\int \! d^{2}b_{1}\, \frac{1}{x^{-}}G^{h_{1}}(x^{+},x^{+}x^{-}s,\left| \vec{b}-\vec{b}_{1}\right| )\nonumber \\
 & \times  & \int ^{x^{\mathrm{targ}}}_{0}\! d\hat{x}^{-}\int ^{1}_{0}\! dz^{+}\int ^{\hat{x}^{-}+x^{-}}_{0}\! dx_{1}^{-}\, G^{h_{2}}(x_{1}^{-},x_{1}^{-}\frac{s_{0}}{x^{-}}z^{+}s,b_{1})\nonumber \label{g3p-(0)-approx} \\
 & \times  & G^{h_{2}}(\hat{x}^{-}+x^{-}-x_{1}^{-},(\hat{x}^{-}+x^{-}-x_{1}^{-})\frac{s_{0}}{x^{-}}(1-z^{+})s,b_{1})\nonumber \label{x} \\
 & \times  & \frac{F_{\mathrm{remn}}\left( x^{\mathrm{targ}}-\hat{x}^{-}\right) }{F_{\mathrm{remn}}\left( x^{\mathrm{targ}}\right) }\label{g3p-(0)-approx} 
\end{eqnarray}
and
\begin{eqnarray}
\widehat{\widehat{G}}_{3\mathrm{I}\! \mathrm{P}-(1)}^{h_{1}h_{2}}\! (x^{+},x^{-},x^{\mathrm{proj}},x^{\mathrm{targ}},s,b) & = & -\frac{r_{3\mathrm{I}\! \mathrm{P}}}{2}\int \! d^{2}b_{1}\int ^{x^{+}}_{s_{0}/(x^{-}s)}\! \frac{dx_{12}^{+}}{x_{12}^{+}}\, G^{h_{1}}(x^{+},s^{+},\left| \vec{b}-\vec{b}_{1}\right| )\nonumber \label{g3p-(1)-approx} \\
 & \times  & \int ^{1}_{0}\! dz^{+}\, G^{h_{2}}(x^{-},x_{12}^{+}z^{+}x^{-}s,b_{1})\nonumber \label{x} \\
 & \times  & \int ^{x^{\mathrm{targ}}}_{0}\! d\hat{x}^{-}G^{h_{2}}(\hat{x}^{-},x_{12}^{+}(1-z^{+})\hat{x}^{-}s,b_{1})\nonumber \\
 & \times  & \frac{F_{\mathrm{remn}}\left( x^{\mathrm{targ}}-\hat{x}^{-}\right) }{F_{\mathrm{remn}}\left( x^{\mathrm{targ}}\right) }\label{g3p-(1)-approx} 
\end{eqnarray}
and
\begin{eqnarray}
\widehat{\widehat{G}}_{3\mathrm{I}\! \mathrm{P}-(2)}^{h_{1}h_{2}}(x^{+},x^{-},x^{\mathrm{proj}},x^{\mathrm{targ}},s,b) & = & \frac{r_{3\mathrm{I}\! \mathrm{P}}}{4}\, \int d^{2}b_{1}\int ^{x^{+}}_{s_{0}/(sx^{-})}\! \frac{dx_{12}^{+}}{x_{12}^{+}}\, G^{h_{1}}(x^{+},s^{+},\left| \vec{b}-\vec{b}_{1}\right| )\label{g3p-(2)-approx} \\
 & \times  & \int ^{1}_{0}\! dz^{+}\int ^{x^{-}}_{0}\! dx_{1}^{-}\, G^{h_{2}}(x_{1}^{-},\hat{s}_{1},b_{1})\, G^{h_{2}}(x^{-}-x_{1}^{-},\hat{s}_{2},b_{1})\nonumber 
\end{eqnarray}
(similarly for \( \widehat{\widehat{G}}_{3\mathrm{I}\! \mathrm{P}+(i)}^{h_{1}h_{2}} \)).
Thus, the elementary process splits into a single elementary scattering contribution
-- with the weight
\begin{equation}
\widehat{\widehat{G}}_{1\mathrm{I}\! \mathrm{P}}^{h_{1}h_{2}}=G_{1\mathrm{I}\! \mathrm{P}}^{h_{1}h_{2}}+\sum _{\sigma }\sum ^{1}_{i=0}\widehat{\widehat{G}}_{3\mathrm{I}\! \mathrm{P}\! \sigma (i)}^{h_{1}h_{2}},
\end{equation}
and the contribution with all three Pomerons being cut -- with the weight 
\begin{equation}
\widehat{\widehat{G}}_{3\mathrm{I}\! \mathrm{P}\! +(2)}^{h_{1}h_{2}}+\widehat{\widehat{G}}_{3\mathrm{I}\! \mathrm{P}\! -(2)}^{h_{1}h_{2}}.
\end{equation}
 Choosing the first one, one proceeds in the usual way, with the functions \( G_{1\mathrm{I}\! \mathrm{P}}^{h_{1}h_{2}}, \)
\( G^{h_{1}h_{2}}_{\mathrm{soft}}, \) \( G^{h_{1}h_{2}}_{\mathrm{sea}-\mathrm{sea}}, \)
\( G^{h_{1}h_{2}}_{\mathrm{val}-\mathrm{sea}}, \) \( G^{h_{1}h_{2}}_{\mathrm{sea}-v\mathrm{al}} \)
in eq.\ (\ref{g-pom-basic}) being replaced by \( \widehat{\widehat{G}}_{1\mathrm{I}\! \mathrm{P}}^{h_{1}h_{2}}, \)
\( \widehat{\widehat{G}}^{h_{1}h_{2}}_{\mathrm{soft}}, \) \( \widehat{\widehat{G}}^{h_{1}h_{2}}_{\mathrm{sea}-\mathrm{sea}}, \)
\( \widehat{\widehat{G}}^{h_{1}h_{2}}_{\mathrm{val}-\mathrm{sea}}, \) \( \widehat{\widehat{G}}^{h_{1}h_{2}}_{\mathrm{sea}-\mathrm{val}} \),
where we introduce the functions \( \widehat{\widehat{G_{J}}}^{h_{1}h_{2}} \)
via
\begin{equation}
\widehat{\widehat{G_{J}}}^{h_{1}h_{2}}=G^{h_{1}h_{2}}_{J}+\widehat{\widehat{G}}_{3\mathrm{I}\! \mathrm{P}+(0)J}^{h_{1}h_{2}}+\widehat{\widehat{G}}_{3\mathrm{I}\! \mathrm{P}-(0)J}^{h_{1}h_{2}}+\widehat{\widehat{G}}_{3\mathrm{I}\! \mathrm{P}-(1)J}^{h_{1}h_{2}}+\widehat{\widehat{G}}_{3\mathrm{I}\! \mathrm{P}-(1)J}^{h_{1}h_{2}},
\end{equation}
 with \( J \) being ``soft'', ``sea-sea'', ``val-sea'' or ``sea-val''.
Here \( \widehat{\widehat{G}}_{3\mathrm{I}\! \mathrm{P}-(0)\mathrm{soft}}^{h_{1}h_{2}} \)
is given by the eq.\ (\ref{g3p-(0)-approx}) with \( G^{h_{1}}(x^{+},x^{+}x^{-}s,\left| \vec{b}-\vec{b}_{1}\right| ) \)
being replaced by its soft contribution \( G^{h_{1}}_{\mathrm{soft}}(x^{+},x^{+}x^{-}s,\left| \vec{b}-\vec{b}_{1}\right| ) \)
(see eq.\ (\ref{g-h-soft-3p})), whereas \( \widehat{\widehat{G}}_{3\mathrm{I}\! \mathrm{P}-(1)\mathrm{soft}}^{h_{1}h_{2}} \)
is given in (\ref{g3p-(1)-approx}) with both \( G^{h_{1}}(x^{+},s^{+},\left| \vec{b}-\vec{b}_{1}\right| ) \)
and \( G^{h_{2}}(x^{-},x_{12}^{+}z^{+}x^{-}s,b_{1}) \) being represented by
the soft contributions \( G^{h_{1}/h_{2}}_{\mathrm{soft}} \). The other functions
\( \widehat{\widehat{G}}_{3\mathrm{I}\! \mathrm{P}+(i)J}^{h_{1}h_{2}} \) are
defined similarly to the ``soft'' case - considering corresponding ``sea''
and ``valence'' contributions in the cut Pomerons \( G^{h_{1}/h_{2}} \) in
eq.\ (\ref{g3p-(0)-approx}-\ref{g3p-(1)-approx}).

For the contribution corresponding to the cases of all three Pomerons being
cut, we split the processes into three separate cut Pomeron pieces. Each piece
is characterized by the function \( \bar{G}_{i\pm }^{h_{1}h_{2}}(\bar{x}^{+},\bar{x}^{-},s,b_{1}) \),
where \( b_{1}\simeq b/2 \) and 
\begin{eqnarray}
 &  & \bar{G}_{1-}^{h_{1}h_{2}}(\bar{x}^{+},\bar{x}^{-},s,b_{1})=G^{h_{1}}(\bar{x}^{+},\bar{x}^{+}\bar{x}^{-}s,b_{1}),\quad \bar{x}^{+}=x^{+},\, \bar{x}^{-}=s_{0}/(x_{12}^{+}s),\\
 &  & \bar{G}_{2-}^{h_{1}h_{2}}(\bar{x}^{+},\bar{x}^{-},s,b_{1})=G^{h_{2}}(\bar{x}^{-},\bar{x}^{+}\bar{x}^{-}s,b_{1}),\quad \bar{x}^{+}=z^{+}x_{12}^{+},\, \bar{x}^{-}=x_{1}^{-},\\
 &  & \bar{G}_{3-}^{h_{1}h_{2}}(\bar{x}^{+},\bar{x}^{-},s,b_{1})=G^{h_{2}}(\bar{x}^{-},\bar{x}^{+}\bar{x}^{-}s,b_{1}),\quad \bar{x}^{+}=(1-z^{+})x_{12}^{+},\, \bar{x}^{-}=x^{-}-x_{1}^{-},
\end{eqnarray}
 where the variables \( x_{12}^{+},z^{+},x_{1}^{-} \) are generated according
to the integrand of eq.\ (\ref{g3p-(2)-approx}) (similar for \( \bar{G}_{i+}^{h_{1}h_{2}} \)).
After that each contribution \( \bar{G}_{i\pm }^{h_{1}h_{2}}(\bar{x}^{+},\bar{x}^{-},s,b_{1}) \)
is treated separately in the usual way, starting from the eq.\ (\ref{g-pom-basic}),
with the functions \( G_{1\mathrm{I}\! \mathrm{P}}^{h_{1}h_{2}}, \) \( G^{h_{1}h_{2}}_{\mathrm{soft}}, \)
\( G^{h_{1}h_{2}}_{\mathrm{sea}-\mathrm{sea}}, \) \( G^{h_{1}h_{2}}_{\mathrm{val}-\mathrm{sea}}, \)
\( G^{h_{1}h_{2}}_{\mathrm{sea}-v\mathrm{al}} \) being replaced by \( \bar{G}_{i\pm }^{h_{1}h_{2}}(\bar{x}^{+},\bar{x}^{-},s,b_{1}) \)
and by the corresponding partial contributions of soft \( G^{h}_{\mathrm{soft}} \),
``sea-sea''-type \( G^{h}_{\mathrm{sea}-\mathrm{sea}} \), and ``valence-sea''-type
parton scattering \( G^{h}_{\mathrm{val}-\mathrm{sea}} \) (see eq.\ (\ref{g-h-soft-3p}-\ref{g-h-val-sea})),
but without ``valence-valence'' contribution. 

\cleardoublepage

\chapter{Calculation of \protect\( \Phi \protect \) and \protect\( H\protect \) }

\section{Calculation of \protect\( \Phi _{pp}\protect \) \label{axd1}}

Here, we present the detailed calculation of \( \Phi  \) for proton-proton
collisions. As shown earlier, \( \Phi _{pp} \) may be written as

\begin{eqnarray}
\Phi _{pp}(x^{+},x^{-}s,b) & = & \sum ^{\infty }_{r_{1}=0}\cdots \, \sum ^{\infty }_{r_{N}=0}\frac{1}{r_{1}!}\cdots \frac{1}{r_{N}!}\, \int \, \prod ^{r_{1}+\ldots +r_{N}}_{\lambda =1}dx^{+}_{\lambda }dx^{-}_{\lambda }\nonumber \\
 & \times  & \prod _{\rho _{1}=1}^{r_{1}}-G_{1\rho _{1}}\ldots \, \prod _{\rho _{N}=r_{1}+\ldots +r_{N-1}+1}^{r_{1}+\ldots +r_{N}}-G_{N\rho _{N}}\nonumber \\
 & \times  & F_{\mathrm{remn}}(x^{+}-\sum _{\lambda }x^{+}_{\lambda })F_{\mathrm{remn}}(x^{-}-\sum _{\lambda }x^{-}_{\lambda }),
\end{eqnarray}
 with
\begin{eqnarray}
F_{\mathrm{remn}}(x) & = & x^{\alpha _{\mathrm{remn}}}\, \Theta (x)\, \Theta (1-x),
\end{eqnarray}
 and with \( G_{i\lambda } \) being of the form
\begin{equation}
G_{i\lambda }(x_{\lambda }^{+},x_{\lambda }^{-},s,b)=\alpha _{i}(x_{\lambda }^{+}x_{\lambda }^{-})^{\beta _{i}},
\end{equation}
 with \( \alpha _{i} \) and \( \beta _{i} \) being \( s \)- and \( b \)-dependent
parameters. We obtain

\begin{eqnarray}
\Phi _{pp}(x^{+},x^{-},s,b) & = & \sum ^{\infty }_{r_{1}=0}\cdots \, \sum ^{\infty }_{r_{N}=0}\frac{(-\alpha _{1})^{r_{1}}}{r_{1}!}\cdots \frac{(-\alpha _{N})^{r_{N}}}{r_{N}!}\nonumber \\
 & \times  & I_{r_{1},\ldots ,r_{N}}(x^{+})I_{r_{1},\ldots ,r_{N}}(x^{-})\label{R-analytique} 
\end{eqnarray}
 with

\begin{equation}
I_{r_{1},\ldots ,r_{N}}(x)=\int \prod ^{r_{1}+\ldots +r_{N}}_{\lambda =1}dx_{\lambda }\prod ^{r_{1}}_{\rho _{1}=1}x^{\beta _{1}}_{\rho _{1}}\, \ldots \, \prod ^{r_{1}+\ldots +r_{N}}_{\rho _{N}=r_{1}+\ldots +r_{N-1}+1}x^{\beta _{N}}_{\rho _{N}}\, F_{\mathrm{remn}}(x-\sum _{\lambda }x_{\lambda }),
\end{equation}
 which amounts to

\begin{equation}
I_{r_{1},\ldots ,r_{N}}(x)=\int \prod ^{r_{1}+\ldots +r_{N}}_{\lambda =1}[dx_{\lambda }x^{\epsilon _{\lambda }}_{\lambda }](x-\sum _{\lambda }x_{\lambda })^{\alpha _{\mathrm{remn}}}\theta (x-\sum _{\lambda }x_{\lambda })\theta \left( 1-(x-\sum _{\lambda }x_{\lambda })\right) ,
\end{equation}
 with

\begin{equation}
\epsilon _{\lambda }=\left\{ \begin{array}{lcl}
\beta _{1} & \mathrm{for} & \lambda \leq r_{1}\\
\beta _{2} & \mathrm{for} & r_{1}<\lambda \leq r_{1}+r_{2}\\
\ldots  &  & \\
\beta _{N} & \mathrm{for} & r_{1}+\ldots +r_{N-1}<\lambda \leq r_{1}+\ldots +r_{N}
\end{array}\right. .
\end{equation}
 We define new variables,

\begin{equation}
\left\{ \begin{array}{rcl}
u_{\lambda } & = & \frac{x_{\lambda }}{x-x_{1}-\ldots -x_{\lambda -1}}\\
 &  & \\
du_{\lambda } & = & \frac{dx_{\lambda }}{x-x_{1}-\ldots -x_{\lambda -1}}
\end{array},\right. 
\end{equation}
 which have the following property,

\begin{equation}
\prod ^{\lambda -1}_{\alpha =1}(1-u_{\alpha })=\prod ^{\lambda -1}_{\alpha =1}\frac{x-\ldots -x_{\alpha }}{x-\ldots -x_{\alpha -1}}=\frac{x-\ldots -x_{\lambda -1}}{x},
\end{equation}
and therefore

\begin{equation}
\left\{ \begin{array}{rcl}
x_{\lambda } & = & xu_{\lambda }\prod ^{\lambda -1}_{\alpha =1}(1-u_{\alpha })\\
 &  & \\
dx_{\lambda } & = & xdu_{\lambda }\prod ^{\lambda -1}_{\alpha =1}(1-u_{\alpha })
\end{array}.\right. 
\end{equation}
This leads to

\begin{equation}
I_{r_{1},\ldots ,r_{N}}(x)=\int \prod ^{r_{1}+\ldots +r_{N}}_{\lambda =1}\left\{ du_{\lambda }u^{\epsilon _{\lambda }}_{\lambda }x^{1+\epsilon _{\lambda }}\prod ^{\lambda -1}_{\alpha =1}(1-u_{\alpha })^{1+\epsilon _{\lambda }}\right\} \left[ x\prod ^{r_{1}+\ldots +r_{N}}_{\lambda =1}(1-u_{\lambda })\right] ^{\alpha _{\mathrm{remn}}}
\end{equation}
Defining

\begin{equation}
\alpha =\alpha _{\mathrm{remn}}+\sum _{\lambda =1}^{r_{1}+\ldots +r_{N}}\tilde{\epsilon }_{\lambda }=\alpha _{\mathrm{remn}}+r_{1}\tilde{\beta }_{1}+\ldots +r_{N}\tilde{\beta }_{N}
\end{equation}
and

\begin{equation}
\gamma _{\lambda }=\alpha _{\mathrm{remn}}+\sum _{\nu =\lambda +1}^{r_{1}+\ldots +r_{N}}\tilde{\epsilon }_{\nu }=\left\{ \begin{array}{lcl}
\alpha _{\mathrm{remn}}+(r_{1}-\lambda )\tilde{\beta }_{1}+r_{2}\tilde{\beta }2+\ldots +r_{N}\tilde{\beta }_{N} & \mathrm{if} & \lambda \leq r_{1}\\
\alpha _{\mathrm{remn}}+(r_{1}+r_{2}-\lambda )\tilde{\beta }_{2}+r_{3}\tilde{\beta }_{3}+\ldots +r_{N}\tilde{\beta }_{N} & \mathrm{if} & r_{1}<\lambda \leq r_{1}+r_{2}\\
\ldots  &  & \\
\alpha _{\mathrm{remn}}+(r_{1}+\ldots +r_{N}-\lambda )\tilde{\beta }_{N} & \mathrm{if} & \lambda >r_{1}+\ldots +r_{N-1}
\end{array}\right. 
\end{equation}
 with
\begin{eqnarray}
\tilde{\beta } & = & \beta +1\, ,\\
\tilde{\epsilon } & = & \epsilon +1\, ,
\end{eqnarray}
we find

\begin{equation}
I_{r_{1},\ldots ,r_{N}}(x)=x^{\alpha }\int \prod ^{r_{1}+\ldots +r_{N}}_{\lambda =1}du_{\lambda }u^{\epsilon _{\lambda }}_{\lambda }(1-u_{\lambda })^{\gamma _{\lambda }}
\end{equation}
 The \( u \)-integration can be done,

\begin{equation}
\int ^{1}_{0}duu^{\epsilon }(1-u)^{\gamma }=\frac{\Gamma (1+\epsilon )\Gamma (1+\gamma )}{\Gamma (2+\epsilon +\gamma )},
\end{equation}
 and we get

\begin{equation}
I_{r_{1},\ldots ,r_{N}}(x)=x^{\alpha }\prod ^{r_{1}+\ldots +r_{N}}_{\lambda =1}\frac{\Gamma (1+\epsilon _{\lambda })\Gamma (1+\gamma _{\lambda })}{\Gamma (2+\epsilon _{\lambda }+\gamma _{\lambda })}
\end{equation}
 Using the relation \( 1+\epsilon _{\lambda }+\gamma _{\lambda }=\gamma _{\lambda -1} \),
we get

\begin{eqnarray}
I_{r_{1},\ldots ,r_{N}}(x) & = & x^{\alpha }\Gamma (1+\beta _{1})^{r_{1}}\ldots \, \Gamma (1+\beta _{N})^{r_{N}}\, \prod ^{r_{1}+\ldots +r_{N}}_{\lambda =1}\frac{\Gamma (1+\gamma _{\lambda })}{\Gamma (1+\gamma _{\lambda -1})}\\
 & = & x^{\alpha }\Gamma (1+\beta _{1})^{r_{1}}\ldots \, \Gamma (1+\beta _{N})^{r_{N}}\frac{\Gamma \left( 1+\gamma _{r_{1}+\ldots +r_{N}}\right) }{\Gamma (1+\gamma _{0})},
\end{eqnarray}
 or
\begin{equation}
\label{I-analytique}
I_{r_{1},\ldots ,r_{N}}(x)=x^{\alpha _{\mathrm{remn}}+r_{1}\tilde{\beta }_{1}+\ldots +r_{N}\tilde{\beta }_{N}}\Gamma (\tilde{\beta }_{1})^{r_{1}}\ldots \, \Gamma (\tilde{\beta }_{N})^{r_{N}}\frac{\Gamma (\tilde{\alpha }_{\mathrm{remn}})}{\Gamma (\tilde{\alpha }_{\mathrm{remn}}+r_{1}\tilde{\beta }_{1}+\ldots +r_{N}\tilde{\beta }_{N})}.
\end{equation}
The final expression for \( \Phi _{pp} \) is therefore

\begin{eqnarray}
\Phi _{pp}(x^{+},x^{-},s,b) & = & x^{\alpha _{\mathrm{remn}}}\sum ^{\infty }_{r_{1}=0}\cdots \, \sum ^{\infty }_{r_{N}=0}\left\{ \frac{\Gamma (1+\alpha _{\mathrm{remn}})}{\Gamma (1+\alpha _{\mathrm{remn}}+r_{1}\tilde{\beta }_{1}+\ldots +r_{N}\tilde{\beta }_{N})}\right\} ^{2}\nonumber \\
 & \times  & \frac{(-\alpha _{1}x^{\tilde{\beta }_{1}}\Gamma ^{2}(\tilde{\beta }_{1}))^{r_{1}}}{r_{1}!}\cdots \frac{(-\alpha _{N}x^{\tilde{\beta }_{N}}\Gamma ^{2}(\tilde{\beta }_{N}))^{r_{N}}}{r_{N}!}\label{r2exact2} 
\end{eqnarray}
with \( x=x^{+}x^{-} \). This is the expression shown in eq.\ (\ref{r2exact}).

\section{Calculation of H\label{axd2}}

The function \( H \) is defined as

\begin{eqnarray}
H(z^{+},z^{-}) & = & \sum ^{\infty }_{m=1}\int \prod ^{m}_{\mu =1}dx^{+}_{\mu }dx^{-}_{\mu }\frac{1}{m!}\prod ^{m}_{\mu =1}G(x^{+}_{\mu },x^{-}_{\mu },s,b)\nonumber \\
 & \times  & \delta (1-z^{+}-\sum ^{m}_{\mu =1}x^{+}_{\mu })\delta (1-z^{-}-\sum ^{m}_{\mu =1}x^{-}_{\mu }).
\end{eqnarray}
 Using the expression

\begin{eqnarray}
G(x^{+}_{\mu },x^{-}_{\mu },s,b) & = & \sum ^{N}_{i=1}\alpha _{i}(x_{\mu }^{+}x_{\mu }^{-})^{\beta _{i}},\label{x} 
\end{eqnarray}
 with the \( \alpha _{i} \) and \( \beta _{i} \) are functions of the impact
parameter \( b \) and the energy squared \( s \), 

\begin{eqnarray}
\alpha _{i} & = & \left( \alpha _{D_{i}}+\alpha ^{*}_{D_{i}}\right) s^{\beta _{D_{i}}+\gamma _{D_{i}}b^{2}}e^{-\frac{b^{2}}{\delta _{D_{i}}}},\\
\beta _{i} & = & \beta _{D_{i}}+\gamma _{D_{i}}b^{2}+\beta ^{*}_{D_{i}}-\alpha _{\mathrm{part}},
\end{eqnarray}
 with \( \alpha ^{*}_{D_{i}}\neq 0 \) and \( \beta ^{*}_{D_{i}}\neq 0 \) only
if \( \alpha _{D_{i}}=0 \), and using the same method as for the calculation
of \( \Phi _{pp}(x^{+},x^{-}) \), one may write

\begin{eqnarray}
H(z^{+},z^{-}) & = & \underbrace{\sum ^{\infty }_{r_{1}=0}\cdots \, \sum ^{\infty }_{r_{N}=0}}^{\infty }_{r_{1}+\ldots +r_{N}=0}\frac{(\alpha _{1})^{r_{1}}}{r_{1}!}\cdots \frac{(\alpha _{N})^{r_{N}}}{r_{N}!}J_{r_{1},\ldots ,r_{N}}(z^{+})J_{r_{1},\ldots ,r_{N}}(z^{-})
\end{eqnarray}
 with 

\begin{equation}
J_{r_{1},\ldots ,r_{N}}(z)=\int \prod ^{r_{1}+\ldots +r_{N}}_{\lambda =1}[dx_{\lambda }x^{\epsilon _{\lambda }}_{\lambda }]\delta \left( 1-z-\sum ^{r_{1}+\ldots +r_{N}}_{\lambda =1}x_{\lambda }\right) 
\end{equation}
 and

\begin{equation}
\epsilon _{\lambda }=\left\{ \begin{array}{lcl}
\beta _{1} & \mathrm{for} & \lambda \leq r_{1}\\
\beta _{2} & \mathrm{for} & r_{1}<\lambda \leq r_{1}+r_{2}\\
\ldots  &  & \\
\beta _{N} & \mathrm{for} & r_{1}+\ldots +r_{N-1}<\lambda \leq r_{1}+\ldots +r_{N}
\end{array}\right. .
\end{equation}
One may use the \( \delta  \) function to obtain

\begin{equation}
J_{r_{1},\ldots ,r_{N}}(z)=\int \prod ^{r_{1}+\ldots +r_{N}}_{\lambda =2}[dx_{\lambda }x^{\epsilon _{\lambda }}_{\lambda }]\left( (1-z)-\sum ^{r_{1}+\ldots +r_{N}}_{\lambda =2}x_{\lambda }\right) ^{\epsilon _{1}}.
\end{equation}
 Introducing \( \tilde{x}_{\alpha }=x_{\alpha +1} \), we define new variables,

\begin{equation}
\left\{ \begin{array}{rcl}
u_{\lambda '} & = & \frac{\tilde{x}_{\lambda '}}{1-z-\tilde{x}_{1}-\ldots -\tilde{x}_{\lambda '-1}}\\
 &  & \\
du_{\lambda '} & = & \frac{d\tilde{x}_{\lambda '}}{1-z-\tilde{x}_{1}-\ldots -\tilde{x}_{\lambda '-1}}
\end{array}\right. ,
\end{equation}
 which have the following property,

\begin{equation}
\prod ^{\lambda '-1}_{\alpha =1}(1-u_{\alpha })=\prod ^{\lambda '-1}_{\alpha =1}\frac{1-z-\ldots -\tilde{x}_{\alpha }}{1-z-\ldots -\tilde{x}_{\alpha -1}}=\frac{1-z-\ldots -\tilde{x}_{\lambda '-1}}{1-z},
\end{equation}
 and therefore

\begin{equation}
\left\{ \begin{array}{rcl}
\tilde{x}_{\lambda '} & = & (1-z)u_{\lambda '}\prod ^{\lambda '-1}_{\alpha =1}(1-u_{\alpha })\\
 &  & \\
d\tilde{x}_{\lambda '} & = & (1-z)du_{\lambda '}\prod ^{\lambda '-1}_{\alpha =1}(1-u_{\alpha })
\end{array}\right. .
\end{equation}
 This leads to

\begin{eqnarray}
J_{r,s,t}(z) & = & \int \prod ^{r_{1}+\ldots +r_{N}-1}_{\lambda '=1}\left\{ du_{\lambda '}u^{\epsilon _{\lambda '+1}}_{\lambda '}(1-z)^{1+\epsilon _{\lambda '+1}}\prod ^{\lambda '-1}_{\alpha =1}(1-u_{\alpha })^{1+\epsilon _{\lambda '+1}}\right\} \nonumber \\
 & \times  & \left[ (1-z)\prod ^{r_{1}+\ldots +r_{N}-1}_{\lambda '=1}(1-u_{\lambda '})\right] ^{\epsilon _{1}}
\end{eqnarray}
 Defining

\begin{equation}
\alpha '=\epsilon _{1}+\sum _{\lambda '=1}^{r_{1}+\ldots +r_{N}-1}\tilde{\epsilon }_{\lambda '+1}=r_{1}\tilde{\beta }_{1}+\ldots +r_{N}\tilde{\beta }_{N}-1
\end{equation}
 and

\begin{eqnarray}
\gamma _{\lambda '} & = & \epsilon _{1}+\sum _{\nu =\lambda '+1}^{r_{1}+\ldots +r_{N}-1}\tilde{\epsilon }_{\nu +1}\\
 & = & \left\{ \begin{array}{lcl}
\epsilon _{1}+(r_{1}-1-\lambda ')\tilde{\beta }_{1}+r_{2}\tilde{\beta }_{2}+\ldots +r_{N}\tilde{\beta }_{N} & \mathrm{if} & \lambda '\leq r_{1}-1\\
\epsilon _{1}+(r_{1}+r_{2}-1-\lambda ')\tilde{\beta }_{2}+r_{3}\tilde{\beta }_{3}+\ldots +r_{N}\tilde{\beta }_{N} & \mathrm{if} & r_{1}-1<\lambda '\leq r_{1}+r_{2}-1\\
\ldots  &  & \\
\epsilon _{1}+(r_{1}+\ldots +r_{N}-1-\lambda ')\tilde{\beta }_{N} & \mathrm{if} & \lambda '>r_{1}+\ldots +r_{N-1}-1
\end{array}\right. ,
\end{eqnarray}
 with

\begin{eqnarray}
\tilde{\beta } & = & \beta +1,\\
\tilde{\epsilon } & = & \epsilon +1,
\end{eqnarray}
 we find

\begin{equation}
J_{r_{1},\ldots ,r_{N}}(z)=(1-z)^{\alpha '}\int \prod ^{r_{1}+\ldots +r_{N}-1}_{\lambda '=1}du_{\lambda '}u^{\epsilon _{\lambda '+1}}_{\lambda '}(1-u_{\lambda '})^{\gamma _{\lambda '}}
\end{equation}
 The \( u \)-integration can be done,

\begin{equation}
\int ^{1}_{0}duu^{\epsilon }(1-u)^{\gamma }=\frac{\Gamma (1+\epsilon )\Gamma (1+\gamma )}{\Gamma (2+\epsilon +\gamma )}
\end{equation}
 and we get

\begin{equation}
J_{r_{1},\ldots ,r_{N}}(z)=(1-z)^{\alpha '}\prod ^{r_{1}+\ldots +r_{N}-1}_{\lambda '=1}\frac{\Gamma (1+\epsilon _{\lambda '+1})\Gamma (1+\gamma _{\lambda '})}{\Gamma (2+\epsilon _{\lambda '+1}+\gamma _{\lambda '})}
\end{equation}
 Using the relation \( 1+\epsilon _{\lambda '+1}+\gamma _{\lambda '}=\gamma _{\lambda '-1} \),
we get, if \( r_{1}\neq 0 \),

\begin{eqnarray}
J_{r_{1},\ldots ,r_{N}}(z) & = & (1-z)^{\alpha '}\Gamma (\tilde{\beta }_{1})^{r_{1}-1}\Gamma (\tilde{\beta }_{2})^{r_{2}}\ldots \, \Gamma (\tilde{\beta }_{N})^{r_{N}}\prod ^{r_{1}+\ldots +r_{N}-1}_{\lambda '=1}\frac{\Gamma (1+\gamma _{\lambda '})}{\Gamma (1+\gamma _{\lambda '-1})}\\
 & = & (1-z)^{\alpha '}\Gamma (\tilde{\beta }_{1})^{r_{1}-1}\Gamma (\tilde{\beta }_{2})^{r_{2}}\ldots \, \Gamma (\tilde{\beta }_{N})^{r_{N}}\frac{\Gamma (1+\gamma _{r_{1}+\ldots +r_{N}-1})}{\Gamma (1+\gamma _{0})}\\
 & = & (1-z)^{\alpha '}\Gamma (\tilde{\beta }_{1})^{r_{1}-1}\Gamma (\tilde{\beta }_{2})^{r_{2}}\ldots \, \Gamma (\tilde{\beta }_{N})^{r_{N}}\frac{\Gamma (\tilde{\beta }_{1})}{\Gamma (r\tilde{\beta }_{1}+\ldots +r_{N}\tilde{\beta }_{N})}.
\end{eqnarray}
 If \( r_{1}=0 \) and \( r_{2}\neq 0 \), we get

\begin{equation}
J_{r_{1},\ldots ,r_{N}}(z)=(1-z)^{\alpha '}\Gamma (\tilde{\beta }_{2})^{r_{2}-1}\Gamma (\tilde{\beta }_{3})^{r_{3}}\ldots \, \Gamma (\tilde{\beta }_{N})^{r_{N}}\frac{\Gamma (\tilde{\beta }_{2})}{\Gamma (r_{2}\tilde{\beta }_{2}+\ldots +r_{N}\tilde{\beta }_{N})},
\end{equation}
 and so on, which corresponds finally to

\begin{equation}
J_{r_{1},\ldots ,r_{N}}(z)=(1-z)^{r_{1}\tilde{\beta }_{1}+\ldots +r_{N}\tilde{\beta }_{N}-1}\frac{\Gamma (\tilde{\beta }_{1})^{r_{1}}\ldots \, \Gamma (\tilde{\beta }_{N})^{r_{N}}}{\Gamma (r_{1}\tilde{\beta }_{1}+\ldots +r_{N}\tilde{\beta }_{N})}.
\end{equation}
 The final expression for \( H \) is therefore
\begin{eqnarray}
H(z^{+},z^{-}) & = & \underbrace{\sum ^{\infty }_{r_{1}=0}...\, \sum ^{\infty }_{r_{N}=0}}_{r_{1}+...+r_{N}\neq 0}\frac{\left[ (1-z^{+})(1-z^{-})\right] ^{r_{1}\tilde{\beta }_{1}+...+r_{N}\tilde{\beta }_{N}-1}}{\Gamma (r_{1}\tilde{\beta }_{1}+...+r_{N}\tilde{\beta }_{N})^{2}}\nonumber \\
 &  & \qquad \qquad \qquad \times \; \frac{(\alpha _{1}\Gamma (\tilde{\beta }_{1})^{2})^{r_{1}}}{r_{1}!}\cdots \frac{(\alpha _{N}\Gamma (\tilde{\beta }_{N})^{2})^{r_{N}}}{r_{N}!}
\end{eqnarray}

\section{Calculation of \protect\( \Phi _{AB}\protect \)\label{axd3}}

The expression for the virtual emissions in case of nucleus-nucleus collisions
is given as 

\begin{eqnarray}
\Phi _{AB}\left( X^{+},X^{-},s,b\right)  & = & \sum ^{\infty }_{l_{1}=0}\ldots \sum ^{\infty }_{l_{AB}=0}\, \int \, \prod _{k=1}^{AB}\left\{ \prod ^{l_{k}}_{\lambda =1}d\tilde{x}_{k,\lambda }^{+}d\tilde{x}_{k,\lambda }^{-}\right\} \; \prod _{k=1}^{AB}\left\{ \frac{1}{l_{k}!}\, \prod _{\lambda =1}^{l_{k}}-G(\tilde{x}_{k,\lambda }^{+},\tilde{x}_{k,\lambda }^{-},s,b)\right\} \nonumber \\
 & \times  & \prod _{i=1}^{A}F_{\mathrm{remn}}\left( x_{i}^{+}-\sum _{\pi (k)=i}\tilde{x}_{k,\lambda }^{+}\right) \, \prod _{j=1}^{B}F_{\mathrm{remn}}\left( x^{-}_{j}-\sum _{\tau (k)=j}\tilde{x}_{k,\lambda }^{-}\right) ,\label{R2AB} 
\end{eqnarray}
 where \( X^{+}=\left\{ x_{1}^{+}\ldots \, x^{+}_{A}\right\}  \), \( X^{-}=\left\{ x_{1}^{-}\ldots \, x^{-}_{B}\right\}  \)
and \( \pi (k) \) and \( \tau (k) \) represent the projectile or target nucleon
linked to pair \( k \). Using the expression

\begin{eqnarray}
G(\tilde{x}_{k,\lambda }^{+},\tilde{x}_{k,\lambda }^{-},s,b) & = & \sum ^{N}_{i=1}\underbrace{\alpha _{i}(\tilde{x}_{k,\lambda }^{+}\, \tilde{x}_{k,\lambda }^{-})^{\beta _{i}}}_{G_{i,k,\lambda }},
\end{eqnarray}
 with the \( \alpha _{i} \) and \( \beta _{i} \) are functions of the impact
parameter \( b \) and the energy squared \( s \), 

\begin{eqnarray}
\alpha _{i} & = & \left( \alpha _{D_{i}}+\alpha ^{*}_{D_{i}}\right) s^{\beta _{D_{i}}+\gamma _{D_{i}}b^{2}}e^{-\frac{b^{2}}{\delta _{D_{i}}}},\\
\beta _{i} & = & \beta _{D_{i}}+\gamma _{D_{i}}b^{2}+\beta ^{*}_{D_{i}}-\alpha _{\mathrm{part}},
\end{eqnarray}
 with \( \alpha ^{*}_{D_{i}}\neq 0 \) and \( \beta ^{*}_{D_{i}}\neq 0 \) only
if \( \alpha _{D_{i}}=0 \). The remnant function \( F_{\mathrm{remn}} \) is
given as

\begin{equation}
F_{\mathrm{remn}}(x)=x^{\alpha _{\mathrm{remn}}}\, \Theta (x)\, \Theta (1-x).
\end{equation}
We have (see calculation of \( \Phi _{pp} \))

\begin{eqnarray}
\ldots \sum ^{\infty }_{l_{k}=0}\ldots \frac{1}{l_{k}!}\prod ^{l_{k}}_{\lambda =1}-G(s,\tilde{x}_{k,\lambda }^{+},\tilde{x}_{k,\lambda }^{-},b) & = & \ldots \sum ^{\infty }_{l_{k}=0}\ldots \frac{1}{l_{k}!}\prod ^{l_{k}}_{\lambda =1}\left( -G_{1,k,\lambda }-\ldots -G_{N,k,\lambda }\right) \\
 & = & \ldots \, \sum ^{\infty }_{r_{1,k}=0}\cdots \sum ^{\infty }_{r_{N,k}=0}\, \ldots \, \frac{1}{r_{1,k}!\ldots r_{N,k}!}\\
 &  & \quad \qquad \times \; \prod ^{r_{1,k}}_{\rho _{1}=1}-G_{1,k,\rho _{1}}\ldots \prod ^{r_{1,k}+\ldots +r_{N,k}}_{\rho _{N}=r_{1,k}+\ldots +r_{N-1,k}+1}-G_{N,k,\rho _{N}}.\nonumber 
\end{eqnarray}
 So eq.\ (\ref{R2AB}) can be written as

\begin{eqnarray}
\Phi _{AB}\left( X^{+},X^{-},s,b\right)  & = & \sum _{r_{1,1}\ldots r_{N,1}}\ldots \sum _{r_{1,AB}\ldots r_{N,AB}}\, \prod _{k=1}^{AB}\frac{1}{r_{1,k}!\ldots r_{N,k}!}\, \int \, \prod _{k=1}^{AB}\left\{ \prod ^{r_{1,k}+\ldots +r_{N,k}}_{\lambda =1}d\tilde{x}_{k,\lambda }^{+}d\tilde{x}_{k,\lambda }^{-}\right\} \nonumber \\
 & \times  & \prod _{k=1}^{AB}\left\{ \prod _{\rho _{1}=1}^{r_{1,k}}-\alpha _{1}(\tilde{x}_{k,\rho _{1}}^{+}\tilde{x}_{k,\rho _{1}}^{-})^{\beta _{1}}\ldots \prod _{\rho _{N}=r_{1,k}+\ldots +r_{N-1,k}+1}^{r_{1,k}+\ldots +r_{N,k}}-\alpha _{N}(\tilde{x}_{k,\rho _{N}}^{+}\tilde{x}_{k,\rho _{N}}^{-})^{\beta _{N}}\right\} \nonumber \\
 & \times  & \prod _{i=1}^{A}F_{\mathrm{remn}}\left( x_{i}^{+}-\sum _{\pi (k)=i}\tilde{x}_{k,\lambda }^{+}\right) \, \prod _{j=1}^{B}F_{\mathrm{remn}}\left( x^{-}_{j}-\sum _{\tau (k)=j}\tilde{x}_{k,\lambda }^{-}\right) \; ,
\end{eqnarray}
 which leads to

\begin{eqnarray}
\Phi _{AB}\left( X^{+},X^{-},s,b\right)  & = & \sum _{r_{1,1}\ldots r_{N,1}}\ldots \sum _{r_{1,AB}\ldots r_{N,AB}}\, \prod _{k=1}^{AB}\frac{\left( -\alpha _{1}\right) ^{r_{1,k}}}{r_{1,k}!}\ldots \frac{\left( -\alpha _{N}\right) ^{r_{N,k}}}{r_{N,k}!}\nonumber \\
 &  & \qquad \qquad \qquad \qquad \qquad \times \; I^{+}_{R}(X^{+})\, I^{-}_{R}(X^{-})\: ,
\end{eqnarray}
 where

\begin{eqnarray}
I^{\sigma }_{R}(X) & = & \int \, \prod _{k=1}^{AB}\prod ^{r_{1,k}+\ldots +r_{N,k}}_{\lambda =1}d\tilde{x}_{k,\lambda }\left( \tilde{x}_{k,\lambda }\right) ^{\epsilon _{k,\lambda }}\; \prod ^{\mathrm{P}^{\sigma }}_{i=1}F_{\mathrm{remn}}\left( x_{i}-\sum _{\kappa ^{\sigma }(k)=i}\tilde{x}_{k,\lambda }\right) \; ,\label{x} 
\end{eqnarray}
 with \( R=\{r_{j,k}\} \) , \( \mathrm{P}^{+}=A \), \( \mathrm{P}^{-}=B \),
\( \kappa ^{+}(k)=\pi (k) \), and \( \kappa ^{-}(k)=\tau (k) \). Using the
property

\begin{equation}
\label{piAB}
\prod ^{AB}_{k=1}=\prod ^{\mathrm{P}^{\sigma }}_{i=1}\, \prod _{\kappa ^{\sigma }(k)=i},
\end{equation}
 one can write

\begin{eqnarray}
I_{R}^{\sigma }(X) & = & \prod _{i=1}^{\mathrm{P}^{\sigma }}\, \left\{ \int \, \prod _{\kappa ^{\sigma }(k)=i}\, \prod ^{r_{1,k}+\ldots +r_{N,k}}_{\lambda =1}d\tilde{x}_{k,\lambda \: }\left( \tilde{x}_{k,\lambda }\right) ^{\epsilon _{k,\lambda }}\right. \nonumber \\
 &  & \left. \qquad \times \; F_{\mathrm{remn}}\left( x_{i}-\sum _{\kappa ^{\sigma }(k)=i}\tilde{x}_{k,\lambda }\right) \right\} \; .
\end{eqnarray}
 Let us rename the \( \tilde{x}_{k,\lambda } \) linked to the remnant \( i \)
as \( \tilde{x}_{1},\, \tilde{x}_{2},\, \ldots ,\, \tilde{x}_{r^{\sigma }_{1,i}+\ldots +r^{\sigma }_{N,i}} \),
where \( r^{\sigma }_{p,i} \) is per definition the number of Pomerons of type
\( p \) linked to remnant \( i \). So we get for the term in brackets

\begin{equation}
\int \, \prod ^{r^{\sigma }_{1,i}+\ldots +r^{\sigma }_{N,i}}_{\nu =1}d\tilde{x}_{\nu }\, \tilde{x}_{\nu }^{\epsilon _{\nu }}F_{\mathrm{remn}}\left( x_{i}-\sum ^{r^{\sigma }_{1,i}+\ldots +r^{\sigma }_{N,i}}_{\nu =1}\tilde{x}_{\nu }\right) .
\end{equation}
 This is exactly the corresponding \( I \) for proton-proton scattering. So
we have

\begin{equation}
I_{R}^{\sigma }(X)=\prod _{i=1}^{\mathrm{P}^{\sigma }}\, x_{i}^{\alpha _{\mathrm{remn}}+r^{\sigma }_{1,i}\tilde{\beta }_{1}+\ldots +r_{N,i}^{\sigma }\tilde{\beta }_{N}}\, \Gamma (\tilde{\beta }_{1})^{r_{1,i}^{\sigma }}\ldots \Gamma (\tilde{\beta }_{N})^{r_{N,i}^{\sigma }}\, g\left( r^{\sigma }_{1,i}\tilde{\beta }_{1}+\ldots +r_{N,i}^{\sigma }\tilde{\beta }_{N}\right) ,
\end{equation}
 with 

\begin{equation}
g(z)=\frac{\Gamma (1+\alpha _{\mathrm{remn}})}{\Gamma (1+\alpha _{\mathrm{remn}}+z)},
\end{equation}
 and

\begin{equation}
\tilde{\beta }=\beta +1.
\end{equation}
 Since we have

\begin{equation}
r^{\sigma }_{p,i}=\sum _{\kappa ^{\sigma }(k)=i}r_{p,k}\: ,
\end{equation}
 we find finally

\begin{eqnarray}
 &  & \Phi _{AB}\left( X^{+},X^{-},s,b\right) =\nonumber \\
 &  & \qquad \sum _{r_{1,1}\ldots r_{N,1}}\cdots \, \sum _{r_{1,AB}\ldots r_{N,AB}}\, \prod _{k=1}^{AB}\frac{\left( -\alpha _{1}\right) ^{r_{1,k}}}{r_{1,k}!}\ldots \frac{\left( -\alpha _{N}\right) ^{r_{N,k}}}{r_{N,k}!}\label{r2abfinal} \\
 &  & \qquad \prod ^{A}_{i=1}\, (x_{i}^{+})^{\alpha _{\mathrm{remn}}}\, \prod _{\pi (k)=i}\left( \Gamma (\tilde{\beta }_{1})(x_{i}^{+})^{\tilde{\beta }_{1}}\right) ^{r_{1,k}}\ldots \left( \Gamma (\tilde{\beta }_{N})(x_{i}^{+})^{\tilde{\beta }_{N}}\right) ^{r_{N,k}}g\left( \sum _{\pi (k)=i}r_{1,k}\tilde{\beta }_{1}+\ldots +r_{N,k}\tilde{\beta }_{N}\right) \nonumber \\
 &  & \qquad \prod ^{B}_{j=1}\, (x_{j}^{-})^{\alpha _{\mathrm{remn}}}\, \prod _{\tau (k)=j}\left( \Gamma (\tilde{\beta }_{1})(x_{j}^{-})^{\tilde{\beta }_{1}}\right) ^{r_{1,k}}\ldots \left( \Gamma (\tilde{\beta }_{N})(x_{j}^{-})^{\tilde{\beta }_{N}}\right) ^{r_{N,k}}g\left( \sum _{\tau (k)=j}r_{1,k}\tilde{\beta }_{1}+\ldots +r_{N,k}\tilde{\beta }_{N}\right) \nonumber 
\end{eqnarray}

\section{Exponentiation of \protect\( \Phi _{AB}\protect \)\label{axd4}}

We first replace in eq.\ (\ref{r2abfinal}) the function \( g(z) \) by the
``exponentiated'' function \( g_{\mathrm{e}}(z) \),

\begin{eqnarray}
g_{\mathrm{e}}\left( \sum _{\kappa ^{\sigma }(k)=i}r_{1,k}\tilde{\beta }_{1}+\ldots +r_{N,k}\tilde{\beta }_{N}\right)  & = & \exp \left\{ -\epsilon _{\mathrm{e}}\left( \sum _{\kappa ^{\sigma }(k)=i}r_{1,k}\tilde{\beta }_{1}+\ldots +r_{N,k}\tilde{\beta }_{N}\right) \right\} \nonumber \\
 & = & \prod _{\kappa ^{\sigma }(k)=i}\left( e^{-\epsilon _{\mathrm{e}}\tilde{\beta }_{1}}\right) ^{r_{1,k}}\ldots \left( e^{-\epsilon _{\mathrm{e}}\tilde{\beta }_{N}}\right) ^{r_{N,k}},
\end{eqnarray}
and we obtain

\begin{eqnarray}
 &  & \Phi _{\mathrm{e}\, _{AB}}\left( X^{+},X^{-},s,b\right) =\nonumber \\
 &  & \qquad \prod ^{A}_{i=1}\, (x_{i}^{+})^{\alpha _{\mathrm{remn}}}\, \prod ^{B}_{j=1}\, (x_{j}^{-})^{\alpha _{\mathrm{remn}}}\\
 &  & \qquad \sum _{r_{1,1}\ldots r_{1,AB}}\, \prod _{k=1}^{AB}\frac{\left( -\alpha _{1}\right) ^{r_{1,k}}}{r_{1,k}!}\left\{ \prod ^{A}_{i=1}\, \prod _{\pi (k)=i}\left( D_{1}\; (x_{i}^{+})^{\tilde{\beta }_{1}}\right) ^{r_{1,k}}\prod ^{B}_{j=1}\, \prod _{\tau (k)=j}\left( D_{1}\; (x_{j}^{-})^{\tilde{\beta }_{1}}\right) ^{r_{1,k}}\right\} \nonumber \\
 &  & \qquad \qquad \ldots \nonumber \\
 &  & \qquad \sum _{r_{N,1}\ldots r_{N,AB}}\, \prod _{k=1}^{AB}\frac{\left( -\alpha _{N}\right) ^{r_{N,k}}}{r_{N,k}!}\left\{ \prod ^{A}_{i=1}\, \prod _{\pi (k)=i}\left( D_{N}\; (x_{i}^{+})^{\tilde{\beta }_{N}}\right) ^{r_{N,k}}\prod ^{B}_{j=1}\, \prod _{\tau (k)=j}\left( D_{N}\; (x_{j}^{-})^{\tilde{\beta }_{N}}\right) ^{r_{N,k}}\right\} \nonumber 
\end{eqnarray}
 with

\begin{equation}
D_{j}=\Gamma (\tilde{\beta }_{j})e^{-\epsilon _{\mathrm{e}}\tilde{\beta }_{j}}.
\end{equation}
 And using again the property (\ref{piAB}), we can write

\begin{eqnarray}
\Phi _{e\, _{AB}}\left( X^{+},X^{-},s,b\right)  & = & \prod ^{A}_{i=1}\, (x_{i}^{+})^{\alpha _{\mathrm{remn}}}\, \prod ^{B}_{j=1}\, (x_{j}^{-})^{\alpha _{\mathrm{remn}}}\nonumber \\
 & \times  & \sum _{r_{1,1}\ldots r_{1,AB}}\, \prod _{k=1}^{AB}\frac{\left( -\alpha _{1}D^{2}_{1}(x_{\pi (k)}^{+}\; x_{\tau (k)}^{-})^{\tilde{\beta }_{1}}\right) ^{r_{1,k}}}{r_{1,k}!}\nonumber \\
 & \times  & \qquad \ldots \\
 & \times  & \sum _{r_{N,1}\ldots r_{N,AB}}\, \prod _{k=1}^{AB}\frac{\left( -\alpha _{N}D^{2}_{N}(x_{\pi (k)}^{+}\; x_{\tau (k)}^{-})^{\tilde{\beta }_{N}}\right) ^{r_{N,k}}}{r_{N,k}!}.\nonumber 
\end{eqnarray}
 Now the sum can be performed and we get the final expression for the ``exponentiated''
\( \Phi  \),

\begin{eqnarray}
\Phi _{\mathrm{e}\, _{AB}}\left( X^{+},X^{-},s,b\right)  & = & \prod ^{A}_{i=1}\, (x_{i}^{+})^{\alpha _{\mathrm{remn}}}\, \prod ^{B}_{j=1}\, (x_{j}^{-})^{\alpha _{\mathrm{remn}}}\prod ^{AB}_{k=1}e^{-\tilde{G}\left( x_{\pi (k)}^{+}\, x^{-}_{\tau (k)}\right) },\nonumber 
\end{eqnarray}
 with

\begin{equation}
\tilde{G}(x)=\sum ^{N}_{i=1}\tilde{\alpha }_{i}x^{\tilde{\beta }_{i}}\: ,
\end{equation}
 where

\begin{eqnarray}
\tilde{\alpha }_{i} & = & \alpha _{i}\Gamma (\tilde{\beta }_{i})^{2}e^{-2\epsilon _{\mathrm{e}}\tilde{\beta }_{i}},\\
\tilde{\beta }_{i} & = & \beta _{i}+1,
\end{eqnarray}
 and
\begin{eqnarray}
\alpha _{i} & = & \left( \alpha _{D_{i}}+\alpha ^{*}_{D_{i}}\right) s^{\beta _{D_{i}}+\gamma _{D_{i}}b^{2}}e^{-\frac{b^{2}}{\delta _{D_{i}}}},\\
\beta _{i} & = & \beta _{D_{i}}+\gamma _{D_{i}}b^{2}+\beta ^{*}_{D_{i}}-\alpha _{\mathrm{part}},
\end{eqnarray}
 with \( \alpha ^{*}_{D_{i}}\neq 0 \) and \( \beta ^{*}_{D_{i}}\neq 0 \) only
if \( \alpha _{D_{i}}=0 \) .

\cleardoublepage

\end{appendix}

\bibliographystyle{pr2}
\bibliography{a}

\end{document}